\documentclass[12pt]{article}

\usepackage{scicite}

\usepackage{times}

\usepackage{graphicx}
\usepackage{color}
\usepackage[usenames,dvipsnames,svgnames,table]{xcolor}
\usepackage{amsthm}
\usepackage{tabularx}
\usepackage{tabulary}
\usepackage{caption}
\usepackage{multirow}
\usepackage{float}
\usepackage{hyperref}

\newcommand\y{0.15}

\newcommand\twofig{0.45\textwidth}
\newcommand\threefig{0.3\textwidth}
\newcommand\fourfig{0.21\textwidth}
\newcommand\fivefig{0.18\textwidth}
\newcommand\sixfig{0.14\textwidth}

\newenvironment{sciabstract}{%
\begin{quote} \bf}
{\end{quote}}

\newcommand{\revision}[1]{\textcolor{black}{#1}} 

\title{Heterogeneous Interventions Reduce the Spread of COVID-19 in Simulations on Real Mobility Data}

\author
{$^{1+}$Haotian Wang,$^{2+}$Abhirup Ghosh,$^{3}$Jiaxin Ding,$^{4\ast}$Rik Sarkar,$^{1+}$Jie Gao \\
\\
\normalsize{$^{1}$Department of Computer Science, Rutgers University, USA}\\
\normalsize{$^{2}$Department of Computing, Imperial College London, UK}\\
\normalsize{$^{3}$}John Hopcroft Center for Computer Science, Shanghai Jiao Tong University, China\\
\normalsize{$^{4}$School of Informatics, University of Edinburgh, UK}\\
\\
\normalsize{$^\ast$To whom correspondence should be addressed; E-mail:  rsarkar@inf.ed.ac.uk.}
\normalsize{$^+$Co-first author.}
}

\date{}

\begin{document} 

\maketitle 

\begin{sciabstract}
Major interventions have been introduced worldwide to slow down the spread of the SARS-CoV-2 virus. Large scale lockdown of human movements are effective in reducing the spread, but they come at a cost of significantly limited societal functions. We show that natural human movements are statistically diverse, and \revision{the} spread of the disease is significantly influenced by a small group of active individuals and gathering venues. We find that interventions focused on these most mobile individuals and popular venues reduce both the peak infection rate and the total infected population while retaining high \revision{social activity levels}. These trends are seen consistently in simulations with real human mobility data of different scales, resolutions, and modalities from multiple cities across the world. The observation implies that compared to broad sweeping interventions, more heterogeneous strategies that are targeted based on the {\em network effects} in human mobility provide \revision{a} better balance between pandemic control and regular social activities.
\end{sciabstract}

\section*{Introduction}

The global pandemic of a novel coronavirus (SARS-CoV-2) has swept through nearly all countries since December 2019. Due to the high infection rate and therefore high demand for medical resources, most countries have adopted large-scale interventions such as business lockdowns and restricted movements. These intervention methods have proven to be effective -- successfully reducing the number of peak daily infected cases and the total number of infected cases so far, as shown by both direct observation from real data~\cite{Flaxman2020-km, Hsiang2020-ab} as well as indirect metapopulation models and simulations~\cite{Chinazzi2020-pw, Kraemer2020-tq}. 

While these intervention methods have helped \revision{in} controlling the pandemic, they are not sustainable over long \revision{periods}. \revision{To address this challenge, we must understand the effect of human movement, and if a less disruptive intervention strategy can control the epidemic.}

The relation between mobility and spread of COVID-19 has been observed in several recent studies~\cite{badr2020association,Kraemer2020-tq,Li2020-ej} that show empirical evidence of correlation between aggregate mobility and number of cases. These studies do not consider the social costs~\cite{Metcalf368} (for example, the reduction in social and economic activities) and are at large scales that do not provide the understanding necessary for more targeted interventions~\cite{Hsiang2020-ab}. Classical epidemiology approaches can model behaviour at scale of individual agents -- these include models such as SIR or SEIR~\cite{wang2020phase} and their variations \cite{lai2020effect, prem2020effect,Chinazzi2020-pw}, data-driven models \cite{covid19health}, and multi-agent models~\cite{Ferguson2020-kz}. By amending the epidemiological models with characteristics of COVID-19 and incorporating the effect of real human movements, we can gain insights into targeted strategies for interventions and the social costs of interventions.

\revision{The objective of our study is to understand the effect of human movements in the spread of the virus.} We use an agent\revision{-}based simulation where a mobile agent can probabilistically infect other agents that are \revision{in} proximity -- such as at the same venue. We consider movements of the real people captured in three types of mobility data -- check-ins at seven different cities, WiFi connection events on a university campus, and GPS traces of electric bikes -- representing different scales, behaviors, and modalities. Across these diverse circumstances, we observe several common features in human movement, \revision{the} spread of the epidemic, and \revision{the} effect of interventions.

\section*{Results}
We found that mobility statistics are heterogeneous across individuals and venues (Fig.~\ref{fig:diversity}), with few agents being highly active and some venues attracting many visitors. This heterogeneity and consequent network effects influence the spread of COVID-19. The common strategy of blanket lockdown is seen to delay the peak but comes at a high social cost (Fig.~\ref{fig:limited_lockdown}). High mobility agents and popular venues attract early infections and play an important role in the early spread of the infection (Fig.~\ref{fig:track_popular_users}). An important observation from our simulations is that interventions that protect these active entities, such as isolating or immunizing the highly active agents, or closing the popular venues, are seen to have a large effect -- both delaying the peak and reducing total infections -- while still incurring a relatively small social cost (Fig.~\ref{fig:compare_strategies}). Classical models (e.g.SEIR~\cite{wang2020phase}) where infection transmission is not related to spatial proximity do not show such heterogeneous effects (Supplementary Fig.~S7), which implies that mobility makes a significant difference compared to homogeneous models. Finally, a social network or {\em contact graph} -- that connects agents to other agents they meet frequently or to venues that they visit frequently -- is seen to have the property that a similar number and group of people are infected in this static model as under the dynamic mobility simulation model (Fig.~\ref{fig:social-network}). \revision{This section elaborates on these results, starting with a study of the baseline lockdown strategy. }

\subsection*{Evaluation of baseline uniform mobility intervention strategy.}

Under a lockdown or stay-at-home order, mobility drops only to a limited extent~\cite{aktay2020google}. We model such a partial lockdown by randomly removing events with a certain probability. This is our baseline intervention strategy. Fig.~\ref{fig:limited_lockdown} studies the realistic scenario where the intervention lasts for a limited duration, with $80\%$ of the events removed.  Thus, during intervention this retains $20\%$ social value (experiments with other probabilities are in Supplementary Fig.~S1 online). The results agree with the observations across the world  -- the peak of the infection is lowered and delayed~\cite{Hsiang2020-ab}. 
However, we discover that this strategy alone does not significantly reduce the total number of infections when applied for a limited duration. Supplementary Fig.~S2 shows the effects of such a strategy applied for longer durations. 

Part I of Fig.~\ref{fig:limited_lockdown} studies the disease spread with \revision{a} varied starting time of the intervention while keeping its duration fixed at $15$ days. In the Foursquare NYC dataset, the intervention starts on \revision{the} day $9$ when $5\%$ of the population gets infected. Compared to the scenario without intervention, this strategy reduces the total number of infected agents by $8\%$, delays the peak of active infections by $11$ days, and lowers it by $6\%$. The peak of new infections is also delayed by $18$ days. The University dataset shows a similar pattern, as the peaks in active and new infection are delayed by $22$ and $19$ days respectively. Other datasets show  trends consistent with the above (Supplementary Fig.~S3 online).

\revision{We note that an earlier lockdown is not necessarily the better strategy when implemented for a fixed duration.} An imperfect lockdown can allow the infection to survive, which then spreads rapidly on \revision{the} removal of the intervention and produces a high eventual peak or second wave. On the other hand, a lockdown implemented somewhat later achieves a lower peak later on, and \revision{a} lower total number of infections.

In Fig.~\ref{fig:limited_lockdown}, Part II, the intervention strategy starts when $10\%$ of the population is infected (day $13$ for the Foursquare NYC dataset and day $36$ for the University dataset). The longer interventions have a more significant effect in reducing both the peak of active cases and total infections. A $30$-days intervention reduces the total infected population by $27\%$ and $5\%$ in the Foursquare NYC dataset, the University dataset respectively. While for shorter interventions (up to $15$ days), the active infection curves remain uni-modal, a longer intervention ($30$ days) produces two peaks and the second peak is delayed by $32$ and $65$ days for the Foursquare NYC and the University dataset respectively. The growth rate, $\lambda_t$ for the longer intervention does not go to zero ($\lambda_t$ reaches zero when no new person is infected). Thus the disease remains in circulation for a longer period. Interventions up to two months (Supplementary Fig.~S4 online, Part II) show similar conclusions: second peaks are observed, and the cumulative number of infected people does not decrease beyond $10\%$ compared to the setting without intervention.

Dividing a population into non-interacting cohorts has been proposed as an intervention strategy~\cite{NJ2020Restart,UNMC2020Covid}, with the idea that preventing transmission across group boundaries will reduce the spread. We find that grouping does act to delay and lower the peaks and to reduce the total number of infections in Fig.~\ref{fig:divide_people} and~S5. The number of infected people is reduced by $37\%$ in NYC and by $53\%$ in the university with $4$ groups. The effect is milder in the Foursquare datasets with larger population size, for example in the Istanbul dataset (Fig.~\ref{fig:divide_people}B). By taking the same sized population as in \revision{the} NYC dataset (Fig.~\ref{fig:divide_people}C), the effect appears -- the total number of infected agents is reduced by $22\%$ with $4$ groups. 

\subsection*{Heterogeneity in  mobility and contagion dynamics.} \revision{We found that the most active individuals are infected significantly earlier (Fig.~\ref{fig:track_popular_users}) than the average population. A person's activity is quantified by the number of check-ins in the Foursquare datasets and the number of meetings in the University and Bike datasets. The popularity of a venue is measured by number of check-ins.} Of the top $5\%$ most active agents, approximately $60\% - 80\%$ get infected at the peak of active infections compared to the peak height about $20\% - 40\%$ for the overall population, and this peak is reached about $10$ days earlier. Thus, highly active people are at a higher risk, and more likely to be propagators of the disease in the early stages. Similar results on other datasets can be seen in Supplementary Fig.~S6.

The heterogeneity creates fundamental differences in infection spread compared to the homogeneous SEIR model. Supplementary Fig.~S7 shows that in comparison to the homogeneous SEIR model - where agents meet all other agents with equal probability - the data\revision{-}driven simulation achieves a higher peak, but a smaller total number of infections. 

The heterogeneity of venues in spreading the contagion is seen in Fig.~\ref{fig:track_popular_users}E~to~G. The number of agents infected from a venue has a heavy\revision{-}tailed distribution for all cities in the Foursquare datasets, and the results for the NYC dataset \revision{are} shown separately. Most of the venues infect a tiny number of individuals, but a few venues infect a large number of agents - In the Foursquare NYC dataset, $50\%$ people are infected from $0.05\%$ ($19$ venues) most popular venues. There are $718$ and $4,979$ individuals who have visited the top $0.01\%$ most popular venues in the Foursquare NYC and Istanbul datasets respectively. \revision{The shape of active infections curve for the active agents remains similar to that for the overall population, but these agents are infected in higher proportions (by at least $12\%$ higher).}

\subsection*{Targeted interventions on the most active individuals and the most popular venues.}  We study heterogeneous  intervention schemes targeted to put a higher level of protection on the most active agents and most popular venues, thus reducing or eliminating their contribution to spreading the virus.

In these simulations, the {\em social cost} of an intervention is defined as the fraction of social events (or potential interactions) lost due to the intervention. For example, in the check-ins dataset, this is measured as the fraction of check-ins lost. In other datasets, \revision{the} social cost is measured as the fraction of co-located pairs of individuals -- representing potential pairwise meetings. The {\em social value} is correspondingly measured as the complement of the social cost - that is, the fraction of events preserved under the intervention. The {\em health value} of an intervention is measured as the fraction of agents who escape infection due to the intervention (but would have been infected otherwise).

Fig.~\ref{fig:compare_strategies}A to D use the social, and health values to compare the targeted interventions against multiple other strategies including the uniform intervention of staying home, protecting a random subset of agents and closing a random subset of venues. In the Foursquare datasets, closing the most popular venues is the most effective strategy, while protecting the most active agents is the most effective strategy in the University and Bike datasets.

In the NYC dataset, to achieve \revision{a} similar $\sim 80\%$ social value, the strategy of closing the most popular venues shows $60\%$ to $72\%$ more health value than other strategies. In \revision{the} Istanbul dataset, to achieve $60\%$ social value closing the most popular venue achieves at least $52\%$ more health value than other strategies. The other cities in the dataset show consistent patterns (Supplementary Fig.~S8 online). In both the University and Bike datasets, at $80\%$ social value, protecting the most active agents achieves $30\%$ or more health value than other strategies.

The strategy to close the most popular venues is analyzed in detail in Fig.~\ref{fig:compare_strategies}E, F, and Supplementary Fig.~S9.
In Foursquare NYC and Istanbul datasets, closing a few ($\sim 1\%$) most popular venues reduces the total number of infected individuals by more than $40\%$, the peak number of active infections by $\sim30\%$ and delays the time of the peak by up to $30$ days. In larger sample populations, this strategy has a smaller impact on total infections and the peak number but serves to delay the peak. In the university dataset, closing $5\%$ of the most popular venues can reduce the total number of infected individuals by $50\%$ and the peak number of active infections by more than $20\%$.

Figure~\ref{fig:compare_strategies}G, H, and Supplementary Fig.~S10 show the effect of protecting the infection of the most active agents. In the University and Bike datasets, protecting the most active $5\%$ to $10\%$ of the population reduces total infection by $20\%$ to $40\%$ and delays the peak by $90$ to $110$ days. Protecting the most active individuals in the Foursquare datasets has similar effects -- in the NYC dataset, the total number of infected people is reduced by $\sim 30\%$ and \revision{the} peak of active cases is reduced by $\sim25\%$ when $20\%$ most active people are protected. However, protecting $20\%$ most active people reduces the social value by $\sim 45\%$. The rest of the Foursquare datasets show similar patterns.

We study other intervention strategies where the agents and venues to protect are selected randomly in Supplementary Fig.~S11,~S12. They are used in the comparison of different interventions in Fig.~\ref{fig:compare_strategies}.

\subsection*{Contact graph: Social network abstraction to estimate mobility based infection spread.}  We designed social network models derived from mobility data such that the spread of infection in a mobility dataset resembles that in the corresponding social network. To model person to person disease transmission as a social network, we define a graph $G$ with agents as nodes and edges connect agents who have met at least once. Each edge has a weight, $w$, as the average number of times they met in a day. For infection spread via venues, the social network is defined using a bipartite graph where the agents and venues constitute the two sets of nodes, and they connect if a person visited a venue. The edges are weighted by the average number of daily visits.

Fig.~\ref{fig:social-network} compares the contagion simulation on a social network against the simulation on time ordered mobility data. The total number of infected people matches between the simulations (Fig.~\ref{fig:social-network}, Part I). Further, nearly the same set of agents get infected starting with independent random seed agents. We measure the similarity of the agents infected in two simulations using Jaccard similarity which is defined for sets $A, B$ as $\frac{|A\cap B|}{|A\cup B|}$. For all datasets, Jaccard similarity between the agents (identified by unique user identifiers in the datasets) infected in the mobility-based simulation and the social network simulation reach above $65\%$. The active cases' peak differs by $1\%$ (in University) to $24\%$ (in Istanbul). Supplementary Fig.~S13 shows similar patterns for other Foursquare datasets.

For the person to person transmission model, Fig.~\ref{fig:social-network}, Part II compares the total number of infected people for three intervention strategies: staying home, protecting the most active agents, and dividing people into groups. 
The strategies are simulated in the social network setting by preprocessing the graph, $G$. The staying home intervention of skipping a meeting with probability $\alpha$ is simulated by removing an edge with weight $w$ with probability $1-(1-\alpha)^w$. \revision{An agent's activity is defined as the total weight of the edges incident on the corresponding node. The strategy to protect the most active agents translates to removing the nodes with the highest activities from $G$. The strategy to divide the population into groups is simulated by randomly assigning the nodes to groups and then removing the edges between nodes belonging to different groups.} The infection curves for mobility and social network simulations are matched for all strategies for both University and Bike datasets -- the difference between \revision{the} two curves is always below $10\%$. The percentage of people infected in total goes to zero in two models with similar intervention parameters.

\subsection*{Robustness of results.}
The conclusions from the simulations are robust to \revision{parameters changes in the simulation}, and has been confirmed in a range of experiments using different parameter values - for infection probability ($\beta$) (Supplementary Fig.~S14 online), number of seeds (Supplementary Fig.~S15 online), and the probability of an agent being asymptomatic (Supplementary Fig.~S16 online). Decreasing $\beta$  reduces the number of infected people and delays the peak of active infections. In the NYC dataset, $68\%$ of the population gets infected when $\beta=0.75$ compared to $34\%$ at $\beta=0.25$. In the Istanbul dataset, the peak is delayed by $10$ days when $\beta$ is reduced to $0.25$ from $0.75$.
Increasing the number of {\em seeds} -- initial infected people -- increase the robustness or predictability of the spread of infection, but otherwise has a relatively small effect: a moderate number of seeds shows \revision{a} similar behavior to a larger number of seeds -- the total number of infected people differs no more than $4\%$ across datasets for using $20$ seeds compared to $1$ seed.
The fraction of asymptomatic carriers increases the rate of infection (since these carriers remain infectious for a longer period), but does not cause a major growth in infection -- in the NYC dataset $14\%$ more people get infected when increasing the percentage of asymptomatic people from $15\%$ to $75\%$. A similar pattern is found in the other datasets.

The conclusions are also sustained when simulations are carried out on sub-samples of data. (Supplementary Supplementary Fig.~S17 online). Here a fraction of the agents \revision{are} randomly sampled for simulation. For the same fraction of the most active agents protected within the subsample, a larger population has a larger fraction of the total population infected. However, this growth appears to be sublinear, and the effect tapers off with the size of the population. Sampling has a similar effect under the intervention of closing popular venues.

\section*{Discussion}

\revision{We have shown that the diversity of human movement influences the spread of the virus,} making the behavior different from homogeneous models. The heterogeneity can be leveraged to devise less disruptive strategies of intervention. Our conclusions are complementary to other non-pharmaceutical interventions such as maintaining person to person distance, wearing masks and avoiding contact. The analysis is based on datasets that are samples of population behavior, and thus, the infection numbers or percentages generated by the simulations should not be regarded as precise representative values but rather as patterns to expect under various interventions. The trends in relative measurements hold across multiple datasets of varying sizes and representing multiple locations, lifestyles, and geographic scales, giving us confidence that they reflect general properties of infection spread due to mobility and corresponding interventions. 

Mobility in a region may be affected by policies and interventions applied to other regions such as neighboring districts~\cite{Haushofer1063}. This spillover effect can be seen as a network effect \revision{on} a larger scale. It has been shown ~\cite{Holtz202009522} that spillover indeed affects aggregate mobility in US counties and thus intervention policies should be coordinated at larger scales. Our datasets do not include mobility across such administrative zones and consequent spillover effects. As more mobility data from the ongoing pandemic becomes available, it may be possible to incorporate and model such larger-scale network effects on agent mobility and infection, as well as incorporate changes in human behavior due to COVID-19~\cite{gao2020mapping,abouk2020immediate}. \revision{The datasets used here are of  pre-COVID behavior between (2012 and 2018) and representative of the spread of infection under ``normal'' behavior.} The effect of changed mobility under social distancing and COVID awareness will require corresponding datasets and separate study.

In simulations, we found that the existing common strategies come with some caveats. The cohort strategy, i.e. partitioning the population into non-interacting groups~\cite{NJ2020Restart,UNMC2020Covid,adams2020supporting} is effective in circumstances such as small populations, a large number of groups, and the specific dataset of University mobility (Fig.~\ref{fig:divide_people}), but not necessarily in other scenarios. The simulations are restricted to certain slices of the society, and in practice, it is infeasible to maintain a consistent partitioning across different communities such as residence, work, and school, which is likely to reduce the impact of this strategy. A larger number of partitions or cohorts will reduce the spread of infection, but a large number of cohorts are difficult to implement and schedule, and they come at a larger social cost of reduced interaction across cohorts. The implementation of a cohort strategy may differ by circumstances, such as separating the groups spatially (different classrooms) versus separating temporally (different time slots), which will influence the person-to-person distancing and infection propagation. \revision{For the other common strategy of a blanket lockdown, which many countries have implemented, the results in Fig.~\ref{fig:limited_lockdown} show that the timing and duration of the action subtly influence the peak and total infections.} Since this strategy is a limited resource - in the sense that it cannot be sustained for long - it needs to be deployed with careful planning, and with consideration of available medical and economic resources. 

We found targeted interventions on the most mobile agents and most popular venues to have a significant effect on pandemic control at a low social cost (Fig.~\ref{fig:compare_strategies}). These results are complementary to prioritizing protections for essential workers such as medical professionals~\cite{chang2020protecting,lancet2020covid}, as our conclusions are for the mobility component of infection spread, which is not addressed in other works~\cite{wang2020reasons}. In this respect, those most at risk and likely spreaders of infection are workers whose professions require frequent movement. While isolating them from work is not practical, better protection strategies such as protective equipment, strict regulations, and early vaccinations can help to reduce infections. Closing the most popular venues such as university cafeteria (Supplementary Fig.~S18) are also unsustainable over long durations, but modified operations and strict social distancing at these venues can help to significantly reduce infections. The influence of these various strategies relative to other non-pharmaceutical interventions remains to be investigated. It has been suggested~\cite{Haushofer1063} that with more data, implicit Randomized Controlled Trials may be carried out by comparing similar localities enforcing different subsets of policies. 

The difference in results for data-driven mobility and those from traditional homogeneous models (Supplementary Fig.~S7) can be attributed to information-structural properties of human mobility, creating a complex {\em network effect}. \revision{The inclusion} of social and economic considerations are likely to add to these complexities, for example, it has been argued that the worldwide food supply chain is being affected by labor shortages, trade restrictions, and factory closures~\cite{laborde2020covid}. The relations between mobility, essential industries, and economics will be important to decipher for the development of long term strategies. 

\revision{In conclusion, from the perspective of actions to control the pandemic, our results suggest interventions on highly mobile people and most popular venues are most likely to be effective.} These interventions can take forms of priority in testing, vaccination, protective equipment, or stricter regulations. \revision{The infection} spread at venues can take the form of limiting large gatherings, reducing the need to visit the venue or different modes of operations (e.g. deliveries and takeaways) -- and should be encouraged more at popular venues. Our model and simulation methodology can be used to identify the points of action in other datasets and environments. 

We found that large scale general lockdowns may not reduce the total number of infections in all cases, but they can be used to delay infections and lower the peak. Thus, it can be used to temporarily avoid overloading hospitals, and to gain time to build up resources. Partitioning a population into cohorts can be useful in small populations (such as small schools). In large groups where they are harder to implement, they are also less likely to effective if implemented. 

We note that our analysis is purely on the mobility dimension. \revision{The spread} of the infection depends on various other factors, such as the nature and architecture of the venue (indoor vs outdoor), awareness of people, and other complex behavioral factors that should ideally be taken into account. Mobility patterns themselves are dependent on economic and industrial aspects that are beyond the scope of our study. More research is needed for detailed models incorporating complex aspects of infection spread.

\captionsetup{labelfont=bf}

\begin{figure}
    \centering
        \includegraphics[width=\textwidth]{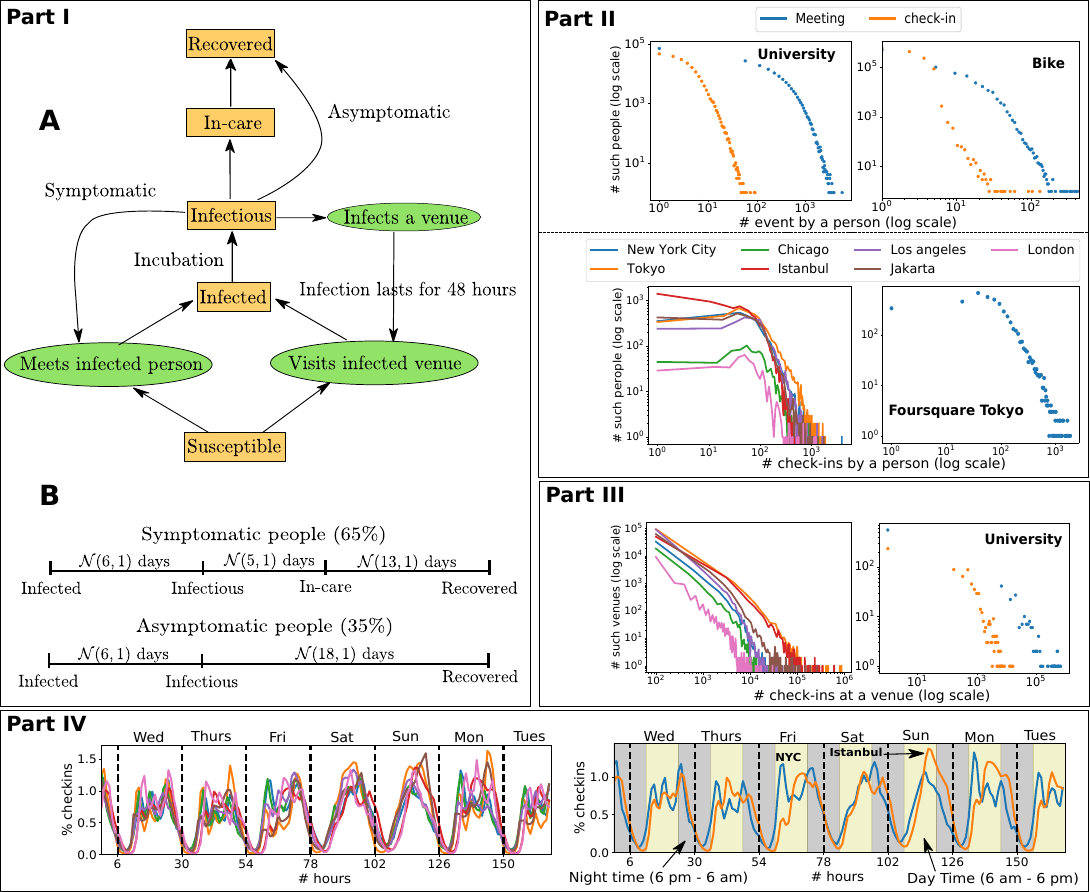}
    \caption{The model for infection spread (Part I) and dataset properties (Part II to IV). The model (A) is adapted from SEIR infection spread model and the parameters (B) are from CDC for COVID-$19$. The model states and actions are represented by boxes and ovals respectively.
    	Part II and III present distributions of the number of check-ins or meetings per person and per venue, depicting the heterogeneity of mobility. 
    	Part IV: Normalized check-in counts aggregated over a week for all Foursquare datasets show daily patterns with increased activity on weekends. Patterns vary across cities.
    }
    \label{fig:model}
    \label{fig:diversity}
\end{figure}

\begin{figure}
    \centering
    \includegraphics[width=\textwidth]{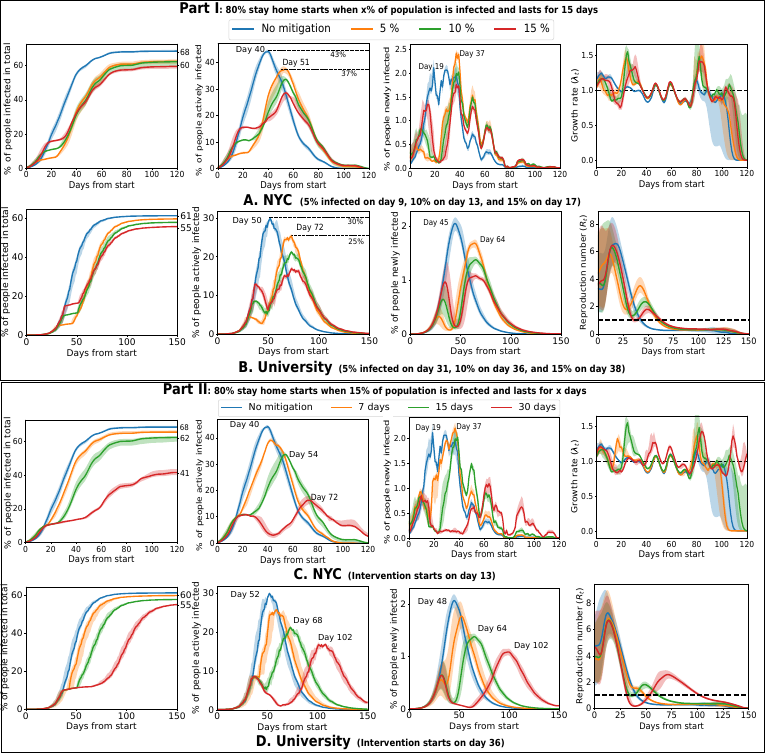}
    \caption{Uniform lockdown for limited duration delays and lowers the peak of the infections, but often has a small effect on total infection.
    	Here, $80\%$ check-in or meeting events are randomly removed. Part I: A limited duration intervention ($15$ days) delays (by $11$ days in NYC and $22$ days in the University) and lowers (by $6\%$ in Foursquare NYC and $5\%$ in the University) the peak of active cases, but does not reduce the total number of infected people, as the second peak is imminent. Part II: With the increasing length of intervention, the number of infected people reduces (by $33\%$ in NYC), but in the University dataset, the effect is milder.
    }
    \label{fig:limited_lockdown}
\end{figure}

\begin{figure}
    \centering
    \includegraphics[width=\textwidth]{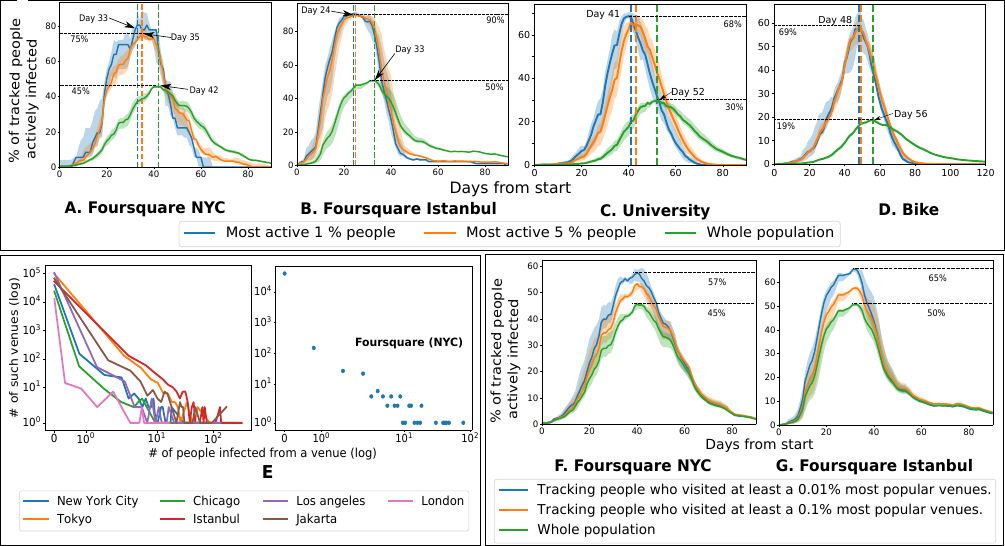}
    \caption{Heterogeneity of mobility in venues and agents results in varied risks to catch and spread the virus. Agents with higher activity get infected with a higher proportion and earlier (A to D).    In all Foursquare datasets, the number of people infected from a venue has heavy-tailed distributions (E). People who visit the most popular venues get infected in higher proportions (F, G). 
    }
    \label{fig:track_popular_users}
\end{figure}


\begin{figure}[htbp]
    \centering
    \includegraphics[width=\textwidth]{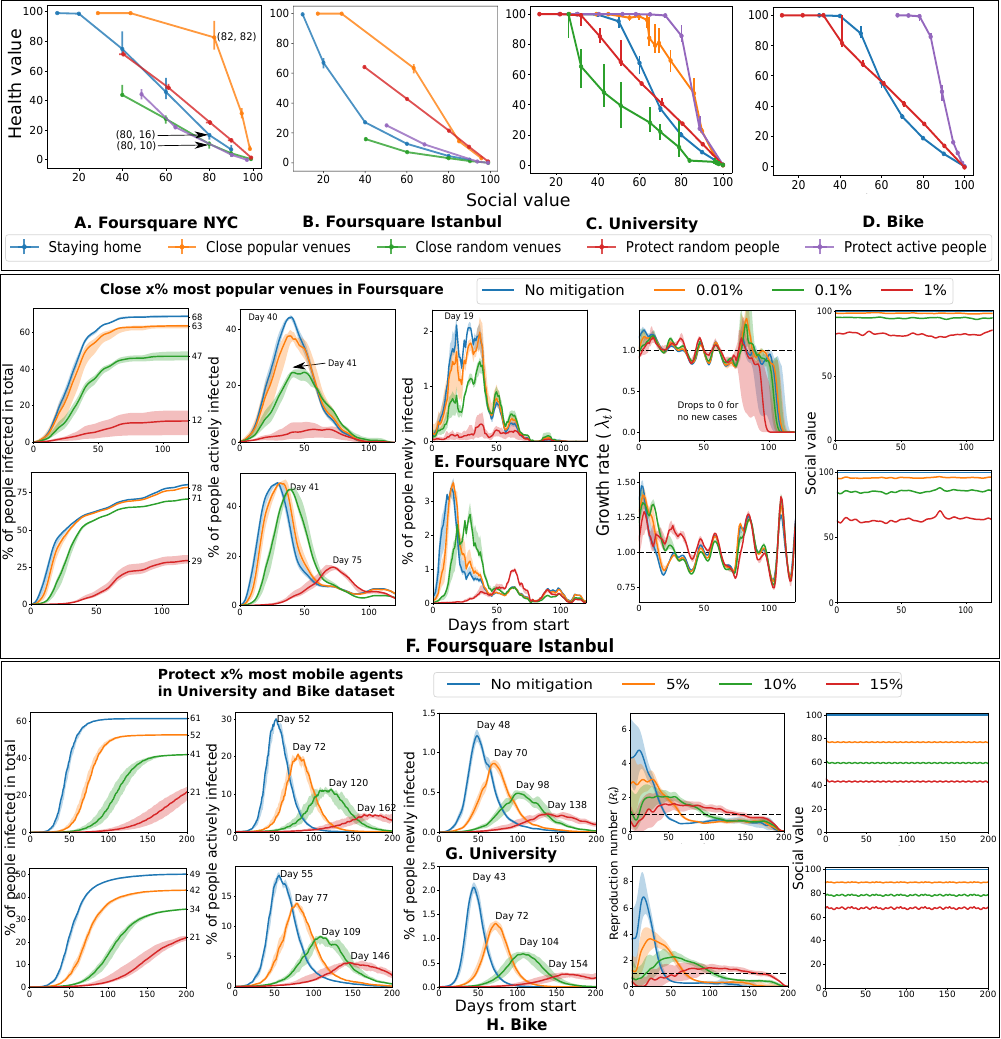}
    \caption{
    }
    \label{fig:compare_strategies}
\end{figure}
\newpage
\noindent \textbf{Figure~\ref{fig:compare_strategies}:} Across all datasets, the most effective strategies in terms of their health and social values consider the heterogeneity of mobility (A to D) and (E to H) present the details of the best strategies. The health value quantifies the percentage of people not infected compared to without intervention scenario, and the social value is the percent of check-ins or meetings that remain under intervention.
    	An ideal strategy would have both high health, and social values.  Across Foursquare datasets, closing the most popular venues is the most effective strategy (A, B); isolating or protecting the most active agents is the best strategy in University (C) and the Bike (D) datasets. The Bike dataset does not have venues and therefore it omits corresponding strategies.
    	E to H show details of the best strategies. They reduce infections to a large extent (closing the most popular $1\%$ venues results in $57\%$ reduction in infection in the NYC dataset), delays and lowers the peak of active infections (protecting the most active $5\%$ agents in the University dataset delays the peak by $20$ days) and maintain high social values ($1\%$ most popular venues contribute to $20\%$ check-ins in the NYC dataset).

\newpage
\begin{figure}[!hp]
    \centering
    \includegraphics[width=0.69\textwidth]{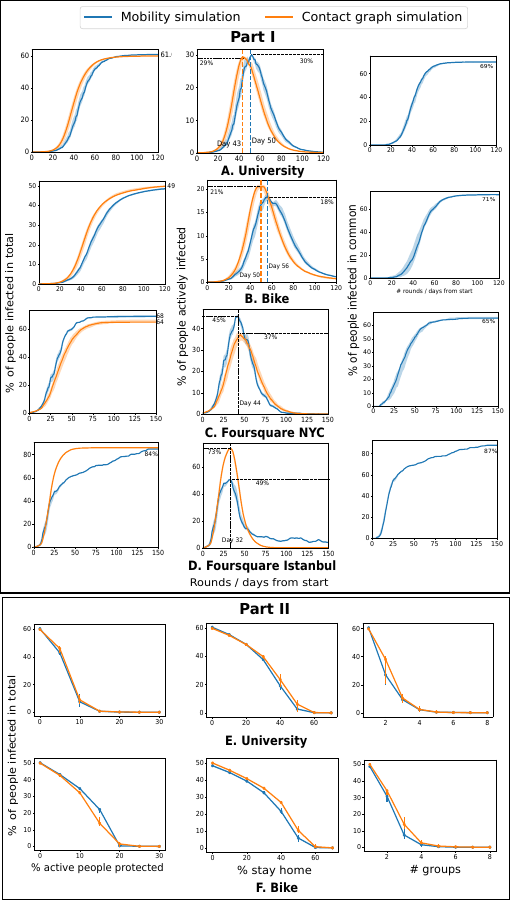}
    \caption{}
    \label{fig:social-network}
\end{figure}
\newpage
\noindent \textbf{Figure~\ref{fig:social-network}:} Contagion simulation in the derived social network model infects a similar number and similar set of agents as the simulation using mobility data in both with and without intervention settings.
Part I: The curves for total infection numbers, actively infected agents, and Jaccard similarity of the infected agents match under no intervention scenario (A to C). Although in the Istanbul dataset (D), two infection number curves do not match after $25$ days, similar sets of people get infected in the two simulations. Part II: In person to person infection spread model, with different intervention parameters, the final infection number is similar between the two models in all three intervention strategies.
\newpage

\newpage
\begin{figure}[!h]
    \centering
    \includegraphics[width=\textwidth]{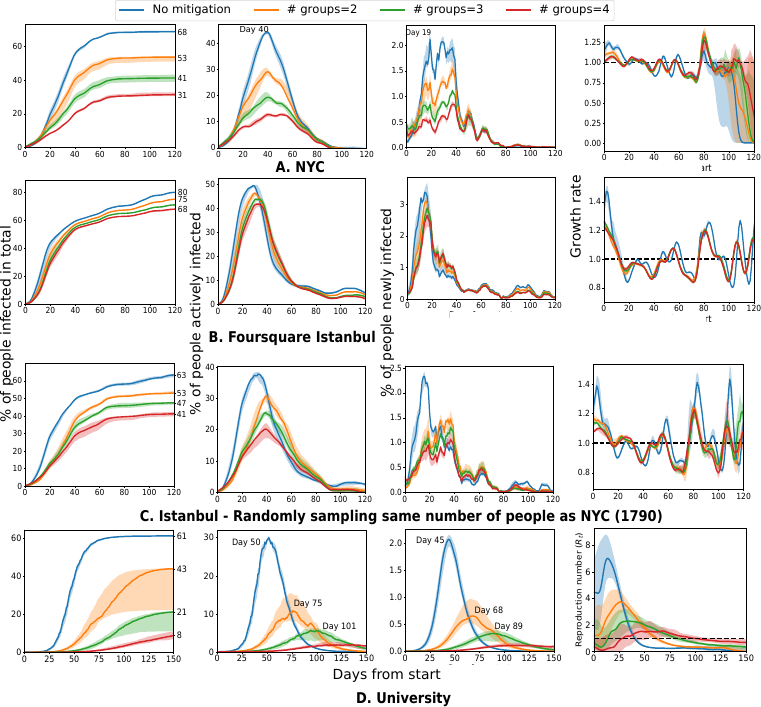}
    \caption{Diving agents into cohorts cuts off the contacts between different cohorts, like separating students into different sessions for courses and acts to reduce the spread of the disease.
    With the more groups, the number of infected people is reduced (by $37\%$ in NYC and by $53\%$ in the university with $4$ groups). The large population of the Istanbul dataset produces densely connected groups that result in substantial intra-group infection spread (B). Thus, when we sample the same number of people as NYC, the effect of separating groups re-appear (C). In addition, the peaks of active cases are also delayed and flatten, especially in the University dataset.}
    \label{fig:divide_people}
\end{figure}

\section*{Method}

Our work takes the approach of multi-agent models but incorporates real mobility traces to study the probable spread of an infectious disease. The study involves three data sets that represent three typical social settings. (Refer to Supplementary material for detailed data description).
\begin{itemize}
	\item \emph{Foursquare check-ins.} We use check-in data~\cite{yang2019revisiting, yang2020lbsn2vec++} from the Foursquare mobile application, which captures snapshots of human mobility in popular public spaces. The dataset includes seven major cities across the world (New York, Chicago, Los Angeles, London, Tokyo, Istanbul, and Jakarta), and contains a total of $2,293,716$ check-ins from $24,068$ individuals in $397,610$ venues over a period of $140$ days. 
	\item \emph{University campus~\cite{sui2016characterizing}:}  WiFi log data on connections of mobile devices with nearby WiFi access points 
	from a large university campus (Tsinghua University) is a representative case of mobility in a large university setting. The dataset contains $106,975$ average daily localization points from $47,359$ individuals.
	
	\item \emph{Personal electric bikes~\cite{wang2020distributed}:} Densely sampled GPS mobility traces of $46,078$ personally owned electric bikes over the period of a month provides a unique dataset as they capture unstructured personal mobility patterns not tied to specific public spaces or institutions.
	
\end{itemize}
All three data sets have fine-grained location resolution of within $10$s of meters, which allow us to investigate inter-personal interactions and interactions of individuals and public venues. These datasets collectively cover three representative slices of human mobility that play important roles in \revision{the} spread of infectious diseases. Infection can propagate person to person at a meeting event (when two agents are \revision{in} proximity), or through a venue (when a susceptible agent visits a venue recently visited by an infectious agent). Note that these datasets were collected pre-COVID between 2012 and 2018, \revision{representing} ``normal'' human movements. We ran simulations on the three datasets with \revision{various} intervention strategies and show their impact in slowing down the spread. We study the total number of individuals infected under various intervention strategies, \revision{the} growth rate of new infections, the maximum number of infected individuals at any time (height of the peak), and the time-varying reproduction number ($R_t$). \revision{The social cost of an intervention strategy is measured as the number of check-ins and meetings lost due to the strategy.}

\paragraph*{Mobility data driven infection simulation model for COVID-$19$.}
We customize the SEIR model~\cite{wang2020phase} to operate on mobility data and incorporate the properties of COVID-$19$ transmission as probabilistic features using the parameters from Center for Disease Control and Prevention~\cite{wu2020characteristics,CDCparameter} (Fig.~\ref{fig:model} Part I).
Like the SEIR model an agent in our model can be in one of the four states: susceptible, infected (or exposed), infectious, and recovered or immune. Unlike the standard model that assumes person-to-person disease transmission through uniformly probable contacts, we obtain person to person contacts from the mobility datasets and extend the model to disease transmission via check-ins to venues. 

The simulation starts by infecting a random set of {\em seed agents} with all other agents as susceptible. The disease transmits from person to person at a \emph{meeting} (in the University and the Bike dataset), or through a venue (in the Foursquare datasets) with a transmission probability, $\beta$. In the University dataset, two agents meet when they are at the same location (such as a classroom or cafeteria) at the same time. \revision{A meeting occurs when two bikes are within $5$ meters for more than $5$ minutes in the Bike dataset. When an infectious agent visits a venue in the Foursquare dataset, we  mark the venue as infected and it remains infected for $48$ hours.} This aligns with the studies on COVID-$19$ lifespan on various surfaces~\cite{van2020aerosol}. 

A susceptible person is infected with probability $\beta$ when meeting an infectious person or visiting an infected venue. The value of $\beta$ is determined for each scenario based on knowledge of existing variables such as $R_0$ (see supplementary material for details). An infected person remains in incubation for $\mathcal{N}(6, 1)$ days followed by the infectious stage gaining the ability to infect other susceptible agents or venues. Here, $\mathcal{N}(\mu,\sigma)$ denotes a sample from a normal distribution with mean $\mu$ and standard deviation $\sigma$. The distribution models variability in infection behavior. For COVID-$19$ there are a significant fraction of asymptomatic carriers~\cite{nishiura2020estimation}\cite{mizumoto2020estimating}. As CDC mentioned, we consider $35\%$ of the infected population (randomly chosen) to be asymptomatic. A symptomatic person remains infectious for $\mathcal{N}(5, 1)$ days before showing symptoms and subsequently enters the {\em in-care} stage such as at home self-quarantine or hospitalization. At this stage, the person ceases to infect others. After $\mathcal{N}(13, 1)$ days in care, the agent assumed to recover and become immune.
On the other hand, asymptomatic agents do not seek care, and thus they skip the in-care state and remain infectious for $\mathcal{N}(18, 1)$ days.

We study temporal dynamics of the infection spread in four dimensions -- the total number of infected people till date, number of active cases, number of new infections, and epidemiological parameters such as the time-varying reproduction number ($R_t$) or growth rate ($\lambda_t$). The active cases include individuals that have been infected till date but not yet recovered. The time-varying reproduction number~\cite{gostic2020practical}, $R_t$ for day $t$ is the expected number of average individuals infected by a single agent who gets infected on day $t$.
The growth rate, $\lambda_t$, \revision{is} the ratio of the total number of agents newly infected on \revision{the} day $t$ to the same number on the day $t-1$.

\section*{Ethics of data and methodology}

In this study, we used secondary datasets collected in the past (between 2012 and 2018) by other researchers, who have published detailed information about the corpora and the collection processes~\cite{yang2019revisiting, yang2020lbsn2vec++,sui2016characterizing}. The project has gone through the ethics review process at the School of Informatics, University of Edinburgh, and has received a waiver for use of these secondary corpora. More specifically, the datasets were completely anonymous with personal identifying information removed. They are either public or collected with consent as follows:

\begin{itemize}
\item The Foursquare data were derived from open public posts on Twitter between 2012 and 2014 by researchers at University of Freibiurg and made available online (\url{https://sites.google.com/site/yangdingqi/home/foursquare-dataset}). Details of the data has been published previously~\cite{yang2019revisiting, yang2020lbsn2vec++}. 

\item The University and Bike datasets were collected from adult participants with informed consent. The University dataset was collected between September 2015 to November 2015 by the research group at (NetMan) AIOps labs at Tsinghua University, for research purposes. The data collection process was published in 2016~\cite{sui2016characterizing}. 
We received the dataset through personal communication with the authors. The Bike dataset was collected between June 2018 and September 2018 by a private company, Zhejiang Tendency Technology Co., Ltd in the city of ZhengZhou, P.R. China. 

\end{itemize}

\section*{Data Availability}

Among the data used in the work, the Foursquare check-ins dataset spanning seven cities are publicly available and has been cited in the text. The Tsinghua WiFi dataset and Zhengzhou bike datasets cannot be made public due to our non-disclosure agreement with the respective data owners. Our contribution is the program code for the simulations and analysis, which is publicly available at \url{https://github.com/SBUhaotian/Mobility_Contagion}~\cite{githubcode} for reproducibility, and use on other datasets.

\section*{Acknowledgements}

Wang and Gao would like to acknowledge supports from NSF CNS-1618391, DMS-1737812, and OAC-1939459. Ghosh and Sarkar acknowledge support from the COVID-19 fund of the Data Driven Innovation program under Edinburgh and South-East Scotland City Region Deal funded by the governments of Scotland and U.K. Ding acknowledges support from Shanghai Sailing Program 20YF1421300.

\section*{Author contributions statement}
HW and AG implemented the simulation and generated figures. JD contributed to data cleaning and literature search. JG and RS contributed to research formulation and data interpretation. All authors contributed to modeling of COVID-19, formulation of intervention strategies and writing of the draft.

\newpage

\section*{Supplementary materials} 
\setcounter{page}{1}

\noindent{\large\bf Title: Heterogeneous Interventions Reduce the Spread of COVID-19 in Simulations on Real Mobility Data}

\noindent{\bf Authors:} Haotian Wang, Abhirup Ghosh, Jiaxin Ding, Rik Sarkar, Jie Gao.
\newline

\textbf{This PDF file includes:}

Materials and Methods

Figure~\ref{figS:probability_stay_home} to~\ref{figS:close_type_venue}

Table S1 and S2

\subsection*{Data}
All our datasets describe individual mobility by timestamped locations. Datasets use anonymized individual identifiers and have assorted temporal and spatial resolutions. Table S1 quantifies the datasets.

\paragraph*{Foursquare dataset.} Foursquare enables users to record their check-in times at venues and thus produce a log of users' mobility to public places. Although the check-ins are accessible to people only within social circles, many people choose to share them on Twitter. Authors in~\cite{yang2019revisiting, yang2020lbsn2vec++} collected the Foursquare check-ins posted on Twitter. In this work, we use a sub-sample of the available dataset containing $7$ cities (New York, Tokyo, Istanbul, Chicago, Jakarta, Los Angeles, and London) with dense data distributions spanning between Apr-$2012$ and Aug-$2012$. Check-ins for a city are selected by first choosing a Geo-location bounding box containing the city, for example, we consider the venues within the box ranging between latitude $[40.5378^{\circ} N, 41.0131^{\circ} N]$ and longitude $[73.4532^{\circ} W, 74.3074^{\circ} W]$ are considered in the New York City dataset. Then we select the check-ins that happened in the selected venues. Essential characterizations of the datasets agree with the findings in~\cite{yang2019revisiting, yang2020lbsn2vec++}. The check-ins have the temporal resolution in seconds.

\paragraph*{University dataset.} The Tsinghua University campus covers an area of $4.4km^2$ with $47,359$ students and $12,000$ faculties. There are $2,346$ wireless access points deployed in $114$ buildings including classrooms, departments, administrative buildings, apartments, gyms, libraries, restaurants, supermarkets, and hotels. The dense deployment allows tracking the connected devices. Each connected device reports its connecting access point every $5$ minutes. The dataset~\cite{sui2016characterizing} is one week long and assuming that the weekly movement stays similar in this setting, we repeat the dataset to generate \revision{more days}. When two agents stay in the same room within a period, it is regarded as a meeting between them. 

\paragraph*{Bike dataset.} Zhengzhou (China) has wireless tracking terminals that can locate and monitor electric bikes in real-time using a variety of IoT sensors, such as satellite positioning module, voltage and current detection module, and temperature sensing module. Our dataset~\cite{wang2020distributed} has timestamped GPS locations of $46,087$ personally owned electrical bikes with the temporal granularity of $10$ seconds. On average, there are more than $41,000$ people traveling by electrical bikes and more than $19$ million data points collected every day for a month and repeat in the simulation. If a person stops moving in one location for more than $5$ minutes, this location with the period is regarded as a stay point of this agent. 
On average, there are $118,337$ daily stay points and $90,144$ meetings every day.  

\paragraph*{Data resolution.} The data we operate with has a high spatial resolution, which allows us to execute detailed agent-based models. Other forms of data often used in movement analysis include cellular data - representing cell tower connections of mobile phones, including errors at a scale of kilometers, which is the range of cell towers. Even in the urban scenarios of dense deployments, cell towers are separated by hundreds of meters implying correspondingly low resolutions. In comparison, in the bike GPS dataset, we have errors at the level of few meters, and similarly, the errors in \revision{the} University WiFi dataset are bounded by a few tens of meters, which is the typical WiFi range, and connection to the same WiFi usually implies agents being in the same building. The Foursquare dataset correspondingly contains precise information of agents visiting the same venue. Compared to metapopulation models that treat the population as a set of communities, this high-resolution simulation incorporates a more dynamic and precise movement pattern. We are able to simulate movement or presence of agents with precision such as presence in the same building, and correspondingly likely infection transmission.

\subsection*{Intervention Strategies}

We have investigated multiple non-pharmaceutical interventions. Some strategies referenced briefly in the main article are described below.

\paragraph*{Intervening random individuals and venues.}
Here, the venues to close (Supplementary Fig.~\ref{figS:close_random_venue} online) and people to protect (Supplementary Fig.~\ref{figS:protect_random_people} online) are chosen uniformly at random. In the Foursquare, NYC dataset, closing $40\%$ random venues results in $\sim 20\%$ reduction in total infected people and $60\%$ reduction in the social value. In other Foursquare datasets of cities such as Istanbul and Tokyo, the effect of this strategy on health value is minimal with \revision{a} reduction \revision{of} not more than $\sim 5\%$. Other datasets show a similar trend. In the University dataset, closing $40$ random venues brings $17\%$ reduction in total infected people. Across all datasets, closing $40\%$ random venues results in more than $\sim 33\%$ reduction in the infected population (Supplementary Fig.~\ref{figS:protect_random_people} online).

These strategies are shown to be less effective than intervening the most active agents or most popular venues. In the Foursquare NYC dataset, to achieve the same social value ($\sim 60\%$), the strategy to close randomly chosen venues achieves $65\%$ less health value compared to the targeted strategy to close the most popular venues. In the University dataset, to achieve the same social value ($\sim 80\%$), the strategy to protect random people obtains $55\%$ less health value compared to the strategy to protect the most active people. 

\subsection*{Methods: Implementation and Experimental setup}
The Foursquare datasets contain time-ordered check-ins along with categorical unique identifiers for venues and agents. In the University and Bike datasets, meetings are computed and stored as data pre-processing steps. The meetings in the University dataset are computed first grouping the contacts by venues and then finding the overlapping staying intervals. In the Bike dataset, the spatial proximity is computed by a kd-tree, a spatial indexing structure to find the location within a given distance. The meetings are stored with the participates, location, starting, and ending time. Meetings are ordered by their starting time.

The simulation progresses sequentially over the events ordered by check-ins or meetings and keeps track of the states of the agents and venues. While the state machine for the agents is described in the main article, a venue can be in one of the two states -- susceptible and infected. Initially, all venues are susceptible and a venue becomes infected once an infectious agent checks-in. Each infected venue keeps a timer that goes off after $48$ hours of being infected and then the venue becomes susceptible again. If an infectious agent visits an infected venue, the timer is reset to the current check-in time. We maintain the timer by operating on it only at check-in event times.

In the person to person infection spread model, the infection probability, $\beta$, used in the simulation is derived for a particular dataset based on existing knowledge of the initial reproduction number, $R_0$ for COVID-19. $R_0$ denotes the expected number of people infected from a single person, i.e., $R_0=c\beta T$, where $c$ is the average number of daily meetings per person, and $T$ is the average number of days an agent remains infectious. While $c$ is estimated from a dataset, model parameters in Fig.~1B give the value of $T$ as $9.55$ days. For the disease transmission through venues, $\beta$ has a different interpretation and does not correspond to usual parameters such as $R_0$. We evaluate a range of $\beta$ value in Supplementary Fig.~\ref{figS:infection_probability} and find consistent patterns across them.

\revision{There are $10$ initial infectious seeds at the beginning of each simulation. The infection probability is set as $0.75$ for the Foursquare dataset and takes $R_0$ as $3$ to obtain the University and Bike dataset's infection probabilities.}

For both the person to person and via venue transmission models, the probabilistic virus transmission is simulated as follows. A uniform random number is sampled in the range $[0,1]$ and compared against the infection probability, $\beta$. The probabilistic experiment is successful, i.e., the agent becomes infected, if the sample is less than $\beta$.

We study dynamics of the infection spread in four dimensions -- the total number of infected people till date, number of active cases, number of new infections, and epidemiological parameters such as the time-varying reproduction number ($R_t$) or growth rate ($\lambda_t$). The active cases include individuals that have been infected till date but not yet recovered. The time-varying reproduction number, $R_t$ for the day $t$ is the expected number of average individuals infected by a single agent who gets infected on the day $t$. The growth rate, $\lambda_t$, is the ratio of the total number of agents newly infected on the day $t$ to the same number on the day $t-1$. When $\lambda_1$ above $1$, the number of infections grows exponentially; when $\lambda_t$ is below $1$, the number of new infections converges.

In the contact graph\revision{-}based study, the contagion simulation uses the same model and parameters as mobility datasets (Fig.~1 Part I). It starts with $10$ initial randomly selected infectious nodes ({\em seeds}) and all other nodes are at \revision{the} susceptible state. It works in synchronous and discrete\revision{-}time rounds -- at each round, an infectious node infects its susceptible neighbors in $G$ with a probability, $p$. As both the meetings and the check-ins to venues are abstracted to counts instead of timestamped events, the contacts are considered to be distributed uniformly over the duration of the dataset, and infection transmission probabilities $\beta$ are adjusted per edge as $p = (1- (1-\beta)^w)$. In the infection transmission model via venue, an infectious individual infects a venue with probability $\min(1, w)$ at each round, and the infected venue in-turn infects an individual with probability $p$. 

Similar to mobility\revision{-}based simulations, in the social network setting we use $\beta=0.75$ for transmission via venues. In the person-to-person transmission model, $\beta$ is derived from $R_0$ from the same expression, $R_0 = c  \beta T$. Here, $c$ is counted as the average total weight from a node and $T$ \revision{remains} the same as $9.55$.

\paragraph*{Computation of social and health values of interventions.} The social value of an intervention strategy is measured as the percentage of activities (check-ins) preserved under the intervention. When a venue is closed or an agent is protected, the corresponding activities are canceled. The social value of a strategy is denoted by $1-x/y$ in percentage, where $x$ and $y$ number of check-ins (for Foursquare datasets) or meetings (for University and Bike datasets) performed under the intervention and without intervention respectively. \revision{The health value of an intervention is correspondingly measured as the percentage of agents who escape infection due to the intervention. If $x$ and $y$ number of agents are infected under intervention and without intervention respectively, then the strategy's health value is the proportion $1-x/y$, usually written as the percentage.}

The simulations are implemented in Python and executed in standard desktop machines with Intel i$7$ cores and $16$GB memory. The simulation is compute-efficient -- all our simulations run under $5$ minutes. The efficiency results from in-memory processing, pre-computation of meetings prior to simulation, and storing infection states in the data structures for venues. All of these reduce the number of operations at each discrete event timestamp. The result of a simulation contains the timestamps of new infections and recoveries.

As our simulation model is stochastic, each experiment is run $10$ times to test their stability. We report the temporal dynamics in the figures with the median of $10$ runs as the solid curve and the shaded regions denote the area between $25$ and $75$ percentiles. The figures are smoothed with $7$ day rolling average. Moreover, to bring about the major trends, the growth rate and reproduction number plots are further smoothed using a standard Gaussian filter with the standard deviation of the Gaussian kernel set as $2$ days.

\section*{Existing Models for infection spread}

There are various existing models for infection spread. We discuss below the ones most relevant to us. 

\begin{itemize}
	\item \emph{Epidemiology models:} Classical models in epidemiology include SIR (Susceptible-Infected-Recovered), SEIR (Susceptible-Exposed-Infected-Recovered) models \cite{wang2020phase} and other variations \cite{lai2020effect, prem2020effect}. \revision{These models make simplistic assumptions of a homogeneous population and any two individuals have the same structure of interactions and dynamics. The model evolves by using a small number of parameters such as an infectious person's chance infecting another one and then deriving ordinary differential equations.} To incorporate heterogeneity in a large population, meta-population models~\cite{Chinazzi2020-pw} include population structures that describe variations in age groups, behaviors, neighborhoods, but in general, these models are coarse-grained.
	\item \emph{Data-driven models:} A few models such as the one from the Institute of Health Metrics (IHME) \cite{covid19health} use a data-driven model to predict the number of new infections, based on data from other countries. This model assumes that the infection process is uniform across different countries, thus it ignores the important parameters such as the discrepancy of culture, weather, the density of population, and the \revision{lifestyle}. 
	
	\item \emph{Multi-agent models:} The models from Imperial College London~\cite{Ferguson2020-kz} are individual-based multi-agent models. These models are fairly complicated with a large number of parameters describing the interactions between the agents. For example, individuals are assumed to reside in high-density residential areas from census data. Contacts with other individuals are assumed to happen within the household, at school, in the workplace, and in the wider community. The parameters of population density in these scenarios are taken as the average in published data. 
	It is a challenge to choose these parameters and validate the choices against real data. 
\end{itemize}

\section*{Comparison with standard SEIR model}
From a dataset, we count the average number of daily contacts for an agent, $c$. With the population size of $N$, there are $Nc/2$ contacts in total per day. The SEIR simulation progresses in synchronous daily rounds and $Nc/2$ contacts are randomly sampled each day. The other parameters remain similar to Fig.~1.

We apply the standard SEIR model to COVID-$19$ parameters, i.e., consider meetings between random pairs of agents. We simulate the person-to-person transmission model in the University and the Bike datasets keeping the model parameters the same as Fig.~1B. Given the dataset, $Nc/2$ contacts between two agents are randomly sampled. Here, we ignore the time-stamped of each meeting from the dataset. Similar to our simulation model, the simulation starts with $10$ initial seeds and proceeds with probabilistic disease transmission using sampled contacts.

Supplementary Fig.~\ref{figS:Compare_RandomMeeting} compares the infection spreads in two models for a setting without intervention. For both University and Bike datasets, a larger population gets infected by \revision{the} SEIR model compared to our mobility based simulation. In both the datasets, the peak of active infections in \revision{the} SEIR model is at least $15\%$ higher delayed by more than $35$ days than the mobility model. 
This is due to heterogeneity of agents -- more active agents get the virus early and infect other susceptible agents early -- resulting in an early peak. Besides a large fraction of agents have a low number of meetings, therefore, have less risk of being infected which leads to a lower total infection number compared with \revision{the} SEIR model. Our observations match with the observations in~\cite{stehle2011simulation}. 

The difference in meeting distributions in the two simulation models results in different distributions for the number of agents infected from an individual. While the mobility\revision{-}based model has a long tail distribution suggesting that the more active agents infect more people, the SEIR model does not have a long tail.

\newpage

\begin{table}[!ht]
    \centering
    \caption*{\textbf{Table S1: The statistical characteristics for the Foursquare dataset.}}
    \begin{tabular}{cccc}
    & Number of check-ins & Number of people & Number of venues \\
    \hline
    New York & $202,599$ & $1,790$ & $39,764$ \\
    Istanbul & $559,966$ & $8,925$ & $53,075$ \\
    Tokyo & $642,687$ & $4,744$ & $96,931$ \\
    Chicago & $103,432$ & $924$ & $23,391$ \\
    Los Angeles & $402,989$ & $3,590$ & $104,629$ \\
    Jakarta & $336,386$ & $3,623$ & $67,047$ \\
    London & $45,657$ & $472$ & $12,773$ \\
    \end{tabular}
    \label{tab:data_desc}
\end{table}

\begin{table}[H]
    \centering
    \caption*{\textbf{Table S2: The statistical result for University dataset and Bike dataset.} 
    }
    \begin{tabular}{ccccc}
    & \# of daily stay points & \# of people & \# of venues & \# of daily meetings \\
    \hline
    University & $106,975$ & $47,359$ & $2,346$ & $2,476,837$\\
    Bike & $118,337$ & $46,087$ &  N.A. & $90,144$ \\
    \end{tabular}
    \label{tab:data_Bike}
\end{table}
\newpage
\setcounter{figure}{0}
\renewcommand{\thefigure}{S\arabic{figure}}

\begin{figure}[H]
    \centering
    \includegraphics[width=0.9\textwidth]{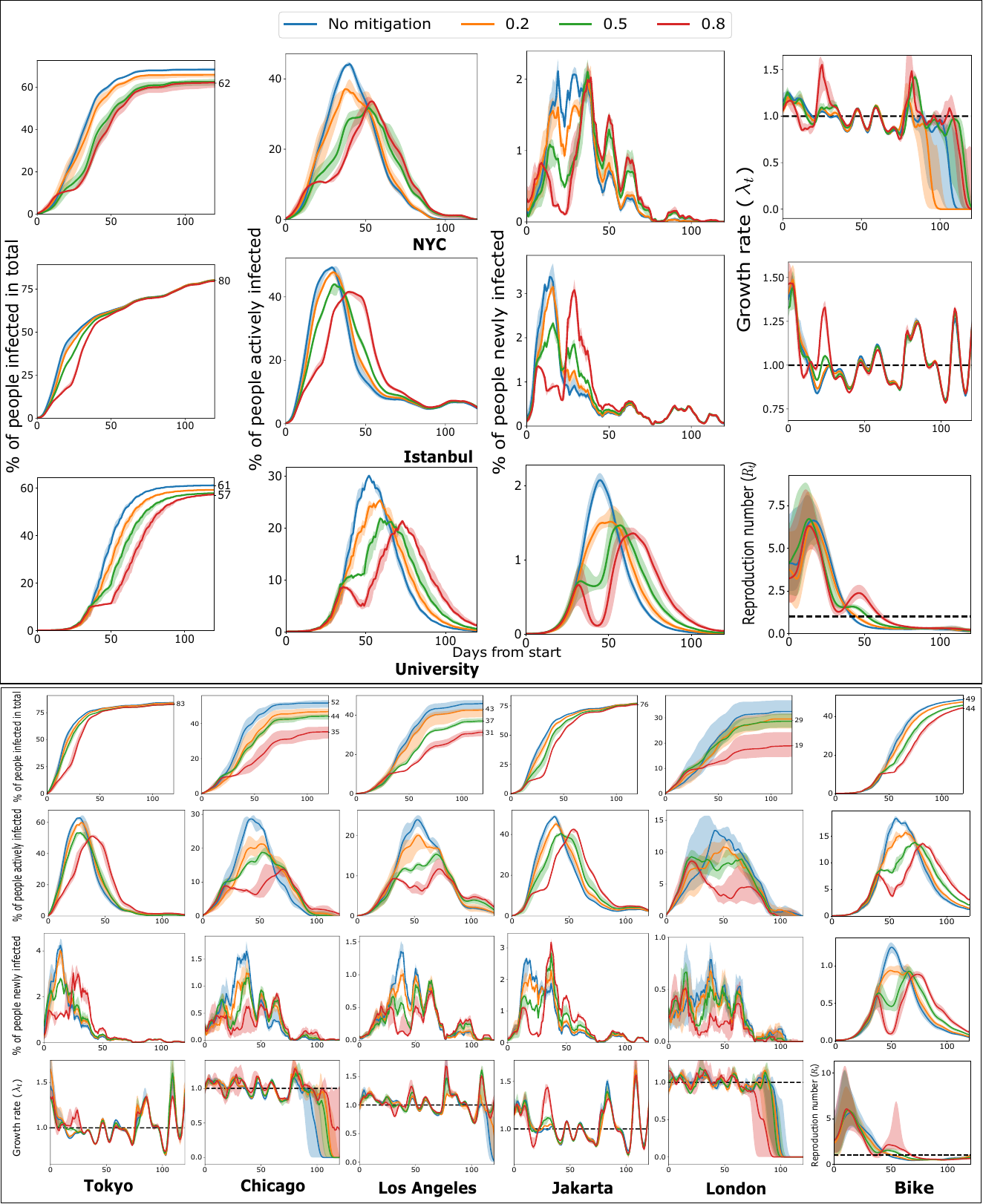}\\
    \caption{Infection spreading with the intervention strategy of varying probability, with which check-in is skipped with random sampling. All interventions start when $10\%$ of the population is infected and it lasts for $15$ days. While with higher probability (i.e., stronger intervention) the peak of active infections gets delayed and lowered, the total number of infected people is independent of the probability (in Istanbul, Tokyo, Jakarta, University, and Bike datasets.}
    \label{figS:probability_stay_home}
\end{figure}

\newpage
\begin{figure}[!hp]
    \centering
    \includegraphics[width=0.9\textwidth]{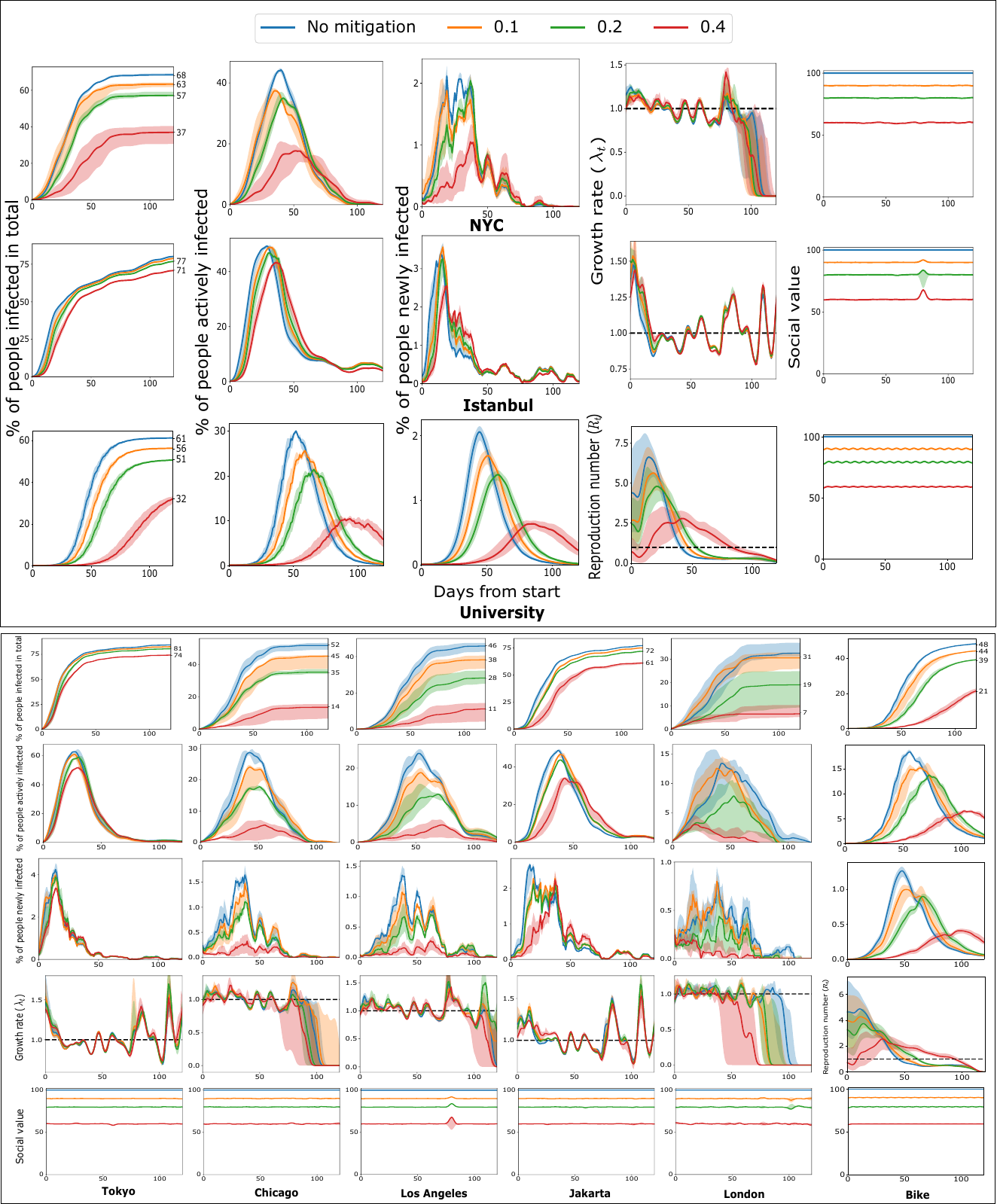}
    \caption{Infection spreading with the intervention strategy of Stay home intervention. This intervention strategy uniformly randomly ignores a check-in with a probability. This represents people reducing their social activities. Naturally, as more check-ins are skipped, the infection is slowed, and the social value is also reduced.}
    \label{figS:stay_home}
\end{figure}

\newpage
\begin{figure}[H]
\centering
\begin{tabular}{m{0.2cm}@{}@{}m{\y\textwidth}@{}@{}m{\y\textwidth}@{}@{}m{\y\textwidth}@{}@{}m{\y\textwidth}@{}@{}m{\y\textwidth}@{}@{}m{\y\textwidth}@{}@{}m{\y\textwidth}@{}}

\rotatebox{90}{\tiny $\%$ of people infected in total} &
\includegraphics[width=\y\textwidth]{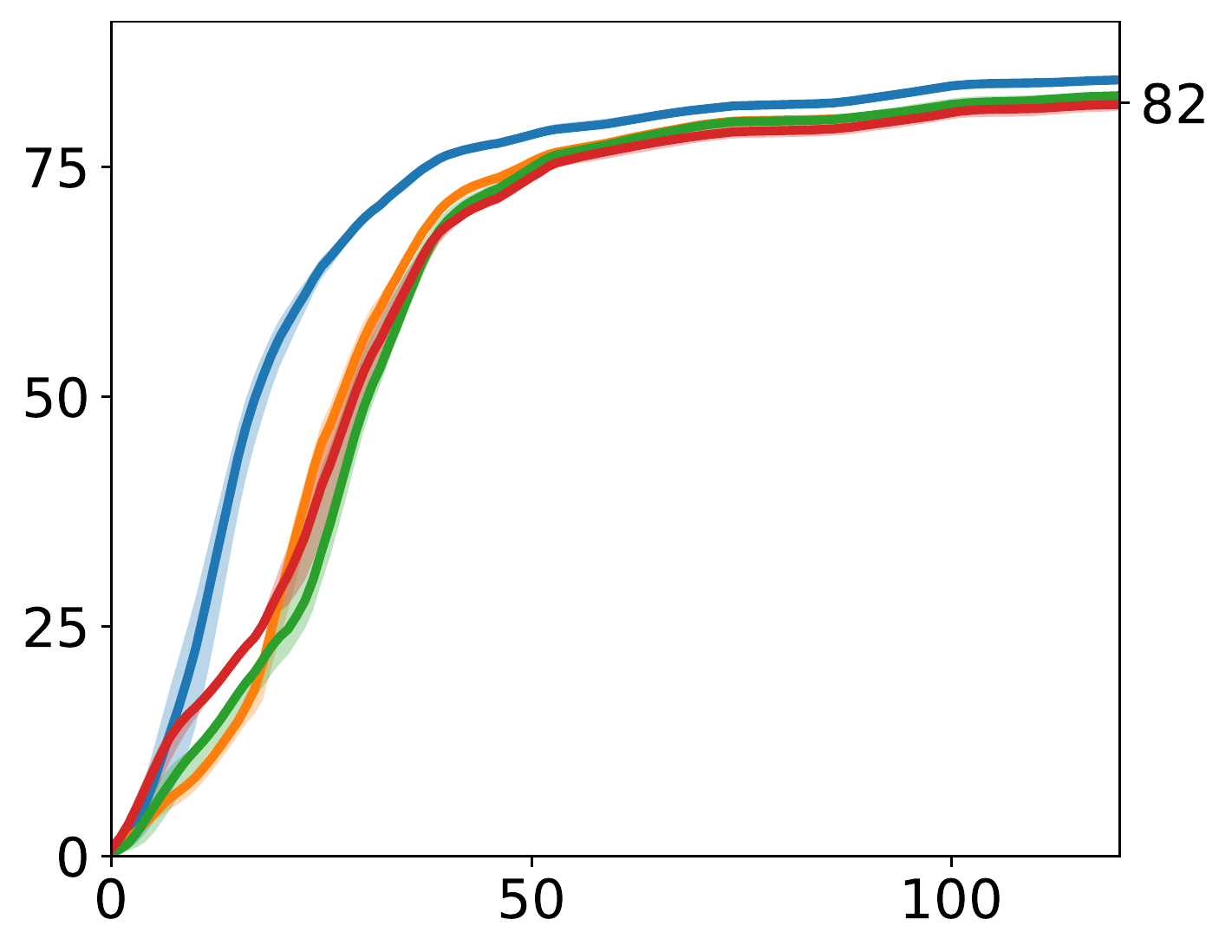}  &
\includegraphics[width=\y\textwidth]{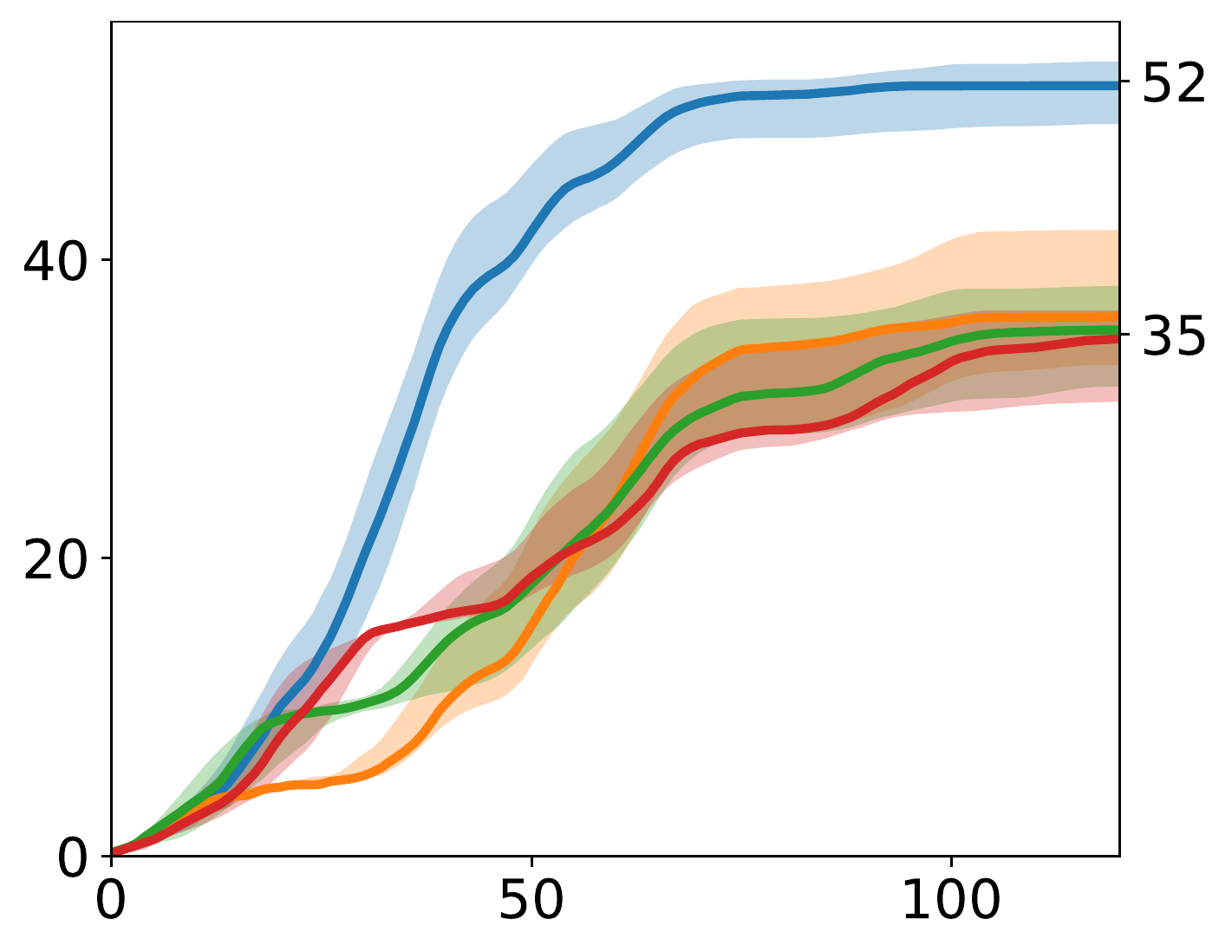}  &
\includegraphics[width=\y\textwidth]{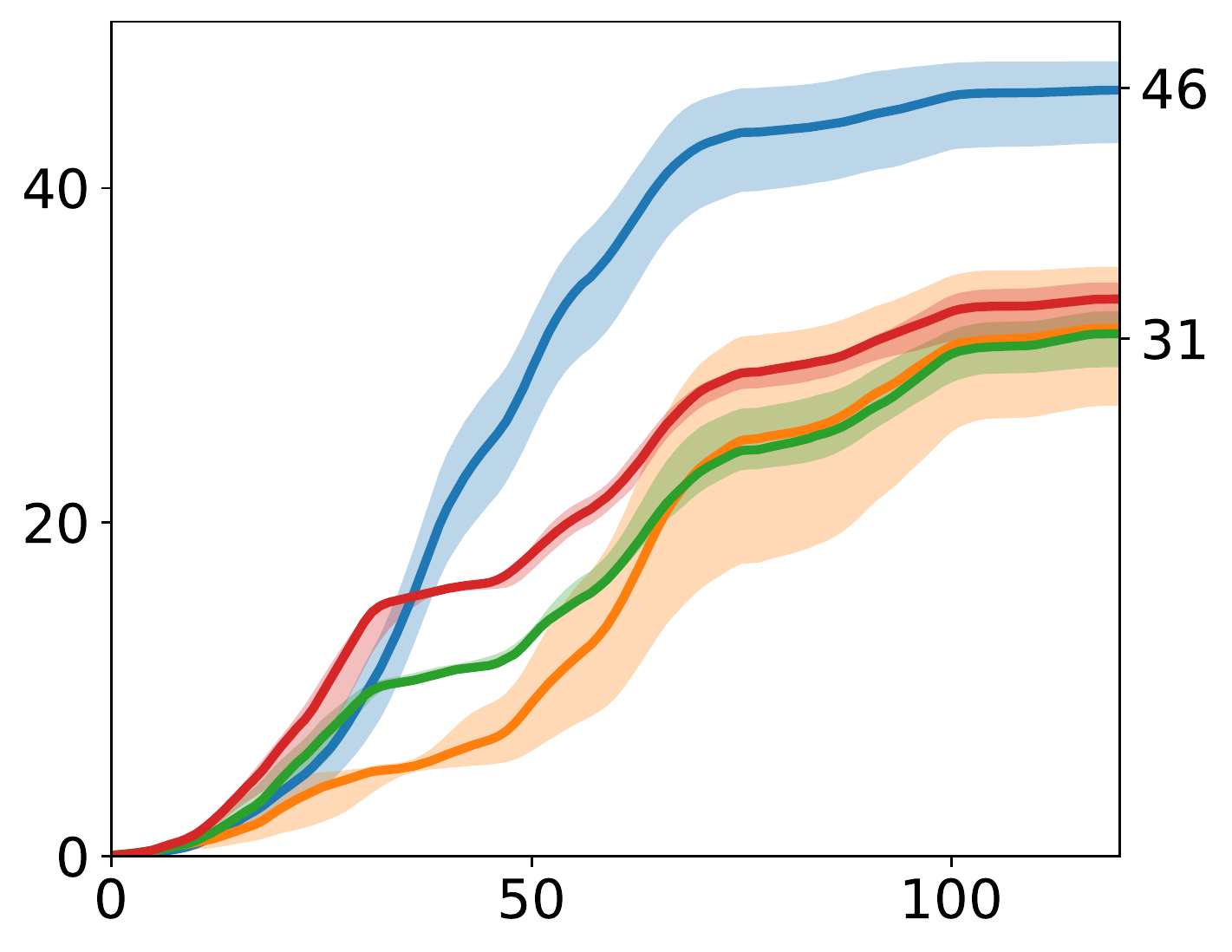}  &
\includegraphics[width=\y\textwidth]{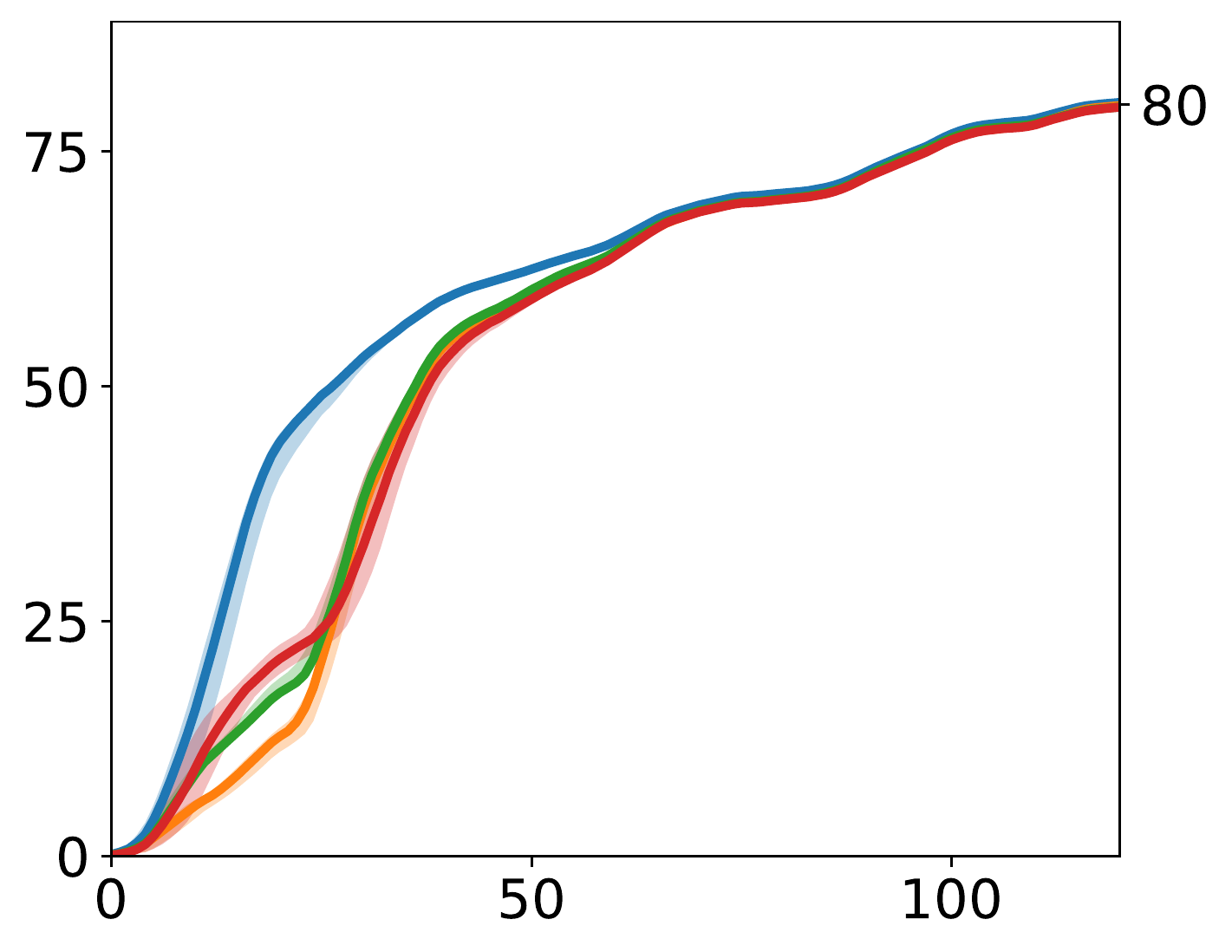}  &
\includegraphics[width=\y\textwidth]{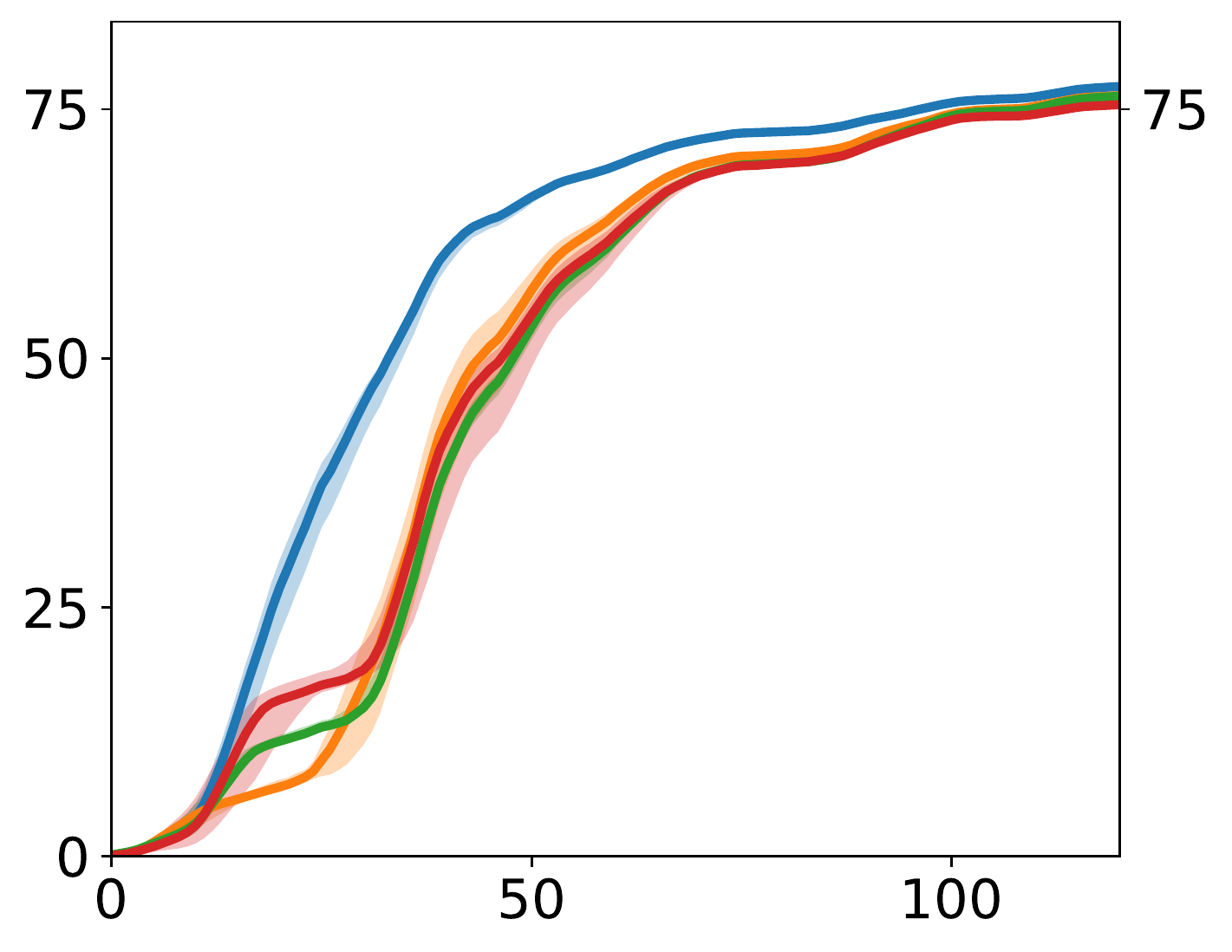}  &
\includegraphics[width=\y\textwidth]{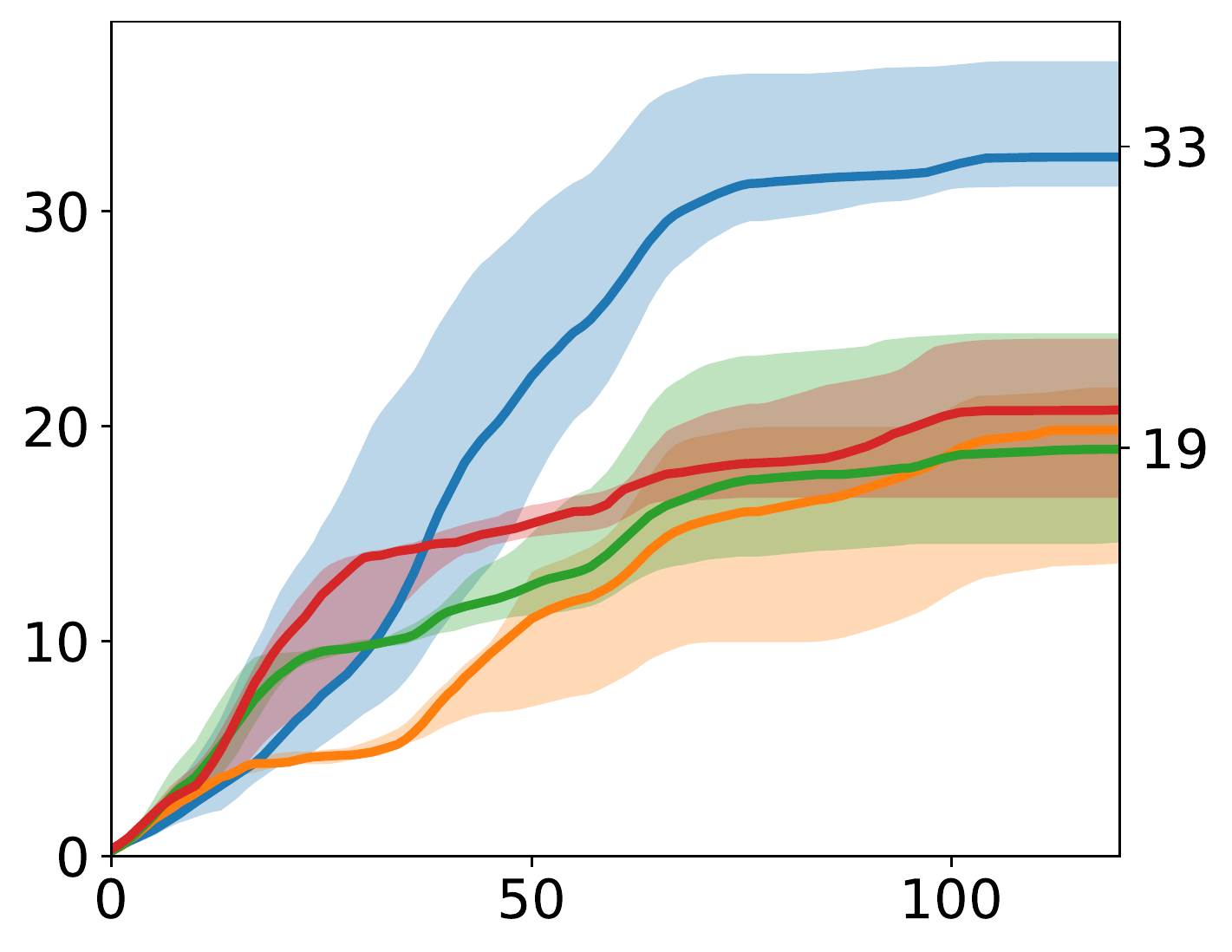} &
\includegraphics[width=\y\textwidth]{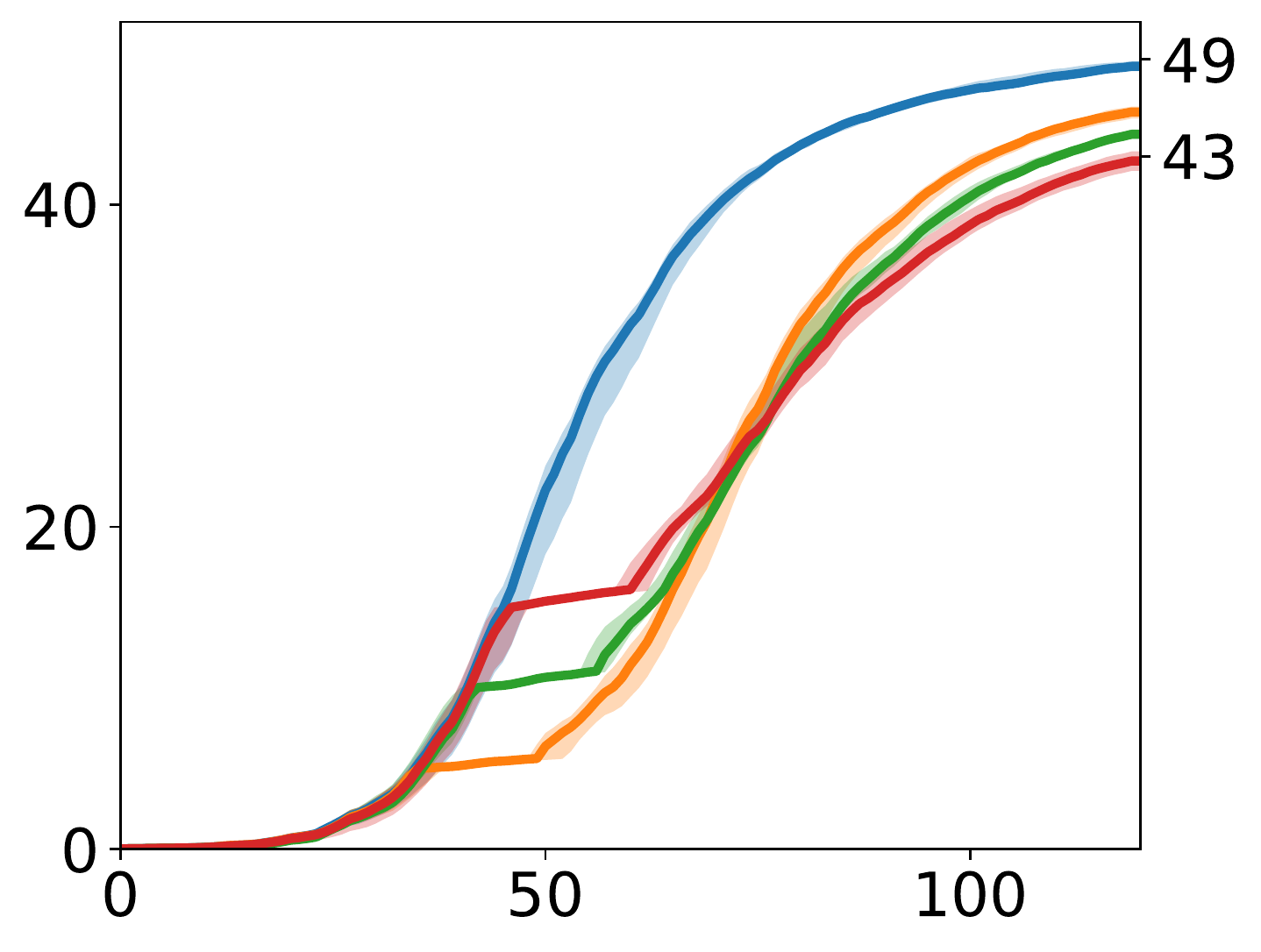}
\\

\rotatebox{90}{\tiny $\%$ of people actively infected} &
\includegraphics[width=\y\textwidth]{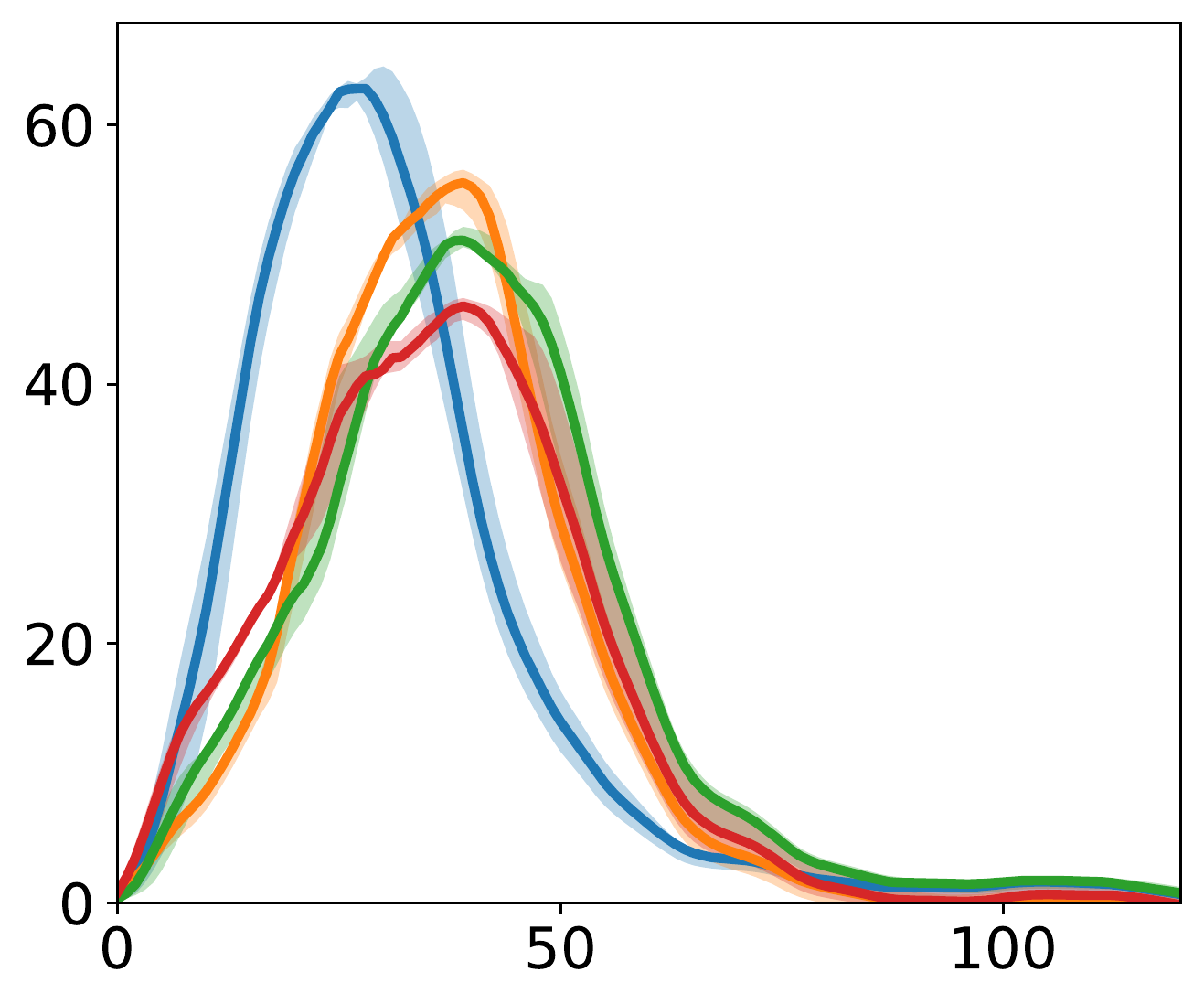}  &
\includegraphics[width=\y\textwidth]{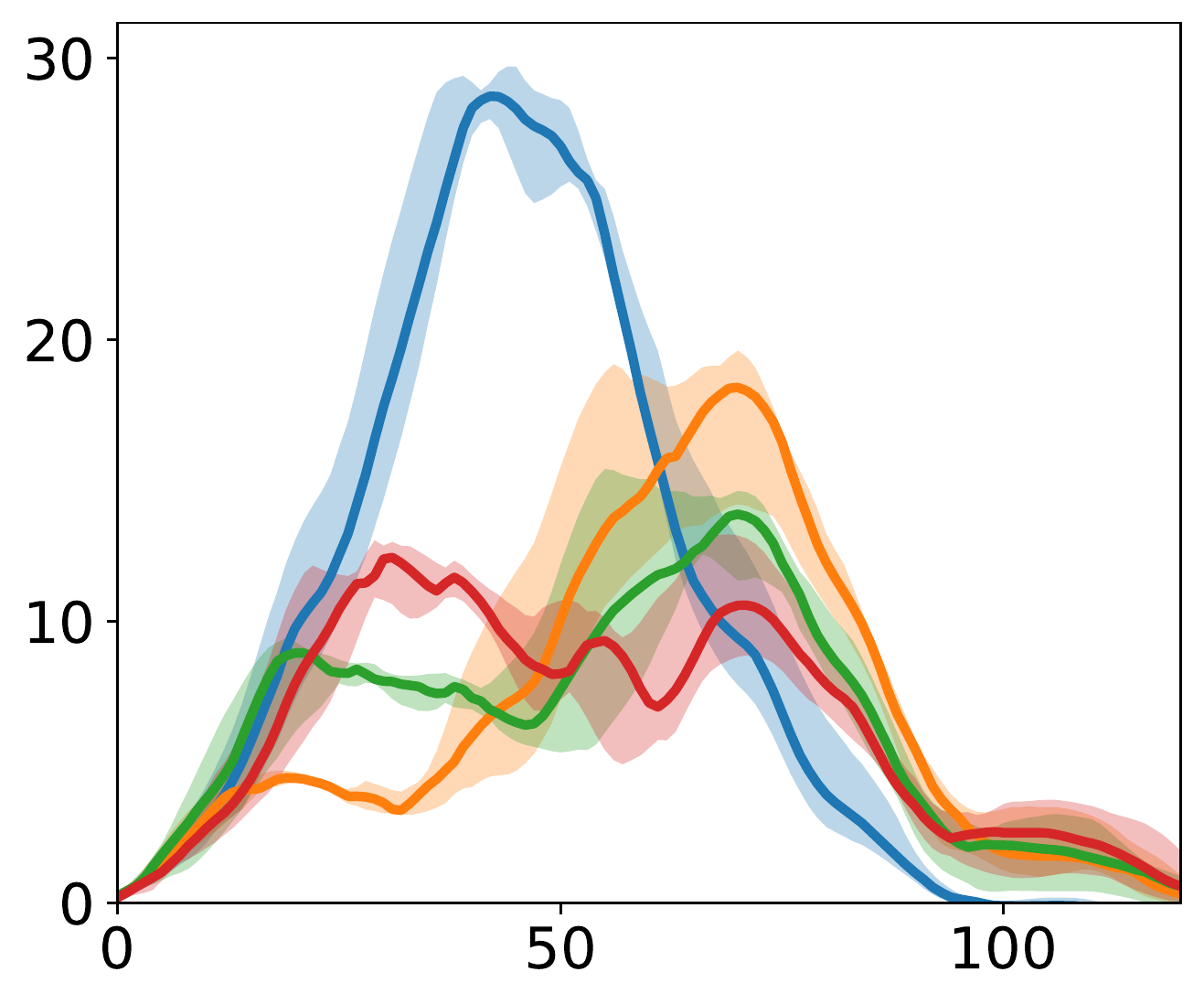}  &
\includegraphics[width=\y\textwidth]{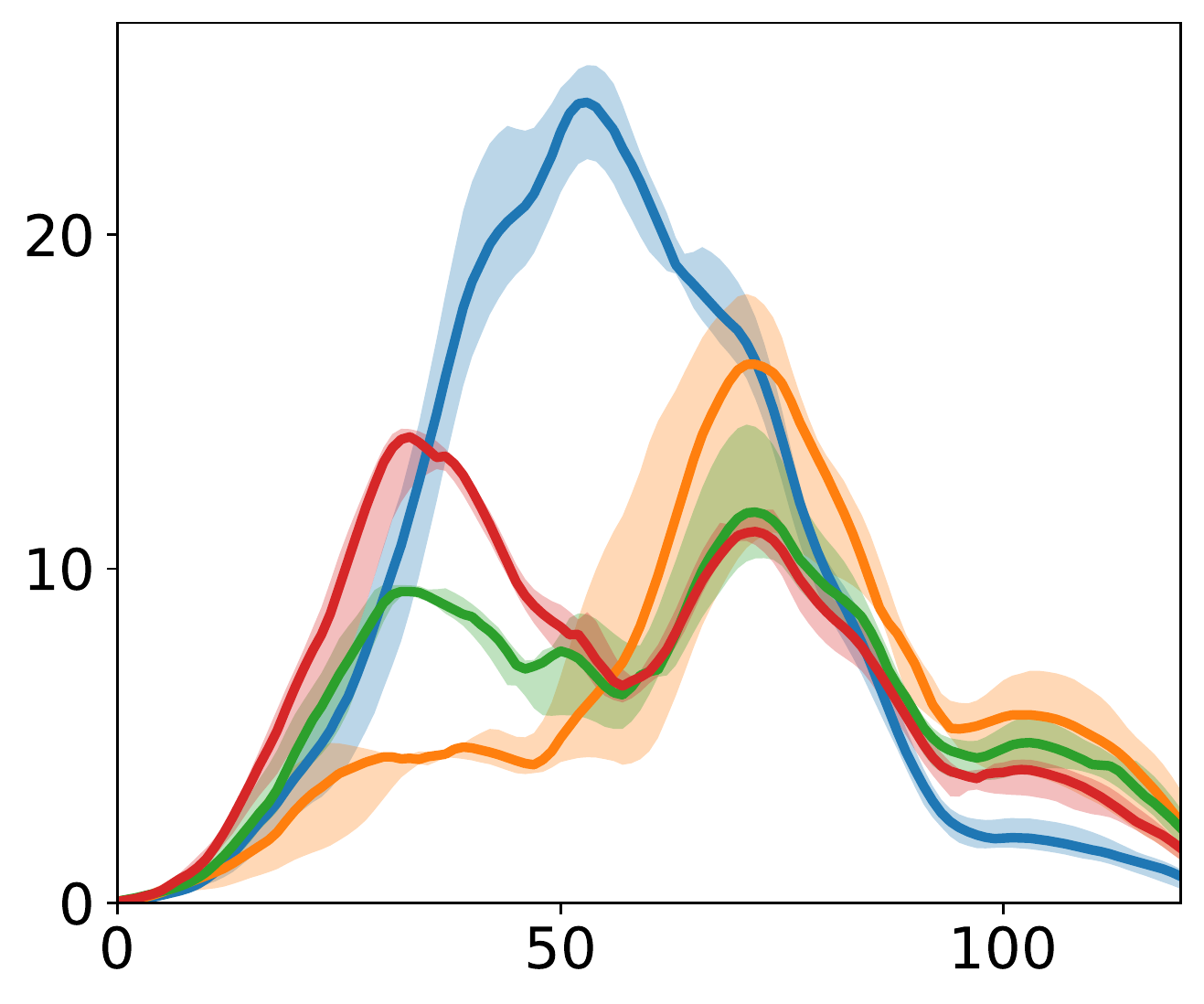}  &
\includegraphics[width=\y\textwidth]{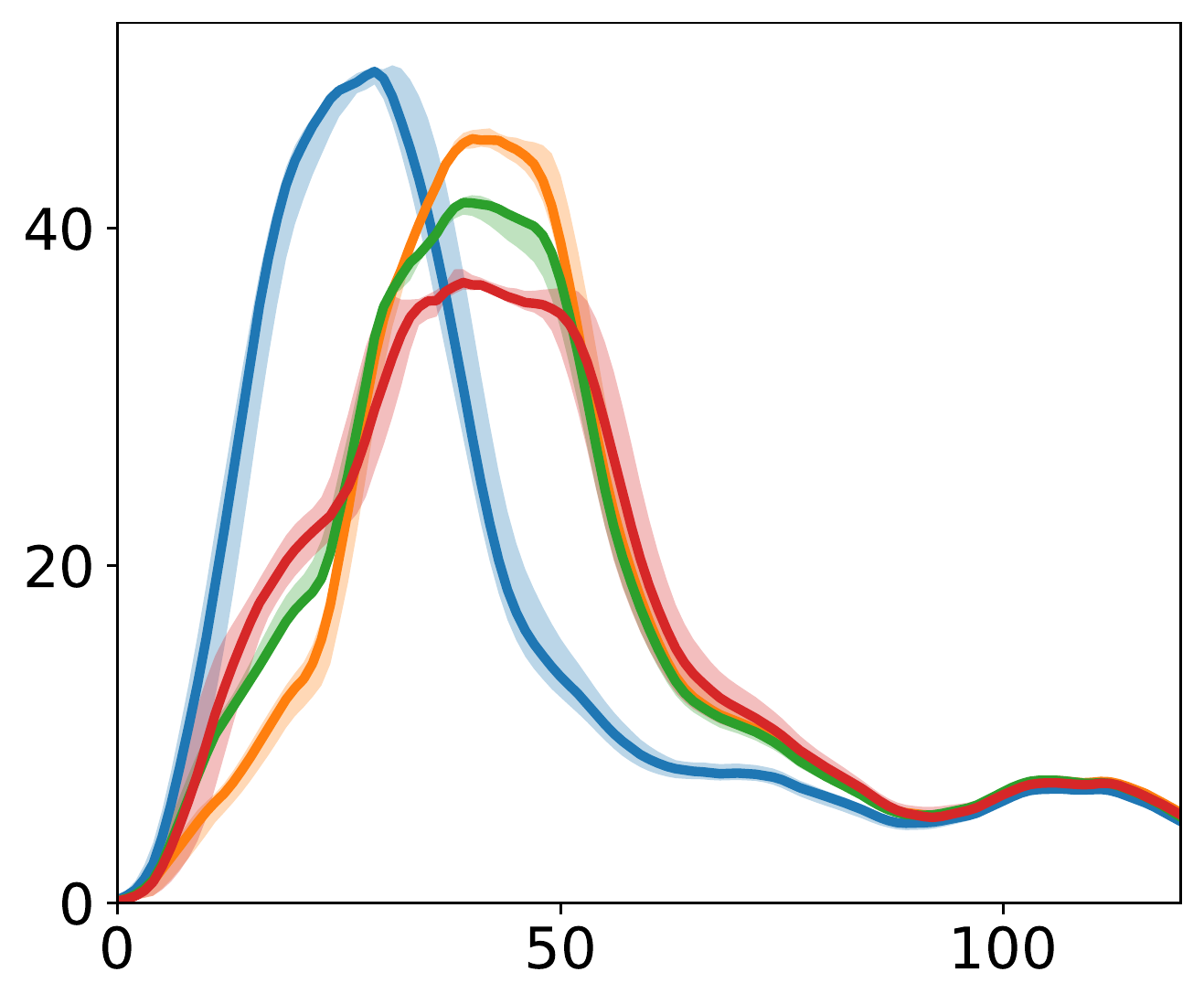}  &
\includegraphics[width=\y\textwidth]{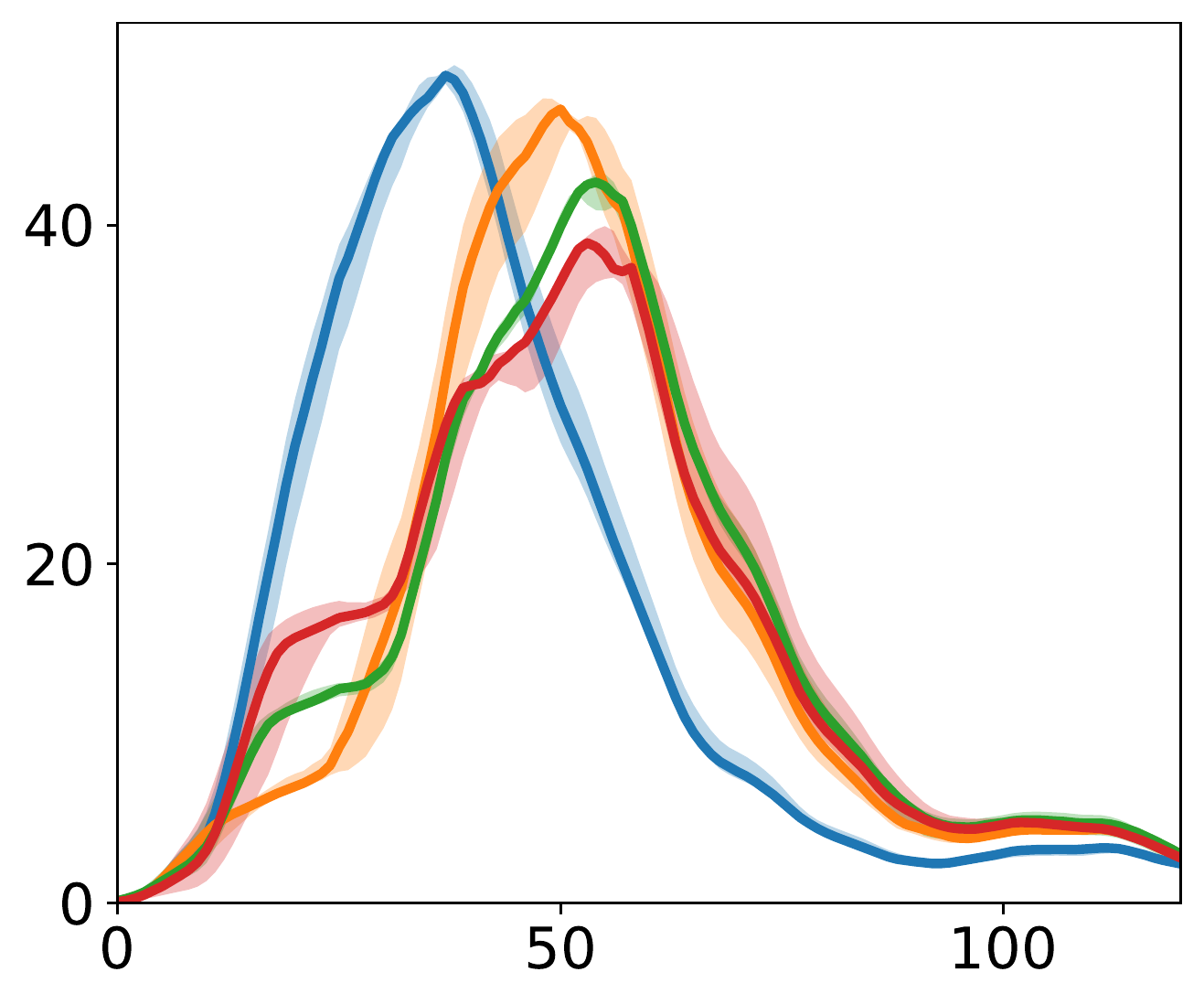}  &
\includegraphics[width=\y\textwidth]{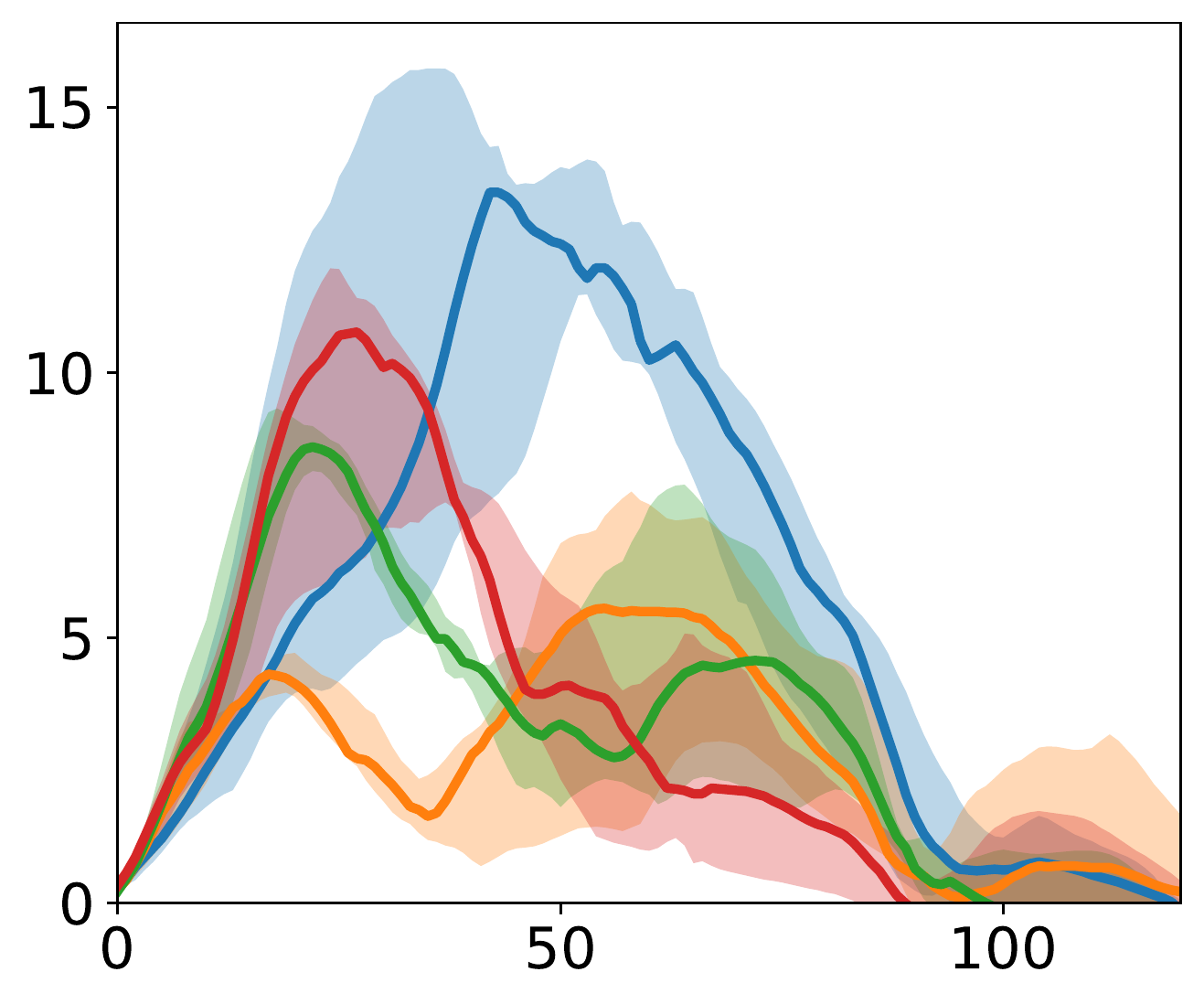} &
\includegraphics[width=\y\textwidth]{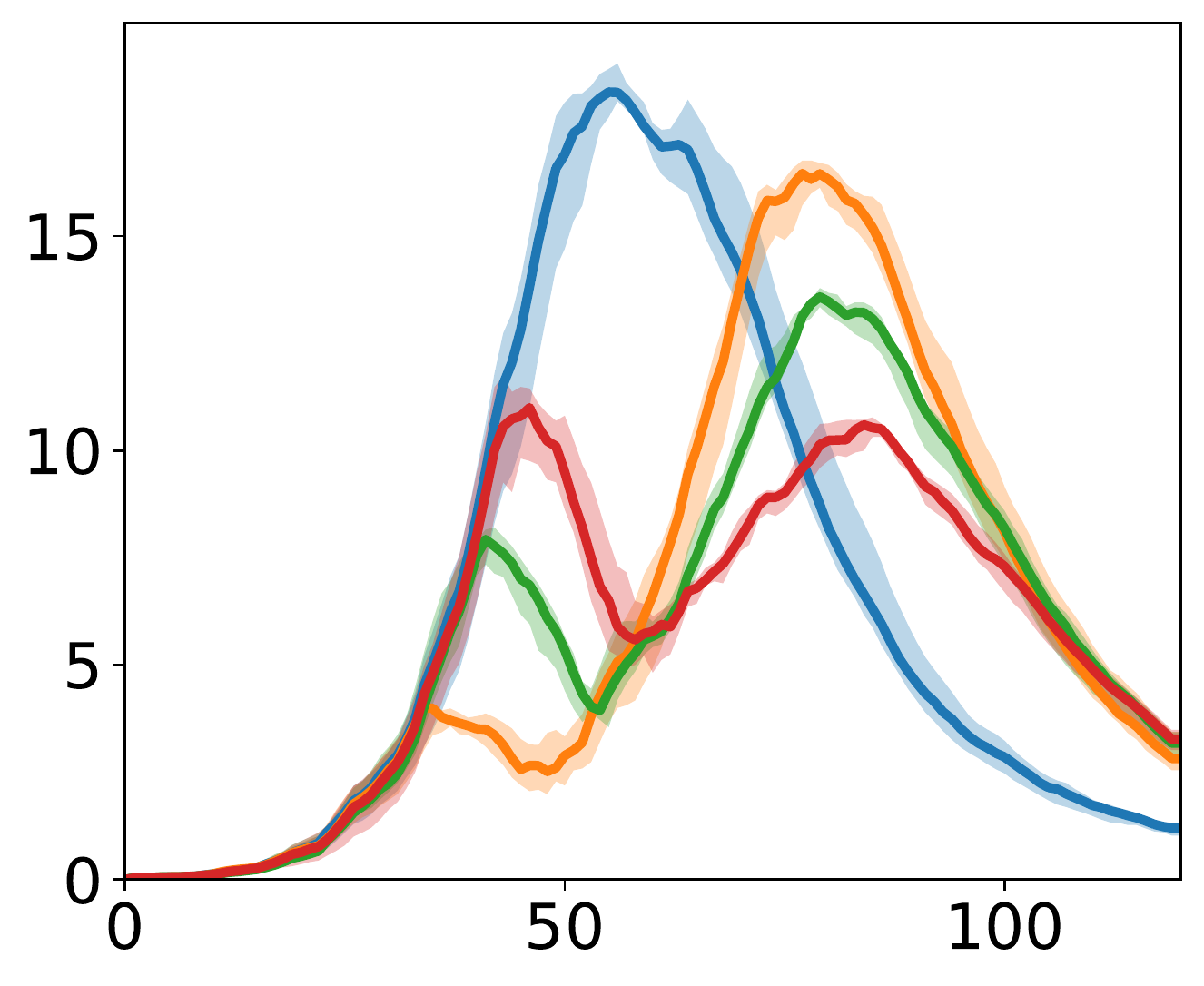}
\\

\rotatebox{90}{\tiny $\%$ of people newly infected} &
\includegraphics[width=\y\textwidth]{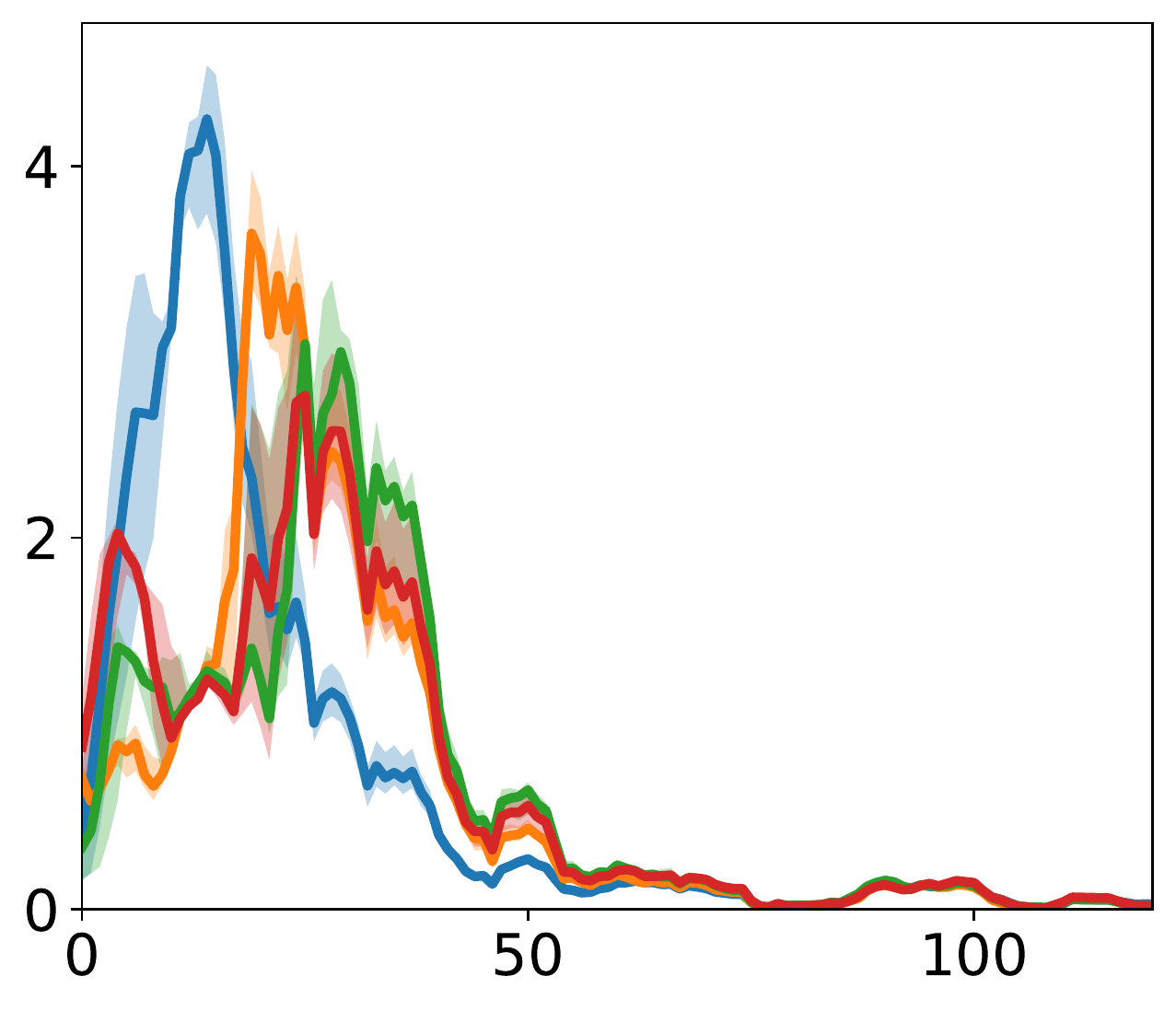}  &
\includegraphics[width=\y\textwidth]{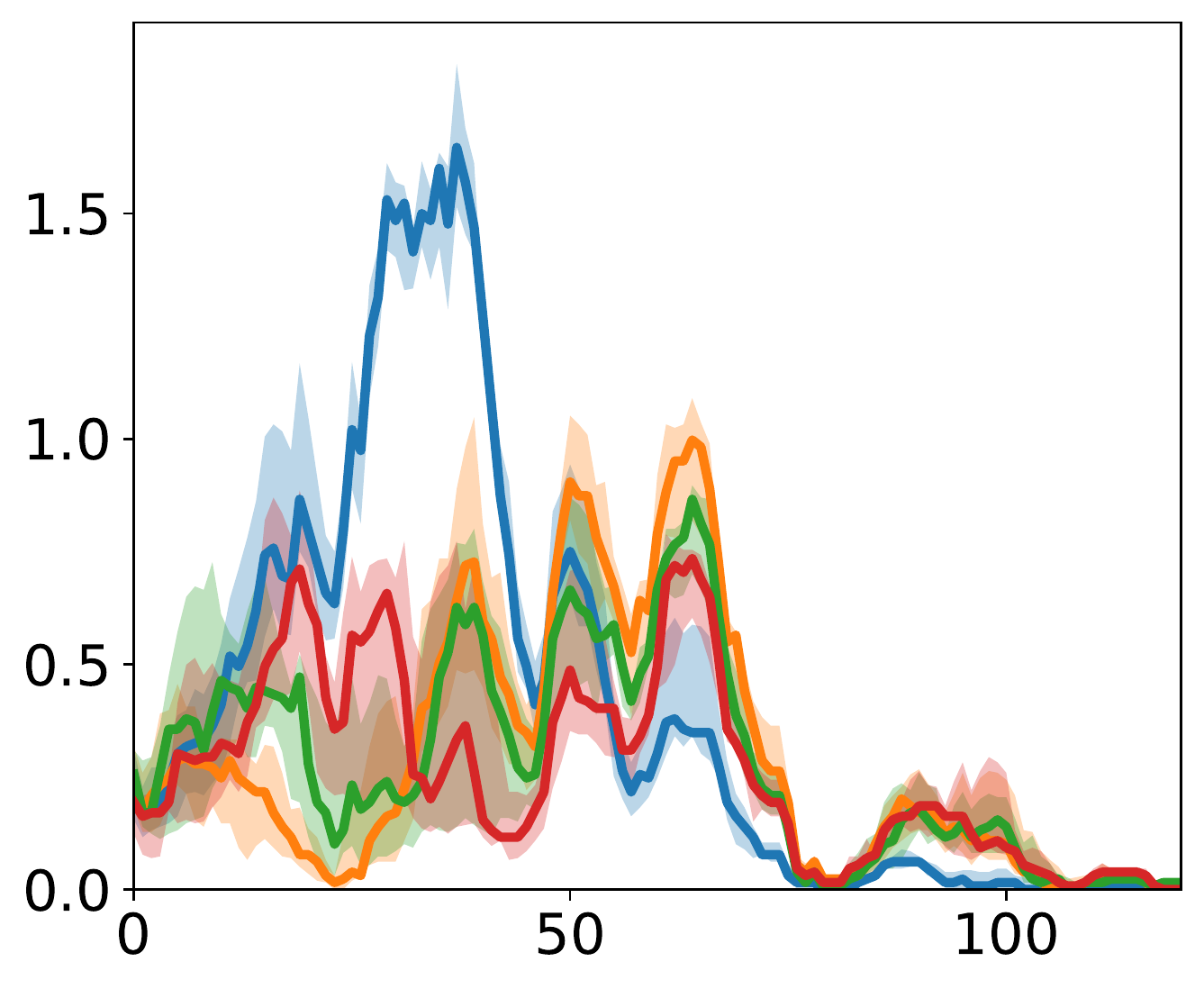}  &
\includegraphics[width=\y\textwidth]{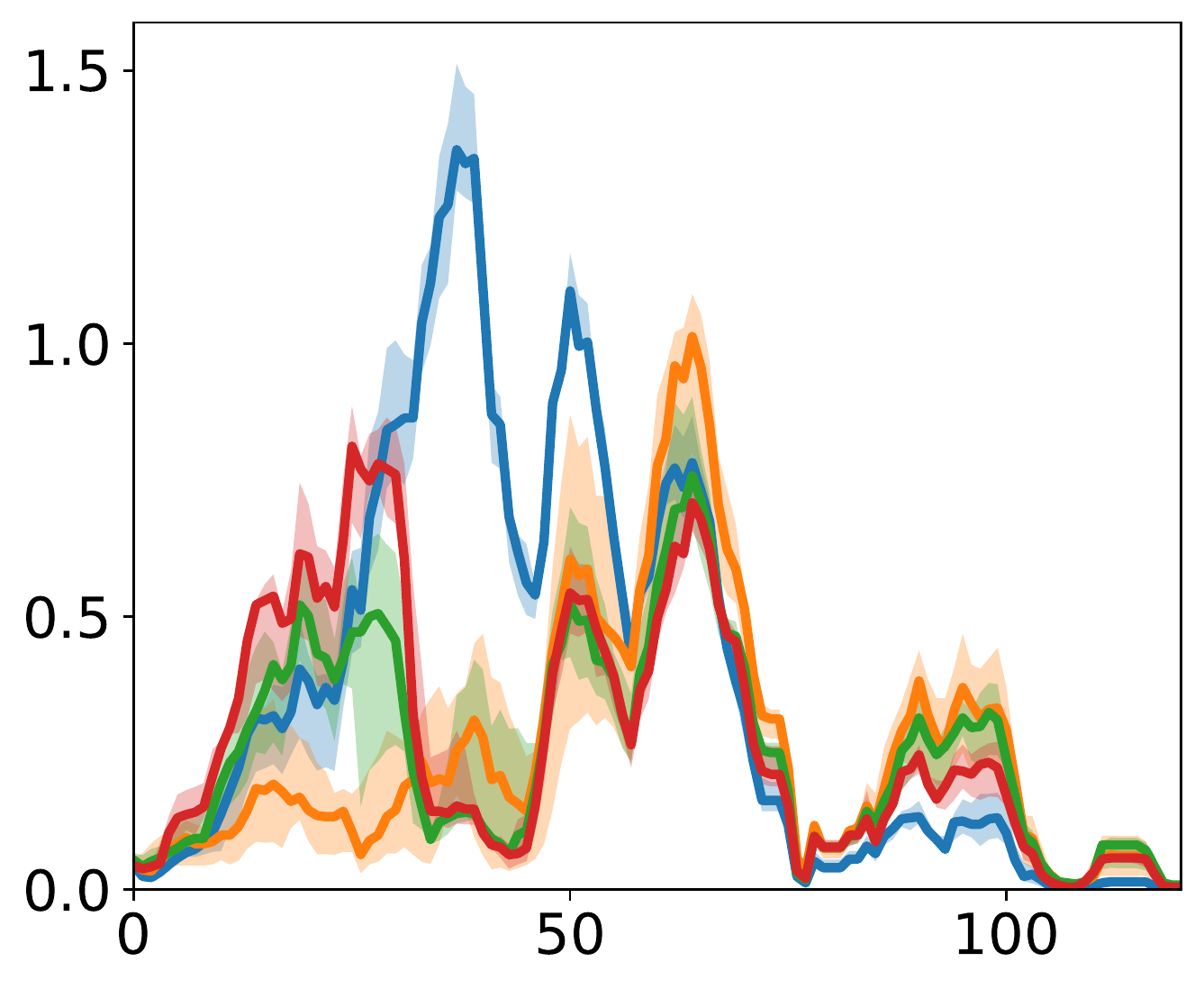}  &
\includegraphics[width=\y\textwidth]{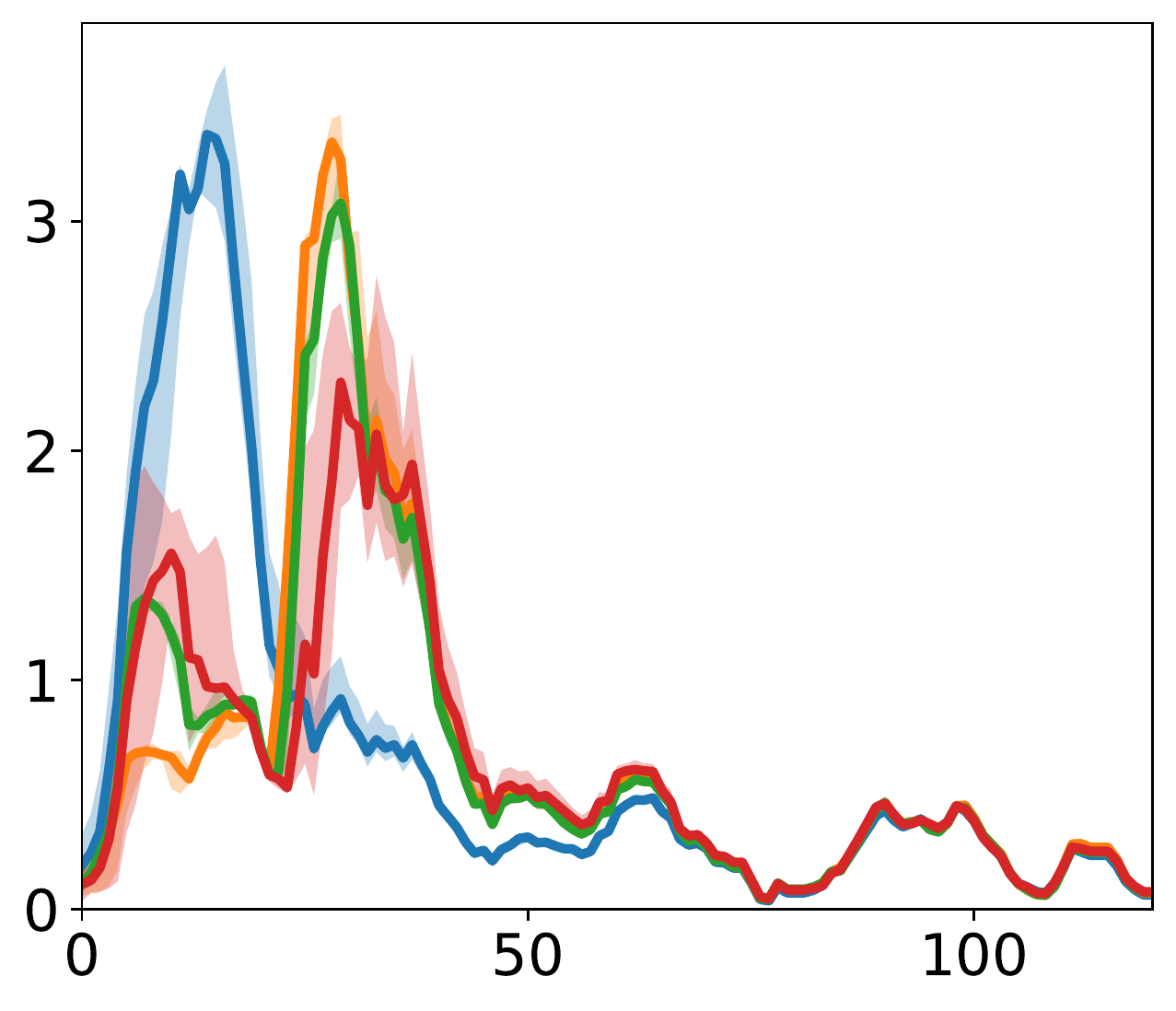}  &
\includegraphics[width=\y\textwidth]{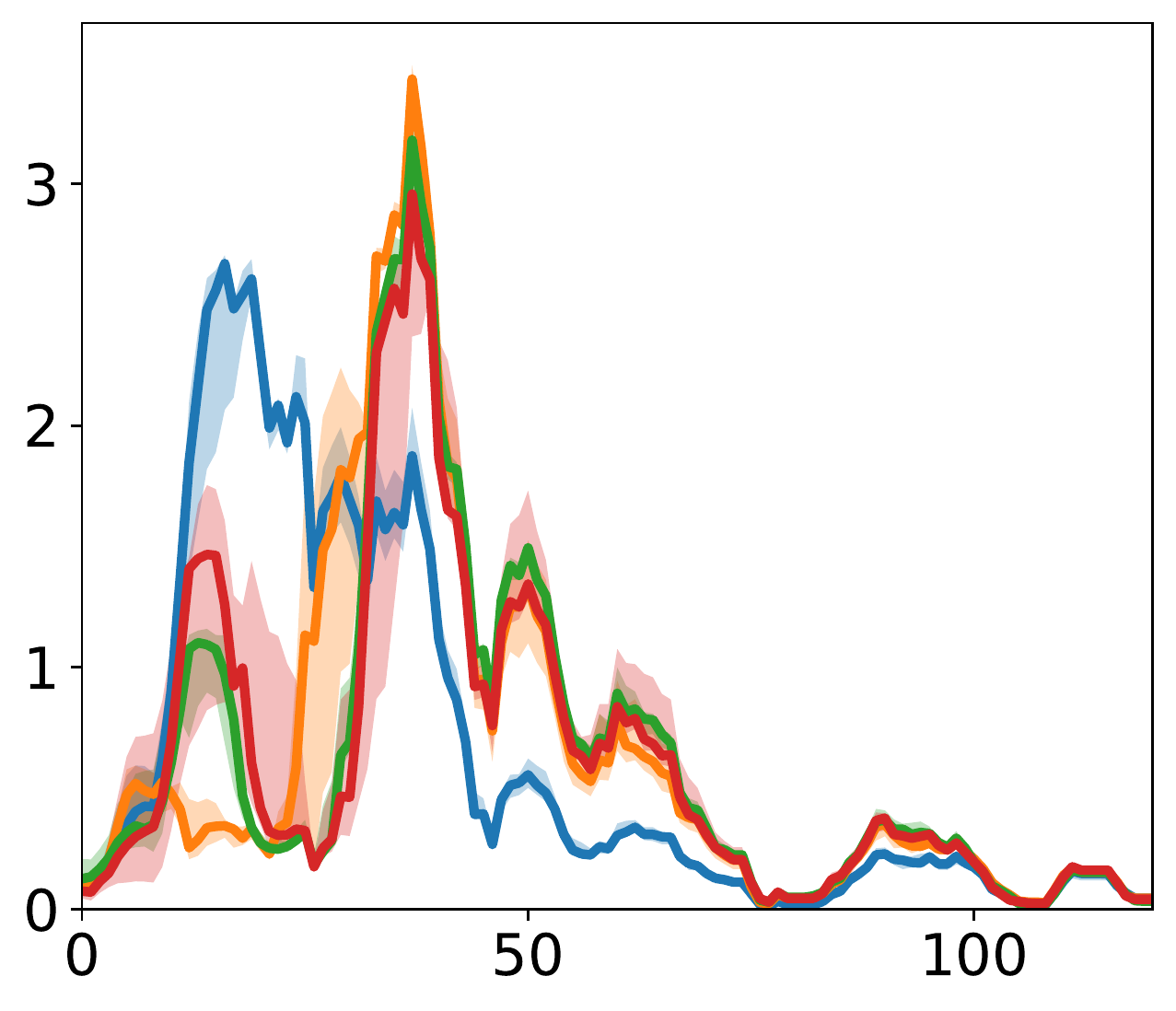}  &
\includegraphics[width=\y\textwidth]{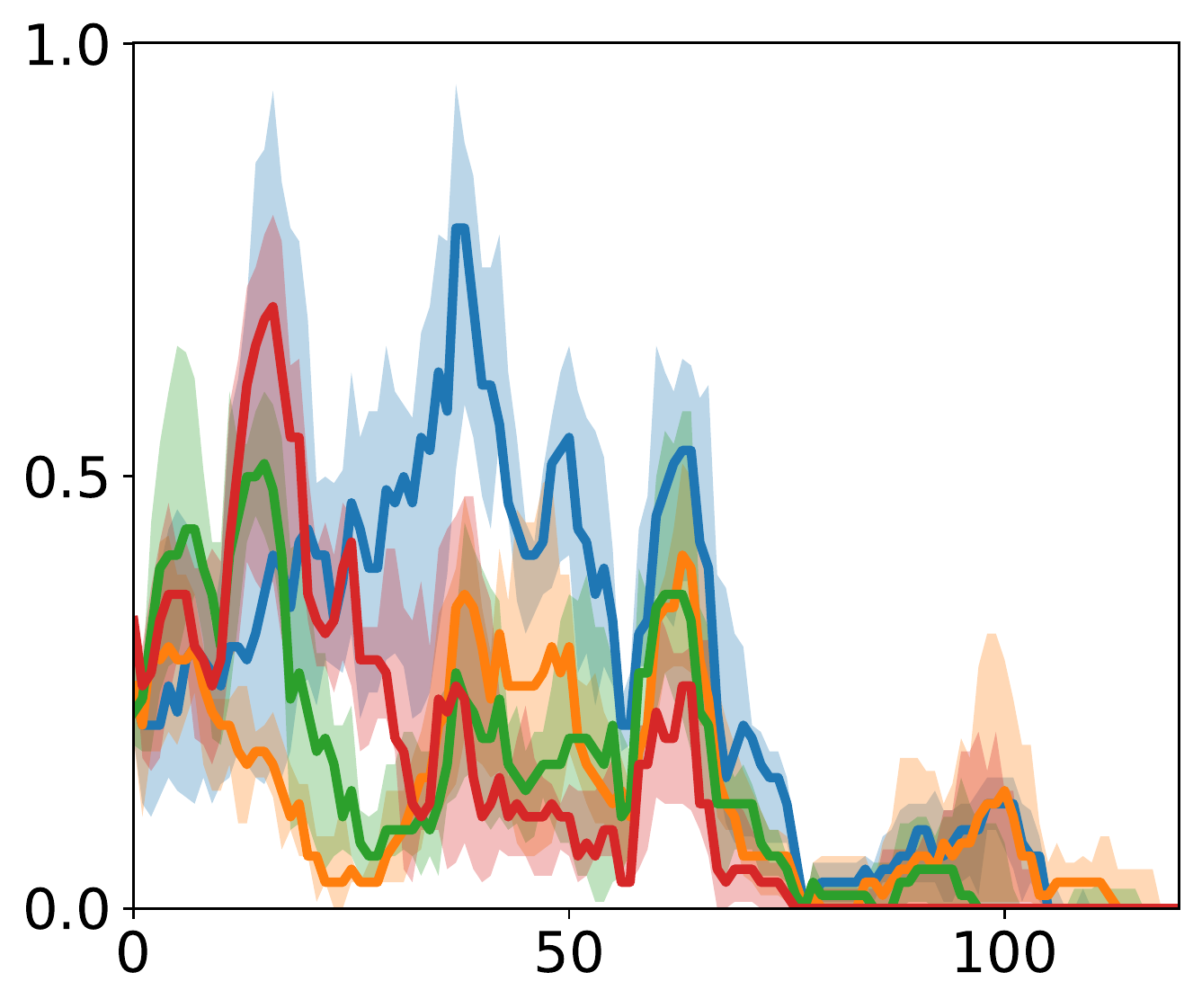} &
\includegraphics[width=\y\textwidth]{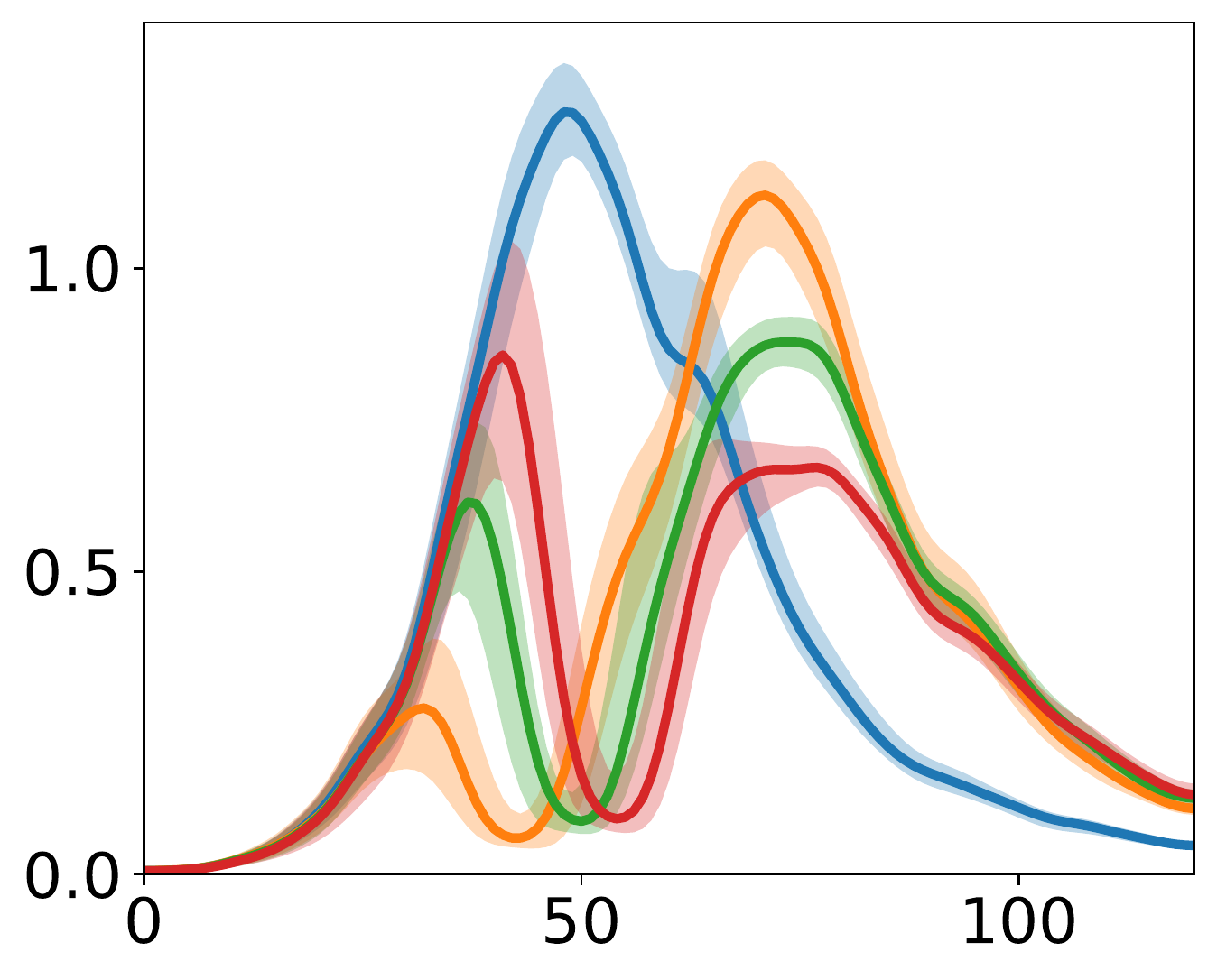}
\\

\rotatebox{90}{\tiny Growth rate ($\lambda_t$)} &
\includegraphics[width=\y\textwidth]{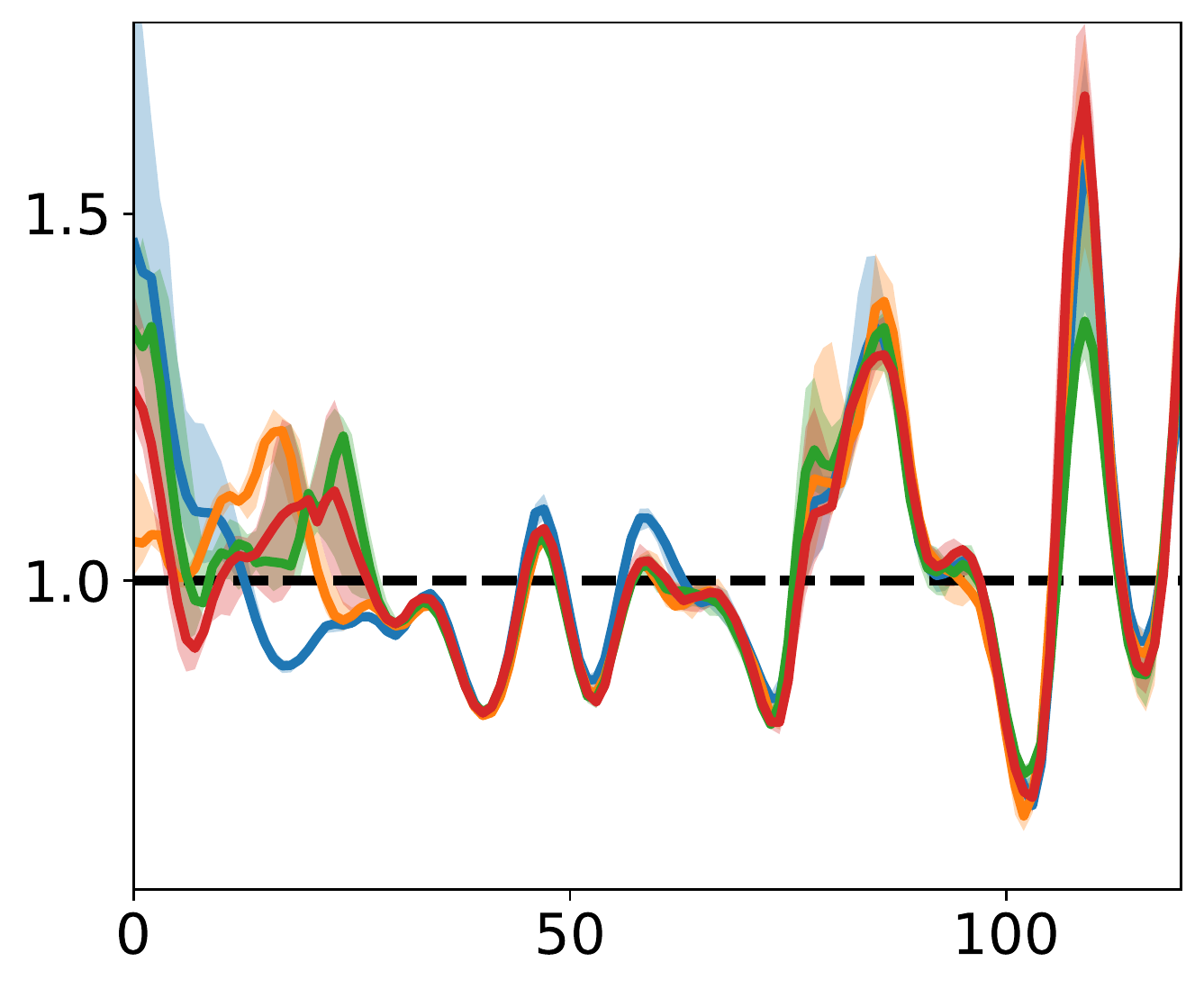}  &
\includegraphics[width=\y\textwidth]{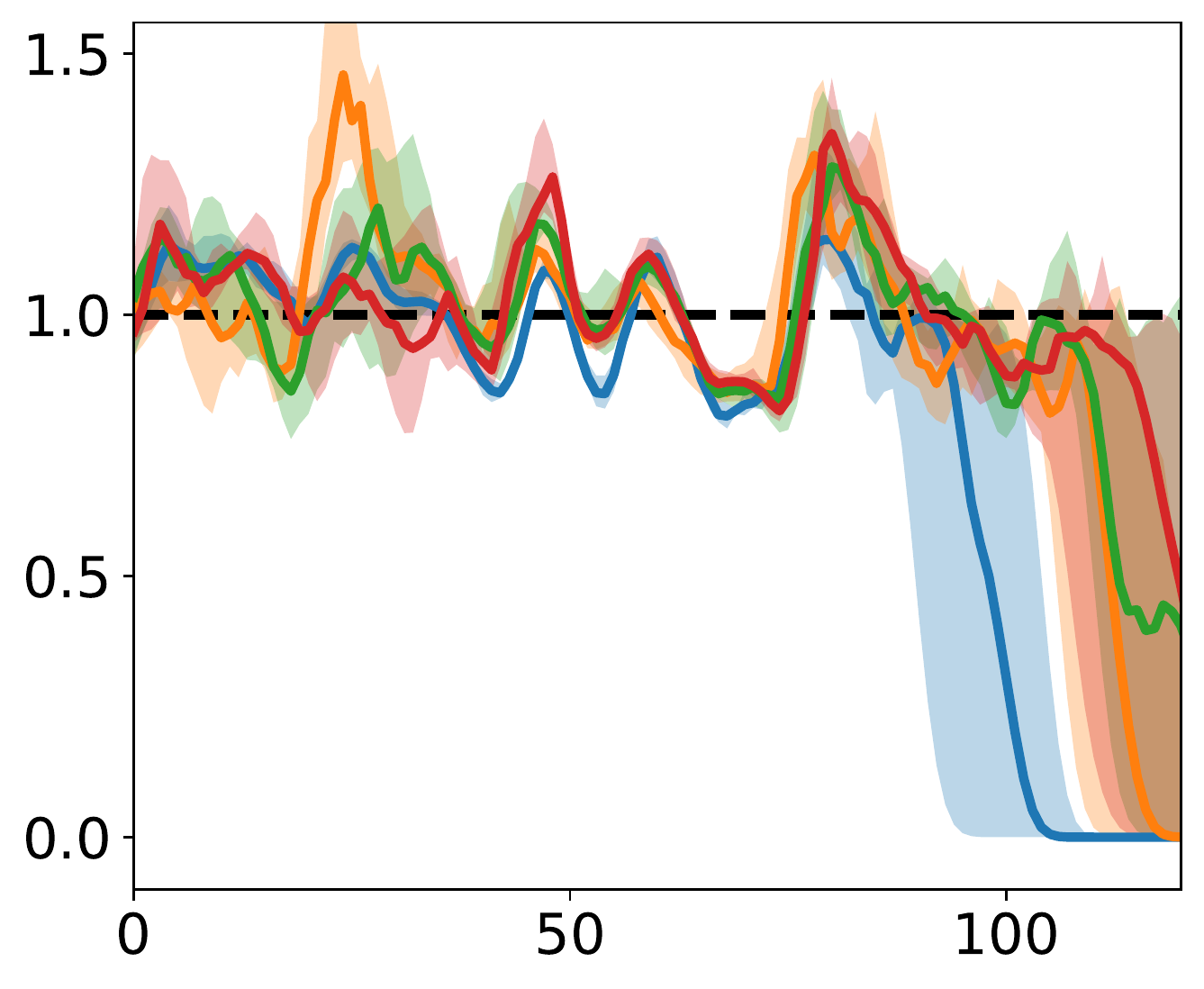}  &
\includegraphics[width=\y\textwidth]{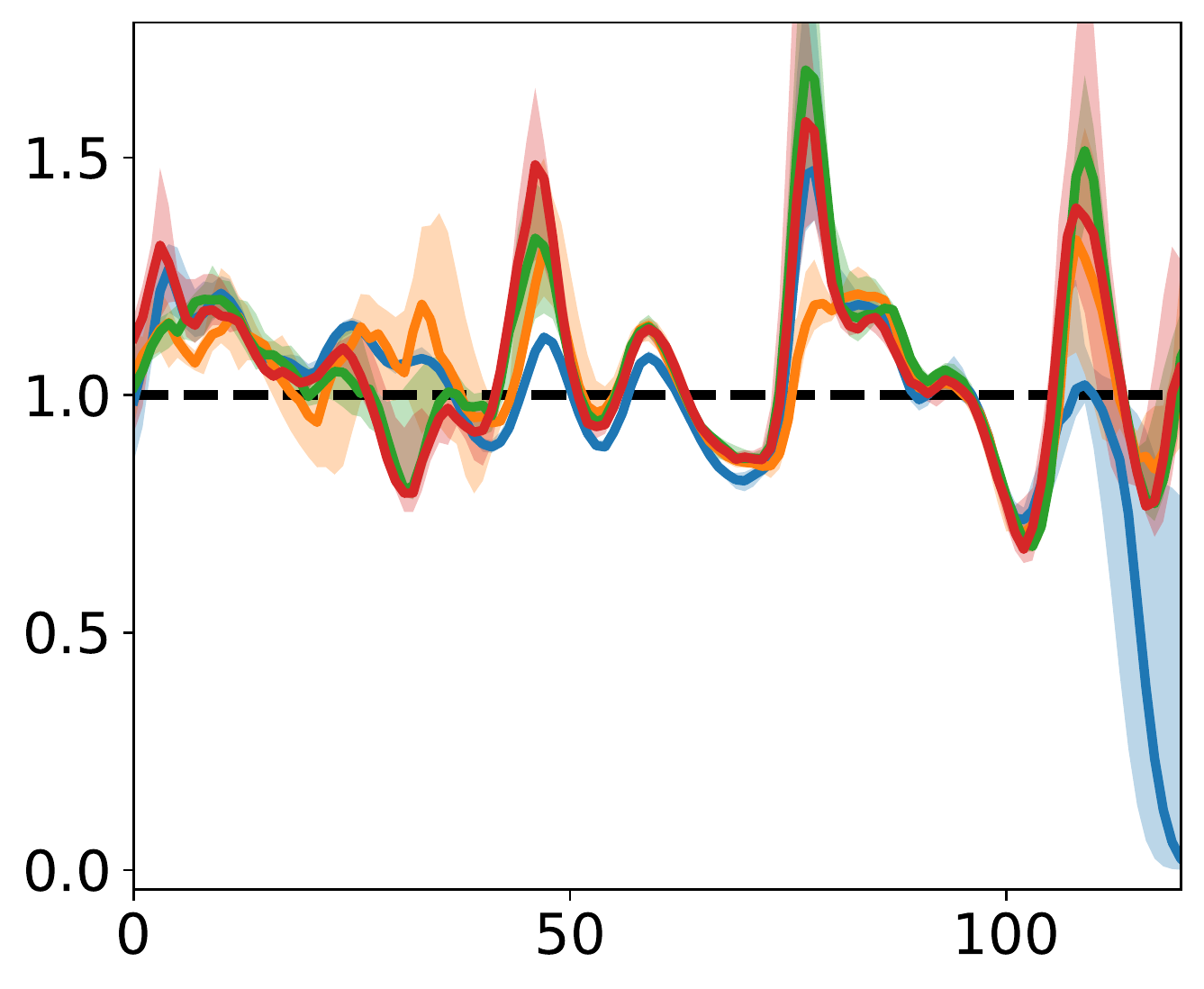}  &
\includegraphics[width=\y\textwidth]{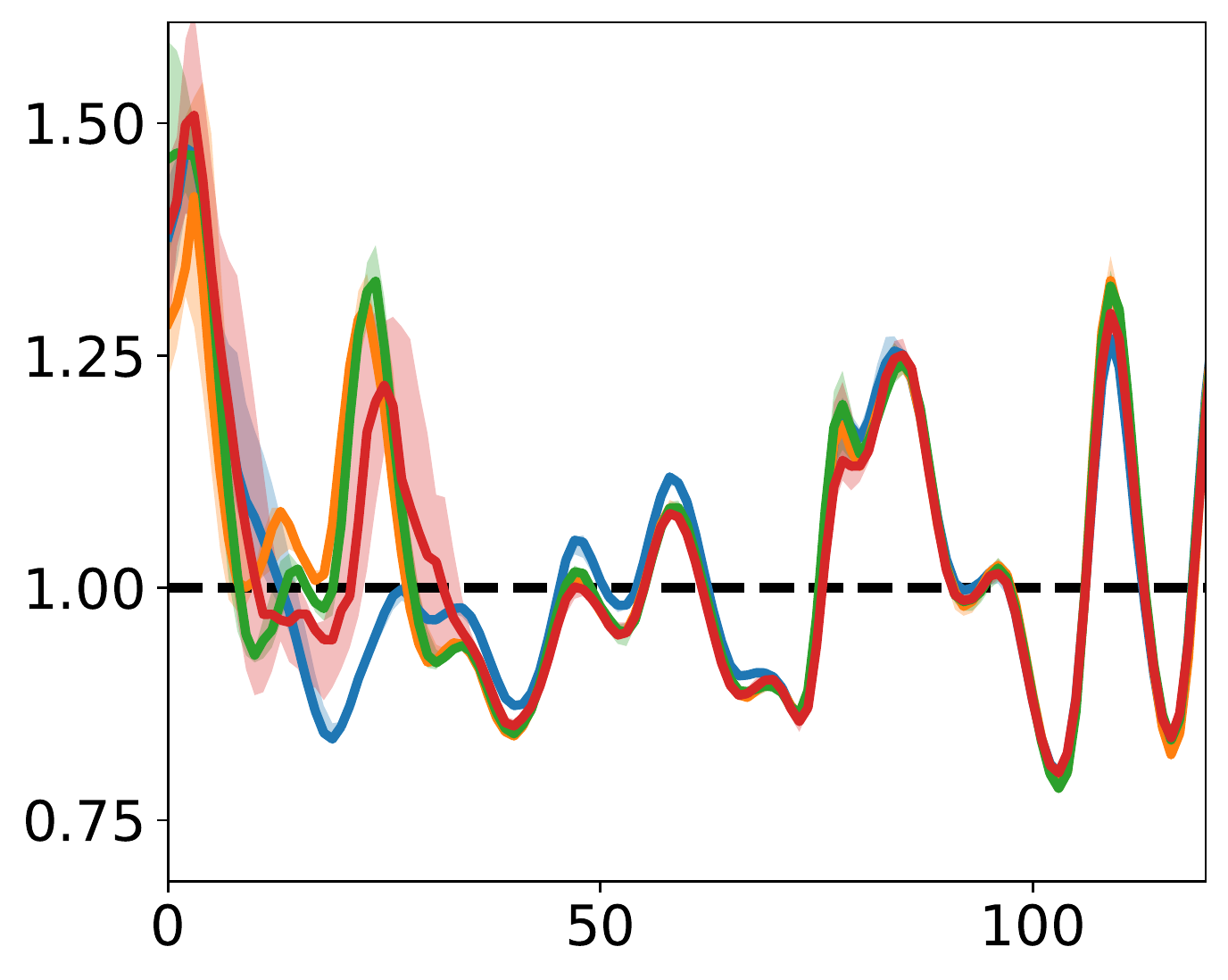}  &
\includegraphics[width=\y\textwidth]{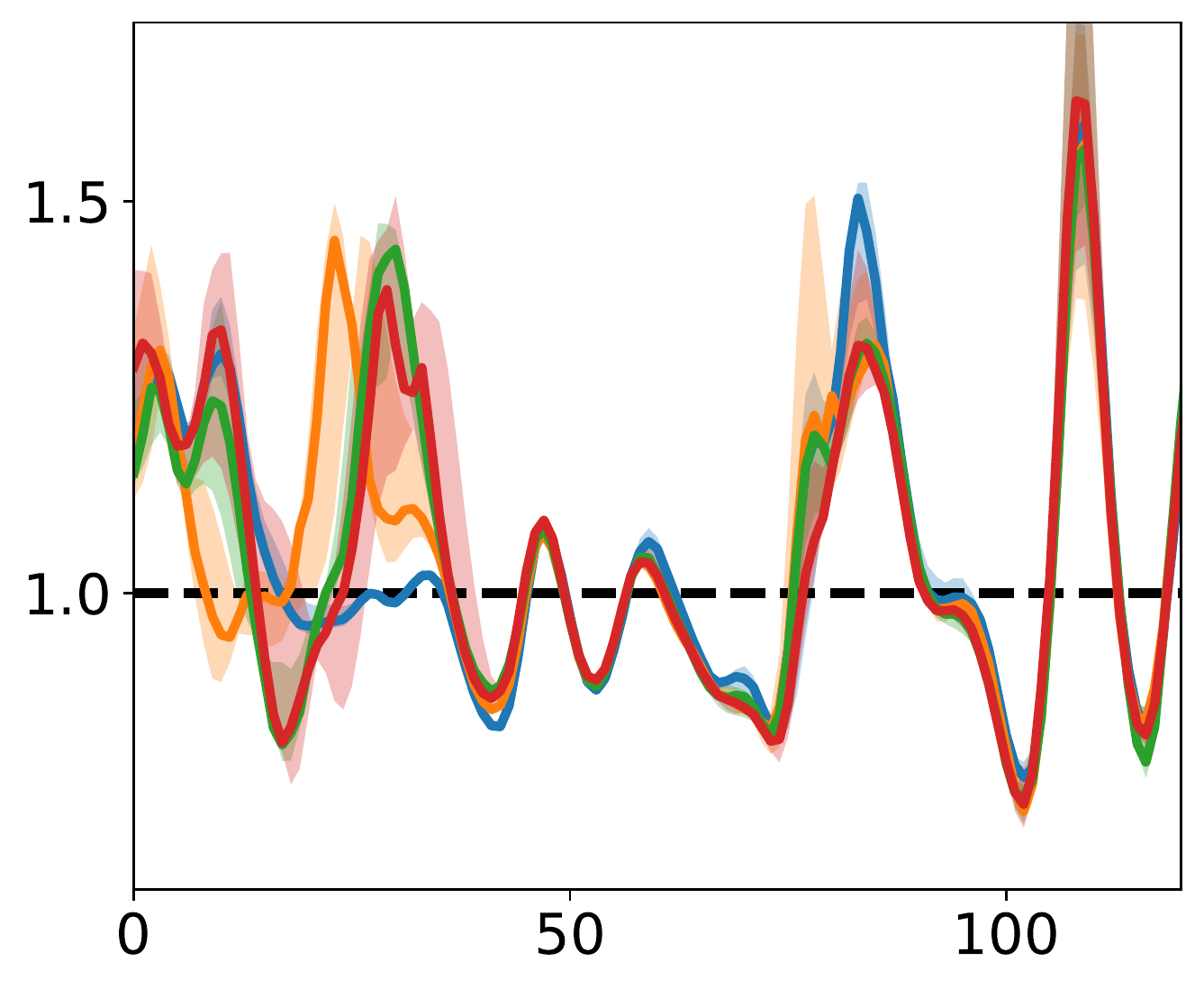}  &
\includegraphics[width=\y\textwidth]{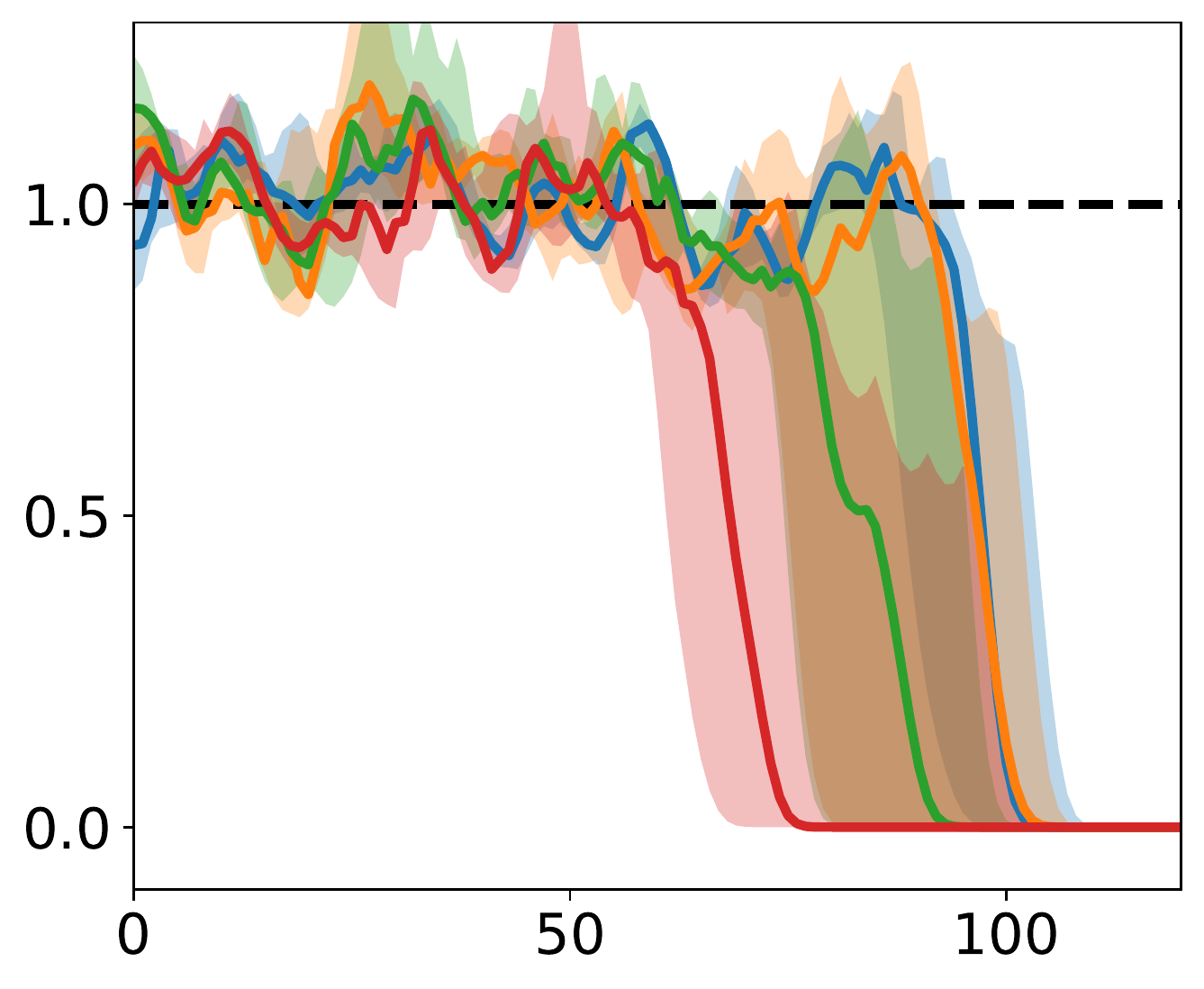} &
\includegraphics[width=\y\textwidth]{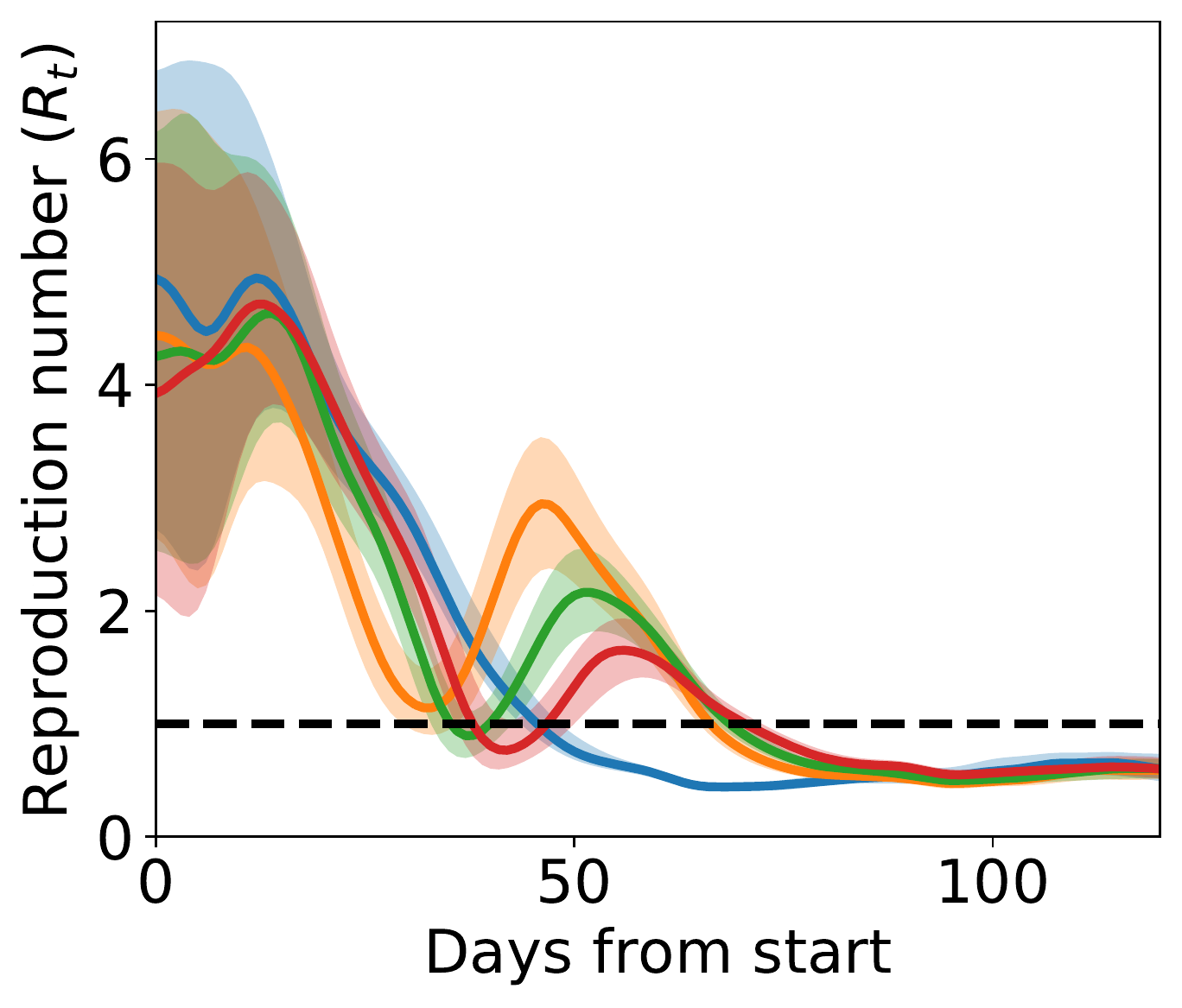}
\\
\multicolumn{8}{c}{Days from Start} \\
\multicolumn{8}{c}{\includegraphics[width=0.6\textwidth]{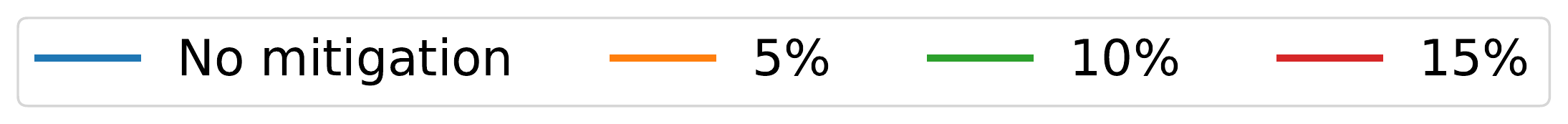}} \\

& \multicolumn{1}{c}{\small Tokyo} & \multicolumn{1}{c}{\small Chicago} & \multicolumn{1}{c}{\small Los Angeles} & \multicolumn{1}{c}{\small Istanbul} & \multicolumn{1}{c}{\small Jakarta} & \multicolumn{1}{c}{\small London} & \multicolumn{1}{c}{\small Bike} 
\end{tabular}
\caption{Infection spreading with a varied start time of uniform intervention as Part I in Figure~2. The intervention strategy uniformly randomly skips $80\%$ of the check-in or meeting events. Intervention starts when $5\%$, $10\%$, and $15\%$ population is infected and lasts for $15$ days. This intervention strategy can reduce the total number of infected agents in some datasets, but the fraction is independent of the starting time of the intervention.}
\label{figS:start_time}
\end{figure}

\newpage
\begin{figure}[H]
\begin{tabular}{m{0.2cm}@{}@{}m{\y\textwidth}@{}@{}m{\y\textwidth}@{}@{}m{\y\textwidth}@{}@{}m{\y\textwidth}@{}@{}m{\y\textwidth}@{}@{}m{\y\textwidth}@{}@{}m{\y\textwidth}@{}}

\multicolumn{8}{c}{\textbf{Part I}}\\
\includegraphics[height=2.1cm]{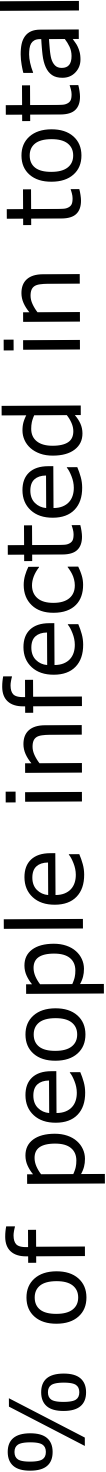}&
\includegraphics[width=\y\textwidth]{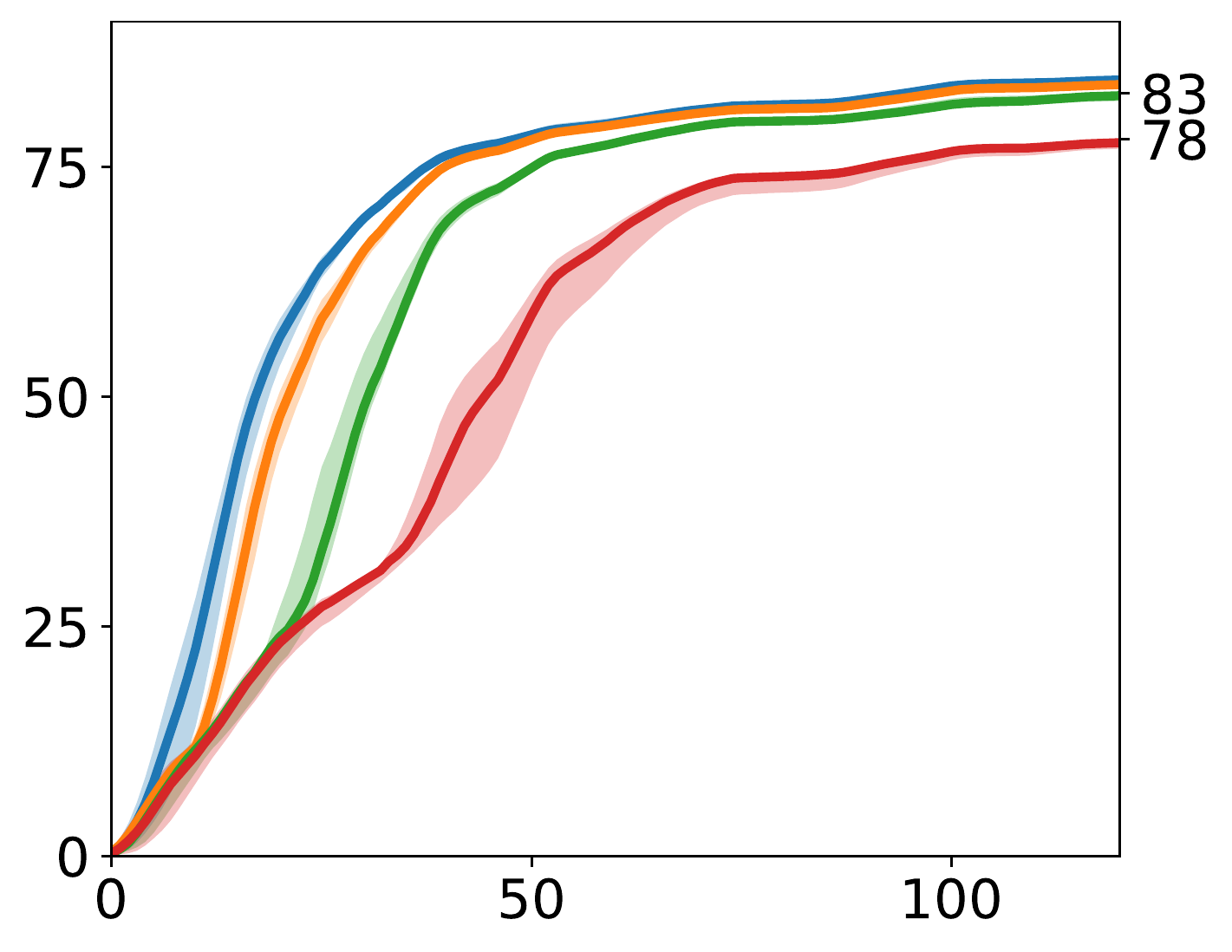}  &
\includegraphics[width=\y\textwidth]{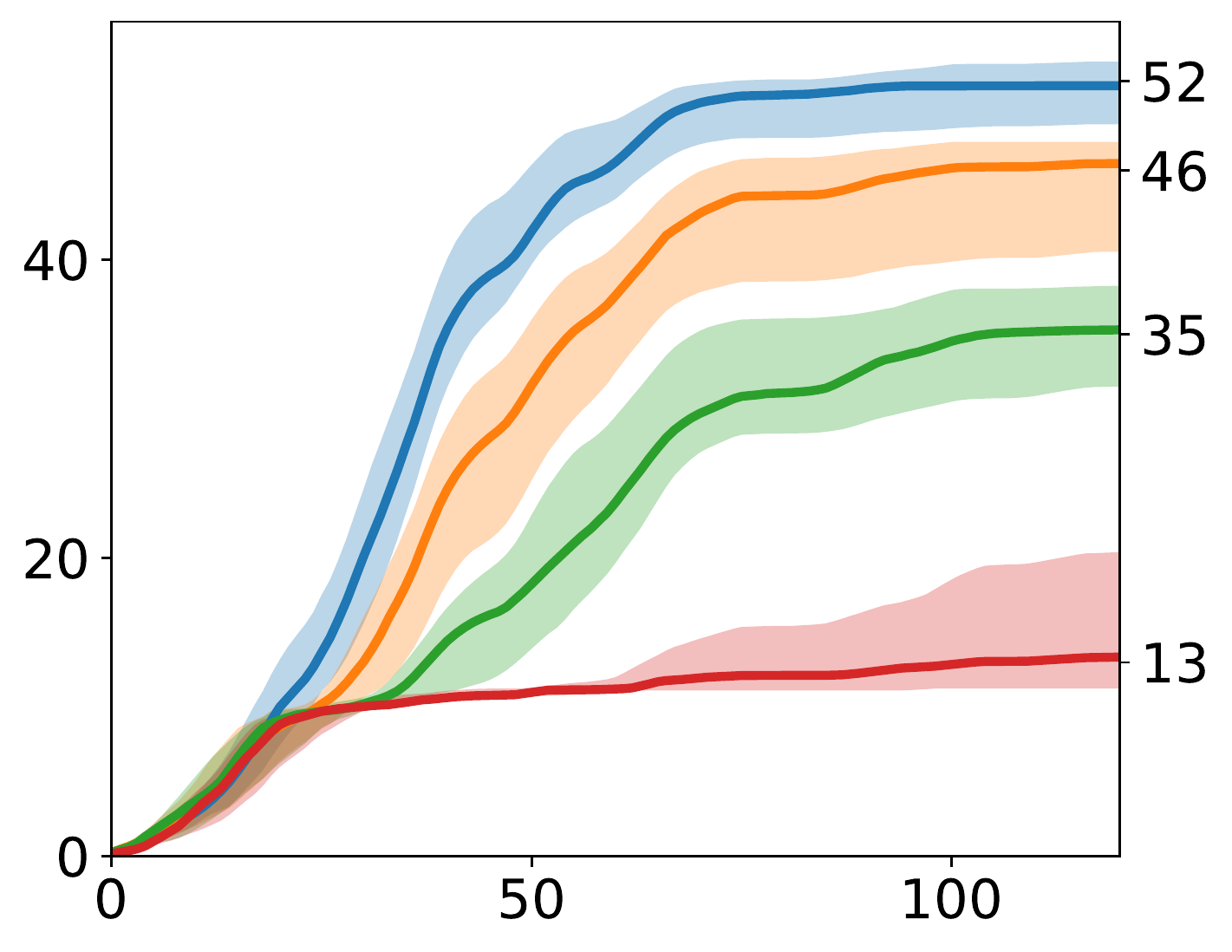}  &
\includegraphics[width=\y\textwidth]{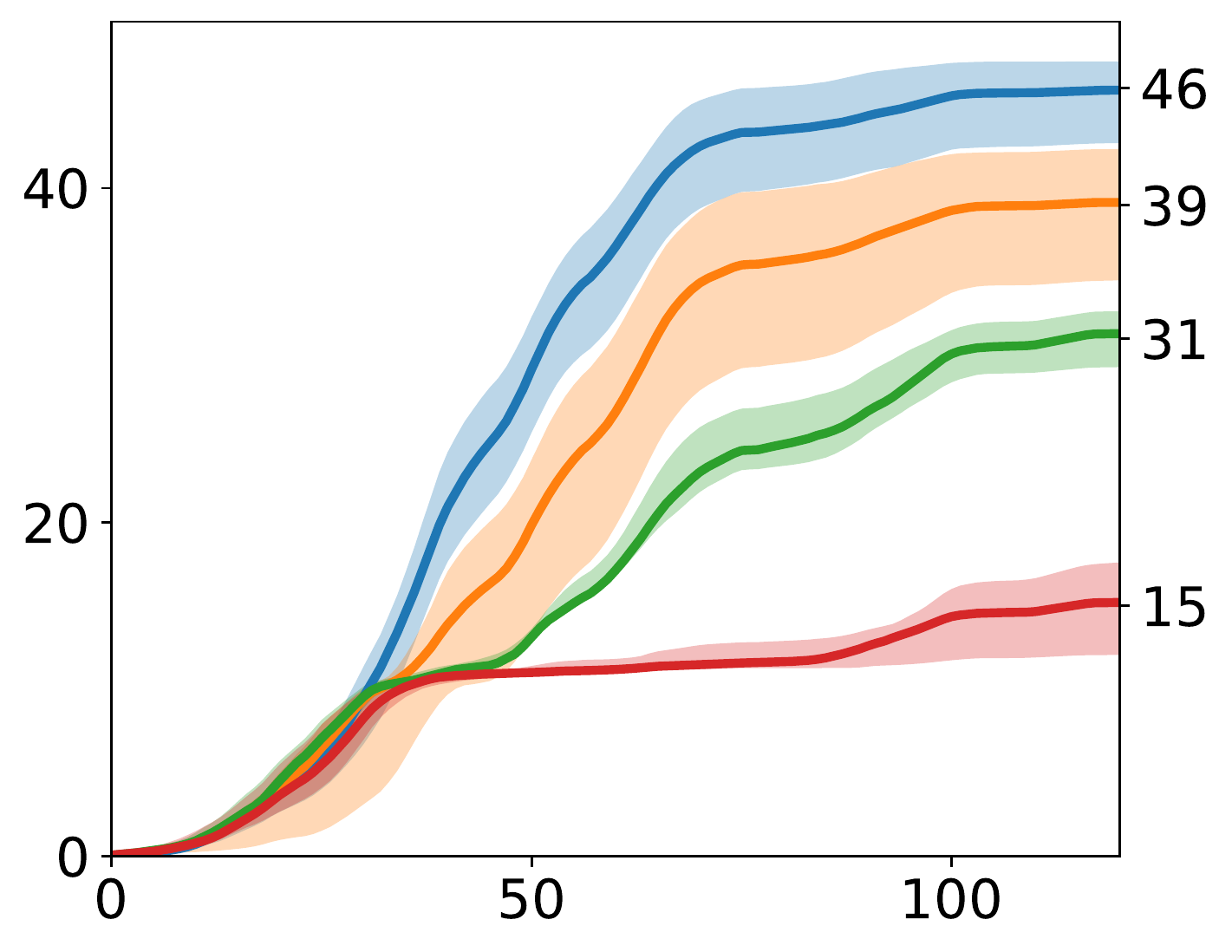}  &
\includegraphics[width=\y\textwidth]{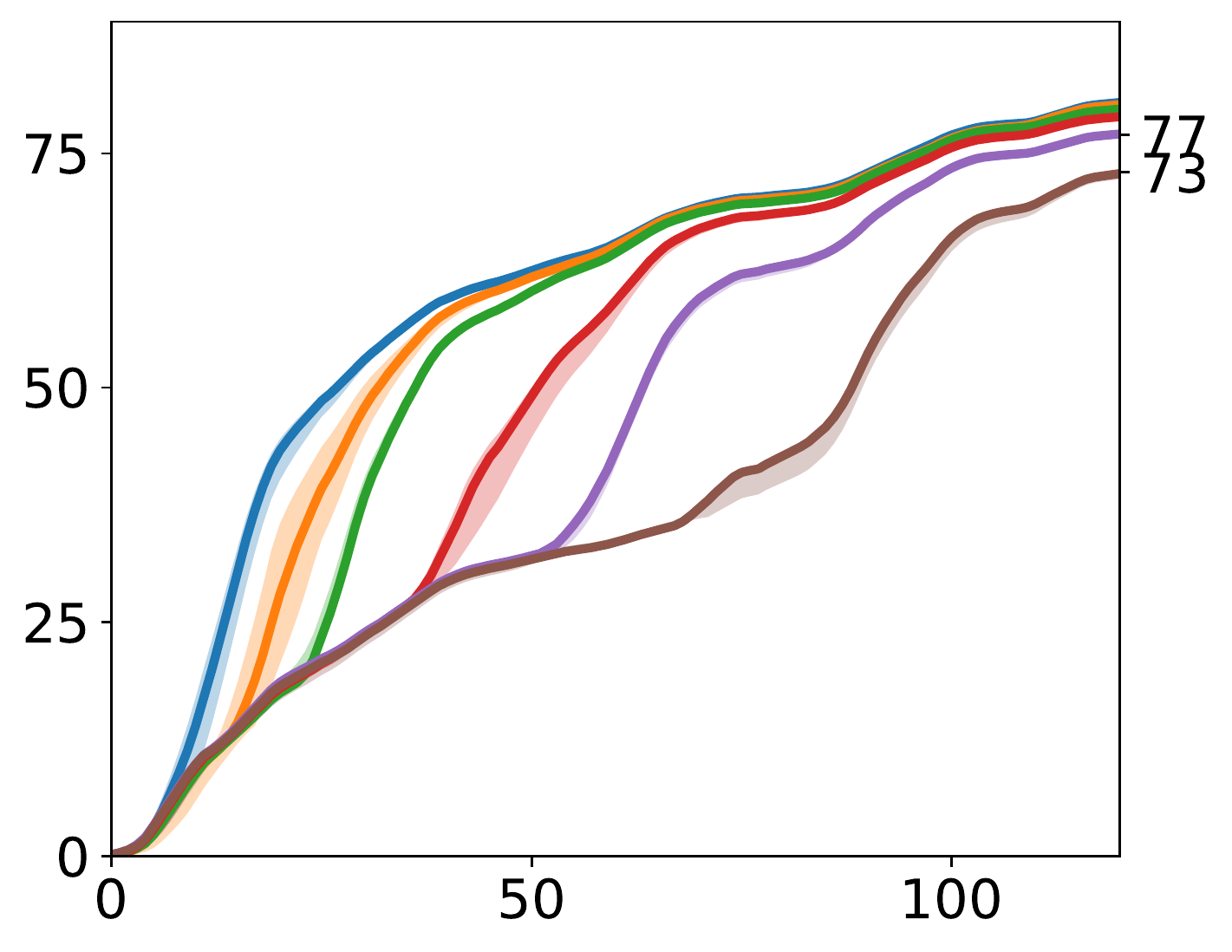}  &
\includegraphics[width=\y\textwidth]{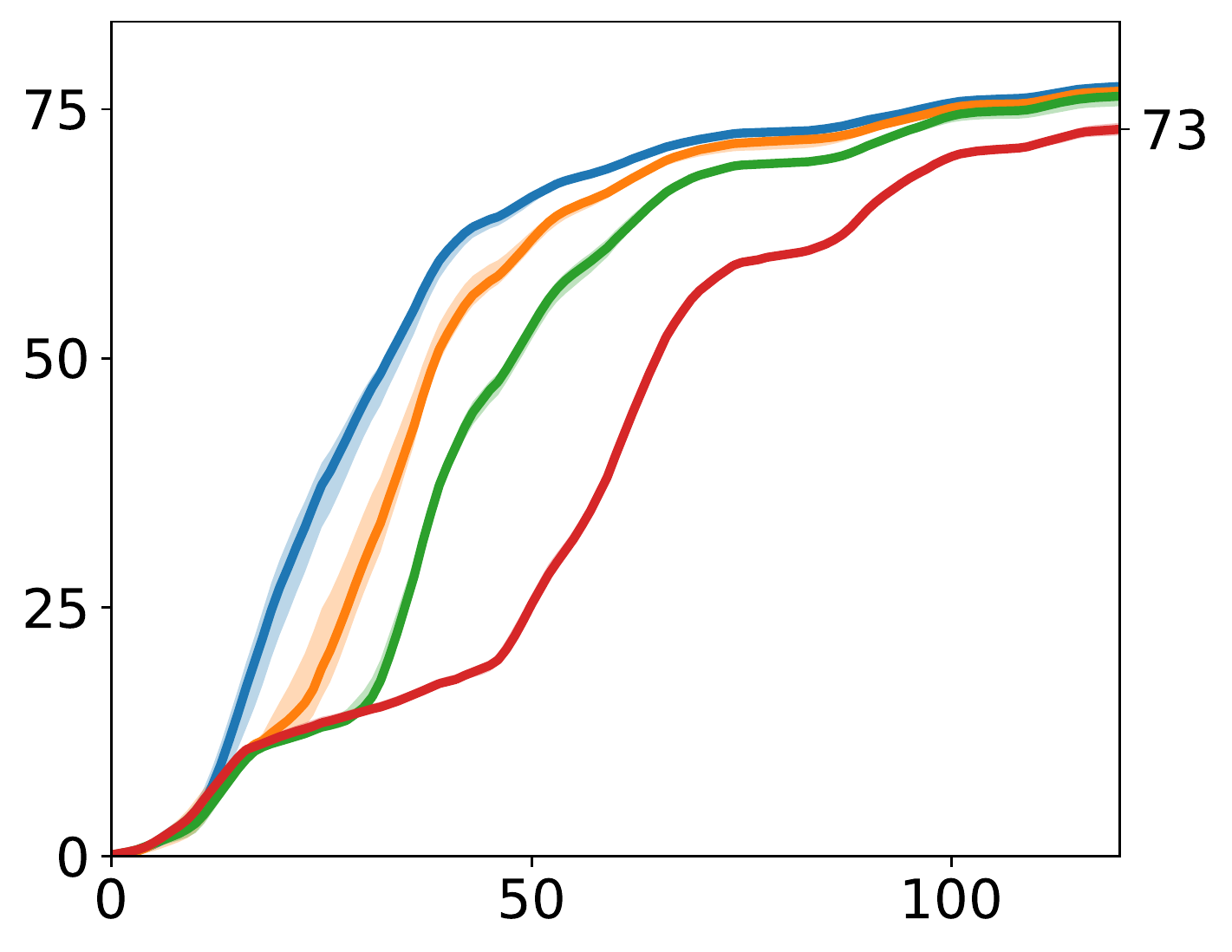}  &
\includegraphics[width=\y\textwidth]{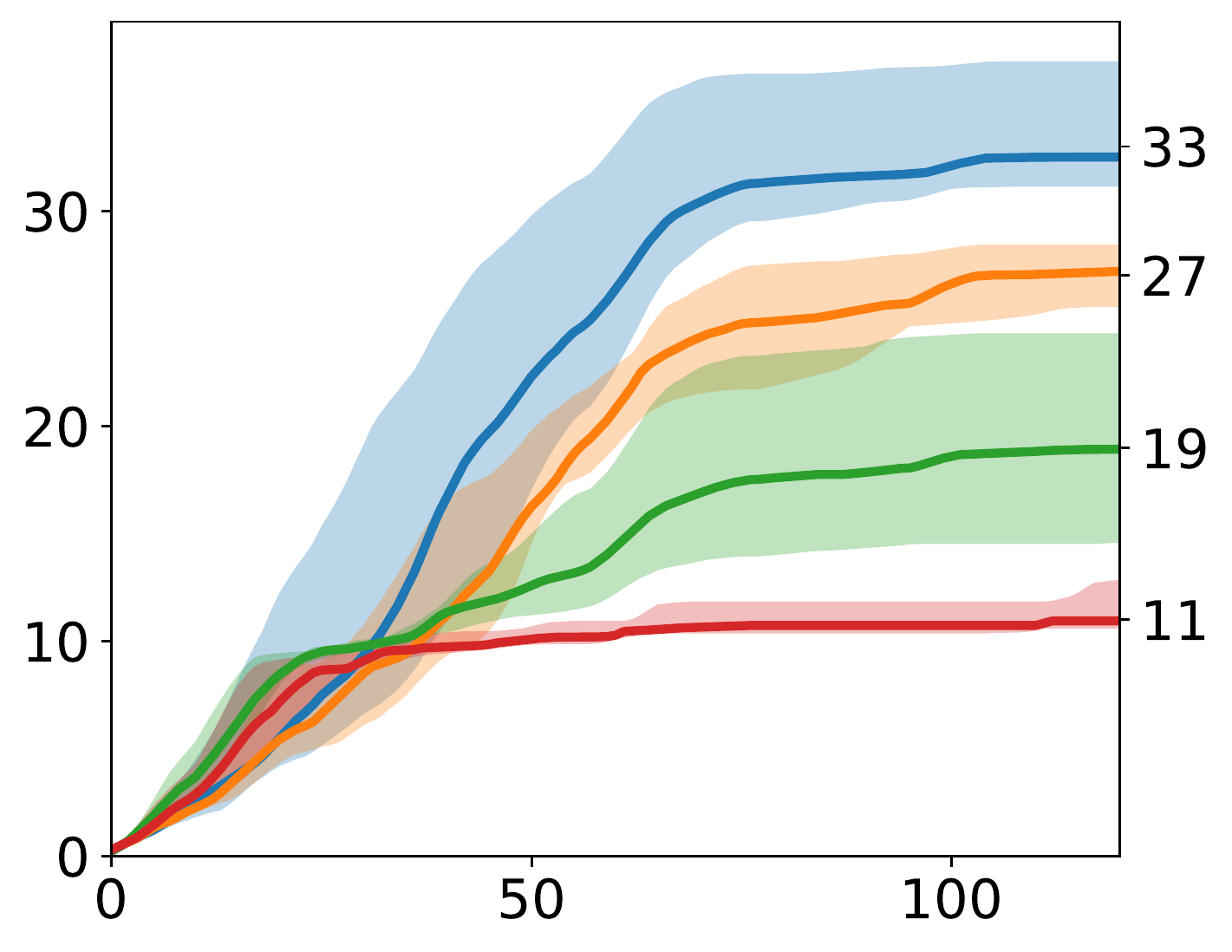} &
\includegraphics[width=\y\textwidth]{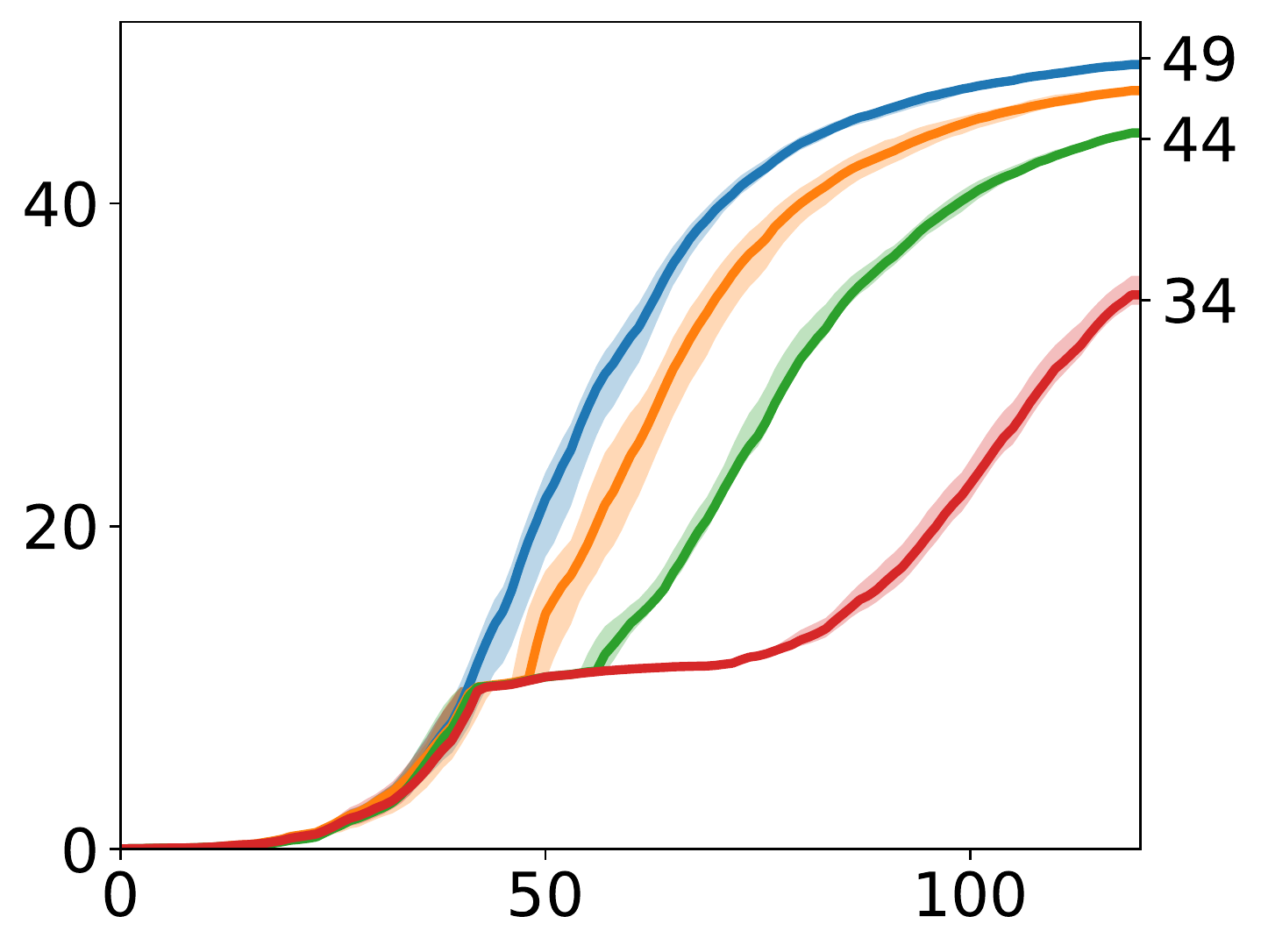}
\\

\includegraphics[height=2.1cm]{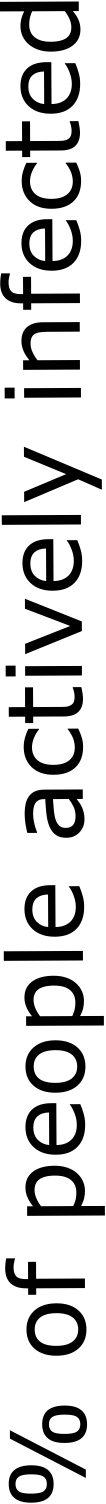}&
\includegraphics[width=\y\textwidth]{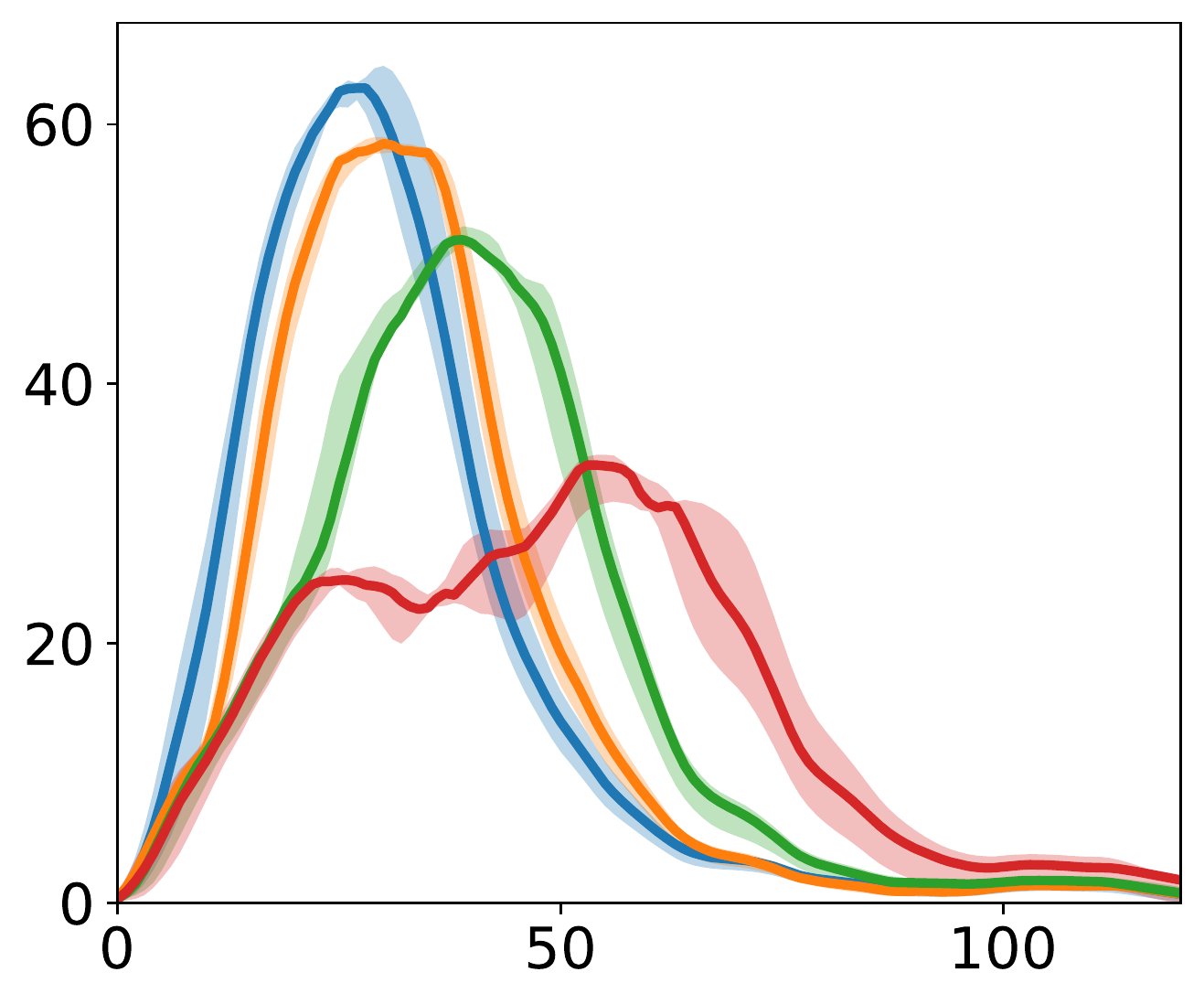}  &
\includegraphics[width=\y\textwidth]{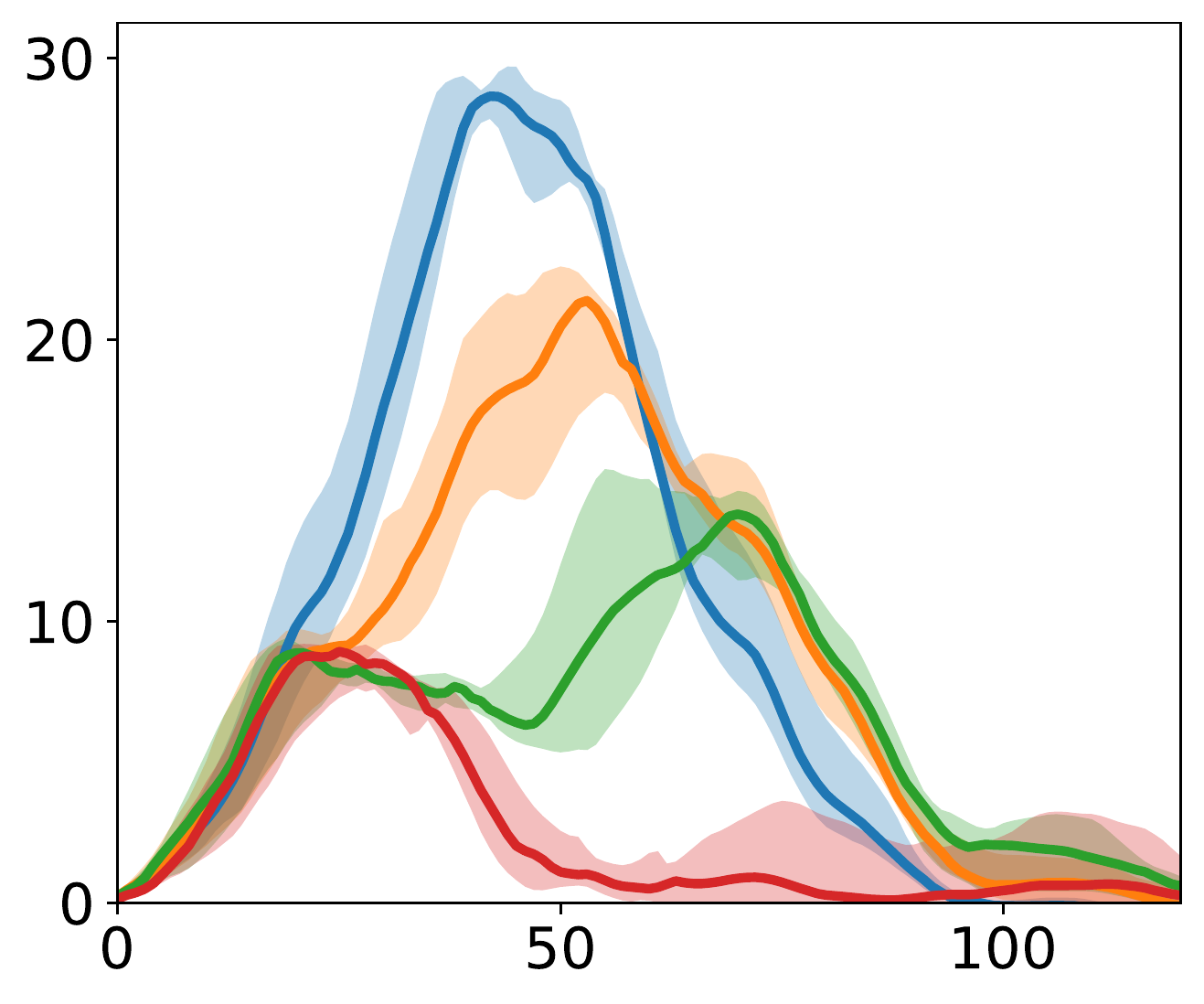}  &
\includegraphics[width=\y\textwidth]{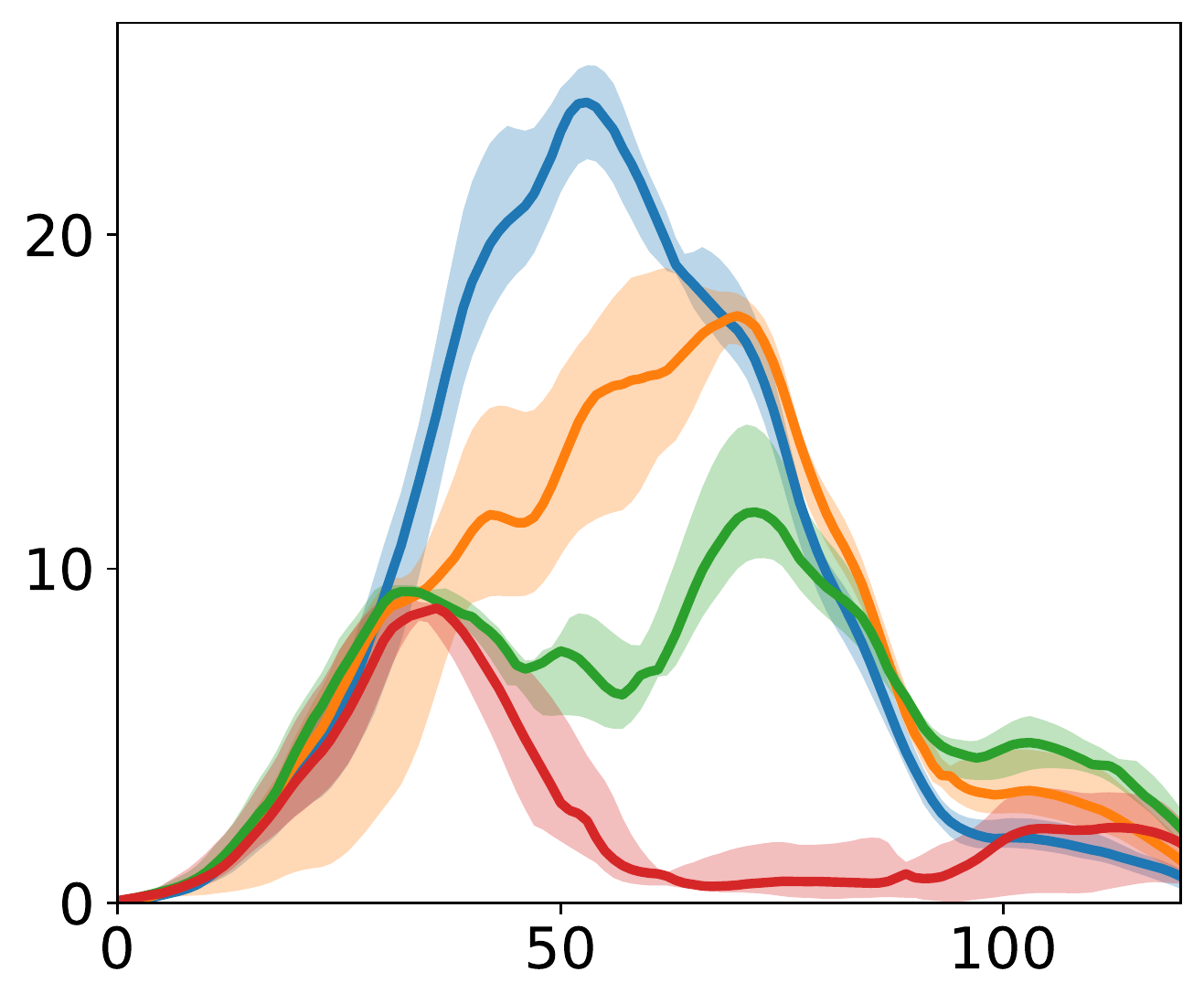}  &
\includegraphics[width=\y\textwidth]{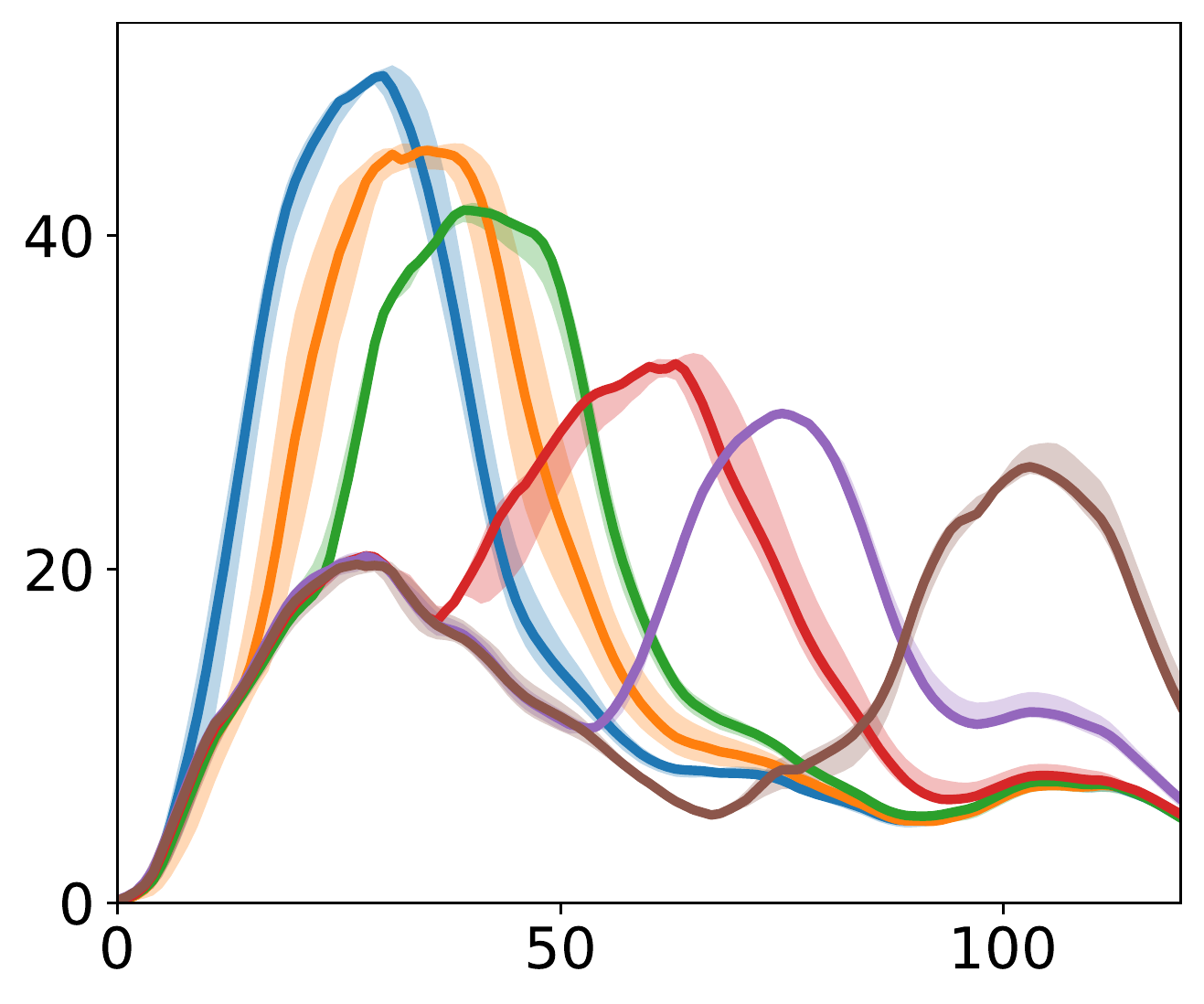}  &
\includegraphics[width=\y\textwidth]{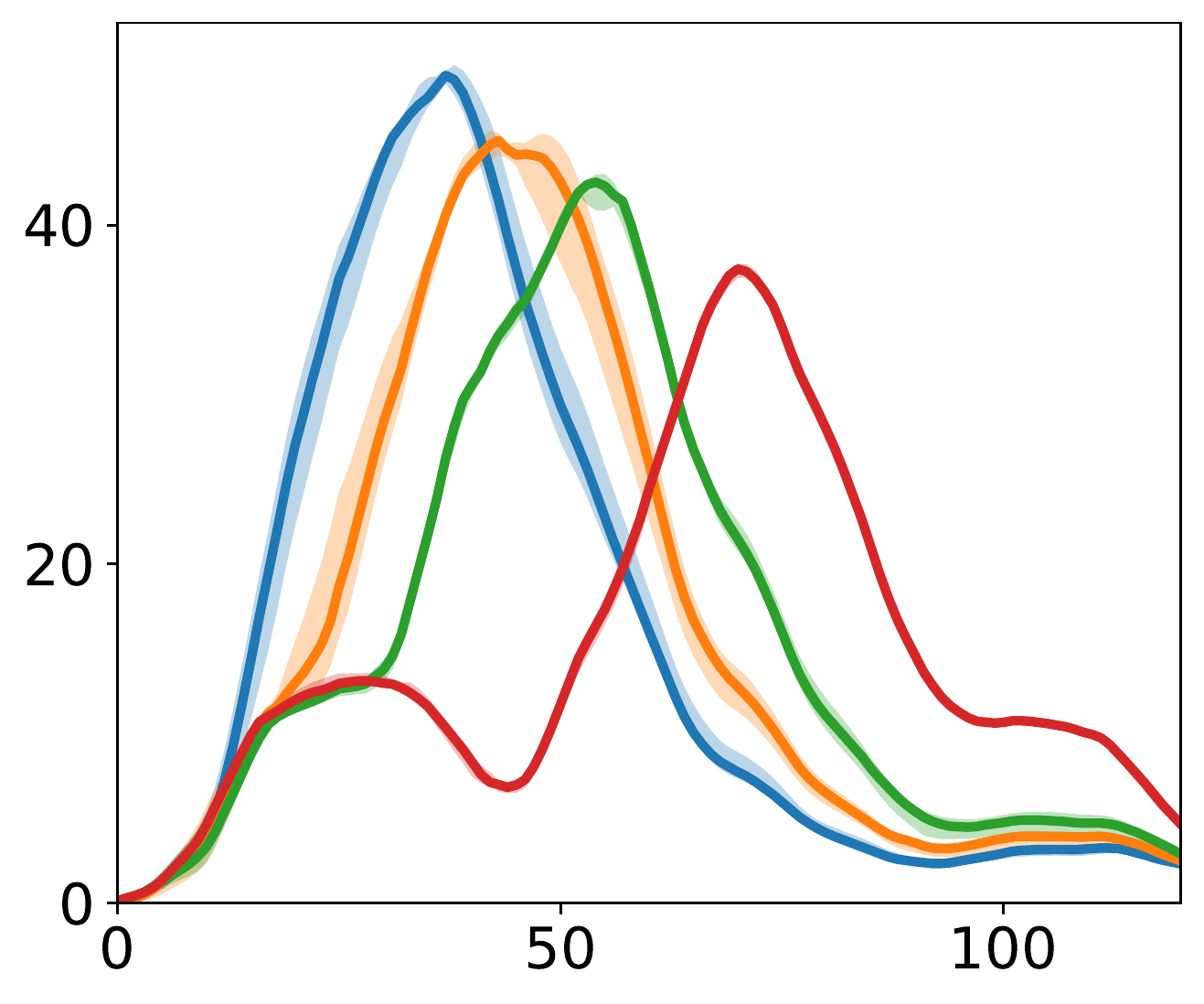}  &
\includegraphics[width=\y\textwidth]{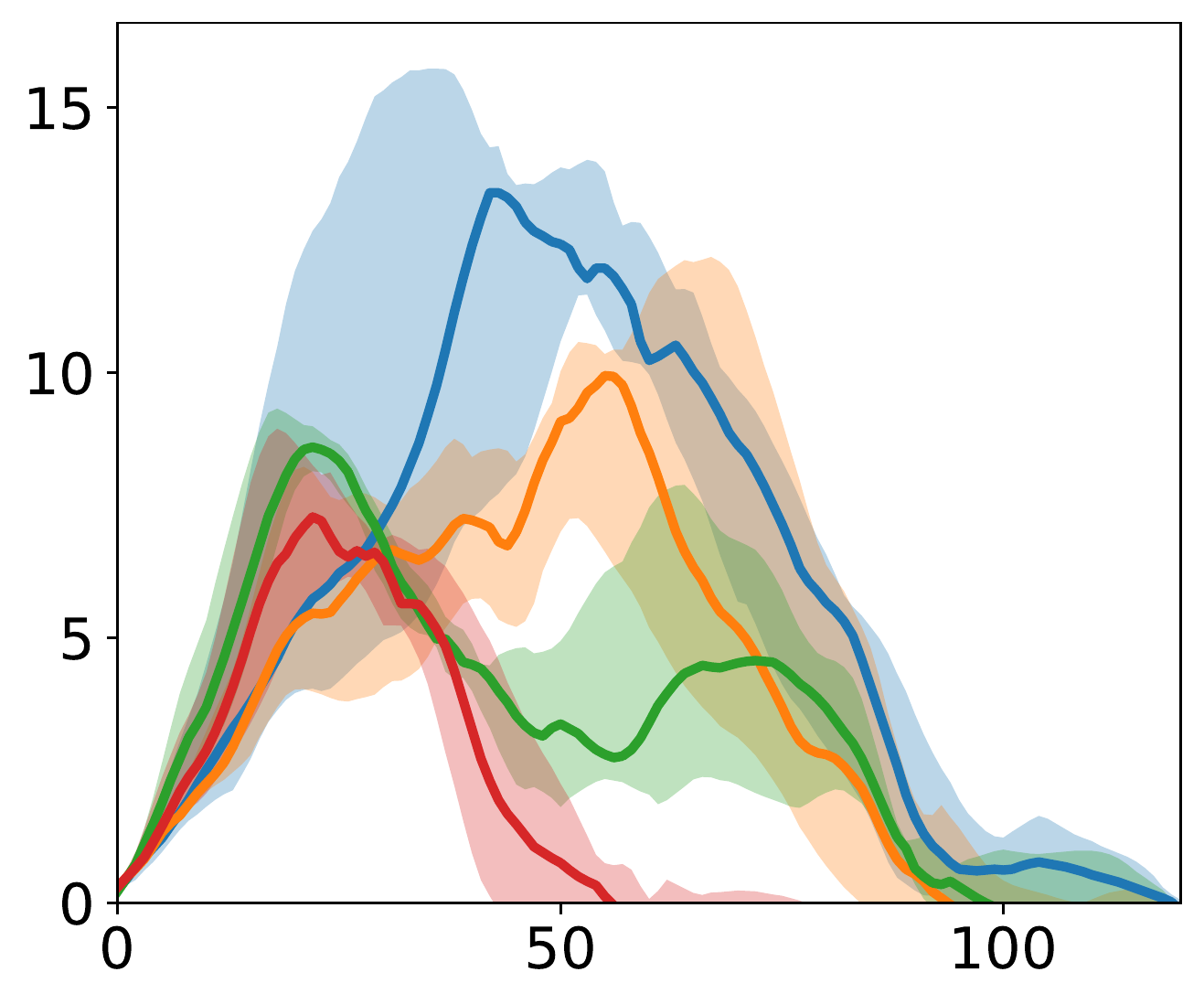} &
\includegraphics[width=\y\textwidth]{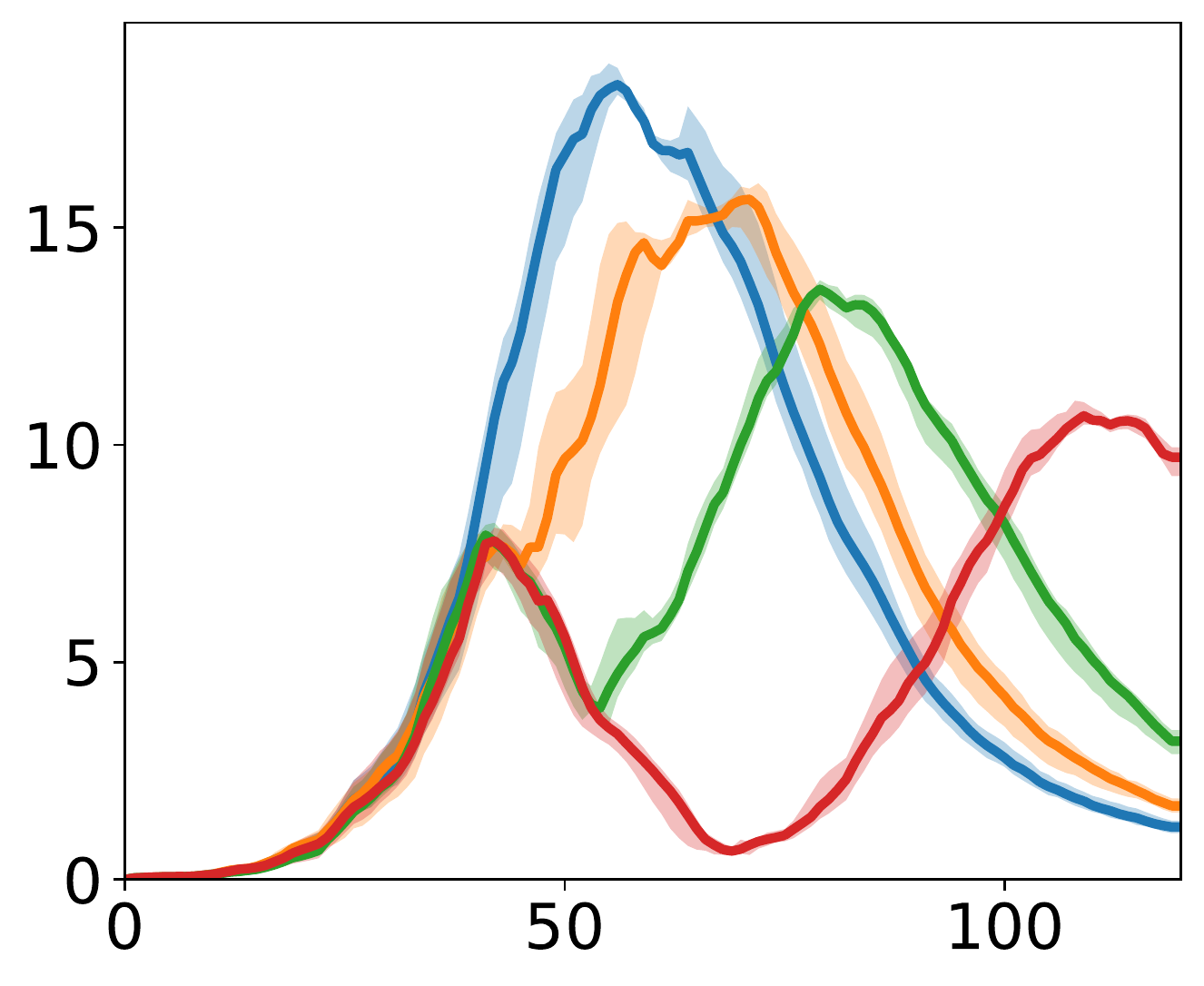}
\\

\includegraphics[height=2cm]{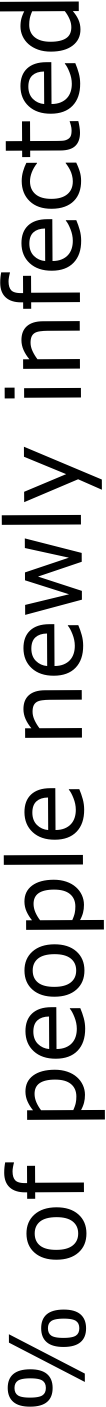}&
\includegraphics[width=\y\textwidth]{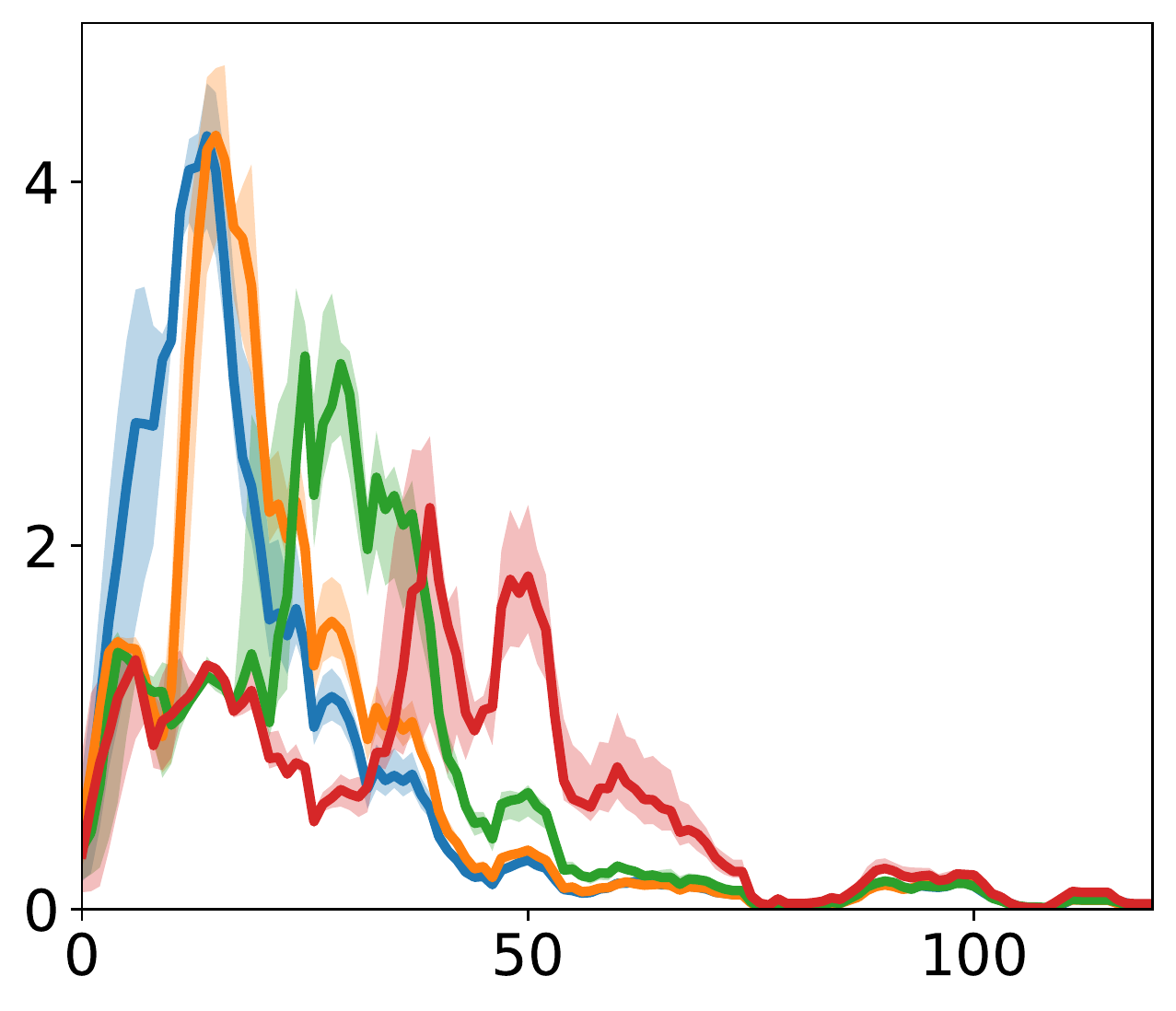}  &
\includegraphics[width=\y\textwidth]{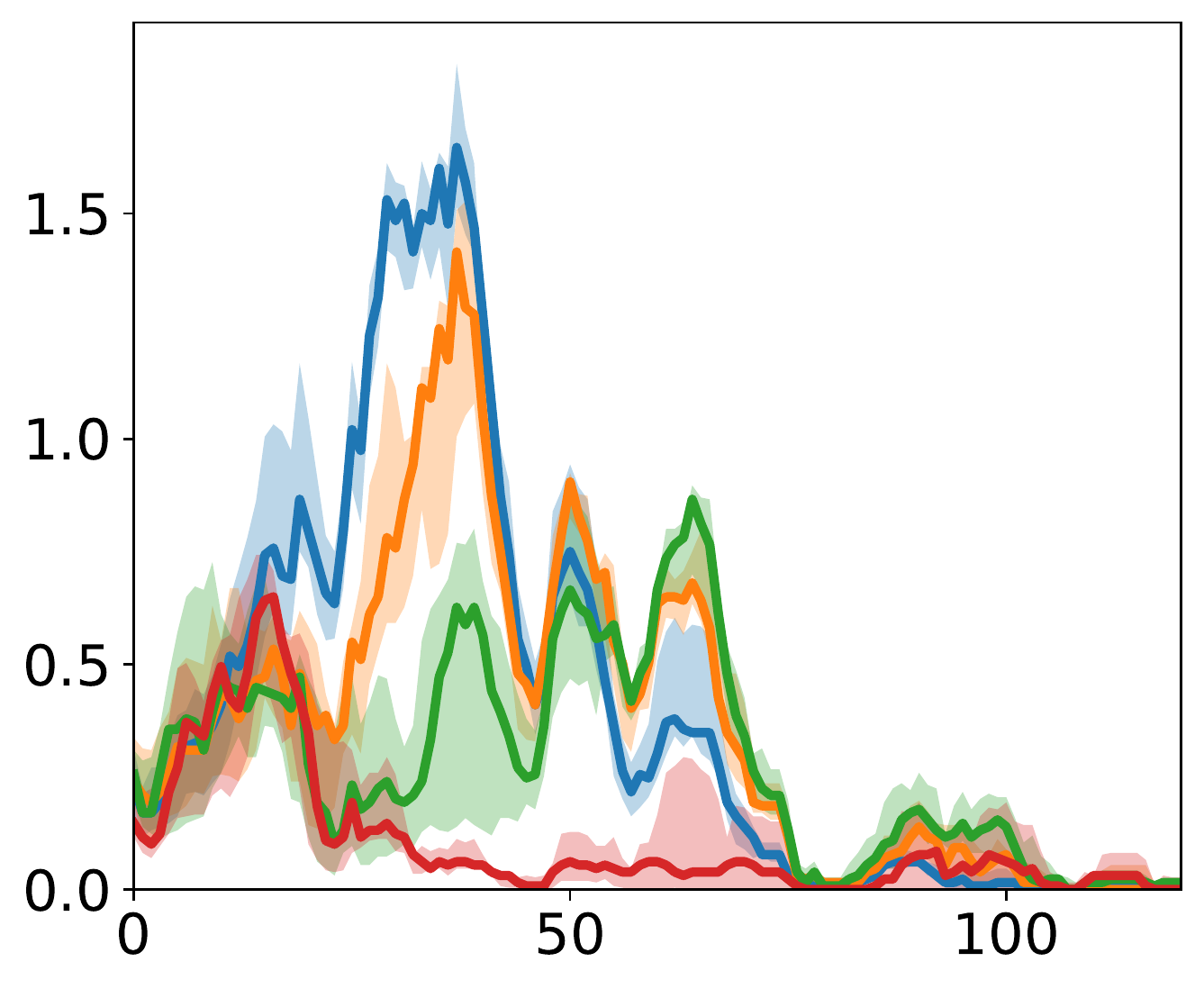}  &
\includegraphics[width=\y\textwidth]{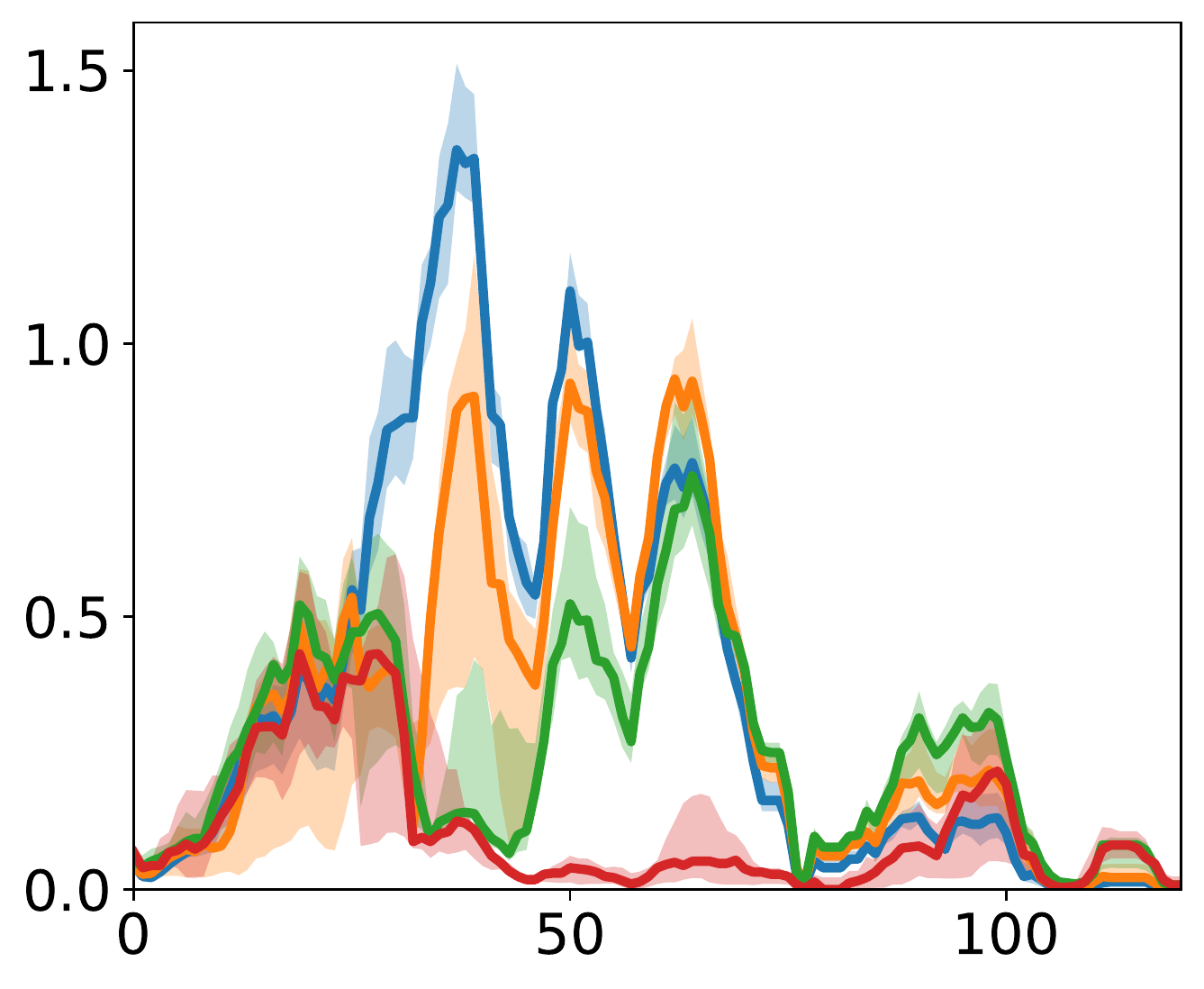}  &
\includegraphics[width=\y\textwidth]{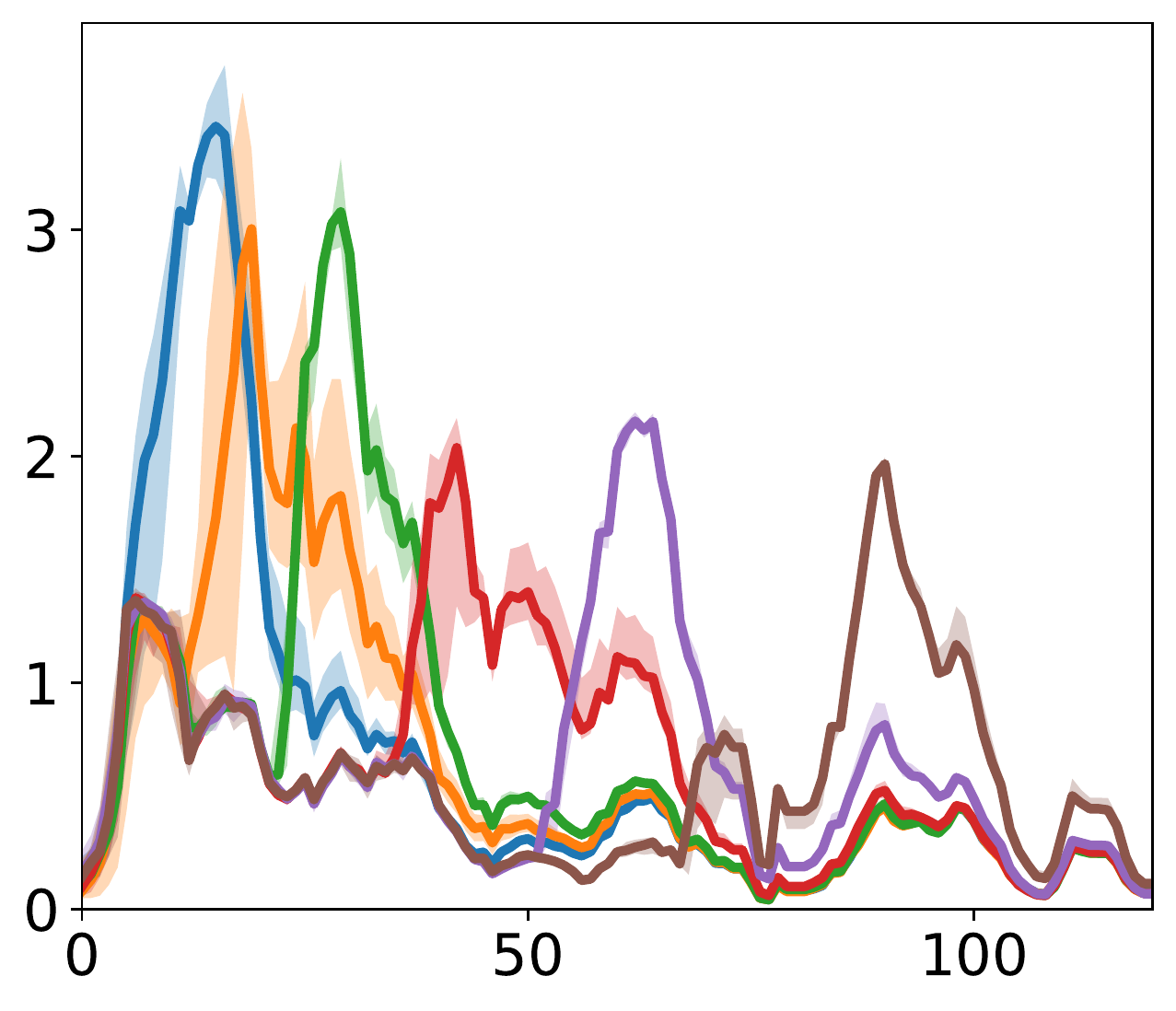}  &
\includegraphics[width=\y\textwidth]{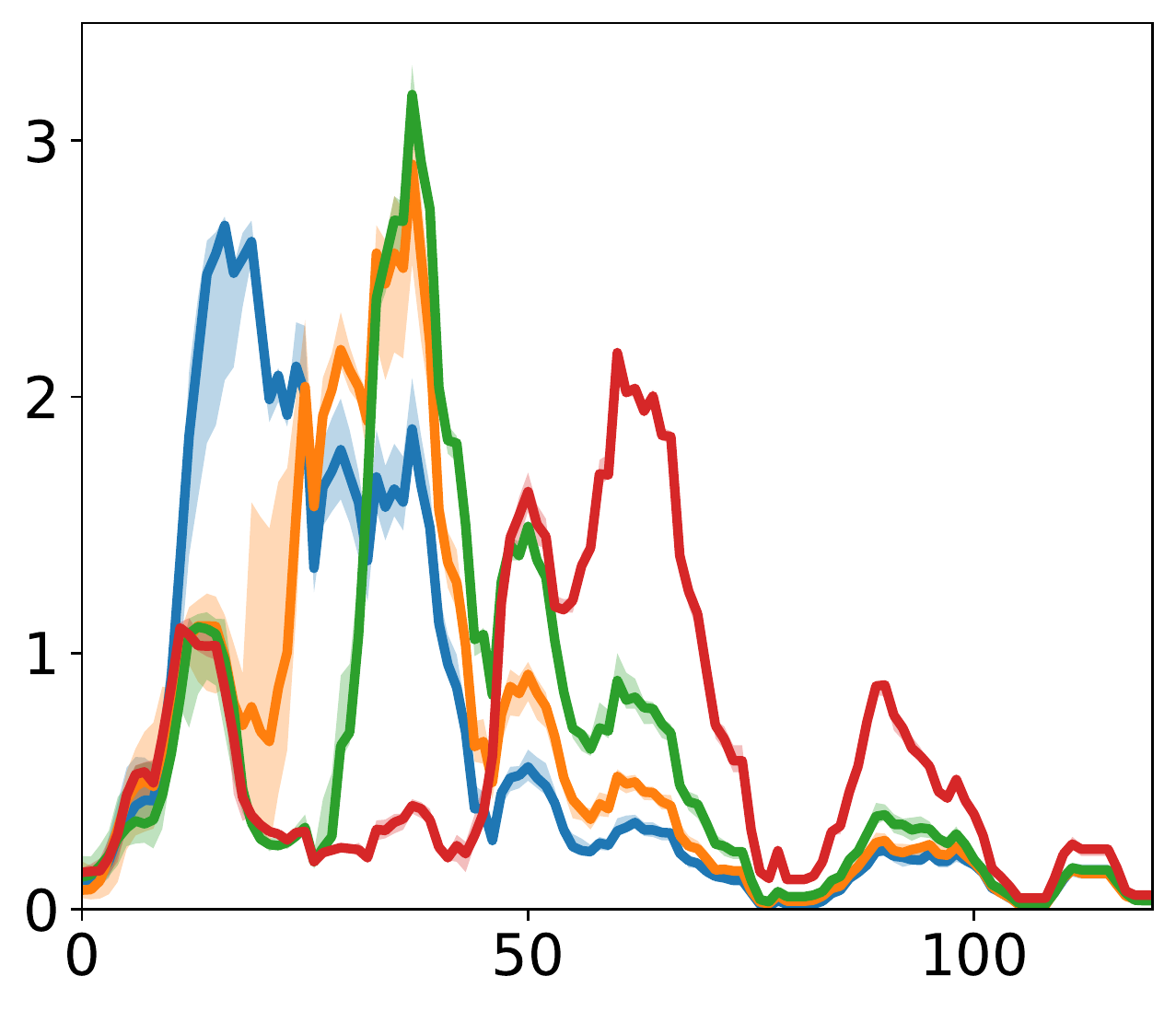}  &
\includegraphics[width=\y\textwidth]{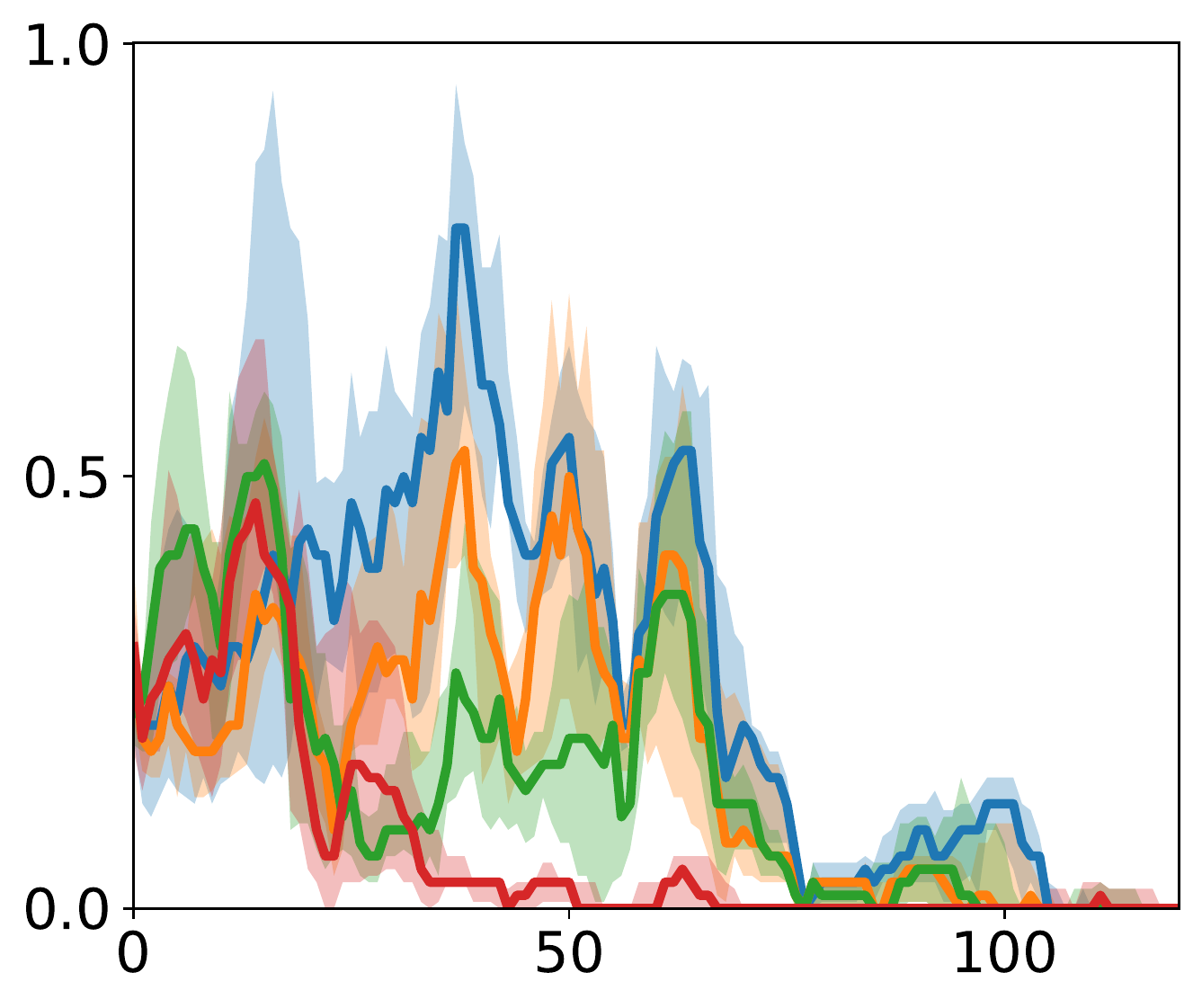} &
\includegraphics[width=\y\textwidth]{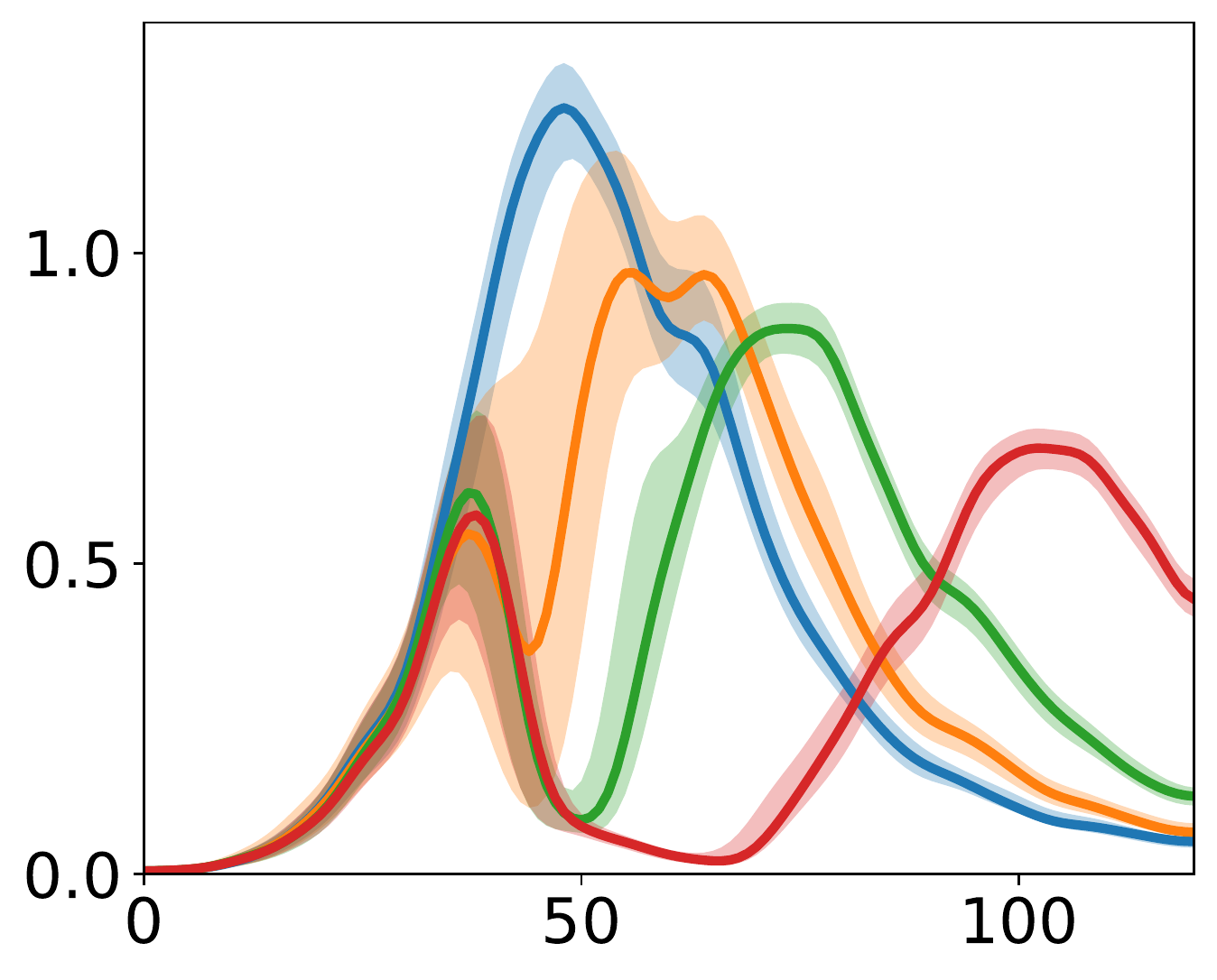}
\\

\includegraphics[height=1.5cm]{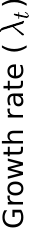}&
\includegraphics[width=\y\textwidth]{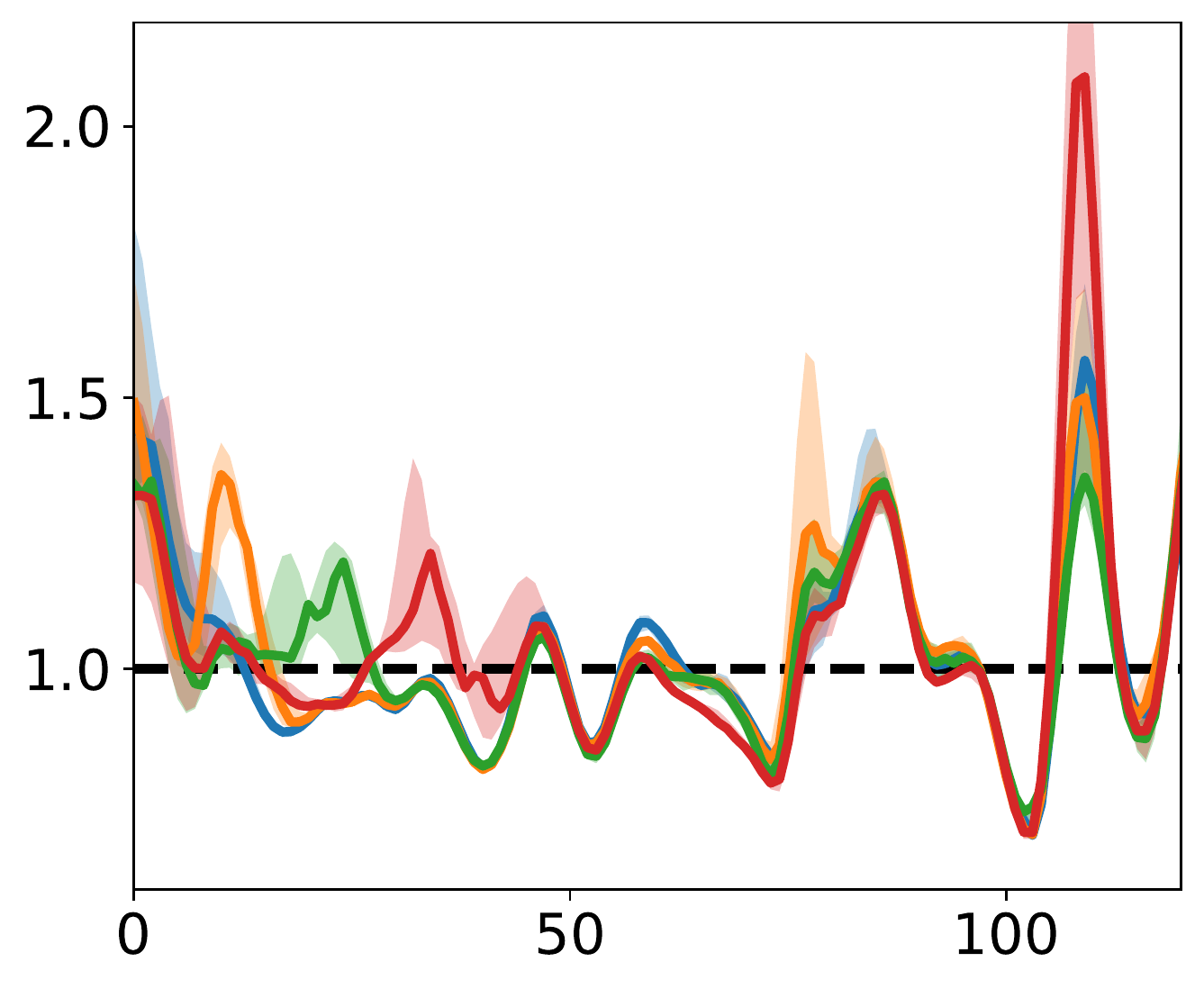}  &
\includegraphics[width=\y\textwidth]{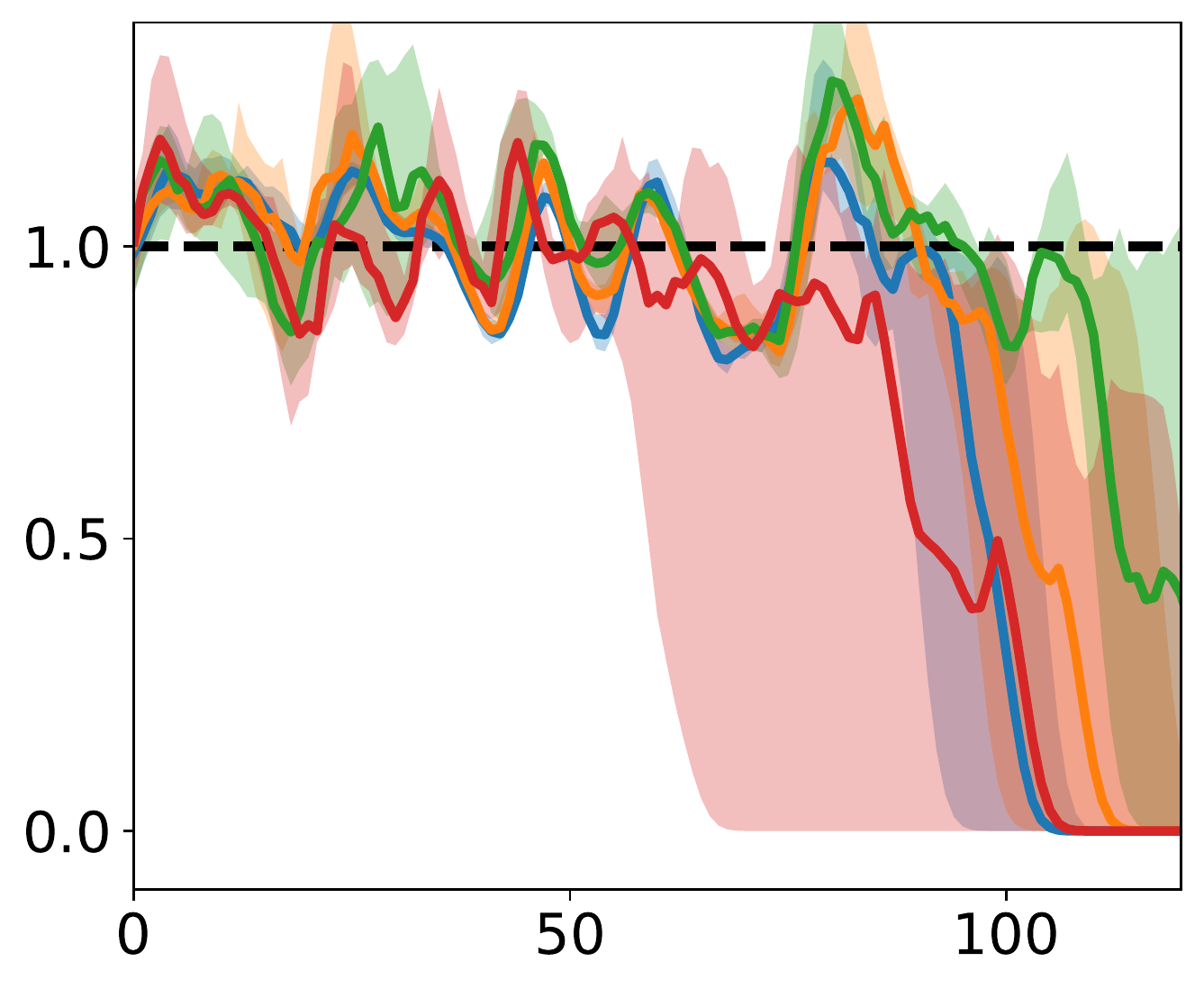}  &
\includegraphics[width=\y\textwidth]{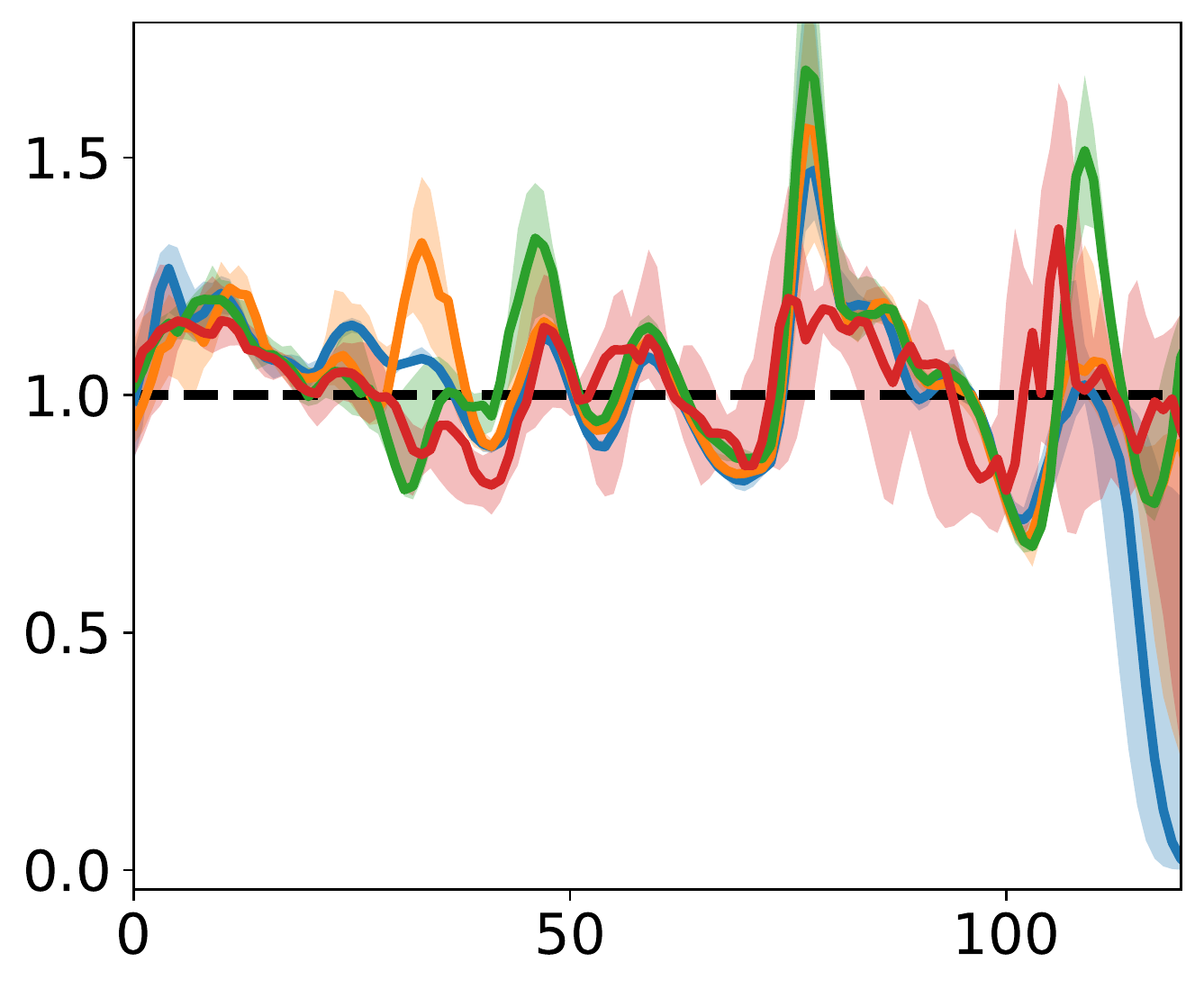}  &
\includegraphics[width=\y\textwidth]{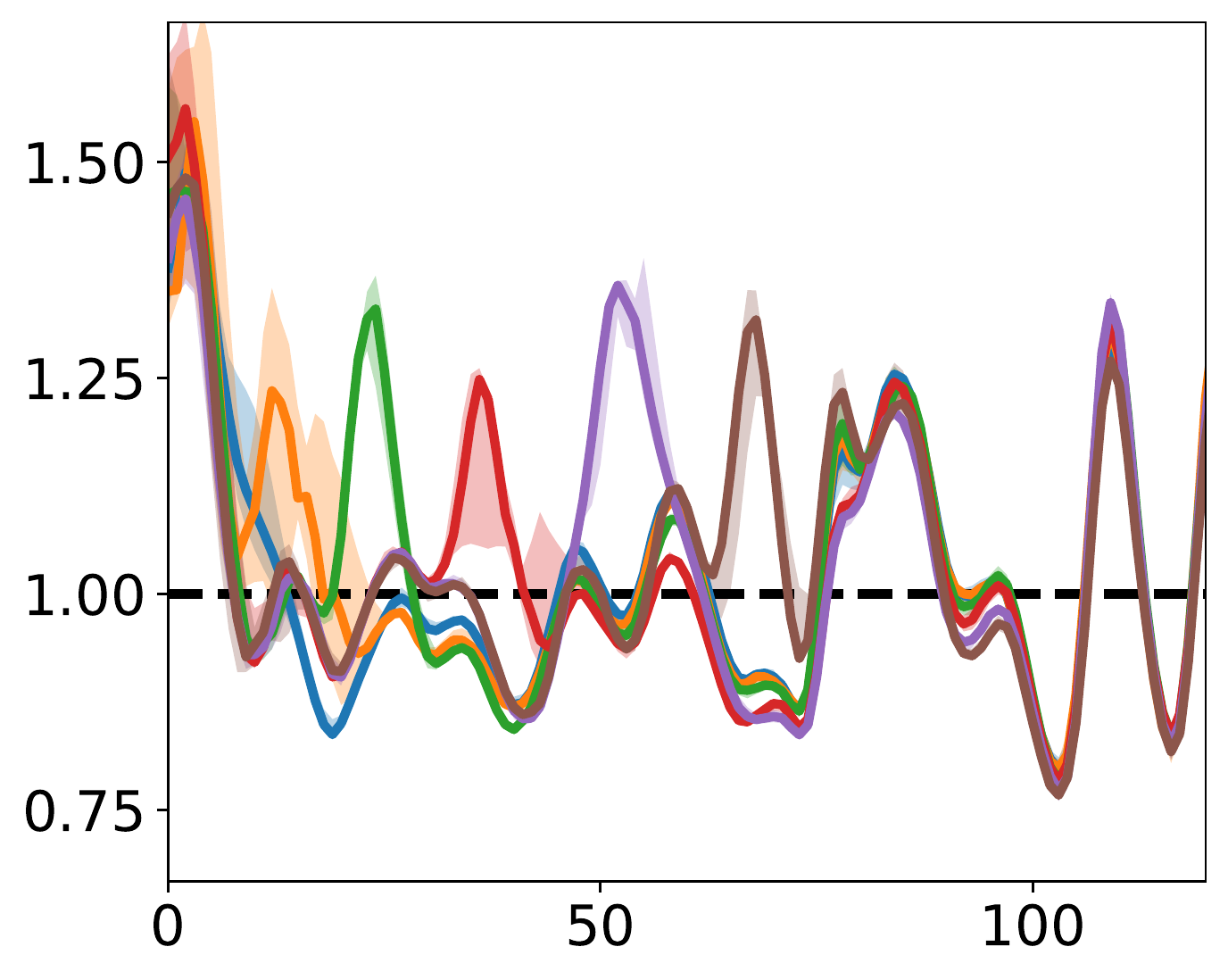}  &
\includegraphics[width=\y\textwidth]{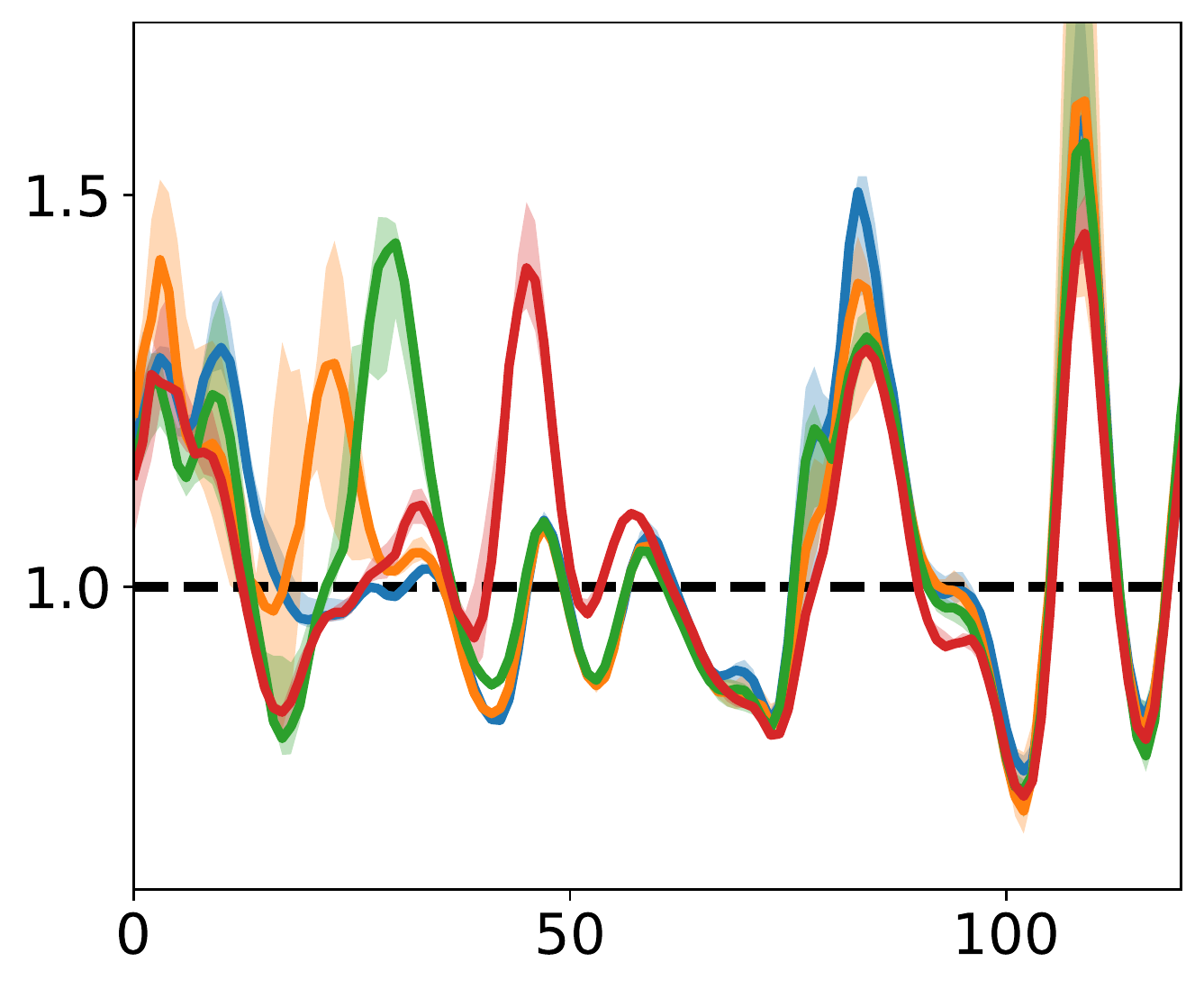}  &
\includegraphics[width=\y\textwidth]{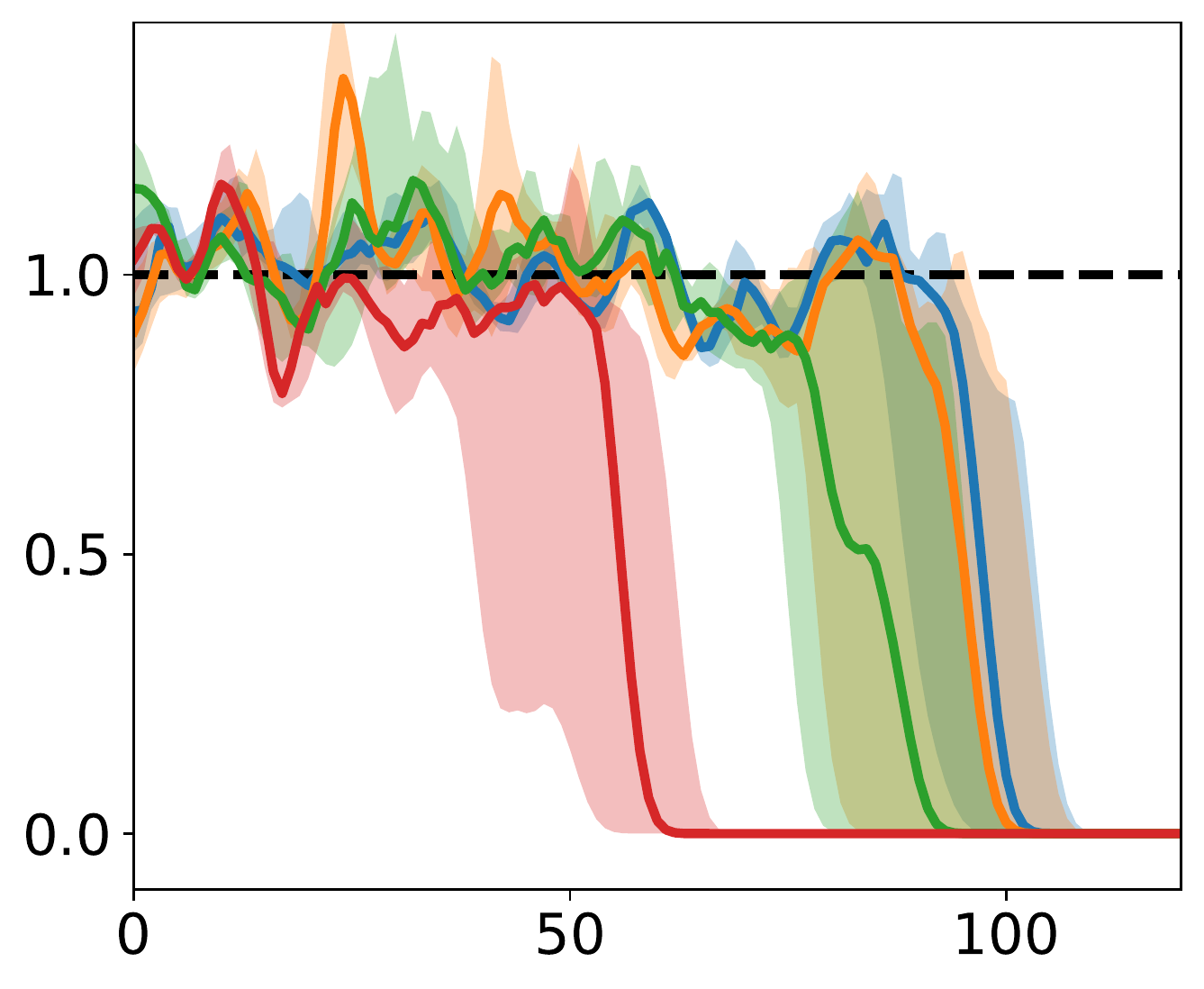} &
\includegraphics[width=\y\textwidth]{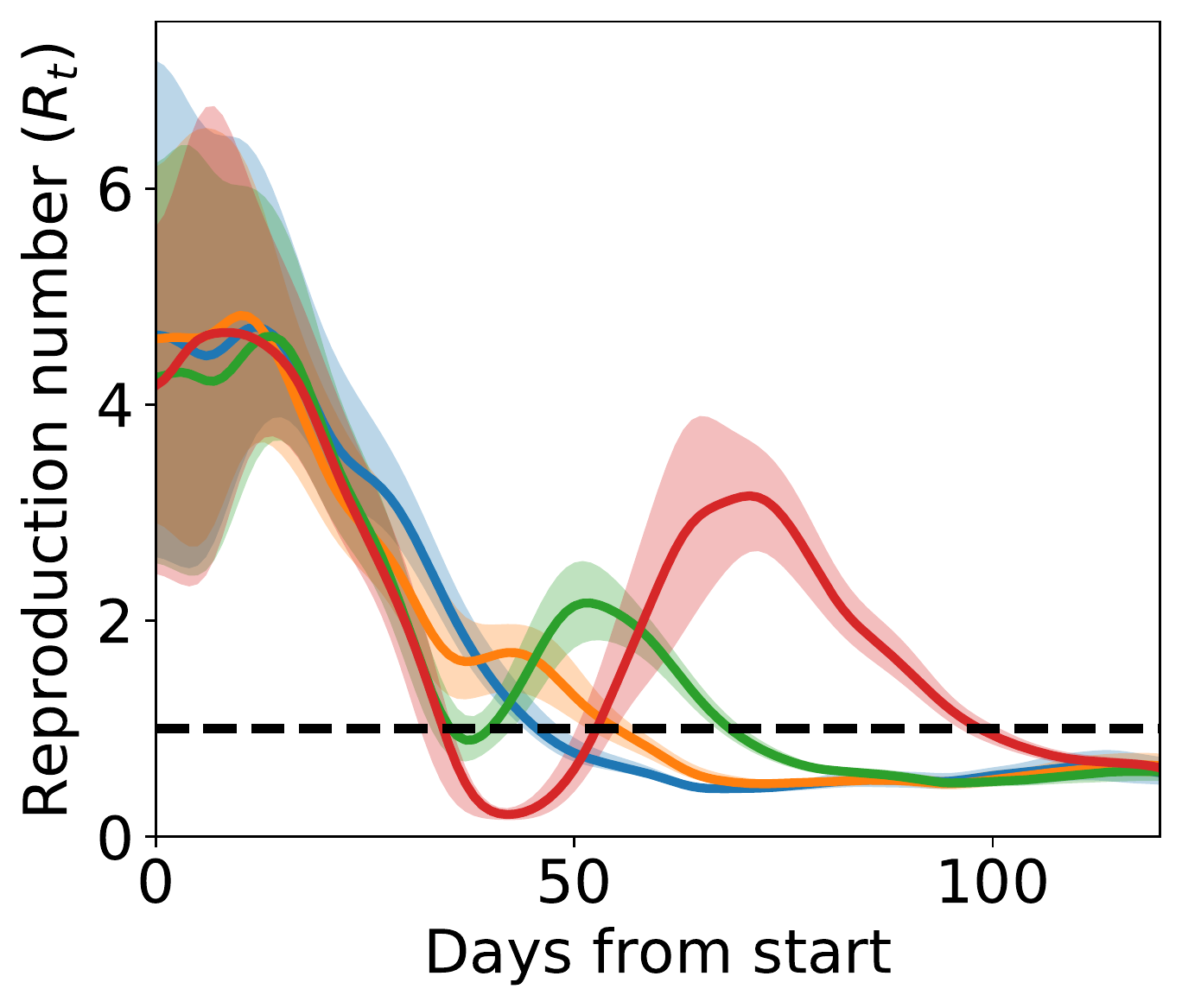}
\\
& \multicolumn{1}{c}{\small Tokyo} & \multicolumn{1}{c}{\small Chicago} & \multicolumn{1}{c}{\small Los Angeles} & \multicolumn{1}{c}{\small Istanbul} & \multicolumn{1}{c}{\small Jakarta} & \multicolumn{1}{c}{\small London} & \multicolumn{1}{c}{\small Bike}\\
\multicolumn{8}{c}{Days from start}\\
\multicolumn{8}{c}{\includegraphics[width=0.6\textwidth]{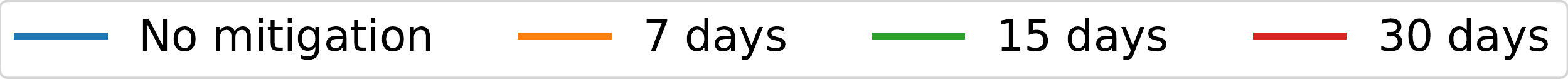}}\\
\multicolumn{8}{c}{\textbf{Part II}}\\
\end{tabular}
\centering
\begin{tabular}{m{0.1cm}m{\fourfig}@{}m{0.1cm}m{\fourfig}@{}m{0.1cm}m{\fourfig}@{}m{0.03cm}m{\fourfig}@{}m{0.1cm}}

\multirow{5}{*}{\includegraphics[height=4cm]{Fig/labels/TotalInfected.pdf}}&
\includegraphics[width=\fourfig]{Fig/foursquare/mobility/istambul/LOCKDOWN_82cdf_infected.pdf} &
\multirow{5}{*}{\includegraphics[height=4cm]{Fig/labels/ActiveInfected.pdf}}&
\includegraphics[width=\fourfig]{Fig/foursquare/mobility/istambul/LOCKDOWN_82active_cases.pdf}  &
\multirow{5}{*}{\includegraphics[height=4cm]{Fig/labels/newInfected.pdf}}&
\includegraphics[width=\fourfig]{Fig/foursquare/mobility/istambul/LOCKDOWN_82new_infected.pdf}  &
\rotatebox{90}{\tiny{Growth rate ($\lambda_t$)}} &
\includegraphics[width=\fourfig]{Fig/foursquare/mobility/istambul/LOCKDOWN_82growth_factor.pdf} & 
\rotatebox{90}{Istanbul}
\\ [-0.25cm]
&
\includegraphics[width=\fourfig]{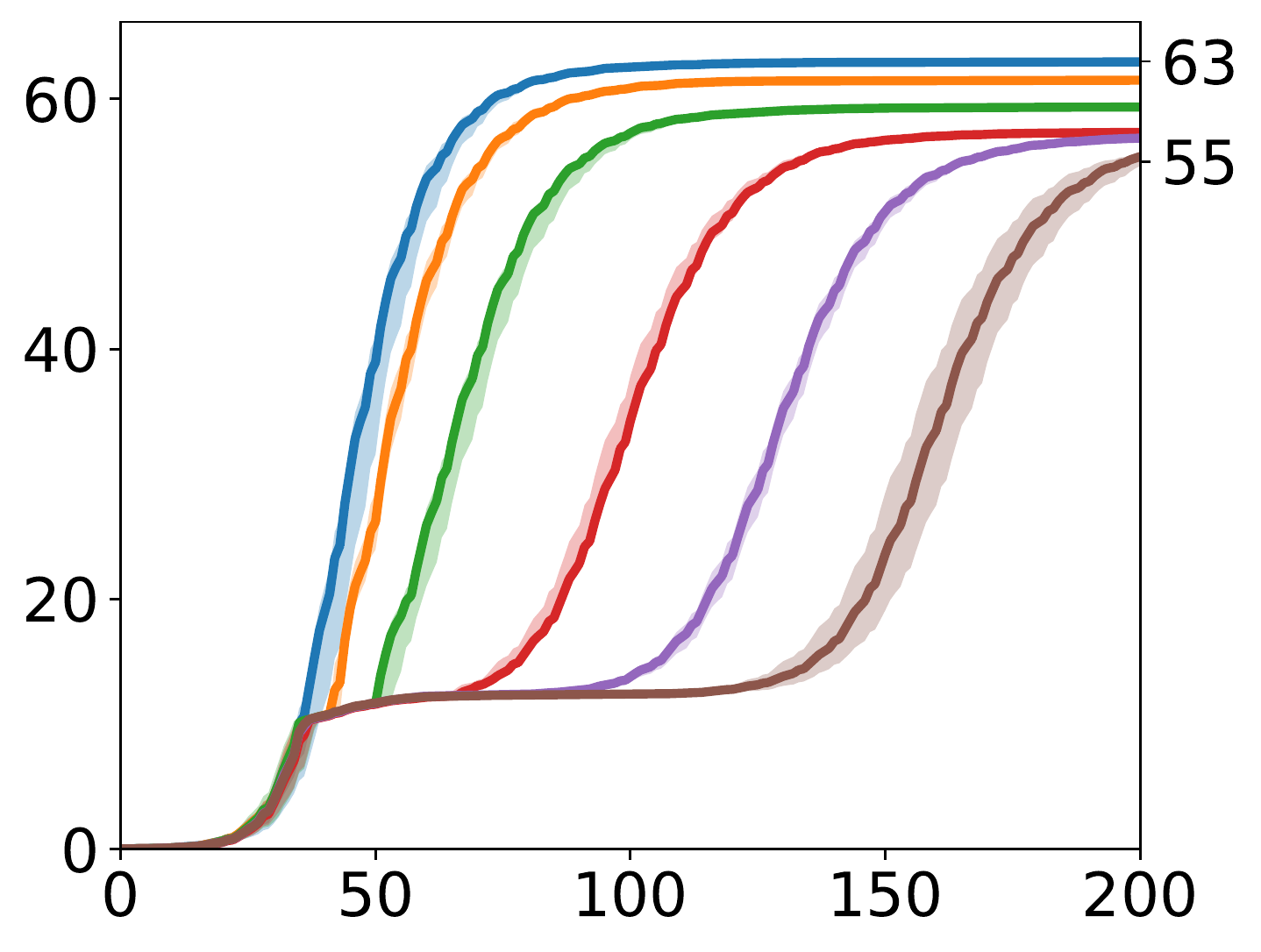}  & &
\includegraphics[width=\fourfig]{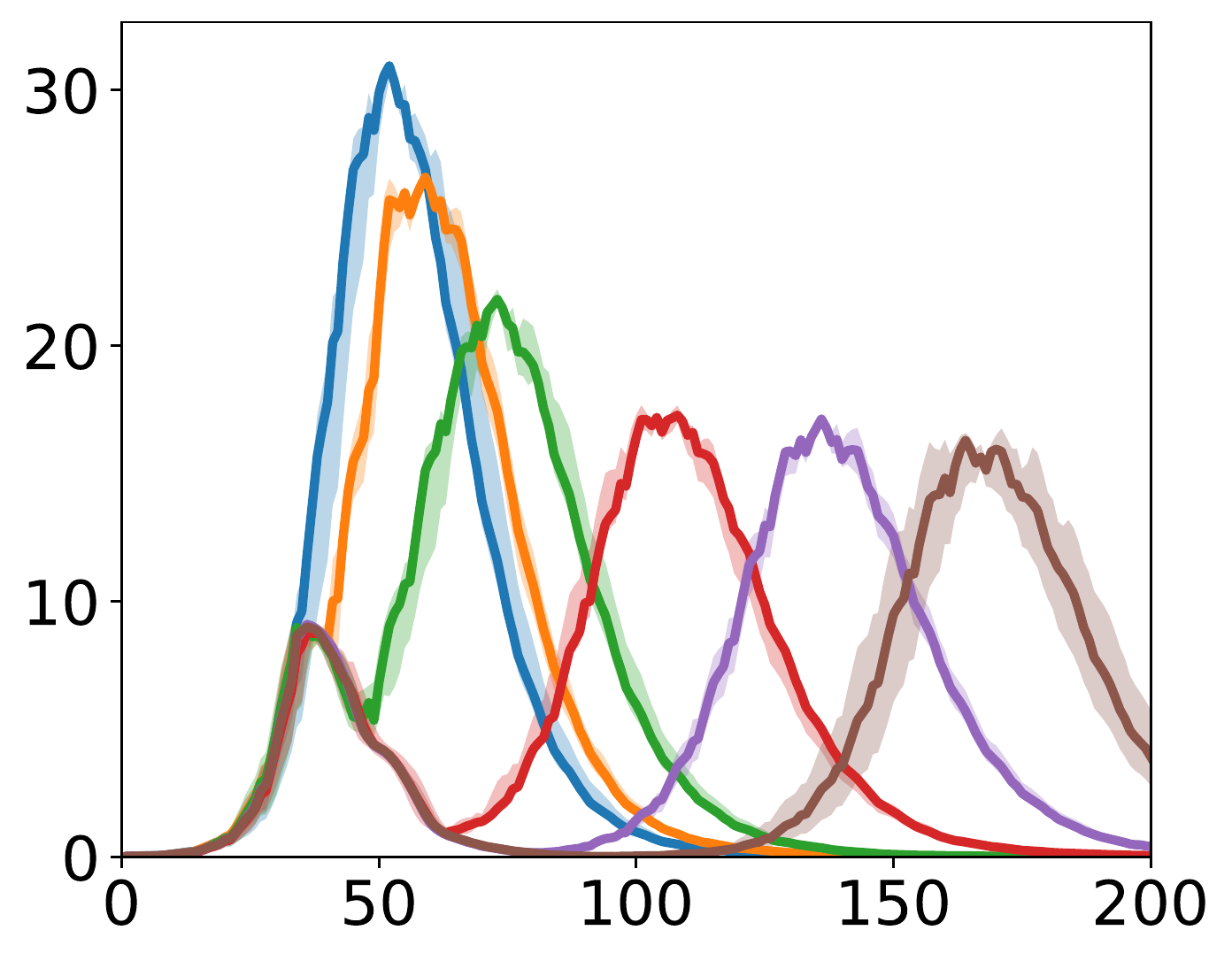}  & &
\includegraphics[width=\fourfig]{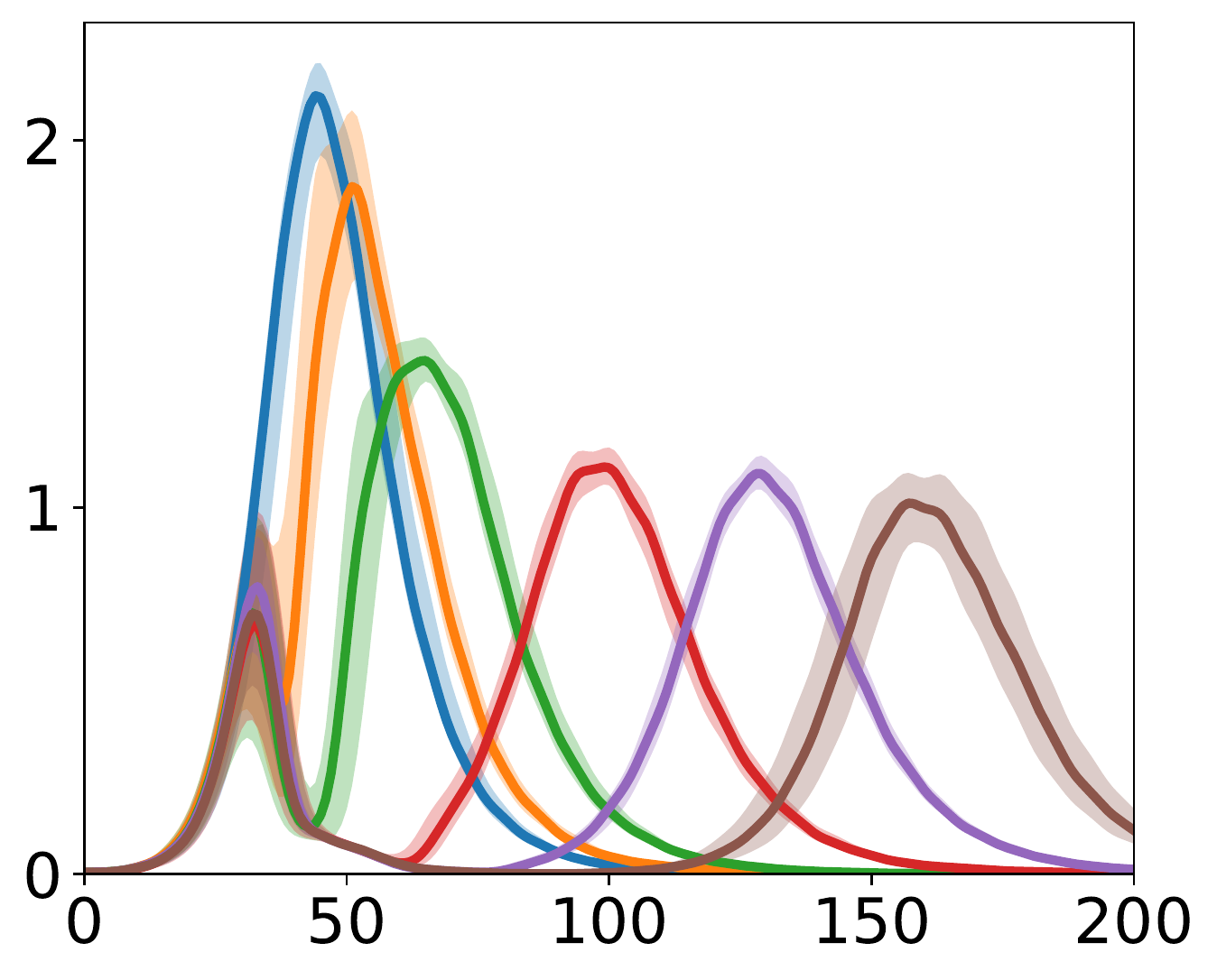}  &
\rotatebox{90}{\tiny{Reproduction number ($R_t$)}} &
\includegraphics[width=\fourfig]{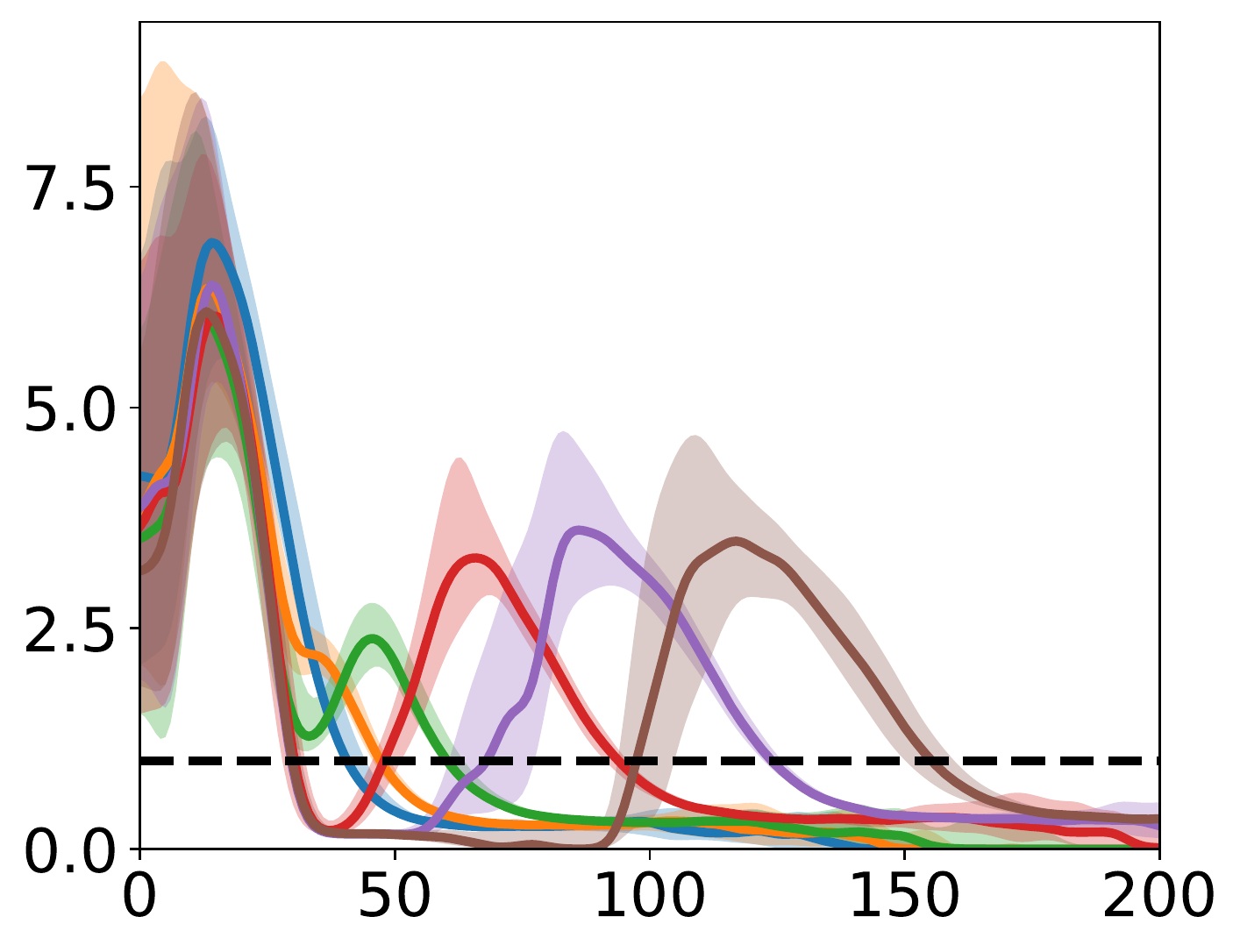}  &
\rotatebox{90}{University}
\\[-0.25cm]
&
\includegraphics[width=\fourfig]{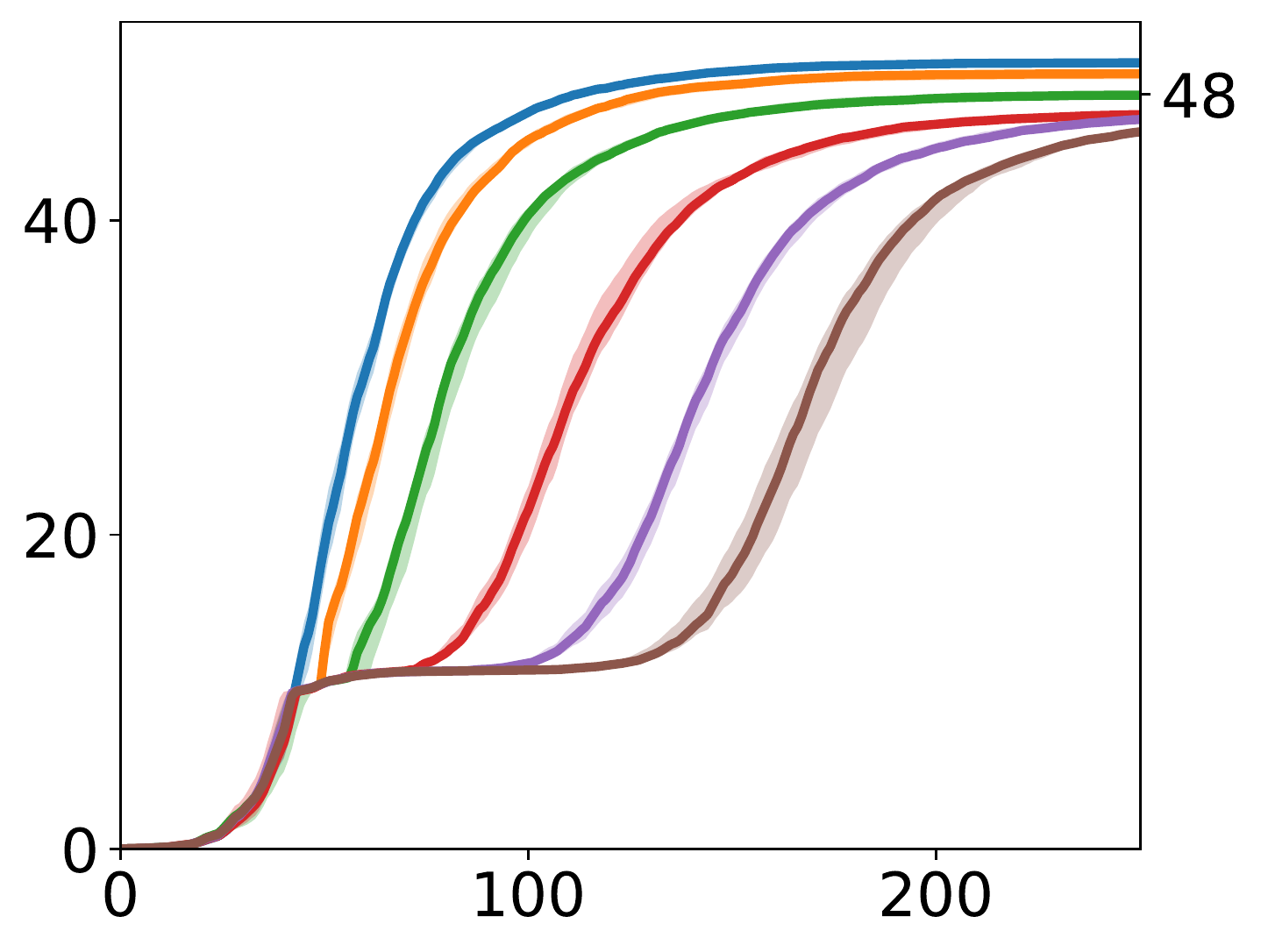}  & &
\includegraphics[width=\fourfig]{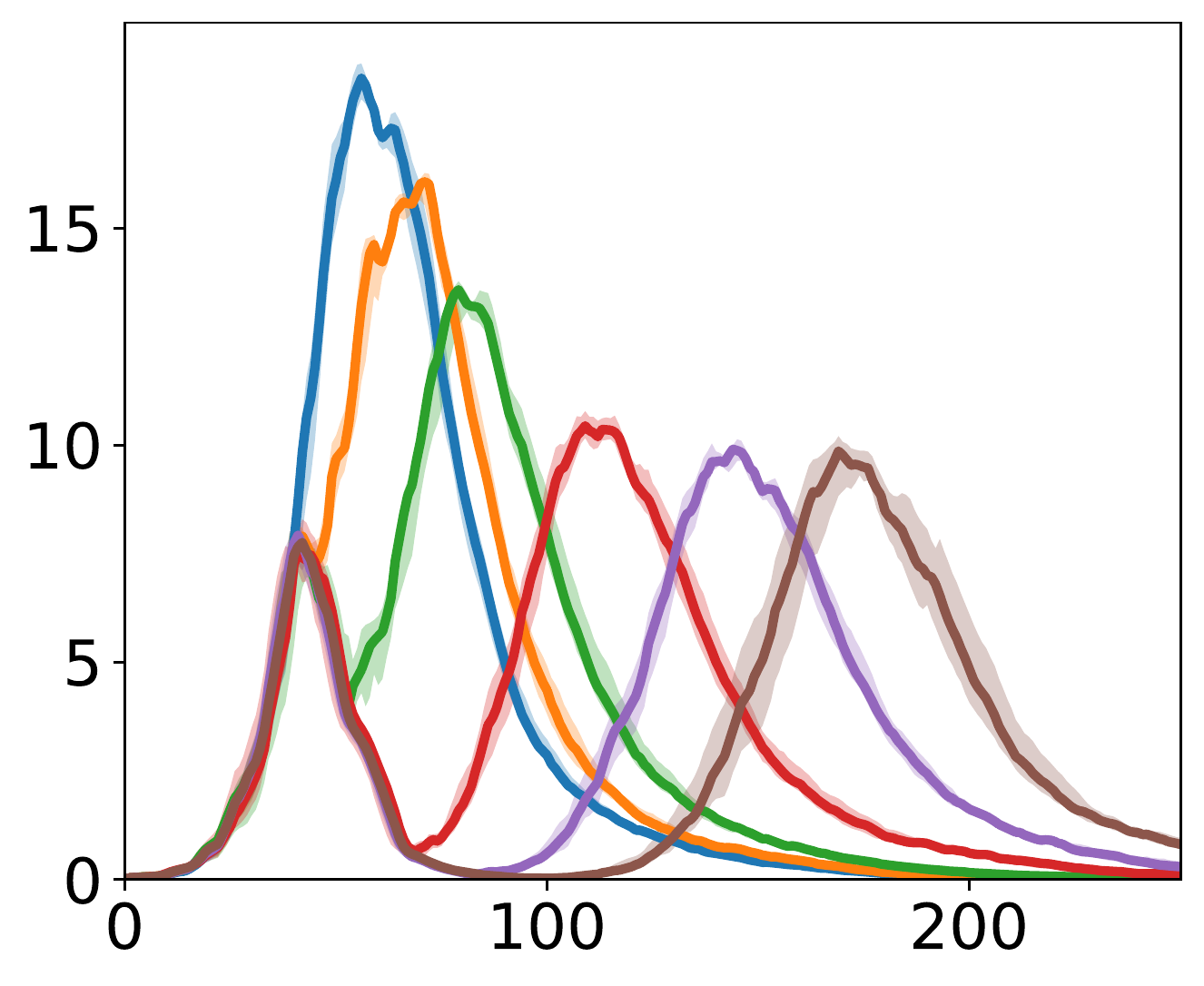}  & &
\includegraphics[width=\fourfig]{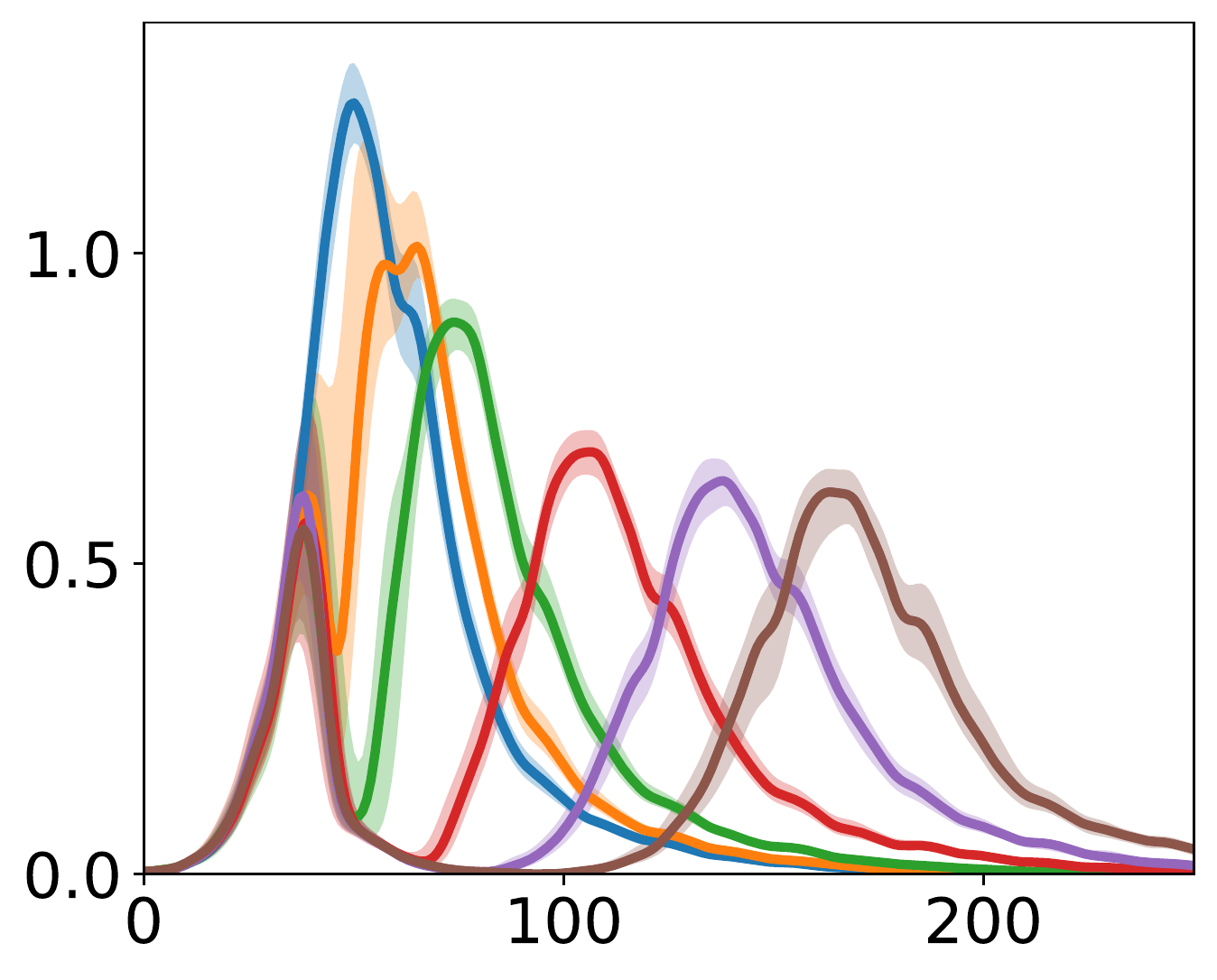}  &
\rotatebox{90}{\tiny{Reproduction Number ($R_t$)}} &
\includegraphics[width=\fourfig]{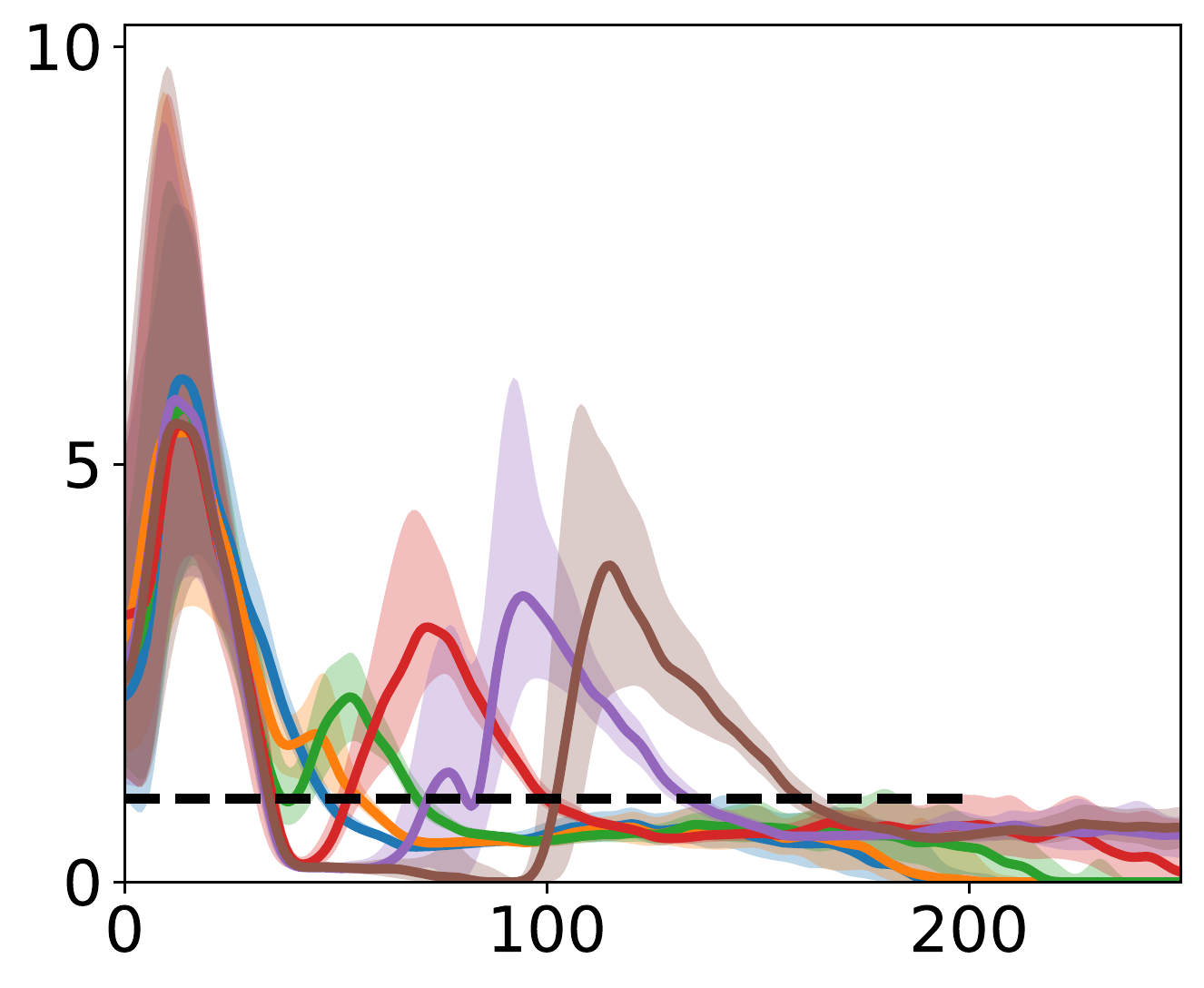}  &
\rotatebox{90}{Bike} \\
\multicolumn{8}{c}{\includegraphics[width=0.7\textwidth]{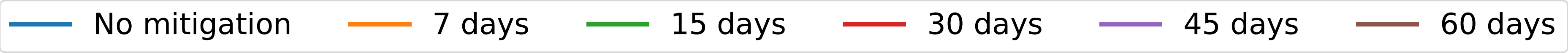}}

\end{tabular}

\caption{}
\label{figS:intervention_length}
\end{figure}
\newpage
\noindent \textbf{Figure~\ref{figS:intervention_length}:}Infection spreading with the intervention strategy of varying intervention length similar to Part II in Figure~2. With long intervention, the total number of infected people is reduced significantly in Chicago, Los Angeles, London, and Bike datasets. Here, $30$ days are enough to have the infected people recovered, and the second wave of infection does not occur. In the other datasets, after the intervention, there is a second wave of infection, leaving a similar number of infected people. We also test longer interventions in Part II in the datasets with larger populations and the patterns in Part I hold.

\newpage
\begin{figure}[H]
    \centering

\begin{tabular}{m{0.1cm}m{\fivefig}@{}m{0.1cm}m{\fivefig}@{}m{0.1cm}m{\fivefig}@{}m{0.1cm}m{\fivefig}@{}m{0.1cm}}

\multirow{14}{*}{\includegraphics[height=4cm]{Fig/labels/TotalInfected.pdf}}&
\includegraphics[width=\fivefig]{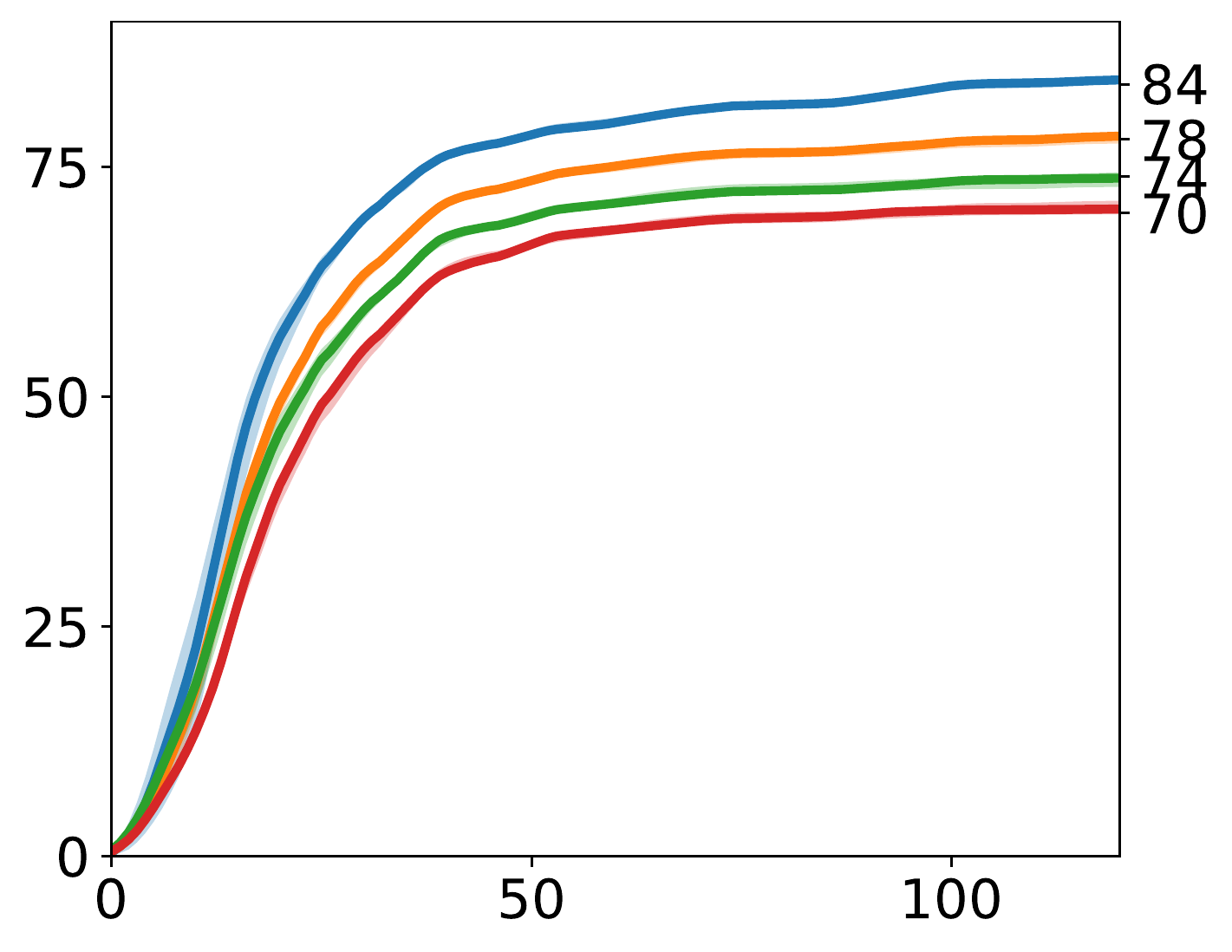} &
\multirow{14}{*}{\includegraphics[height=4cm]{Fig/labels/ActiveInfected.pdf}}&
\includegraphics[width=\fivefig]{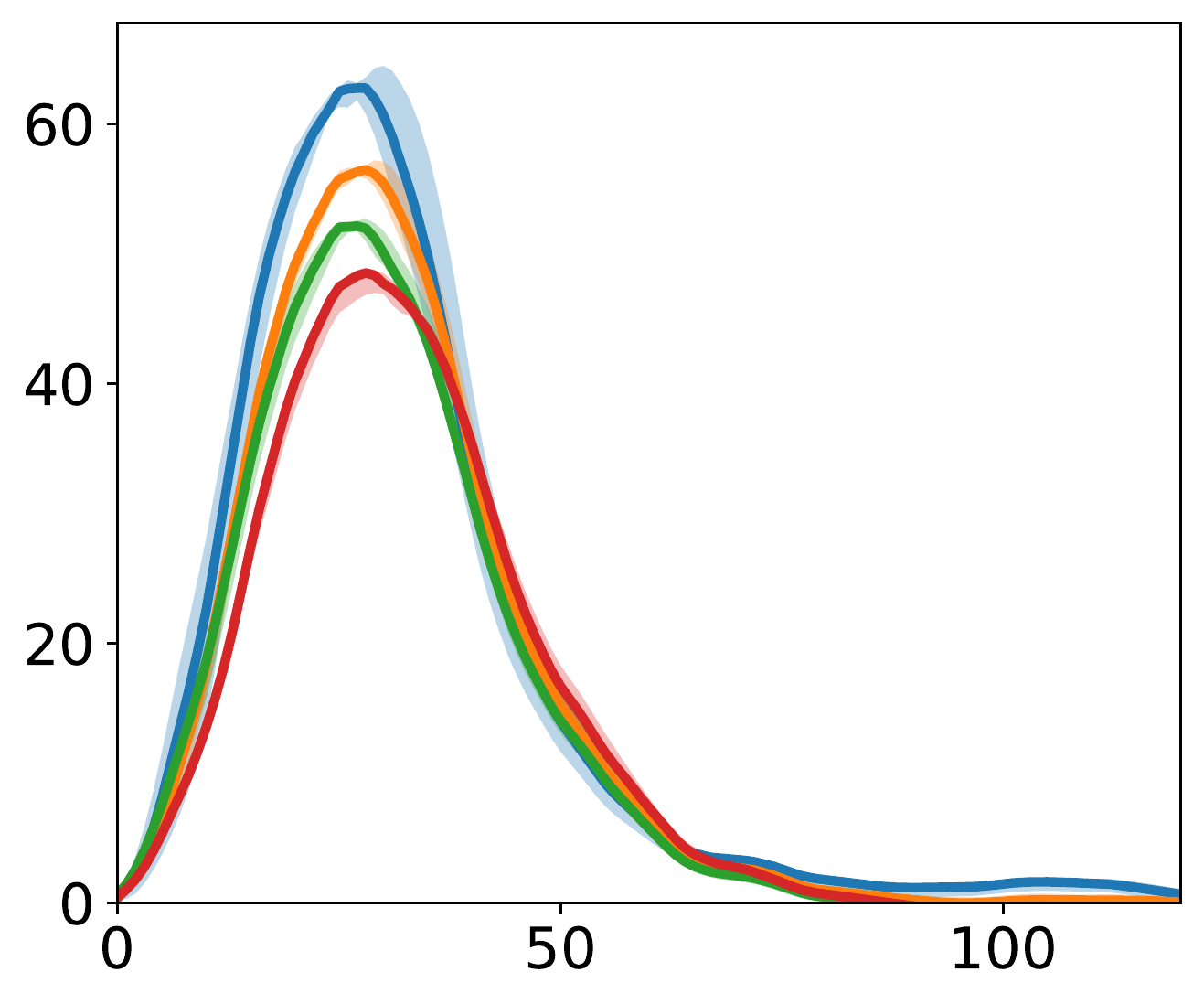}  &
\multirow{14}{*}{\includegraphics[height=4cm]{Fig/labels/newInfected.pdf}}&
\includegraphics[width=\fivefig]{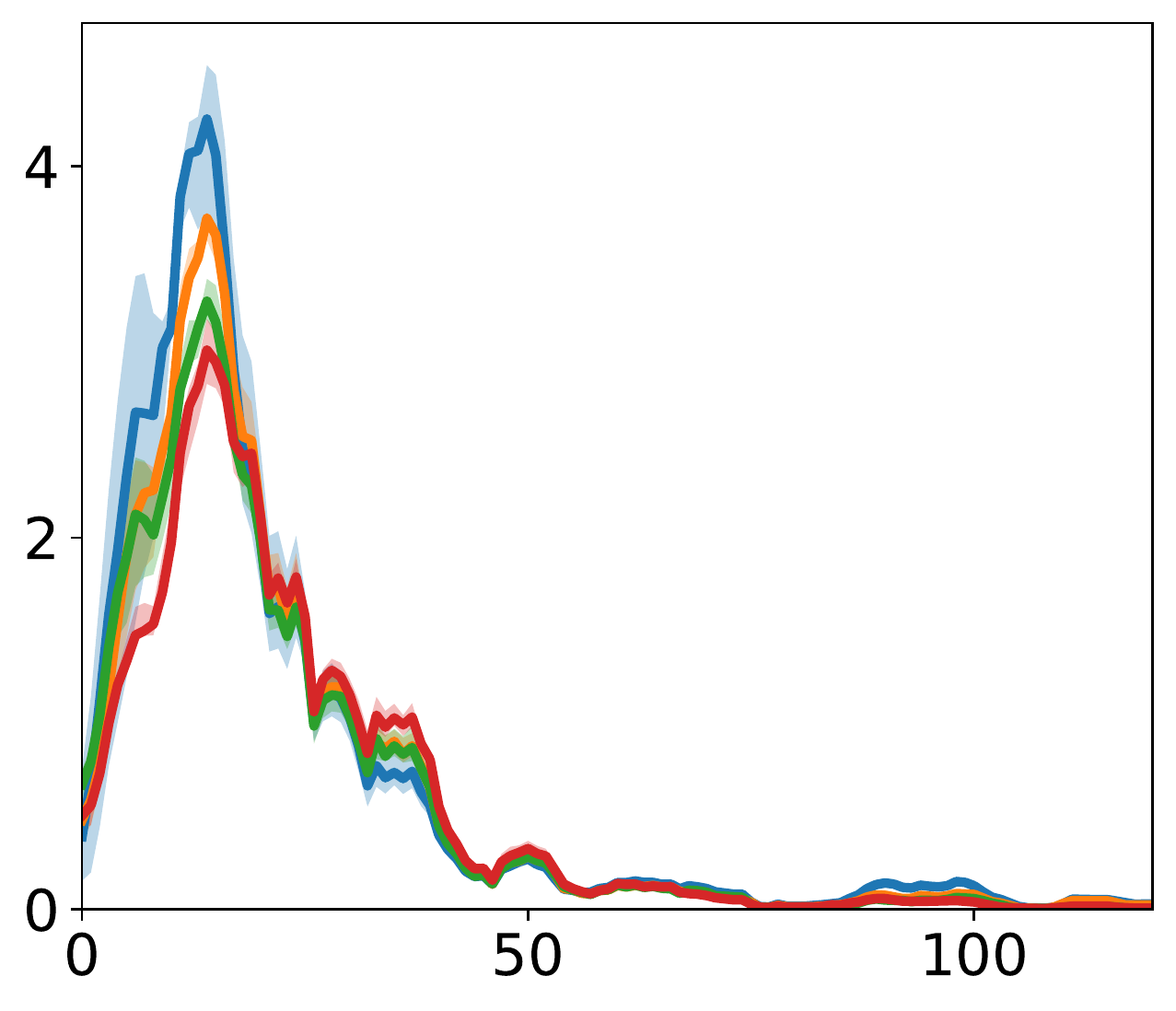}  &
\multirow{14}{*}{\includegraphics[height=2.5cm]{Fig/labels/growthrate.pdf}}&
\includegraphics[width=\fivefig]{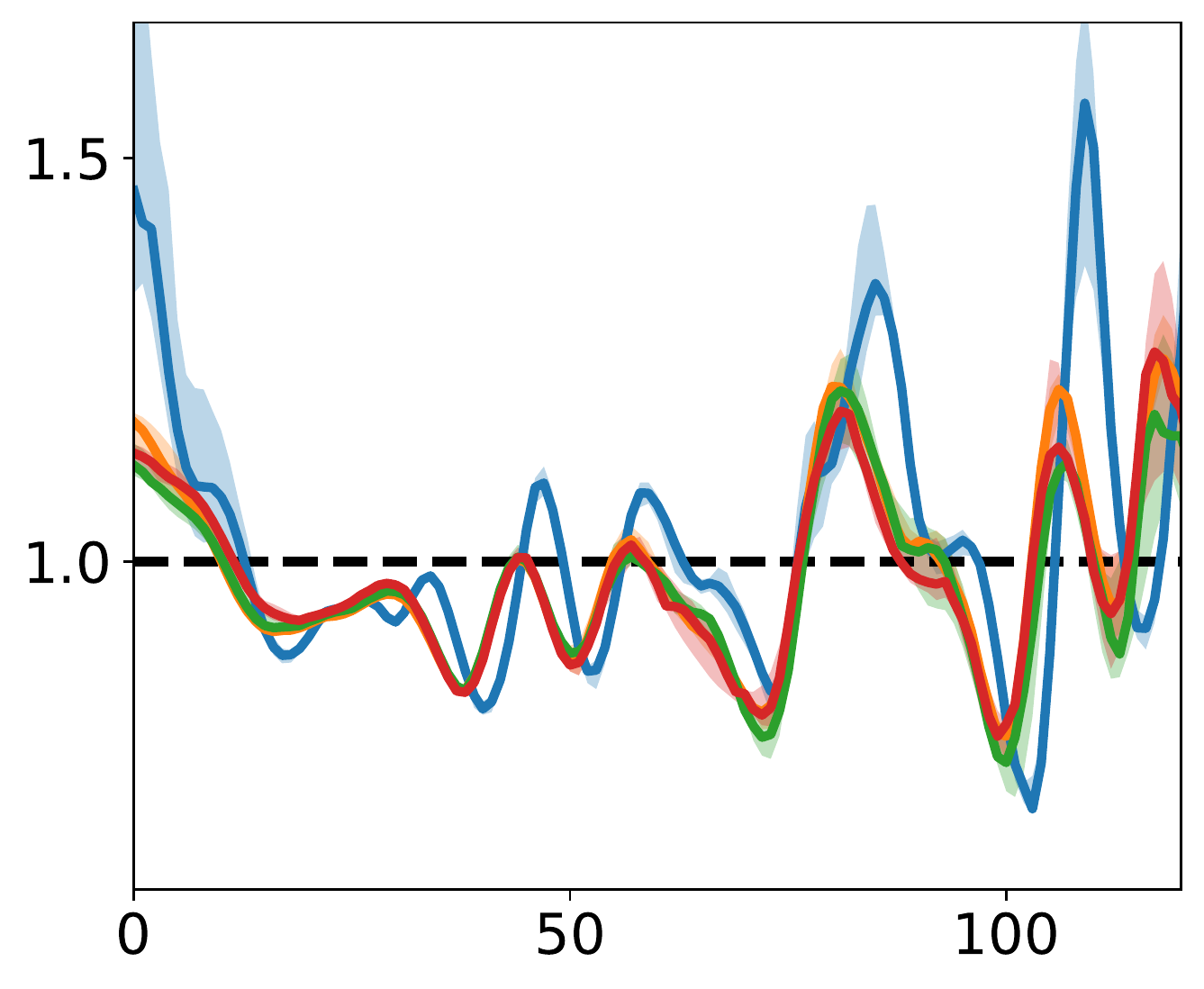} & 
\rotatebox{90}{Tokyo}
\\ [-0.25cm]
&

\includegraphics[width=\fivefig]{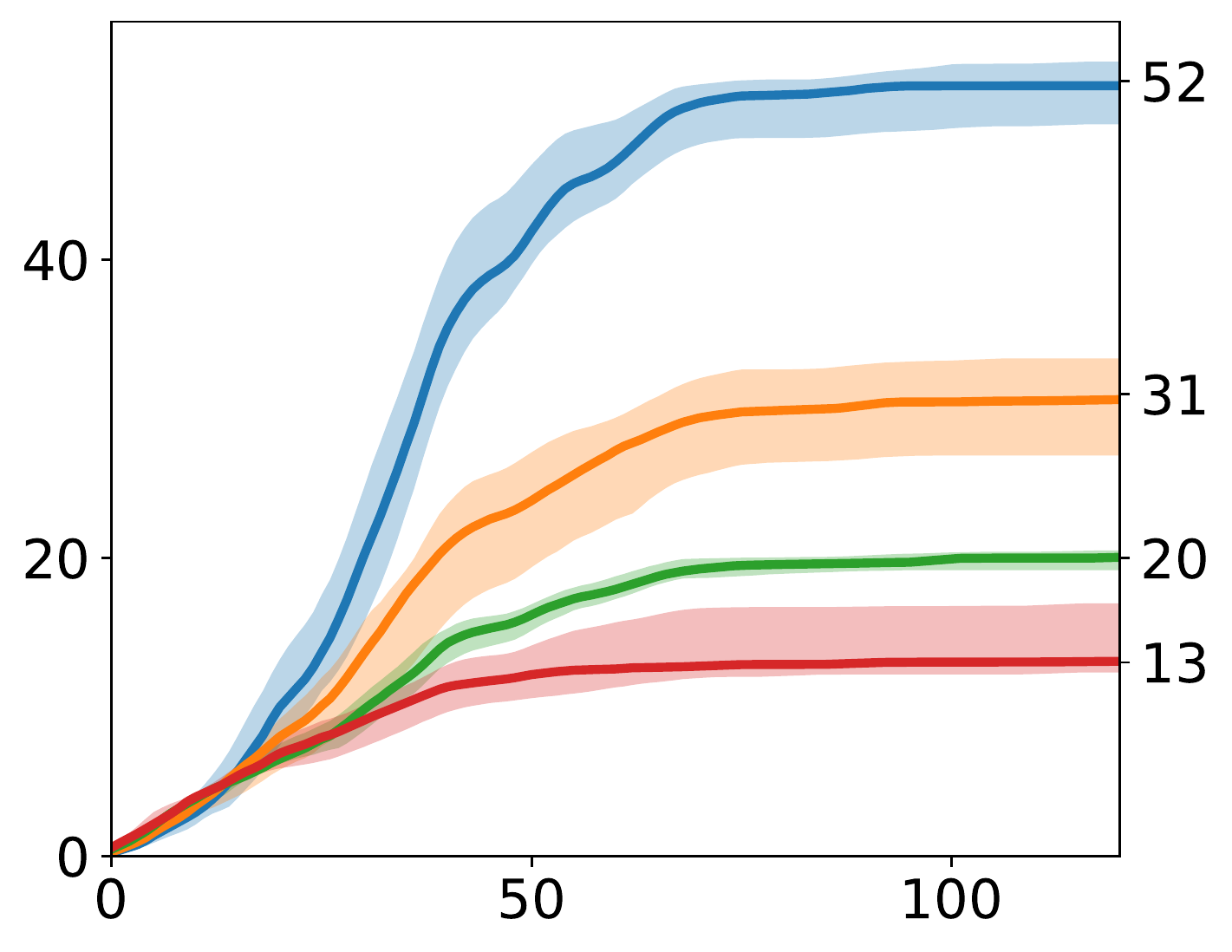} & &
\includegraphics[width=\fivefig]{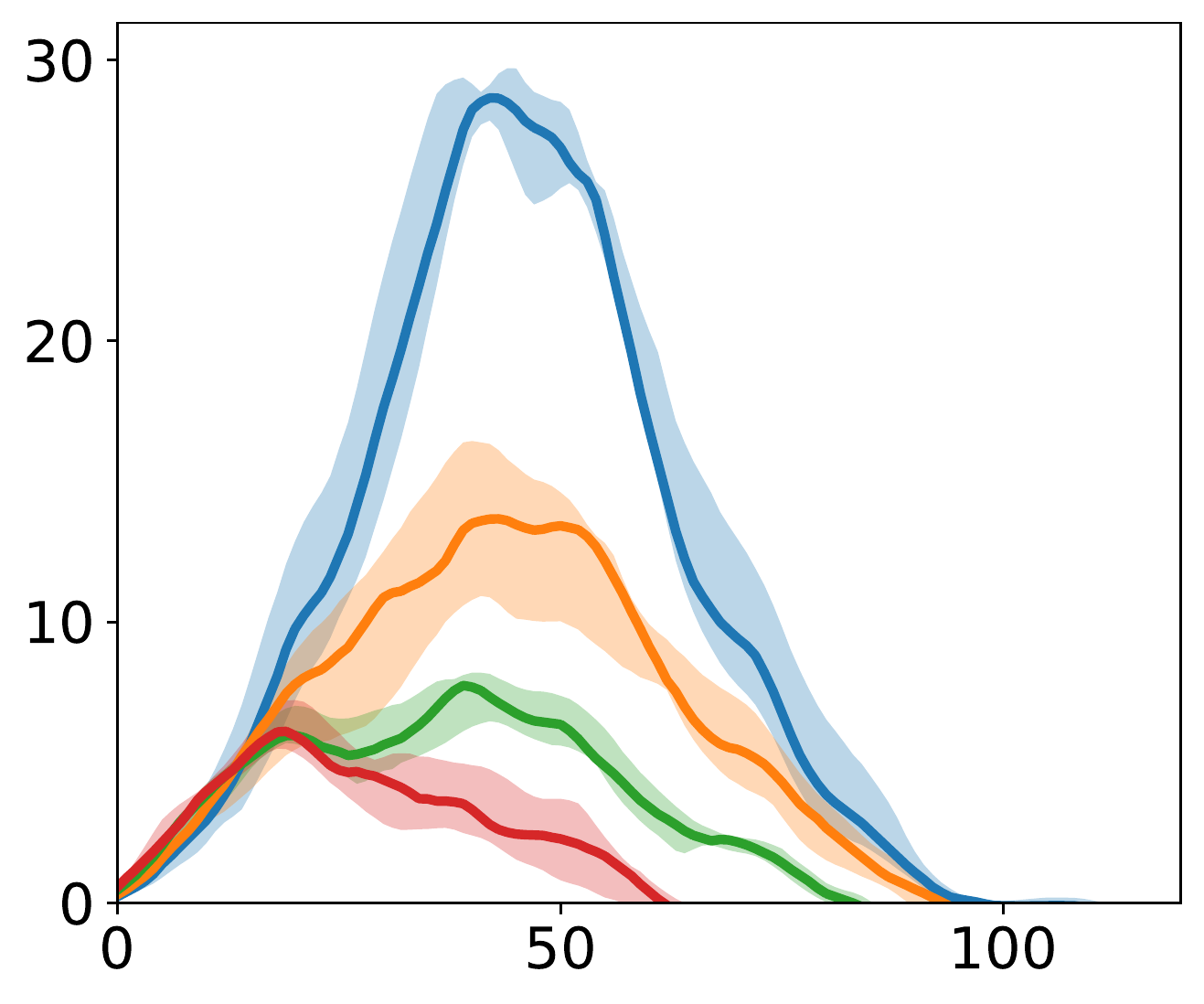}  & &
\includegraphics[width=\fivefig]{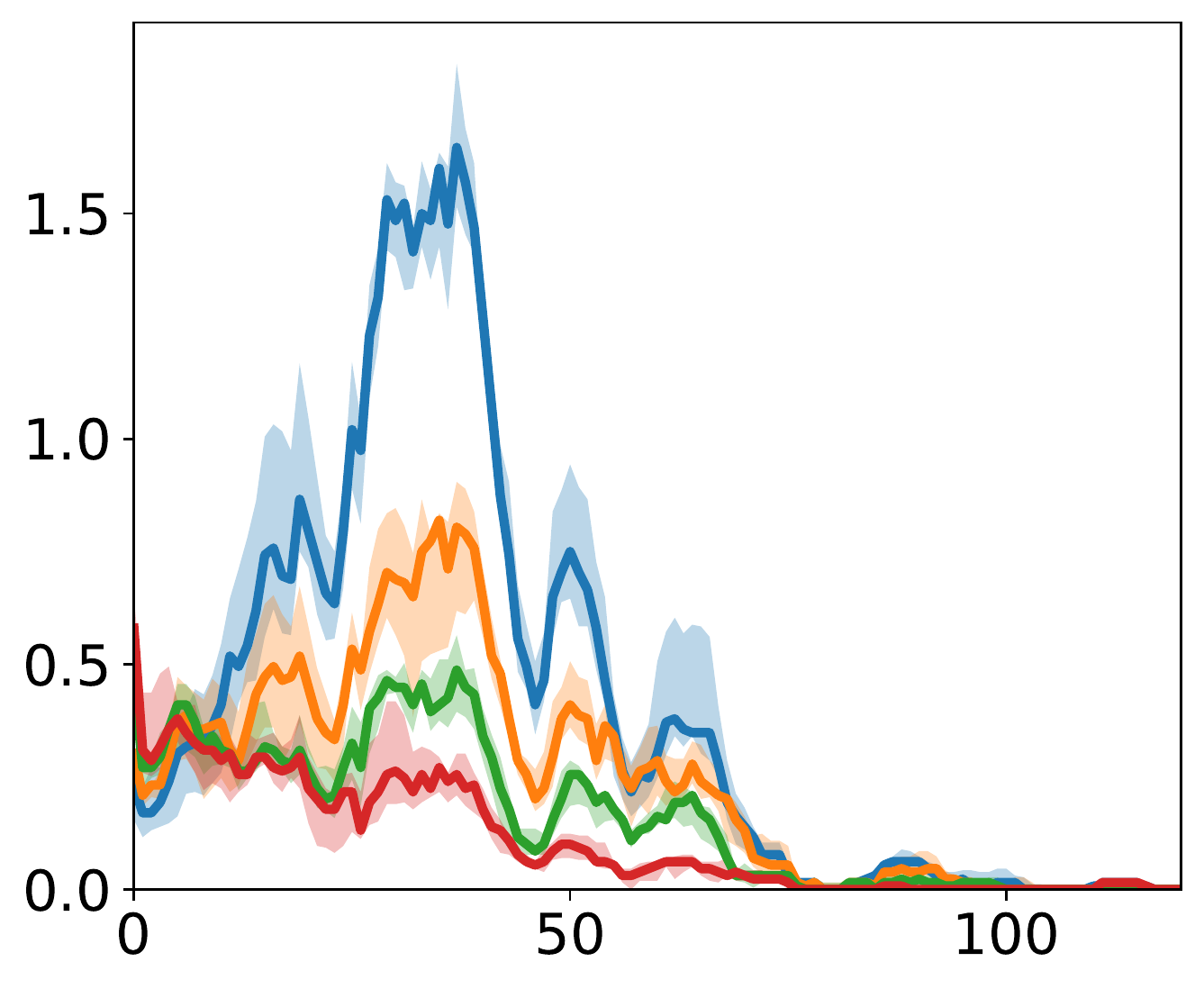}  & &
\includegraphics[width=\fivefig]{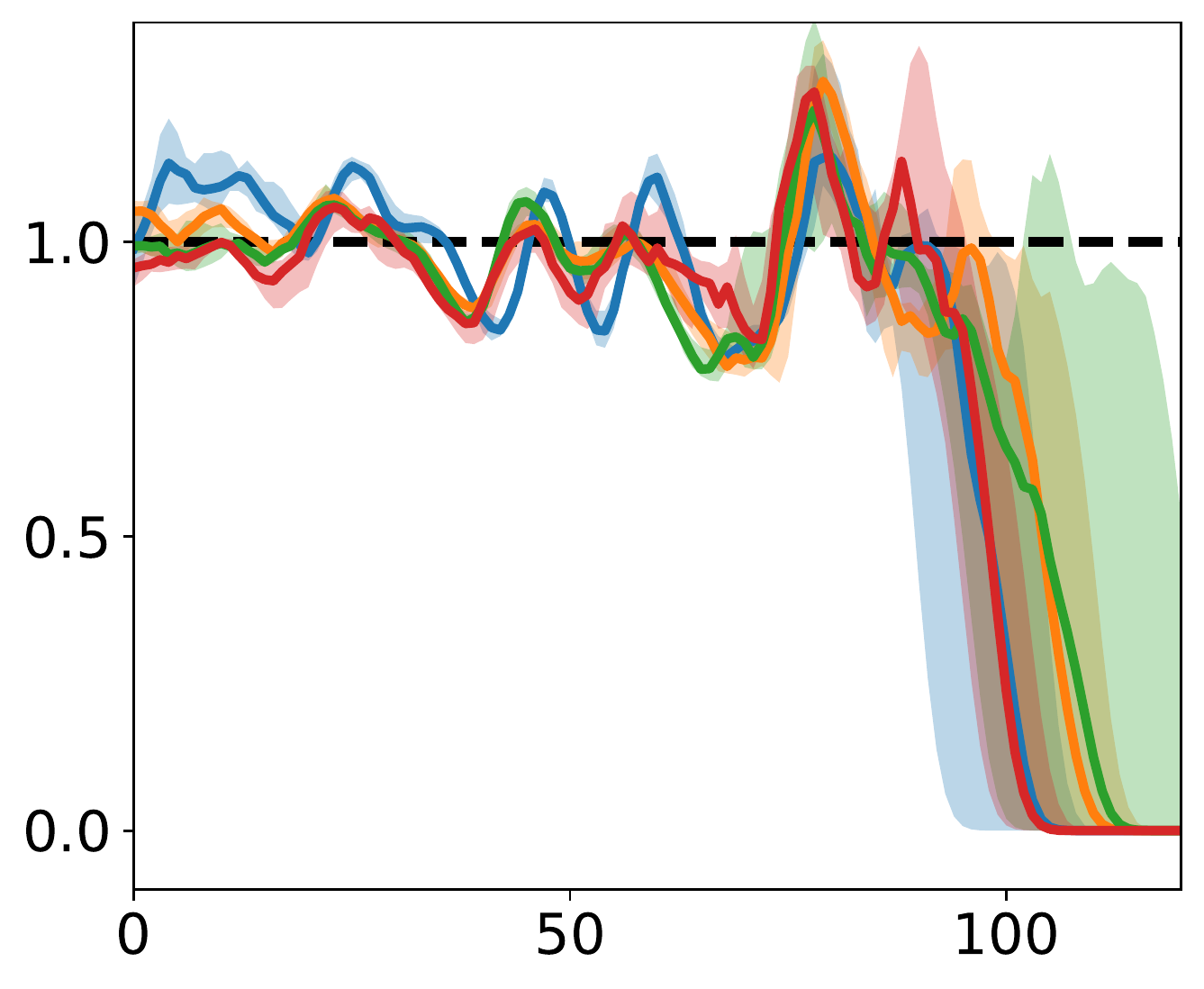}  &
\rotatebox{90}{Chicago}
\\ [-0.25cm]

&
\includegraphics[width=\fivefig]{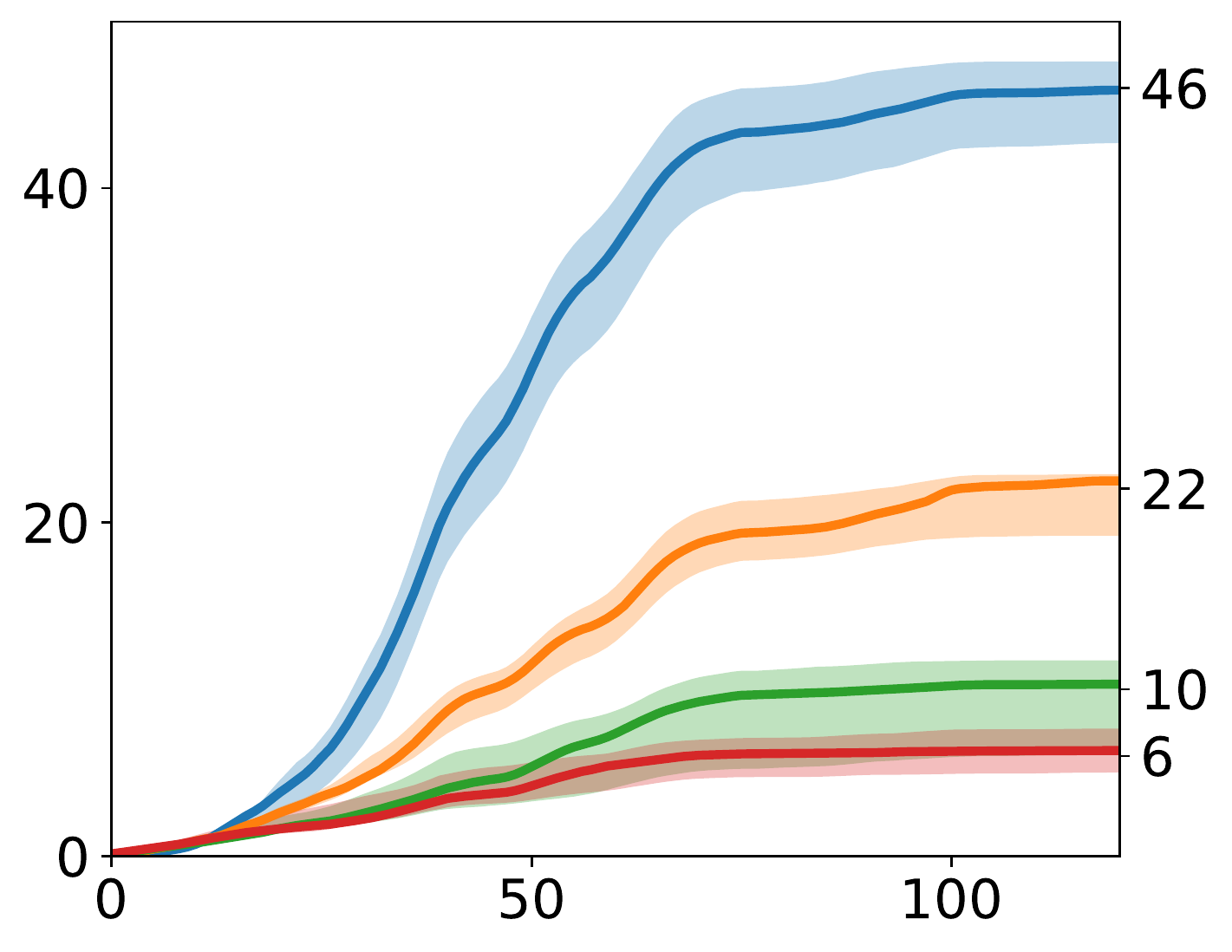} & &
\includegraphics[width=\fivefig]{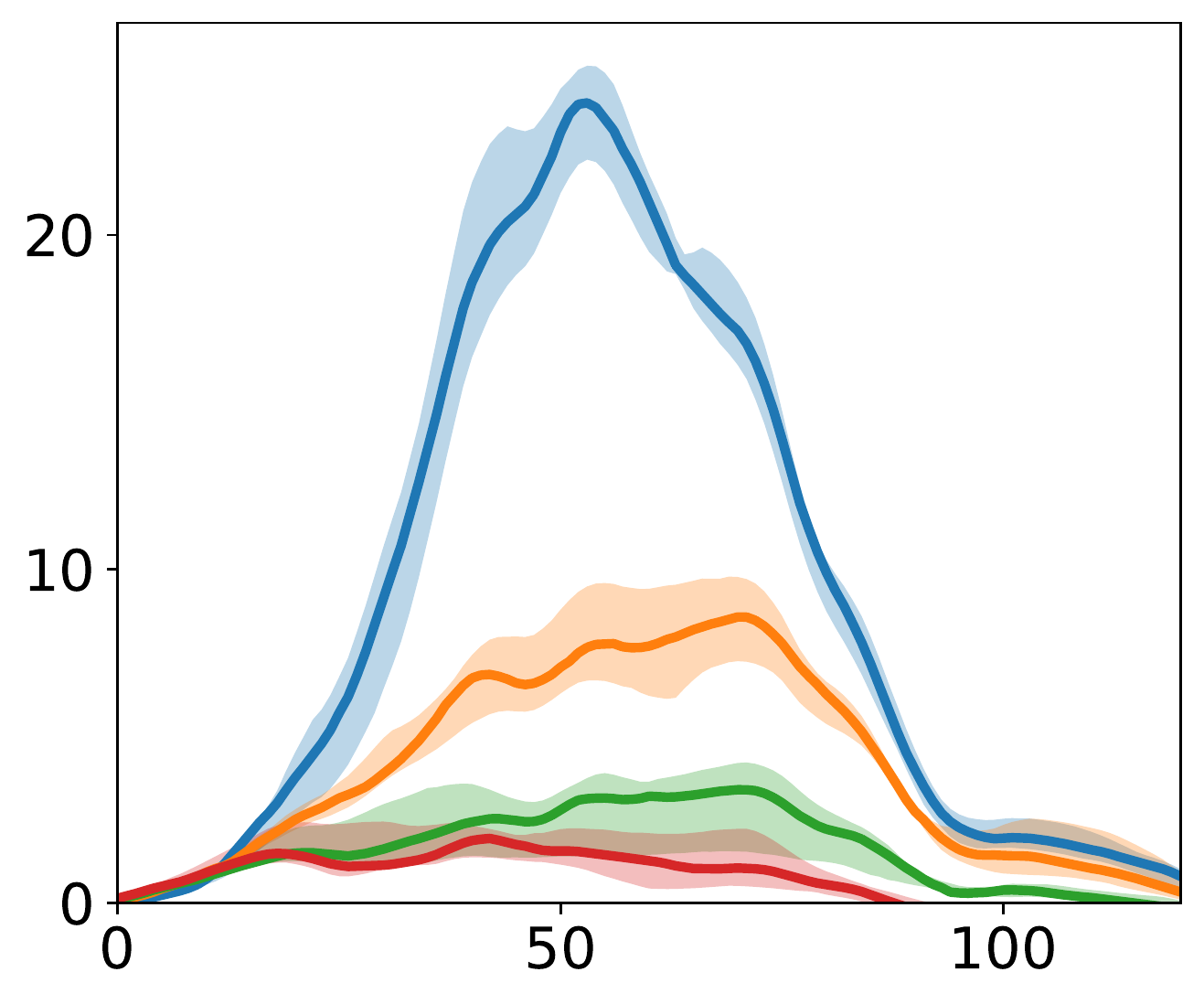} & &
\includegraphics[width=\fivefig]{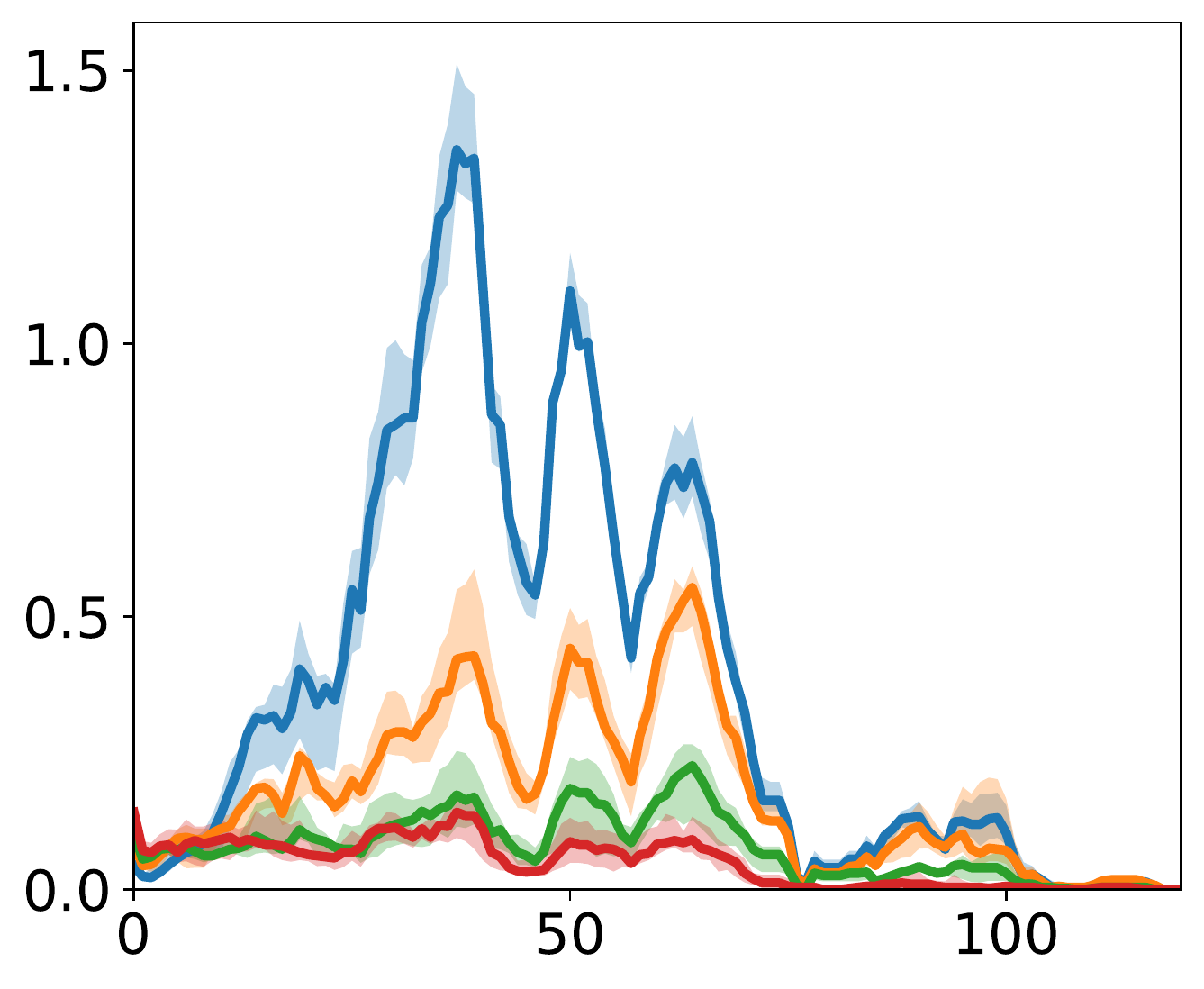}  & &
\includegraphics[width=\fivefig]{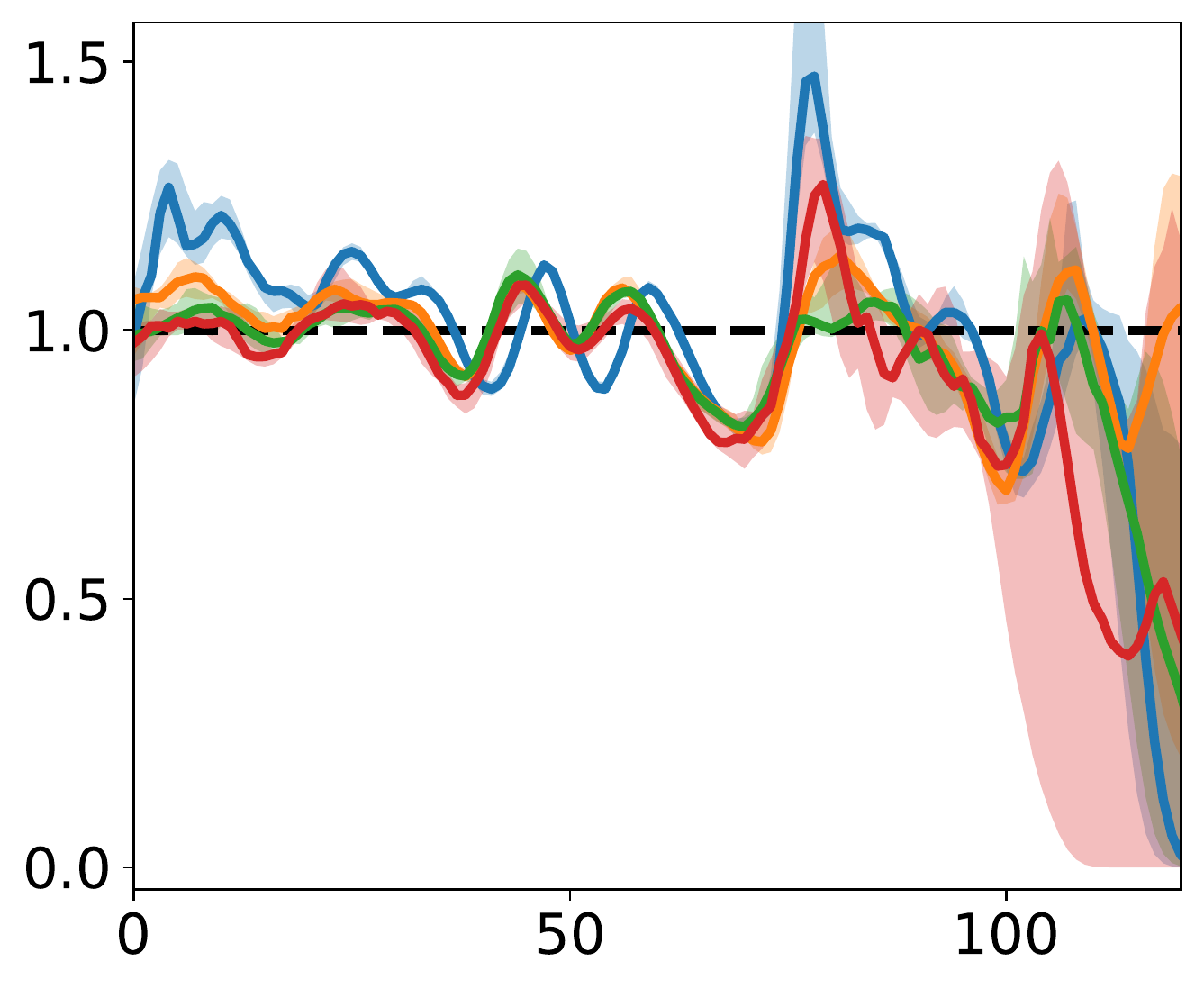} &
\rotatebox{90}{Los Angeles}
\\ [-0.25cm]

&
\includegraphics[width=\fivefig]{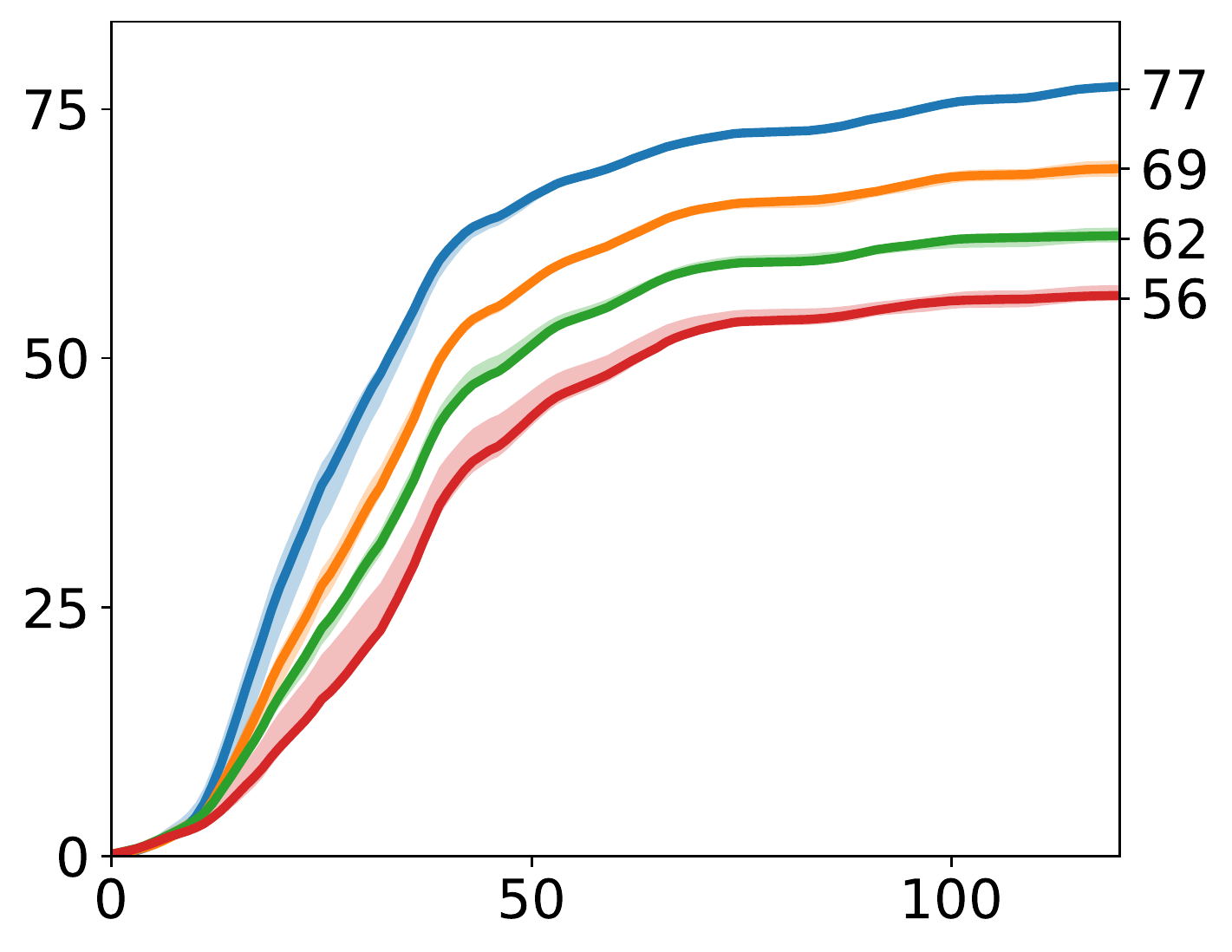} &&
\includegraphics[width=\fivefig]{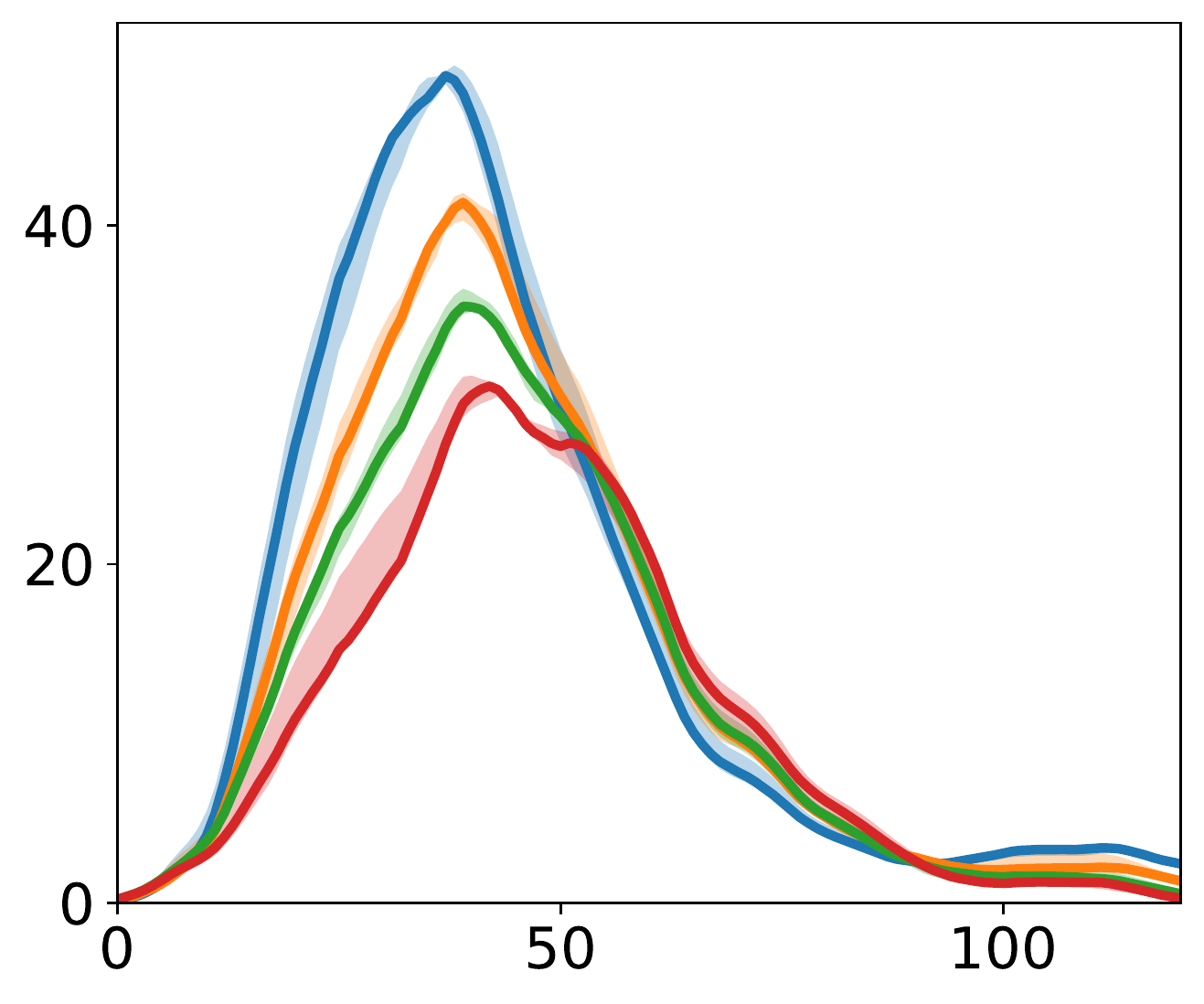}  &&
\includegraphics[width=\fivefig]{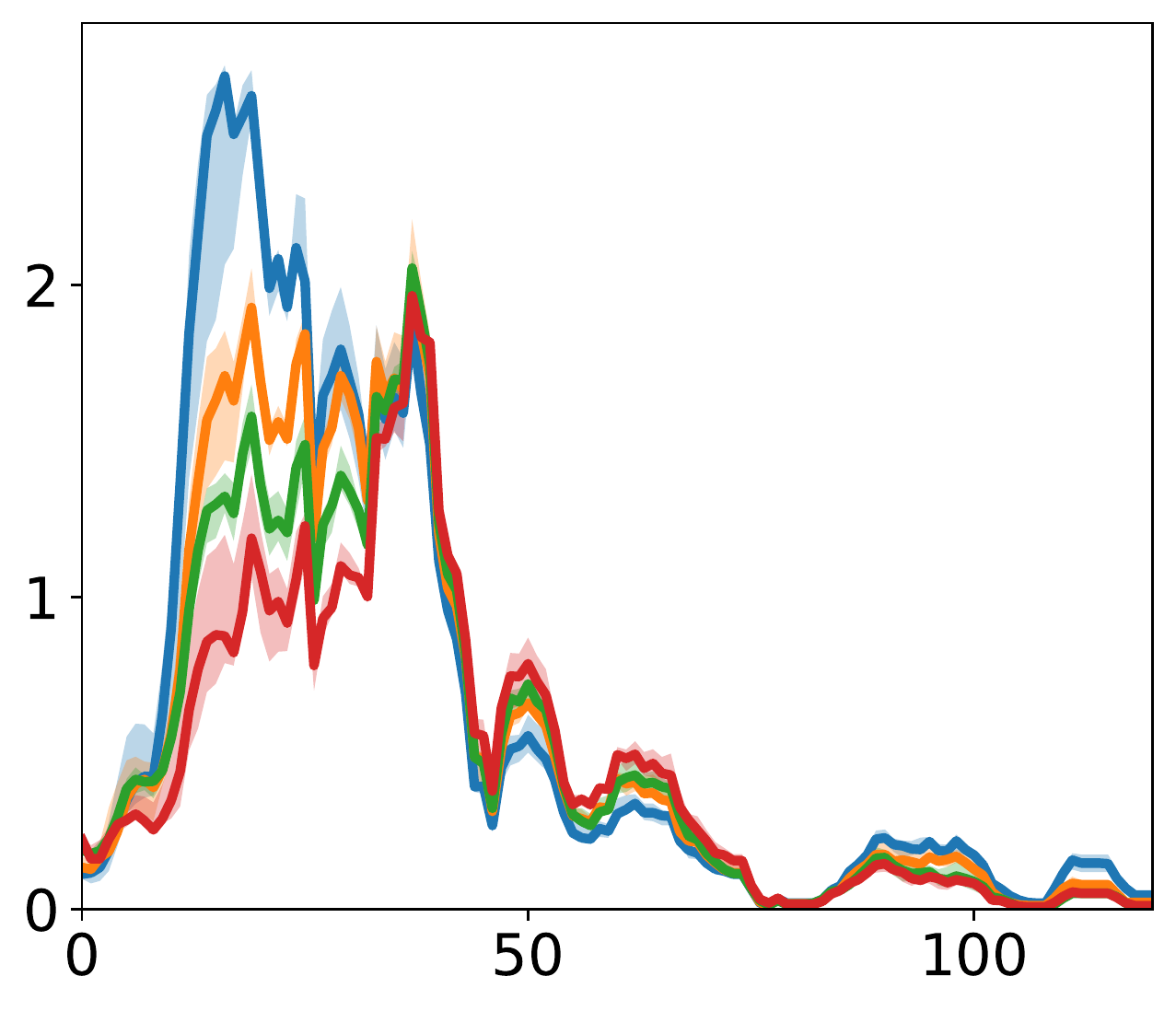}  &&
\includegraphics[width=\fivefig]{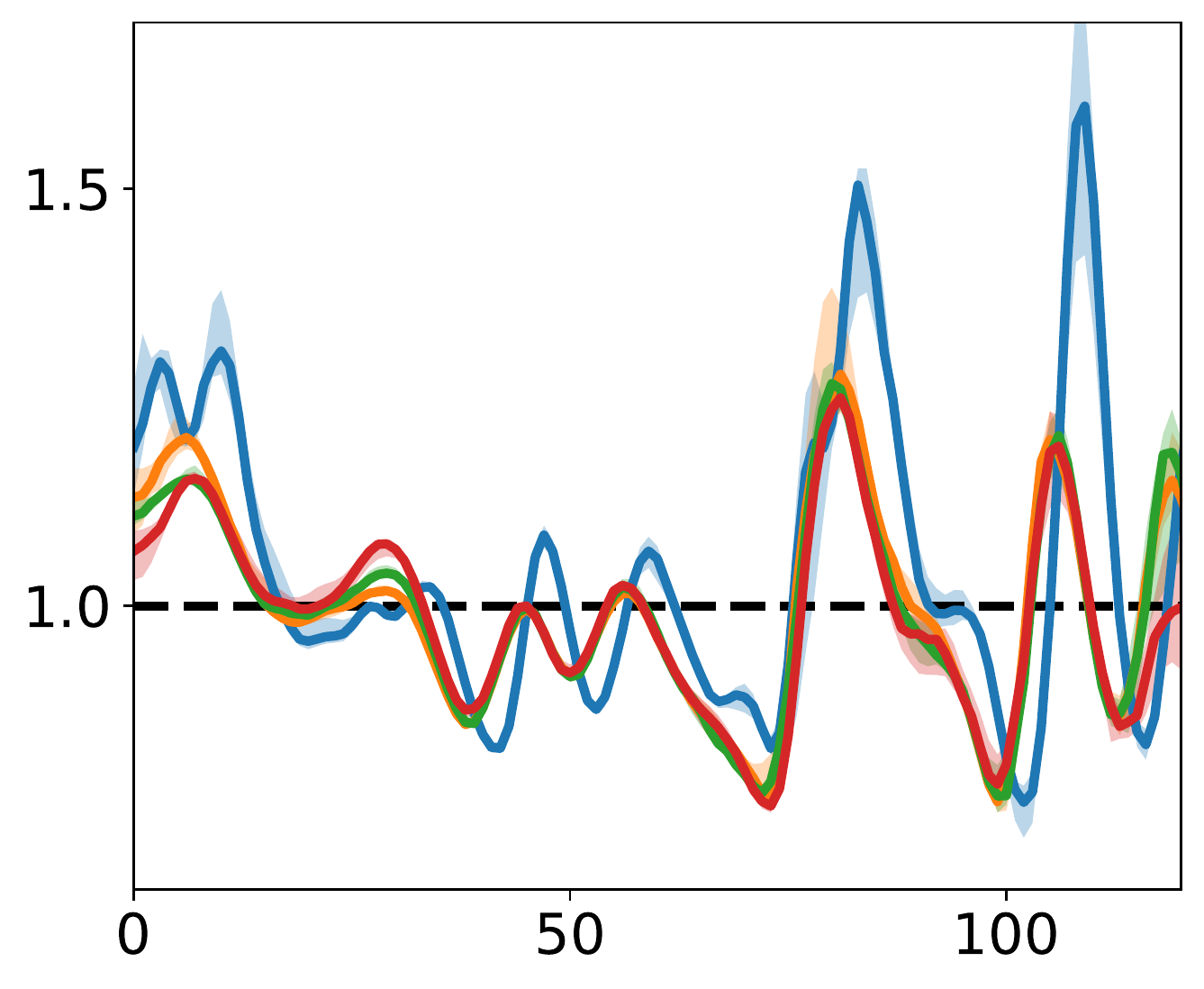} &
\rotatebox{90}{Jakarta}
\\ [-0.25cm]

&
\includegraphics[width=\fivefig]{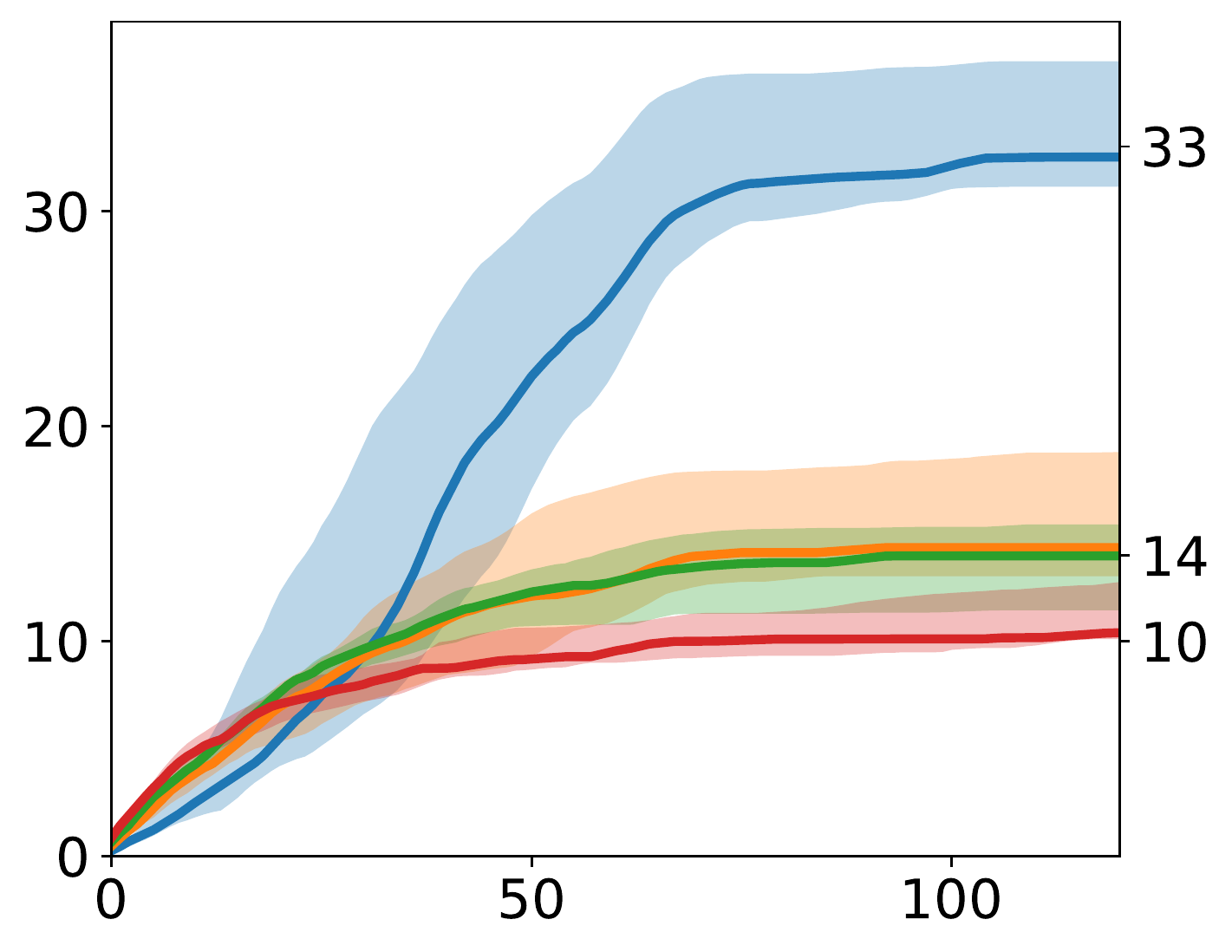} &&
\includegraphics[width=\fivefig]{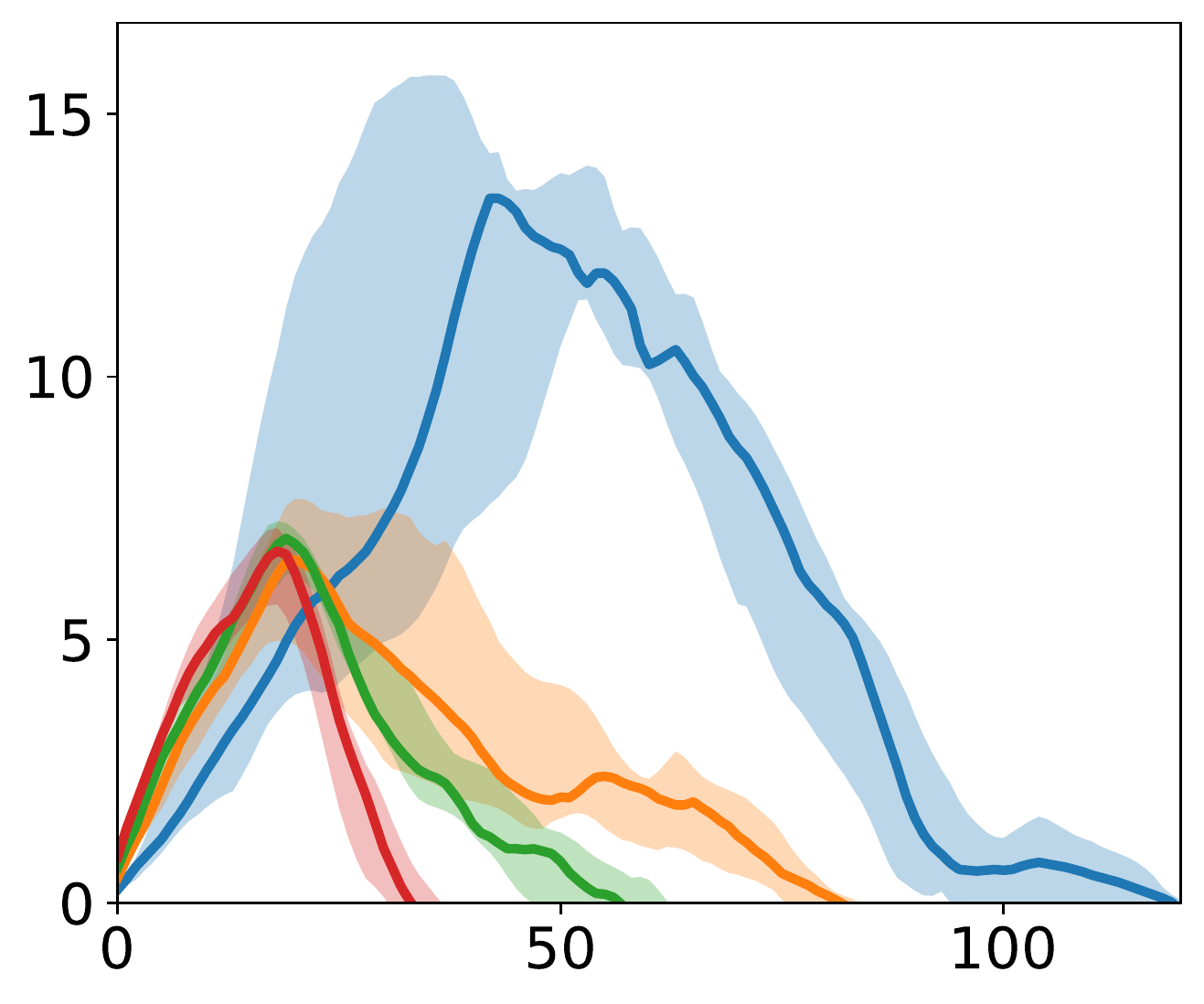}  &&
\includegraphics[width=\fivefig]{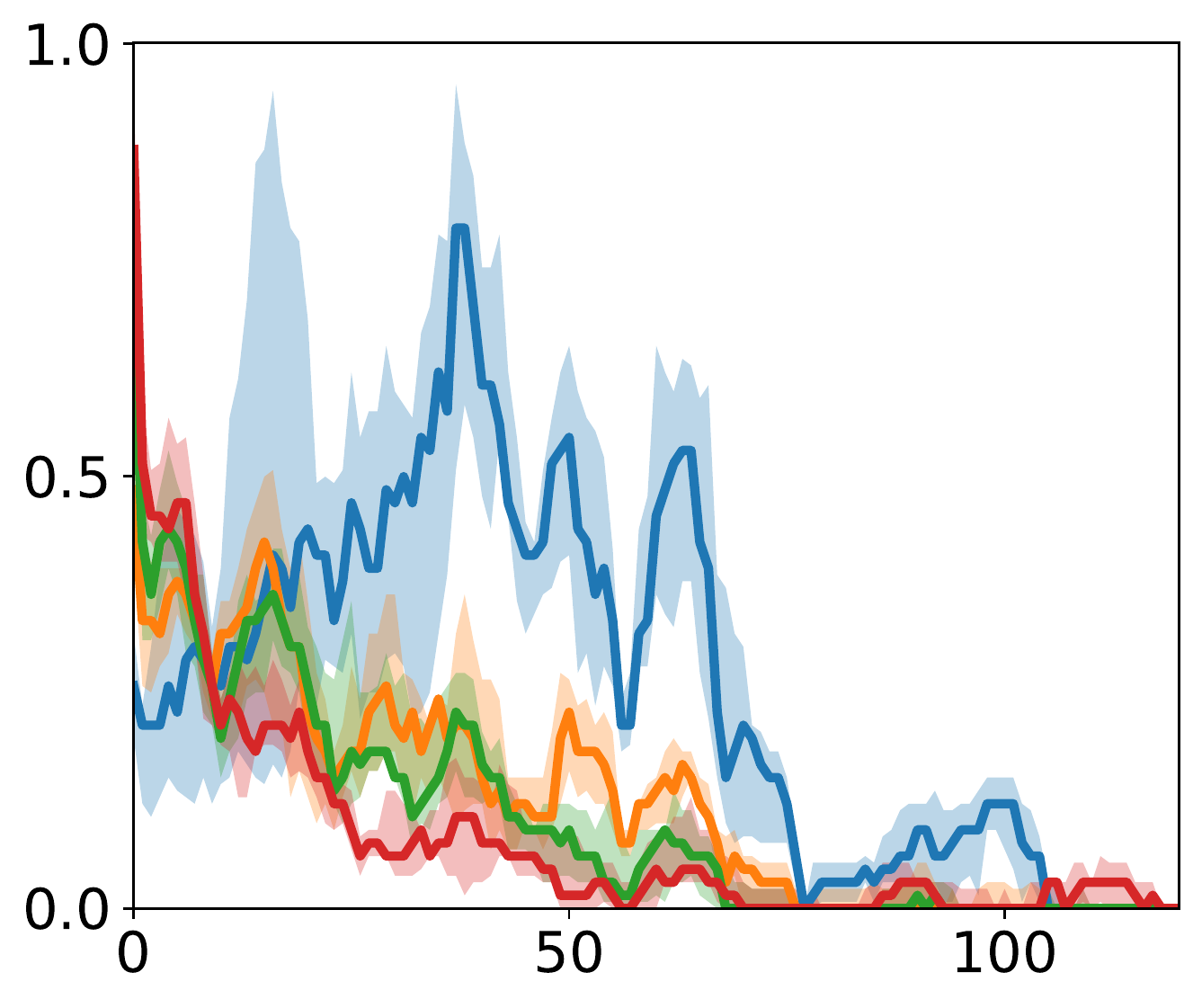}  &&
\includegraphics[width=\fivefig]{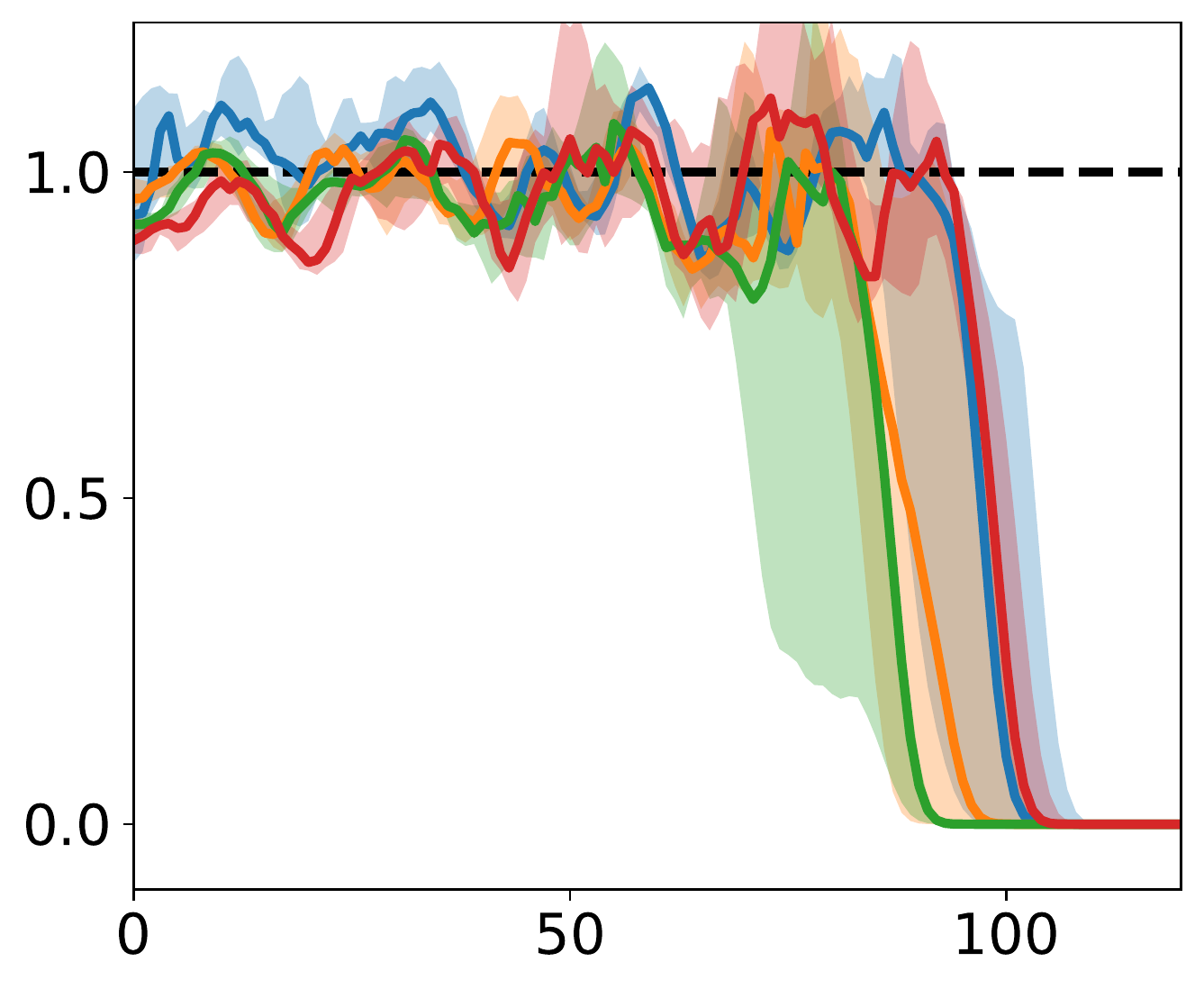} &
\rotatebox{90}{London}
\\ [-0.25cm]

&
\includegraphics[width=\fivefig]{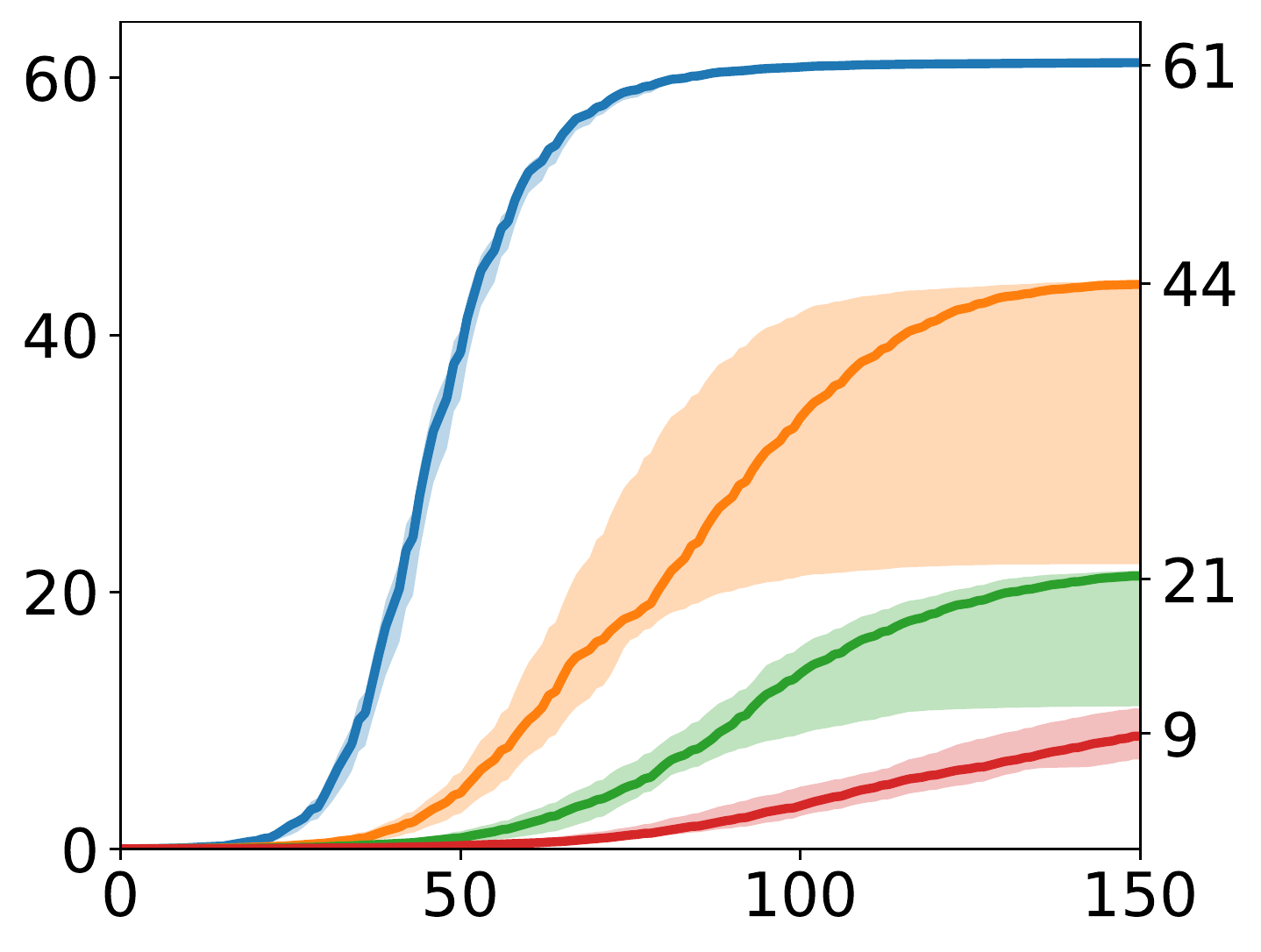} &&
\includegraphics[width=\fivefig]{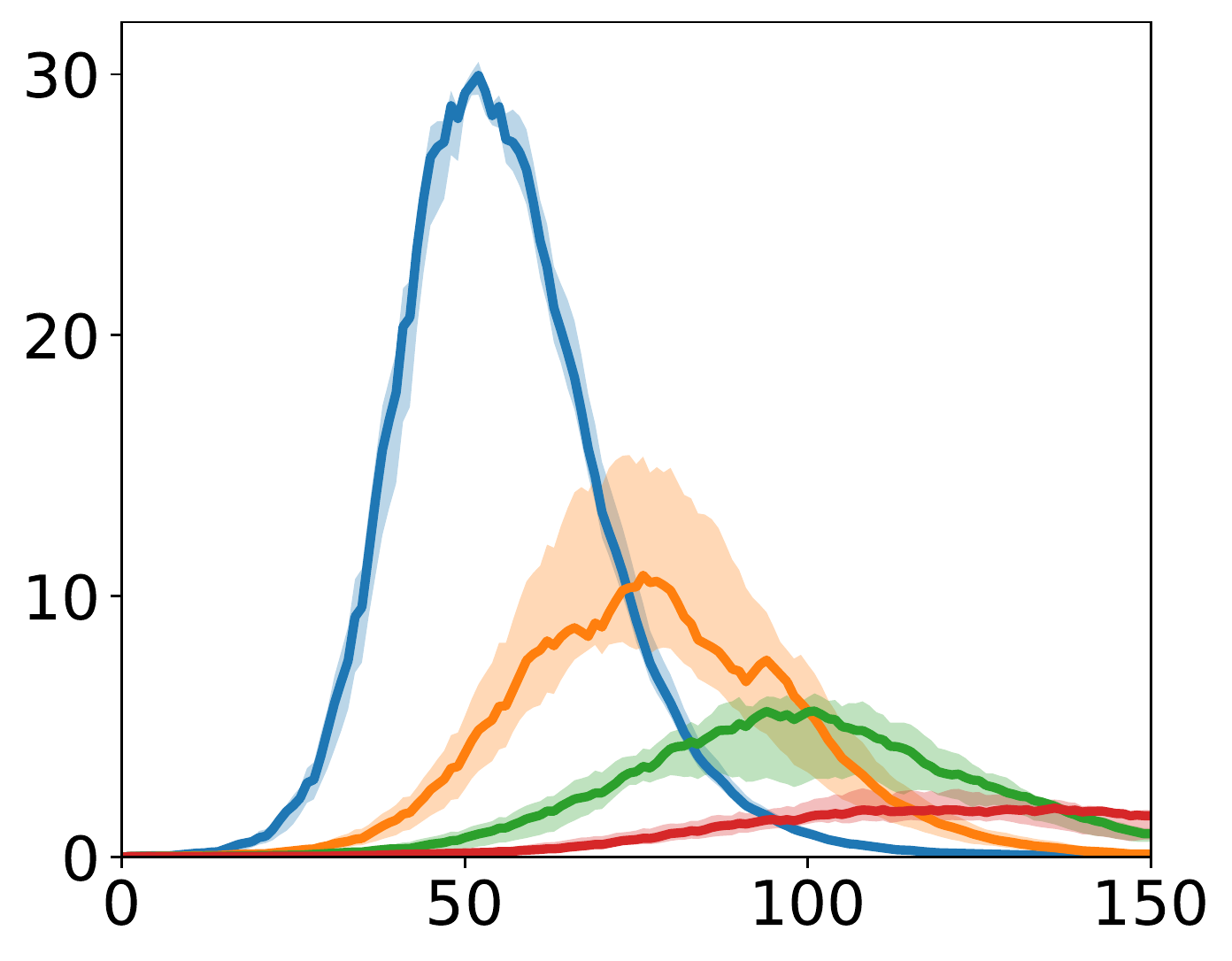} &&
\includegraphics[width=\fivefig]{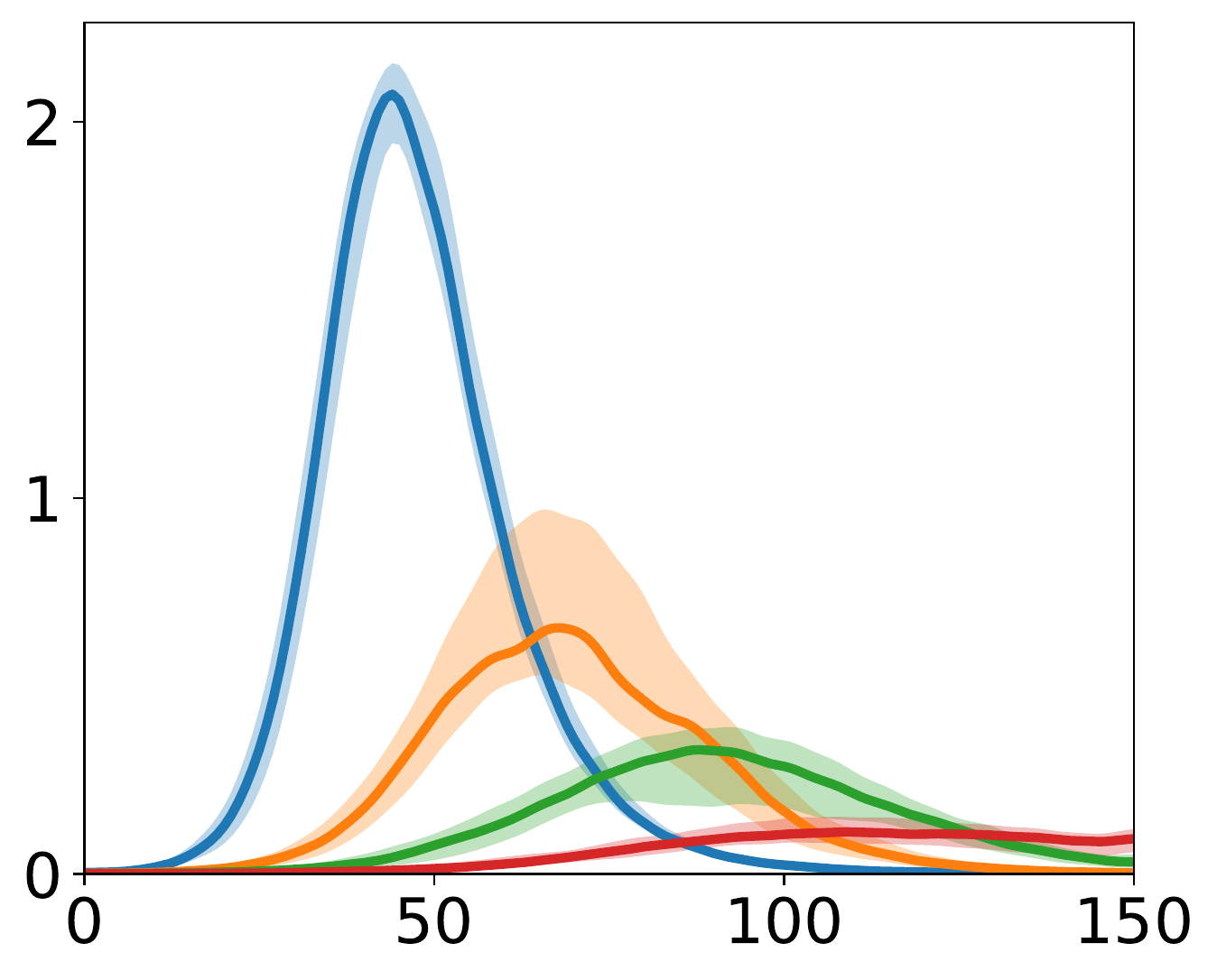} &&
\includegraphics[width=\fivefig]{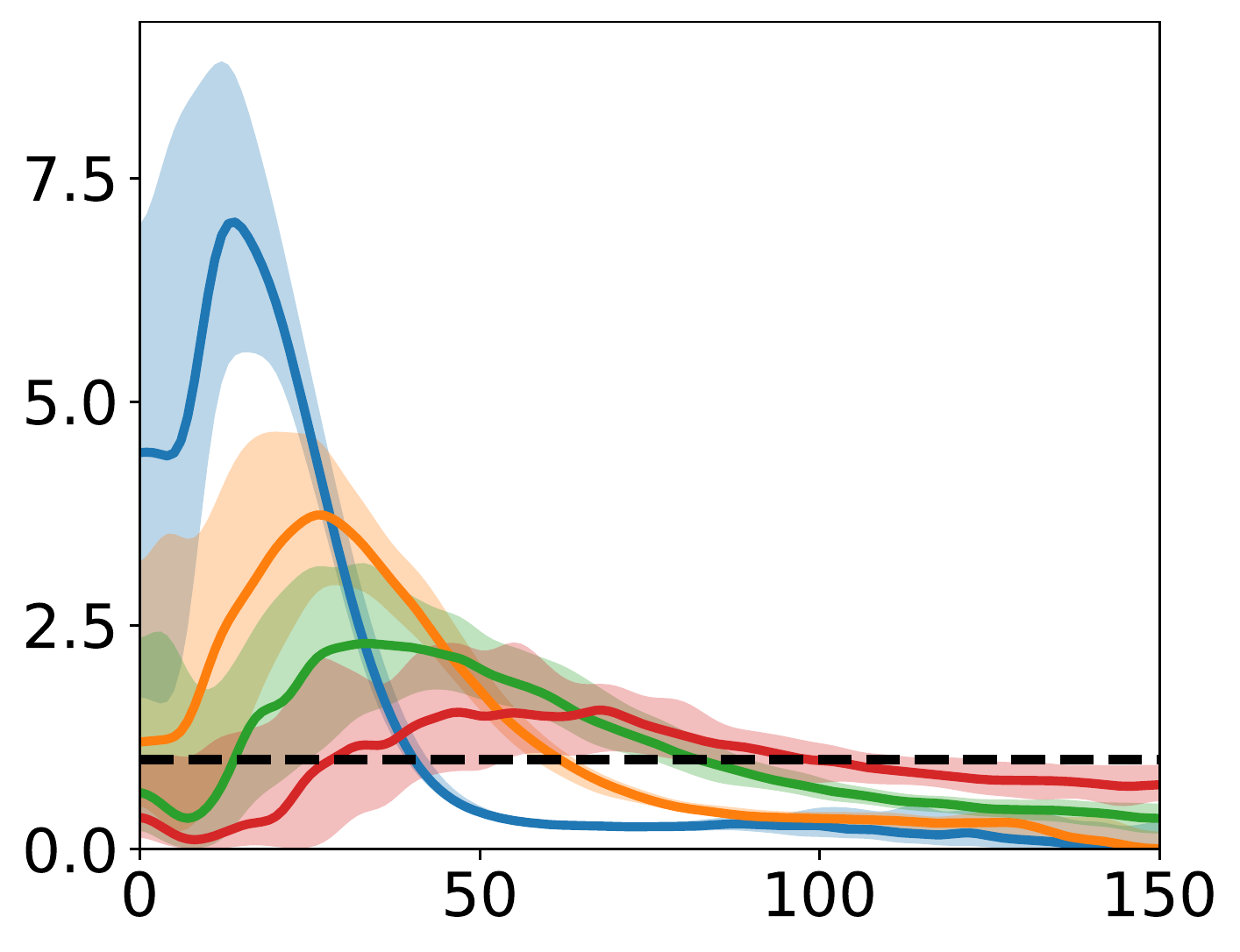} &
\rotatebox{90}{University}\\ [-0.25cm] 
&
\includegraphics[width=\fivefig]{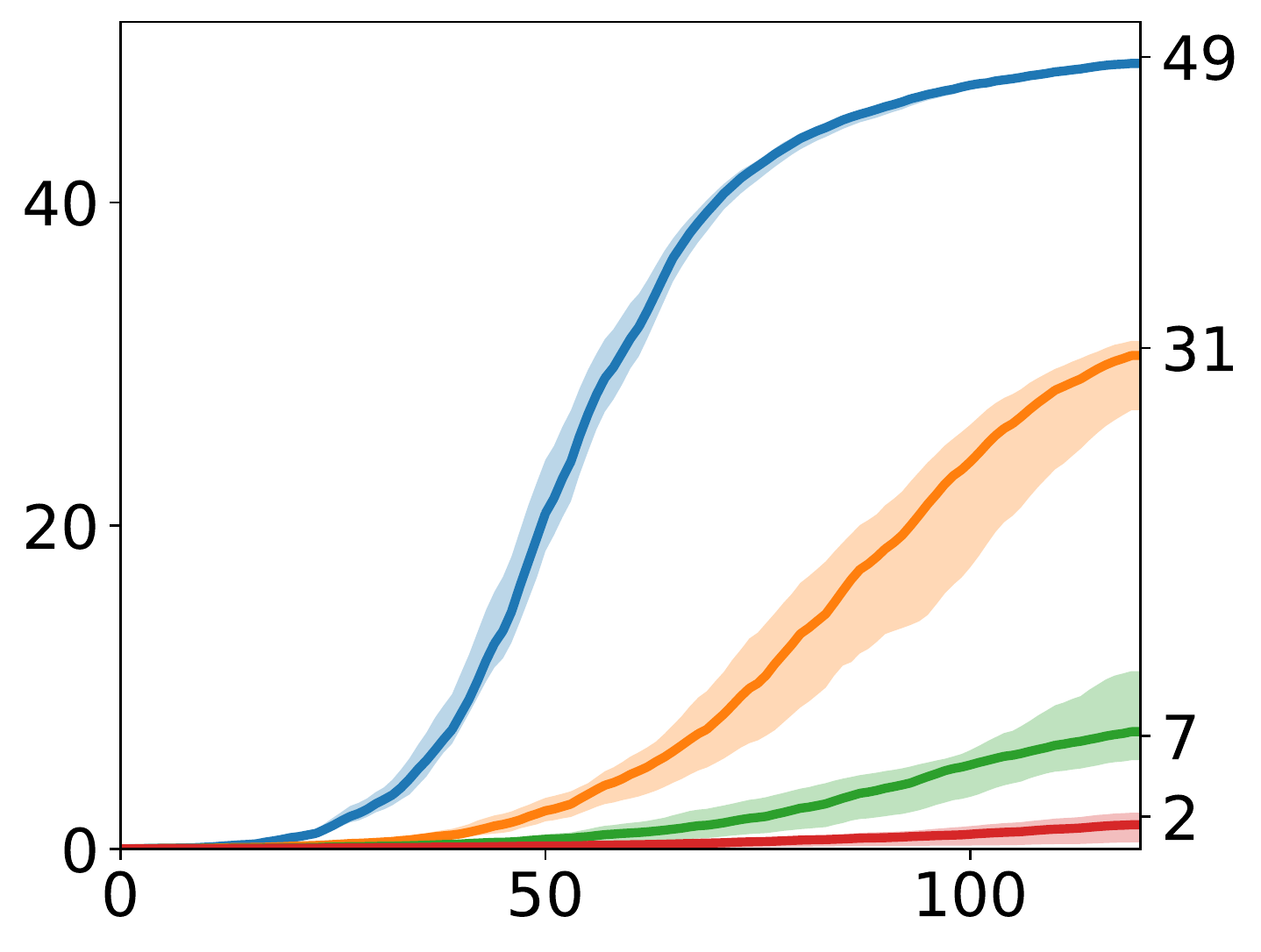} &&
\includegraphics[width=\fivefig]{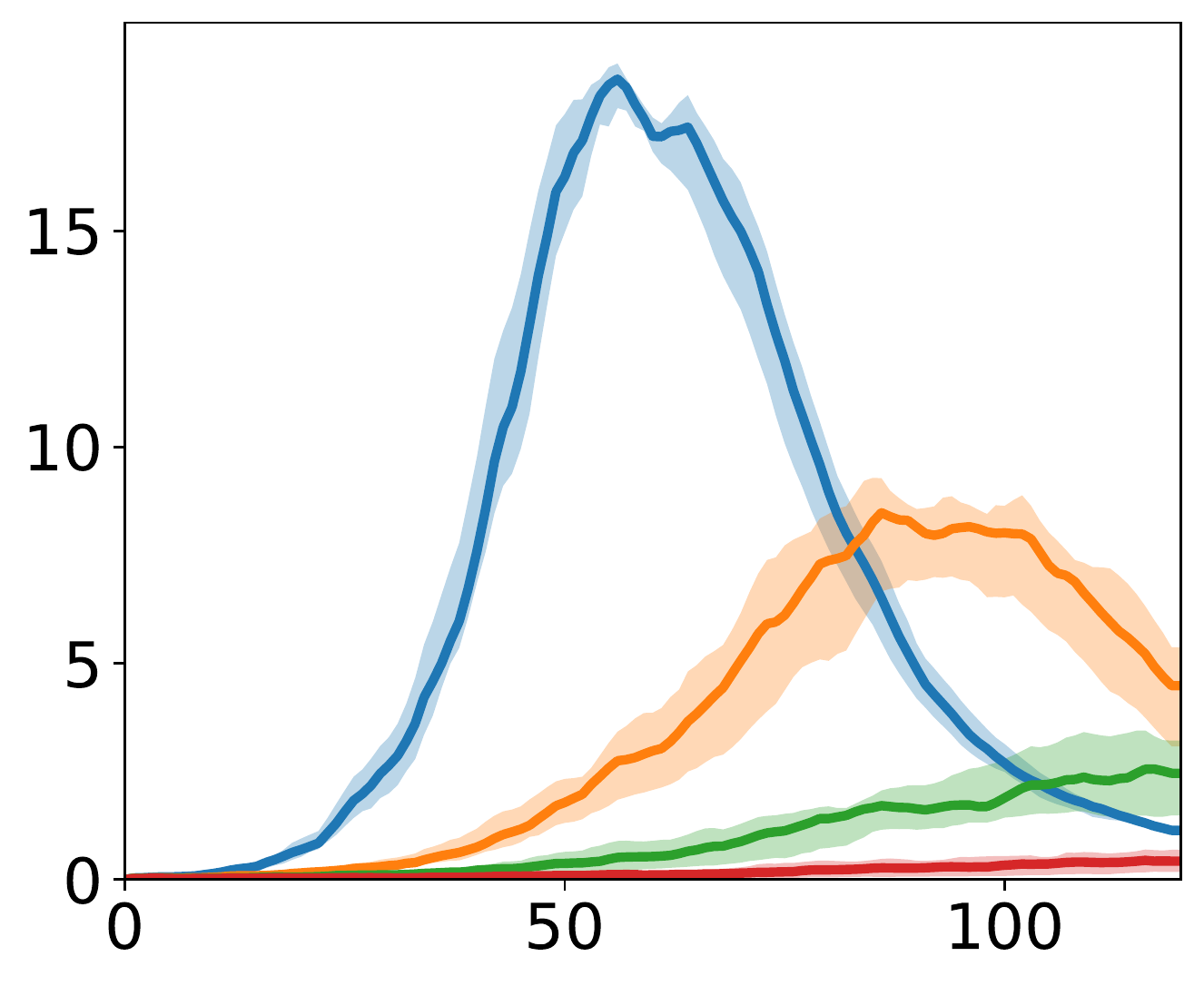} &&
\includegraphics[width=\fivefig]{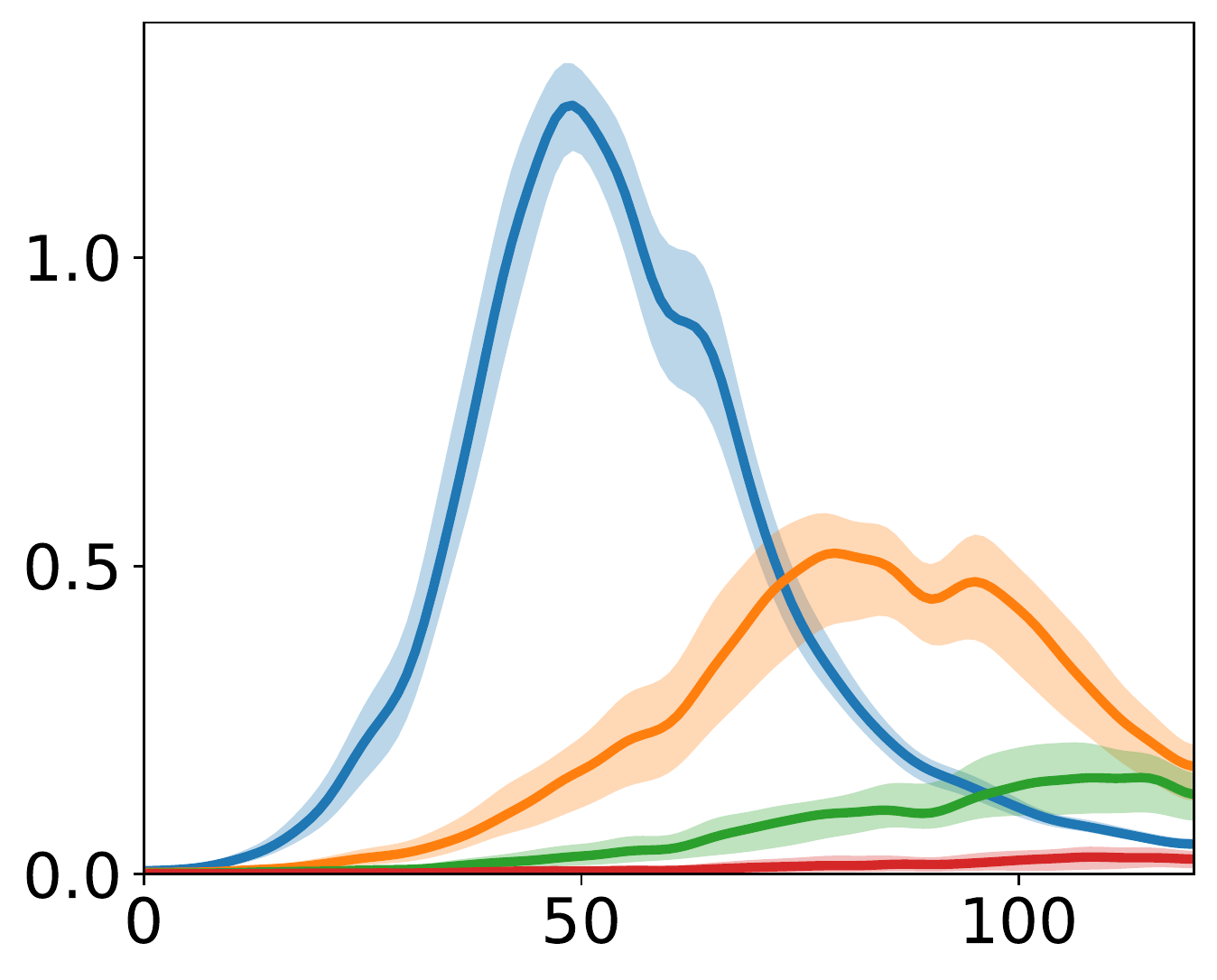} &&
\includegraphics[width=\fivefig]{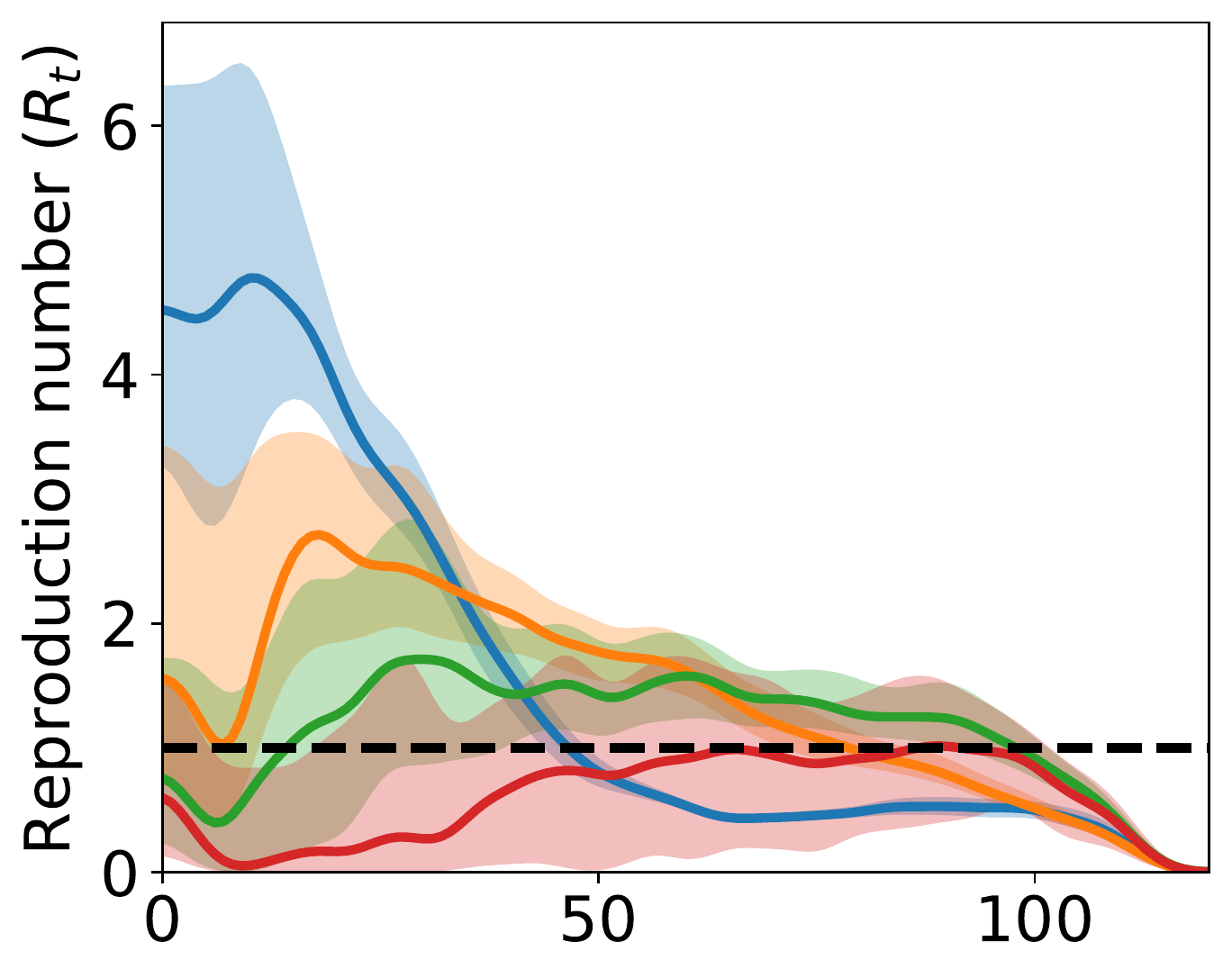} &
\rotatebox{90}{Bike}
\\ [-0.25cm]

\multicolumn{9}{c}{\includegraphics[width=0.15\textwidth]{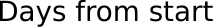}}\\
\multicolumn{9}{c}{\includegraphics[width=0.7\textwidth]{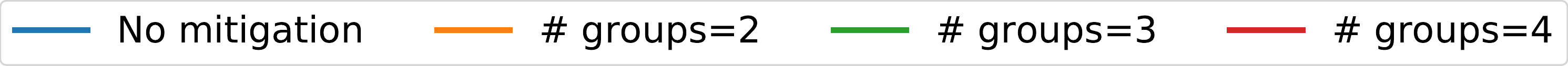}}\\

\end{tabular}
\caption{Infection spreading with the intervention strategy of dividing people into groups. With more groups, each group has fewer agents, and the connections between them become sparser. In most datasets, dividing people into $4$ groups can reduce the total number of infected people significantly.
}
\label{figS:divide_people}
\end{figure}

\newpage
\begin{figure}[!hp]
    \centering
    \begin{tabular}{ccc}
    \includegraphics[width=0.3\textwidth]{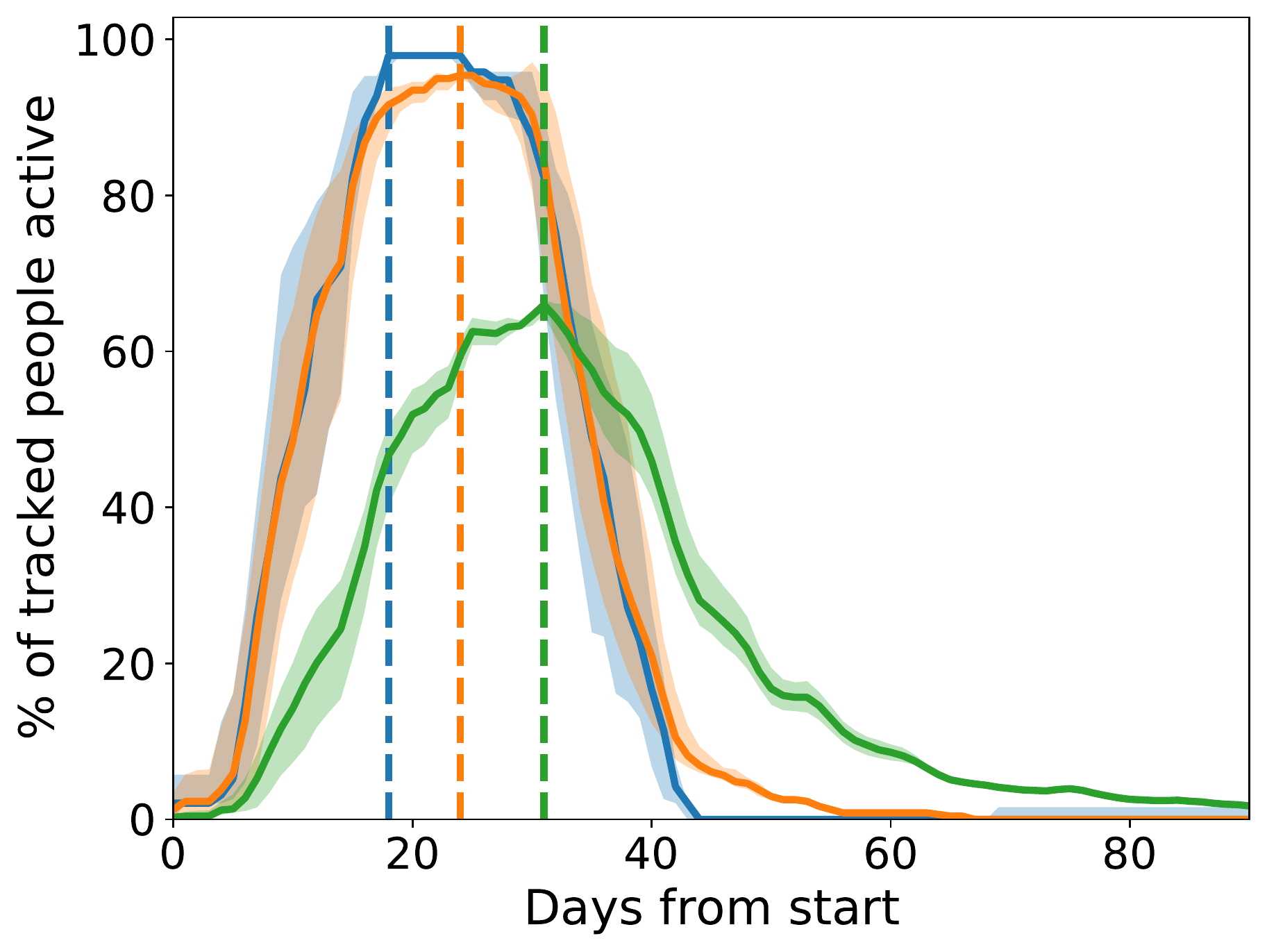}  &
    \includegraphics[width=0.3\textwidth]{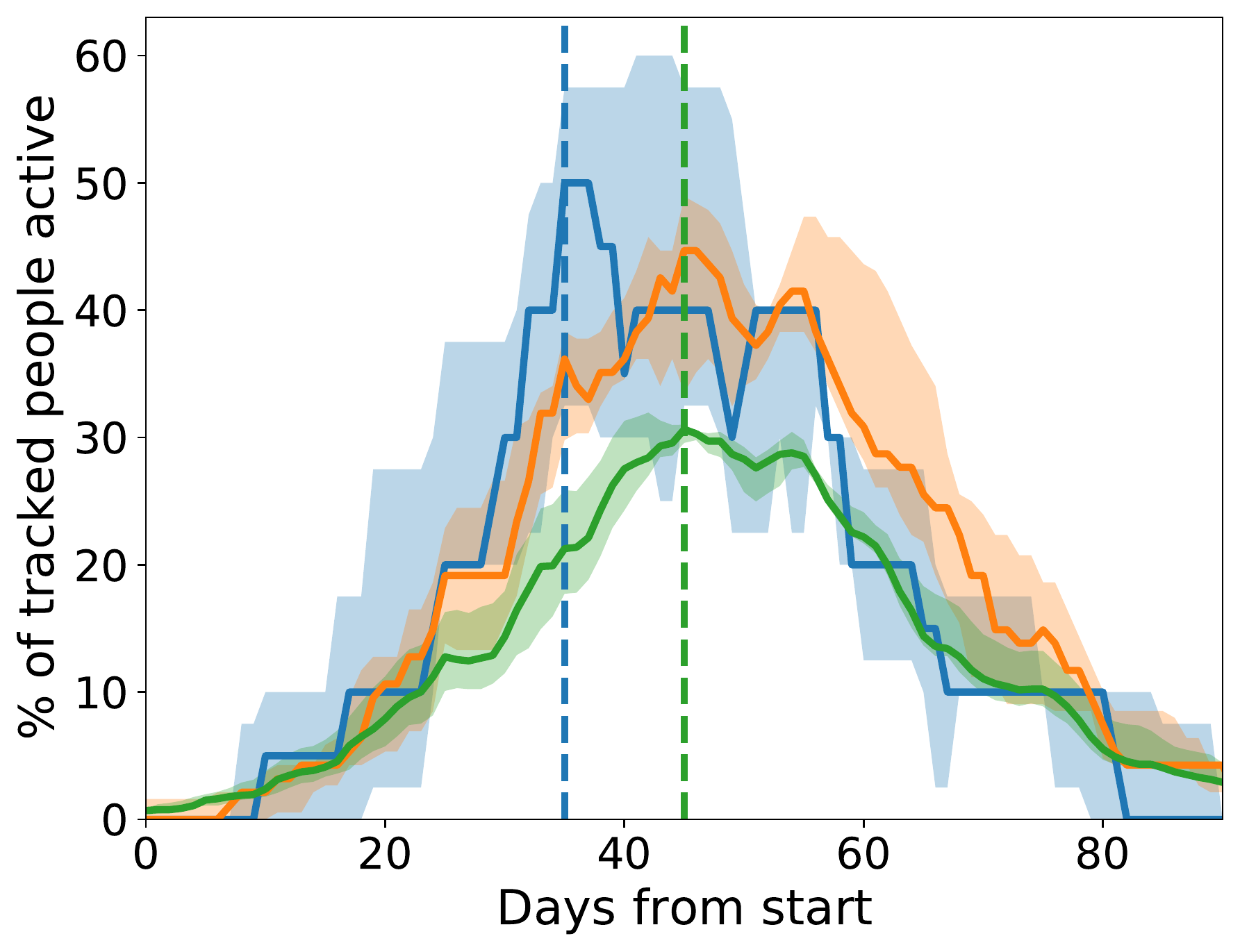}  &
    \includegraphics[width=0.3\textwidth]{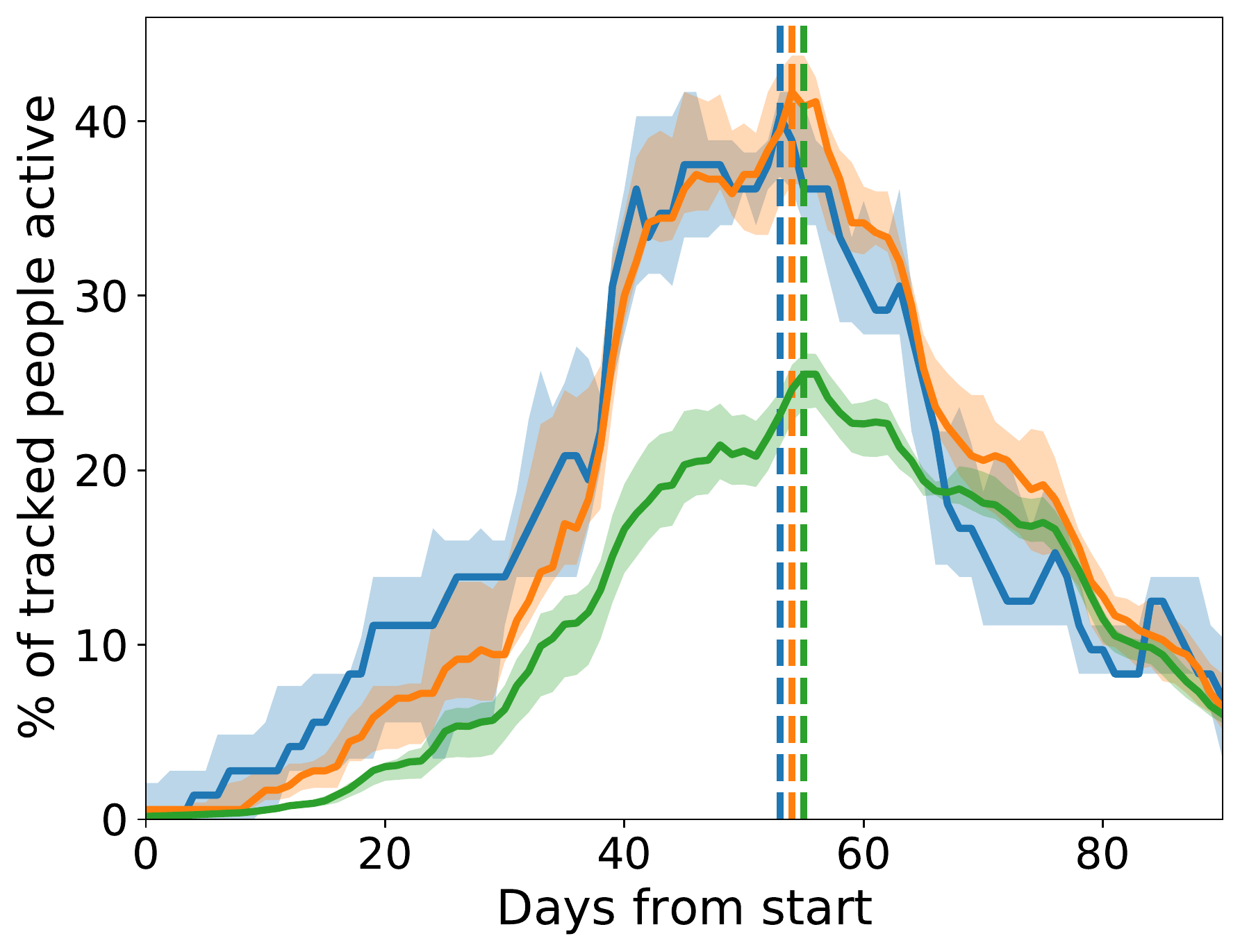}  \\
    Tokyo & Chicago & Los Angeles \\
    \end{tabular}
    \begin{tabular}{cc}
    \includegraphics[width=0.3\textwidth]{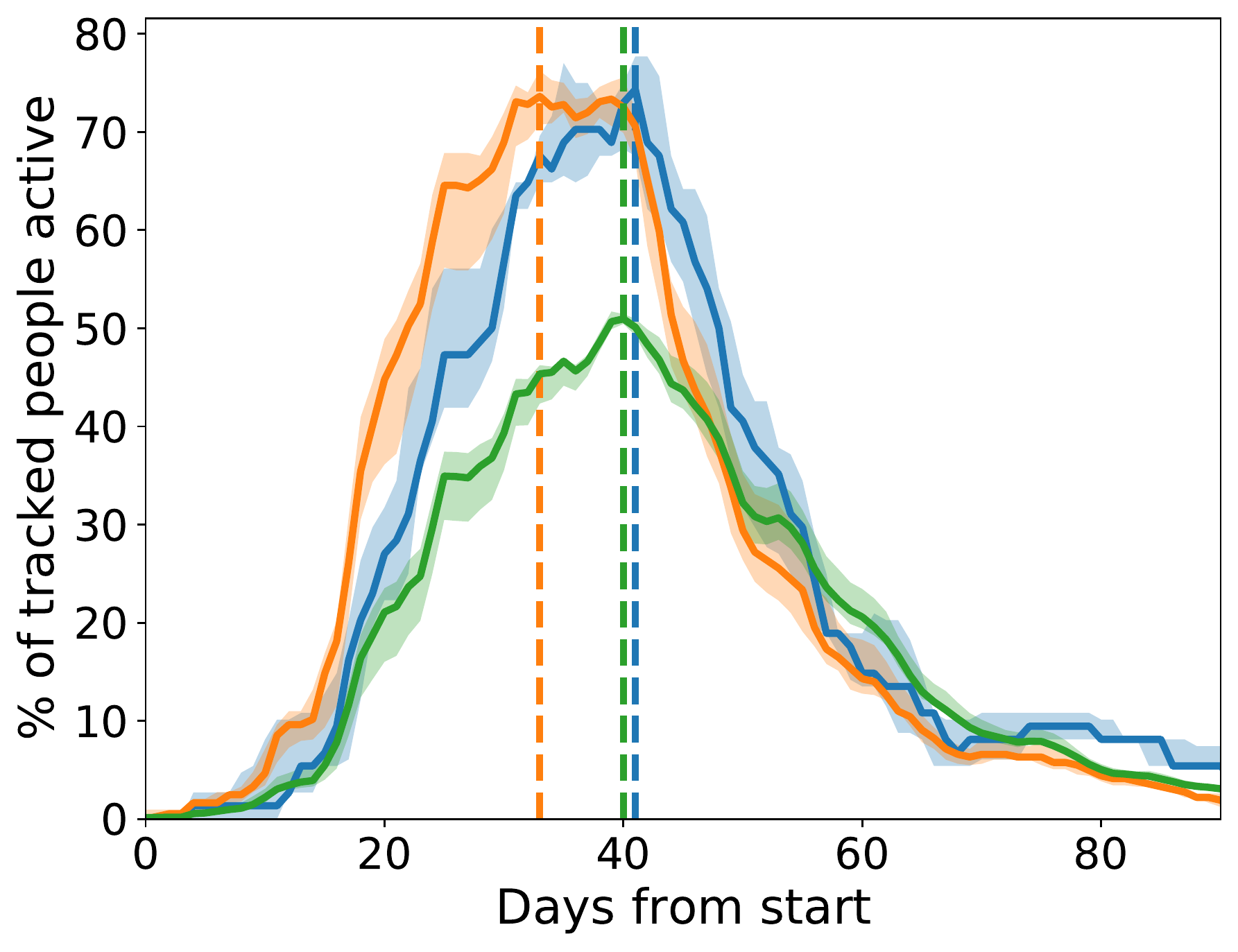}  &
    \includegraphics[width=0.3\textwidth]{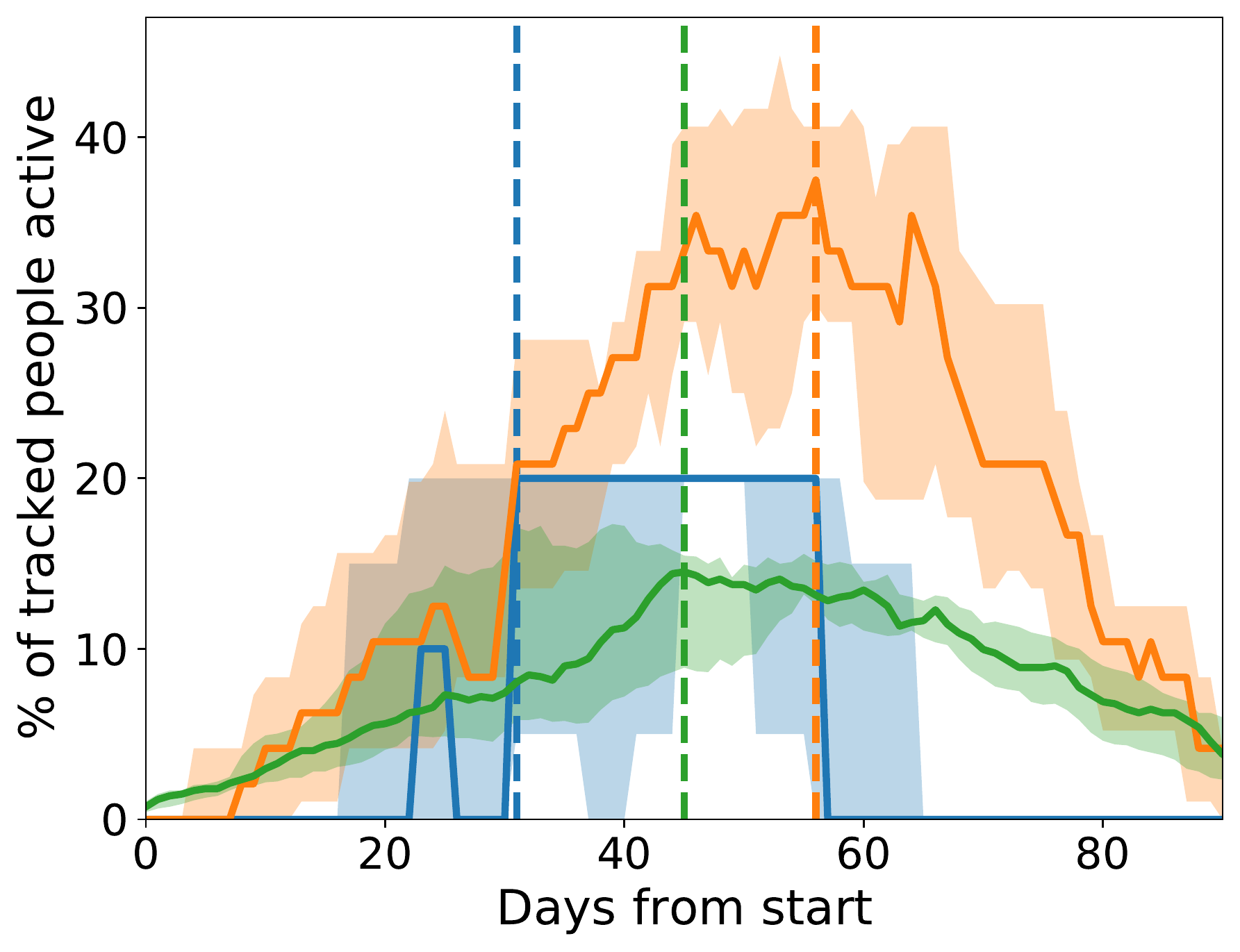}  \\
     Jakarta & London \\
    
	\multicolumn{2}{c}{\includegraphics[width=0.4\textwidth]{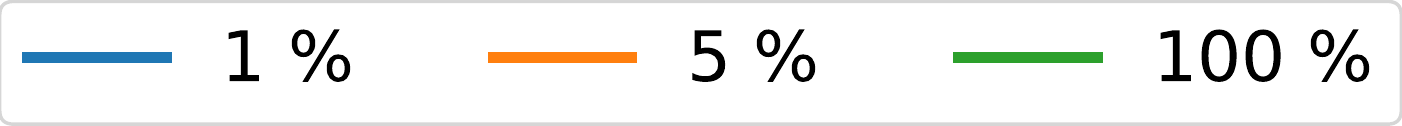}} \\
    
    \end{tabular}
    \caption{Tracking the more active agents similar to Figure~3. The vertical lines denote the days of the peak. All the datasets show consistent trends that the most active agents get infected with a higher proportion and earlier than the whole population.}
    \label{figS:track_popular_users_more}
\end{figure}

\newpage
\begin{figure}[H]
    \centering

\begin{tabular}{m{\fourfig}m{\fourfig}m{\fourfig}m{\fourfig}m{0.1cm}}

\includegraphics[width=\fourfig]{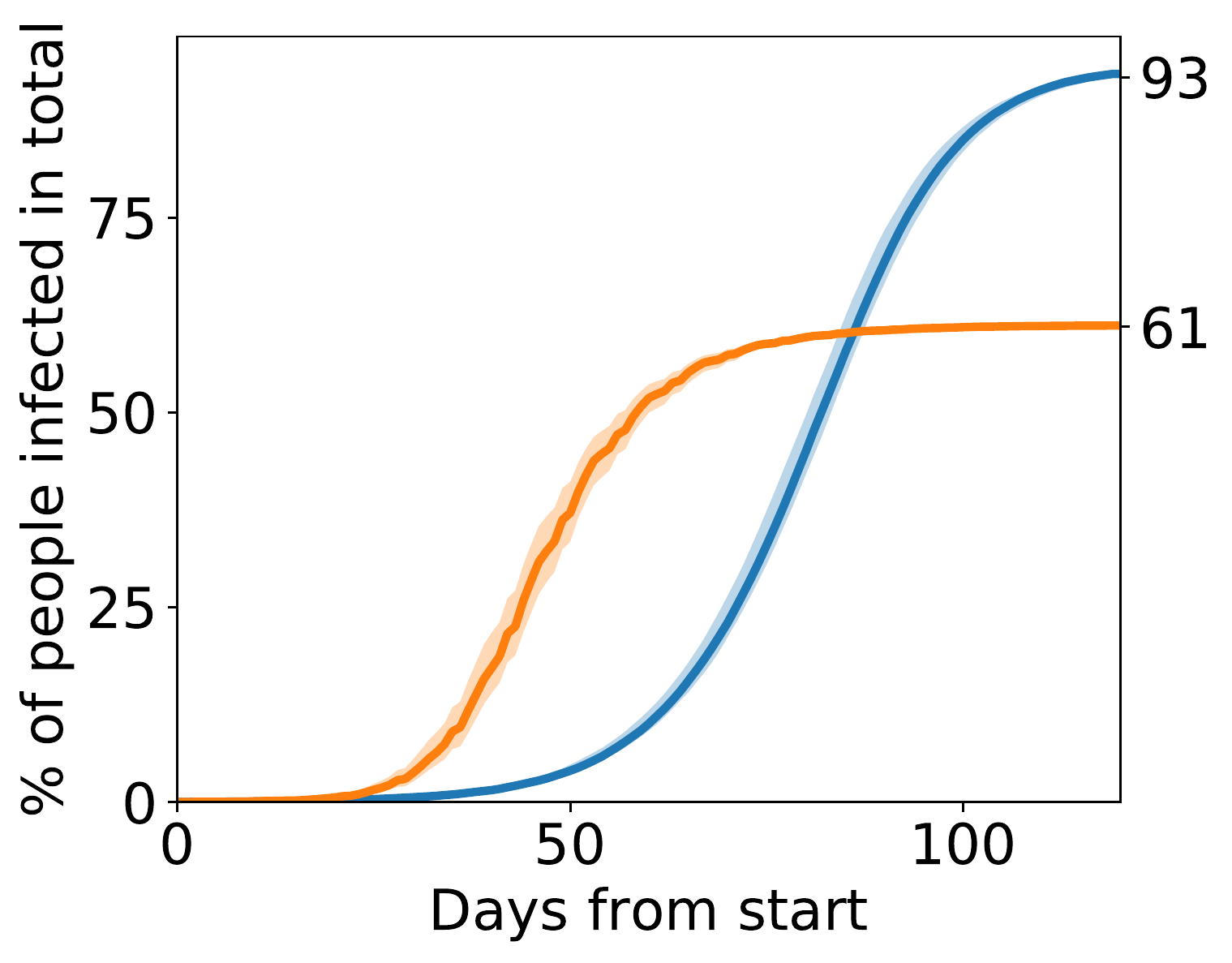} &
\includegraphics[width=\fourfig]{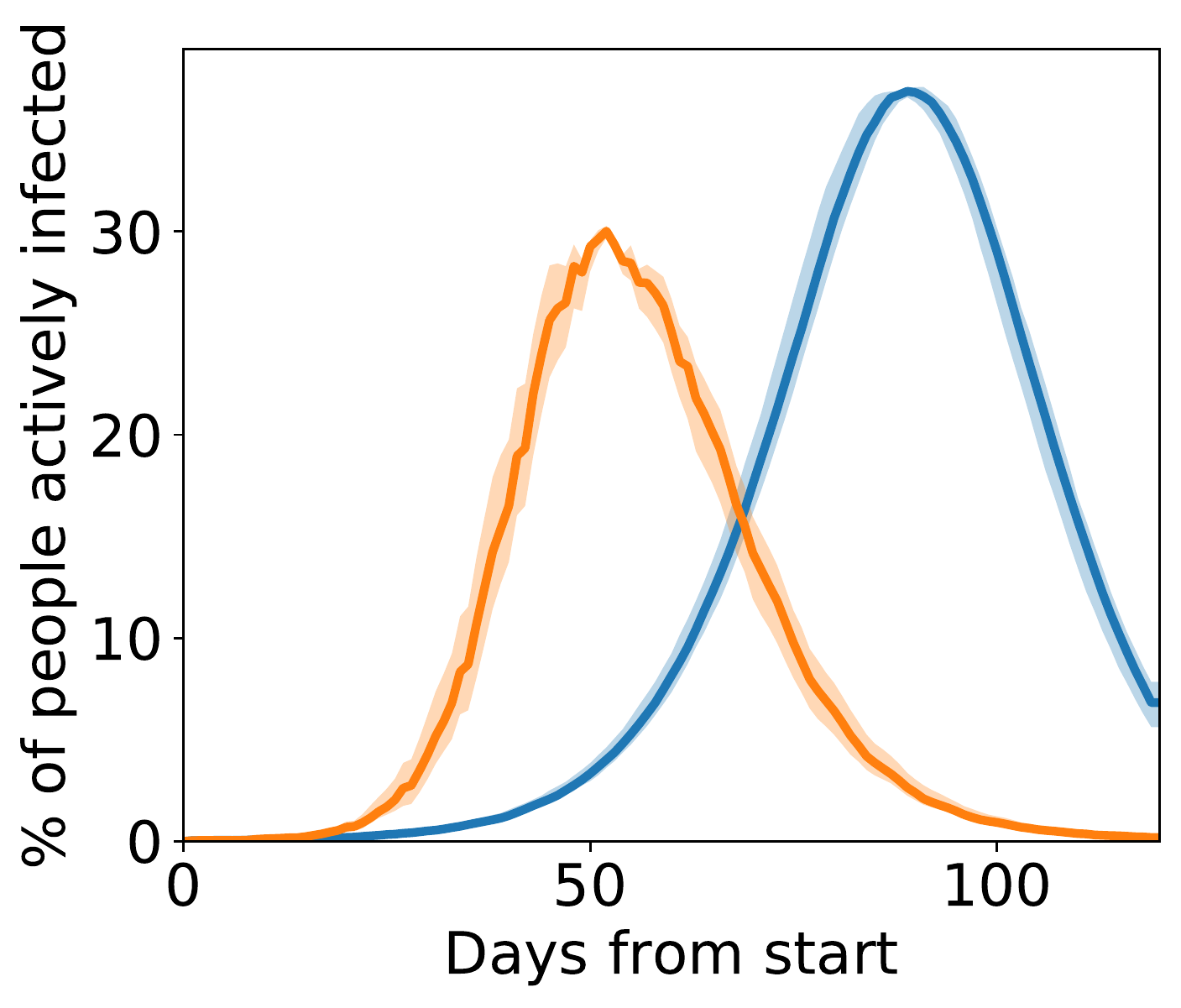} &
\includegraphics[width=\fourfig]{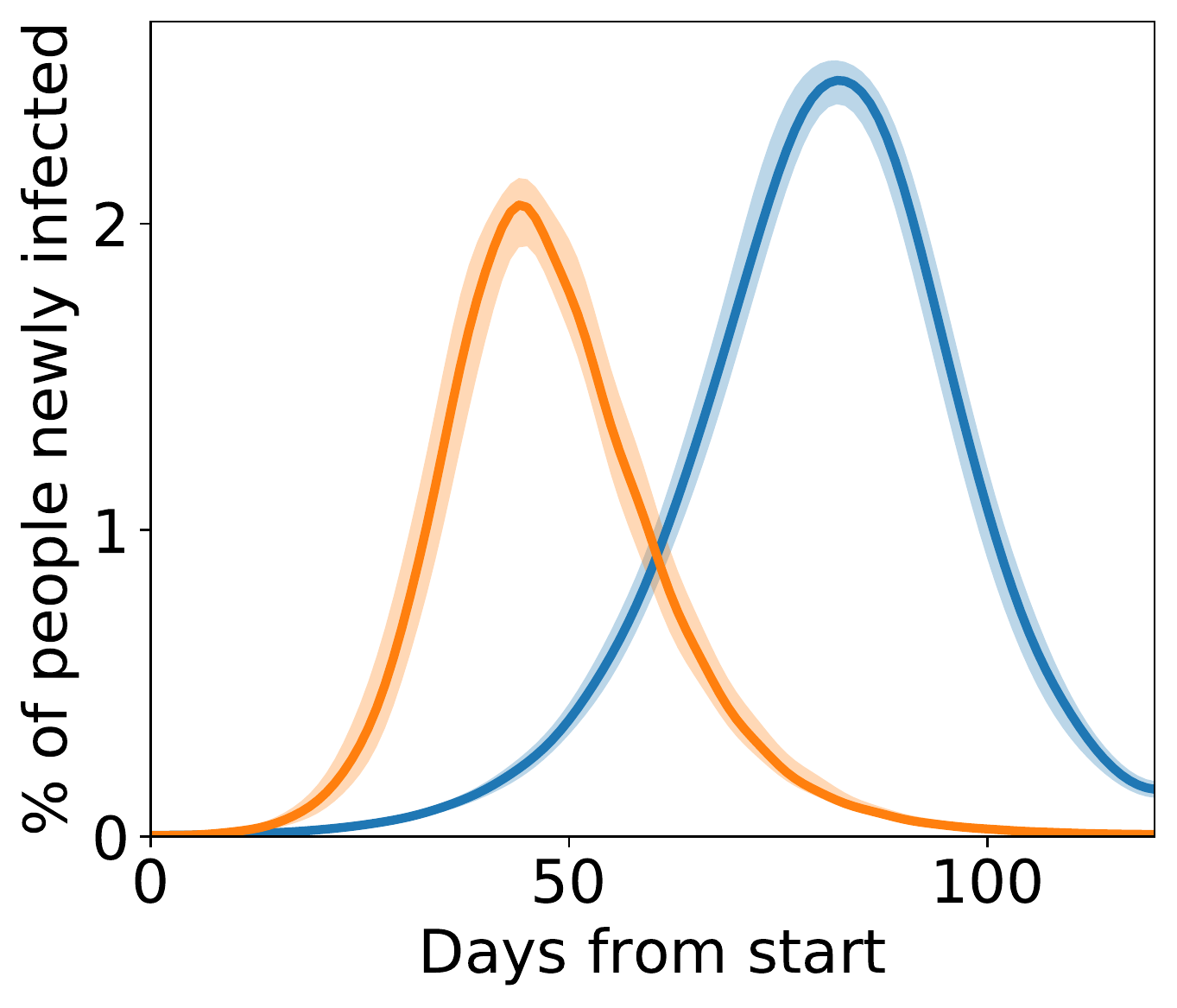} &
\includegraphics[width=\fourfig]{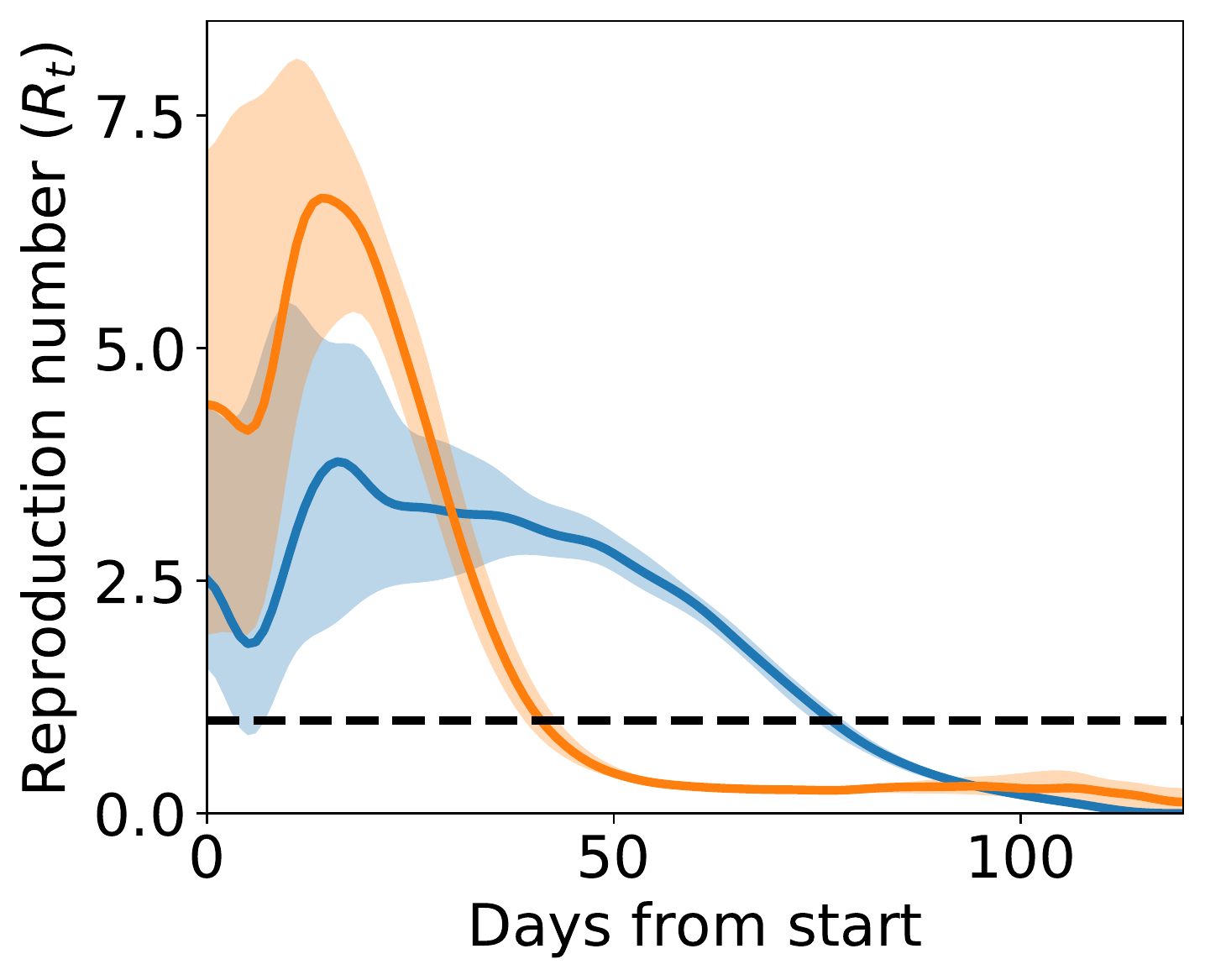} &
\rotatebox{90}{University}\\

\includegraphics[width=\fourfig]{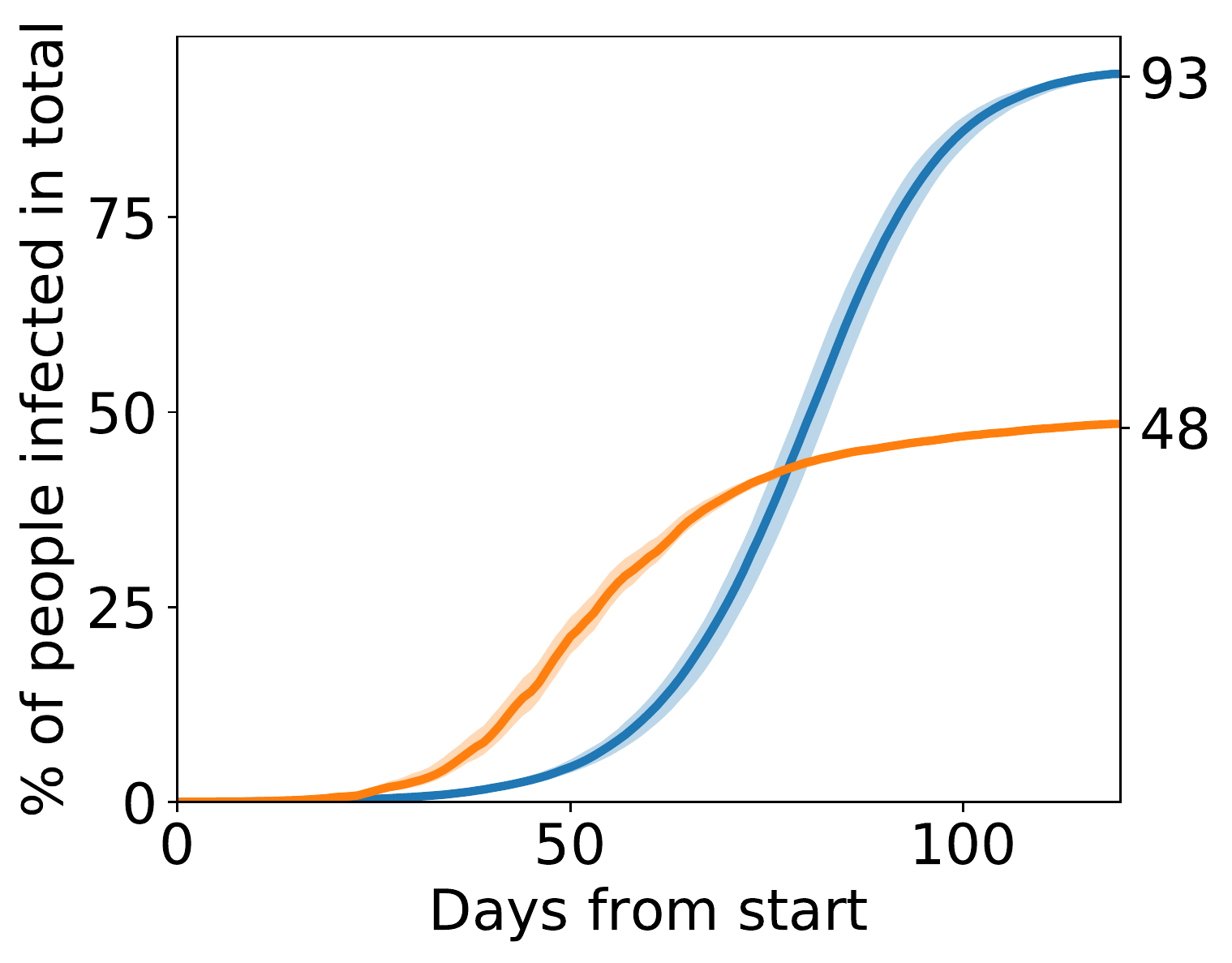} &
\includegraphics[width=\fourfig]{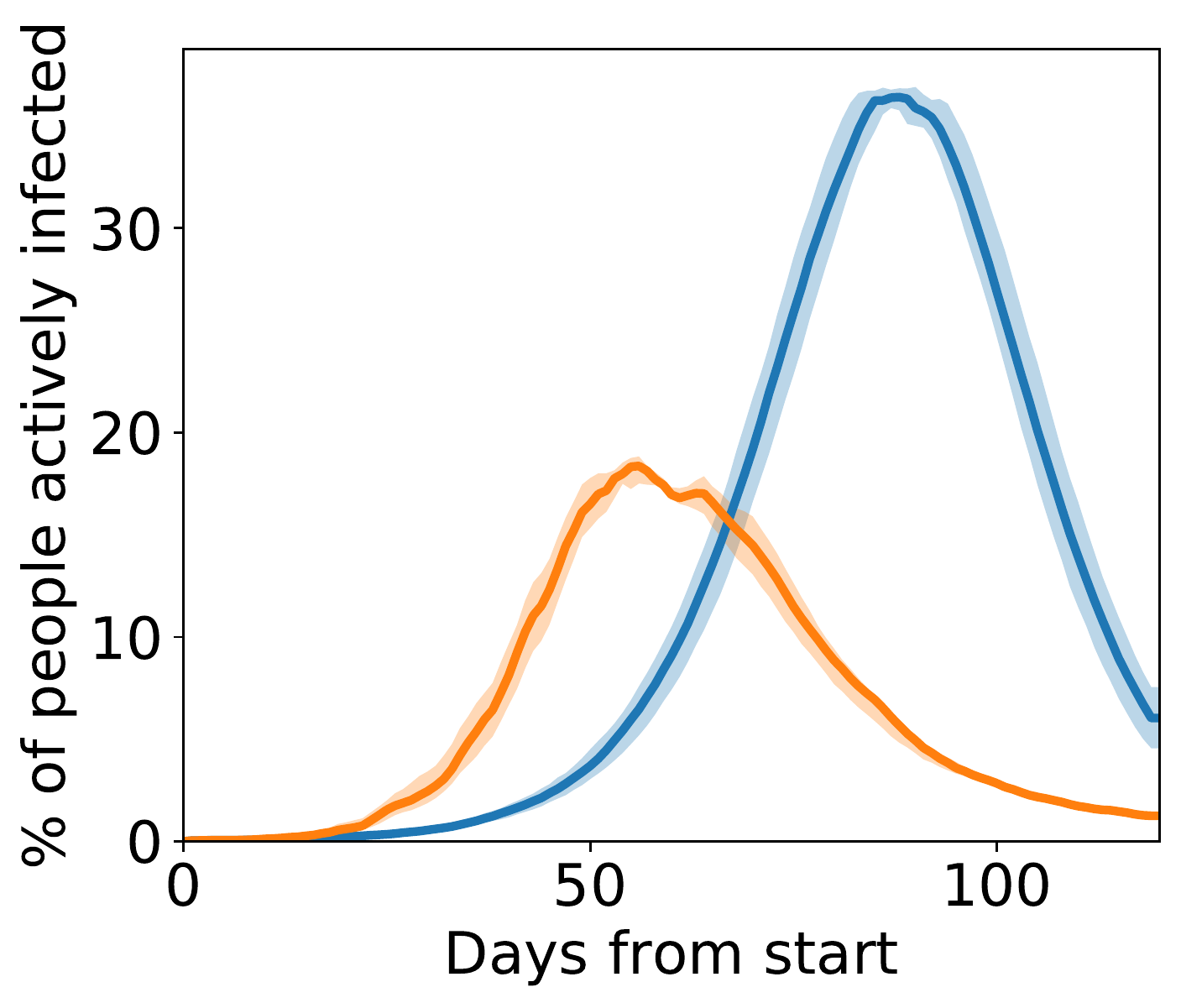} &
\includegraphics[width=\fourfig]{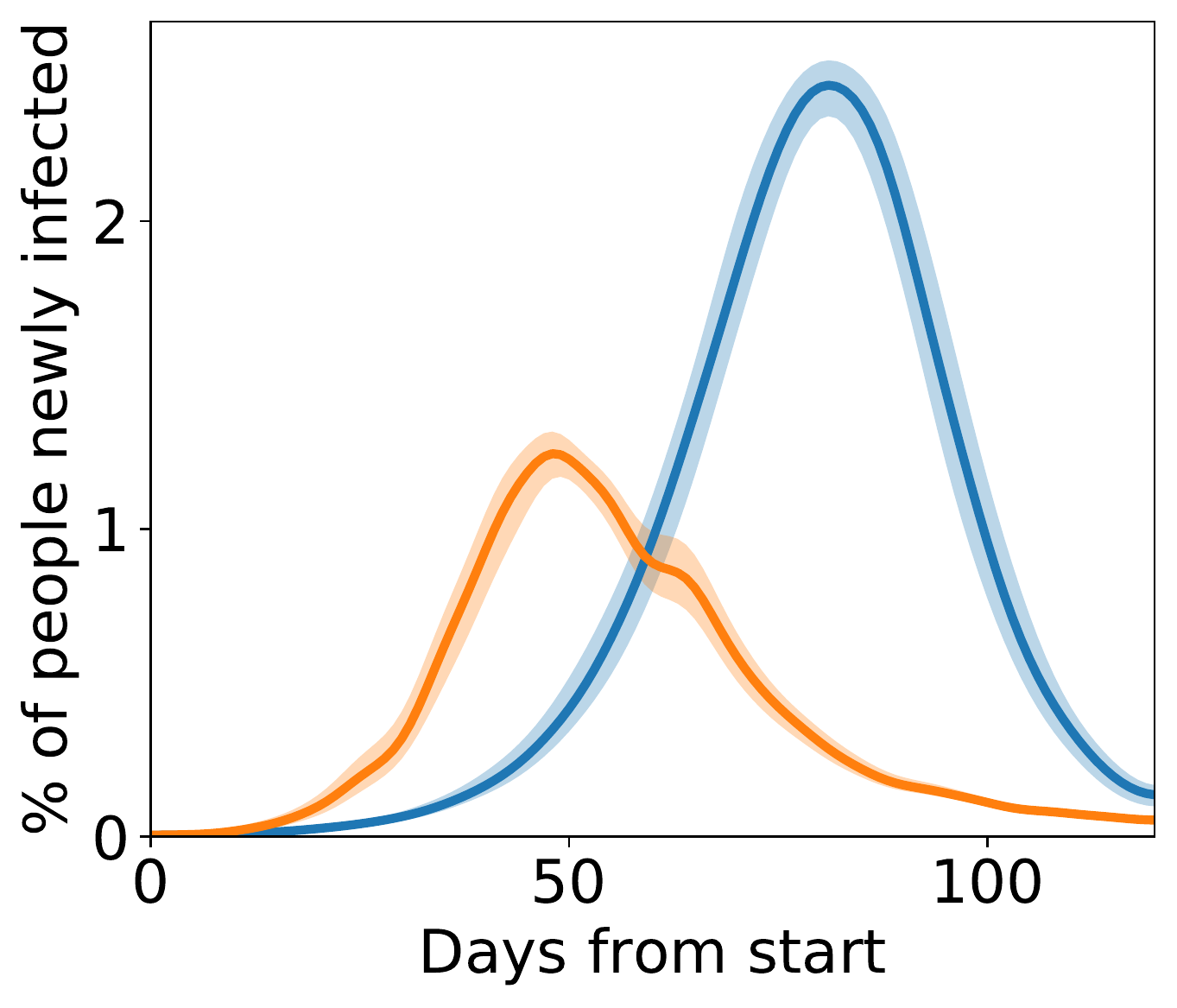} &
\includegraphics[width=\fourfig]{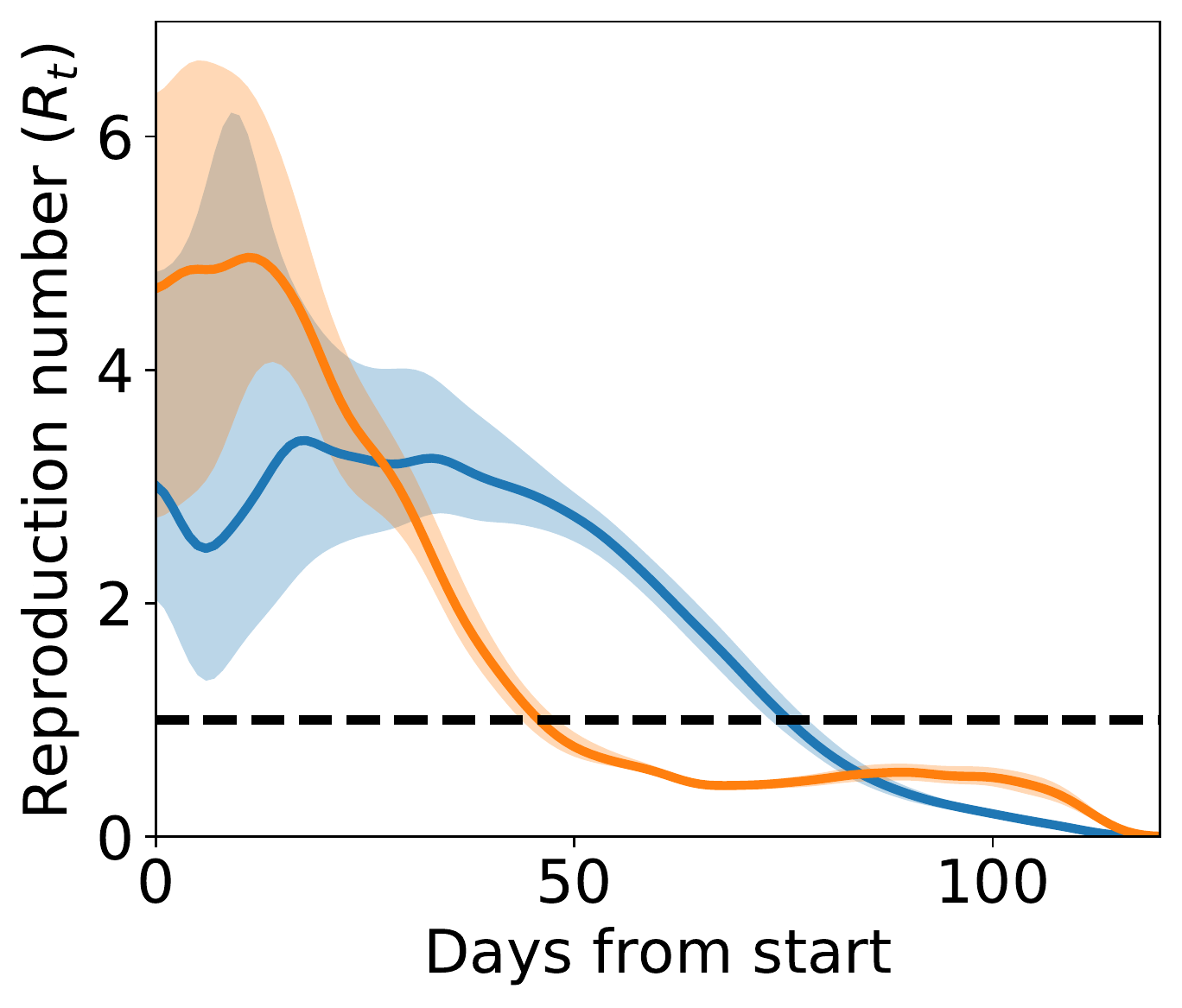} &
\rotatebox{90}{Bike}\\

\multicolumn{2}{c}{\includegraphics[width=\twofig]{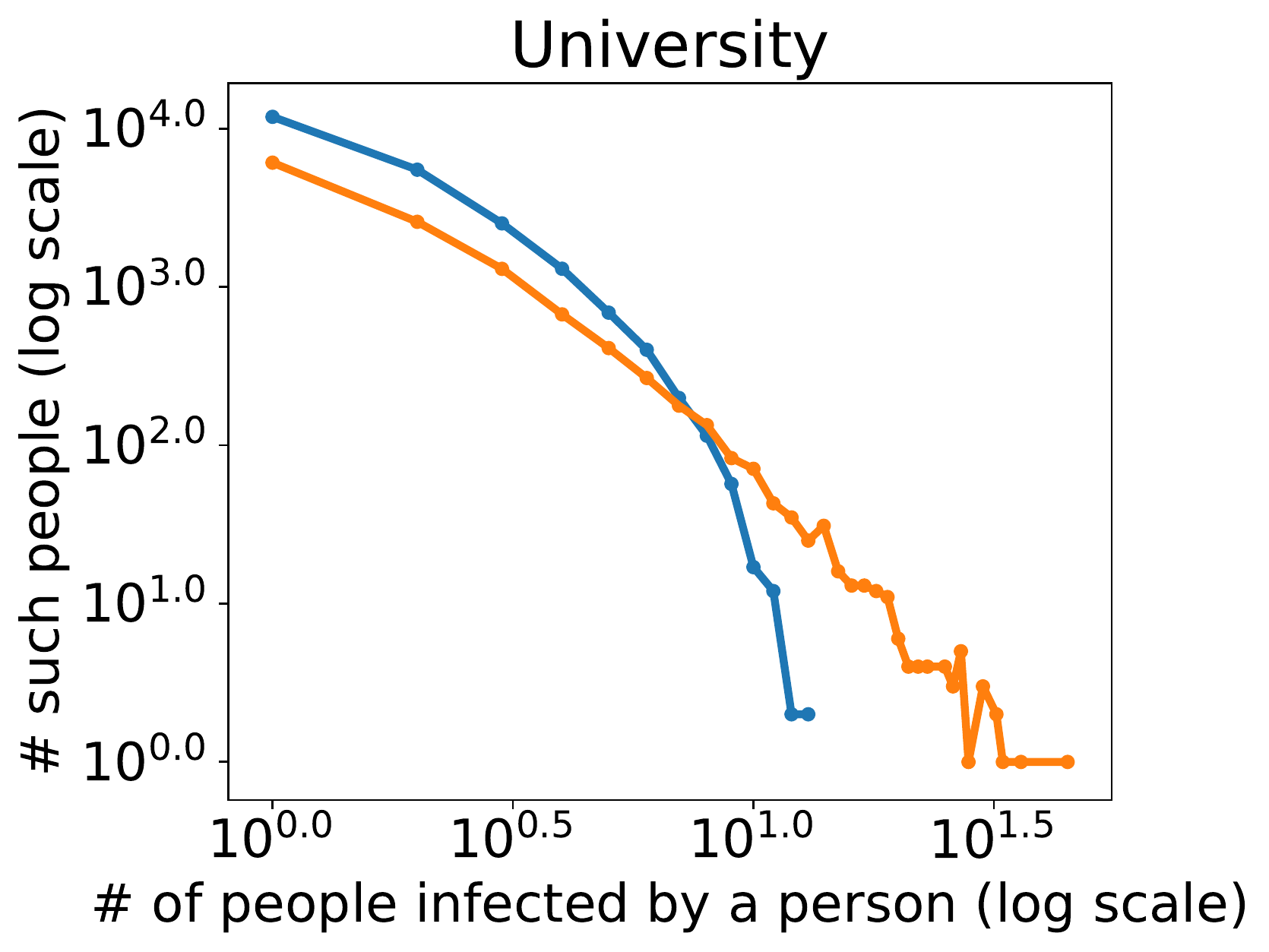}} &
\multicolumn{2}{c}{\includegraphics[width=\twofig]{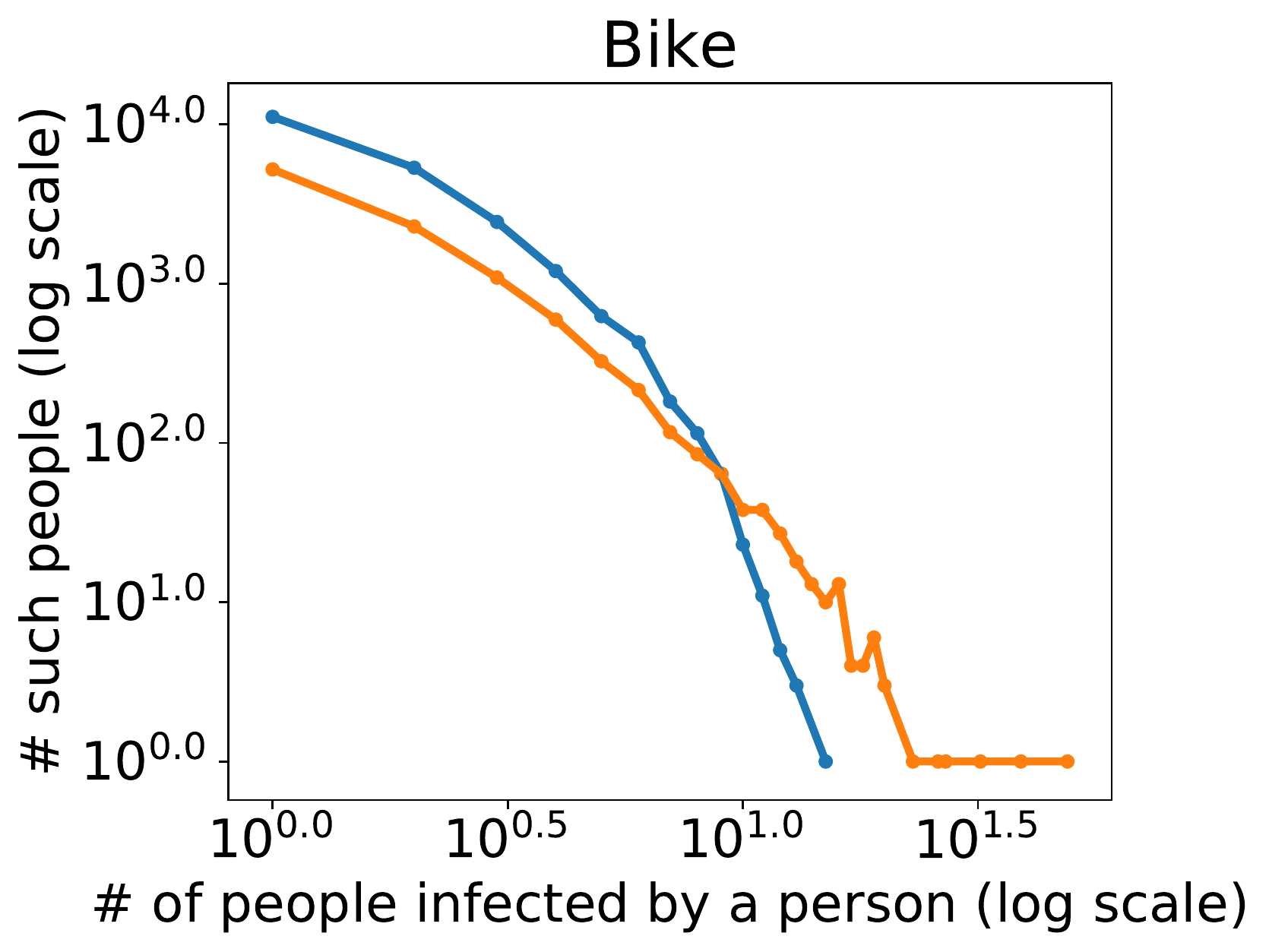}} &
\\

\multicolumn{5}{c}{\includegraphics[width=0.5\textwidth]{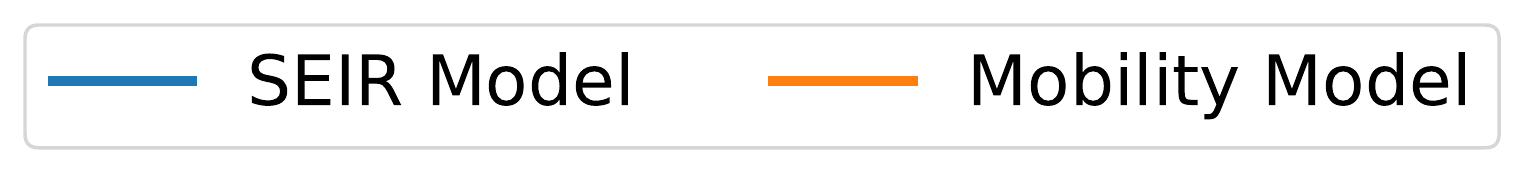}}\\

\end{tabular}
\caption{Infection spreading with random meeting information and real mobility data. In the SEIR model, almost all people get infected in the end, which is a much higher percentage than in the mobility model. At the same time, the peaks of people actively infected in the SEIR model are about $35$ days delayed, compared with the mobility model in two datasets. The third row shows the distribution of the number of people infected by a person. In the mobility model, an agent can infect at most $50$ susceptible agents, while in the SEIR model, at most only $15$ susceptible agents can be infected from one infectious agent. Thus, there is a long tail in the mobility model because there are some more active agents, who spread the disease quickly.}
\label{figS:Compare_RandomMeeting}
\end{figure}

\newpage
\begin{figure}[!hp]
    \centering
    \begin{tabular}{ccc}
    
    \includegraphics[width=0.3\textwidth]{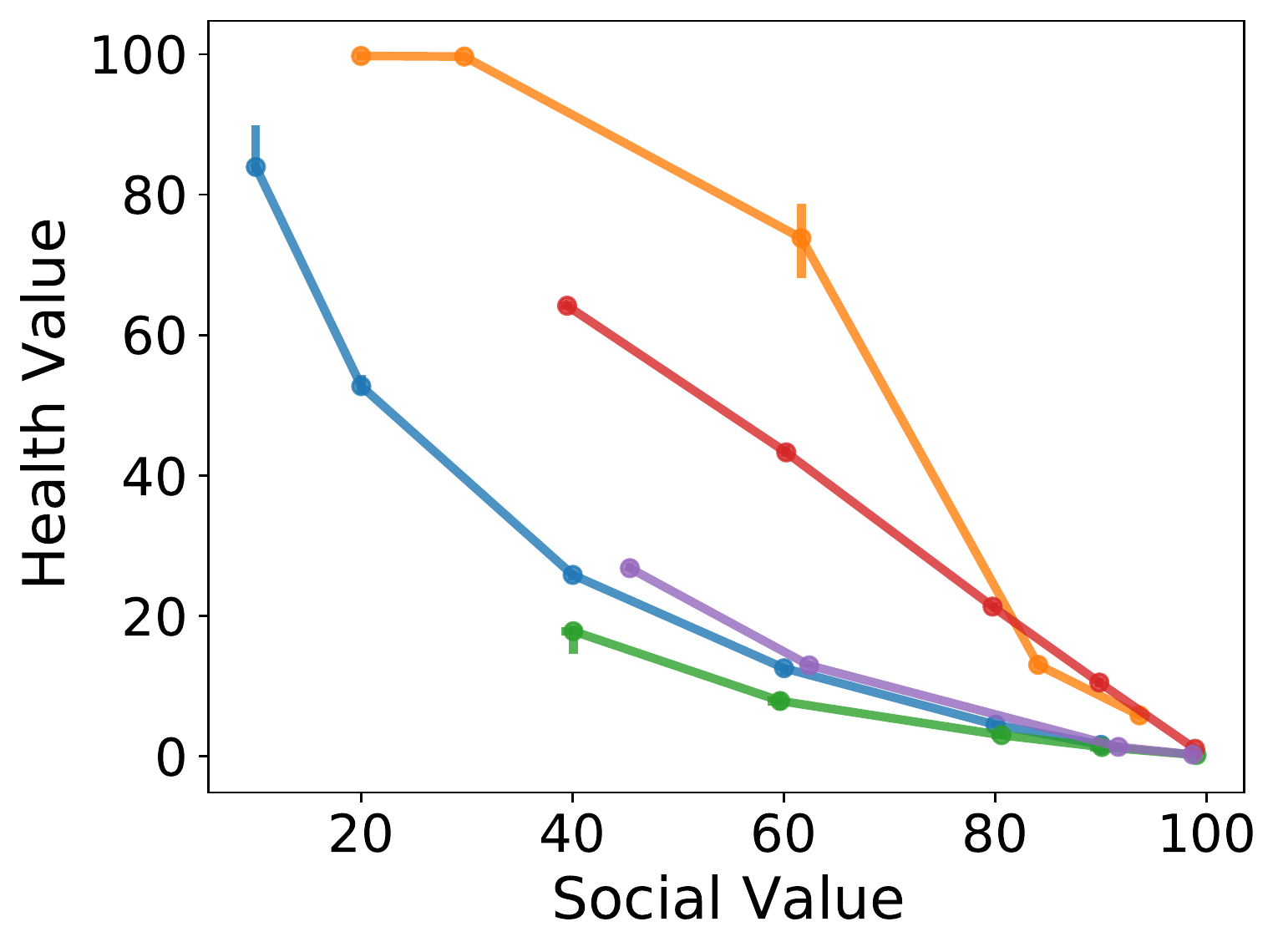}  &
    \includegraphics[width=0.3\textwidth]{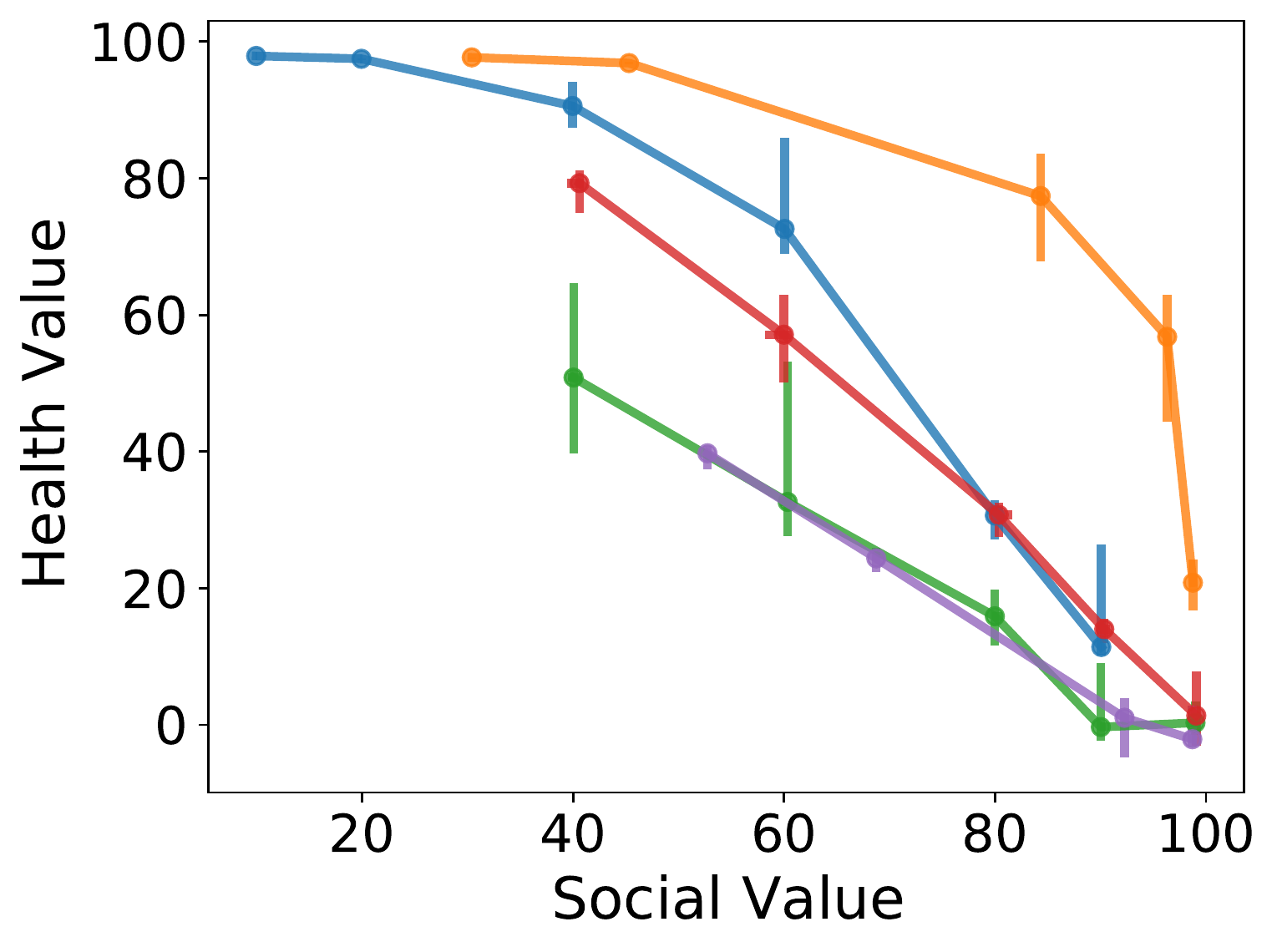}  &
    \includegraphics[width=0.3\textwidth]{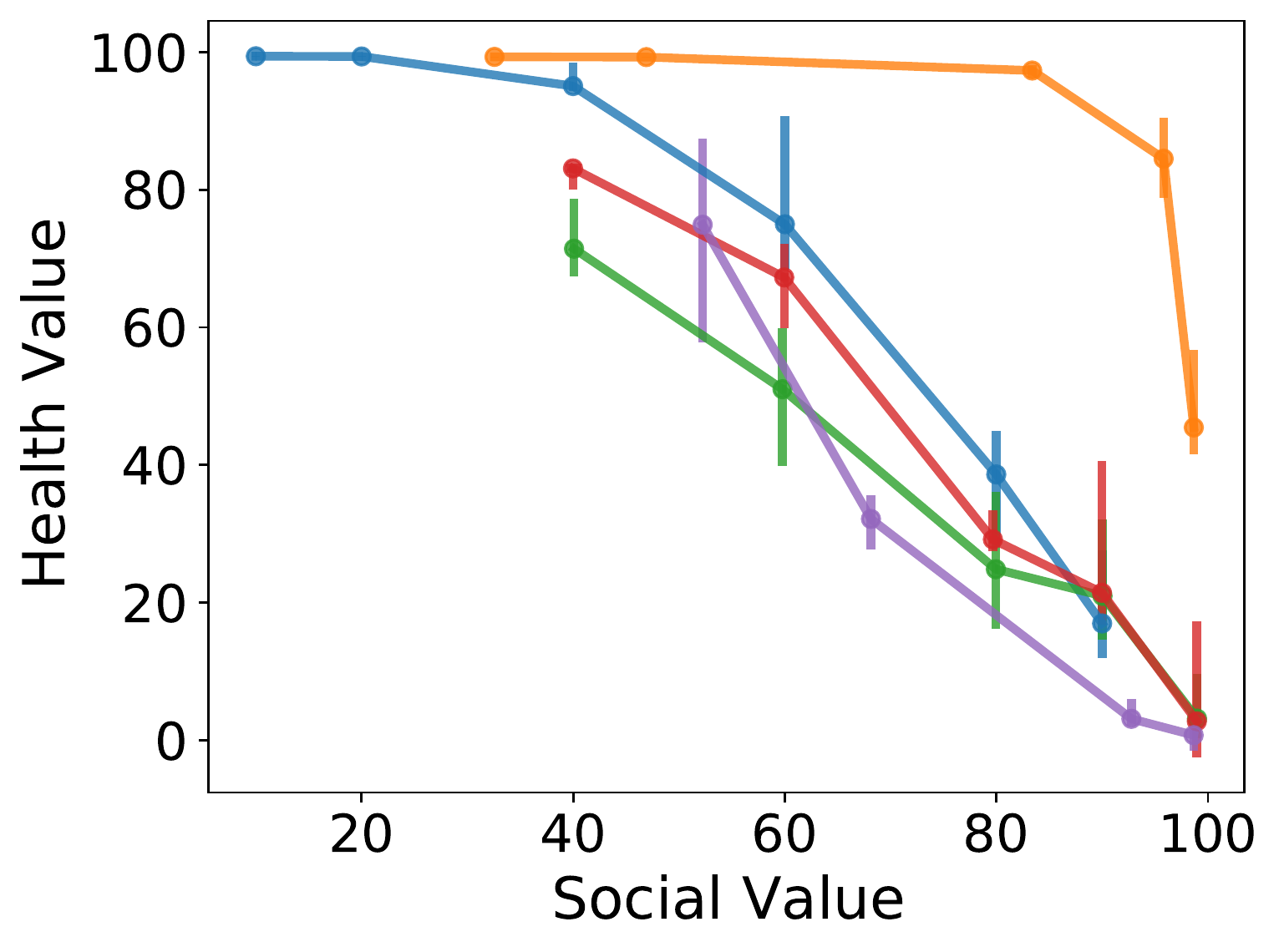}  \\
    Tokyo & Chicago & Los Angeles \\
    \includegraphics[width=0.3\textwidth]{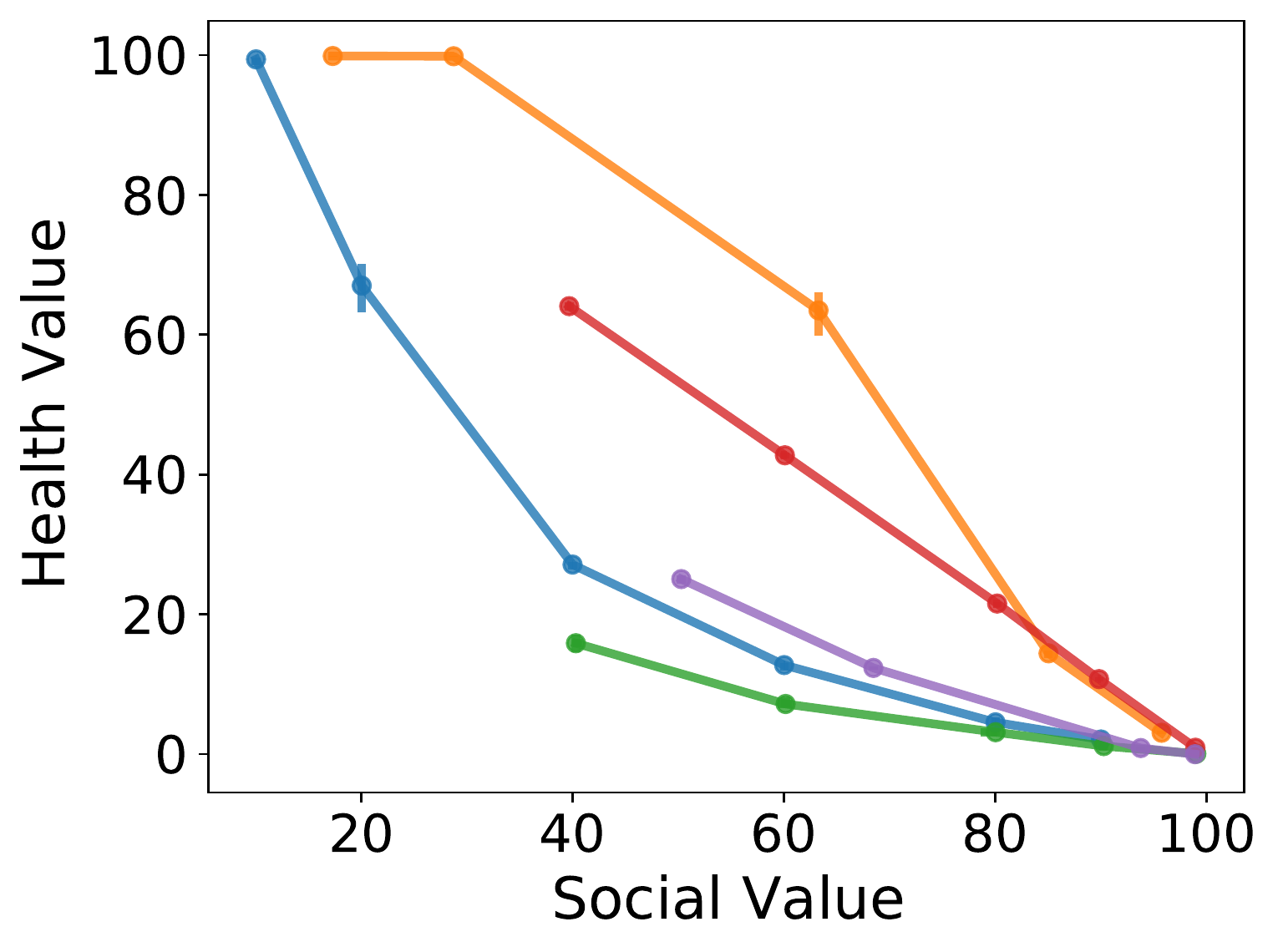}  &
    \includegraphics[width=0.3\textwidth]{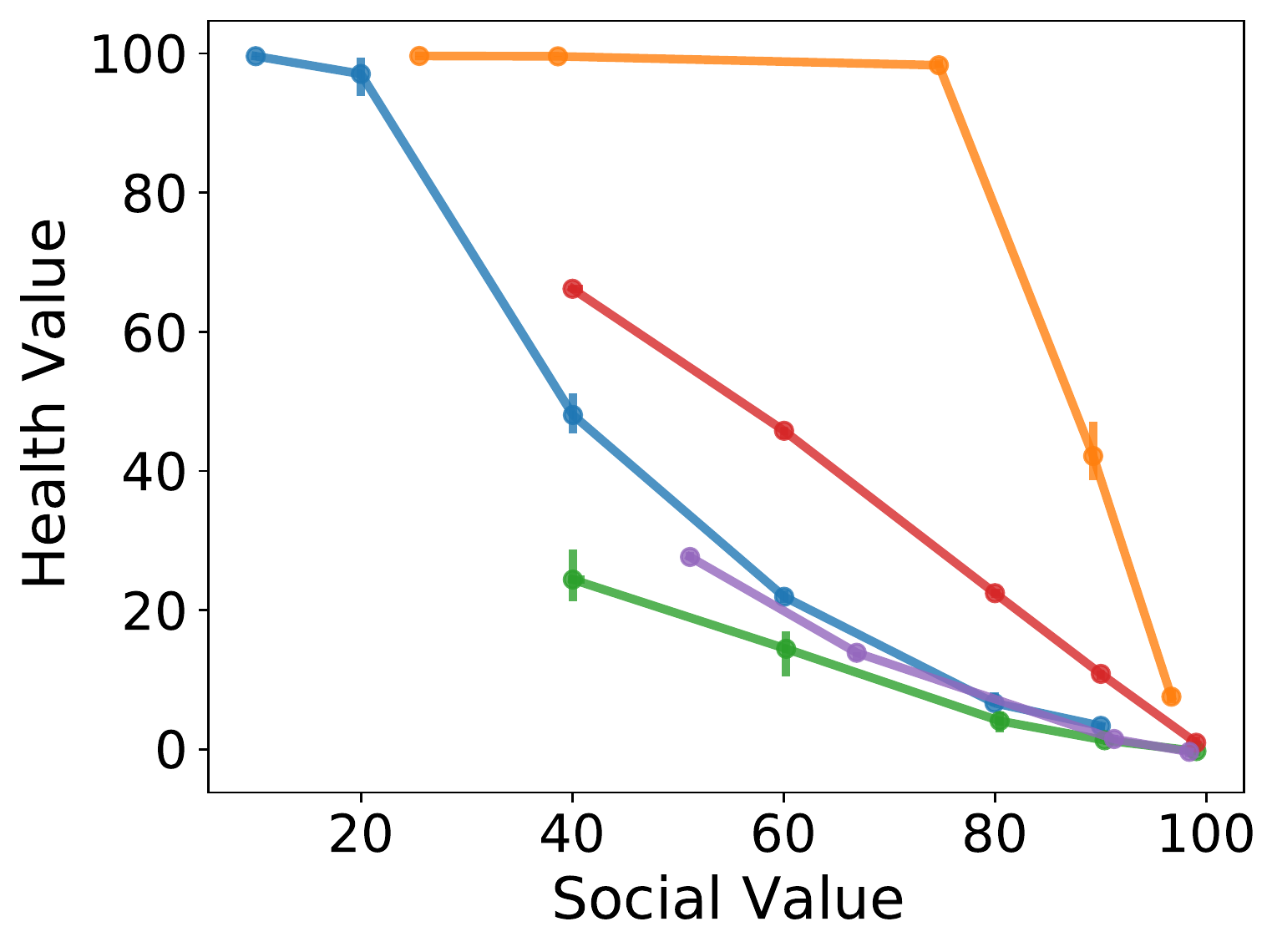}  &
    \includegraphics[width=0.3\textwidth]{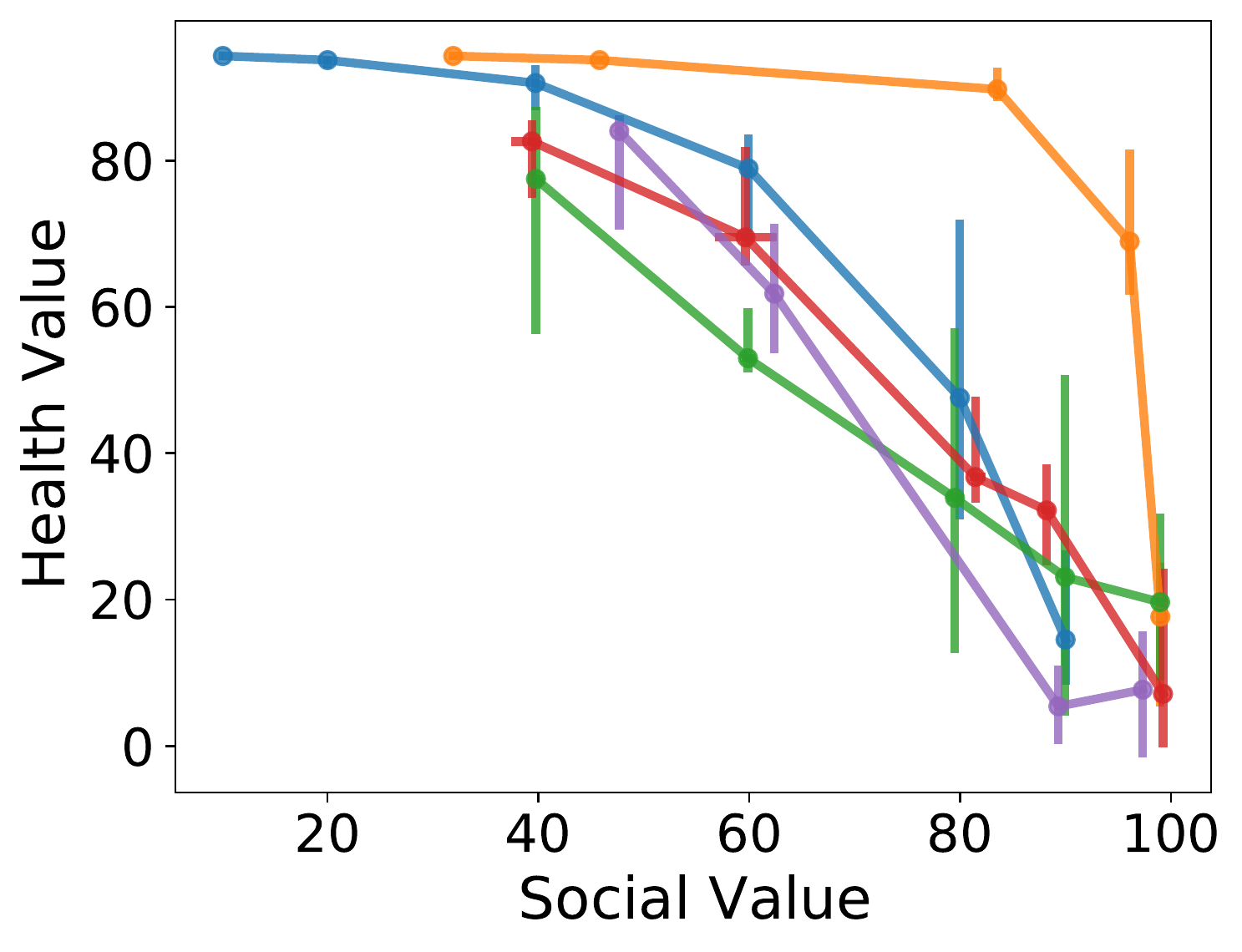}  \\
	Istanbul & Jakarta & London \\
    \multicolumn{3}{c}{\includegraphics[width=\textwidth]{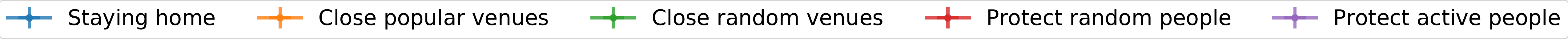}}  \\
    
    \end{tabular}
    \caption{Comparison of different intervention strategies in all other Foursquare datasets similar to Figure~4. Closing popular venues is the most advantageous from both health and social value perspectives.}
    \label{figS:compare_strategies}
\end{figure}

\newpage
\begin{figure}[!hp]
\begin{tabular}{m{0.2cm}@{}>{\centering\arraybackslash}m{\sixfig}>{\centering\arraybackslash}m{\sixfig}>{\centering\arraybackslash}m{\sixfig}>{\centering\arraybackslash}m{\sixfig}>{\centering\arraybackslash}m{\sixfig}|>{\centering\arraybackslash}m{\sixfig}}

\includegraphics[height=2.2cm]{Fig/labels/TotalInfected.pdf}&
\includegraphics[width=\sixfig]{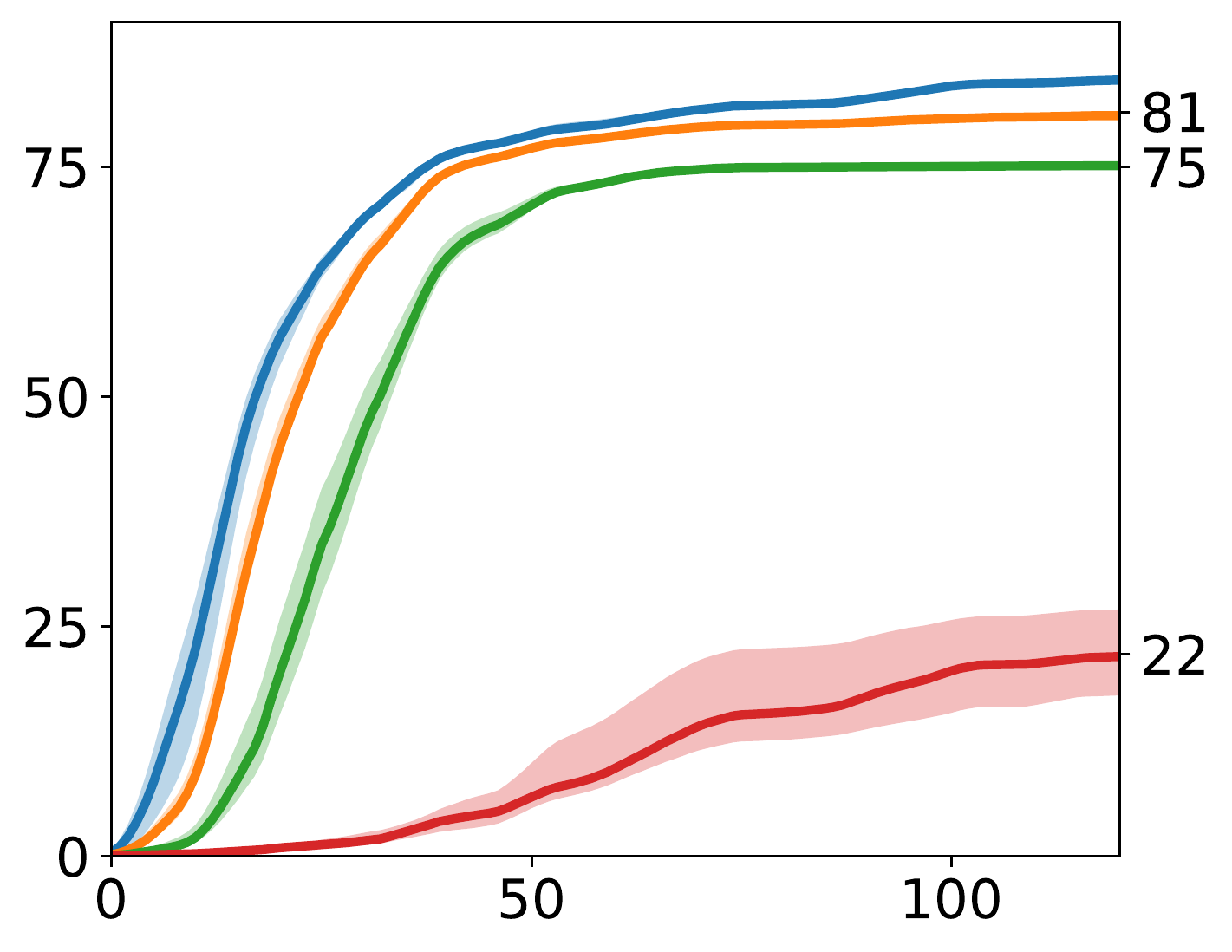}  &
\includegraphics[width=\sixfig]{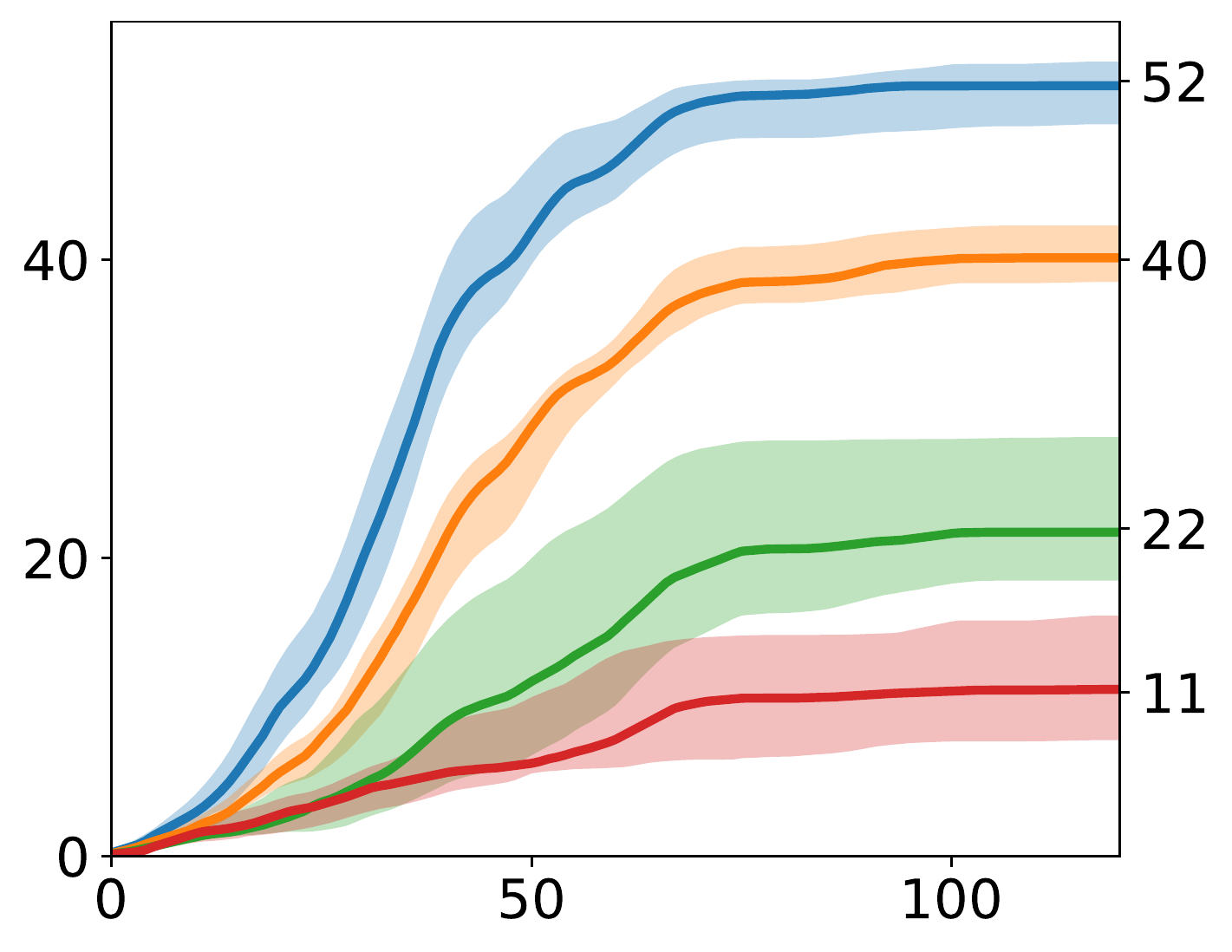}  &
\includegraphics[width=\sixfig]{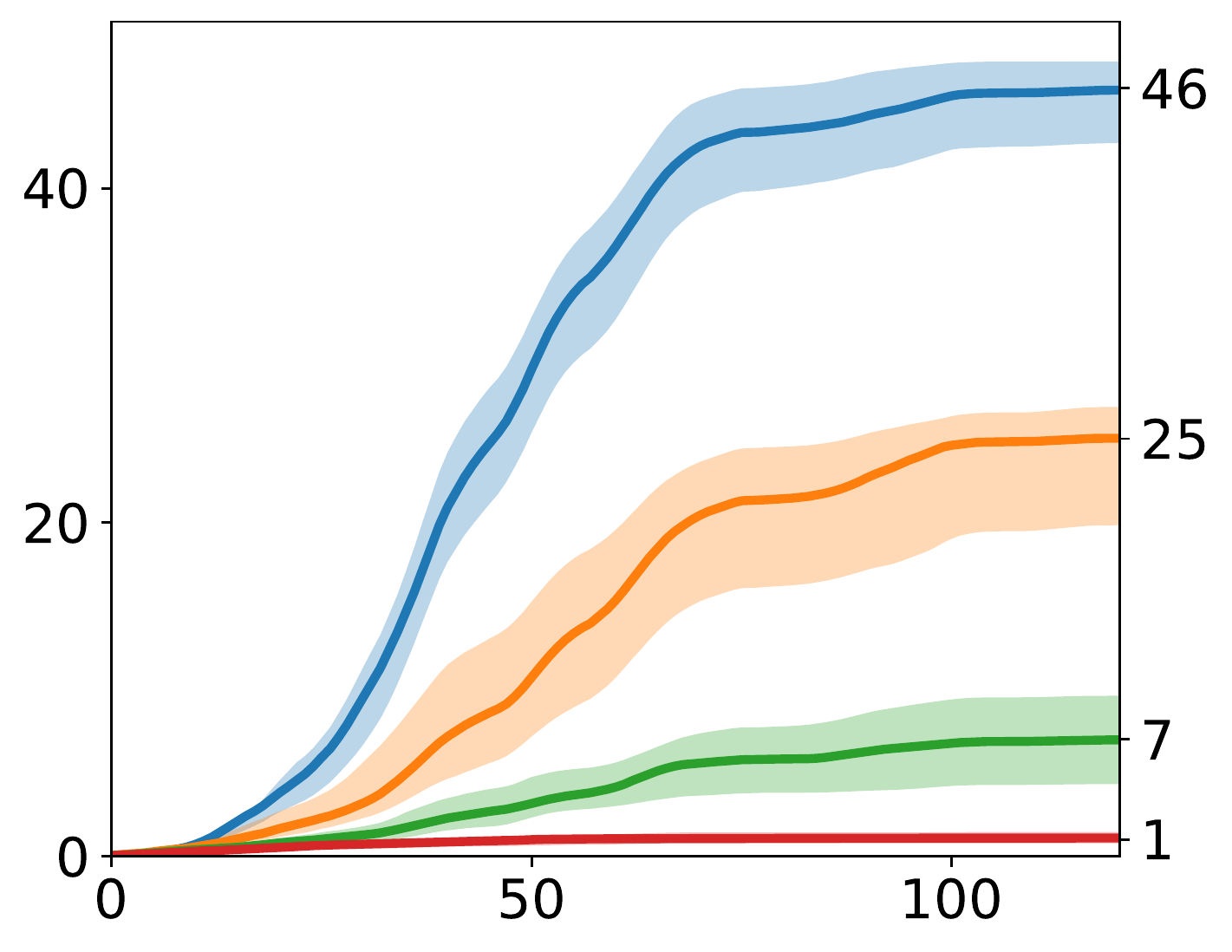}  &
\includegraphics[width=\sixfig]{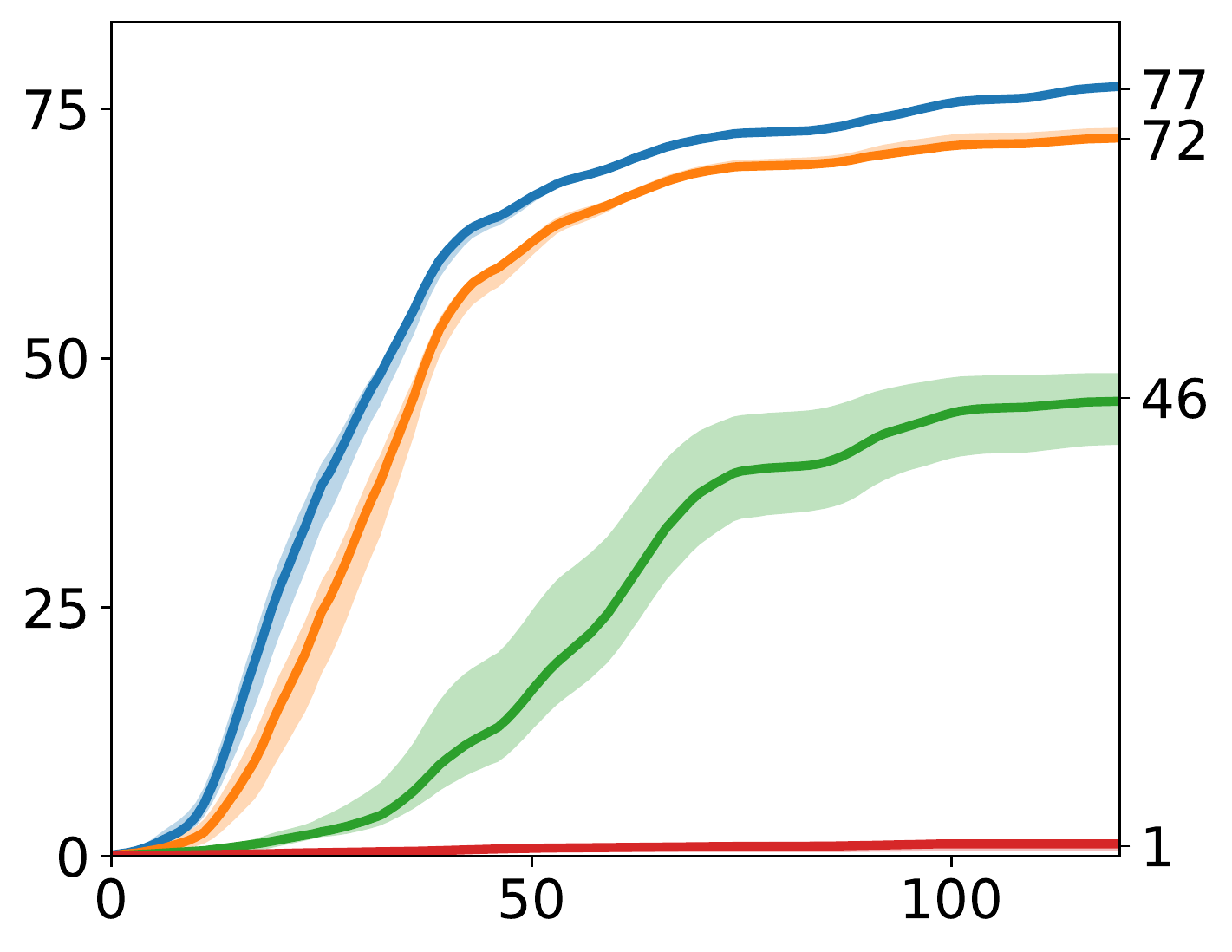}  &
\includegraphics[width=\sixfig]{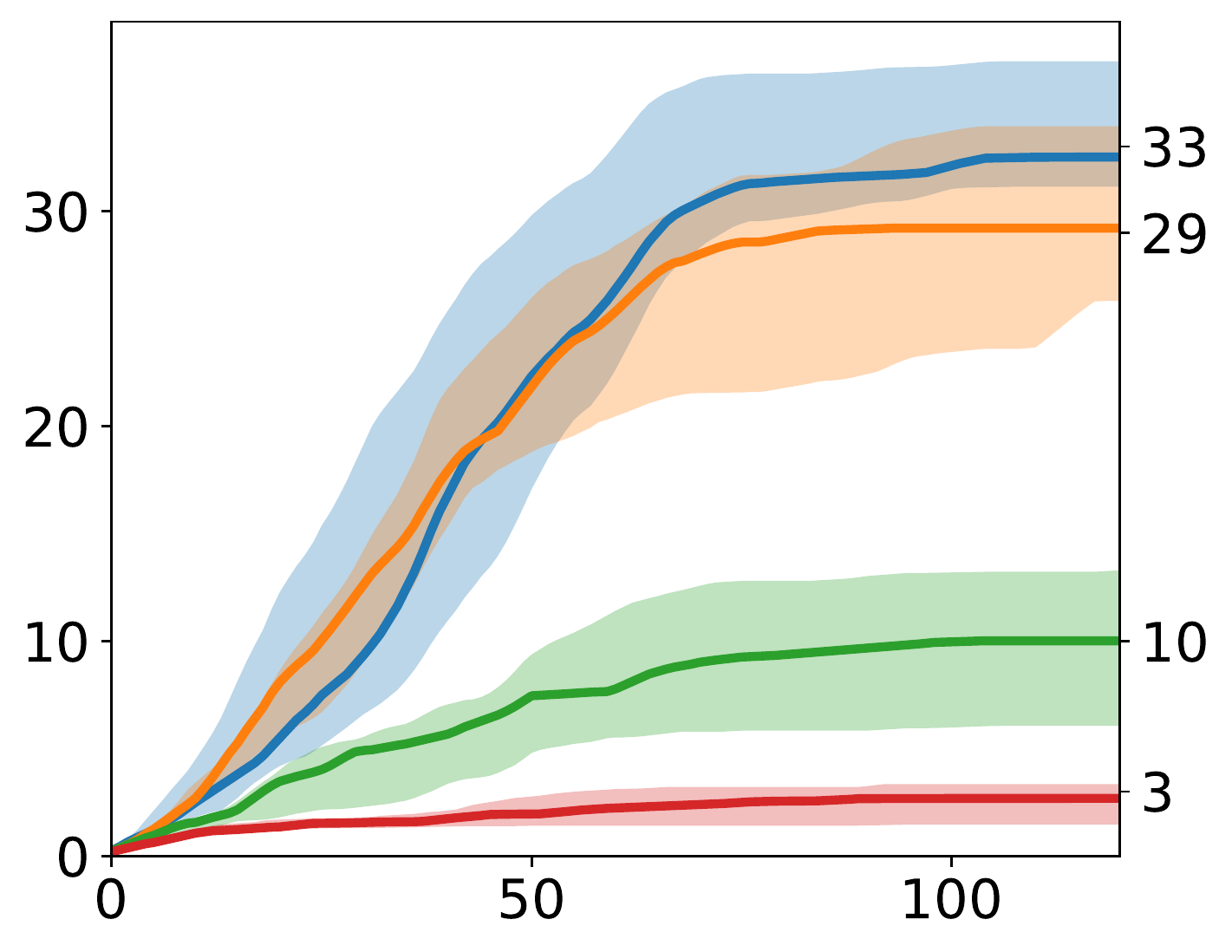} &
\includegraphics[width=\sixfig]{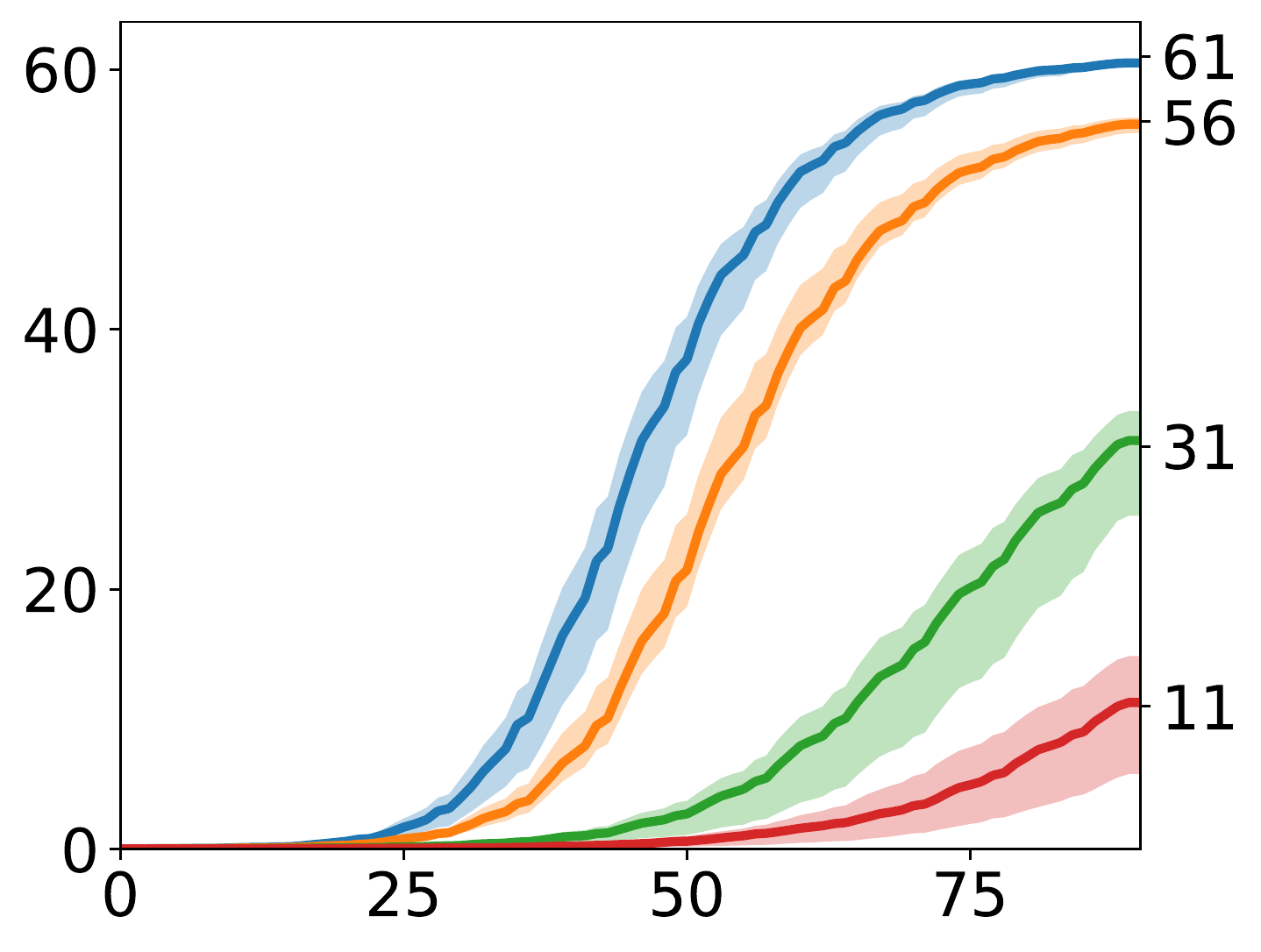} 
\\

\includegraphics[height=2.2cm]{Fig/labels/ActiveInfected.pdf}&
\includegraphics[width=\sixfig]{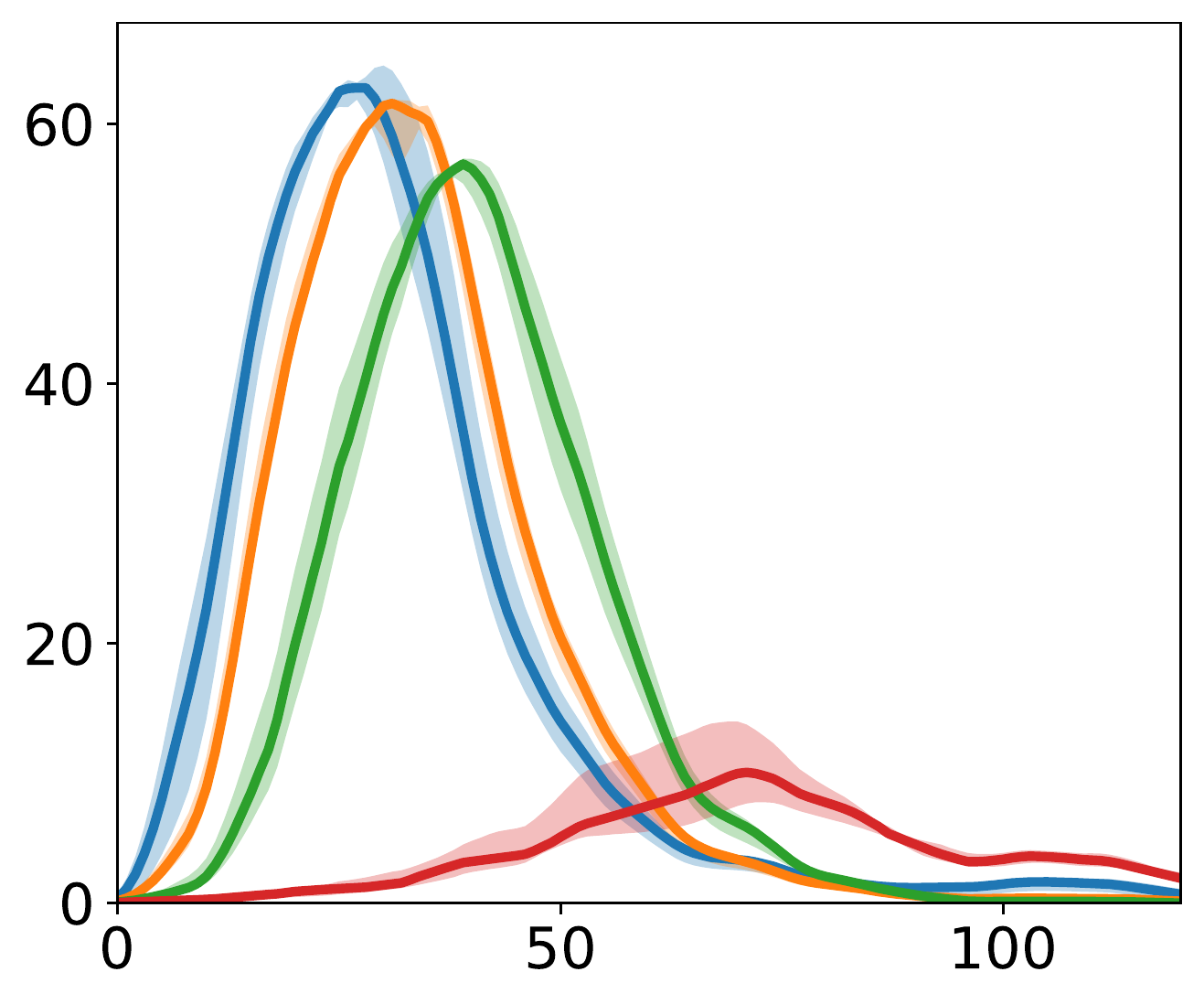}  &
\includegraphics[width=\sixfig]{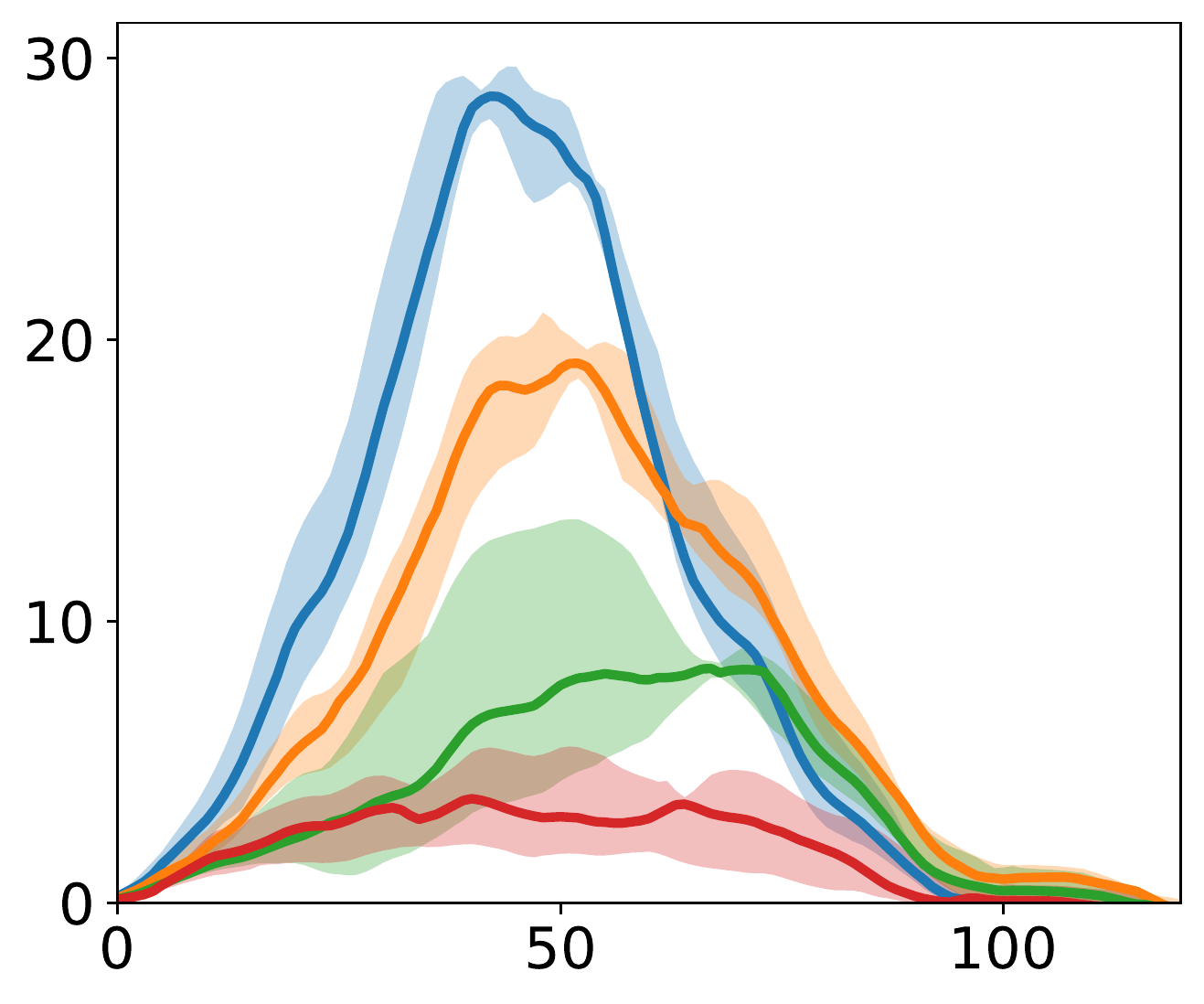}  &
\includegraphics[width=\sixfig]{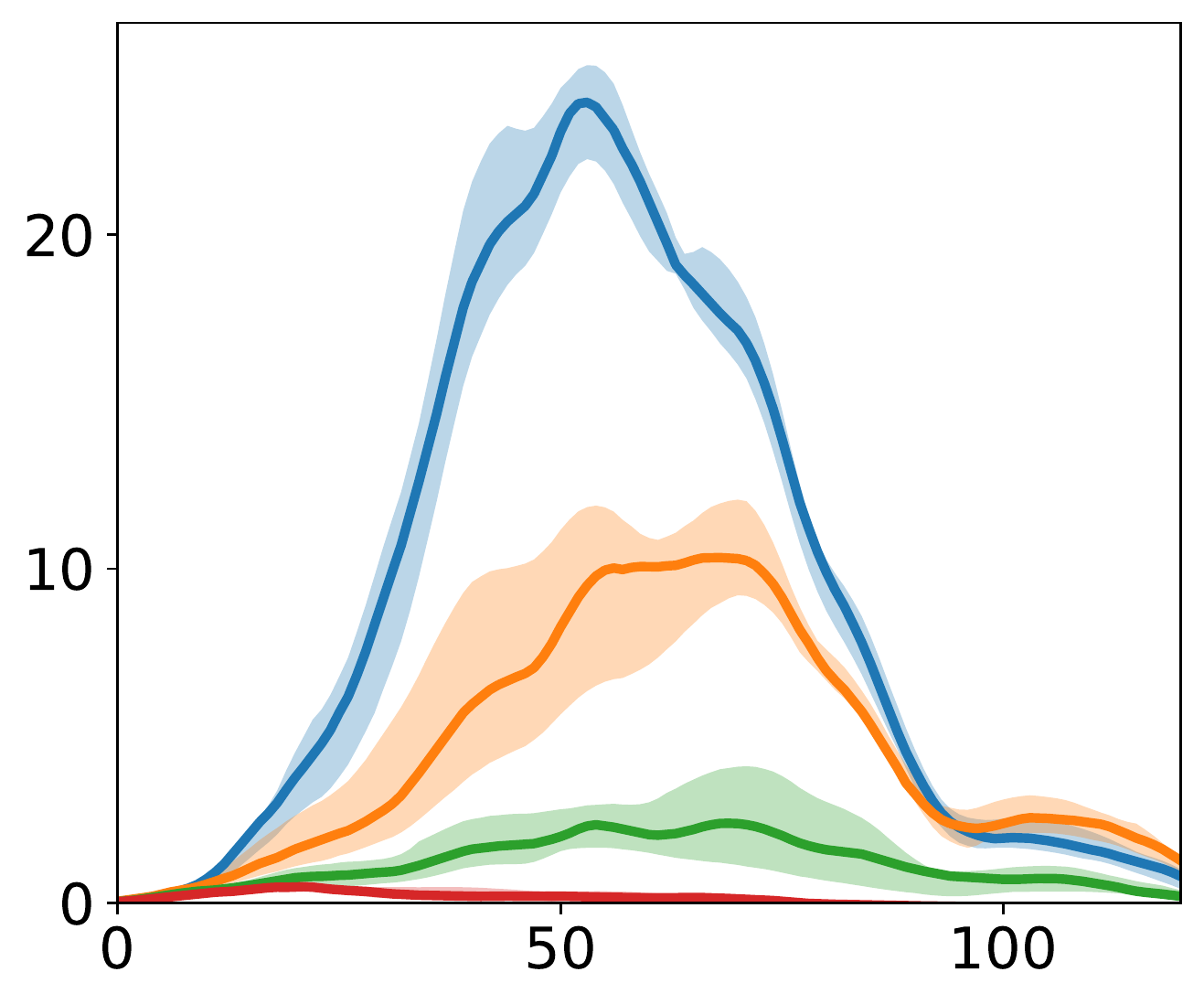}  &
\includegraphics[width=\sixfig]{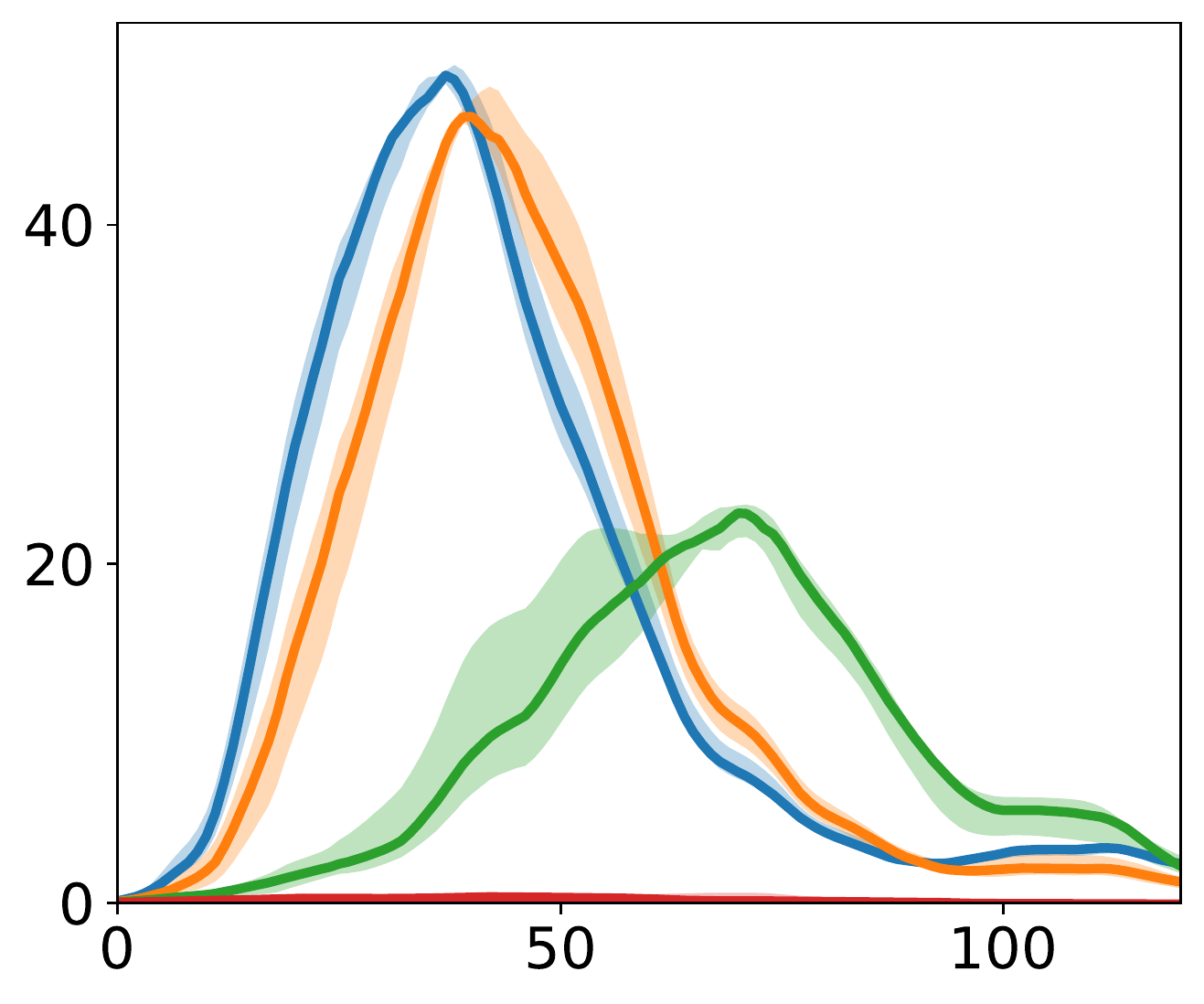}  &
\includegraphics[width=\sixfig]{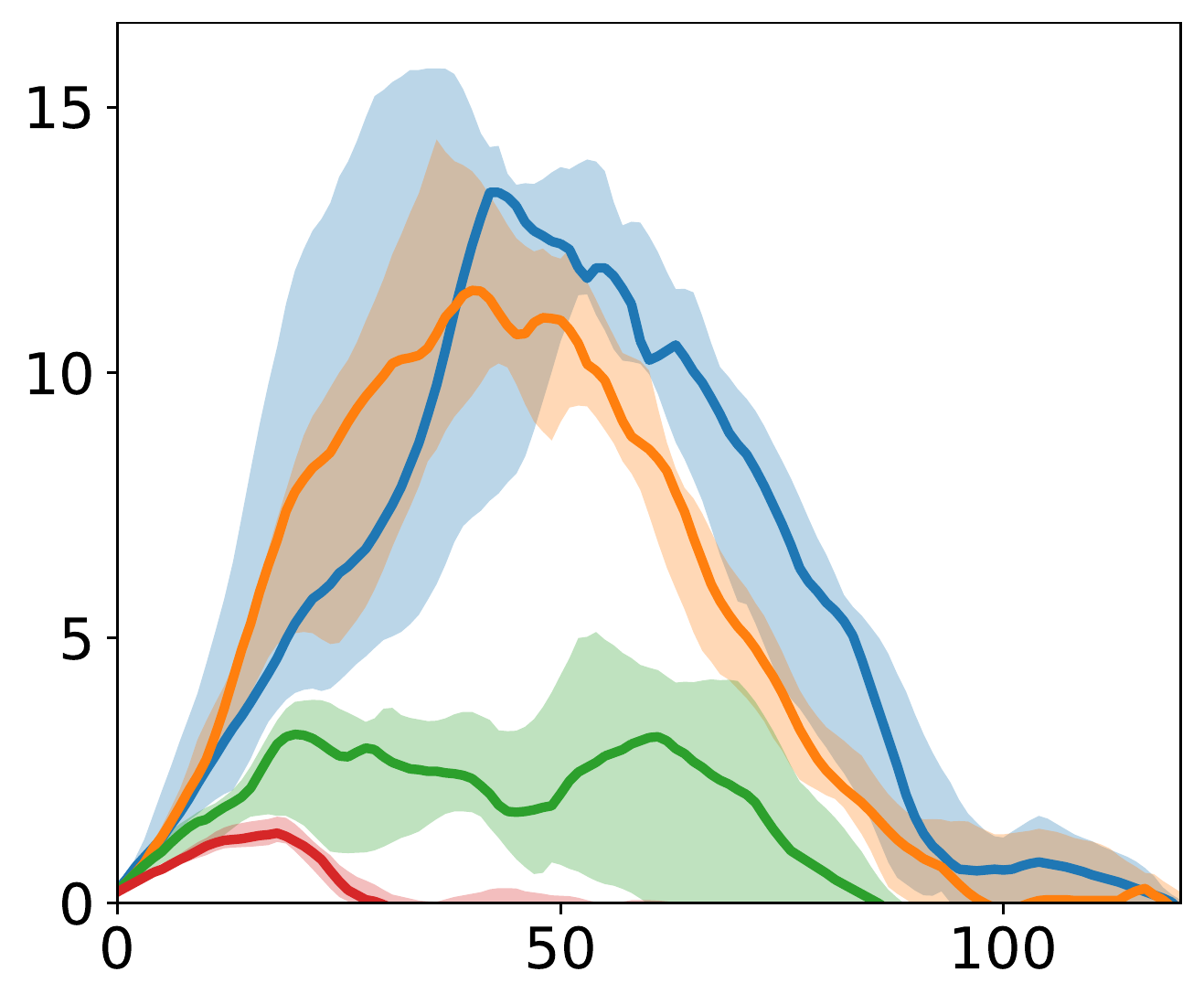} &
\includegraphics[width=\sixfig]{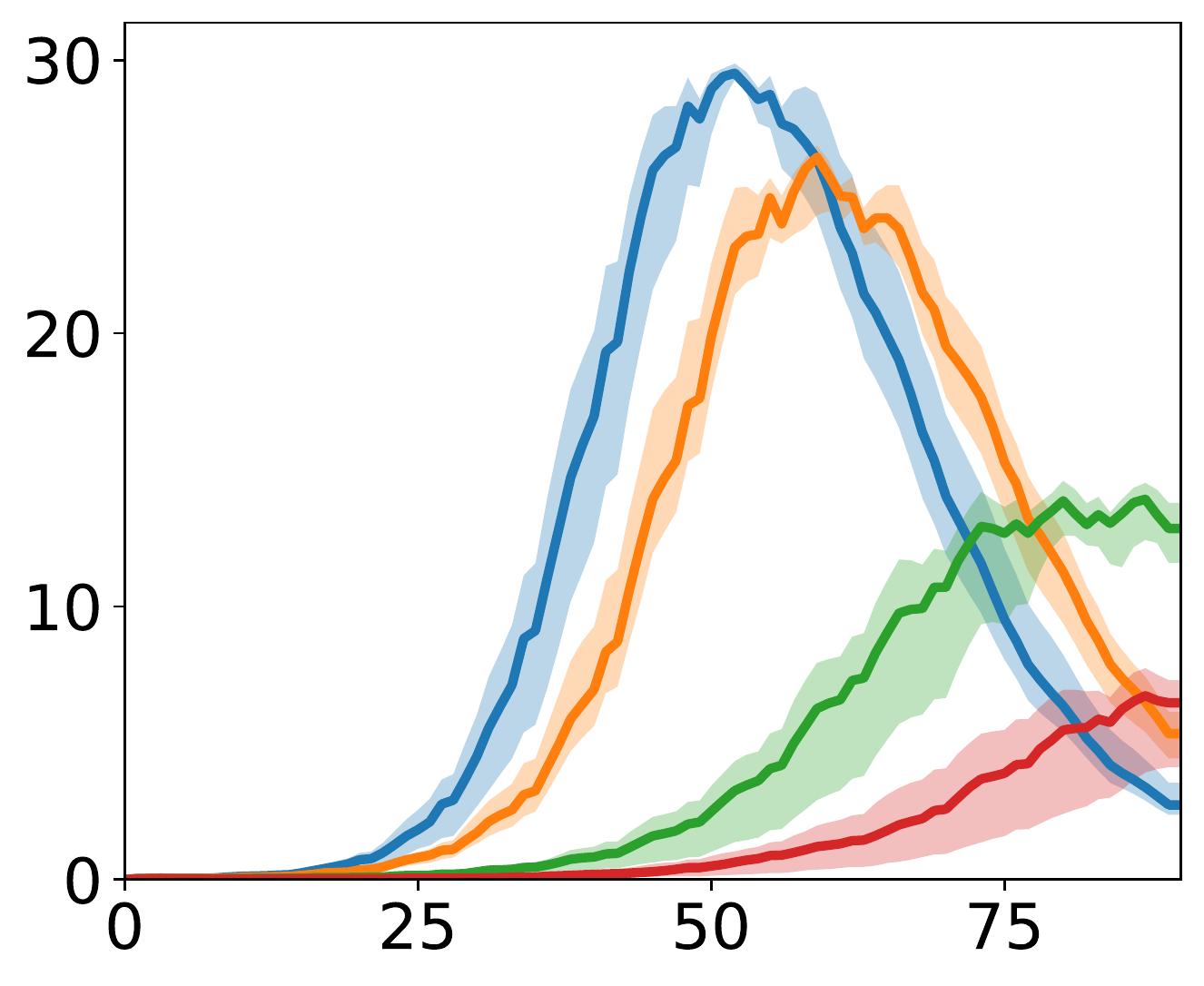} 
\\

\includegraphics[height=2.2cm]{Fig/labels/newInfected.pdf}&
\includegraphics[width=\sixfig]{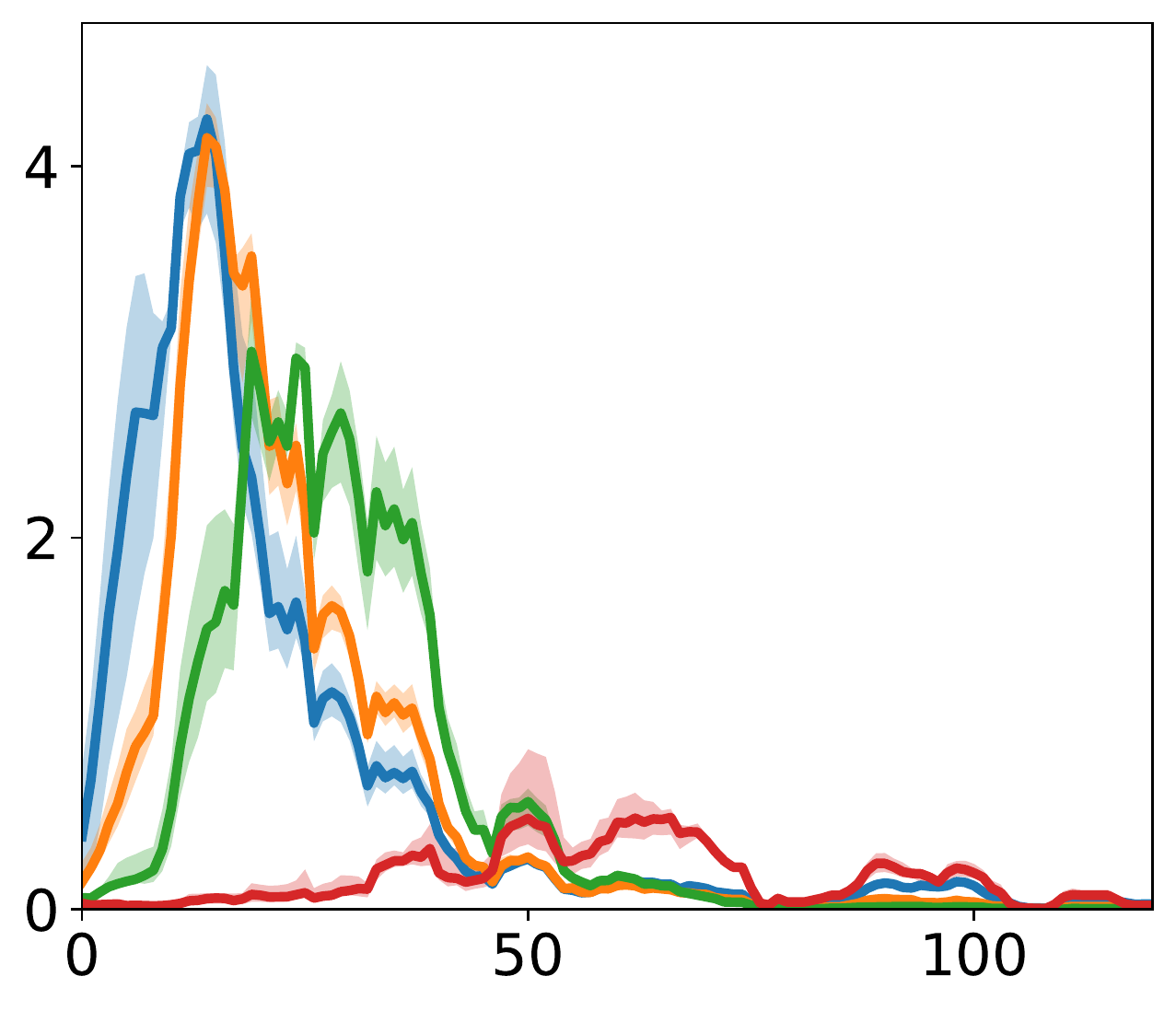}  &
\includegraphics[width=\sixfig]{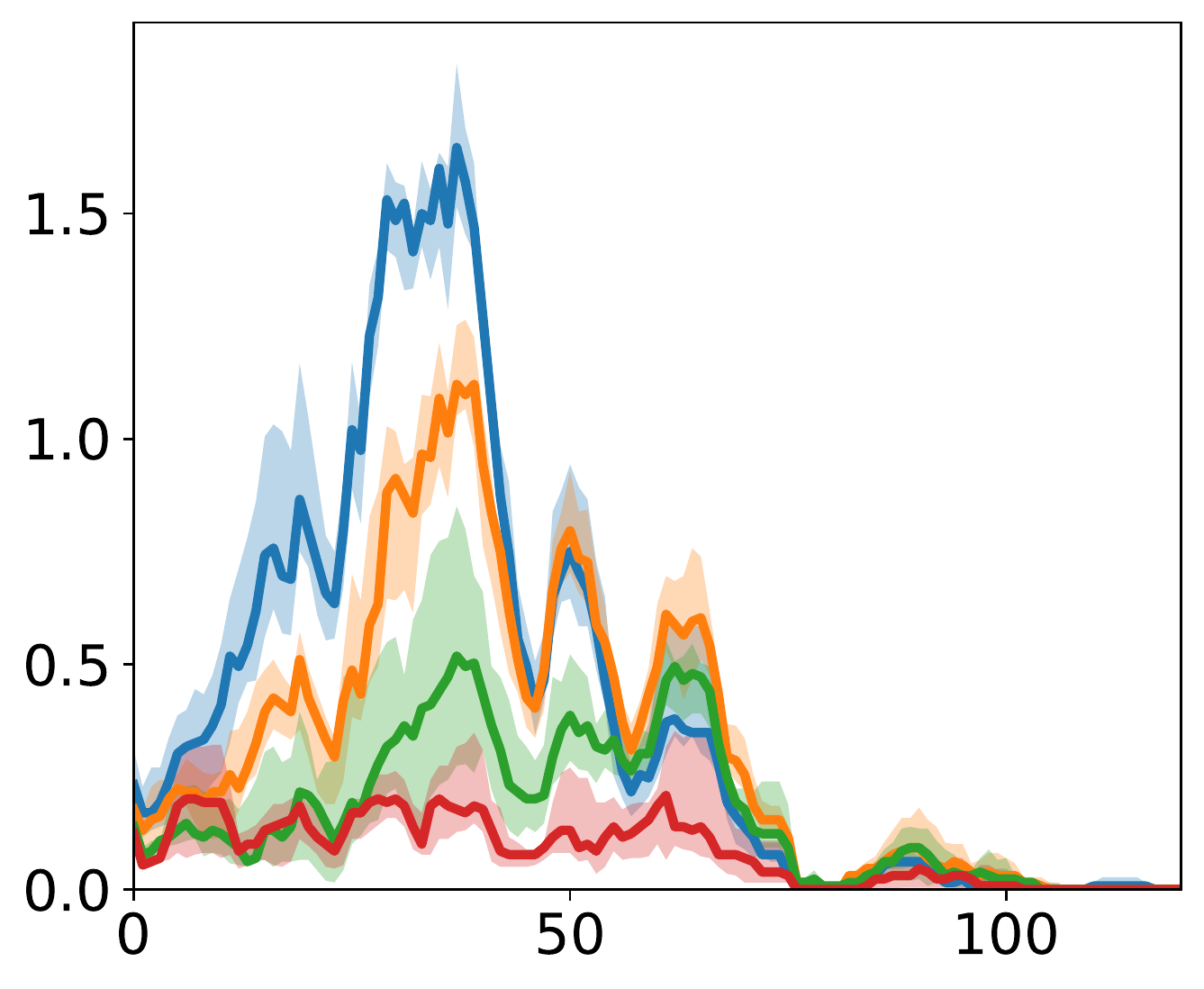}  &
\includegraphics[width=\sixfig]{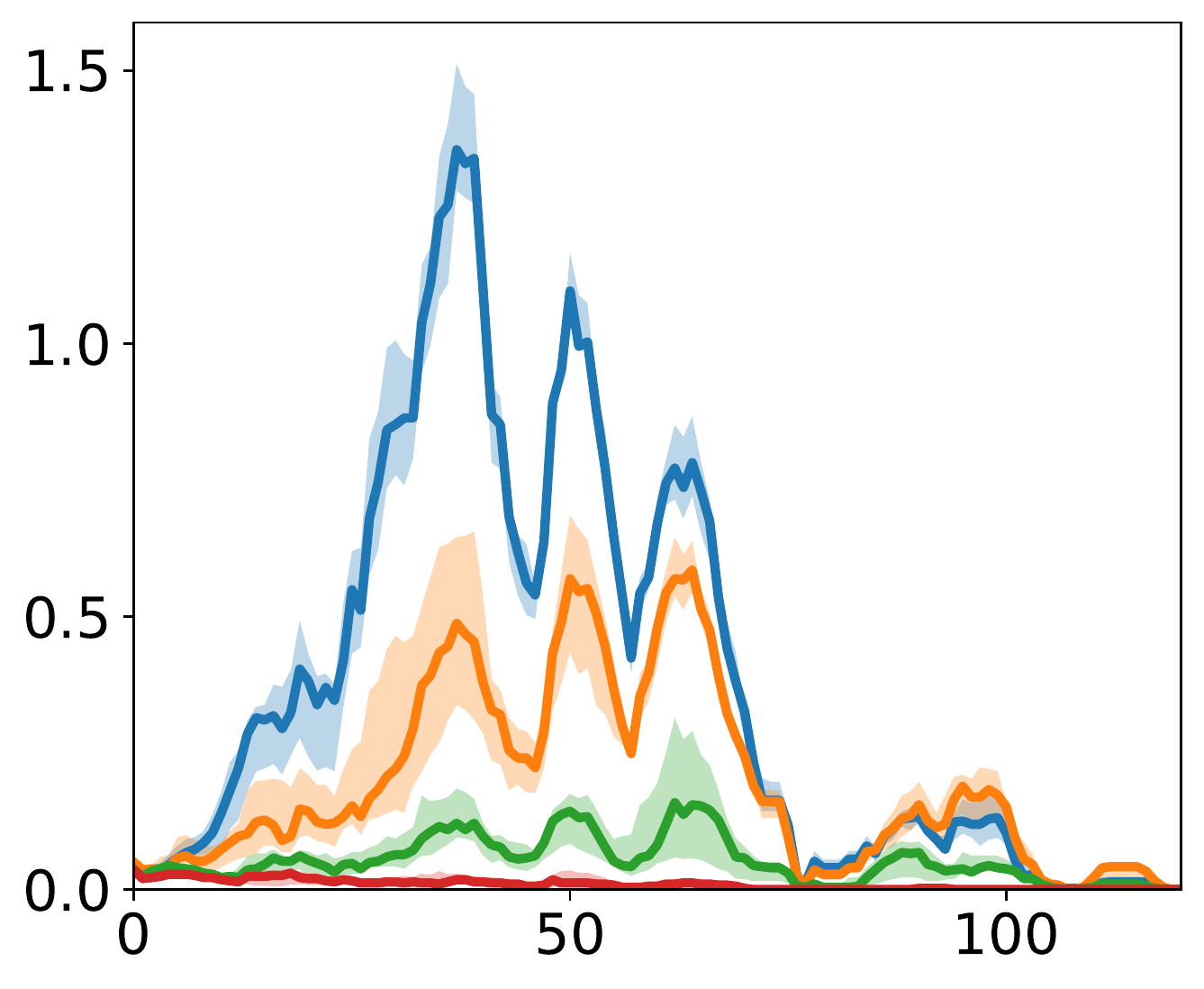}  &
\includegraphics[width=\sixfig]{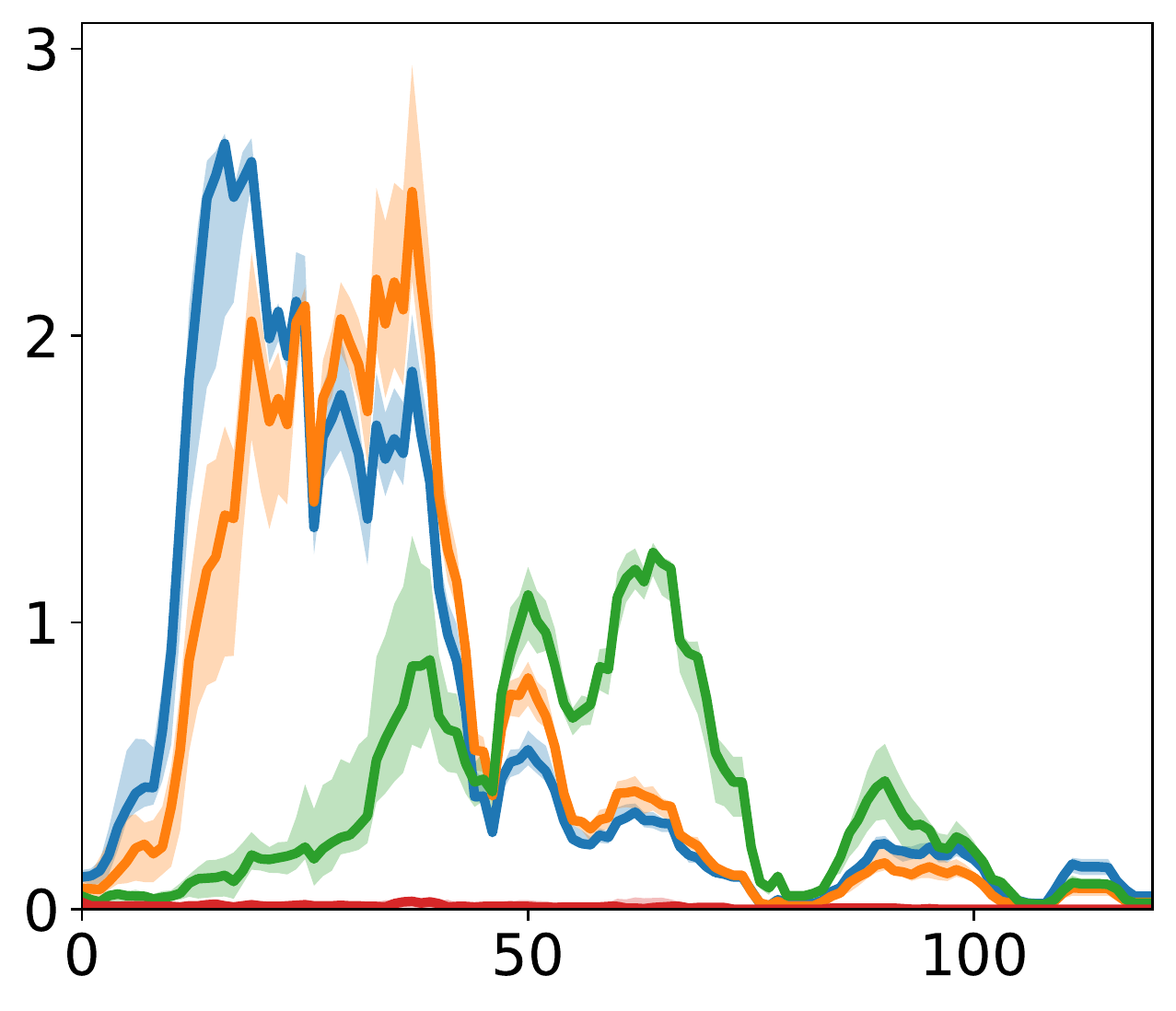}  &
\includegraphics[width=\sixfig]{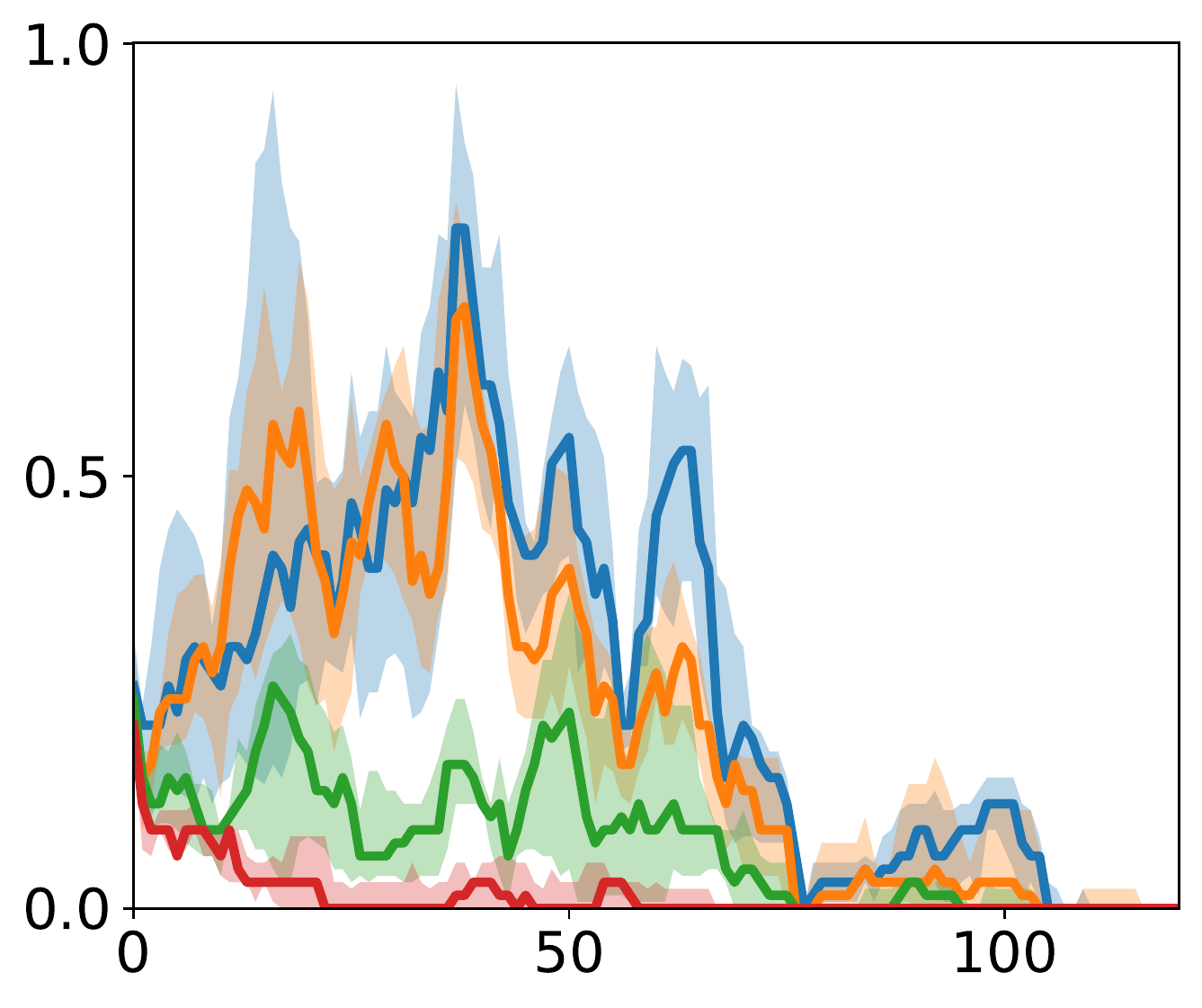} &
\includegraphics[width=\sixfig]{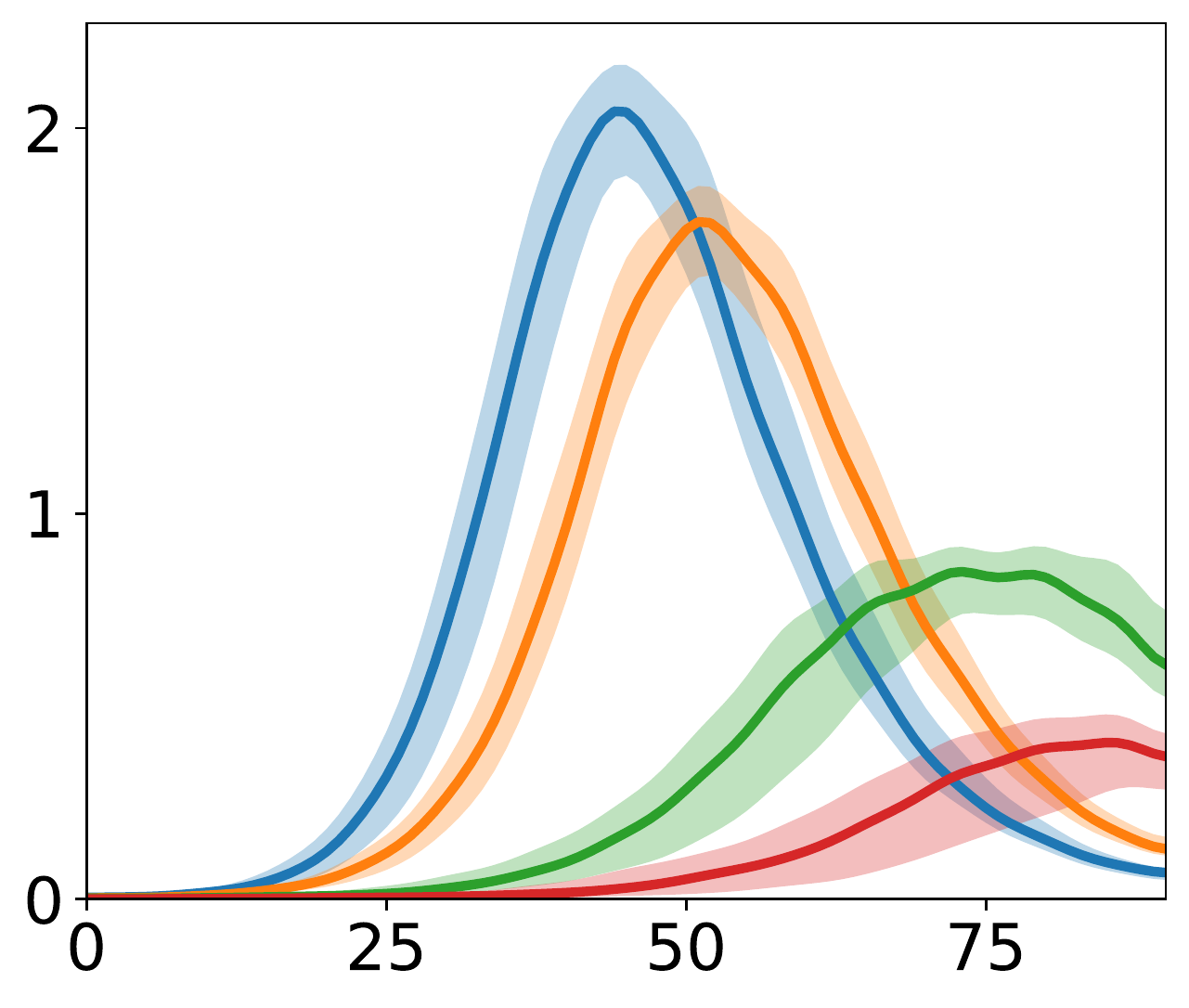} 
\\

\includegraphics[height=1.6cm]{Fig/labels/growthrate.pdf}&
\includegraphics[width=\sixfig]{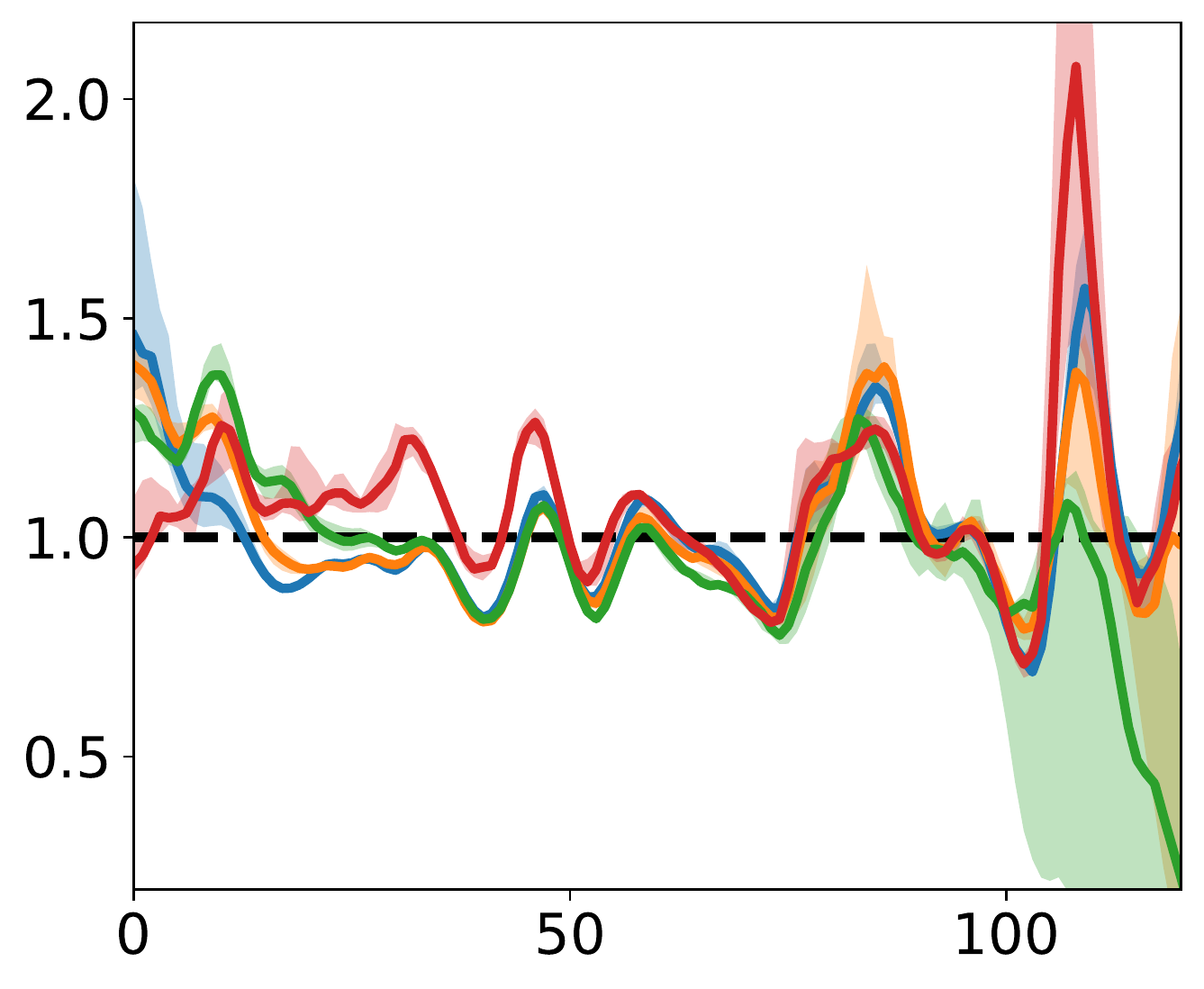}  &
\includegraphics[width=\sixfig]{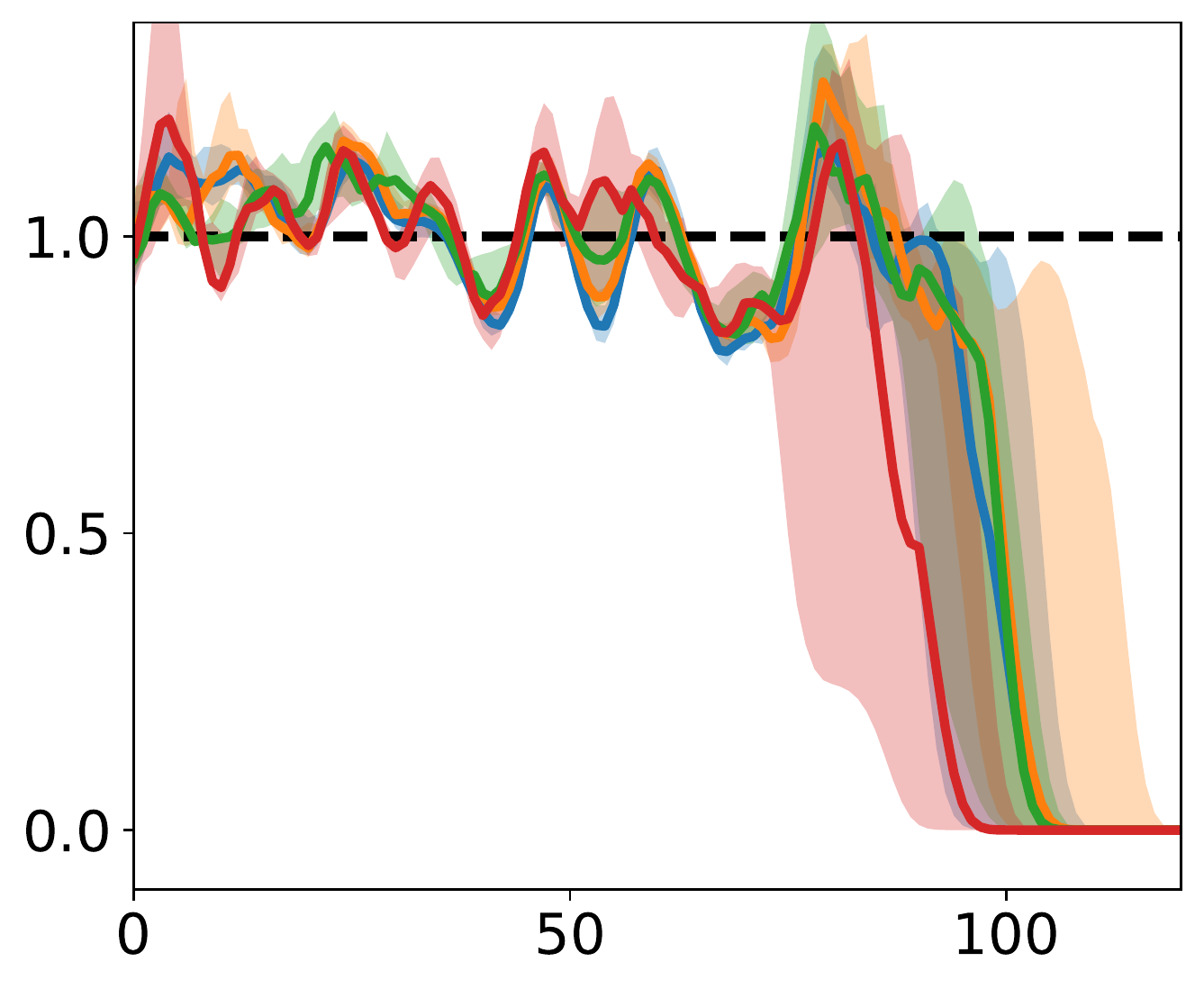}  &
\includegraphics[width=\sixfig]{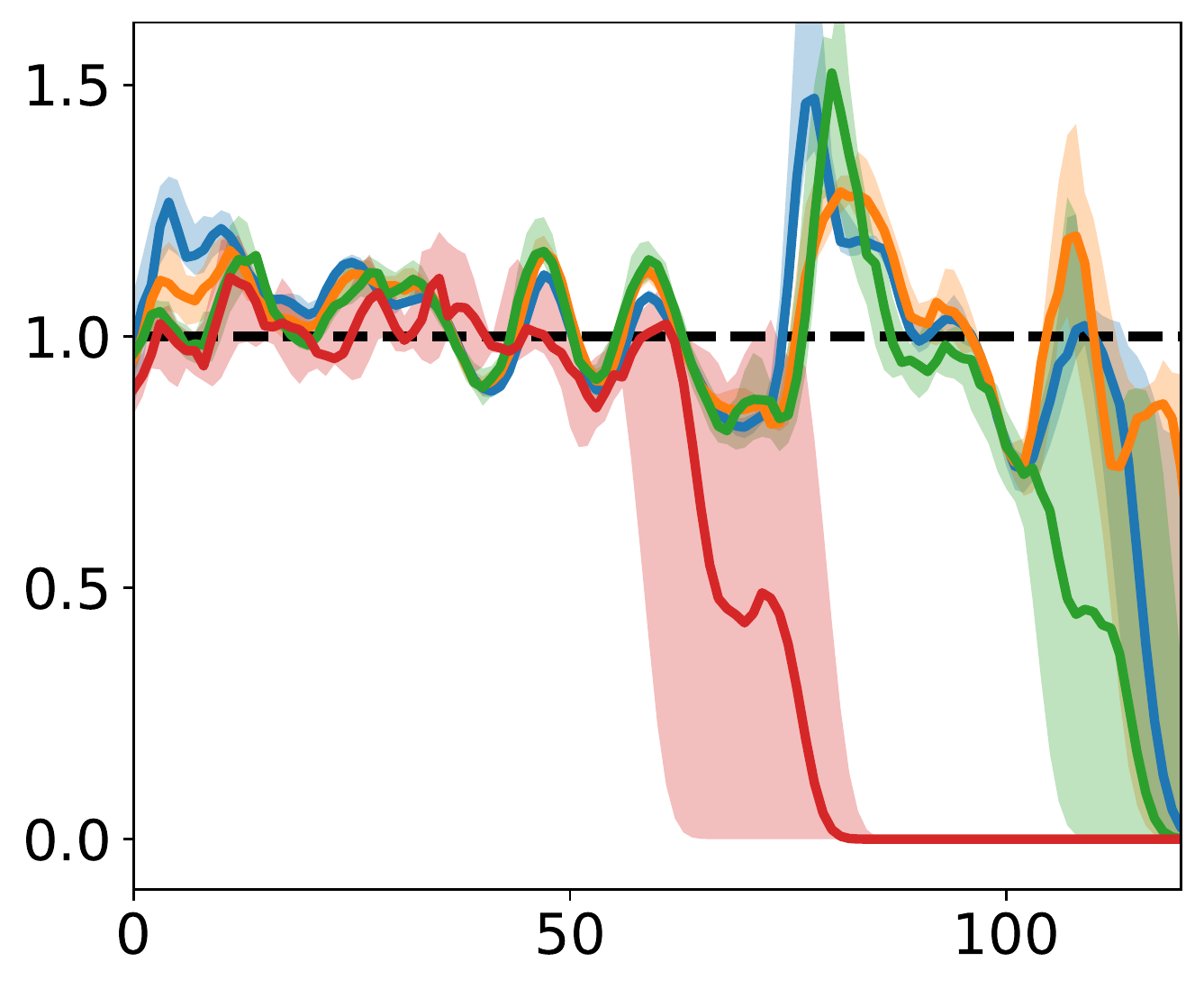}  &
\includegraphics[width=\sixfig]{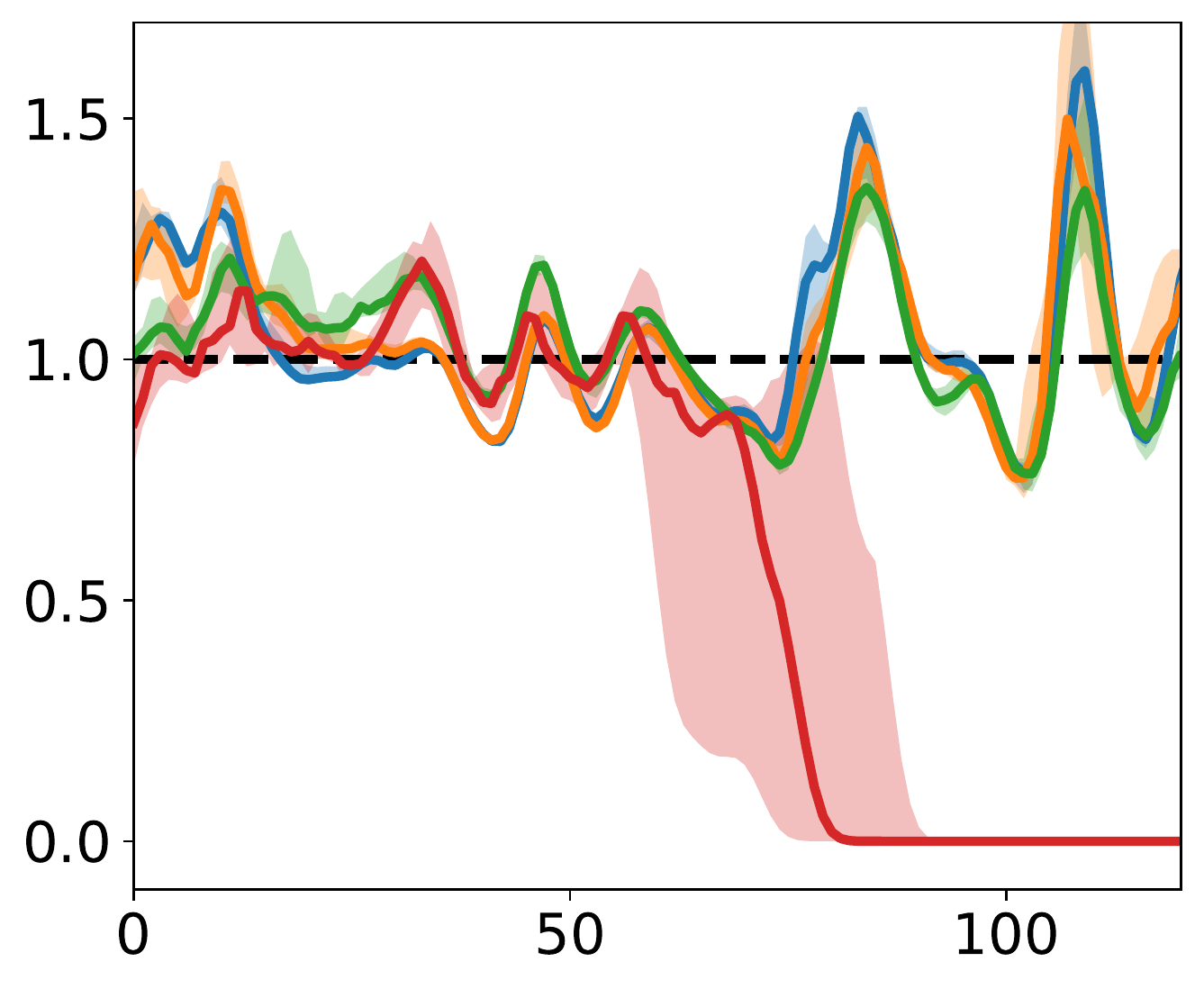}  &
\includegraphics[width=\sixfig]{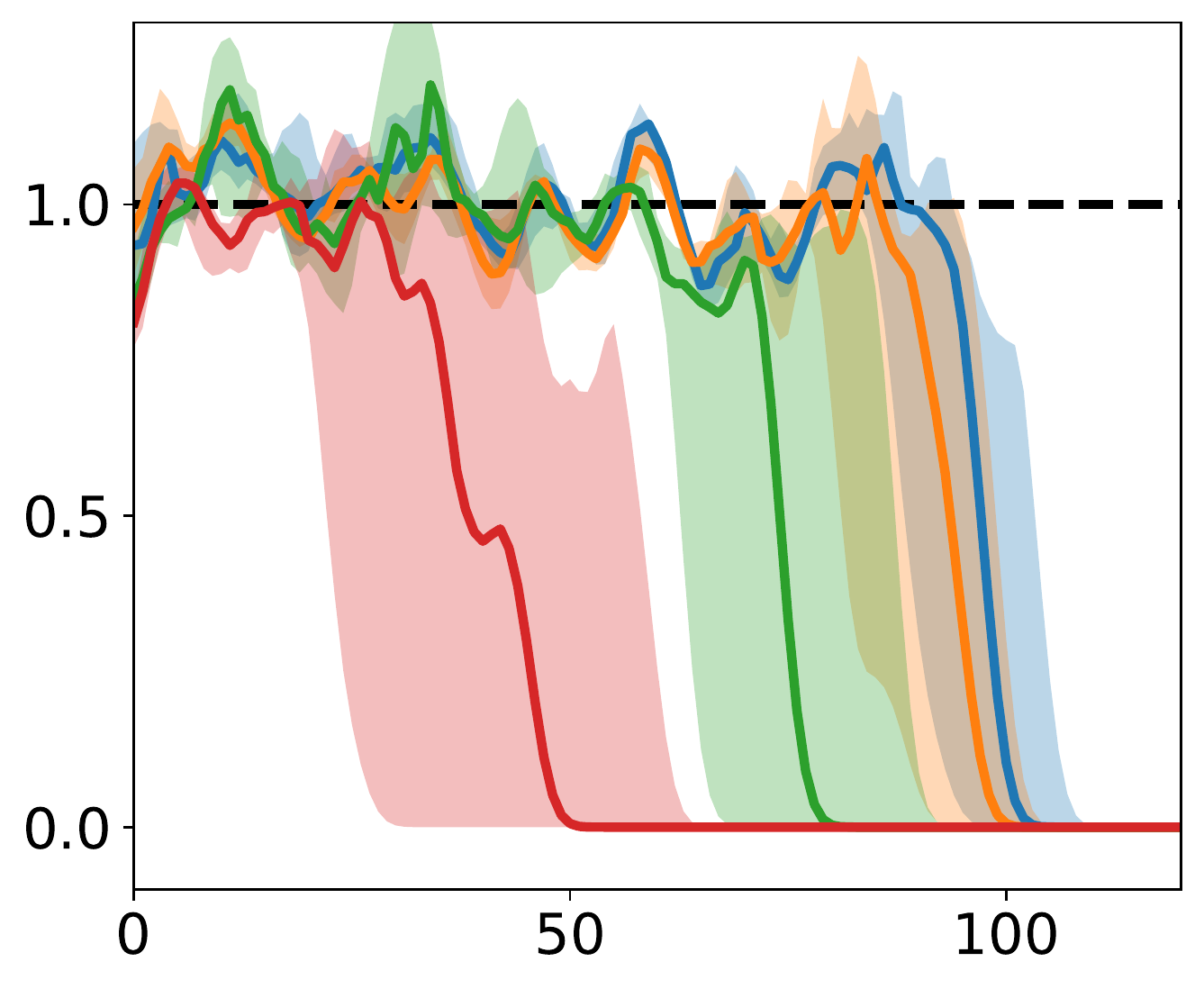} &
\includegraphics[width=\sixfig]{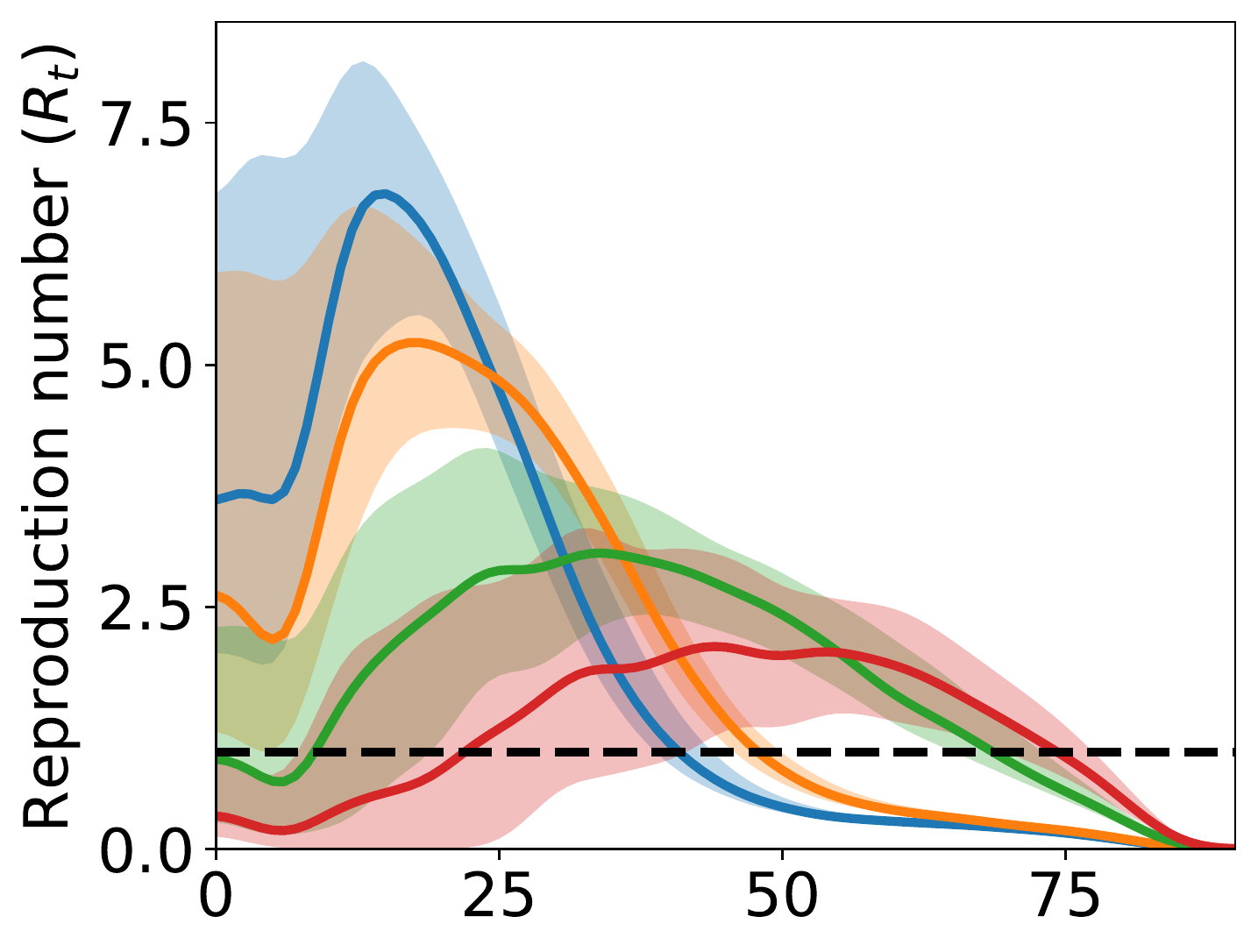} 
\\
\includegraphics[height=1.1cm]{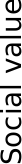}&
\includegraphics[width=\sixfig]{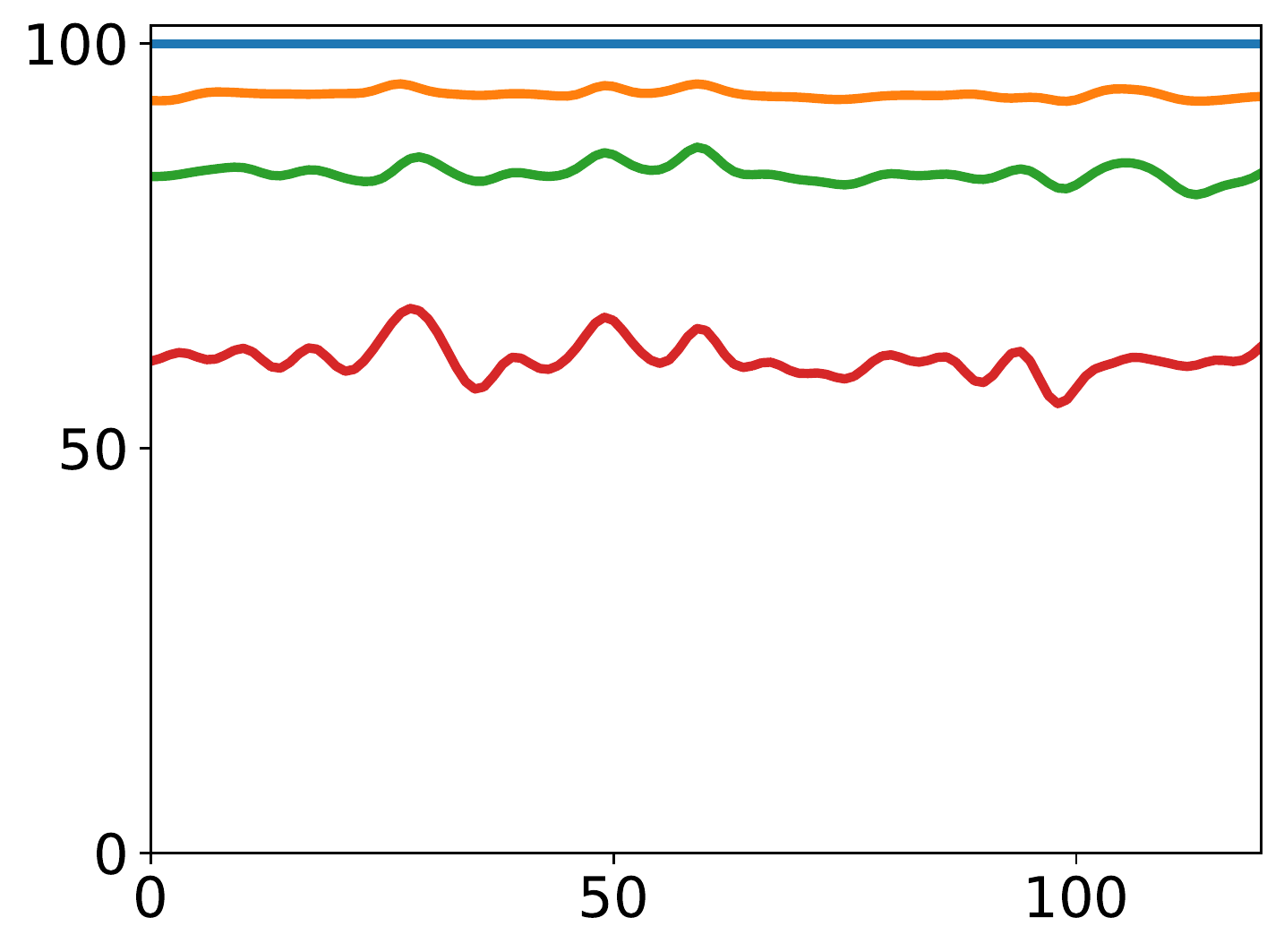}  &
\includegraphics[width=\sixfig]{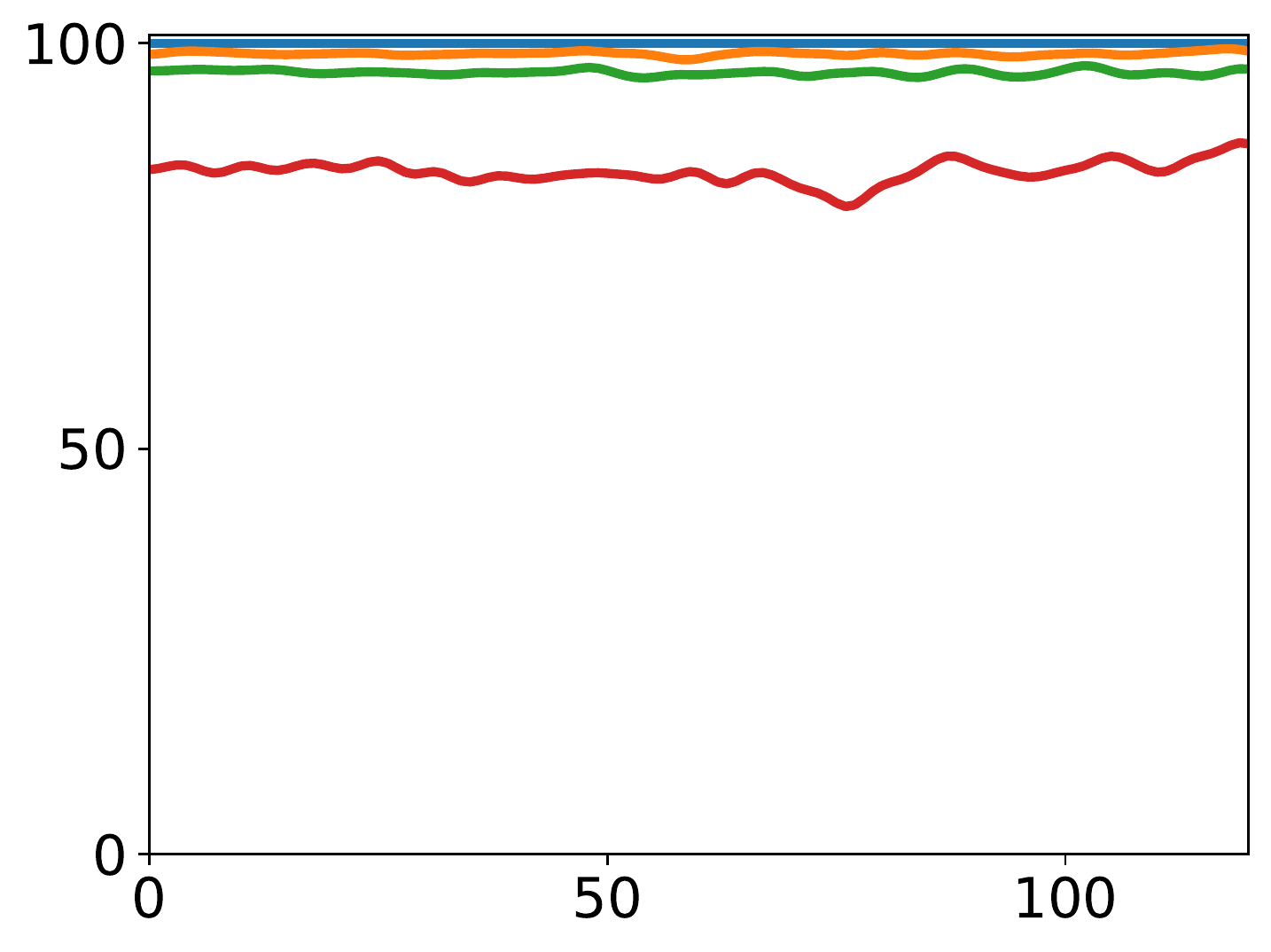}  &
\includegraphics[width=\sixfig]{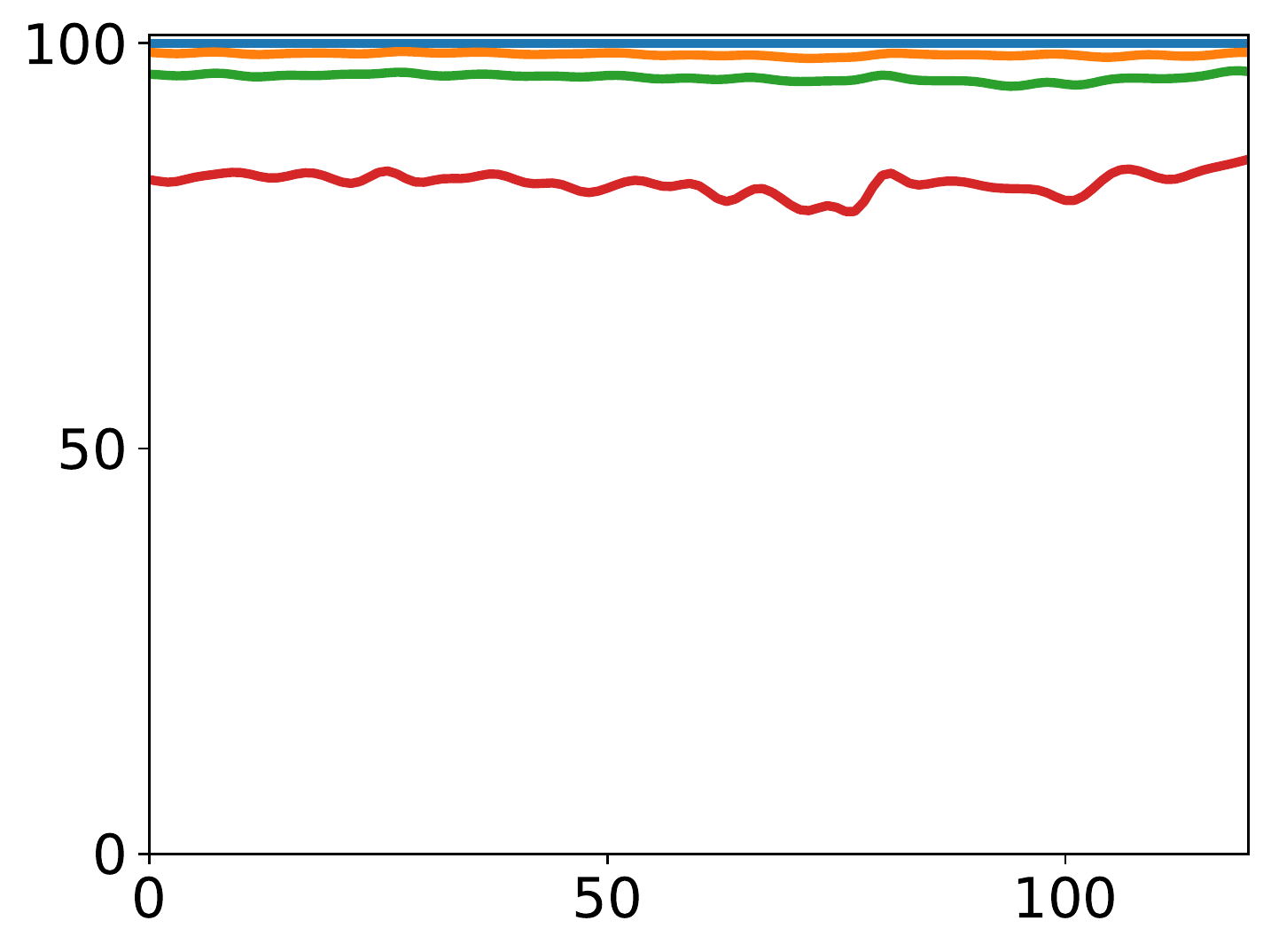}  &
\includegraphics[width=\sixfig]{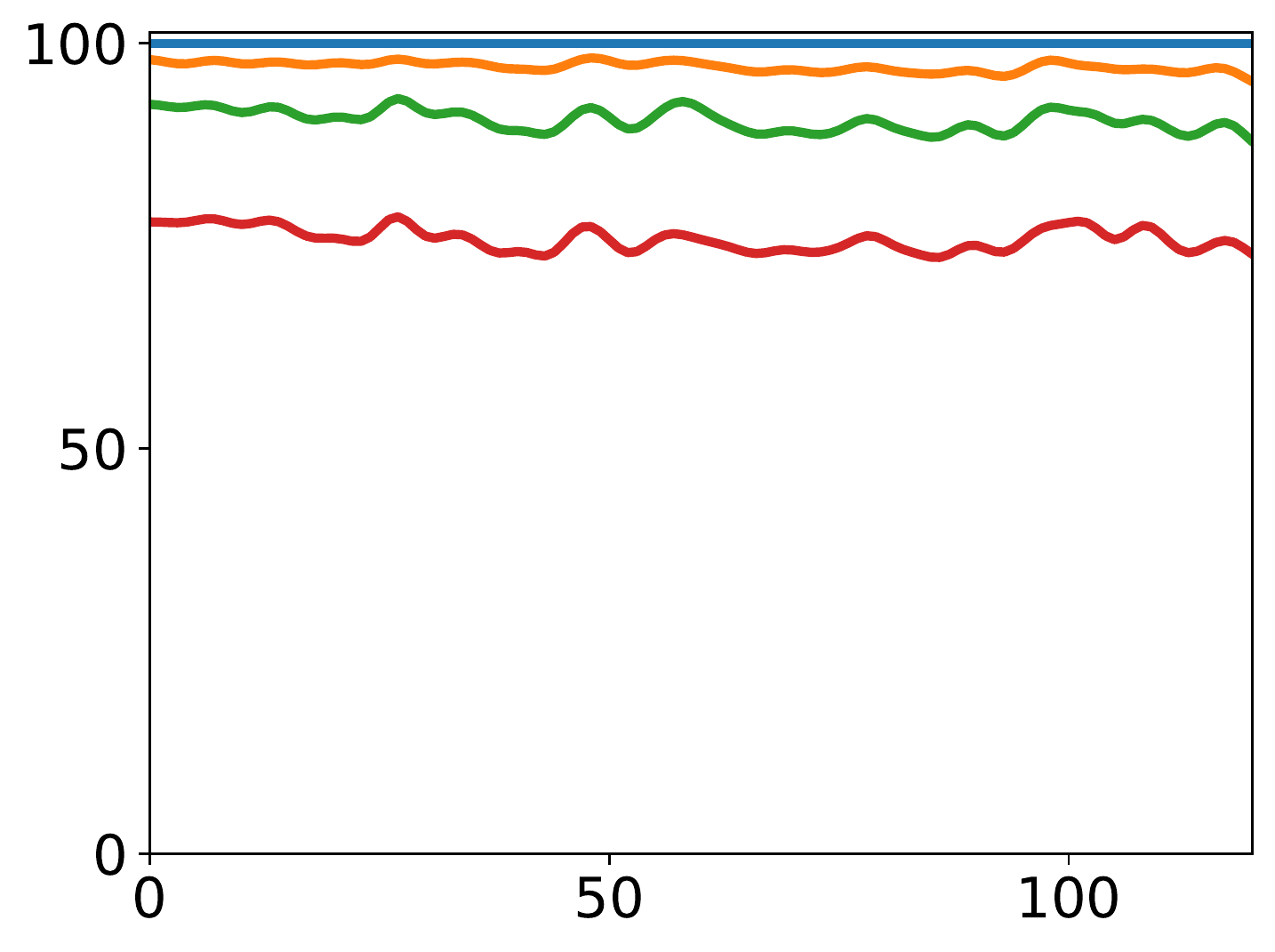}  &
\includegraphics[width=\sixfig]{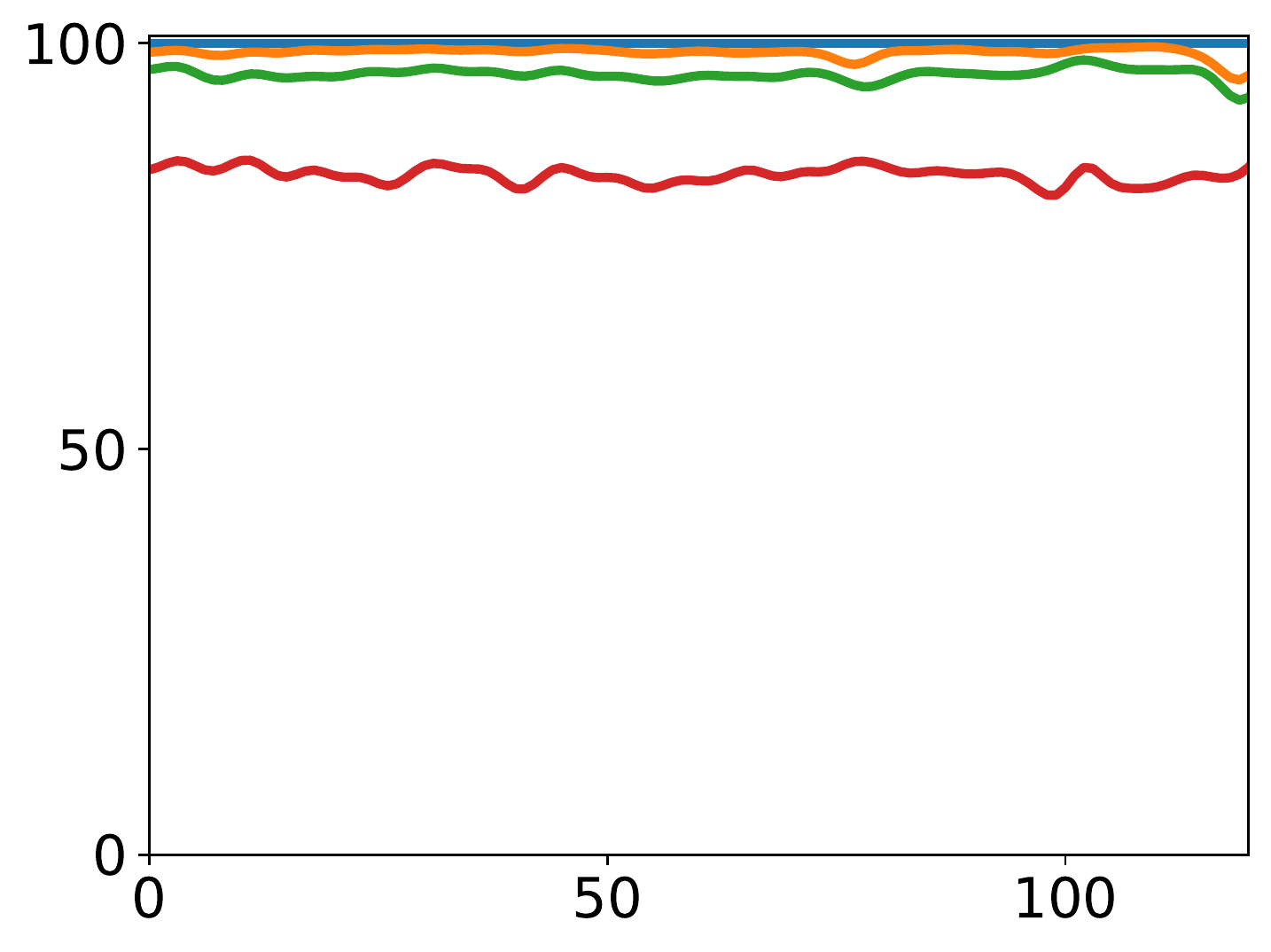}&
\includegraphics[width=\sixfig]{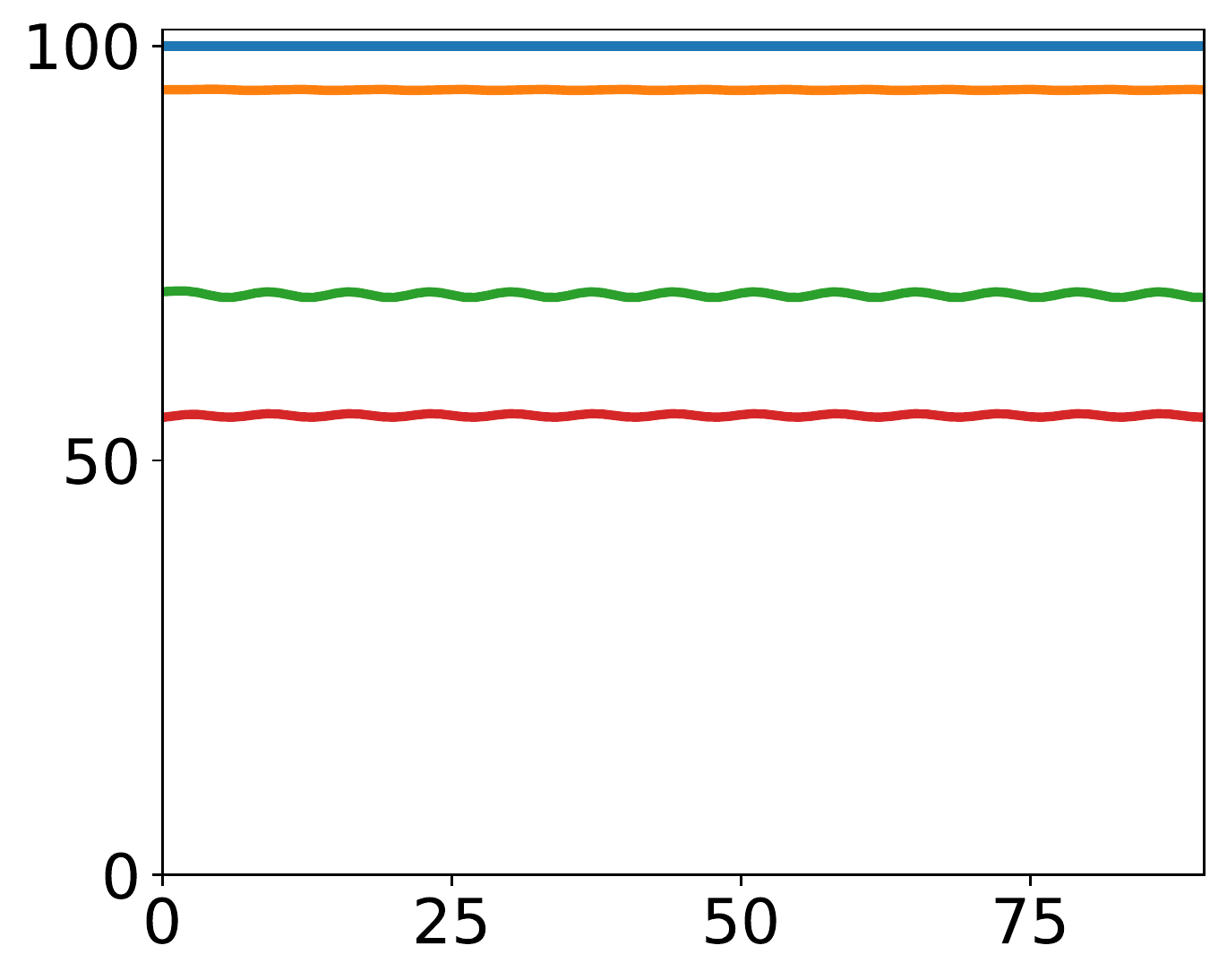}
\\

\multicolumn{6}{c}{\includegraphics[width=0.15\textwidth]{Fig/labels/daysfromstart.pdf}}\\
& {\small Tokyo} & {\small Chicago} & {\small Los Angeles} & {\small Jakarta} & {\small London} & {\small University}\\

\multicolumn{6}{c}{\includegraphics[width=0.6\textwidth]{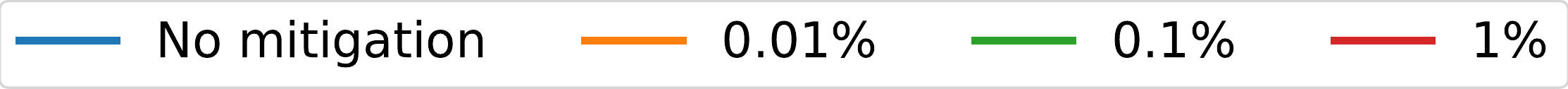}}&
\includegraphics[width=\sixfig]{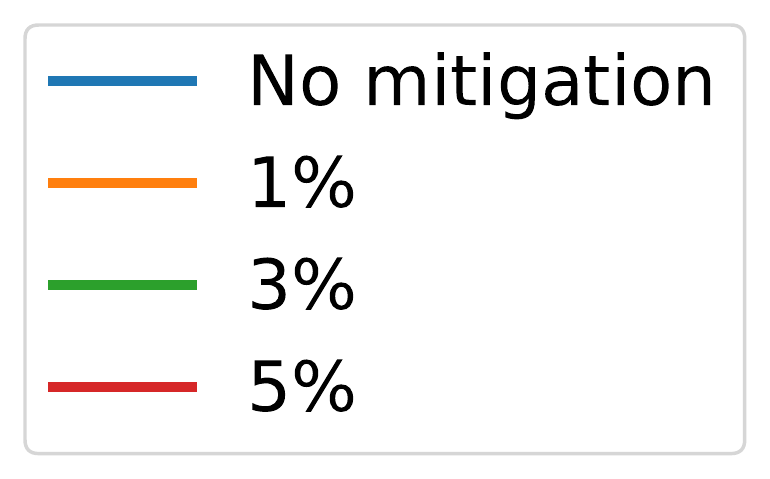}
\\

\end{tabular}
\caption{Infection spreading with the intervention strategy of closing popular venues. With more popular venues closed, the total number of infected people is reduced. The peaks of people actively infected are delayed and become lower. In addition, closing these popular venues has a comparatively smaller influence on the social value than uniform intervention.}
\label{figS:close_popular_venue}
\end{figure}

\newpage
\begin{figure}[!hp]
    \centering
\begin{tabular}{m{0.1cm}m{\fivefig}@{}m{0.1cm}m{\fivefig}@{}m{0.1cm}m{\fivefig}@{}m{0.1cm}m{\fivefig}@{}m{0.1cm}m{\fivefig}@{}m{0.1cm}}
\multirow{14}{*}{\includegraphics[height=4cm]{Fig/labels/TotalInfected.pdf}}&
\includegraphics[width=\fivefig]{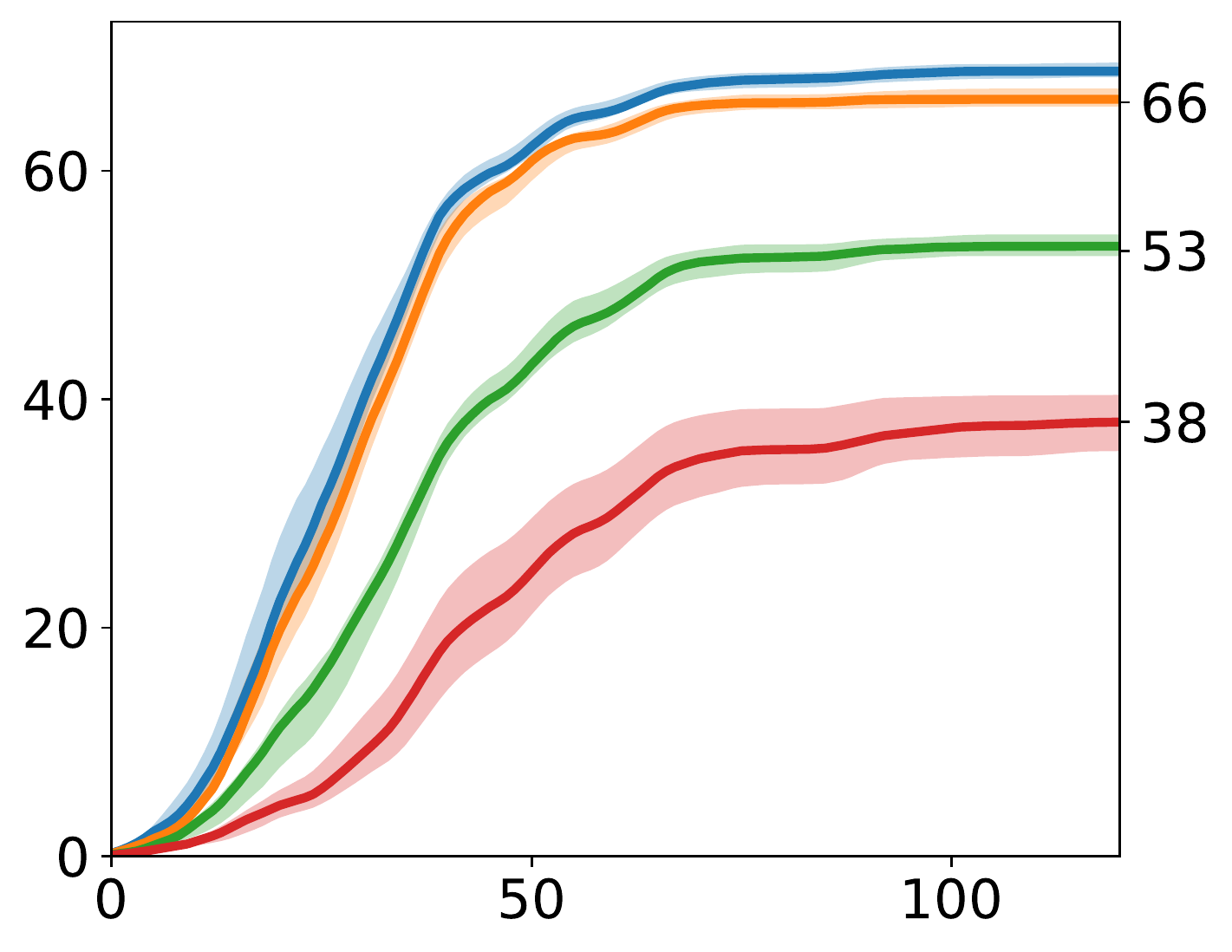} &
\multirow{14}{*}{\includegraphics[height=4cm]{Fig/labels/ActiveInfected.pdf}}&
\includegraphics[width=\fivefig]{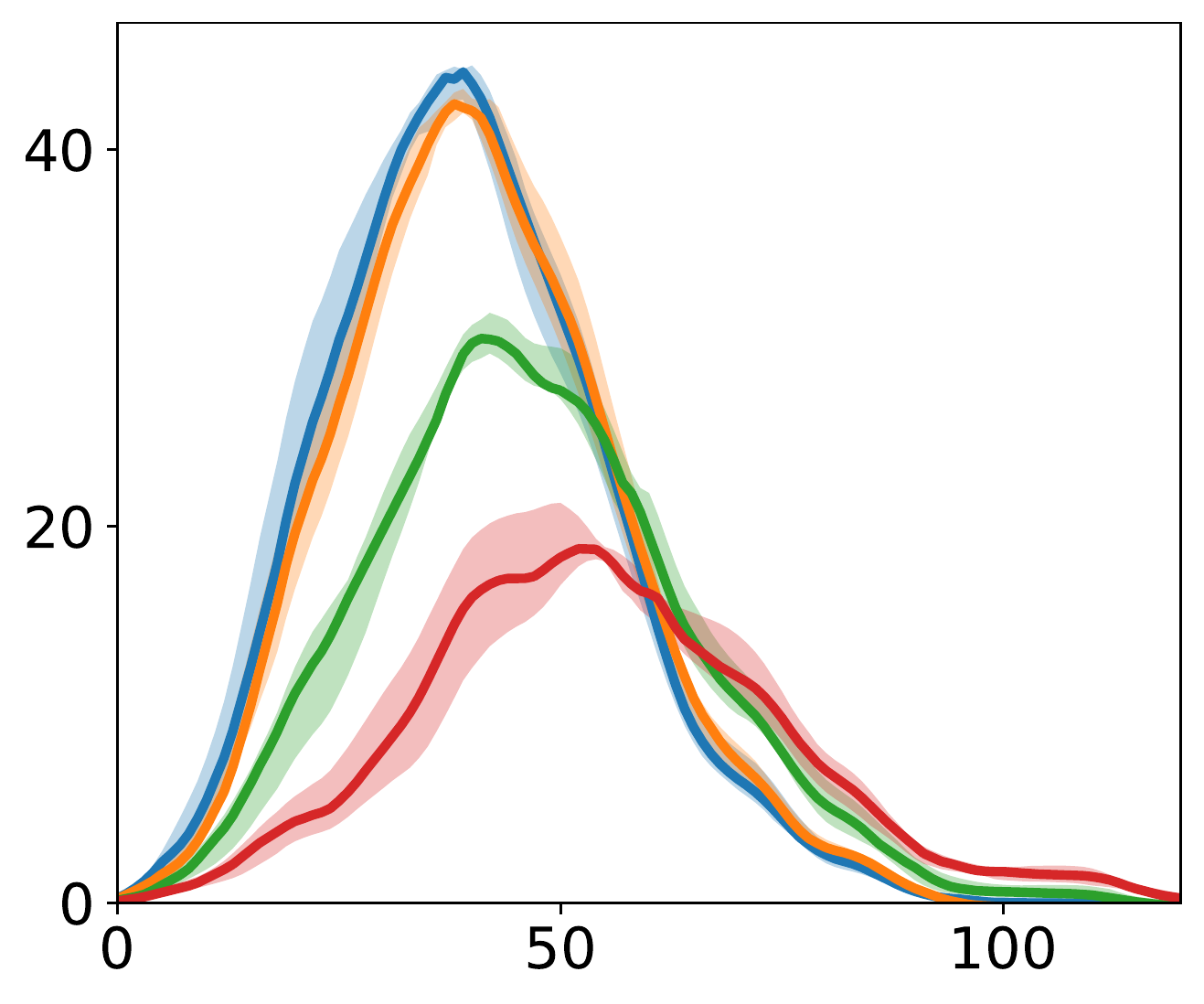}  &
\multirow{14}{*}{\includegraphics[height=4cm]{Fig/labels/newInfected.pdf}}&
\includegraphics[width=\fivefig]{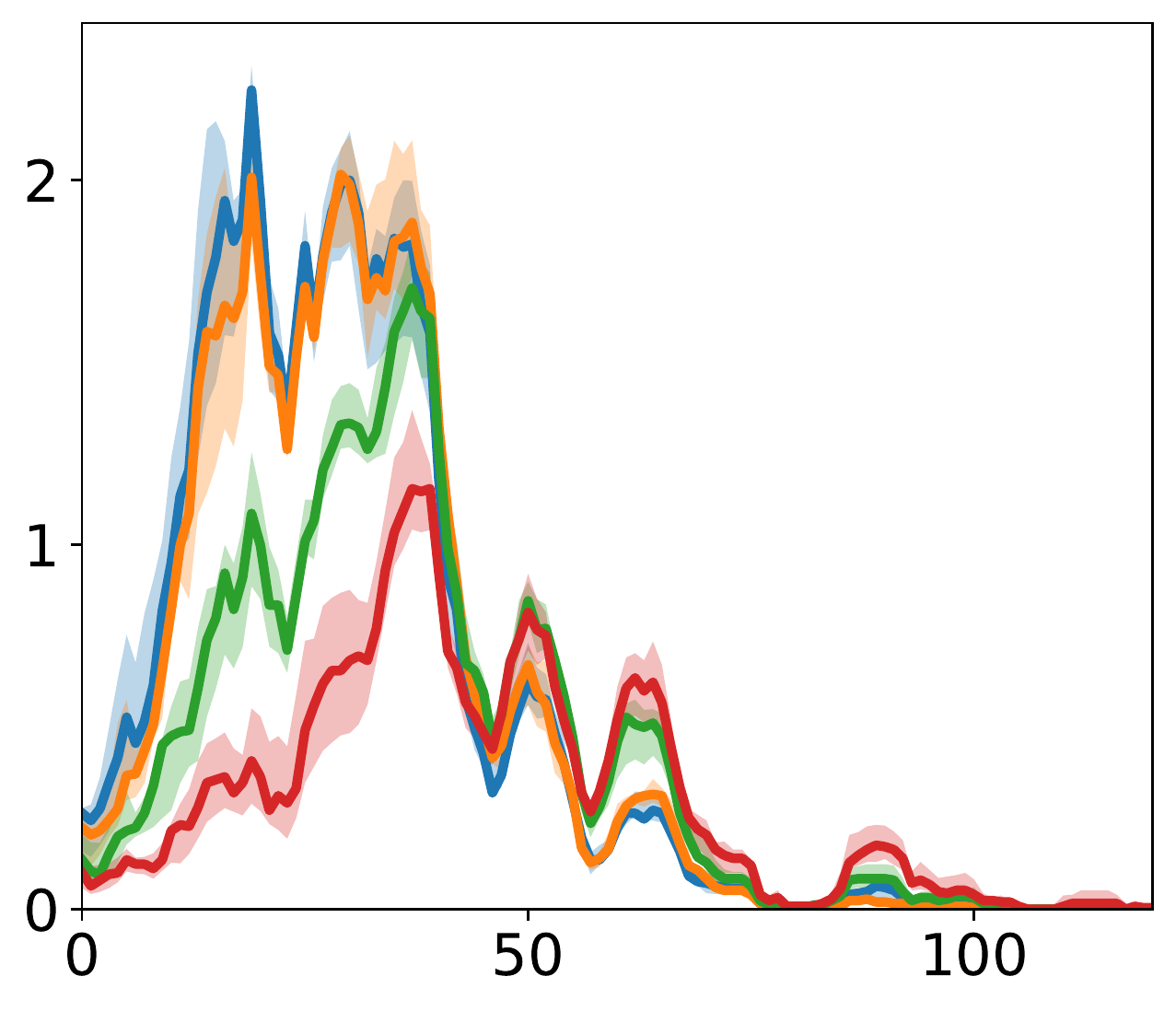}  &
\multirow{14}{*}{\includegraphics[height=2cm]{Fig/labels/growthrate.pdf}}&
\includegraphics[width=\fivefig]{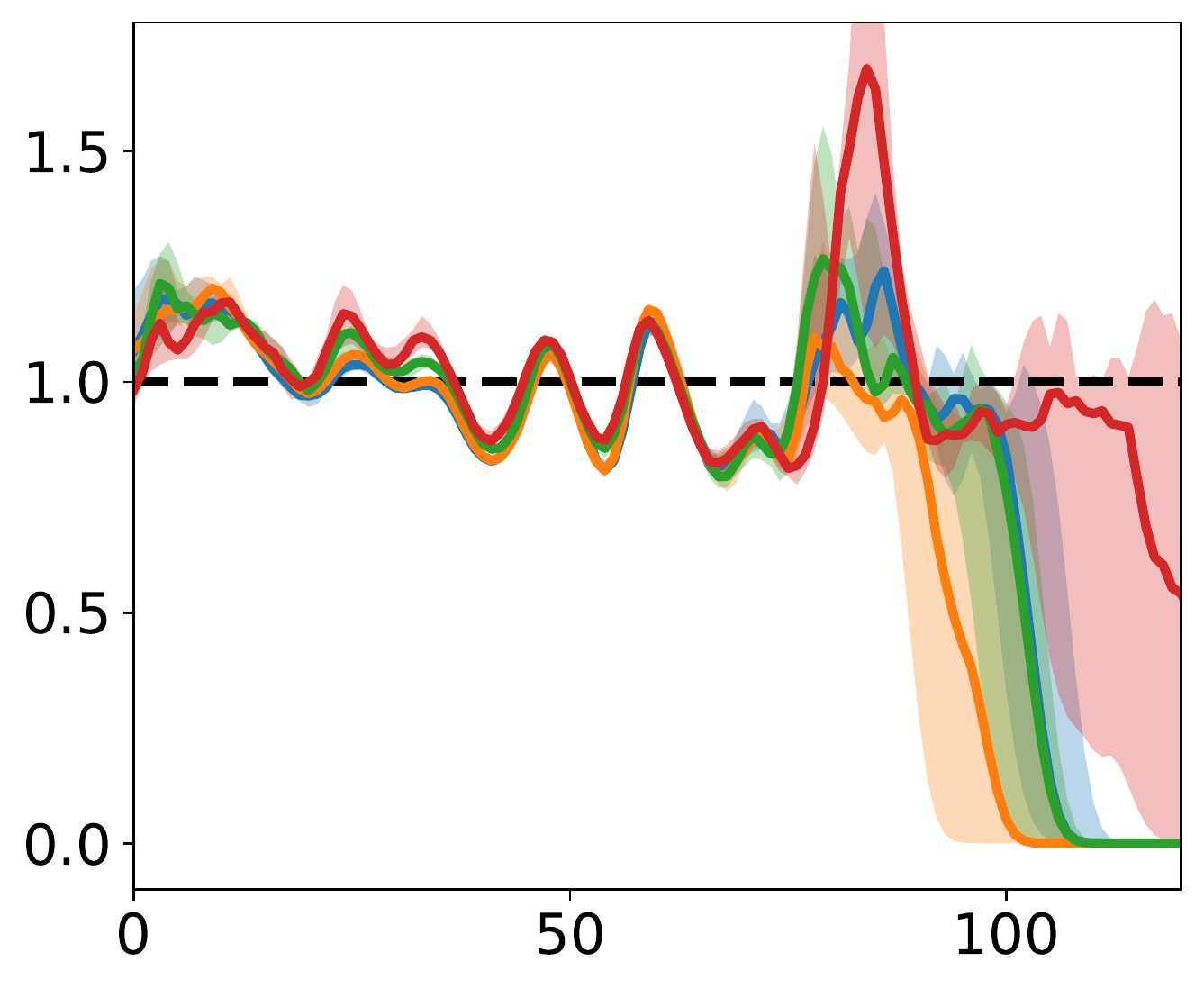} &
\multirow{14}{*}{\includegraphics[height=2cm]{Fig/labels/socialvalue.pdf}}&
\includegraphics[width=\fivefig]{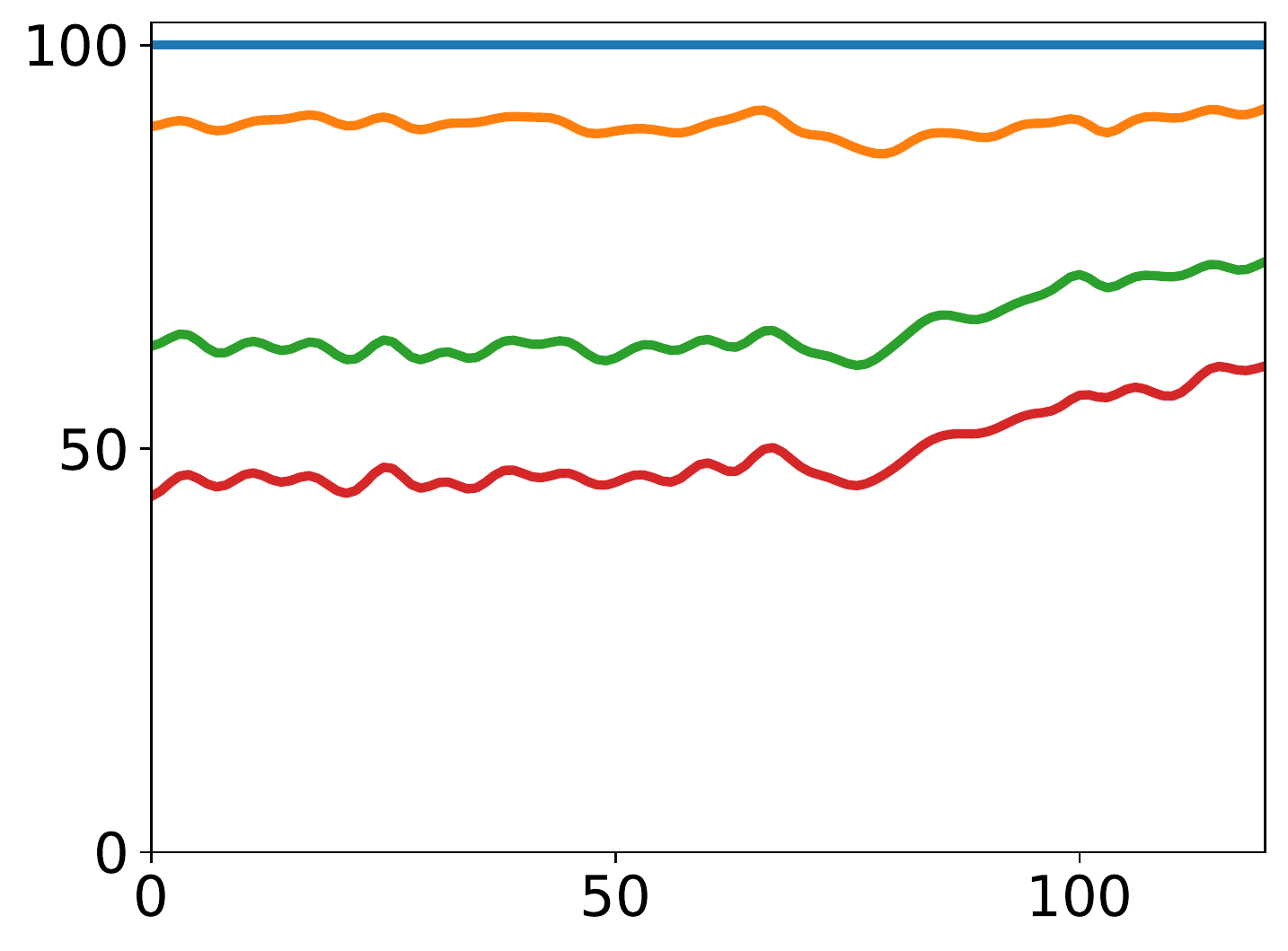} &
\rotatebox{90}{NYC}
\\ [-0.25cm]
&
\includegraphics[width=\fivefig]{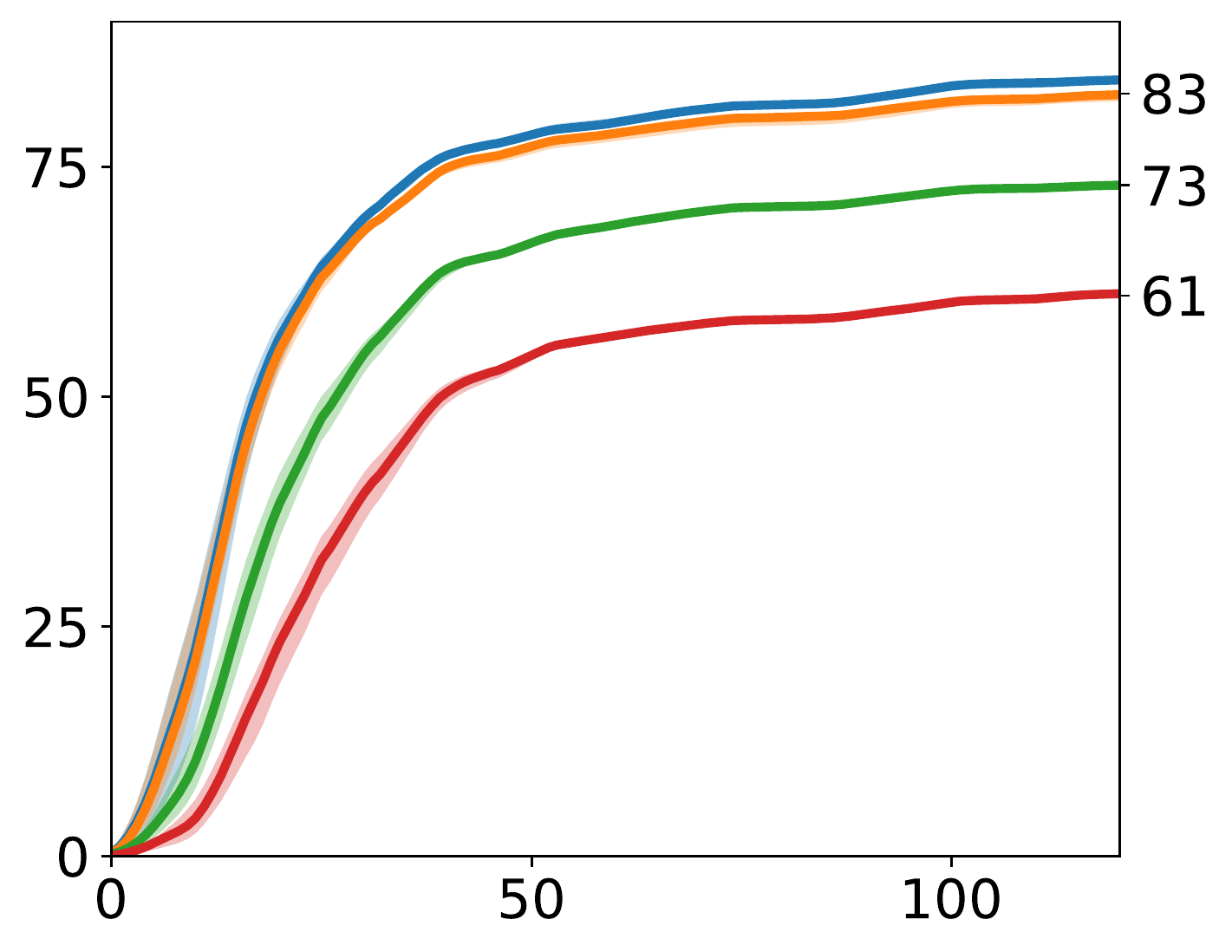} & &
\includegraphics[width=\fivefig]{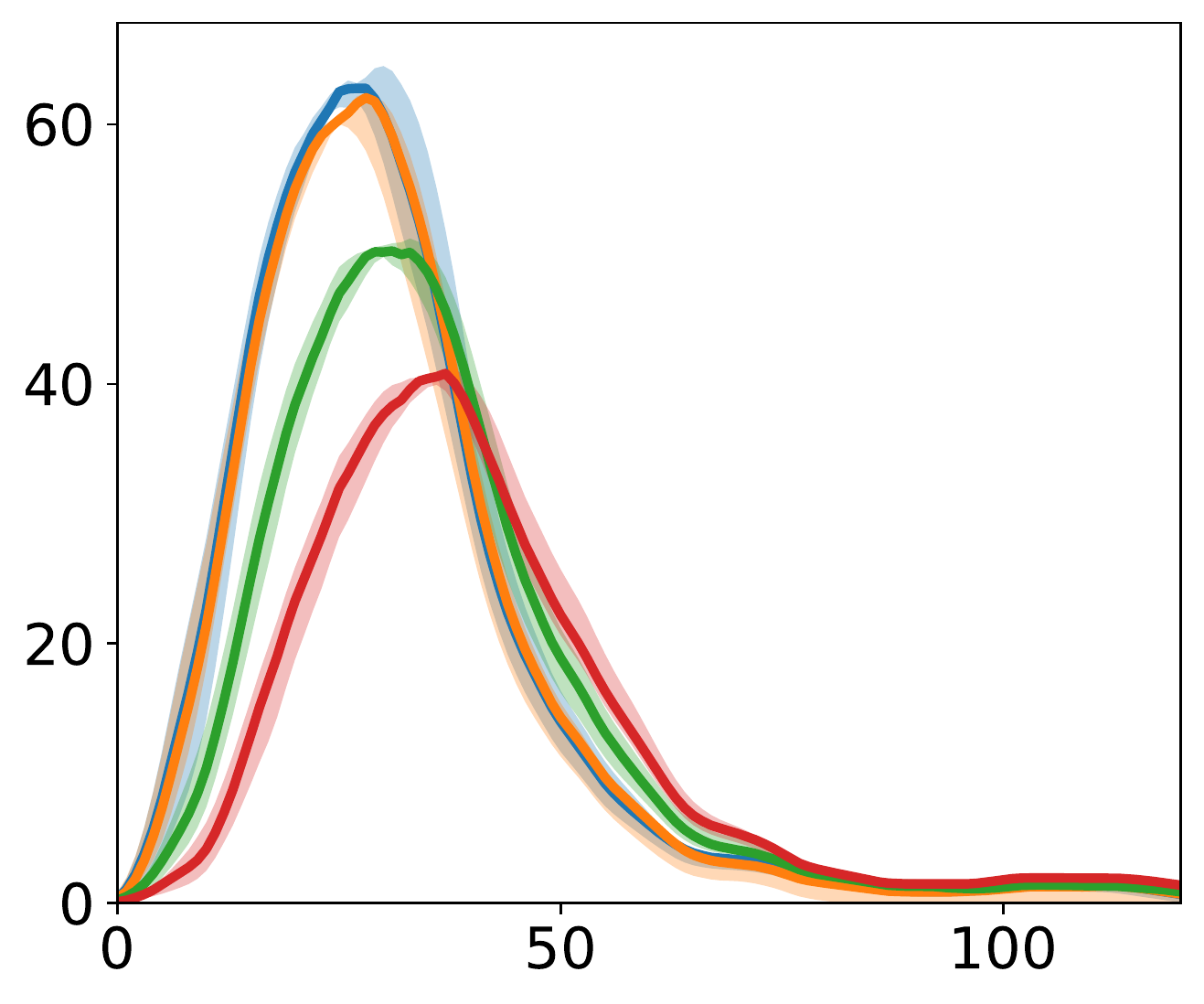}  & &
\includegraphics[width=\fivefig]{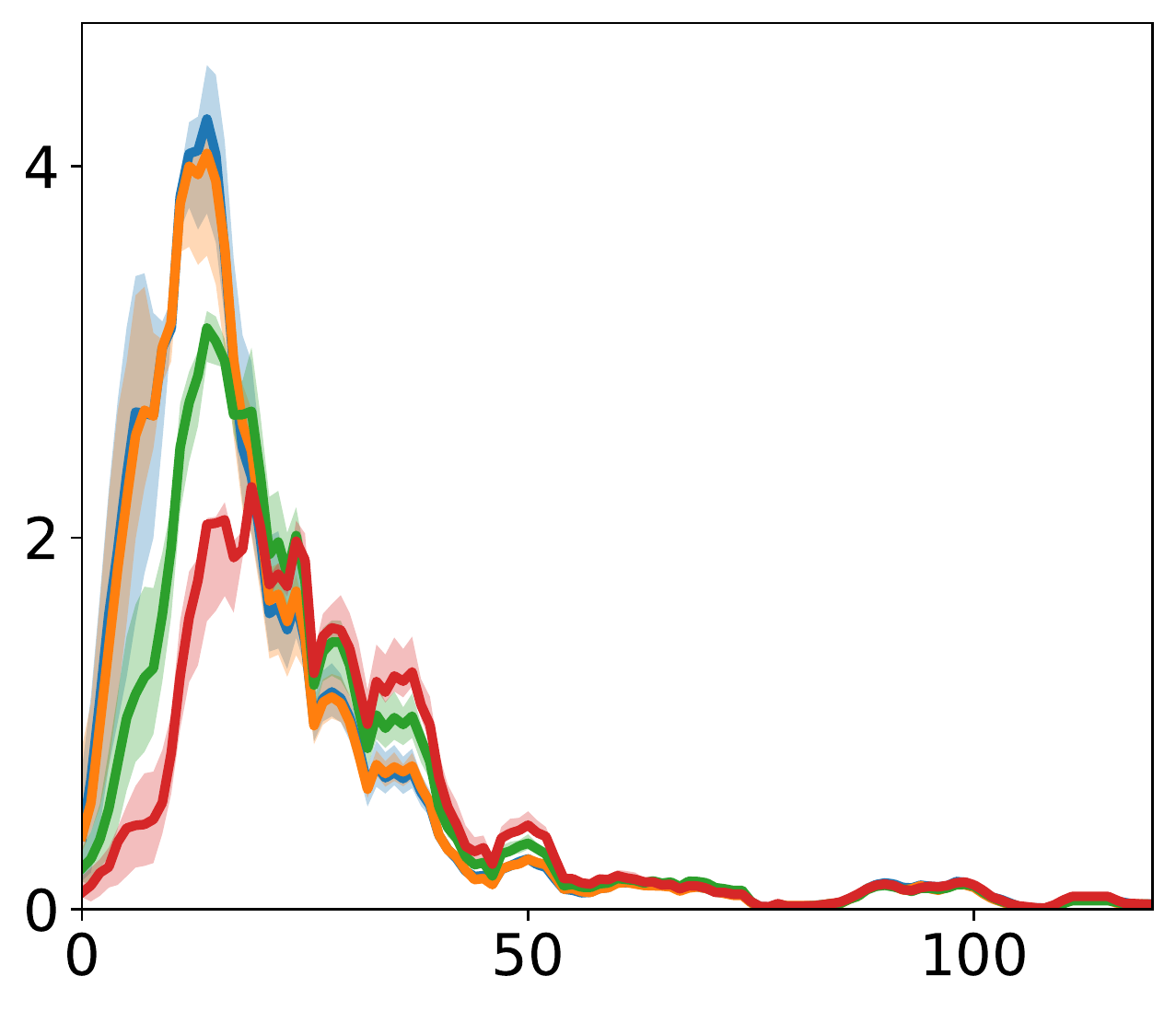}  & &
\includegraphics[width=\fivefig]{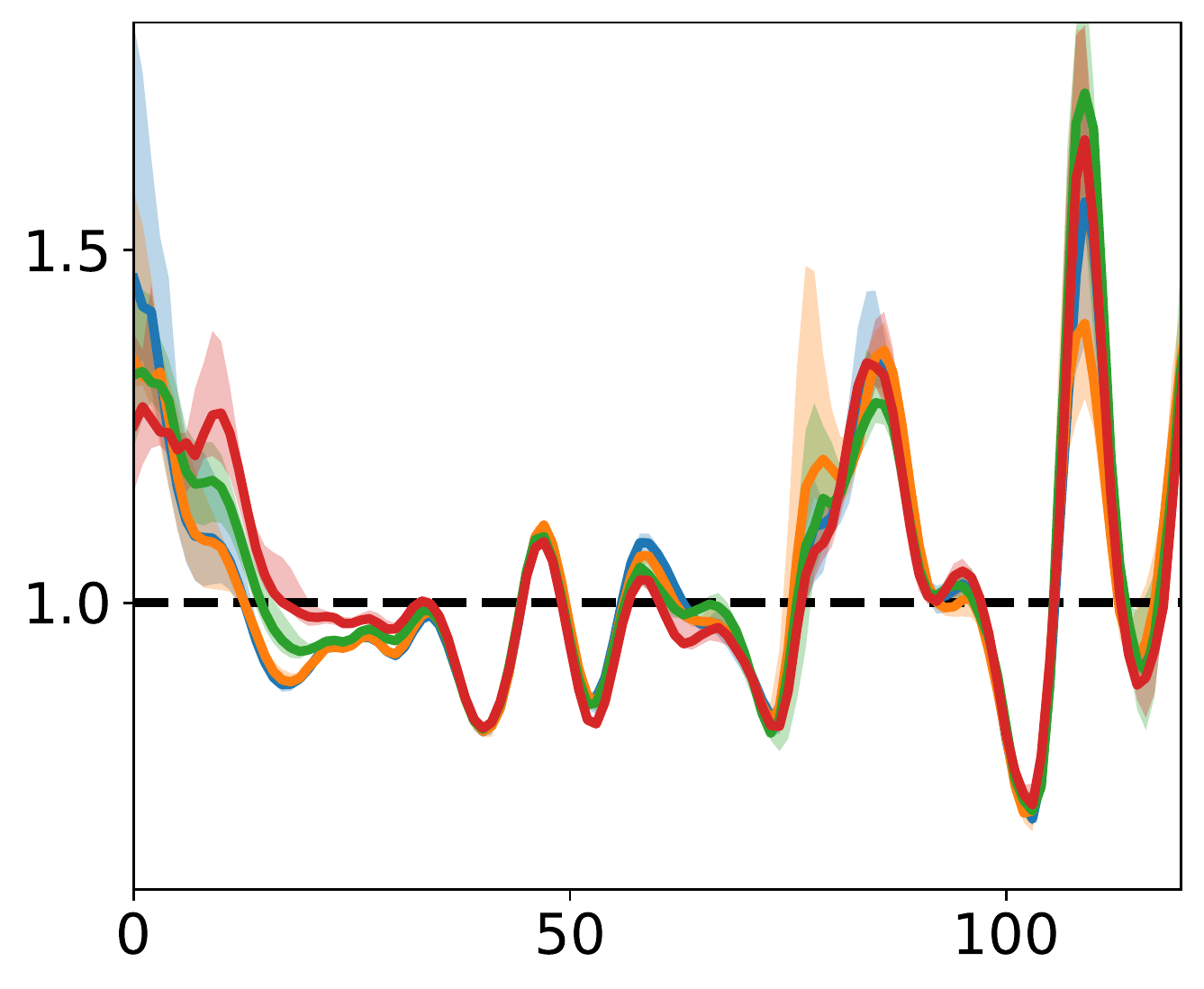} & &
\includegraphics[width=\fivefig]{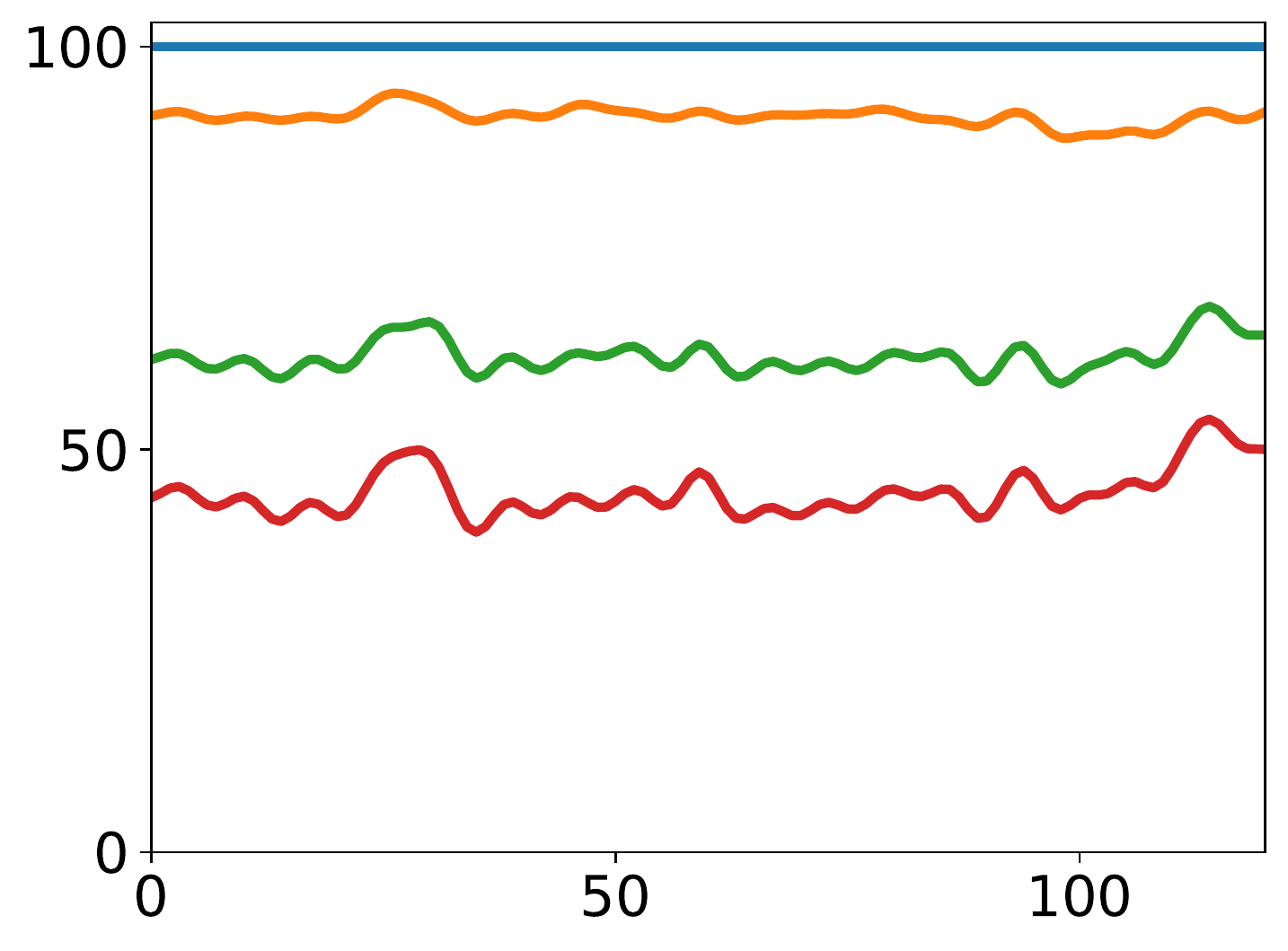} &
\rotatebox{90}{Tokyo}
\\[-0.25cm]
&
\includegraphics[width=\fivefig]{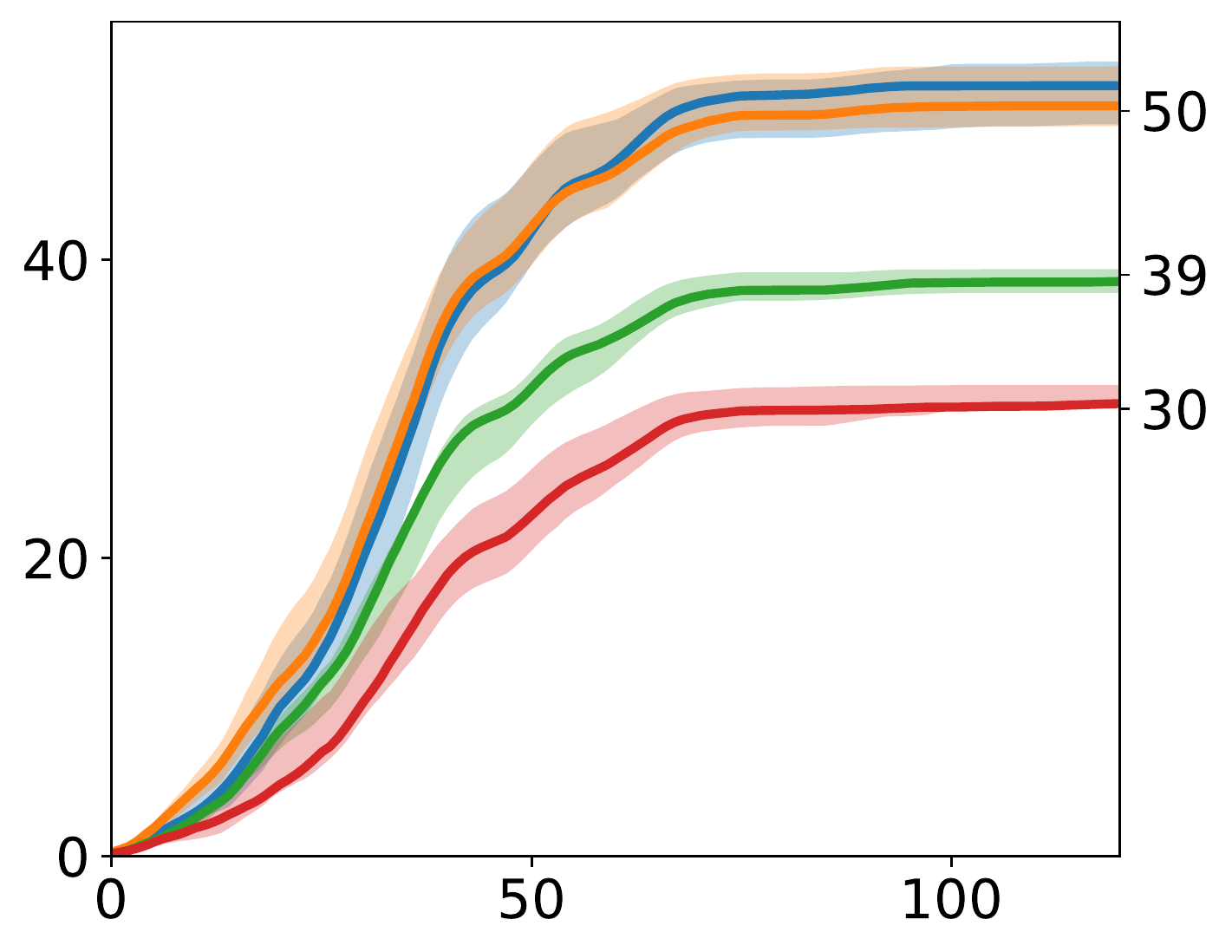} & &
\includegraphics[width=\fivefig]{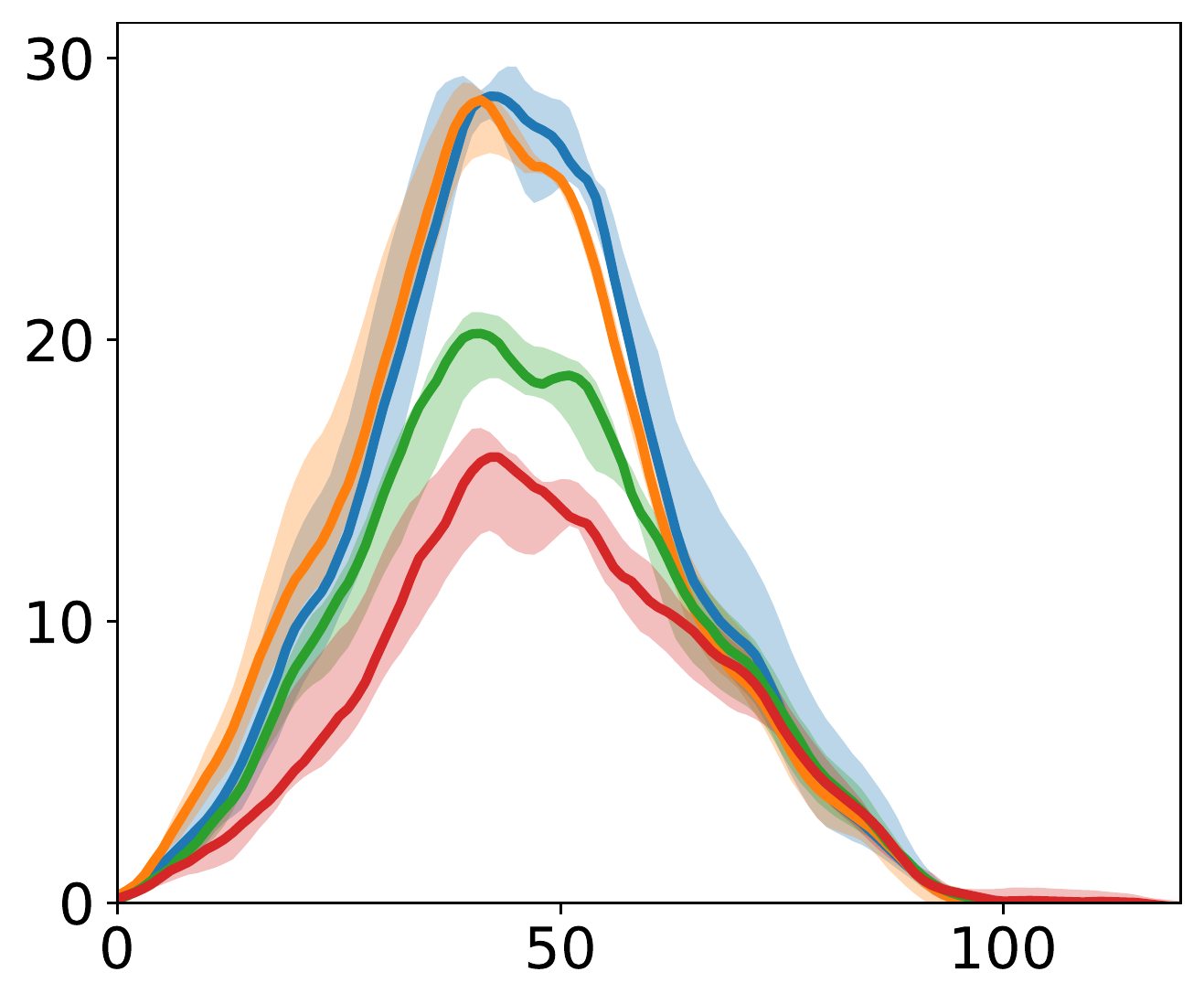}  & &
\includegraphics[width=\fivefig]{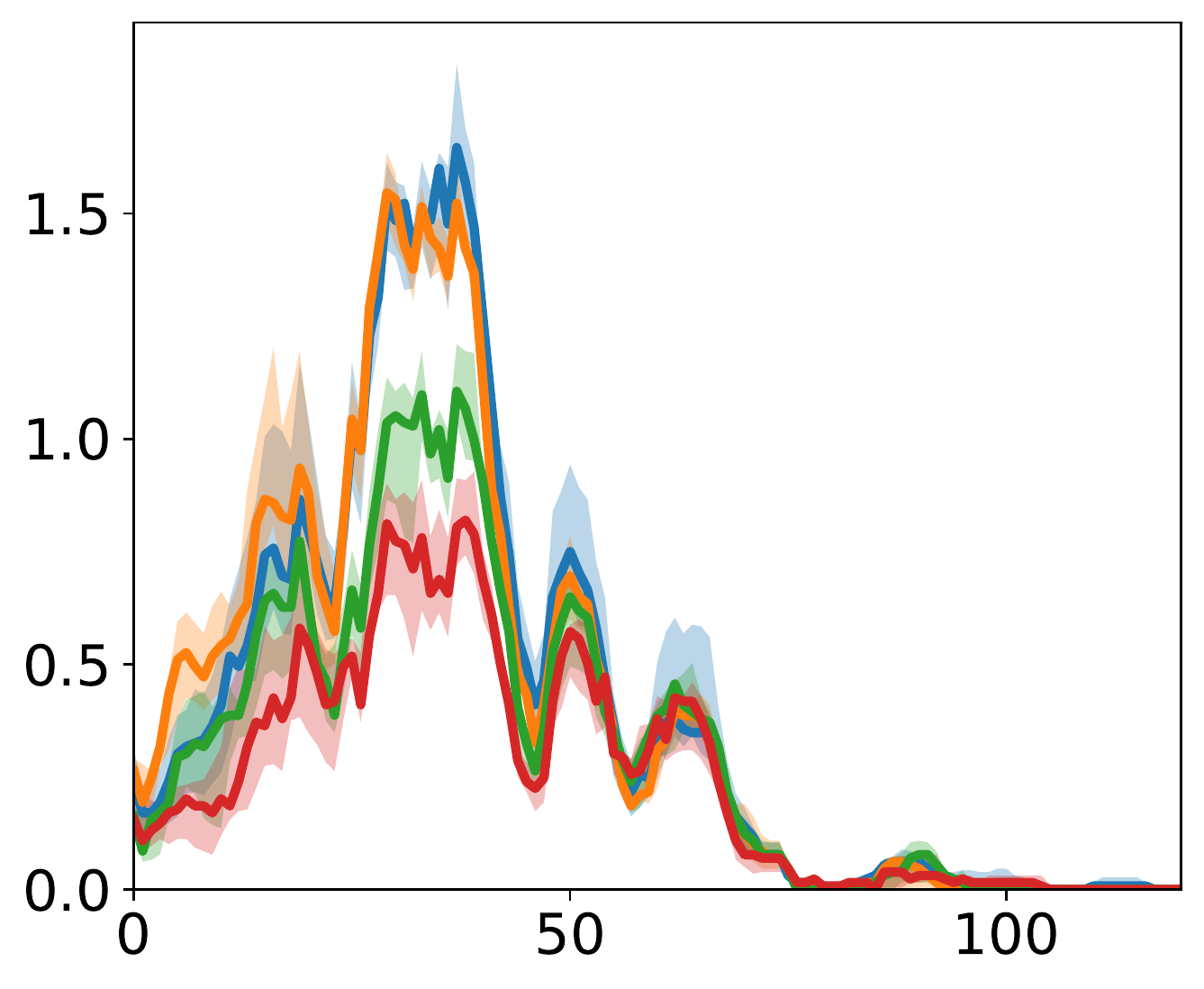}  & &
\includegraphics[width=\fivefig]{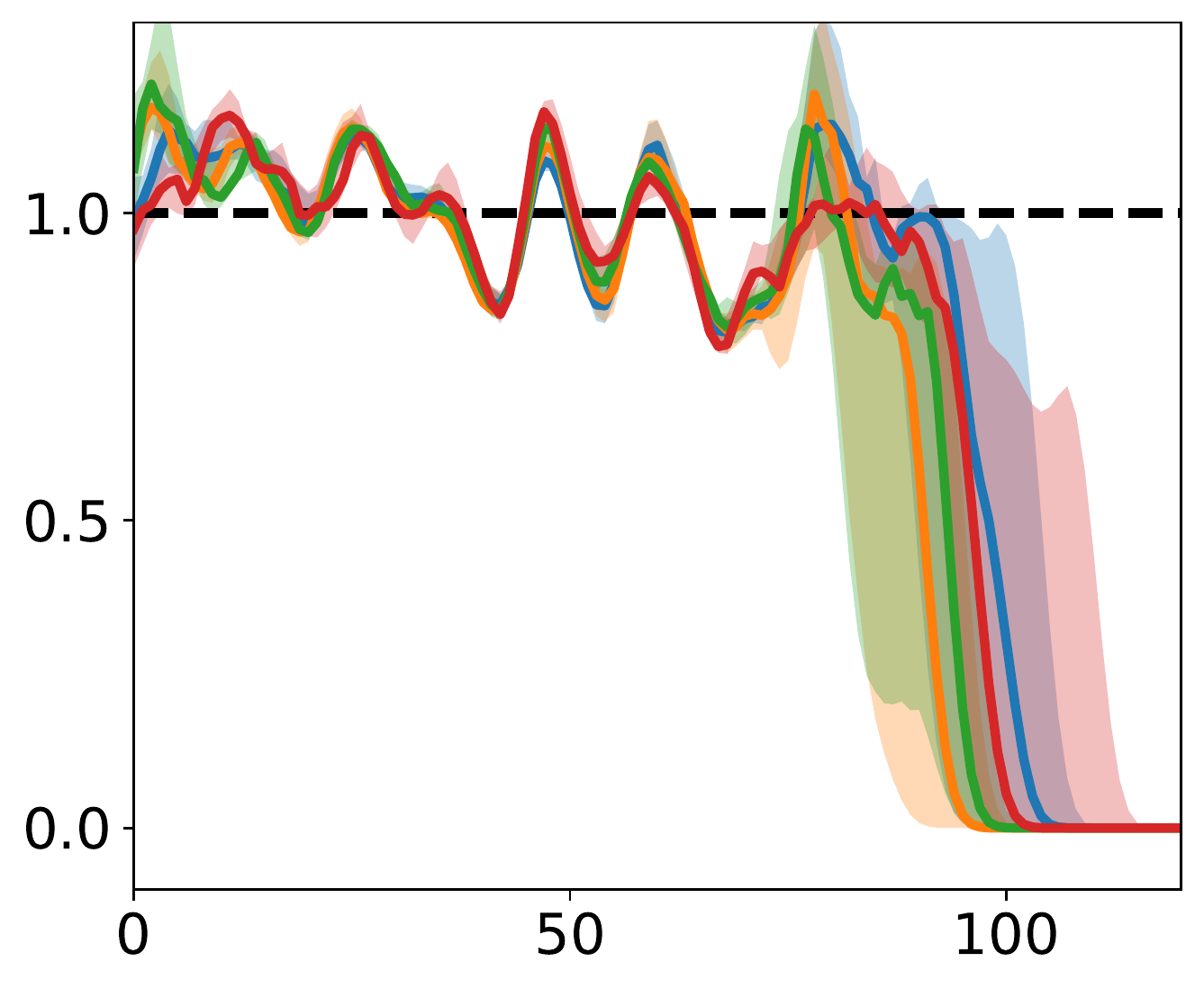}  & &
\includegraphics[width=\fivefig]{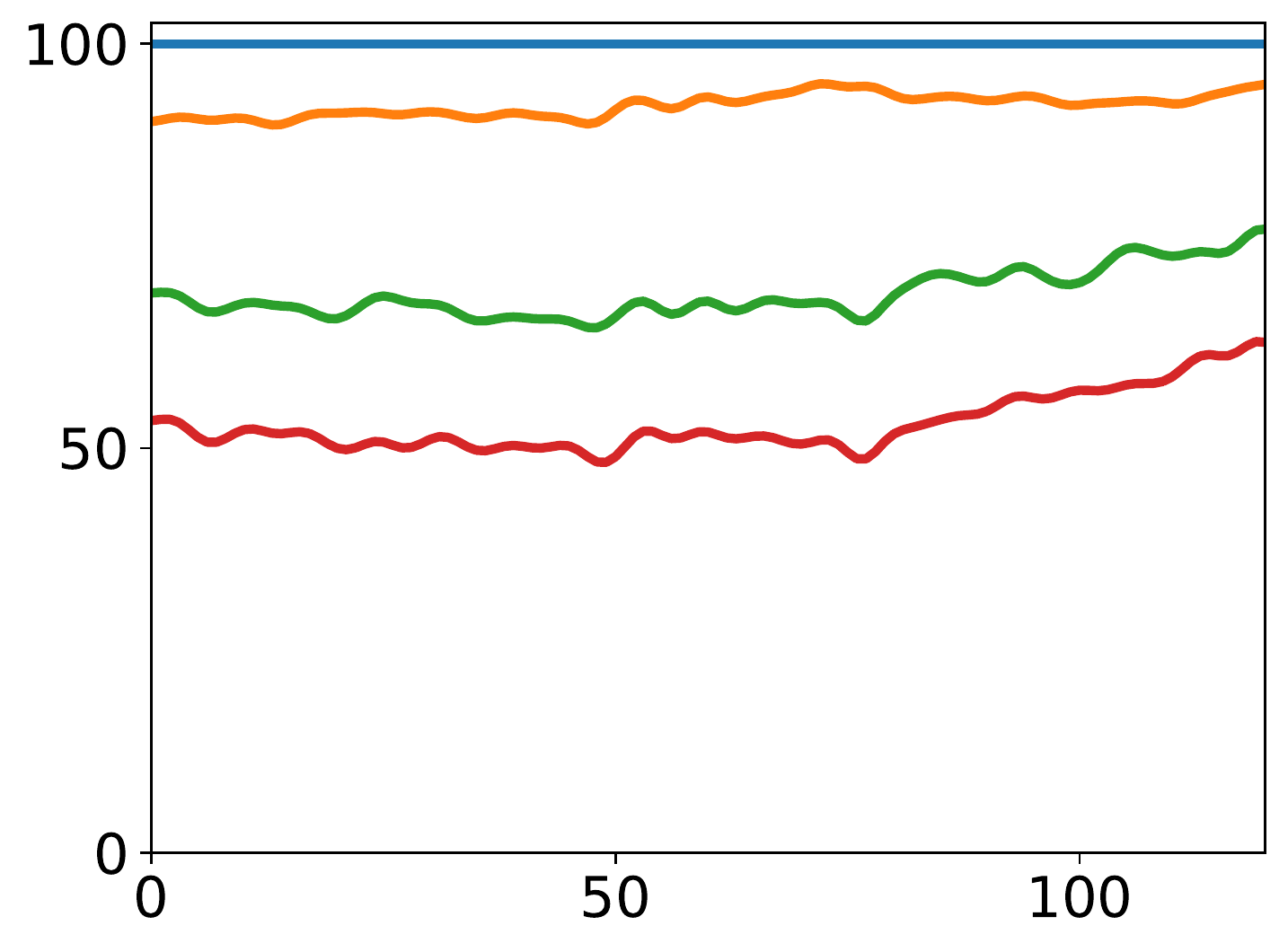} &
\rotatebox{90}{Chicago}
\\ [-0.25cm]

&
\includegraphics[width=\fivefig]{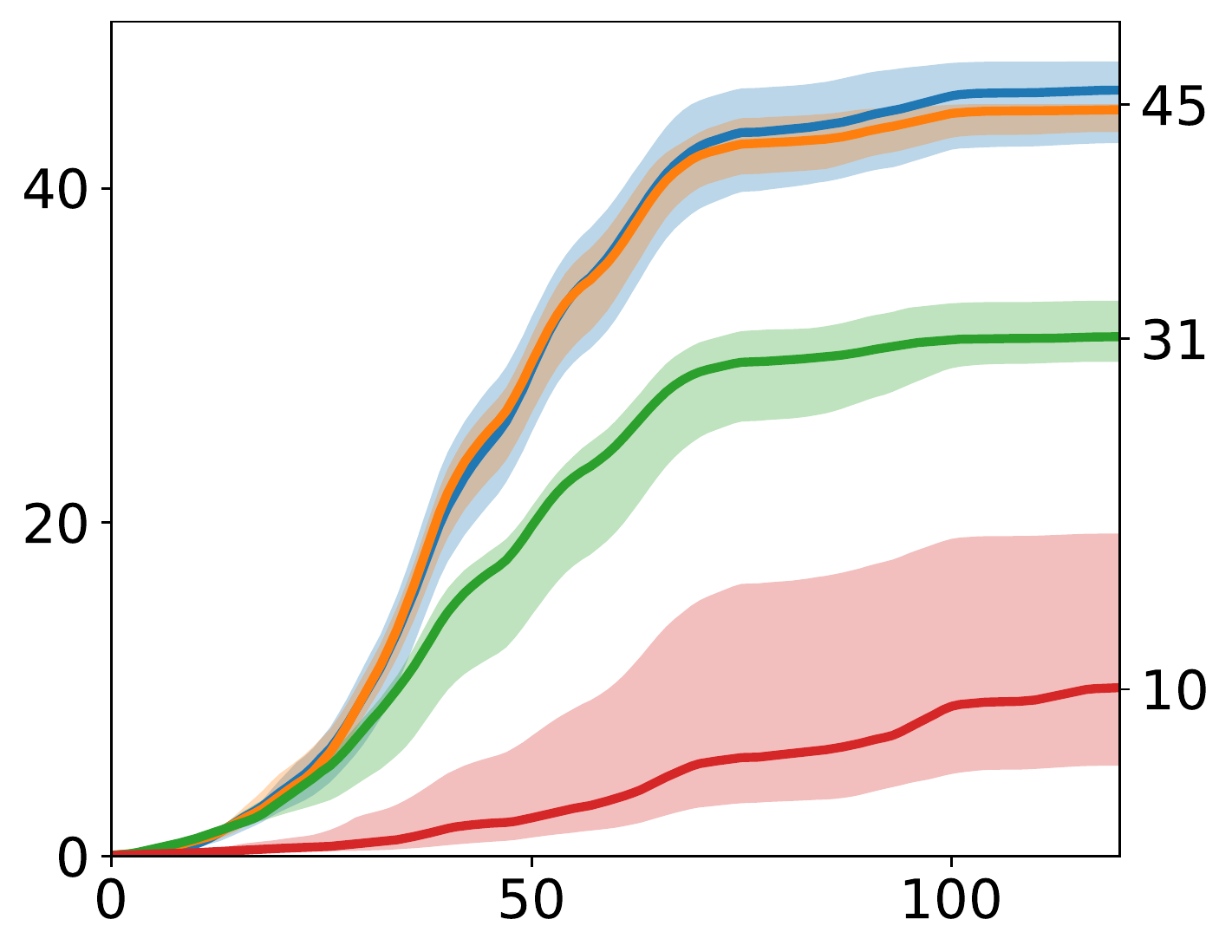} & &
\includegraphics[width=\fivefig]{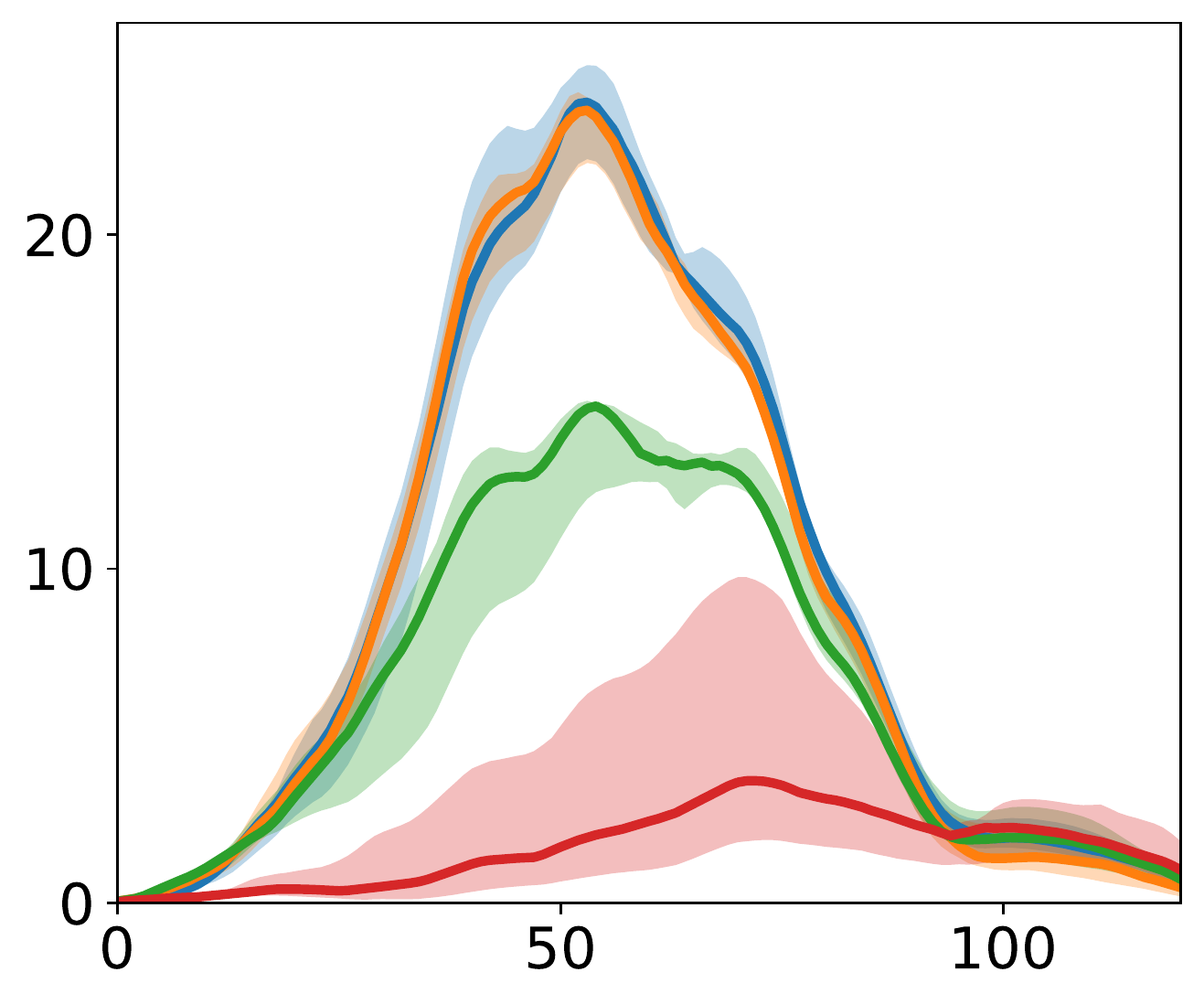} & &
\includegraphics[width=\fivefig]{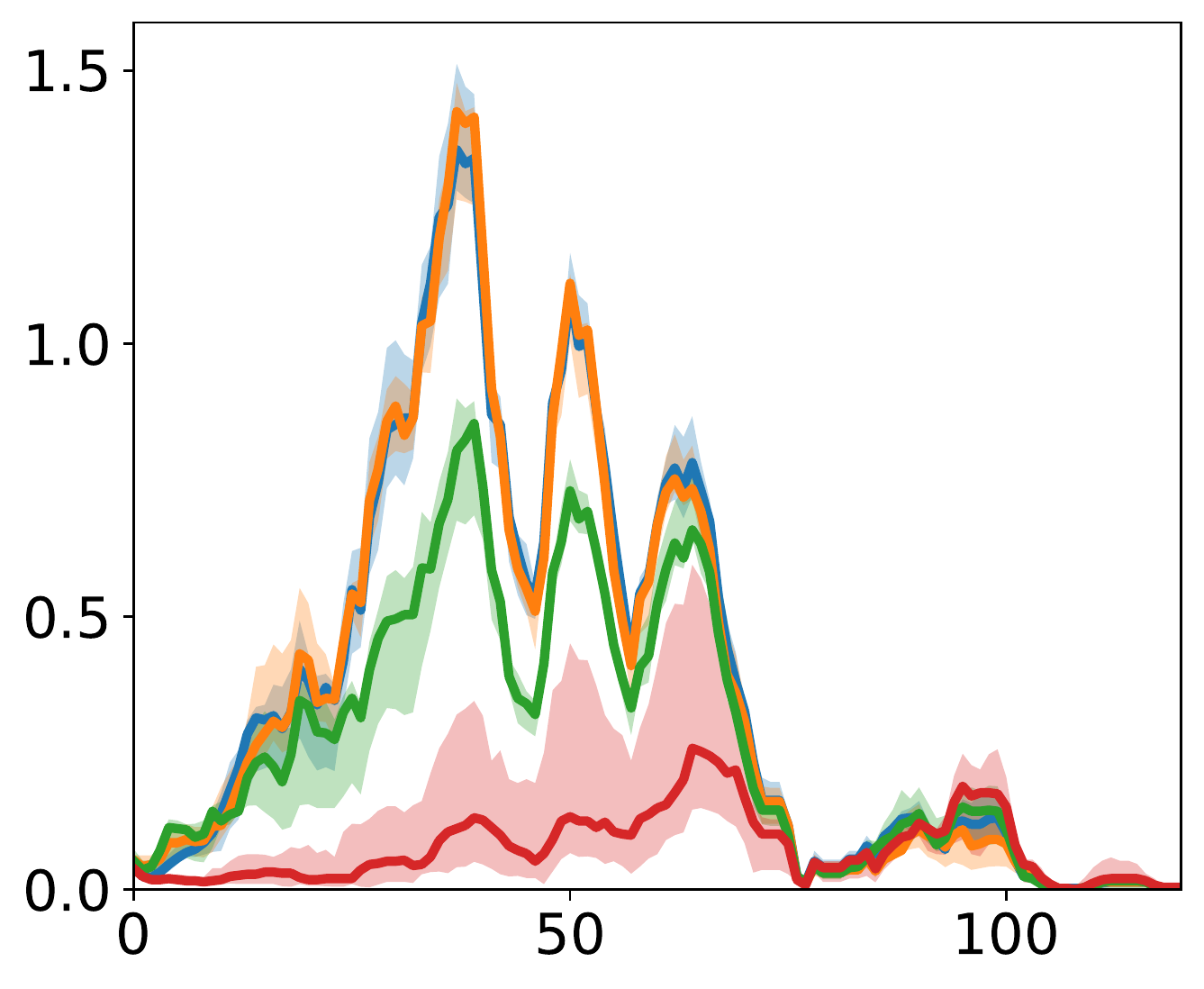}  & &
\includegraphics[width=\fivefig]{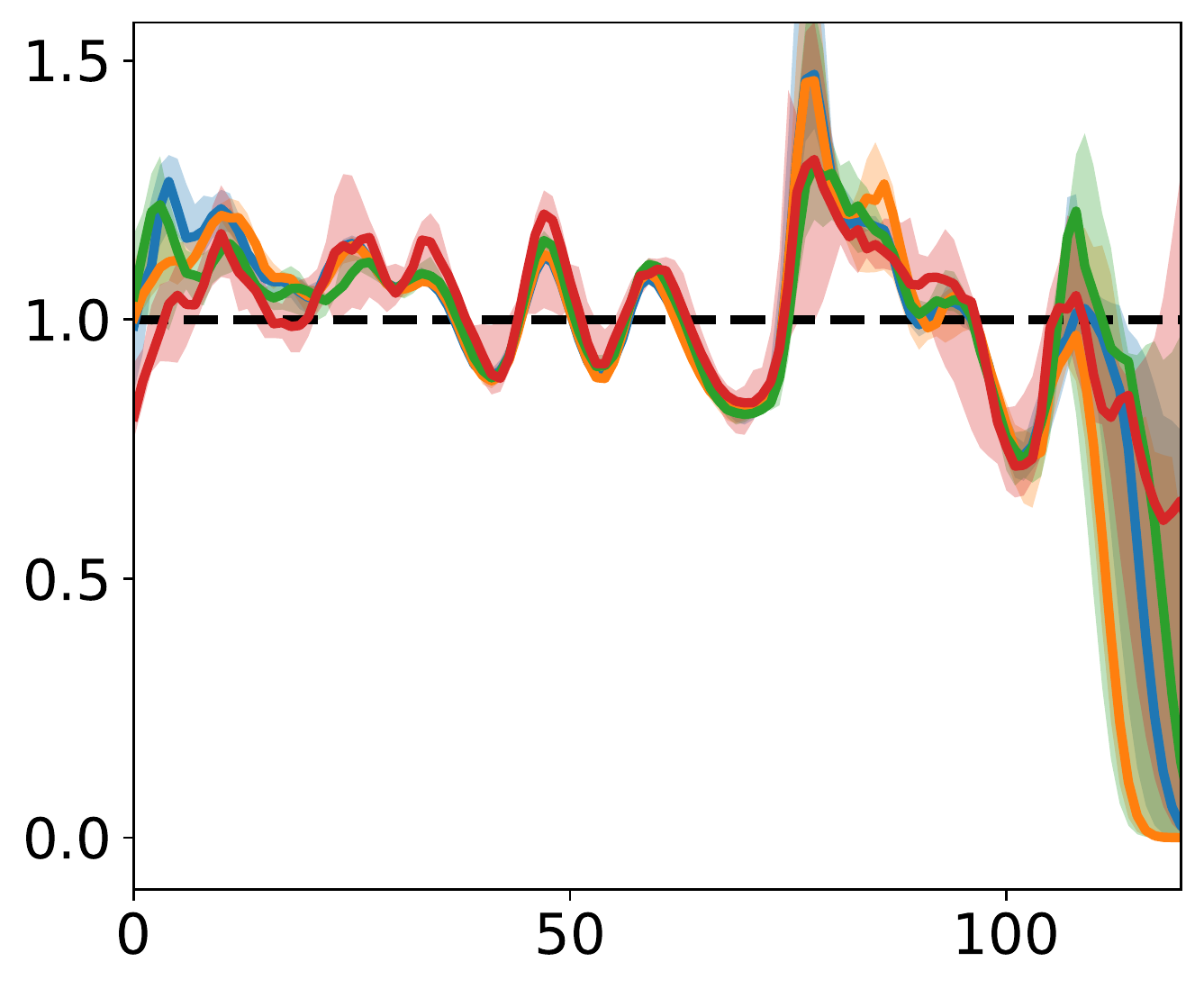} &&
\includegraphics[width=\fivefig]{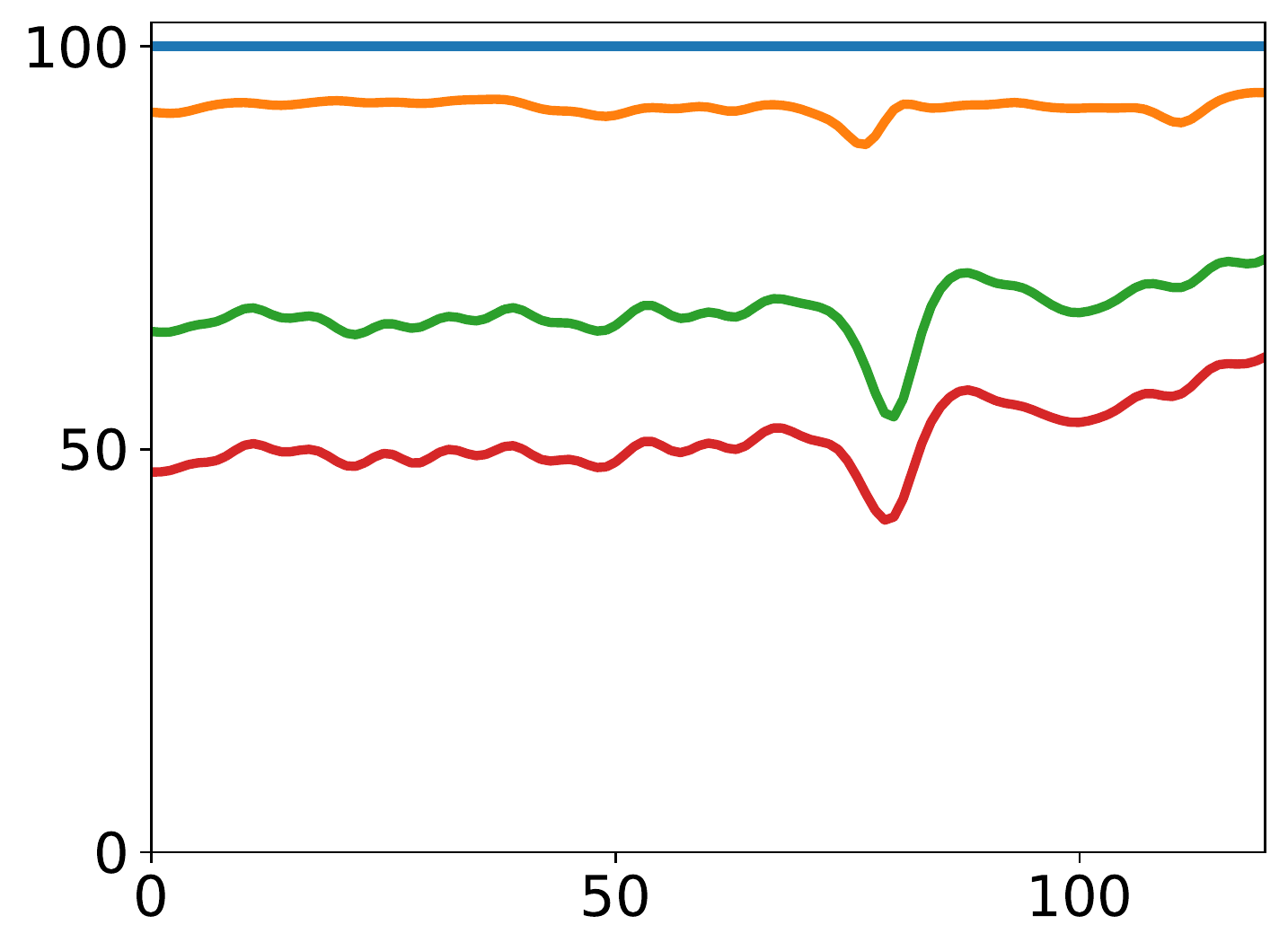} &
\rotatebox{90}{Los Angeles}
\\ [-0.25cm]

&
\includegraphics[width=\fivefig]{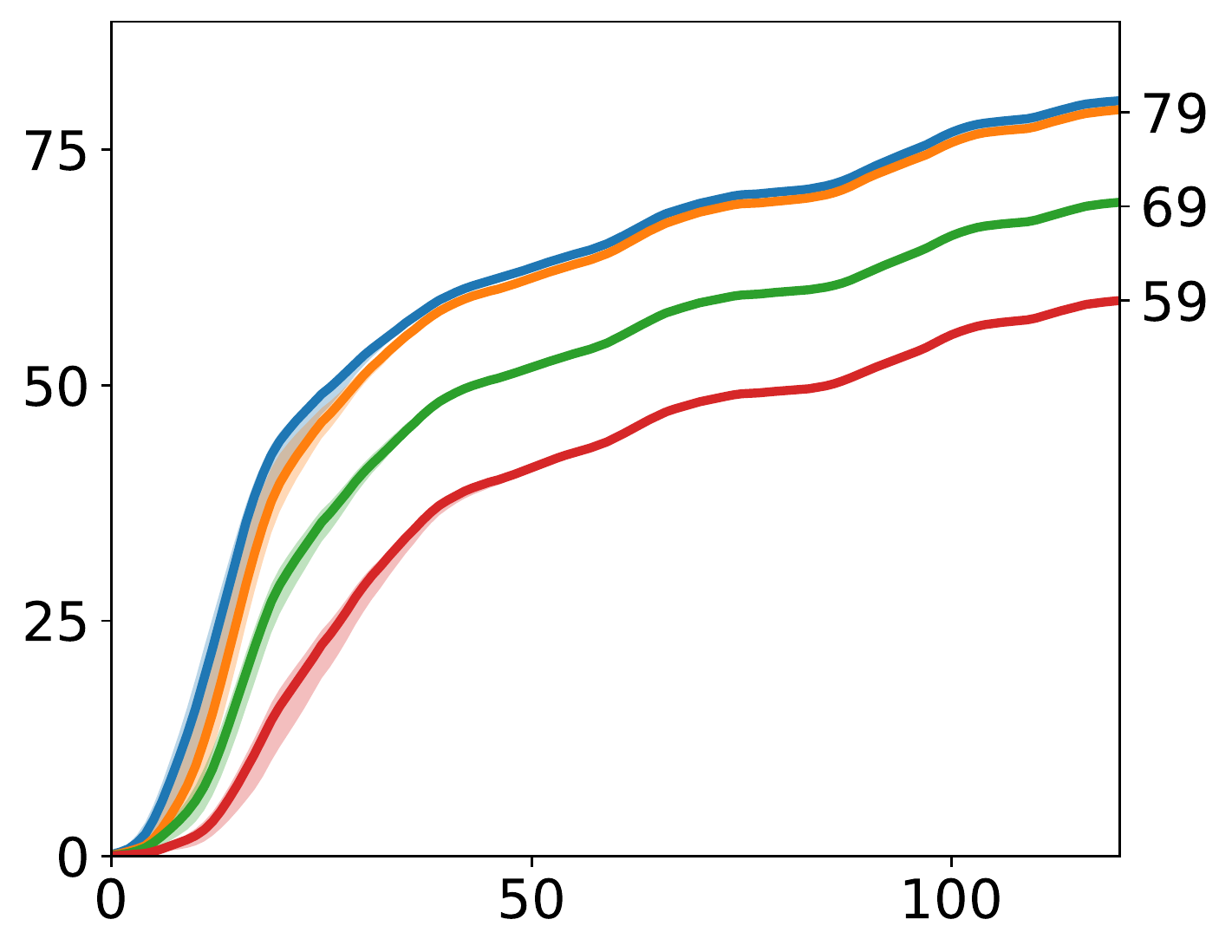} & &
\includegraphics[width=\fivefig]{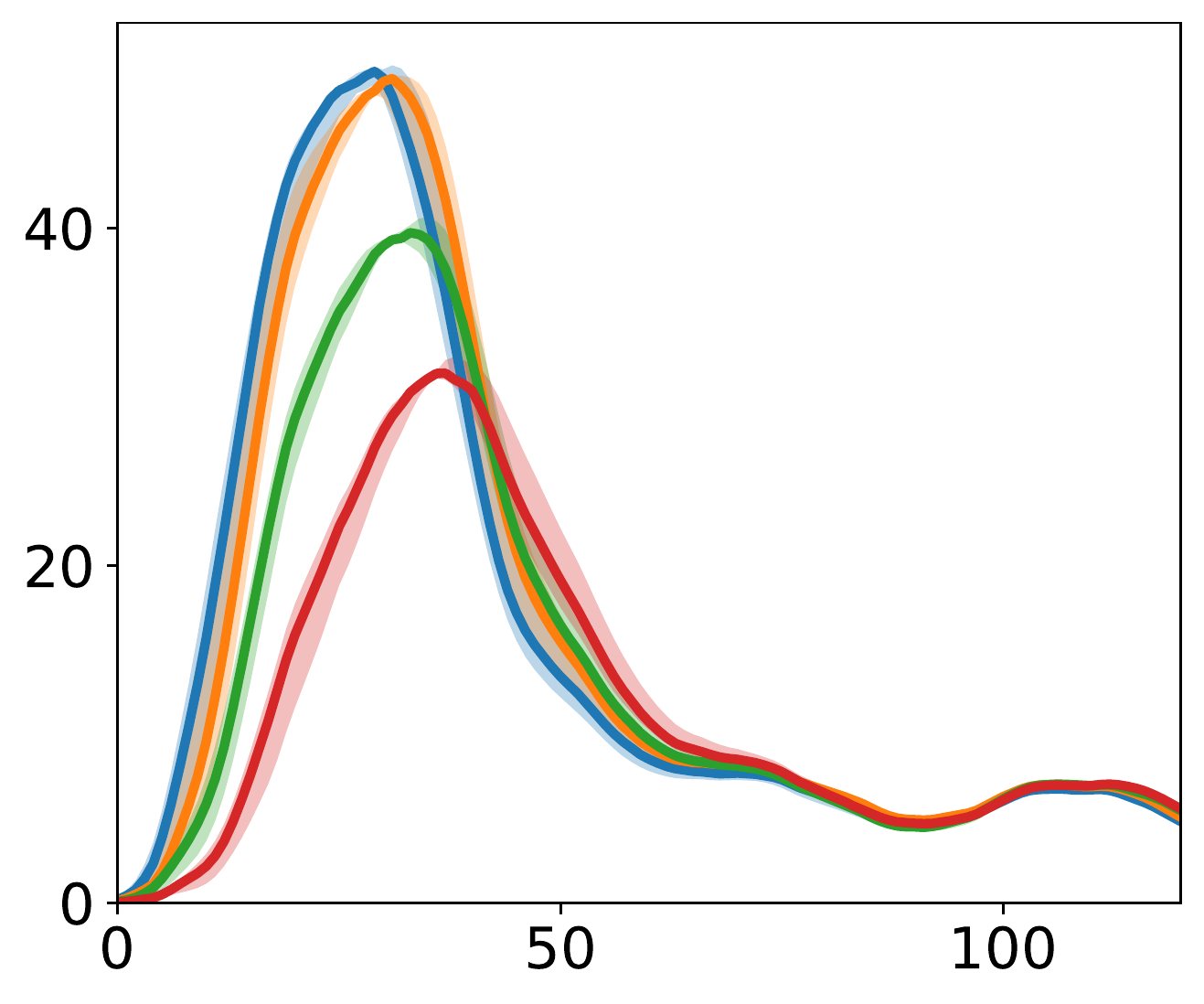}  &&
\includegraphics[width=\fivefig]{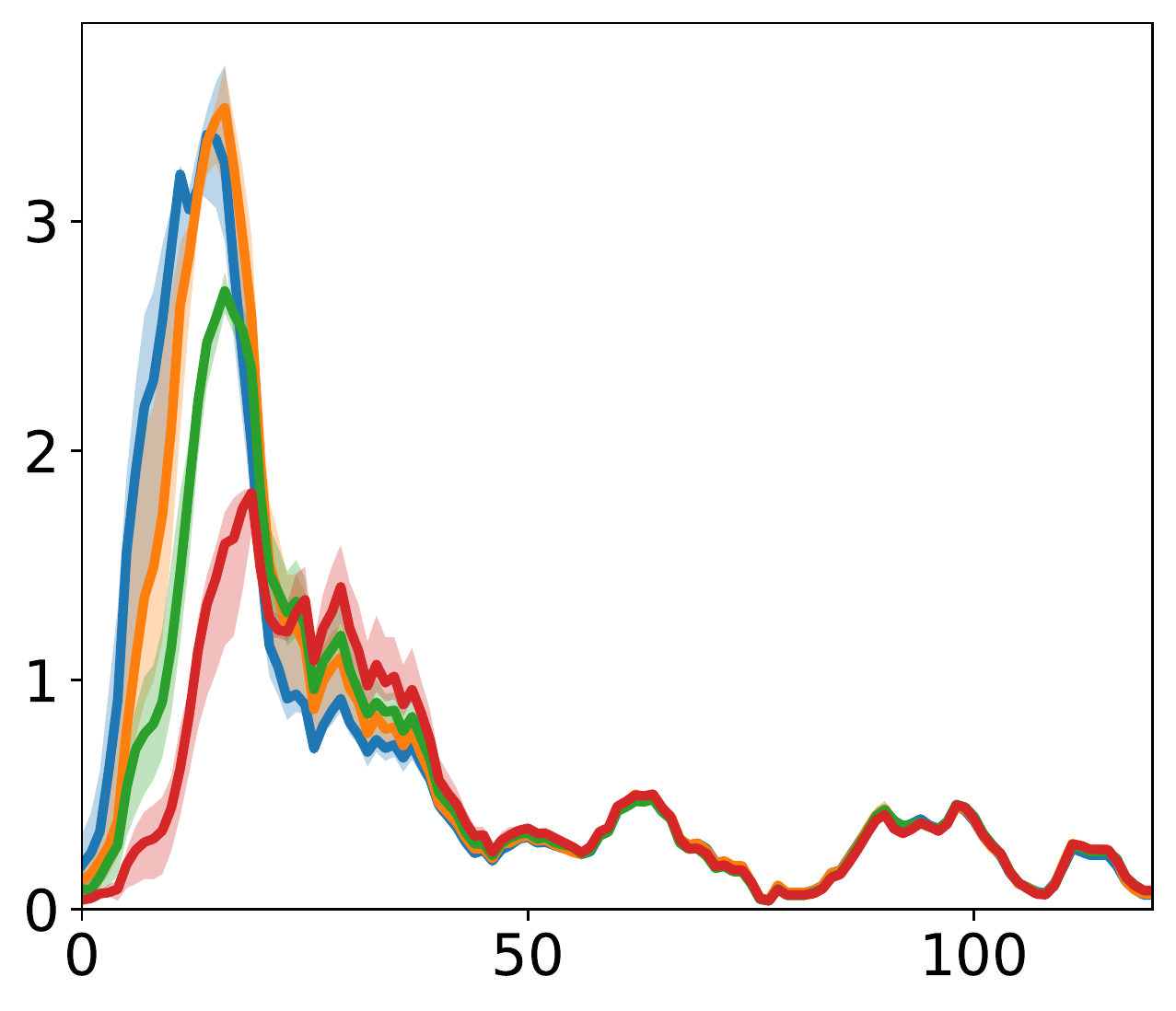}  &&
\includegraphics[width=\fivefig]{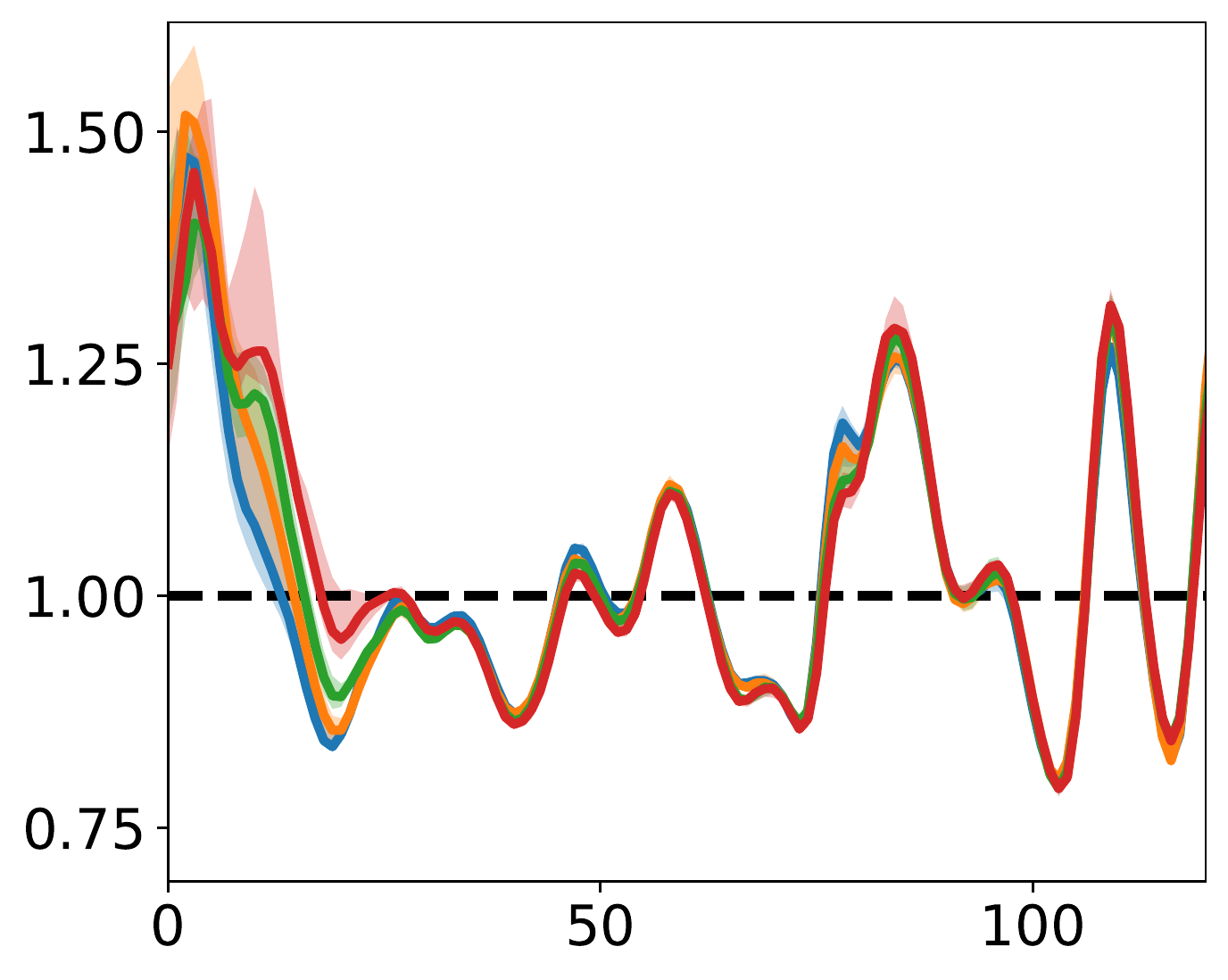} &&
\includegraphics[width=\fivefig]{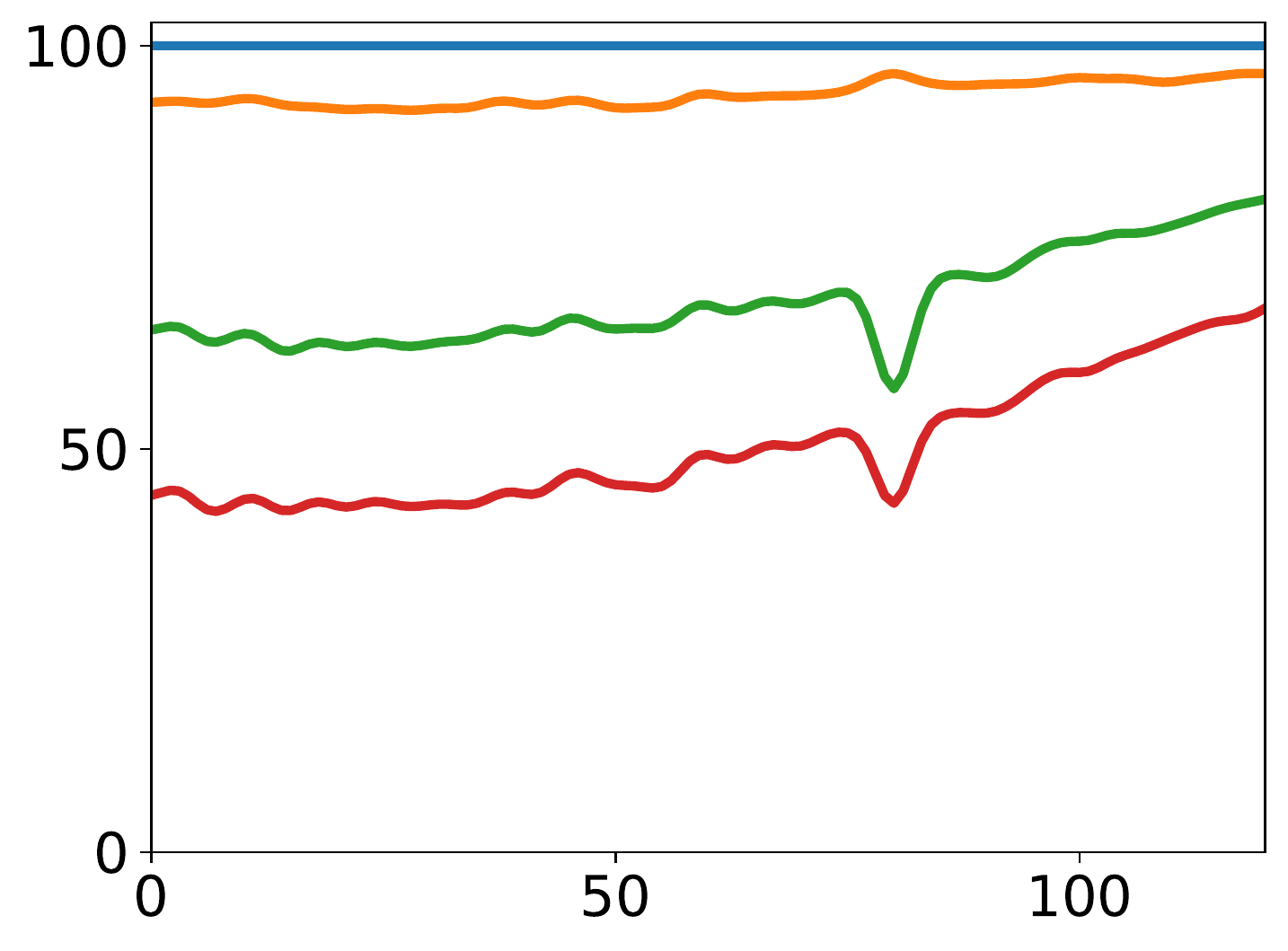} &
\rotatebox{90}{Istanbul}
\\ [-0.25cm]

&
\includegraphics[width=\fivefig]{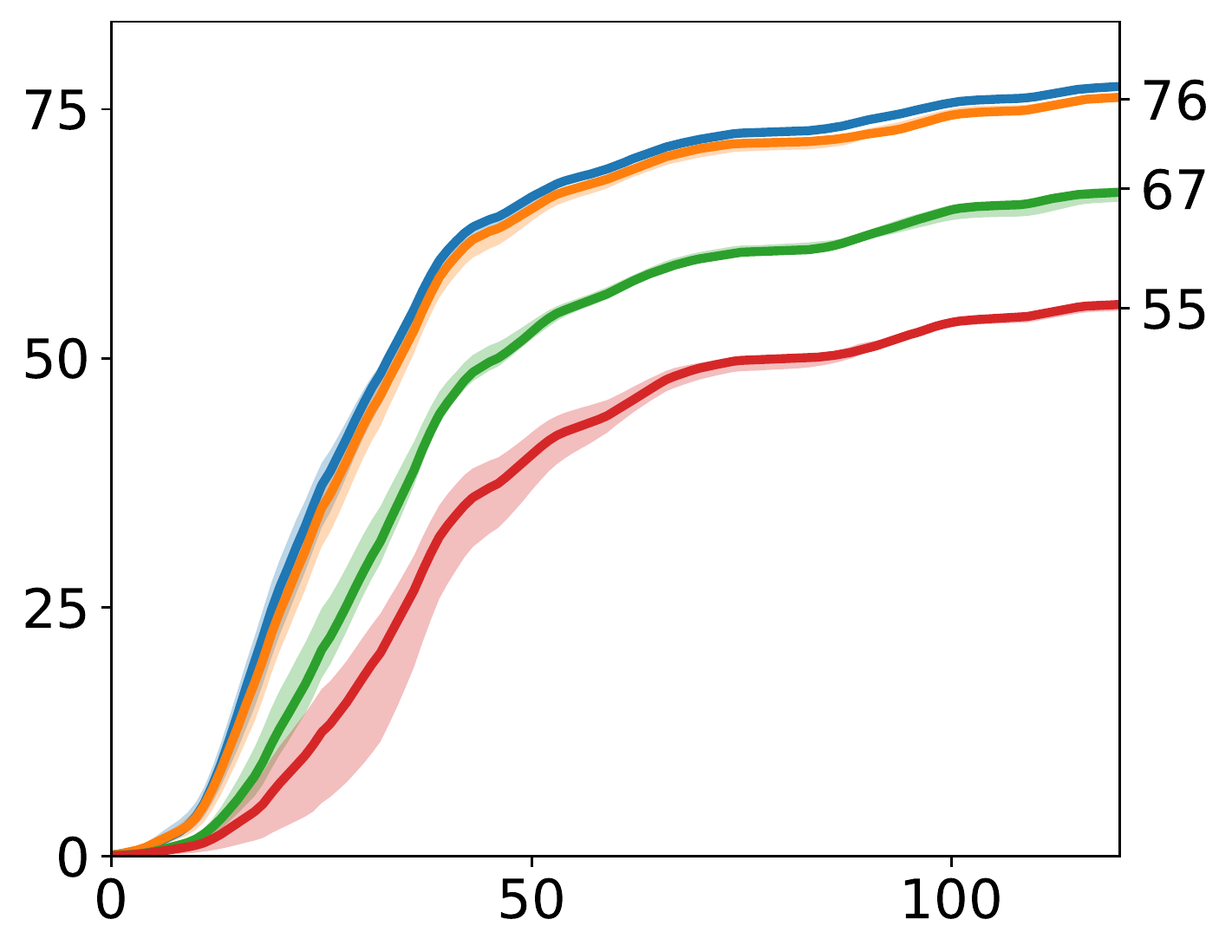} &&
\includegraphics[width=\fivefig]{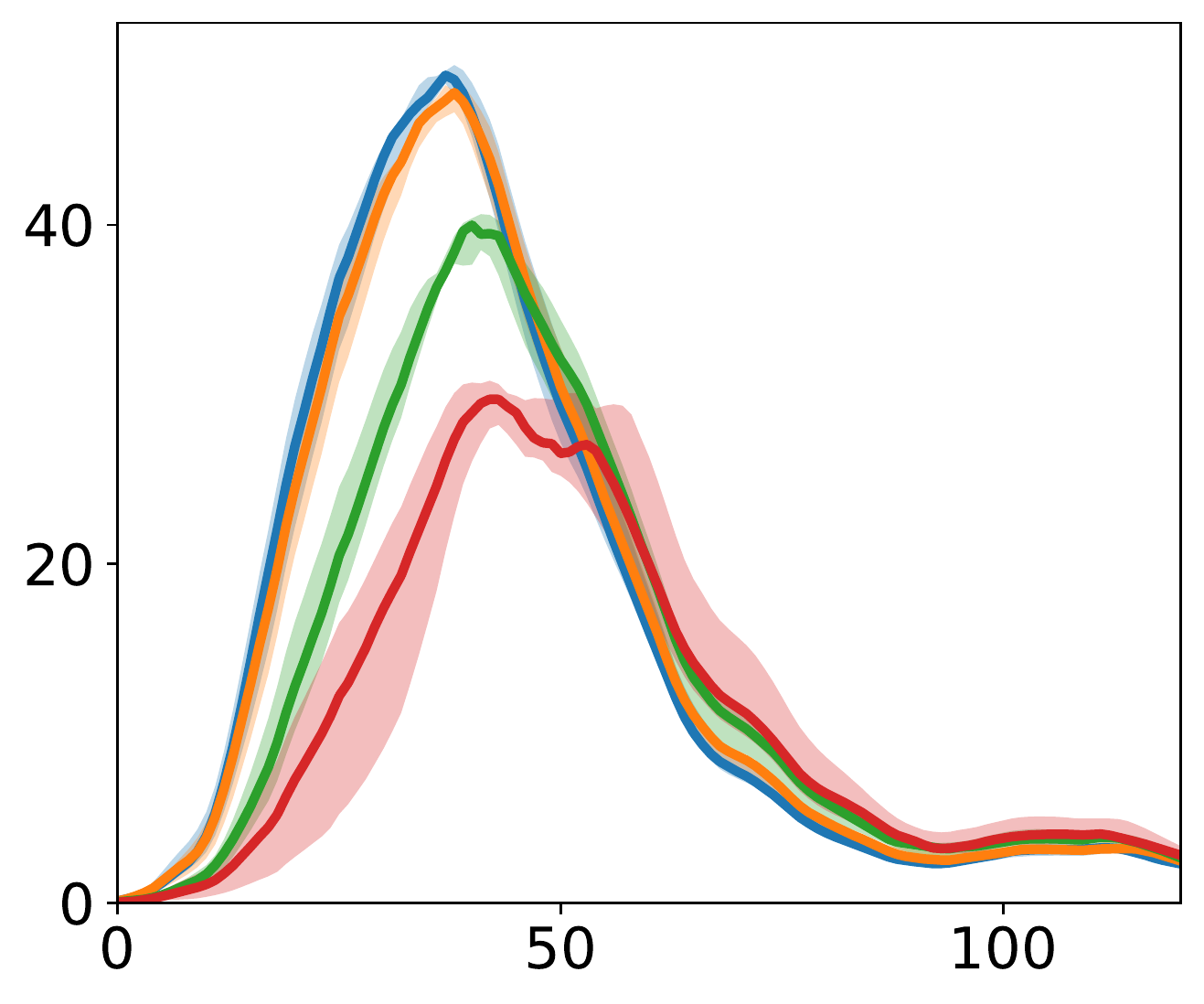}  &&
\includegraphics[width=\fivefig]{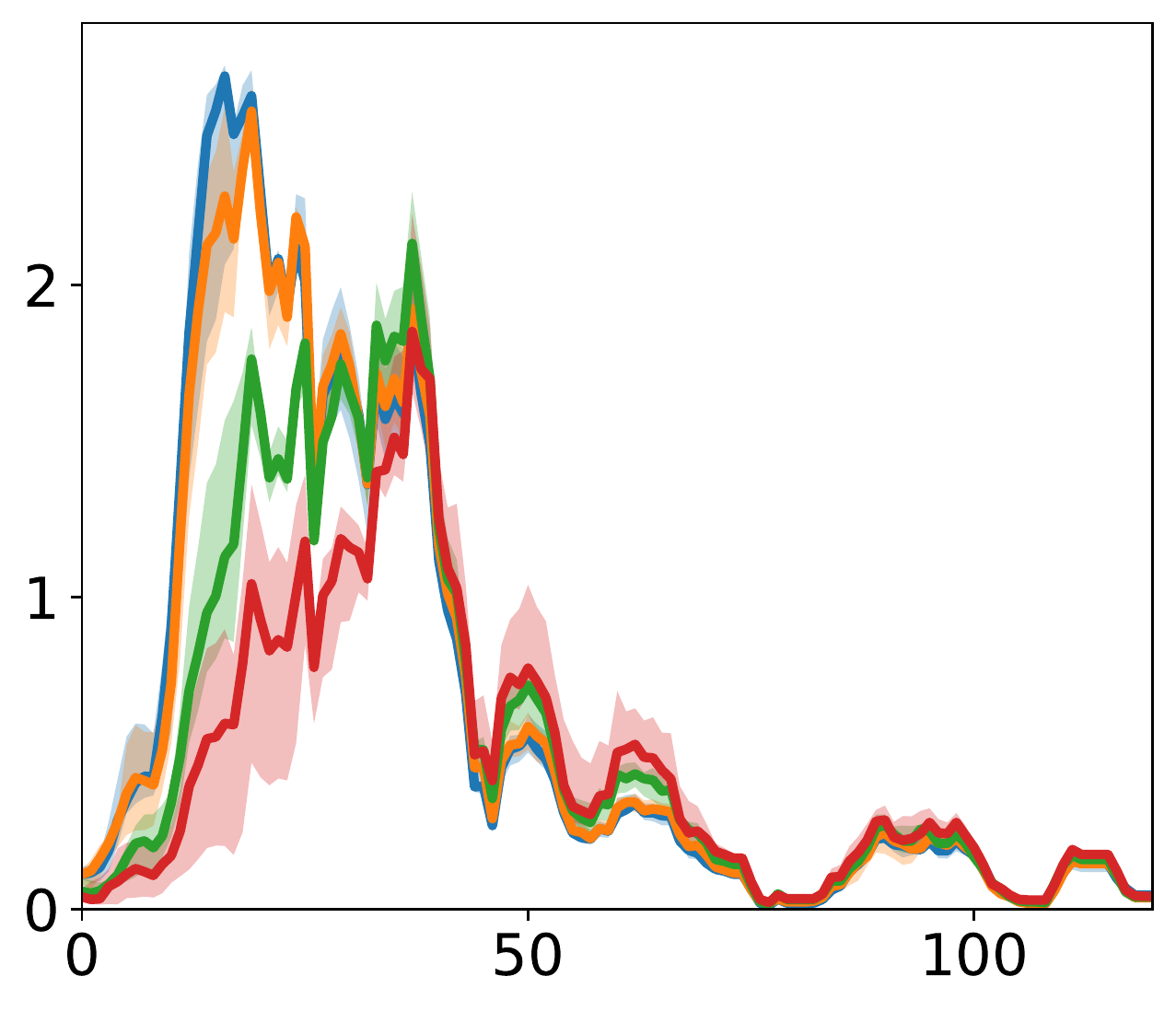}  &&
\includegraphics[width=\fivefig]{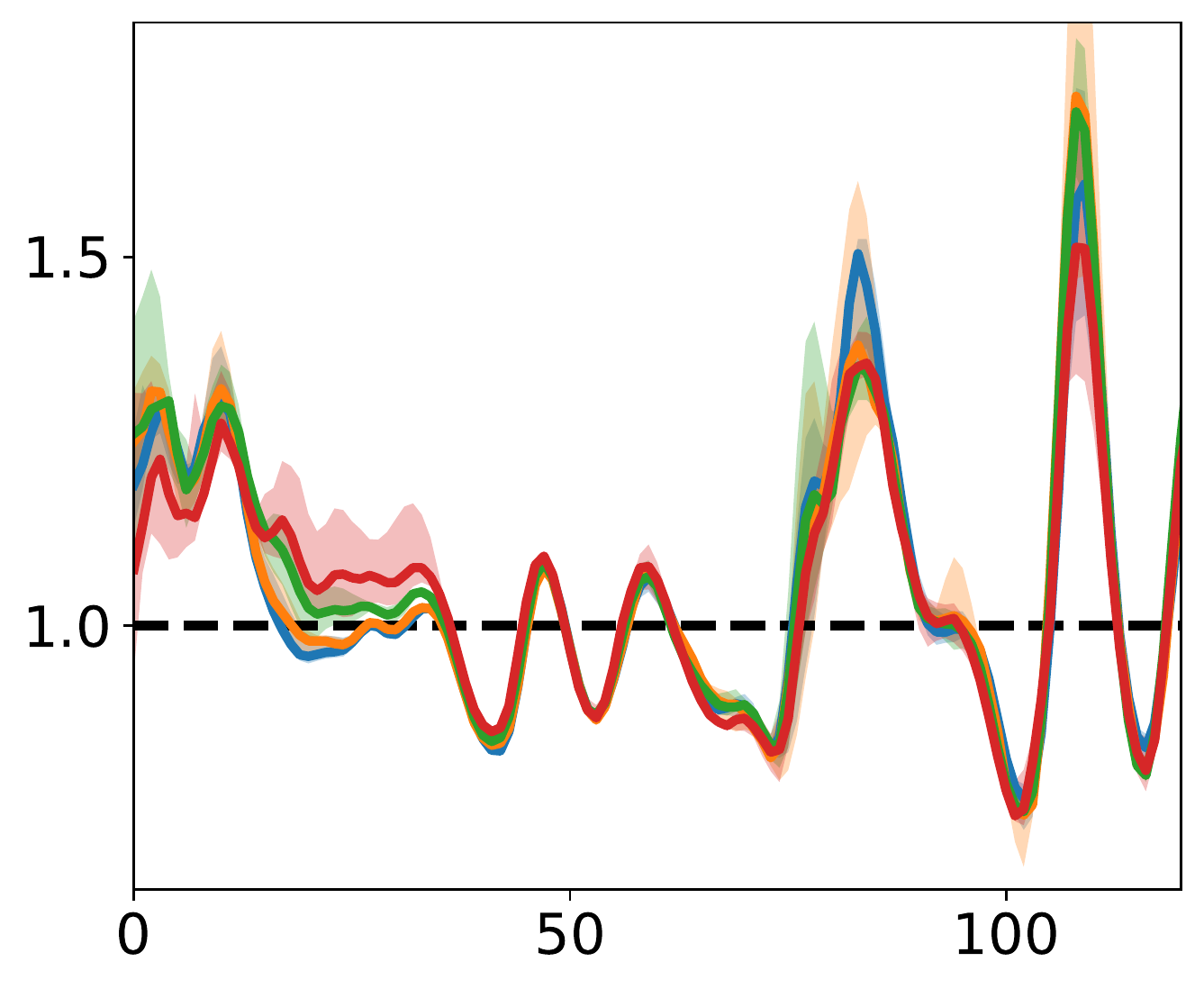} &&
\includegraphics[width=\fivefig]{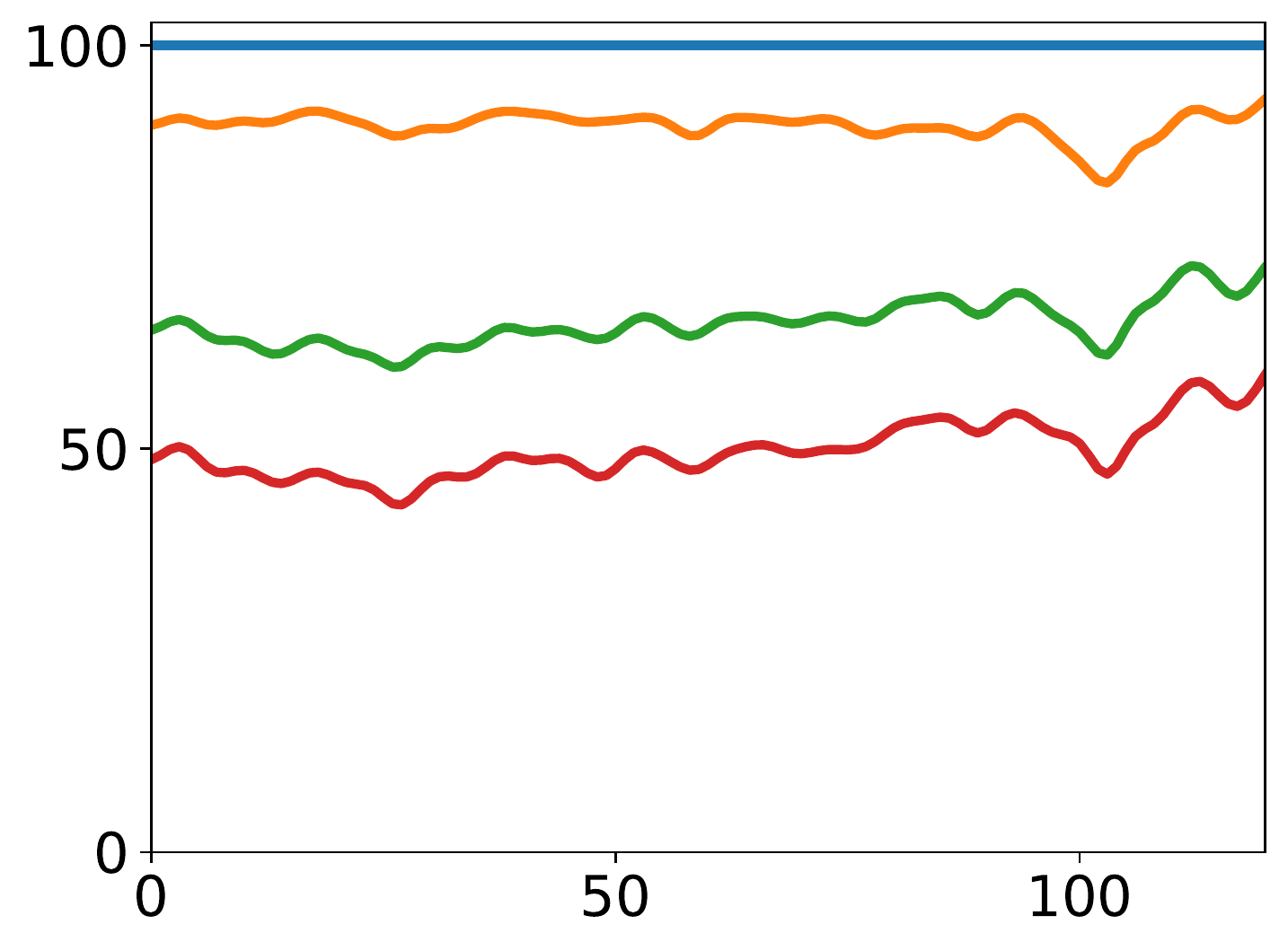} &
\rotatebox{90}{Jakarta}
\\ [-0.25cm]

&
\includegraphics[width=\fivefig]{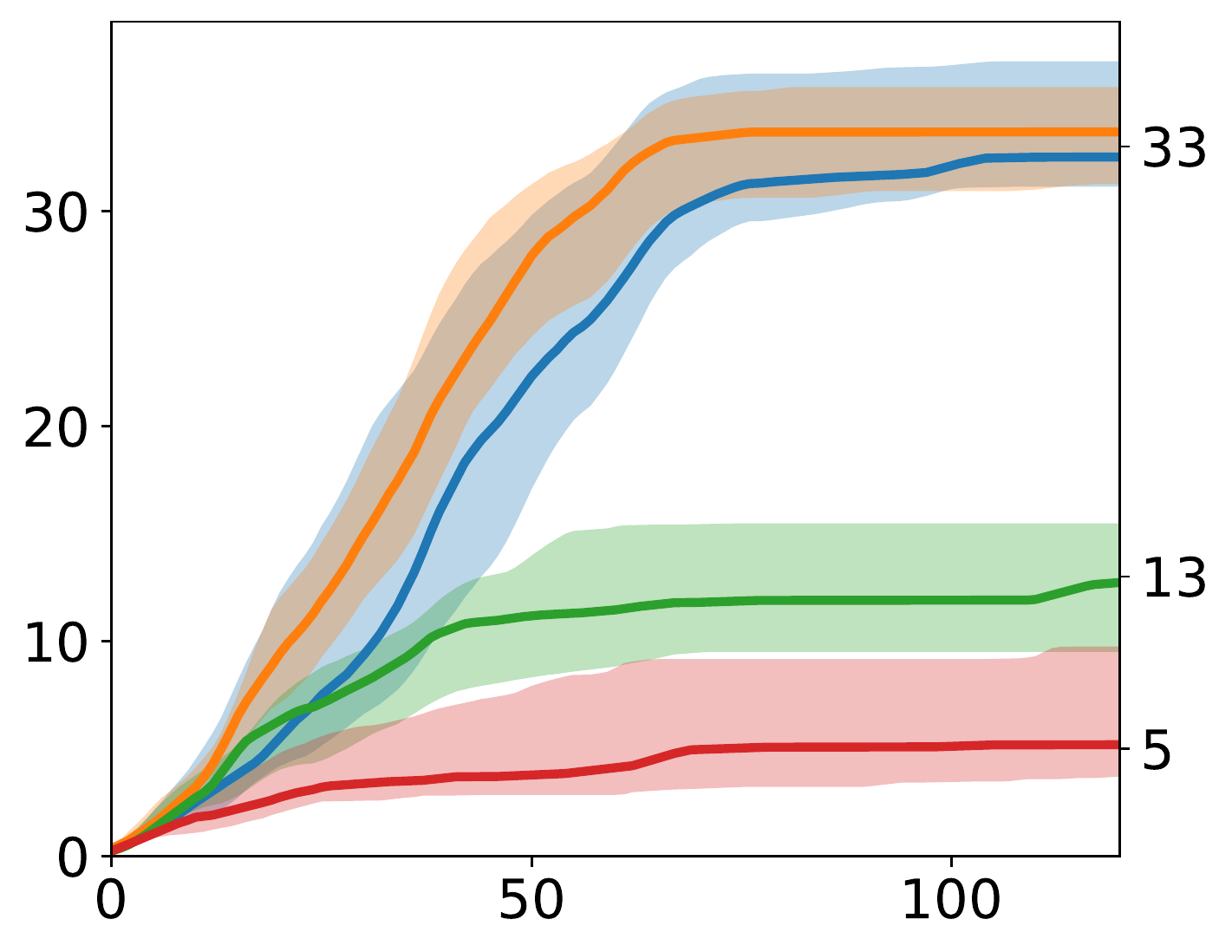} &&
\includegraphics[width=\fivefig]{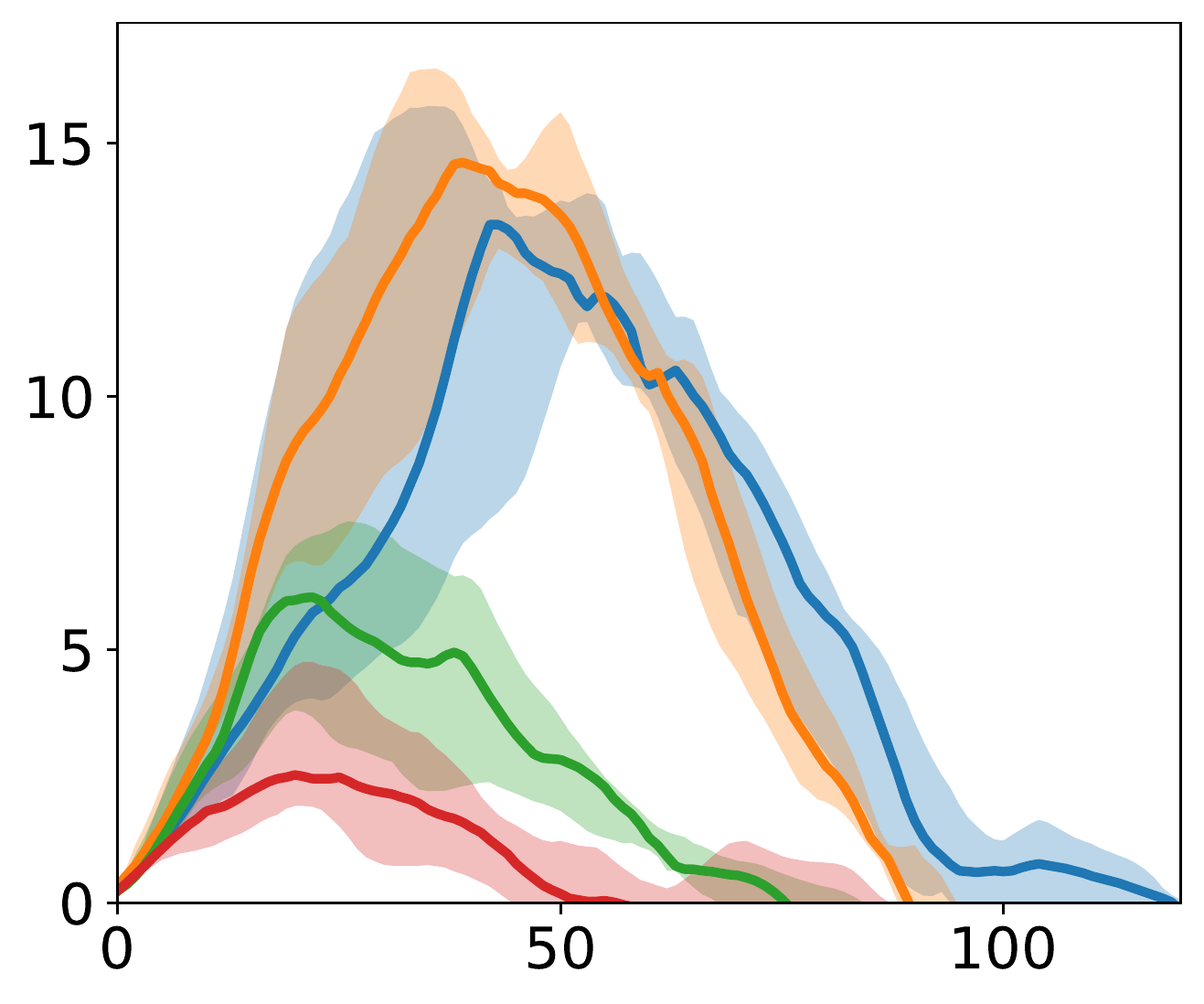}  &&
\includegraphics[width=\fivefig]{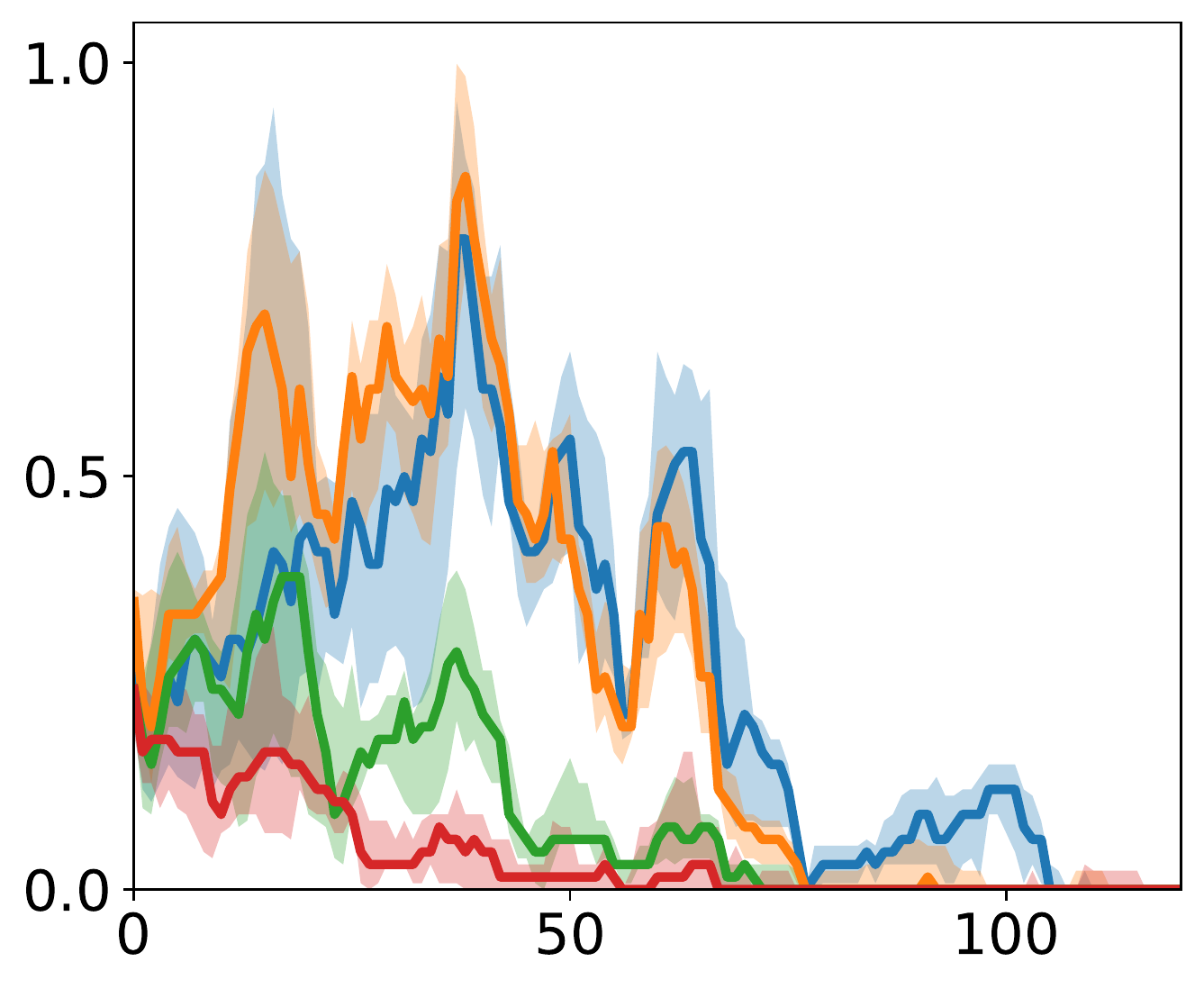}  &&
\includegraphics[width=\fivefig]{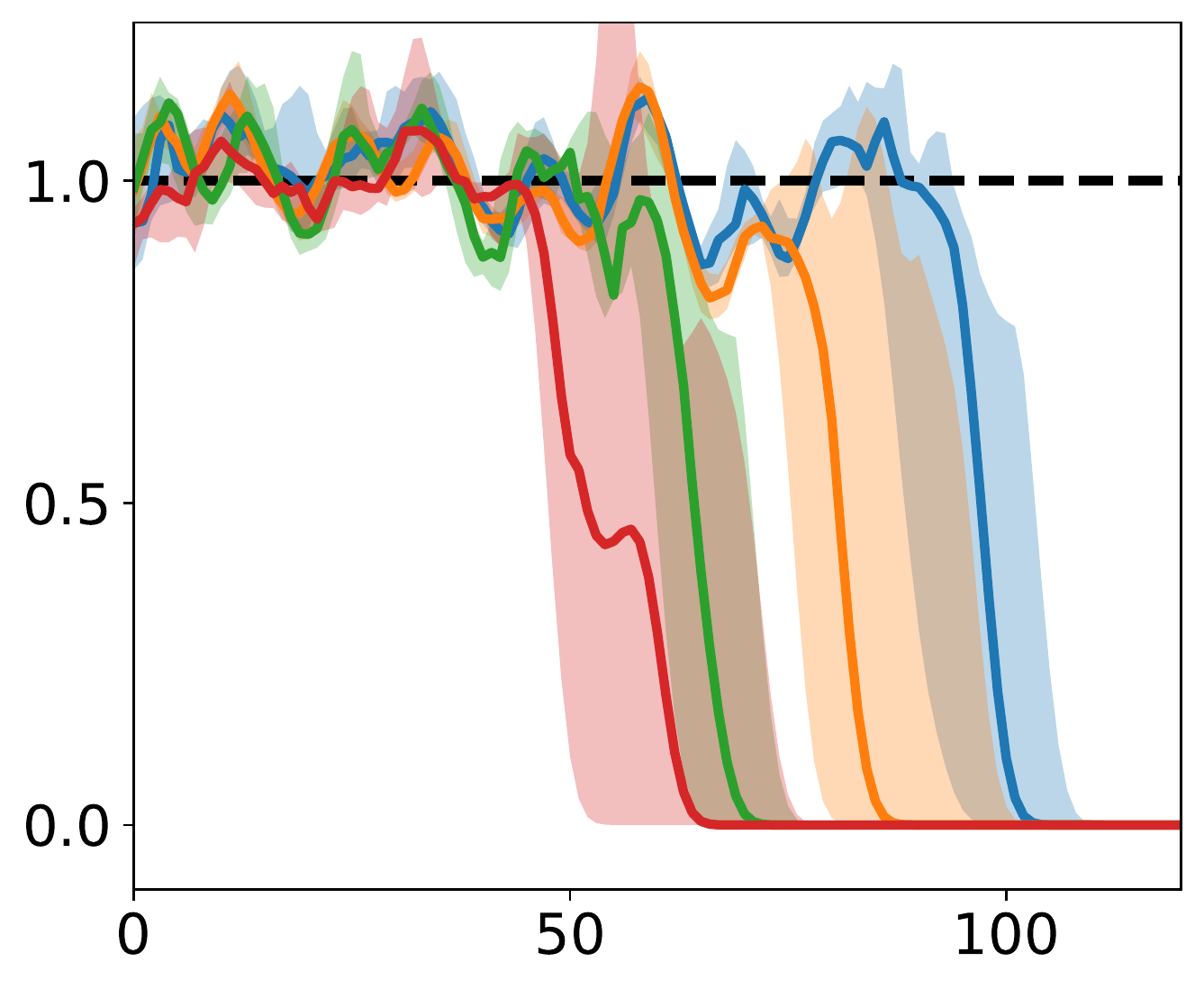} &&
\includegraphics[width=\fivefig]{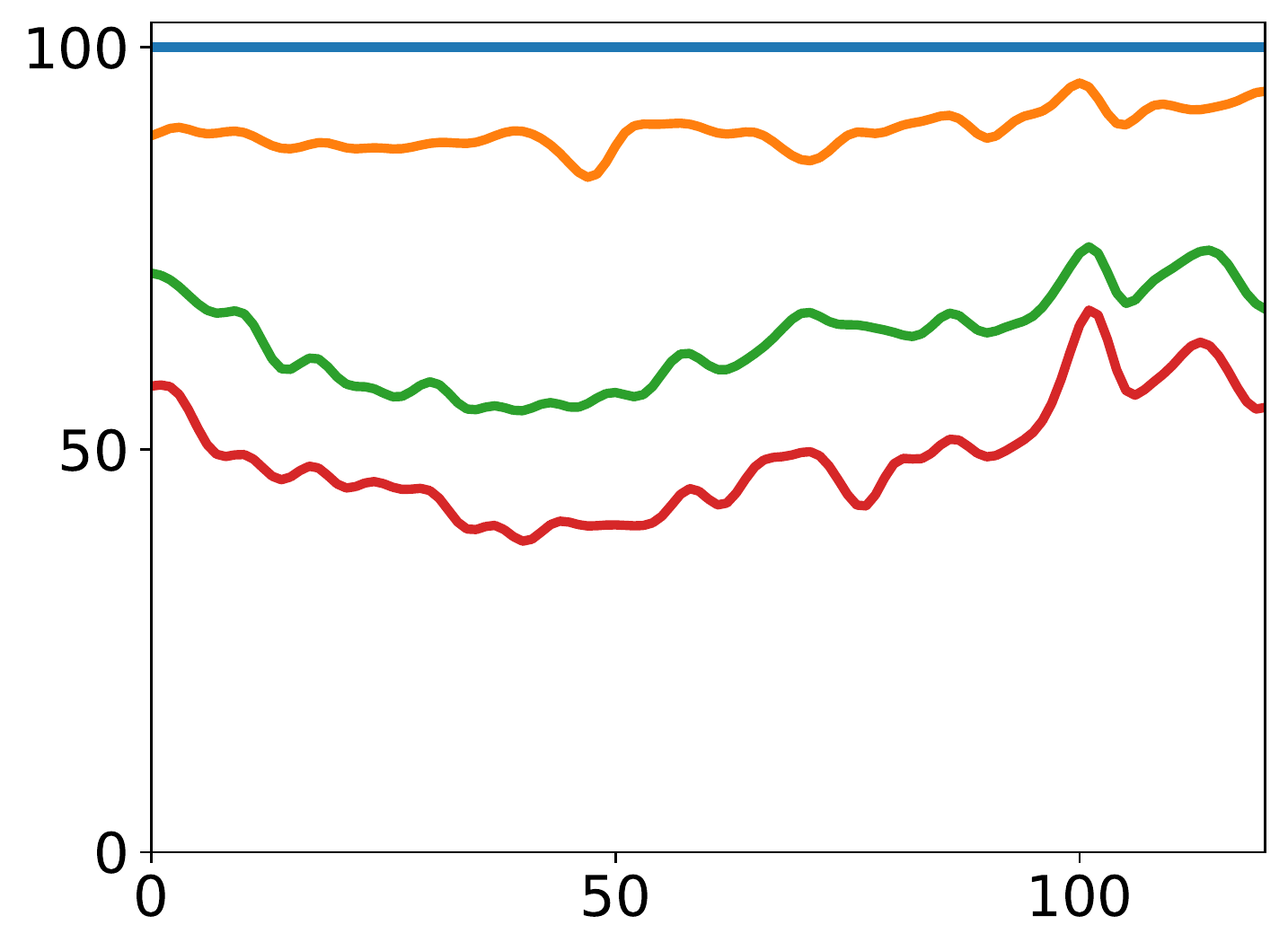} &
\rotatebox{90}{London}
\\ [-0.25cm]

\multicolumn{11}{c}{\includegraphics[width=0.15\textwidth]{Fig/labels/daysfromstart.pdf}}\\
\multicolumn{11}{c}{\includegraphics[width=0.6\textwidth]{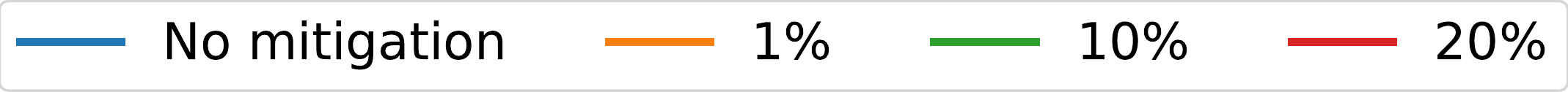}}\\
\end{tabular}
\caption{Infection spreading when protecting most active agents. The total number of infected people is reduced and the peaks of active infections are delayed and lowered with more active agents being protected.
}
\label{figS:protect_acitve_people}
\end{figure}

\begin{figure}[!hp]
    \centering

\begin{tabular}{m{0.1cm}m{\fivefig}@{}m{0.1cm}m{\fivefig}@{}m{0.1cm}m{\fivefig}@{}m{0.1cm}m{\fivefig}@{}m{0.1cm}m{\fivefig}@{}m{0.1cm}}

\multirow{14}{*}{\includegraphics[height=4cm]{Fig/labels/TotalInfected.pdf}}&
\includegraphics[width=\fivefig]{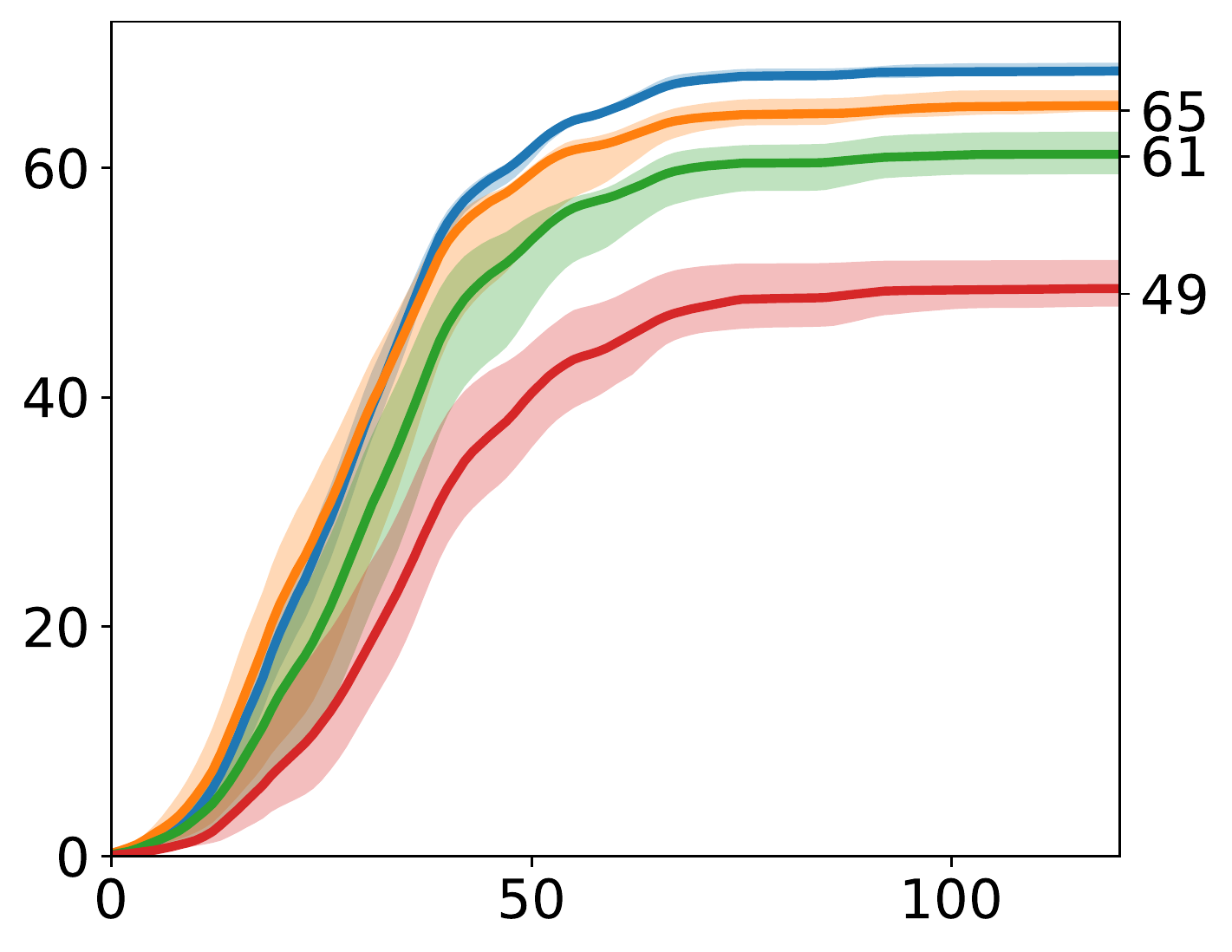} &
\multirow{14}{*}{\includegraphics[height=4cm]{Fig/labels/ActiveInfected.pdf}}&
\includegraphics[width=\fivefig]{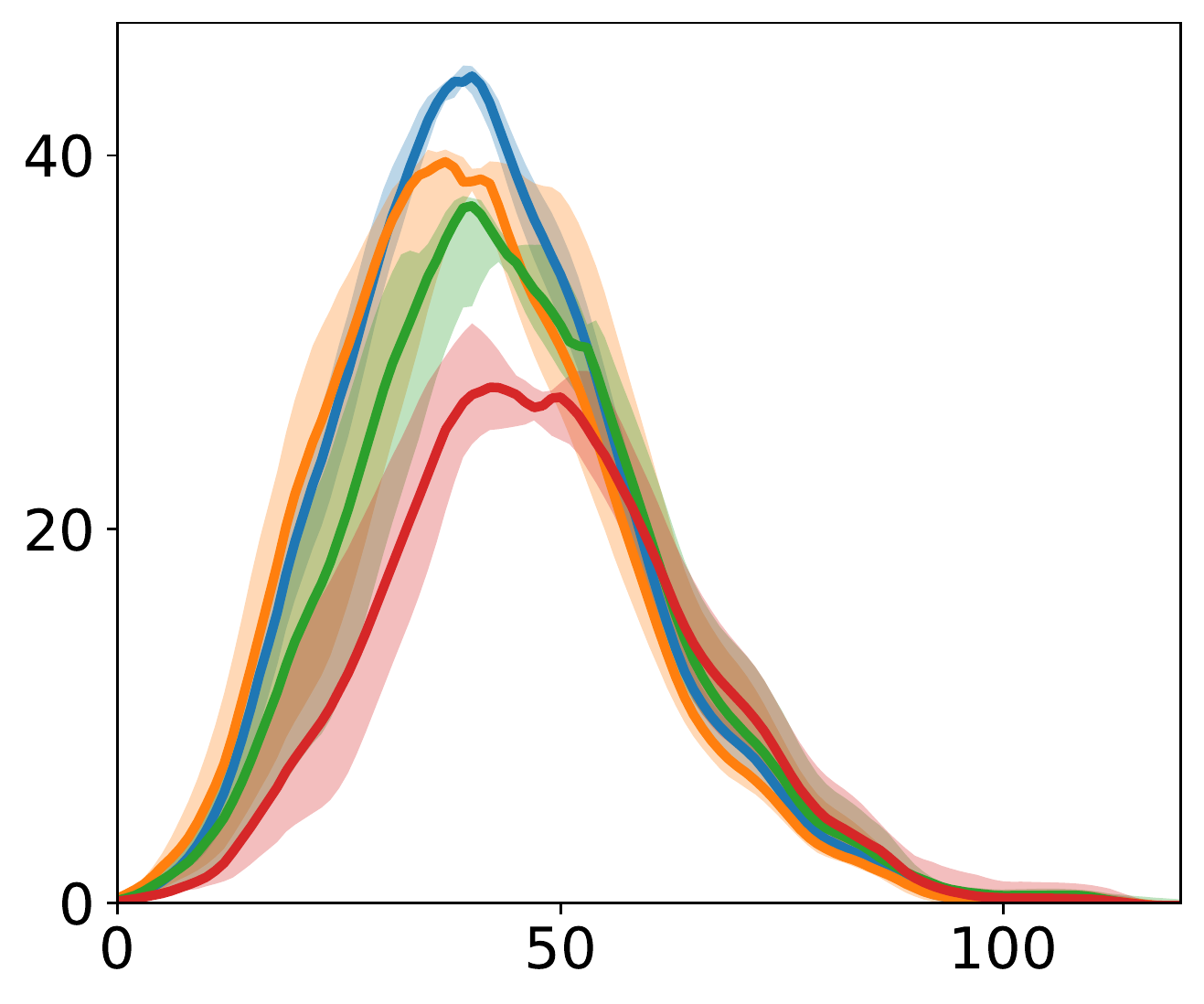}  &
\multirow{14}{*}{\includegraphics[height=4cm]{Fig/labels/newInfected.pdf}}&
\includegraphics[width=\fivefig]{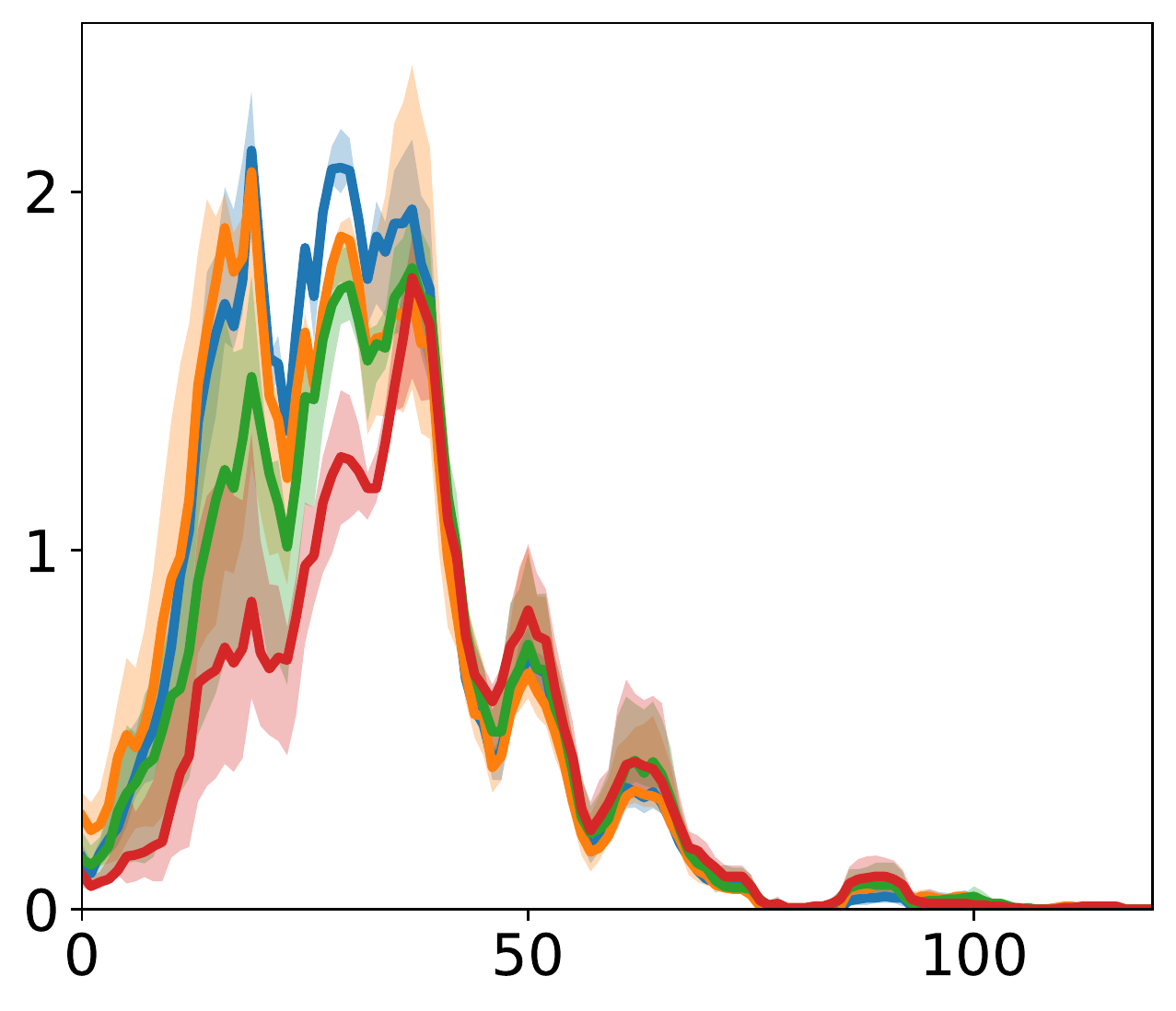}  &
\multirow{14}{*}{\includegraphics[height=2cm]{Fig/labels/growthrate.pdf}}&
\includegraphics[width=\fivefig]{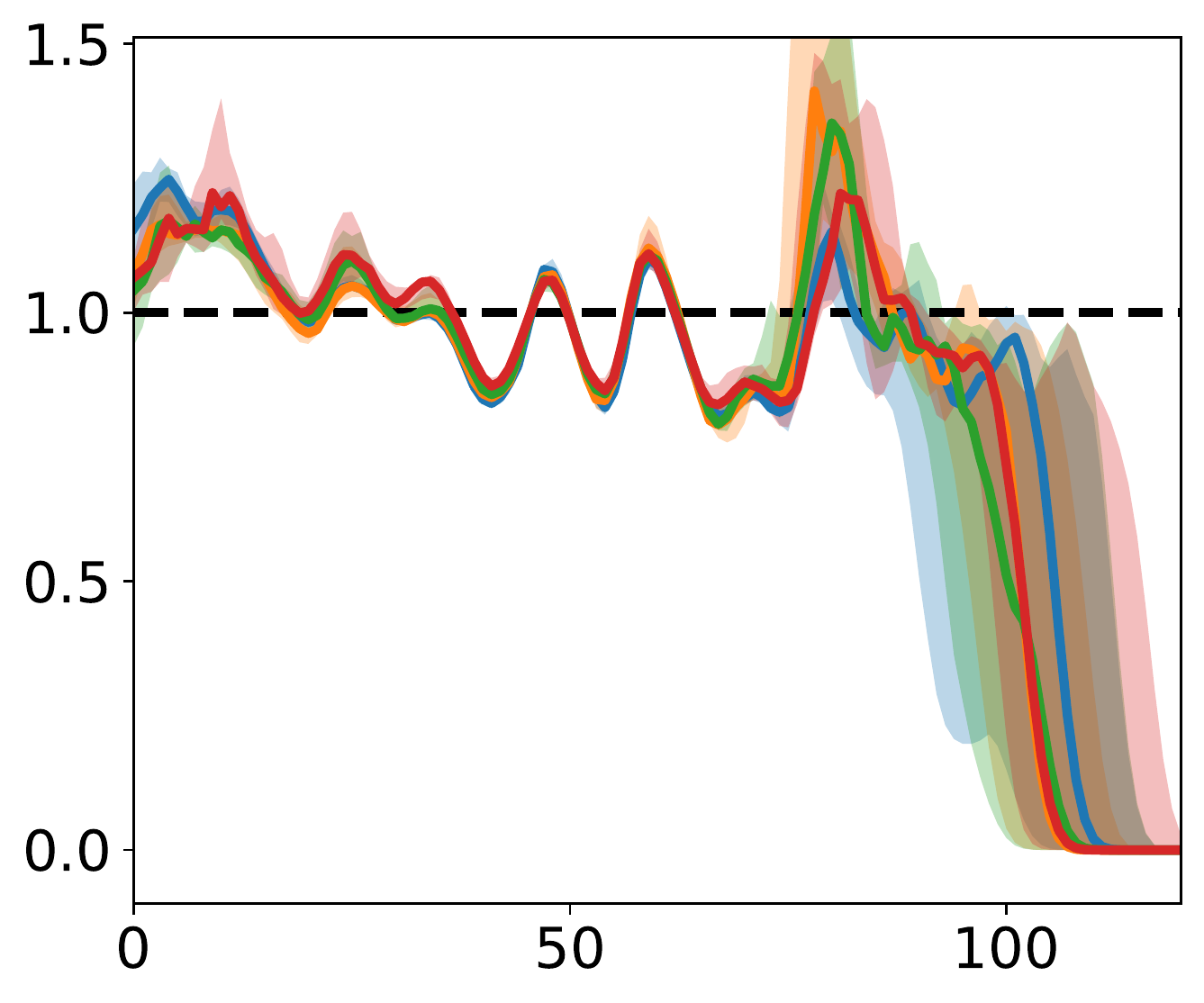} &
\multirow{14}{*}{\includegraphics[height=2cm]{Fig/labels/socialvalue.pdf}}&
\includegraphics[width=\fivefig]{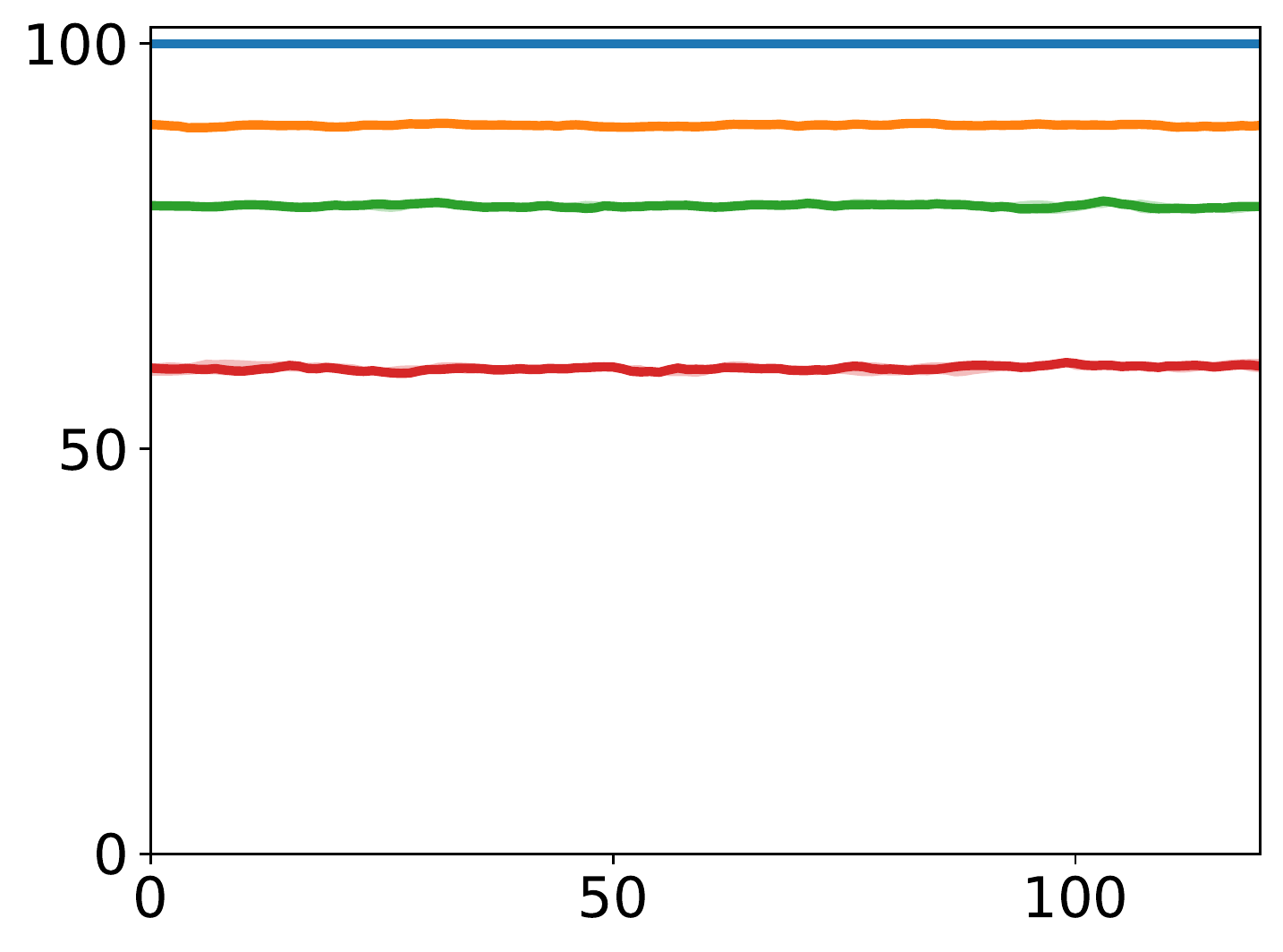} &
\rotatebox{90}{NYC}
\\ [-0.25cm]
&
\includegraphics[width=\fivefig]{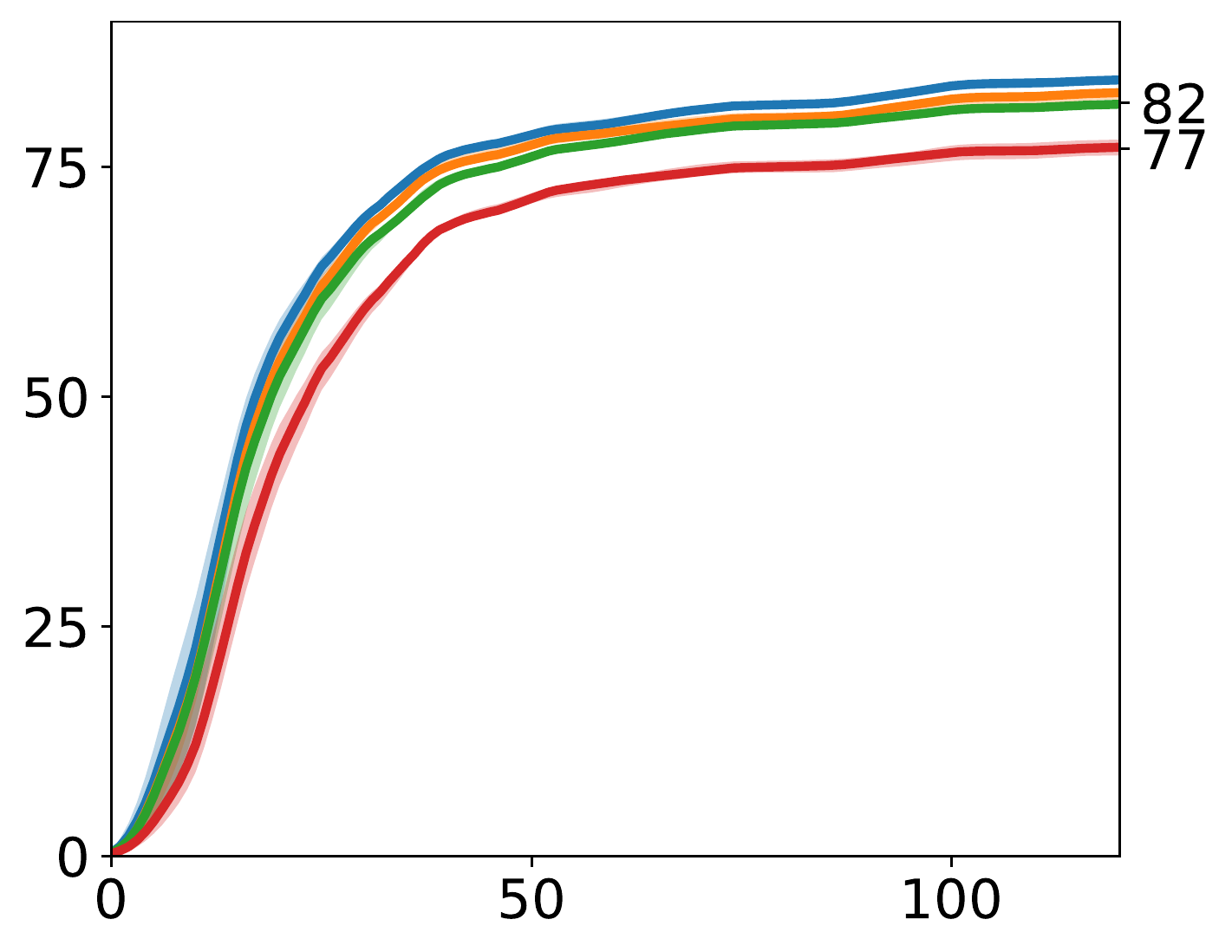} & &
\includegraphics[width=\fivefig]{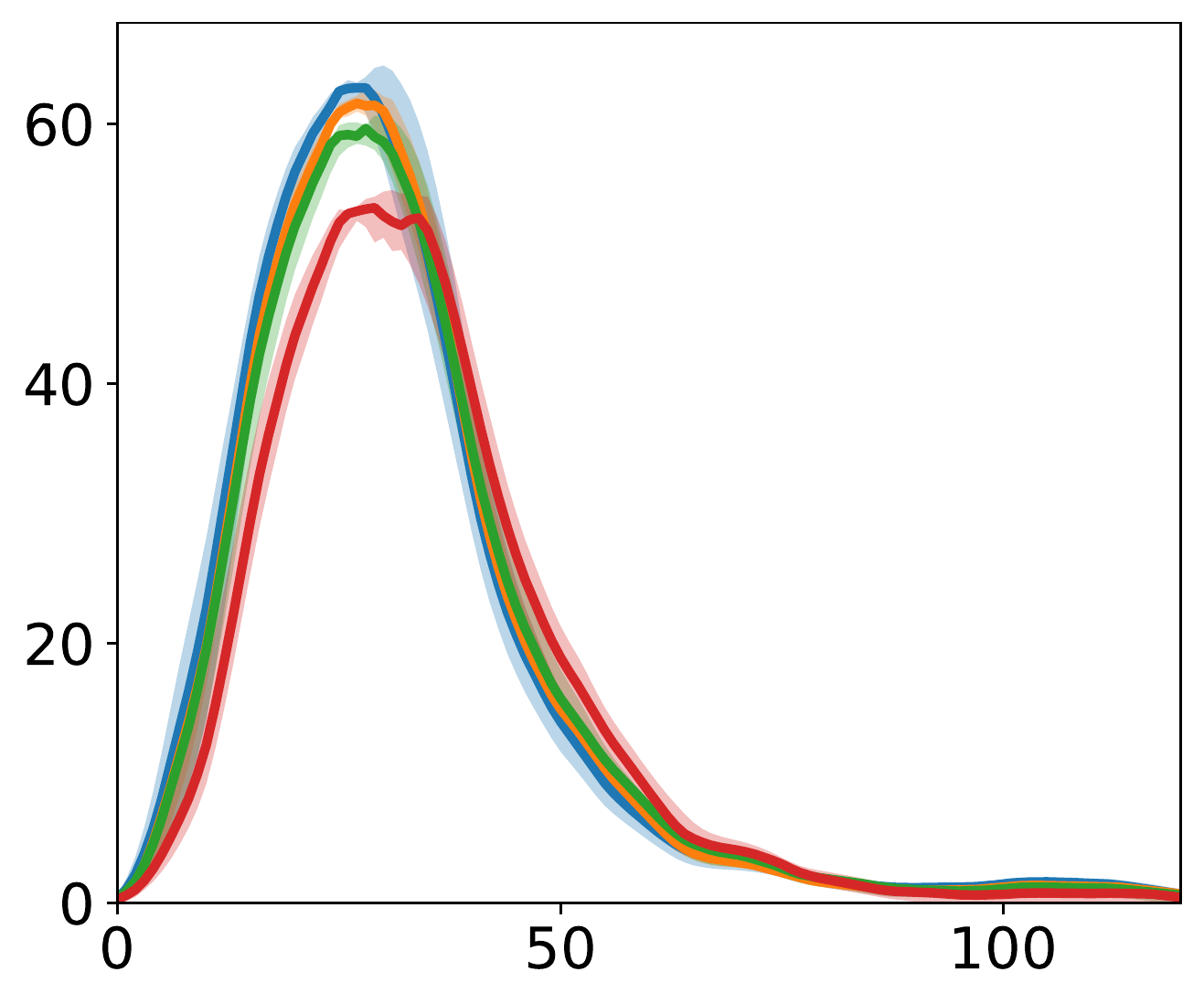}  & &
\includegraphics[width=\fivefig]{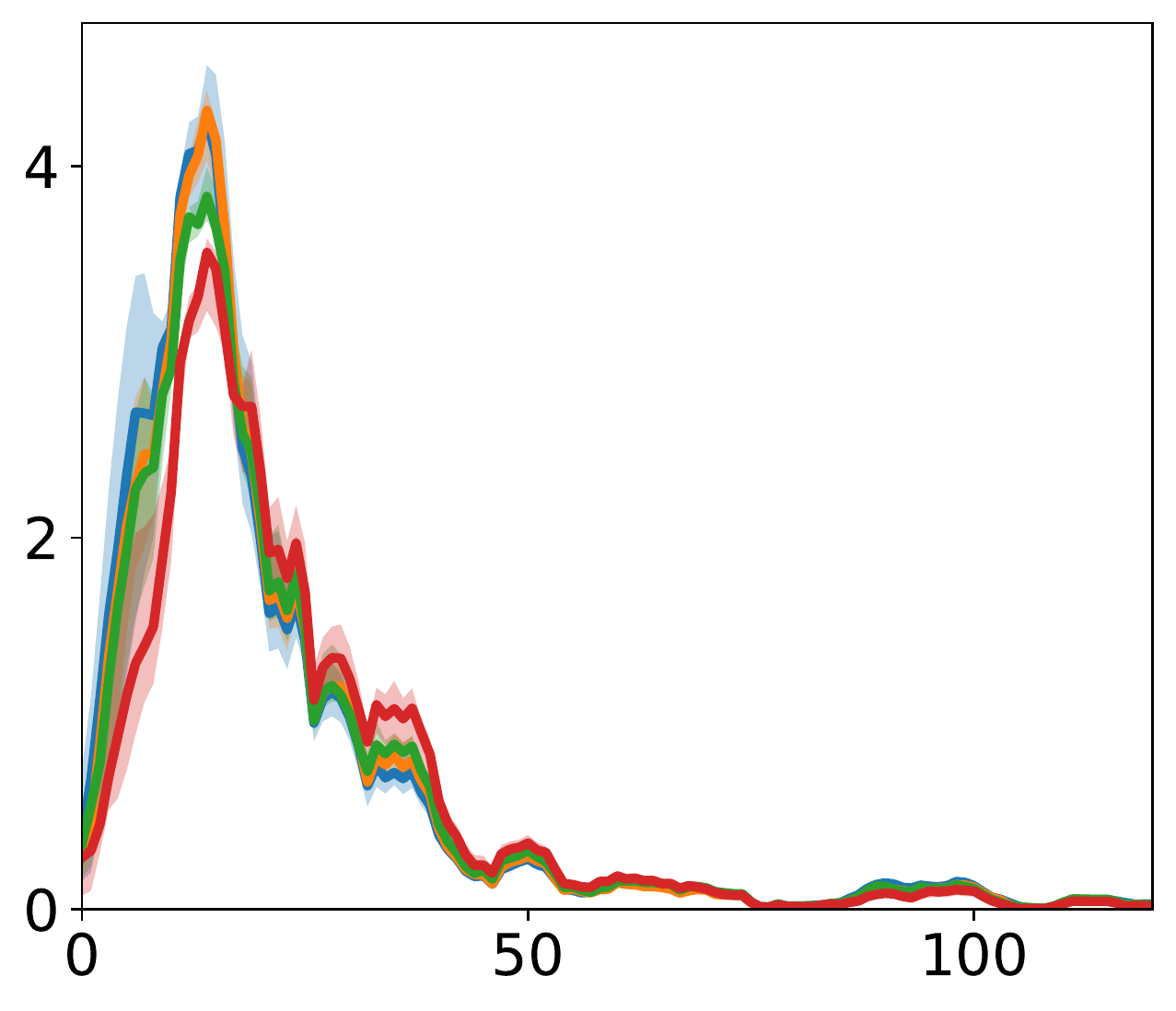}  & &
\includegraphics[width=\fivefig]{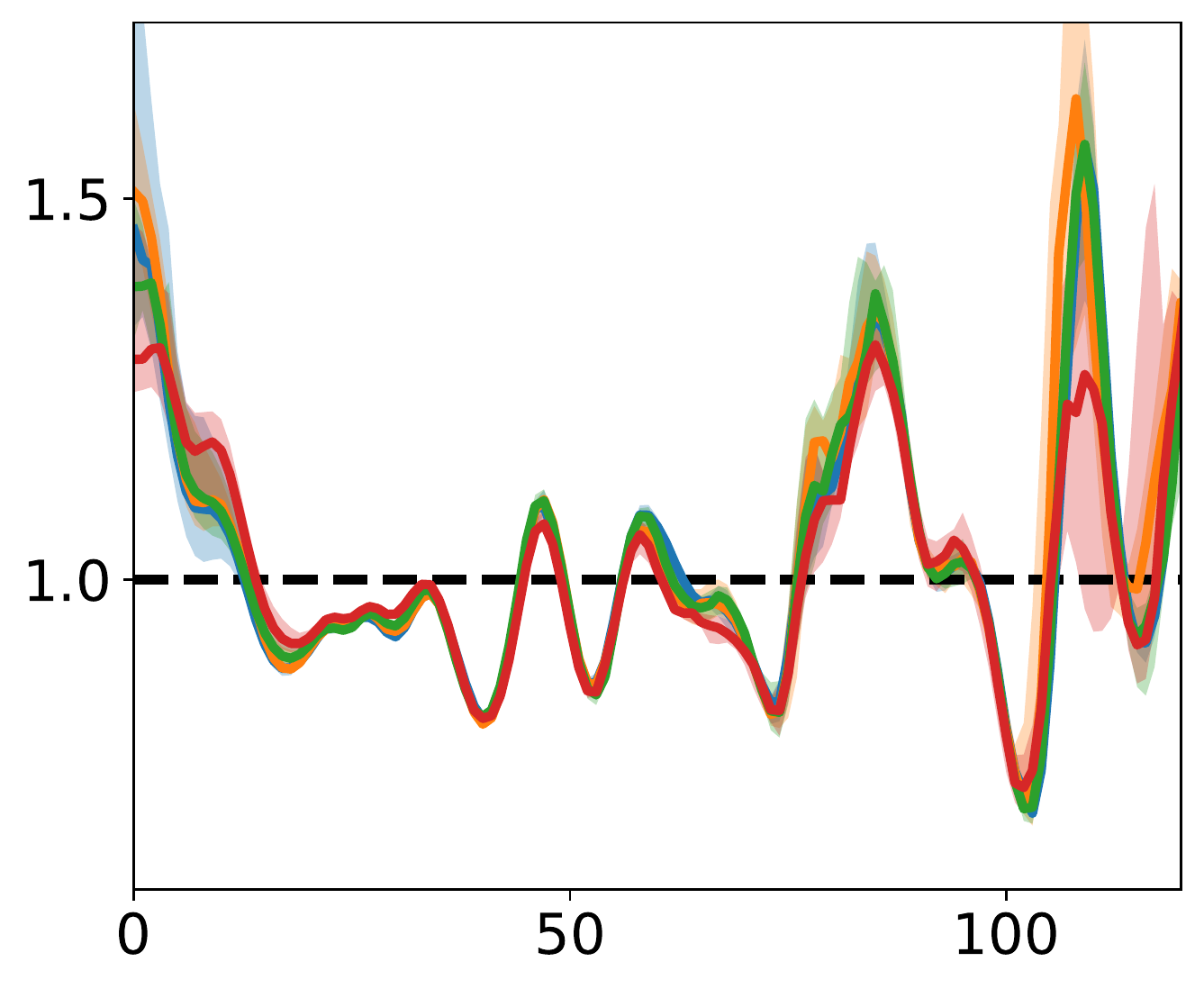} & &
\includegraphics[width=\fivefig]{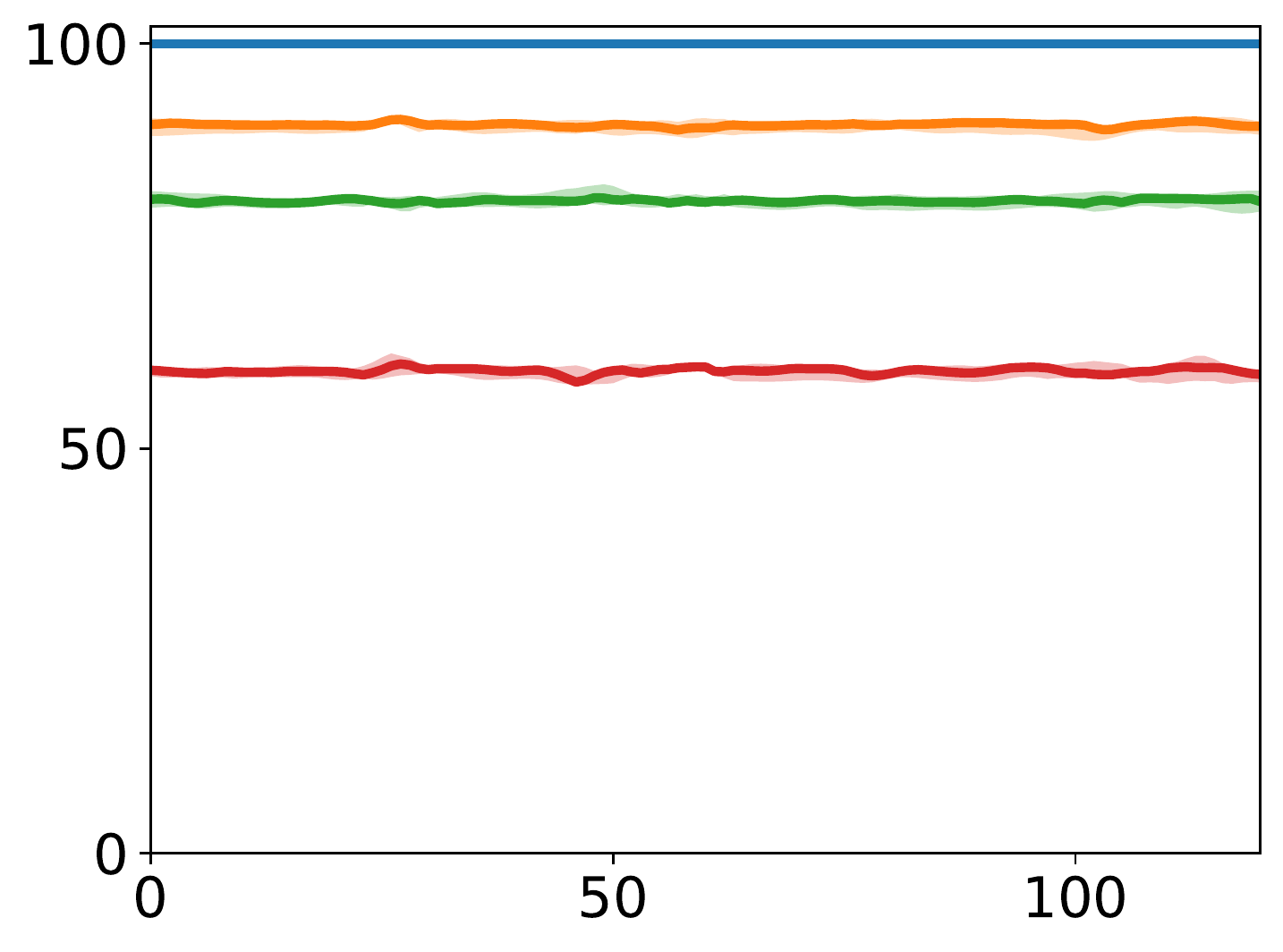} &
\rotatebox{90}{Tokyo}
\\[-0.25cm]
&
\includegraphics[width=\fivefig]{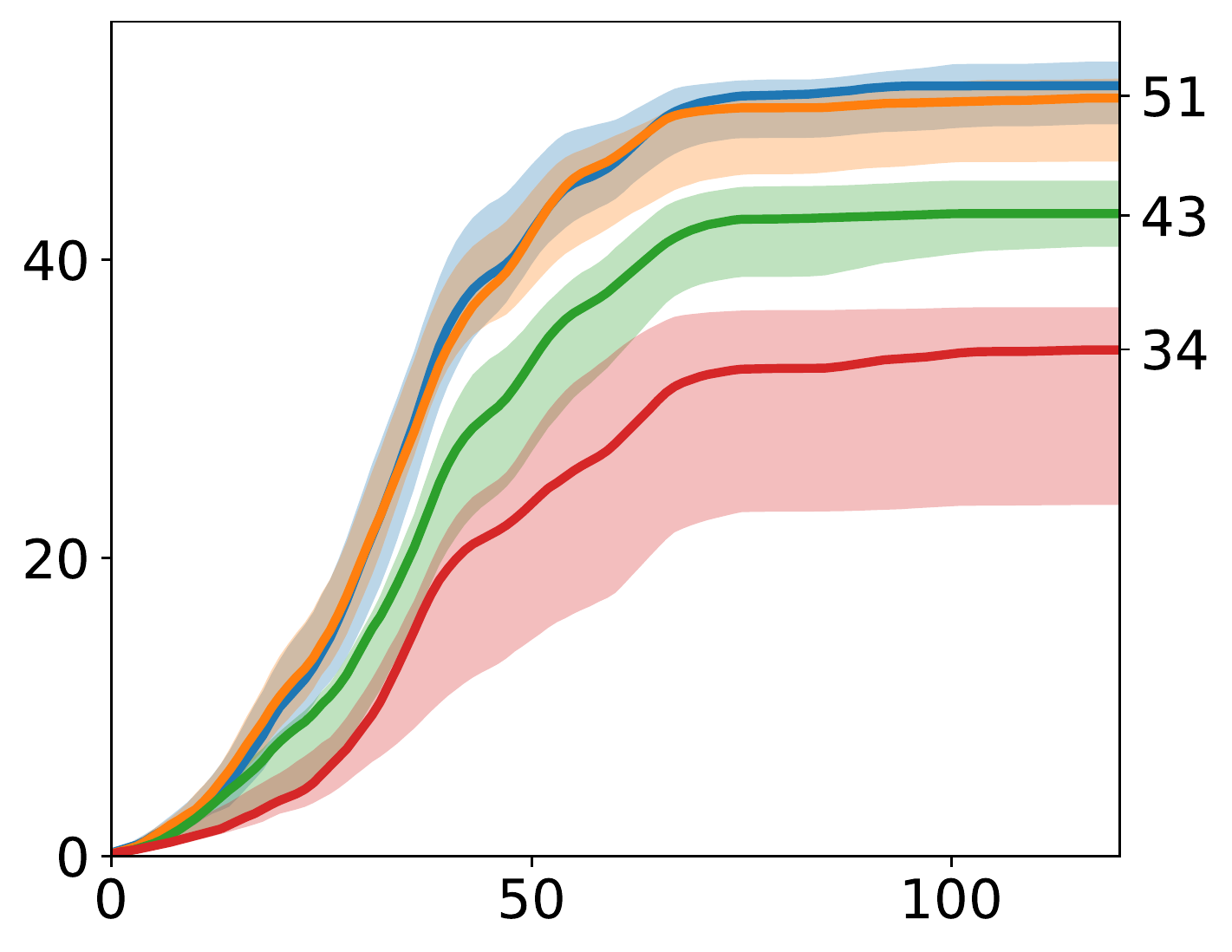} & &
\includegraphics[width=\fivefig]{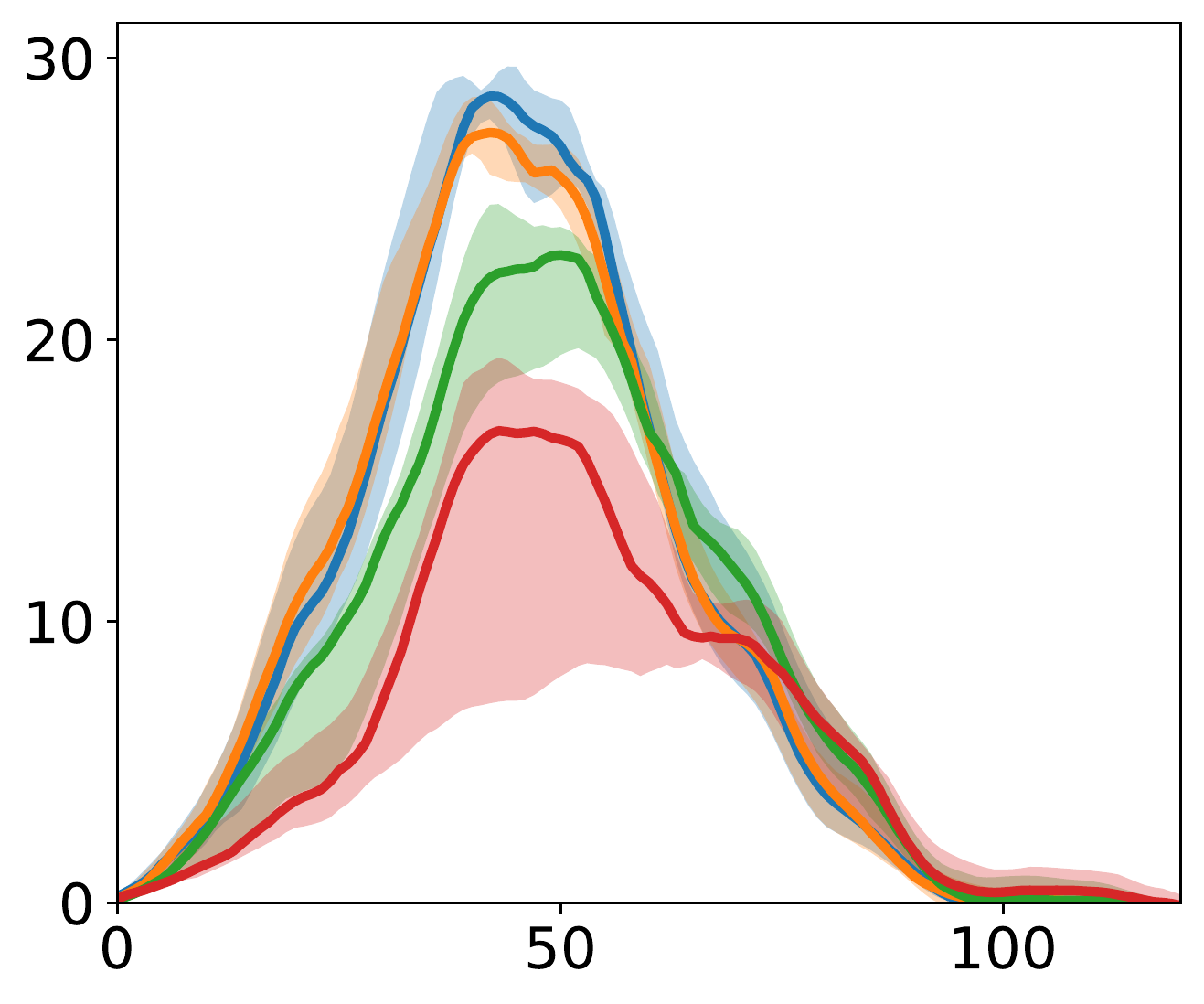}  & &
\includegraphics[width=\fivefig]{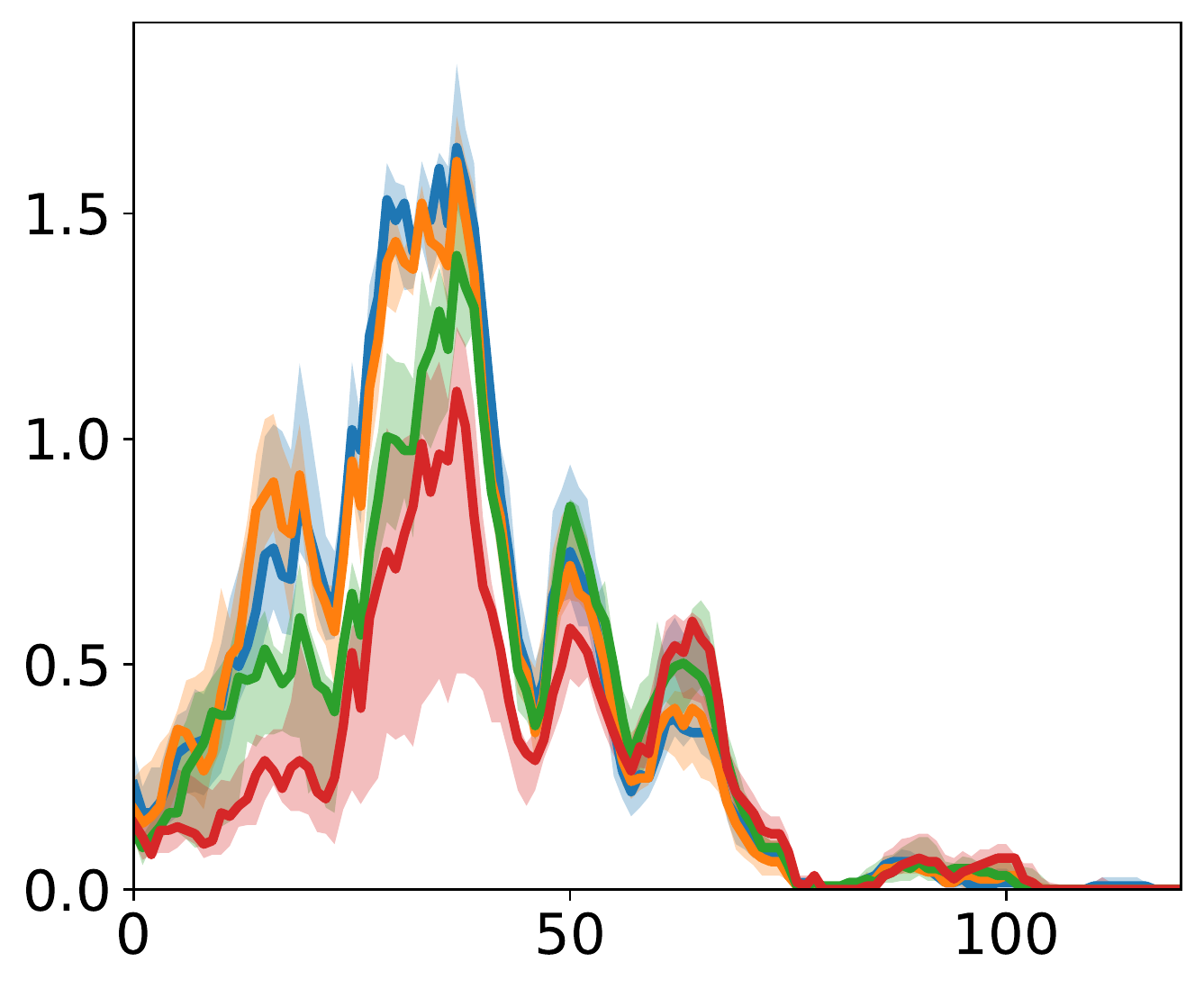}  & &
\includegraphics[width=\fivefig]{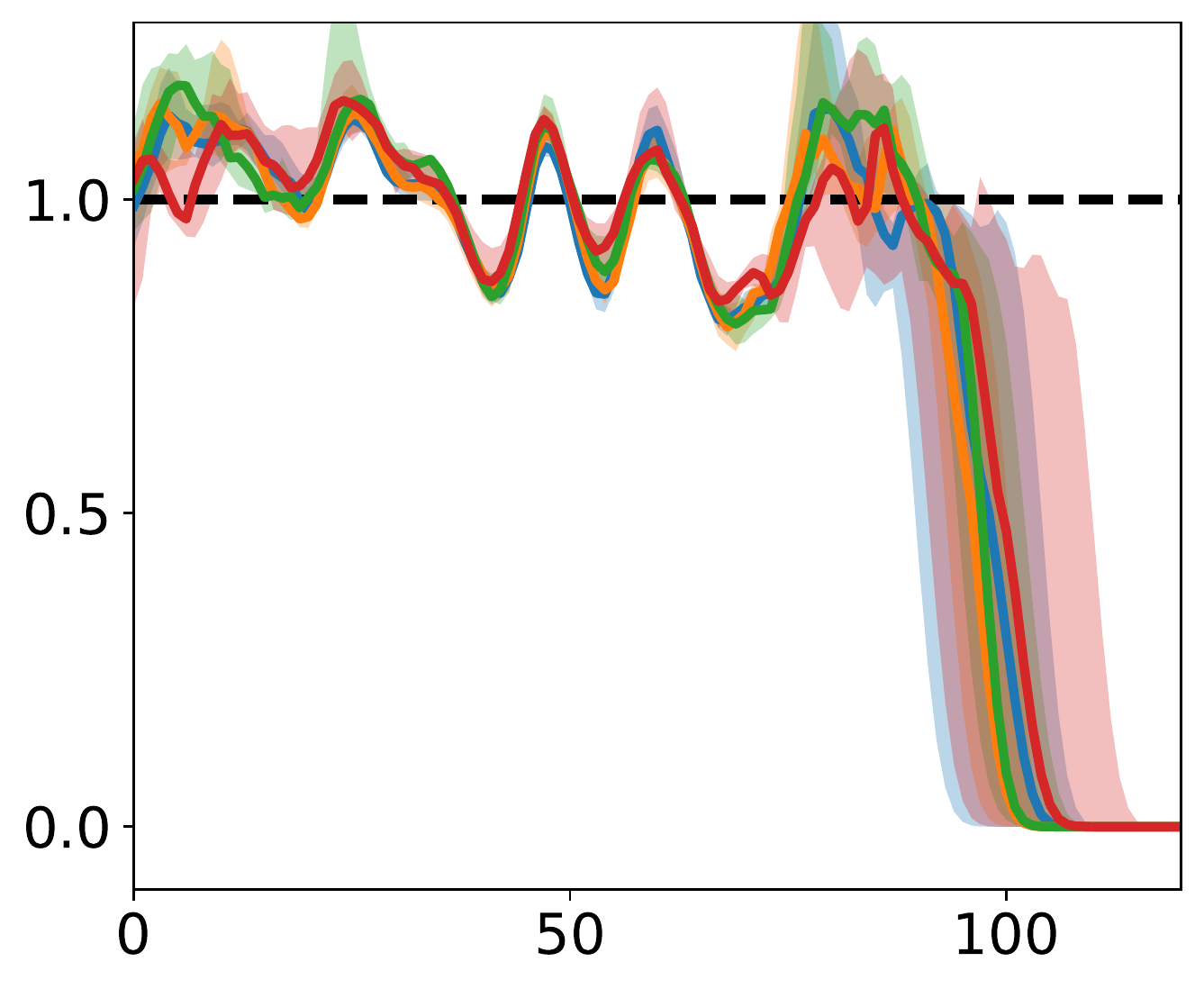}  & &
\includegraphics[width=\fivefig]{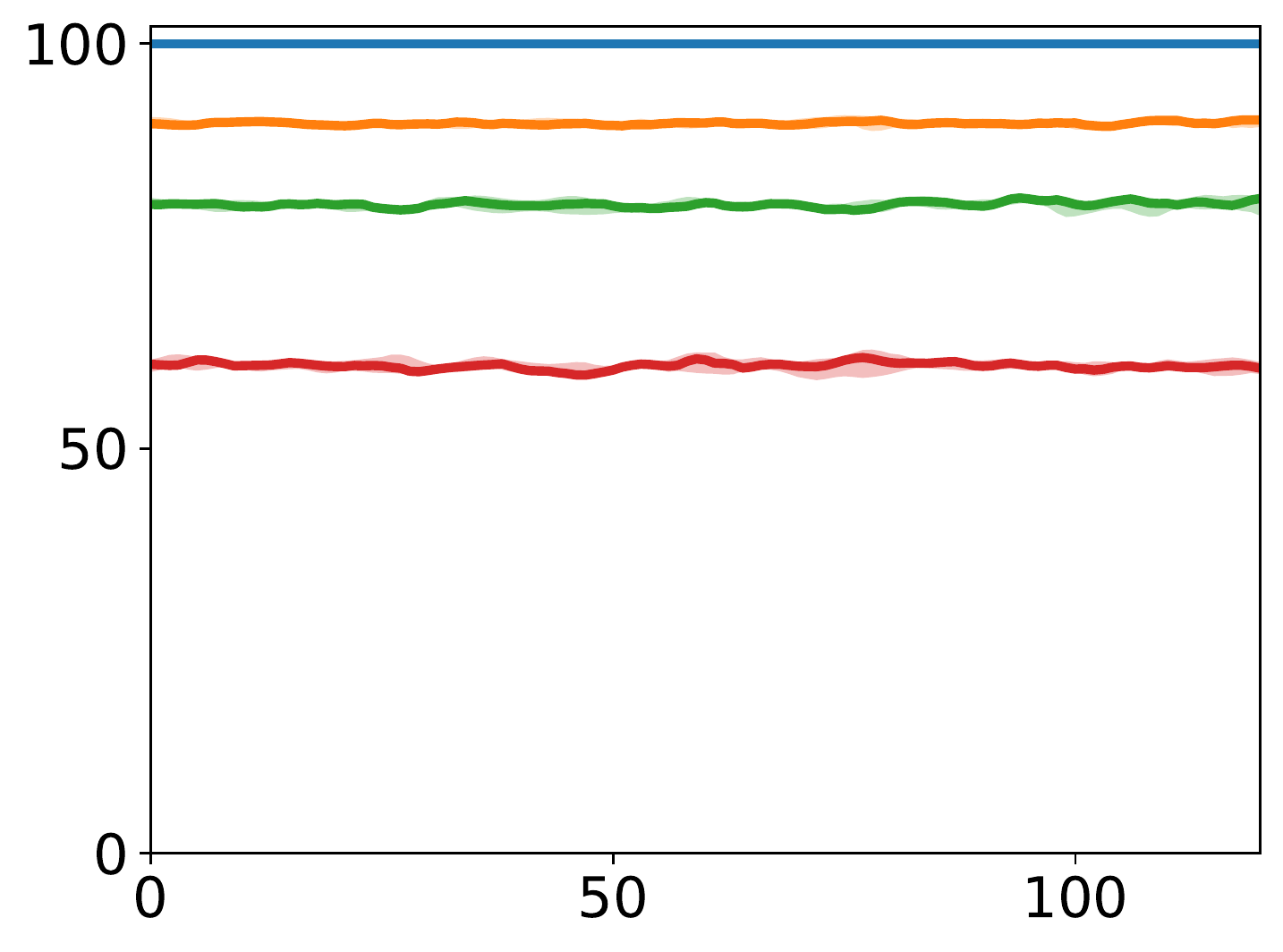} &
\rotatebox{90}{Chicago}
\\ [-0.25cm]

&
\includegraphics[width=\fivefig]{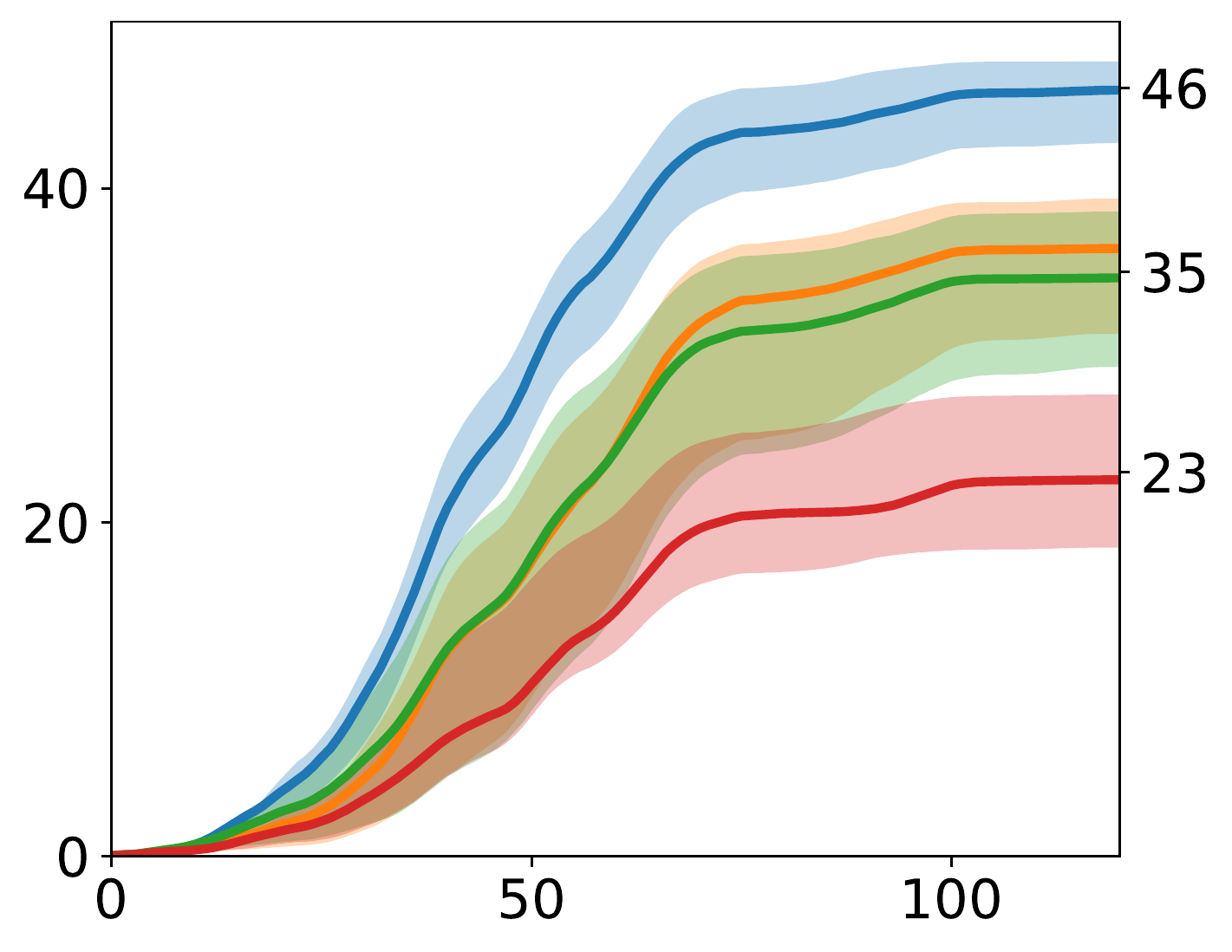} & &
\includegraphics[width=\fivefig]{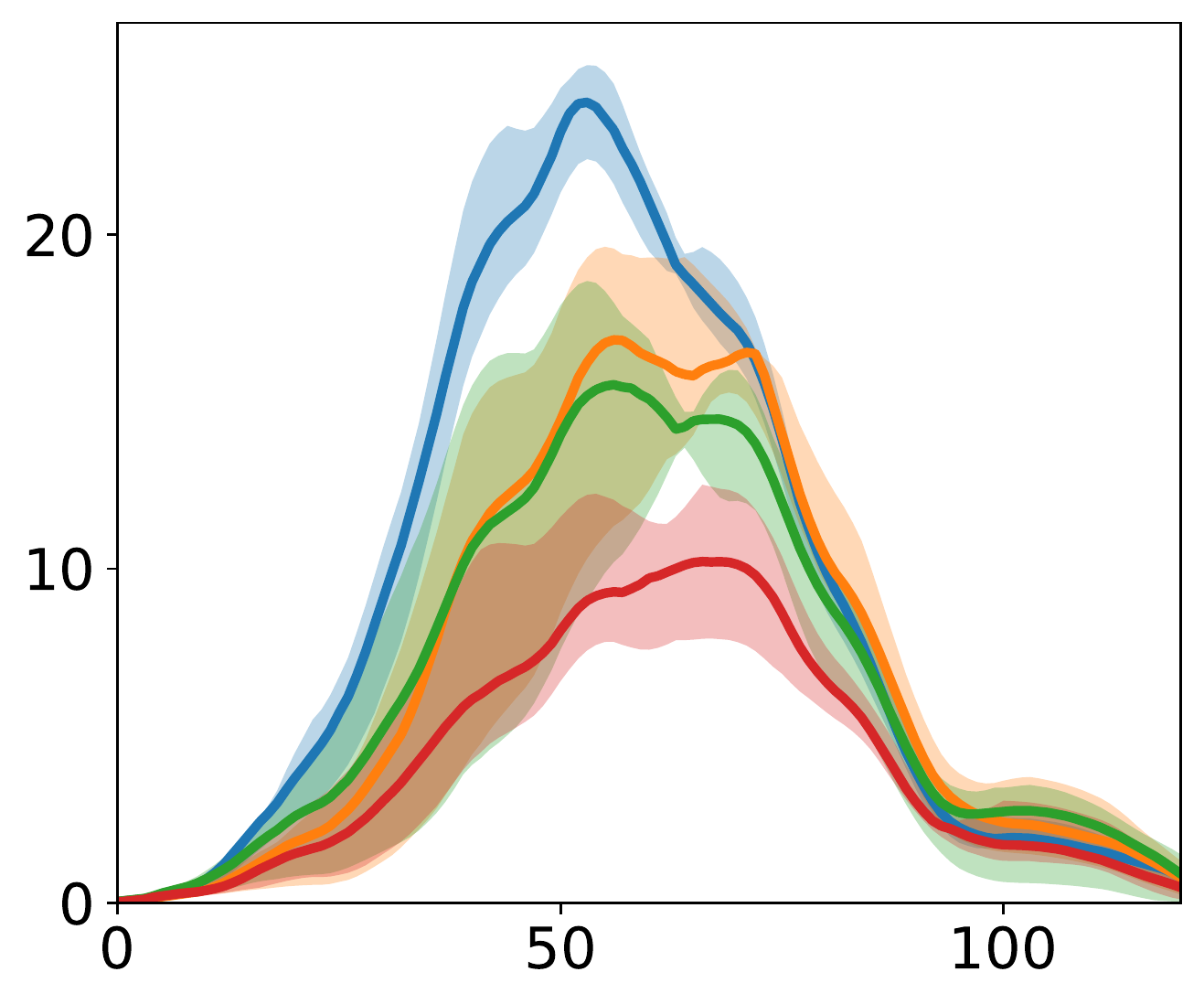} & &
\includegraphics[width=\fivefig]{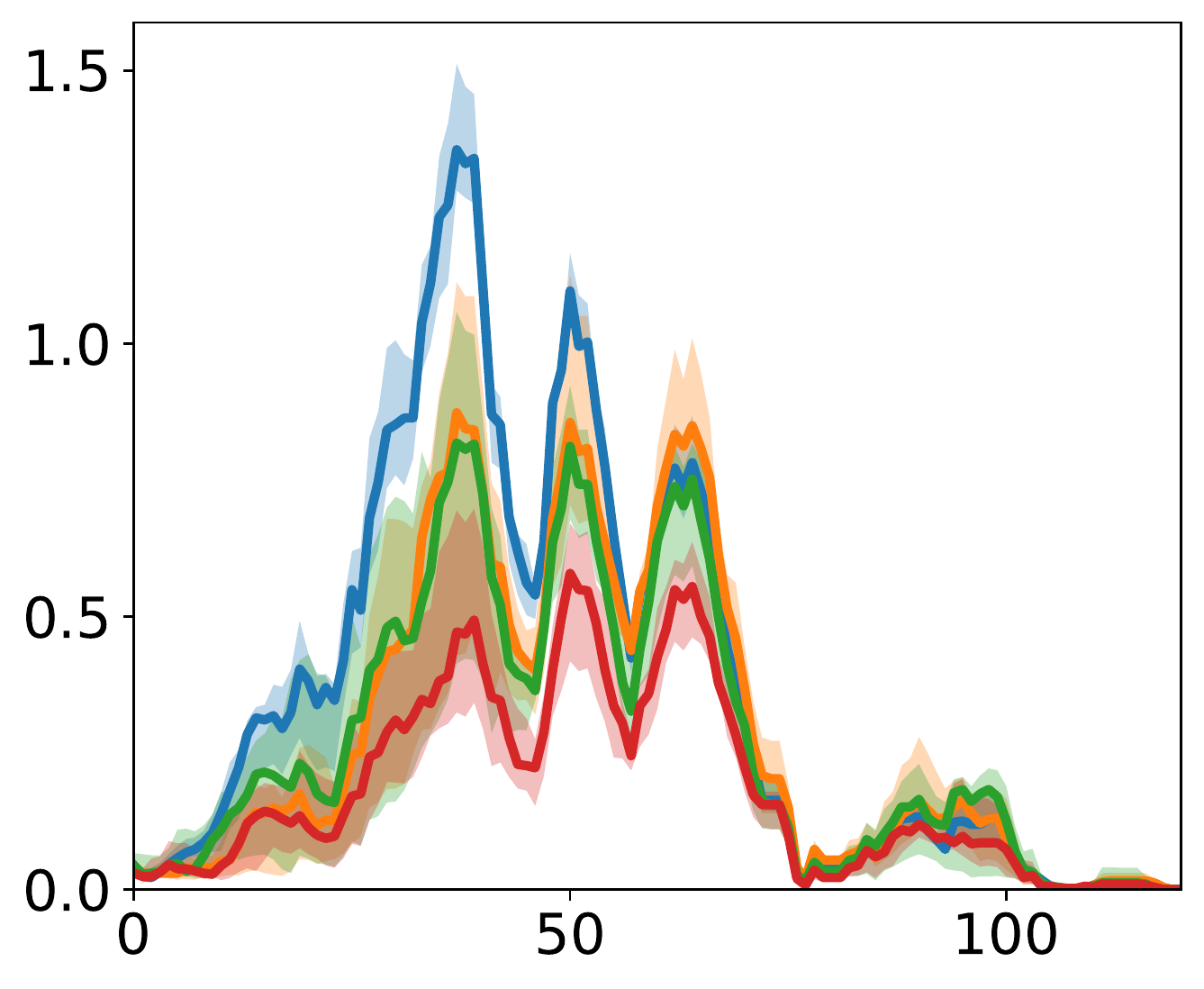}  & &
\includegraphics[width=\fivefig]{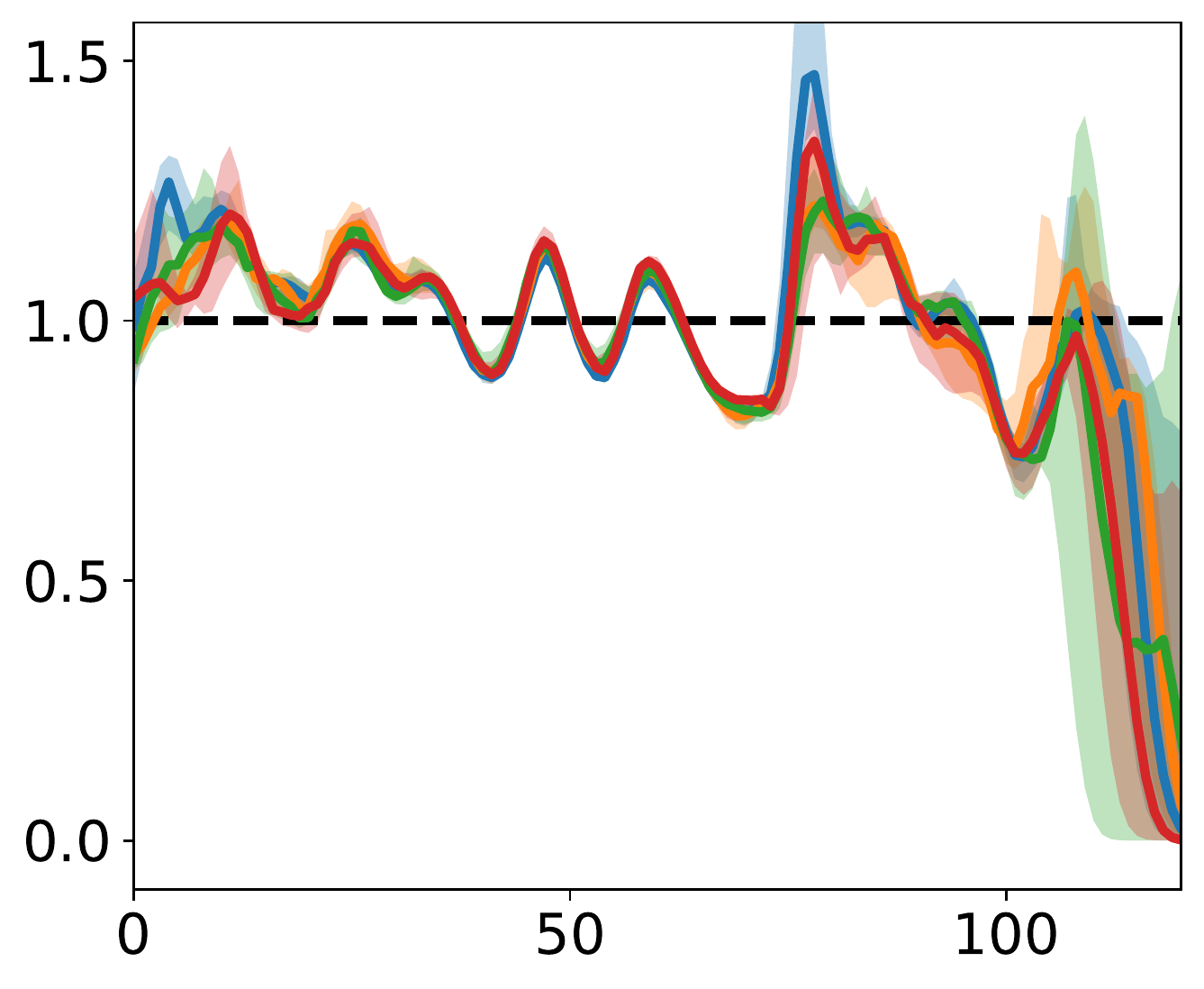} &&
\includegraphics[width=\fivefig]{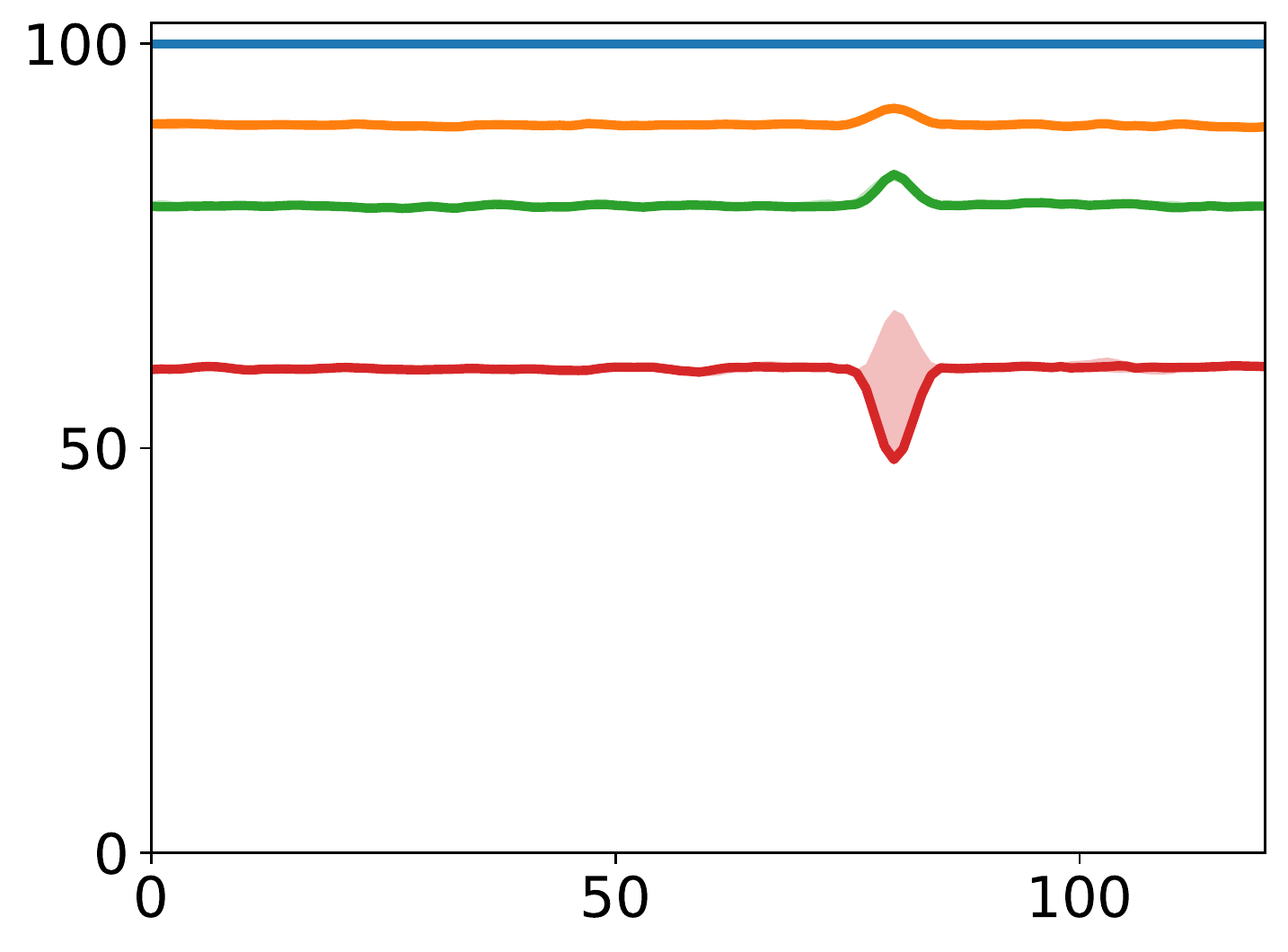} &
\rotatebox{90}{Los Angeles}
\\ [-0.25cm]

&
\includegraphics[width=\fivefig]{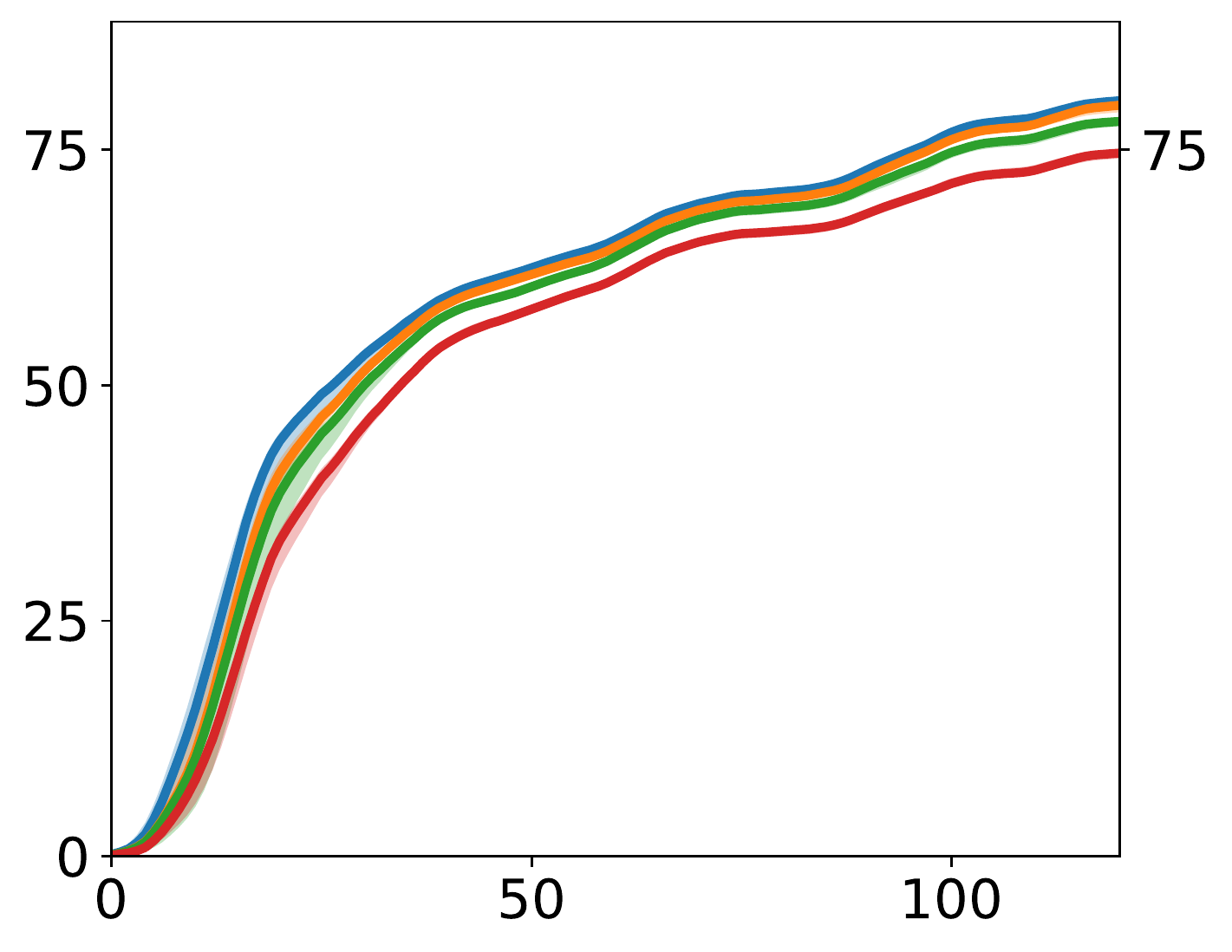} & &
\includegraphics[width=\fivefig]{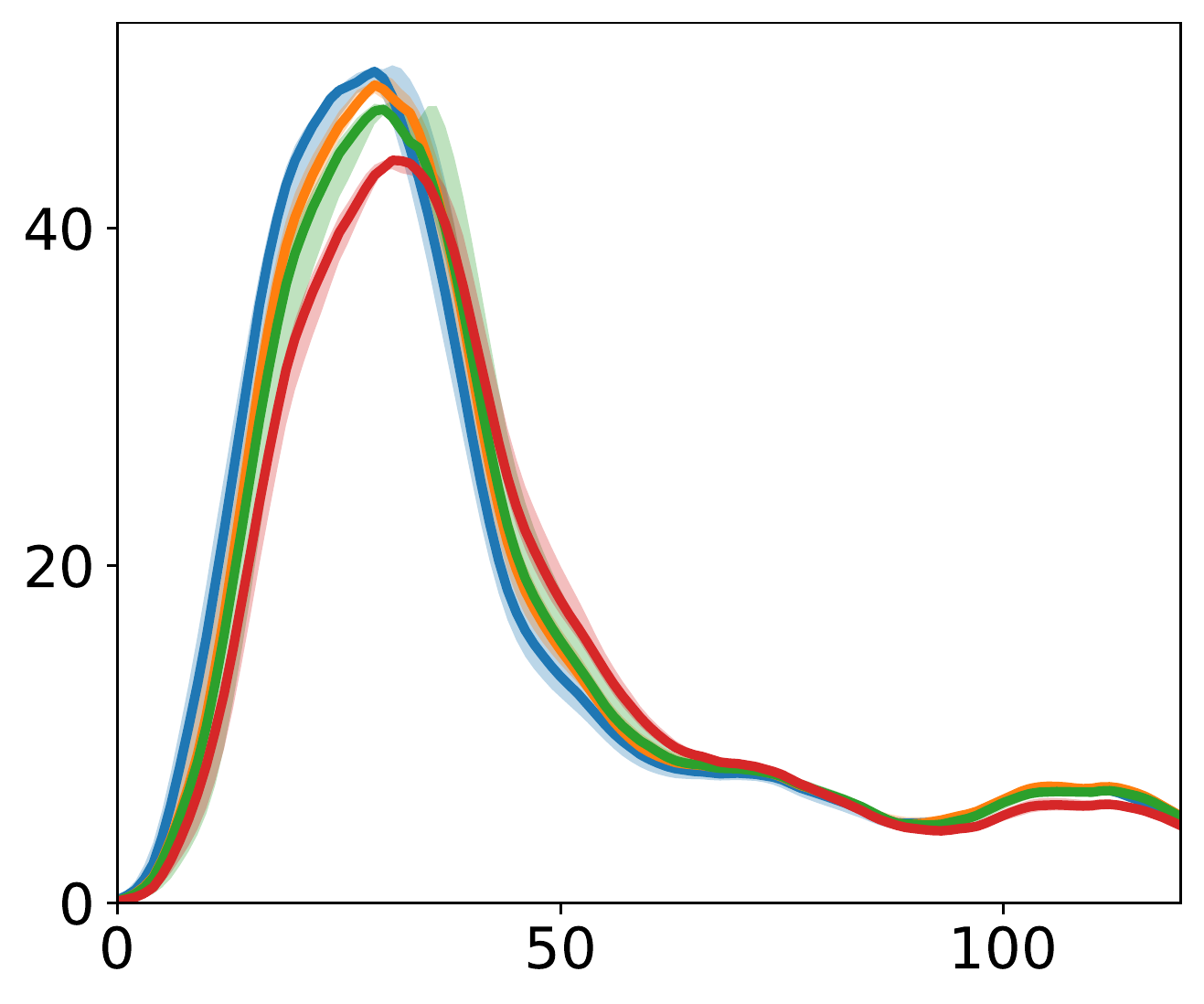}  &&
\includegraphics[width=\fivefig]{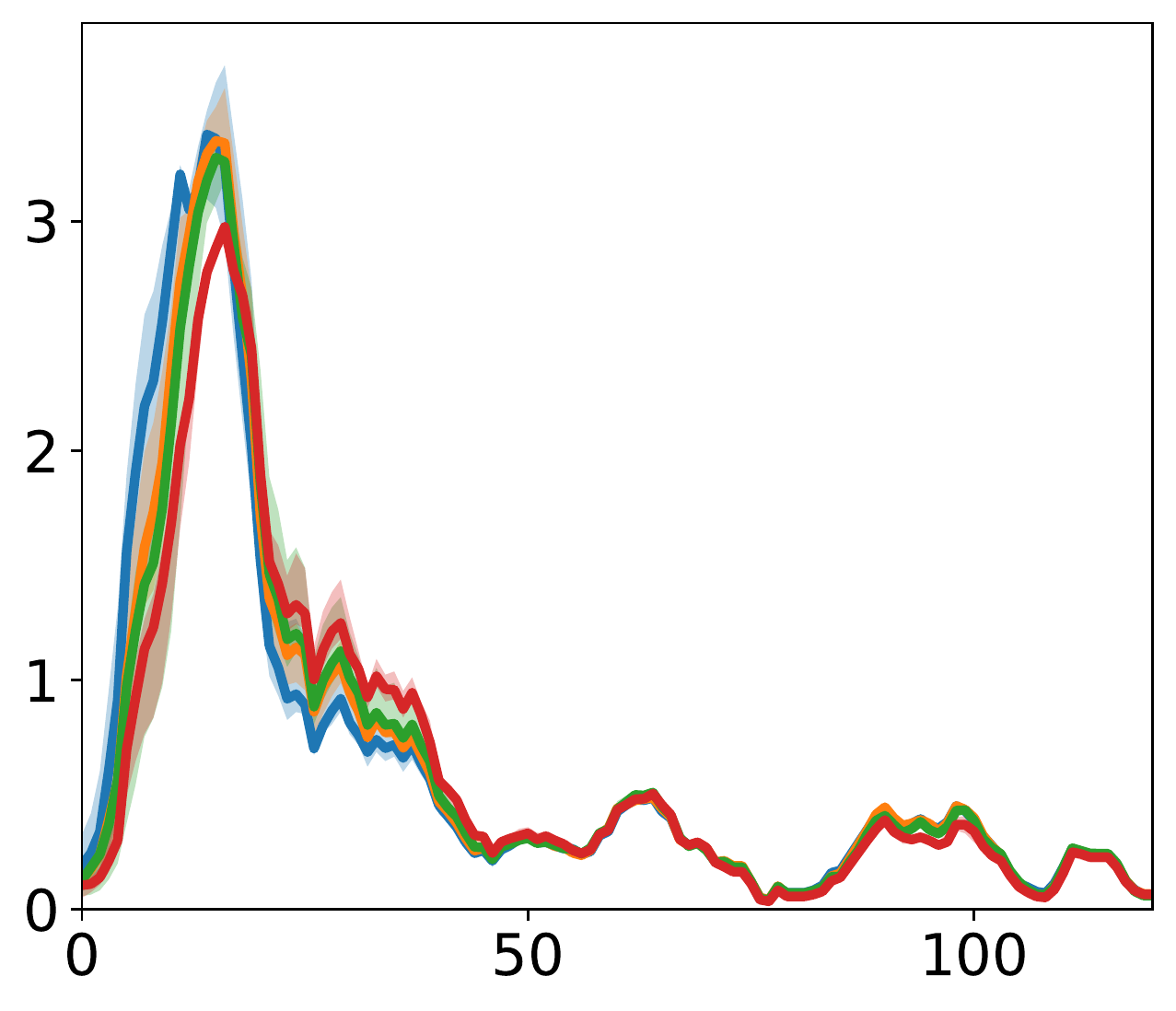}  &&
\includegraphics[width=\fivefig]{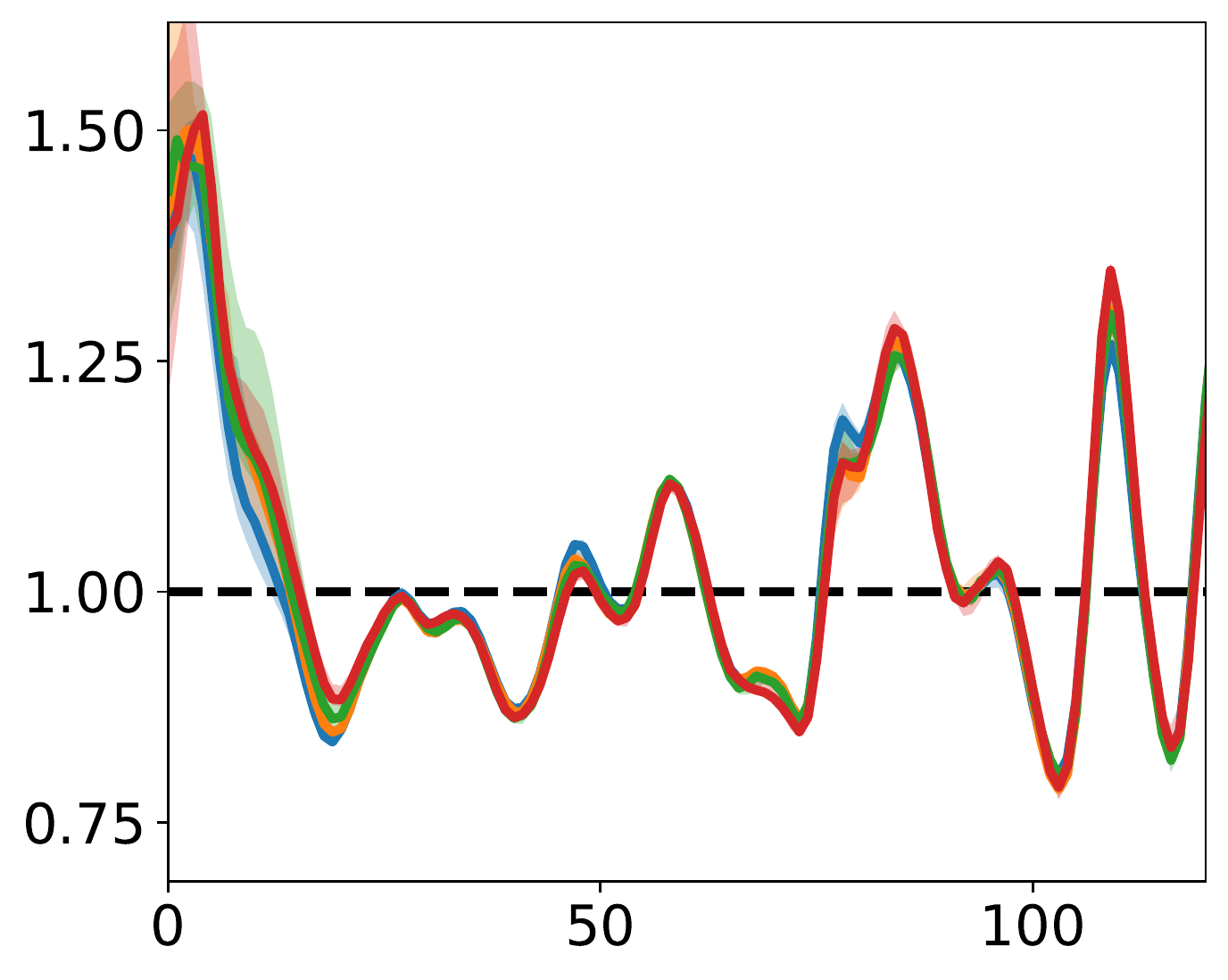} &&
\includegraphics[width=\fivefig]{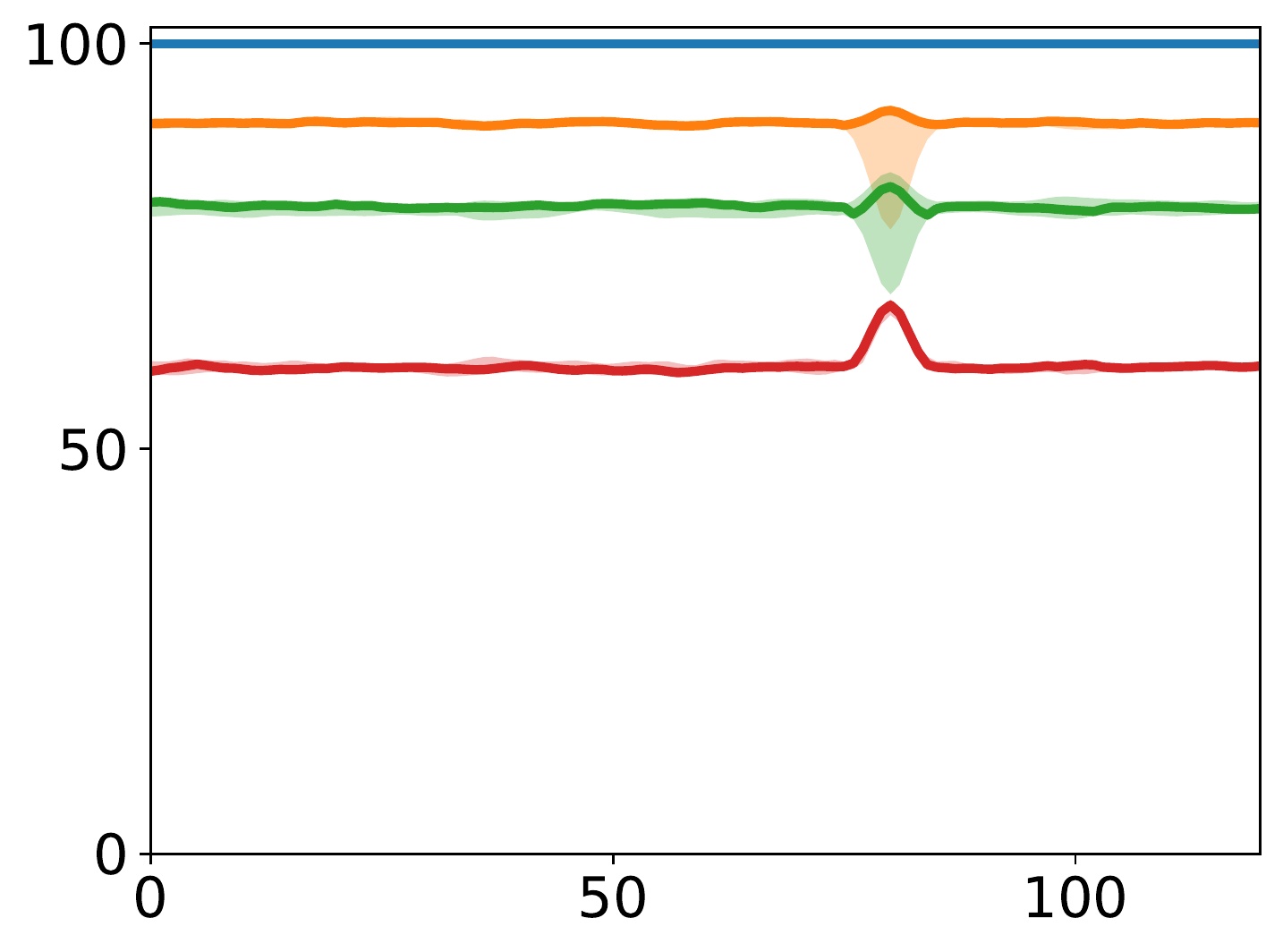} &
\rotatebox{90}{Istanbul}
\\ [-0.25cm]

&
\includegraphics[width=\fivefig]{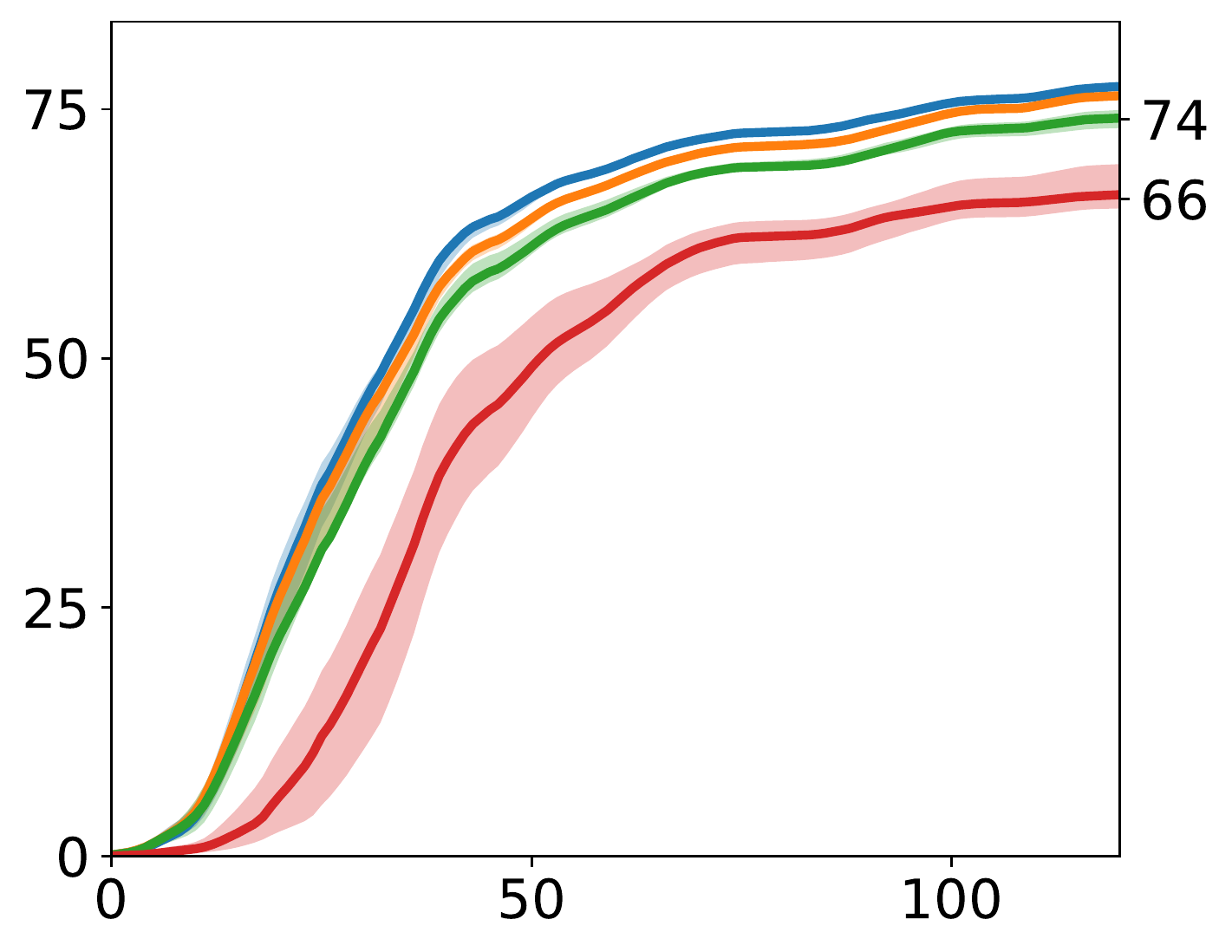} &&
\includegraphics[width=\fivefig]{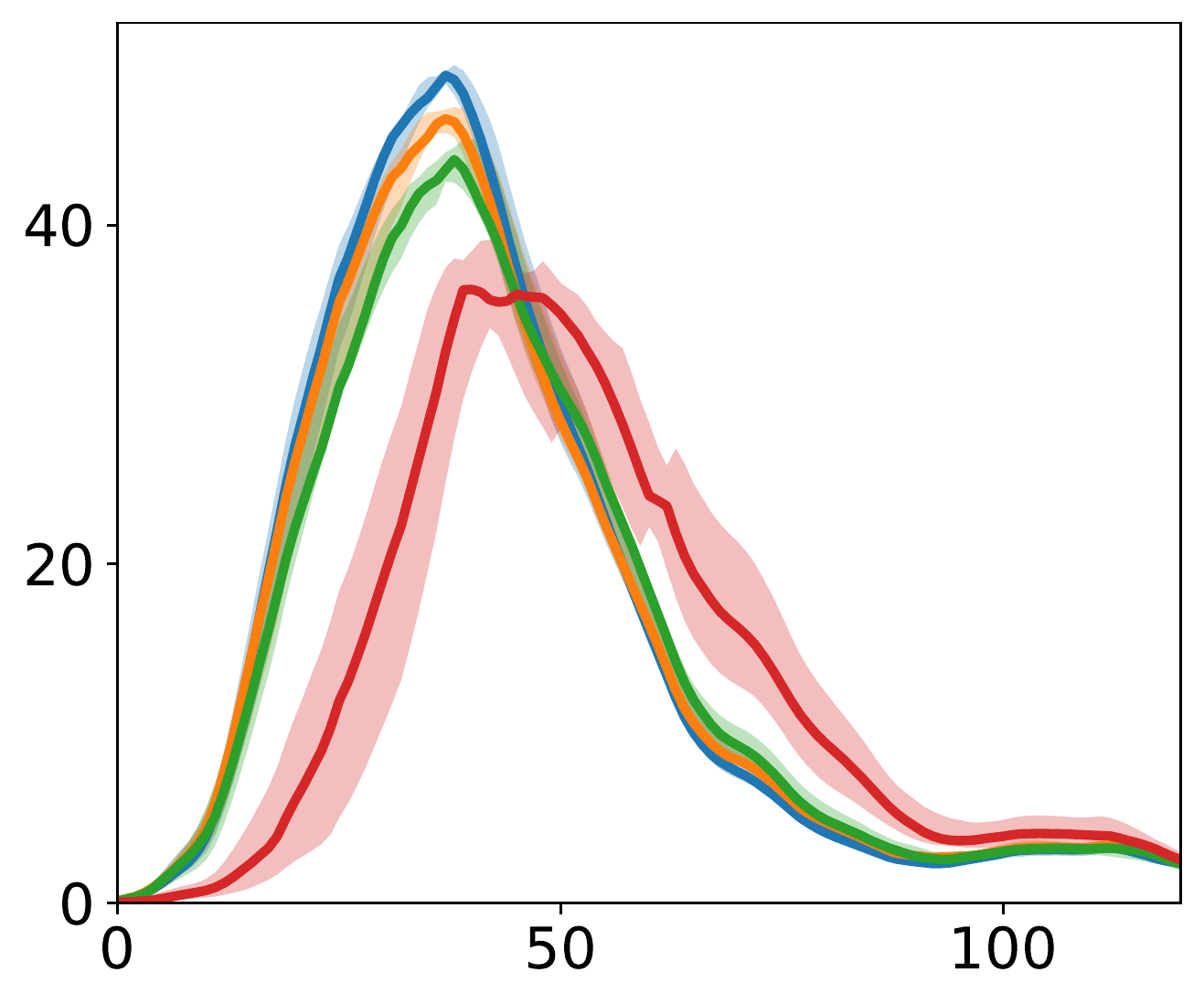}  &&
\includegraphics[width=\fivefig]{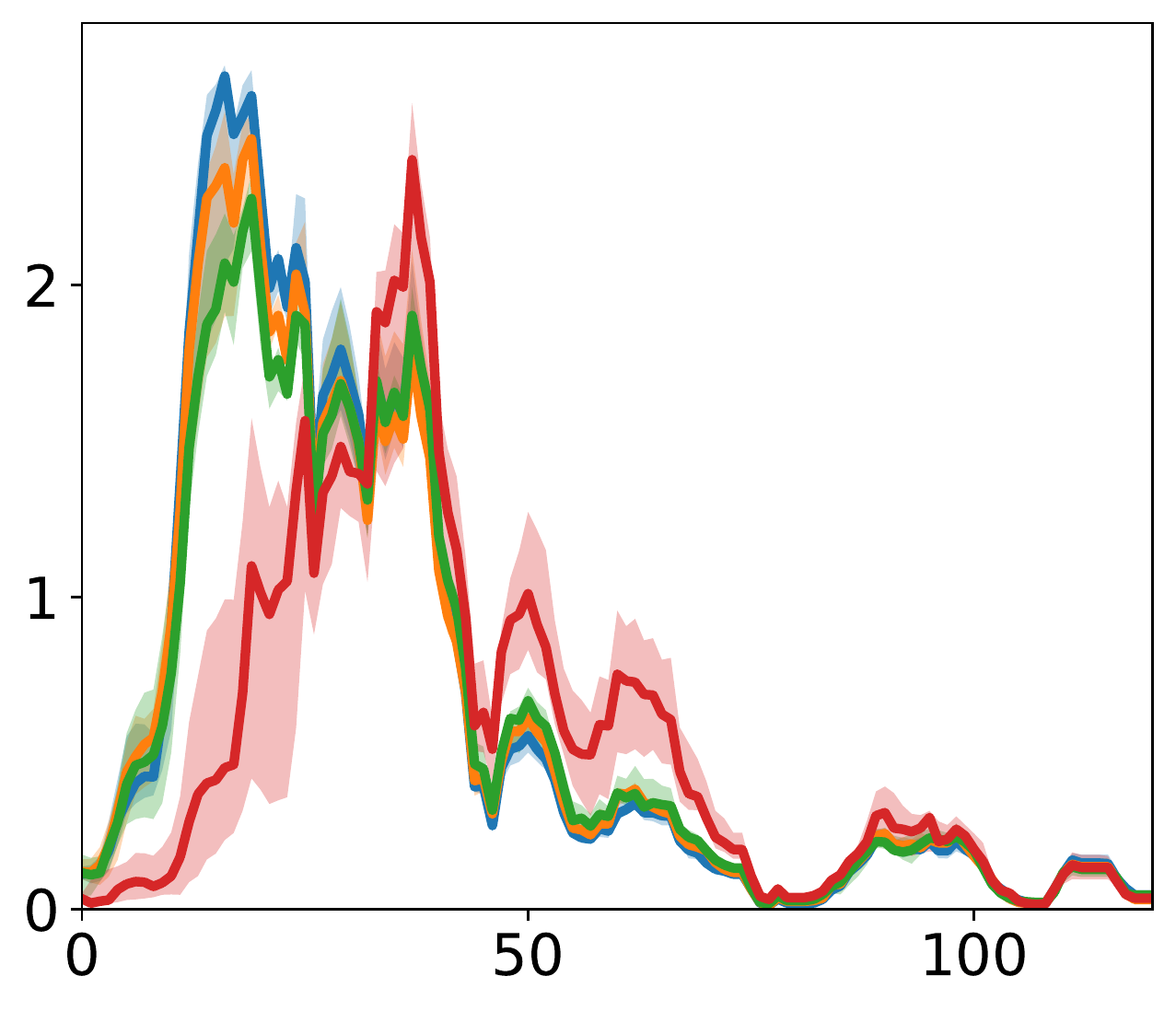}  &&
\includegraphics[width=\fivefig]{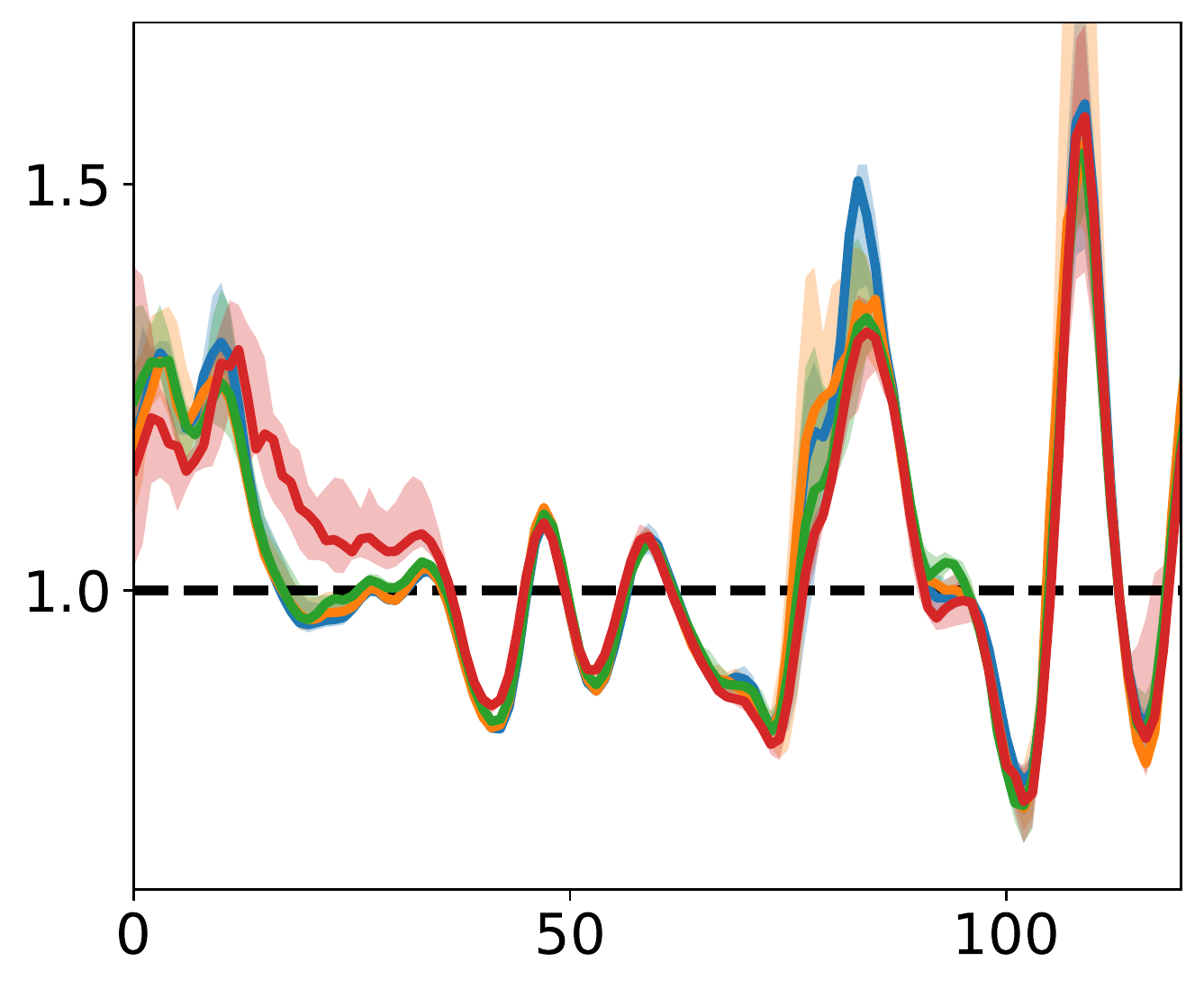} &&
\includegraphics[width=\fivefig]{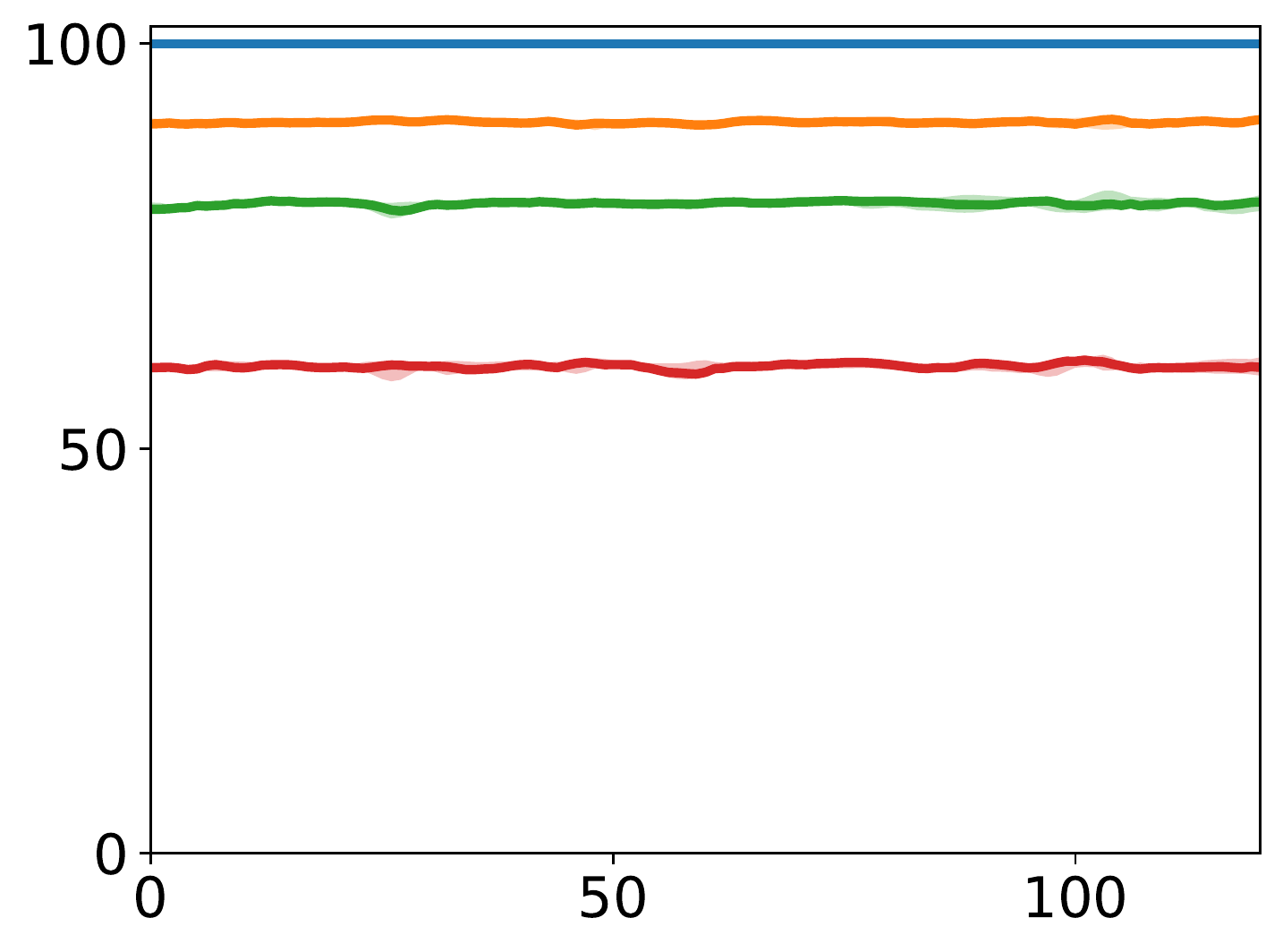} &
\rotatebox{90}{Jakarta}
\\ [-0.25cm]

&
\includegraphics[width=\fivefig]{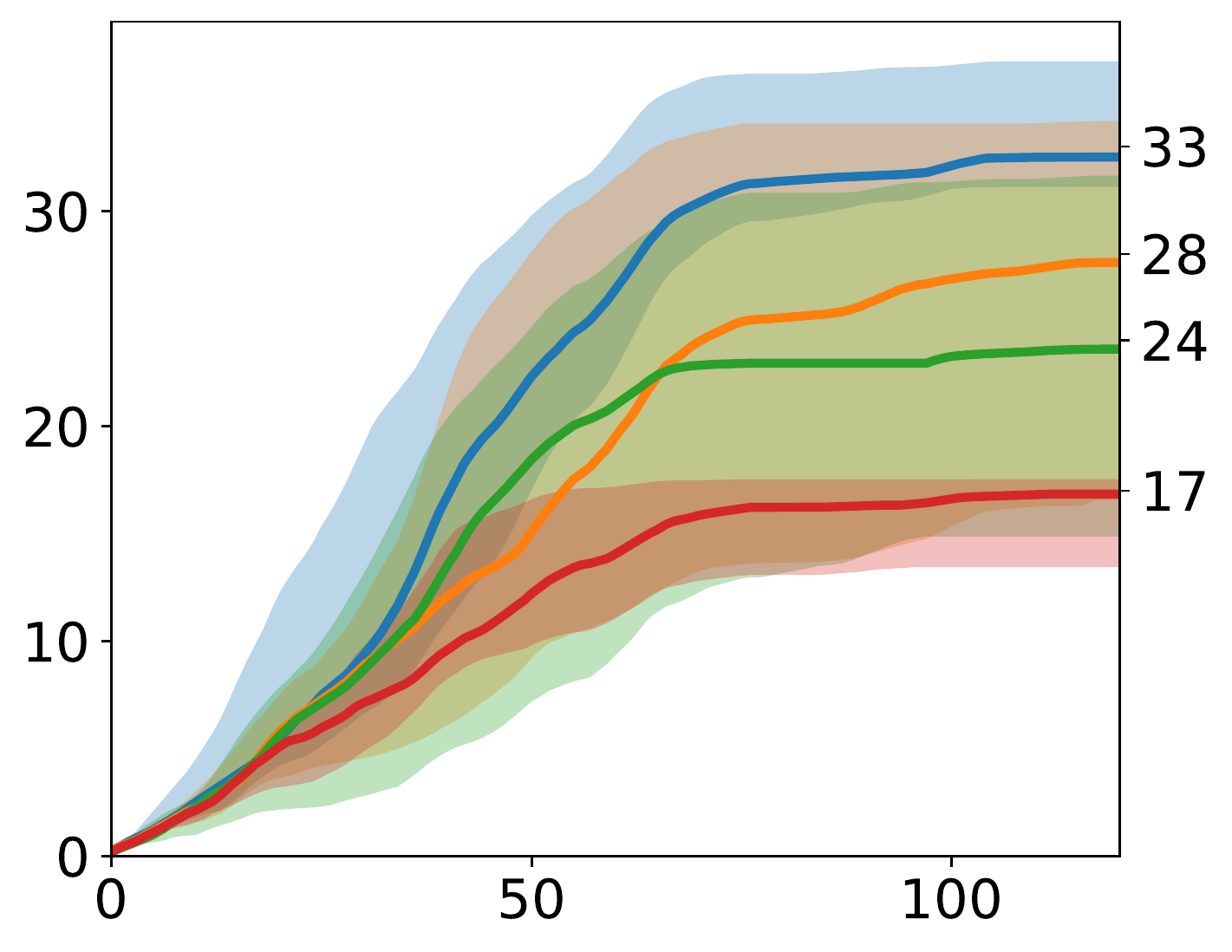} &&
\includegraphics[width=\fivefig]{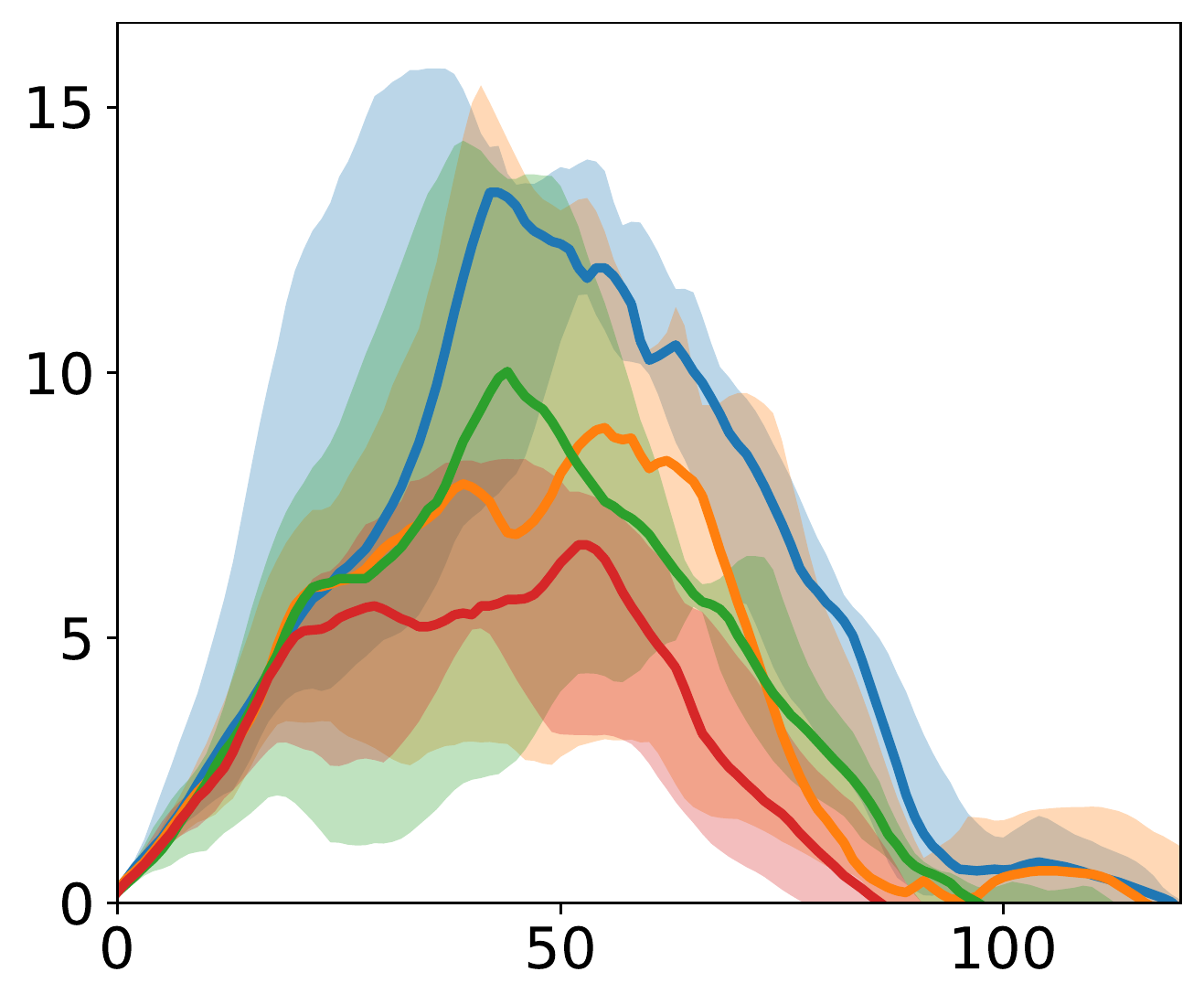}  &&
\includegraphics[width=\fivefig]{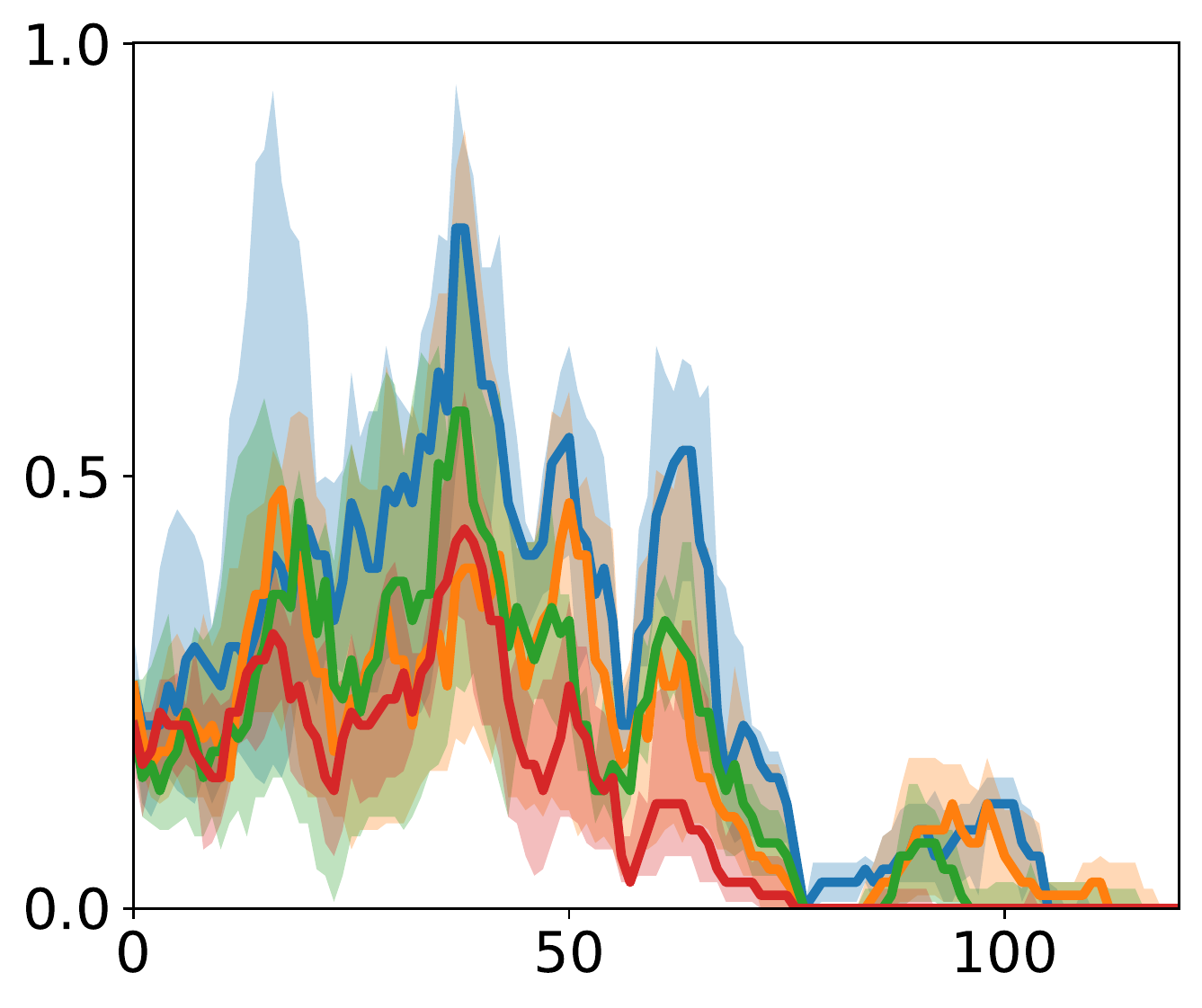}  &&
\includegraphics[width=\fivefig]{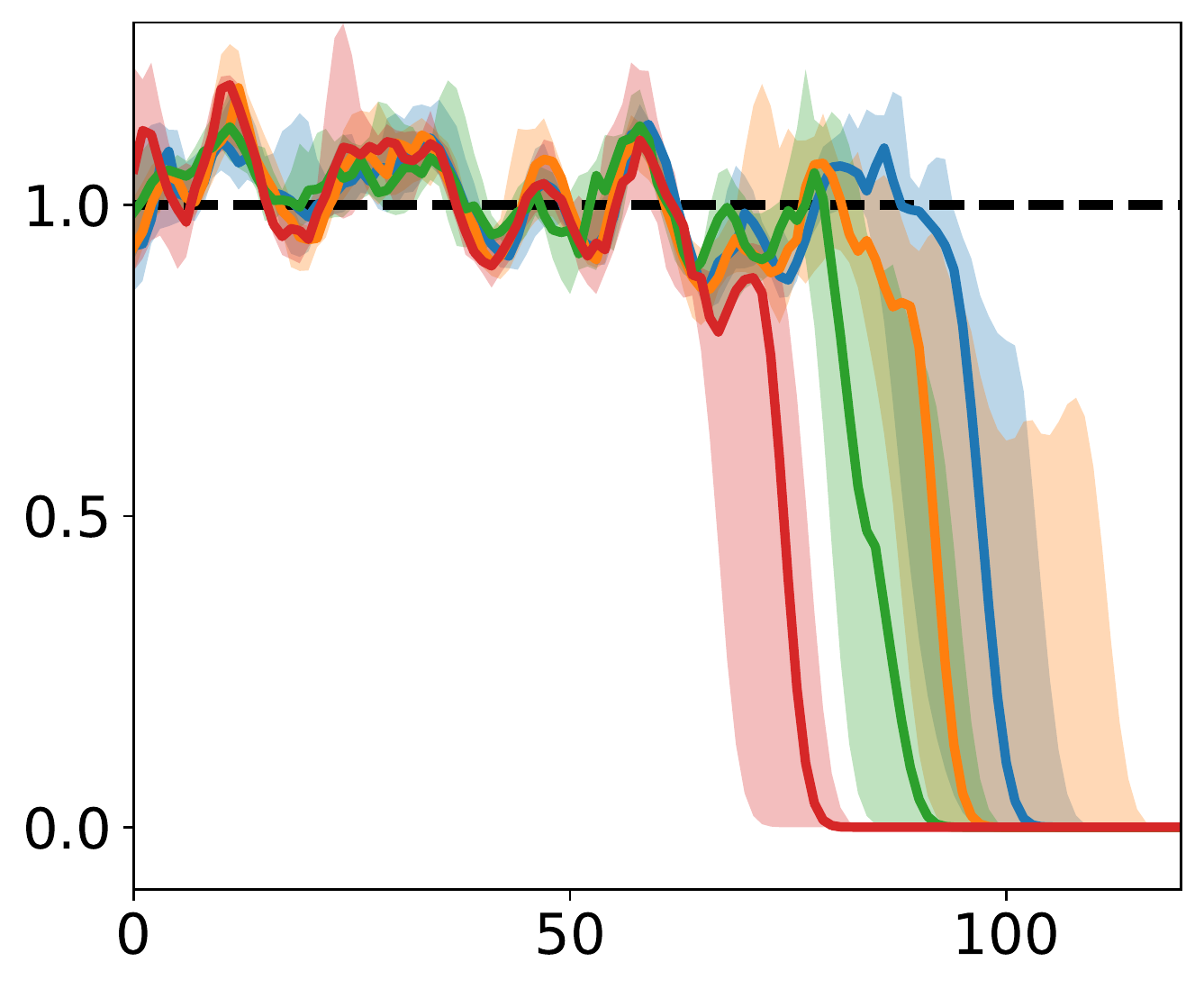} &&
\includegraphics[width=\fivefig]{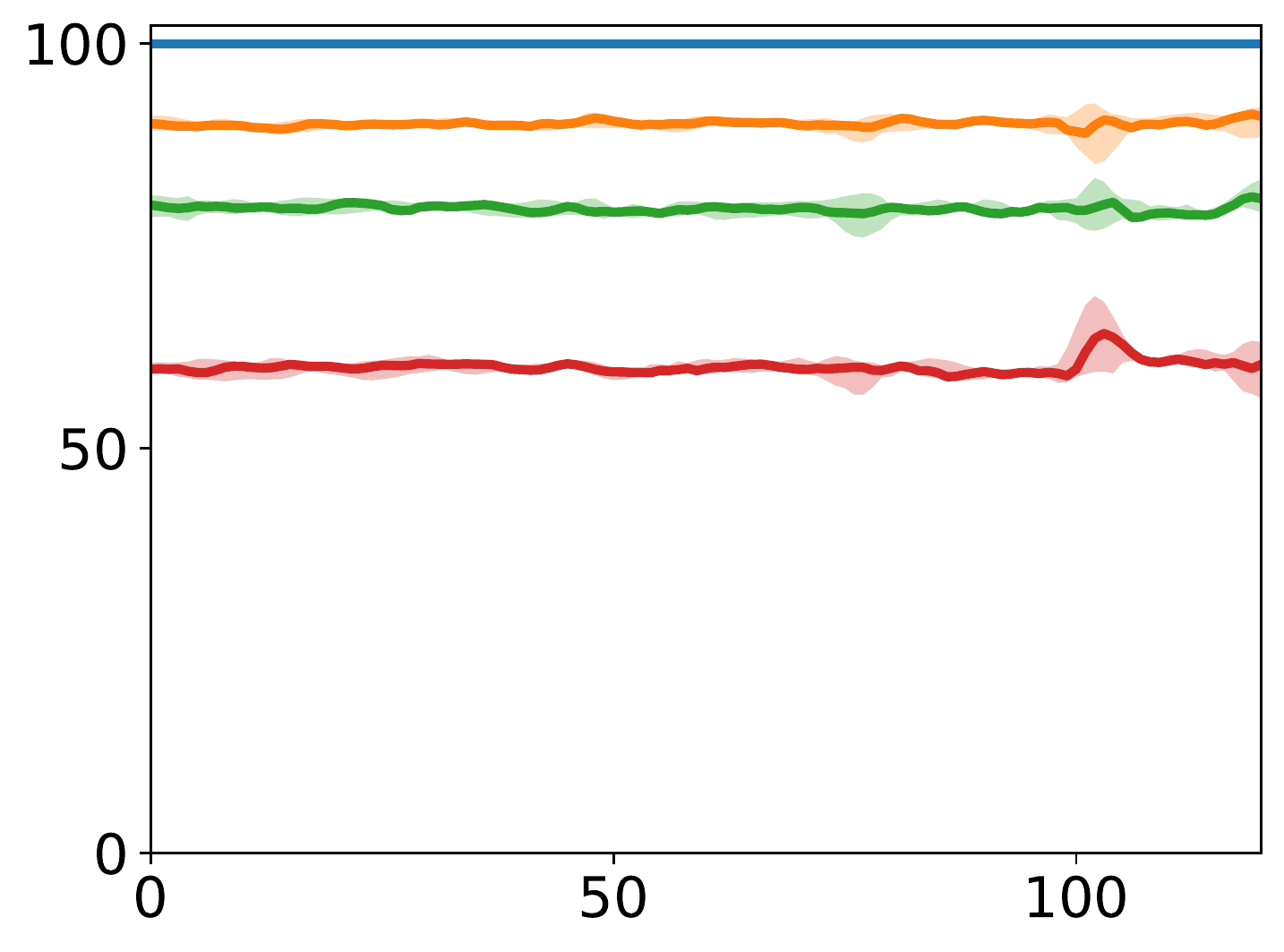} &
\rotatebox{90}{London}
\\ [-0.25cm]

&
\includegraphics[width=\fivefig]{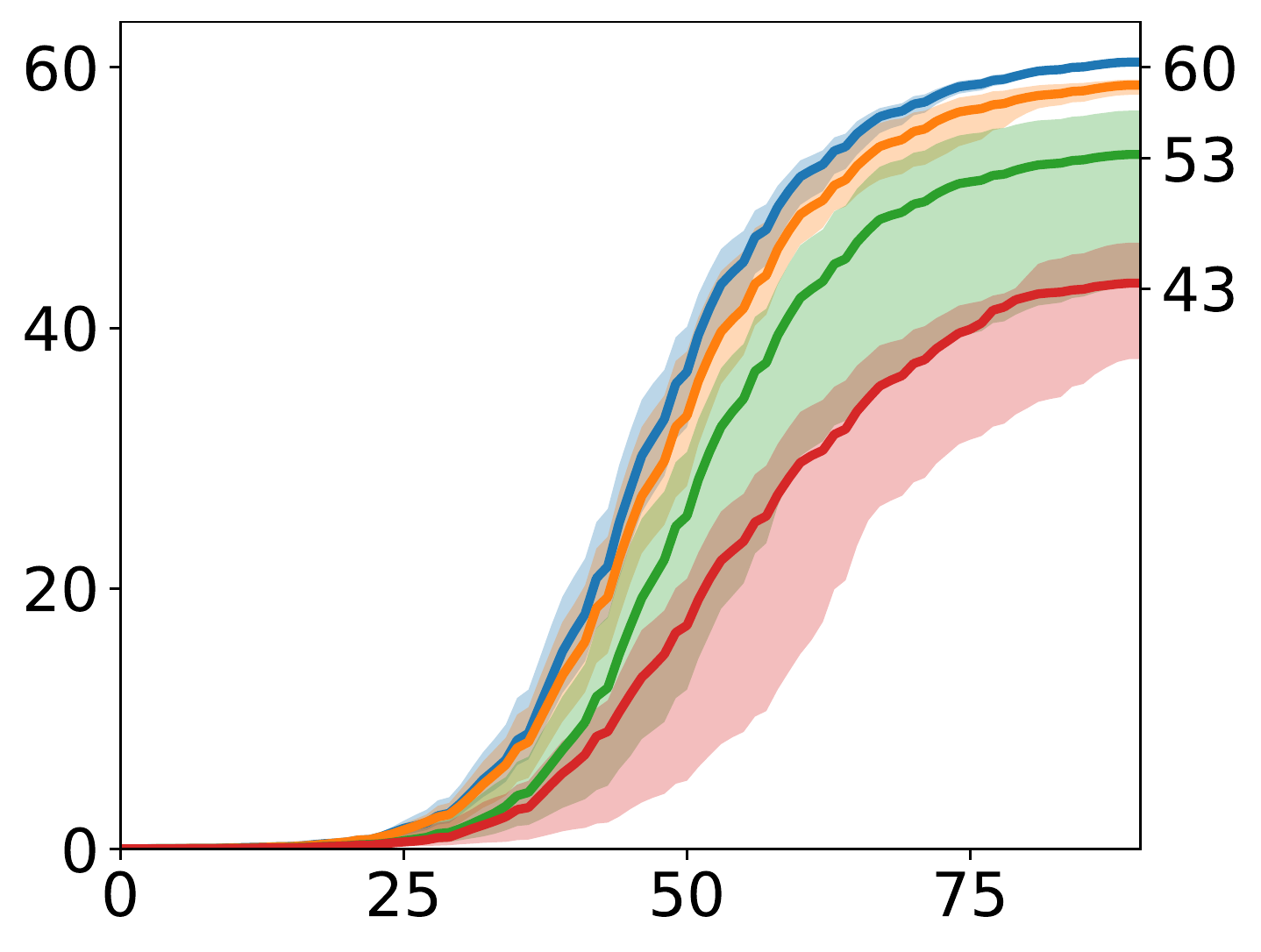} &&
\includegraphics[width=\fivefig]{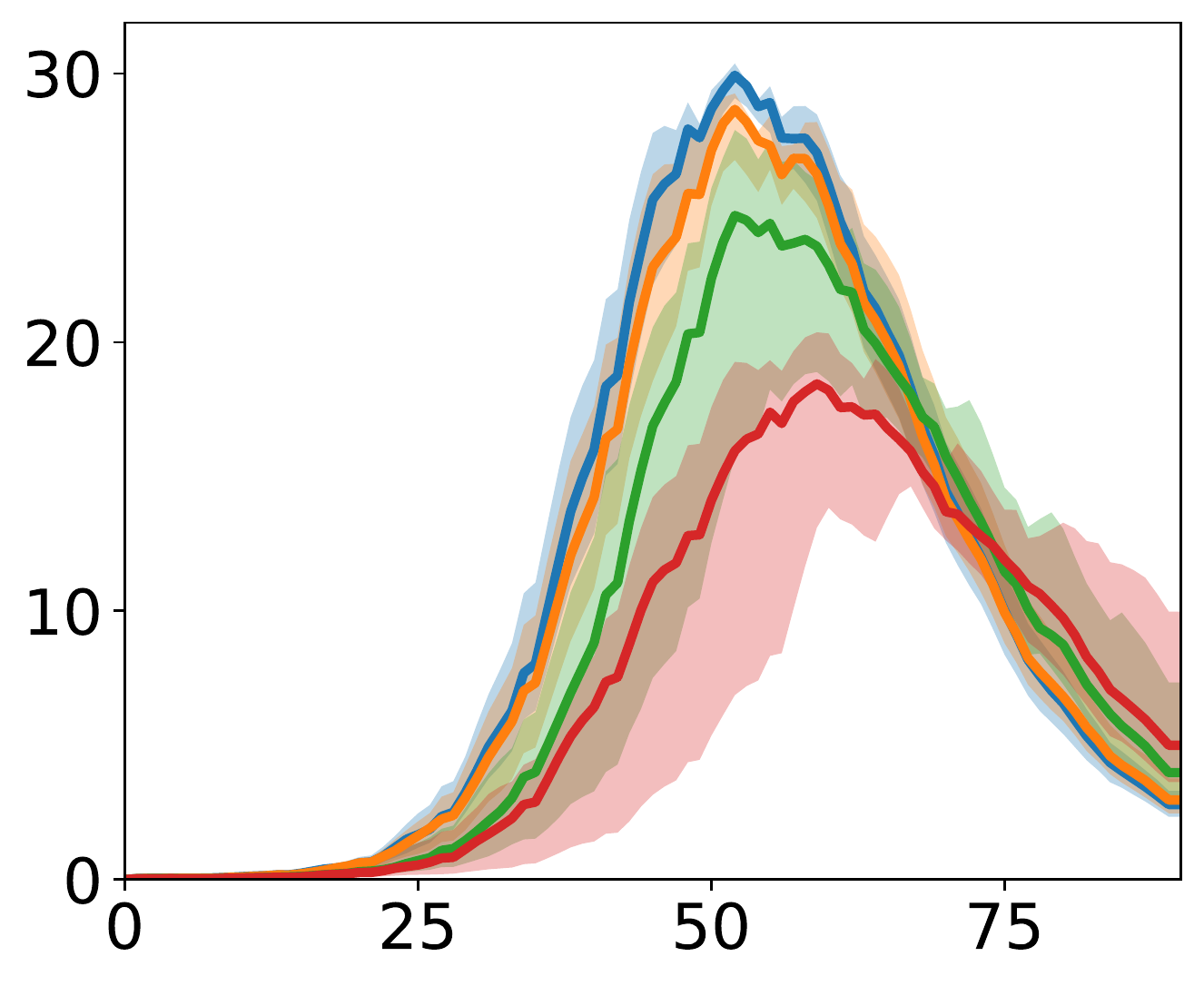}  &&
\includegraphics[width=\fivefig]{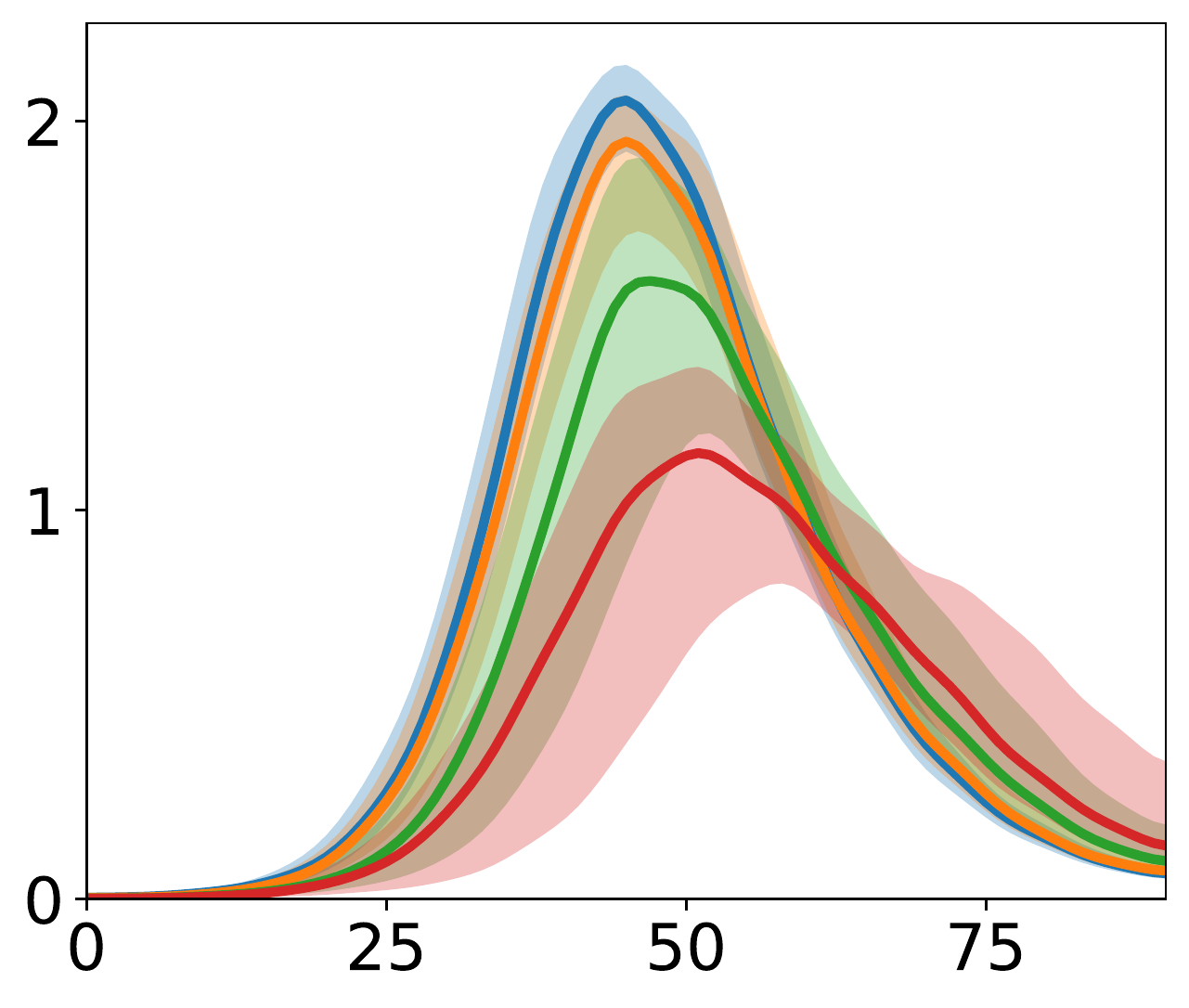}  &&
\includegraphics[width=\fivefig]{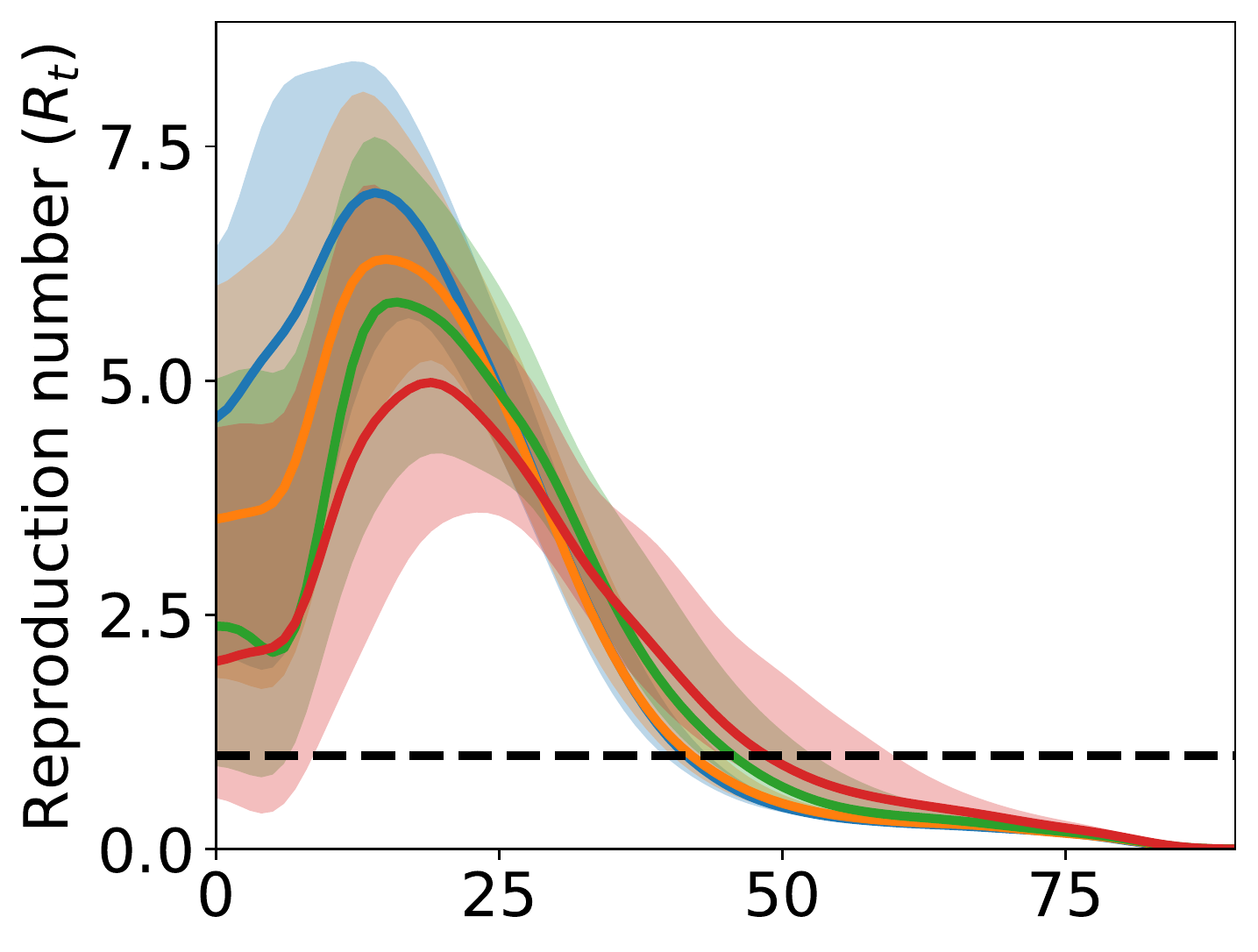} &&
\includegraphics[width=\fivefig]{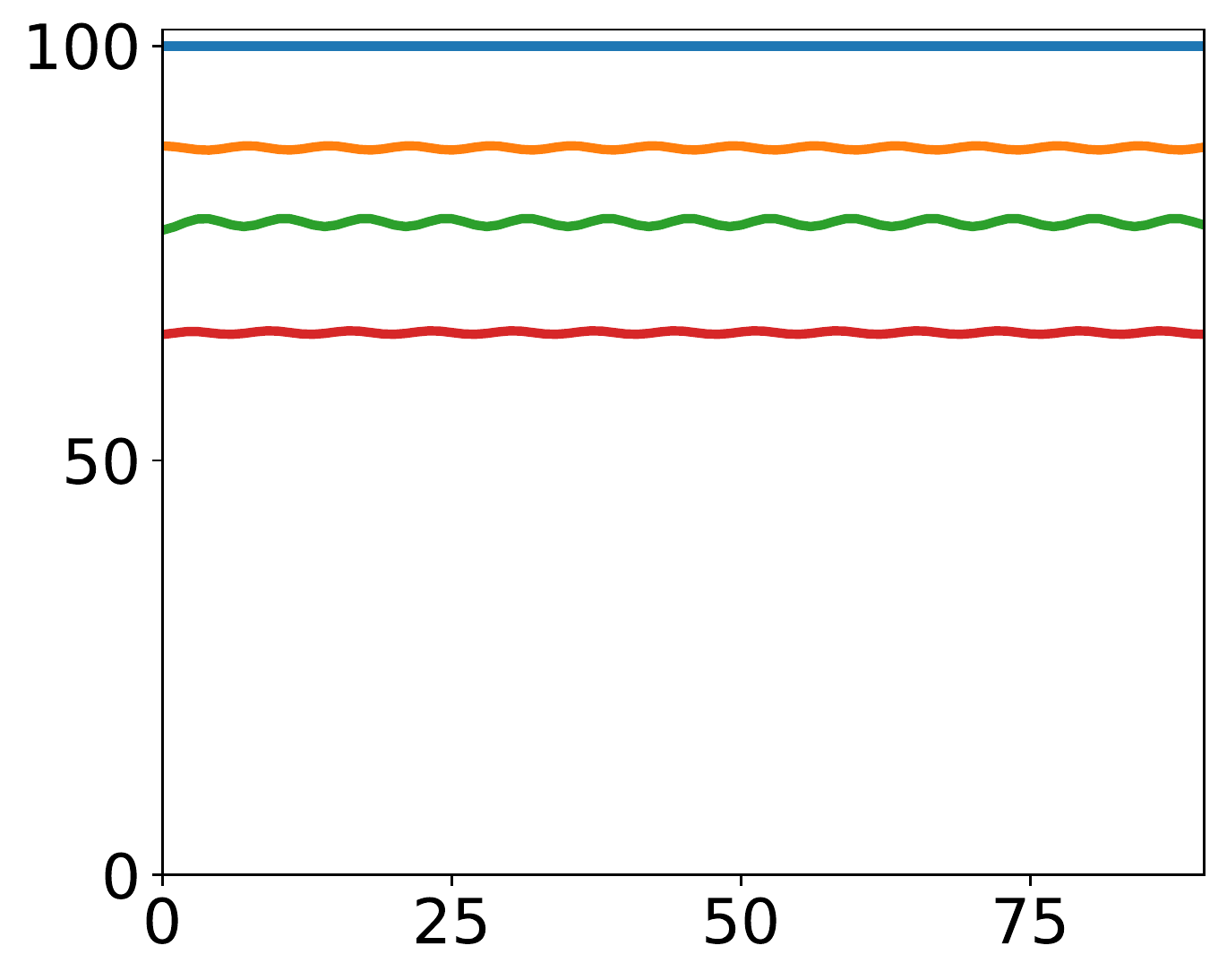} &
\rotatebox{90}{University}
\\ [-0.25cm]

\multicolumn{11}{c}{\includegraphics[width=0.15\textwidth]{Fig/labels/daysfromstart.pdf}}\\
\multicolumn{11}{c}{\includegraphics[width=0.6\textwidth]{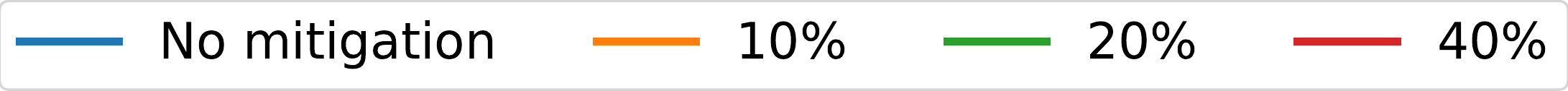}}\\

\end{tabular}
\caption{}
\label{figS:close_random_venue}
\end{figure}
\newpage
\noindent \textbf{Figure~\ref{figS:close_random_venue}:} Infection spreading with the intervention strategy of closing random venues. Compared with closing popular venues, this strategy does not have a strong intervention effect. In some cities like Istanbul, when $40\%$ venues are closed. the total number of infected people is reduced a little, less than $10\%$, but the social value is reduced by about $40\%$.
\clearpage

\newpage
\newcommand\fivesmallfig{0.16\textwidth}
\begin{figure}[H]
    \centering
\begin{tabular}{m{0.1cm}m{\fivesmallfig}@{}m{0.1cm}m{\fivesmallfig}@{}m{0.1cm}m{\fivesmallfig}@{}m{0.1cm}m{\fivesmallfig}@{}m{0.1cm}m{\fivesmallfig}@{}m{0.1cm}}

\multirow{14}{*}{\includegraphics[height=4cm]{Fig/labels/TotalInfected.pdf}}&
\includegraphics[width=\fivesmallfig]{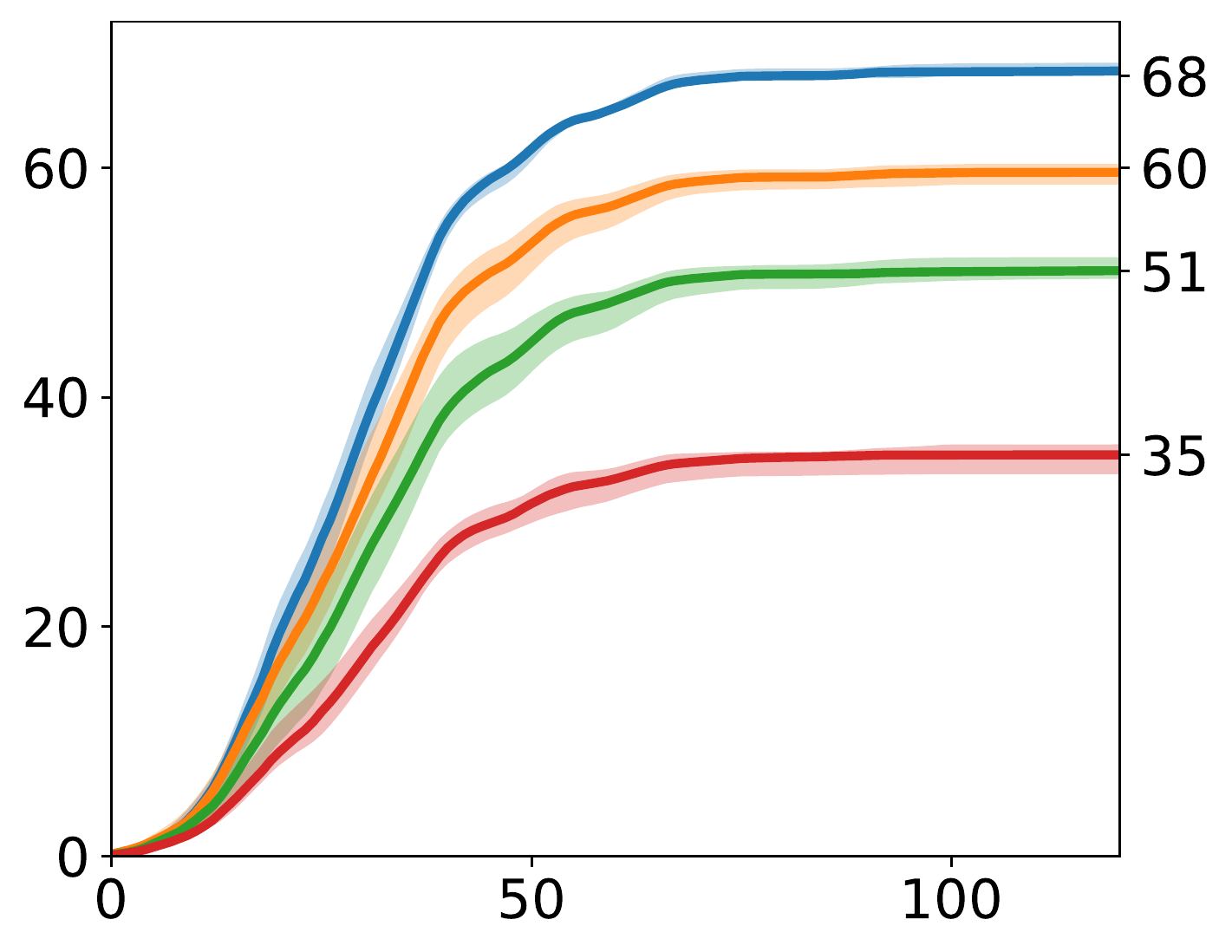} &
\multirow{14}{*}{\includegraphics[height=4cm]{Fig/labels/ActiveInfected.pdf}}&
\includegraphics[width=\fivesmallfig]{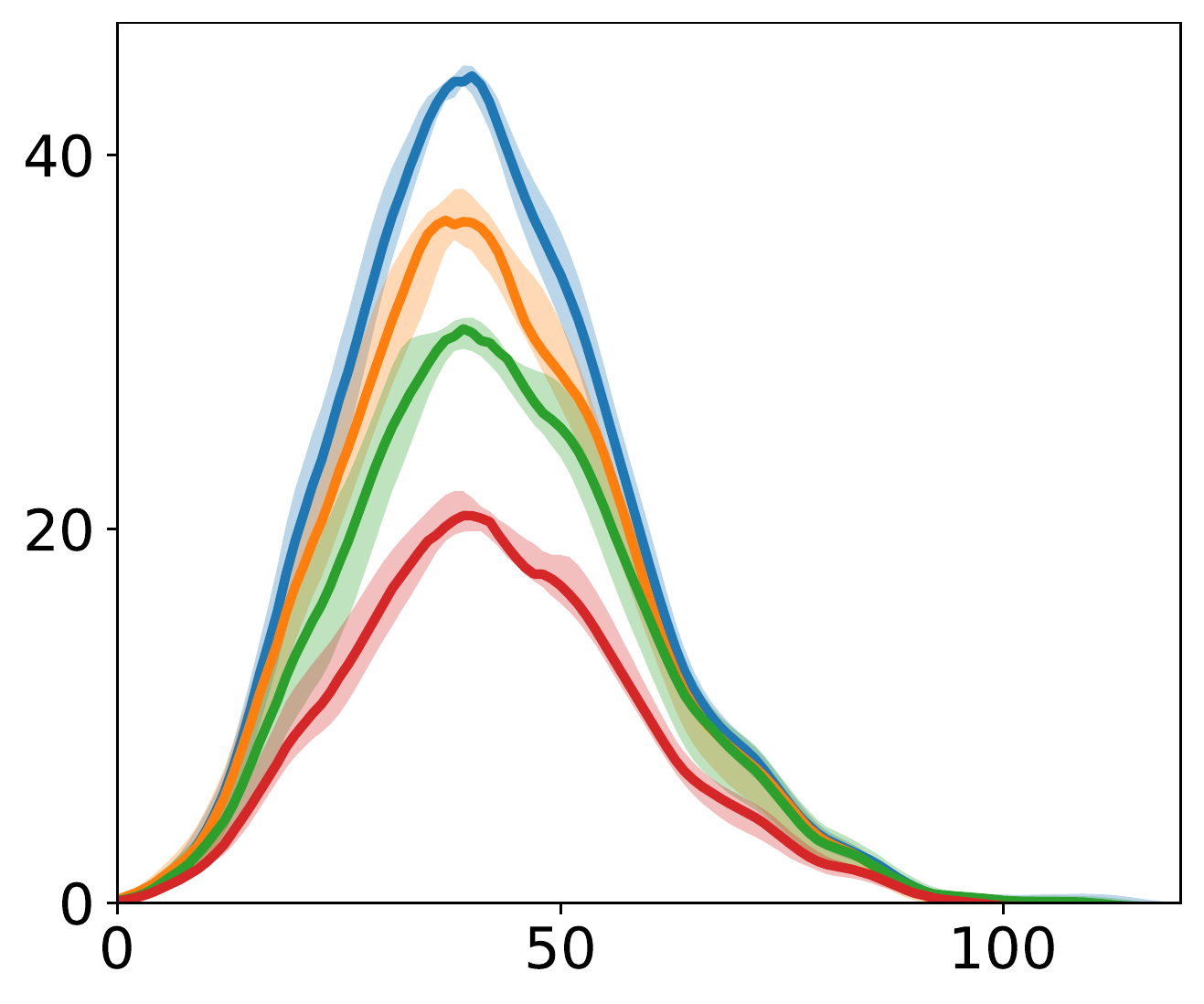}  &
\multirow{14}{*}{\includegraphics[height=4cm]{Fig/labels/newInfected.pdf}}&
\includegraphics[width=\fivesmallfig]{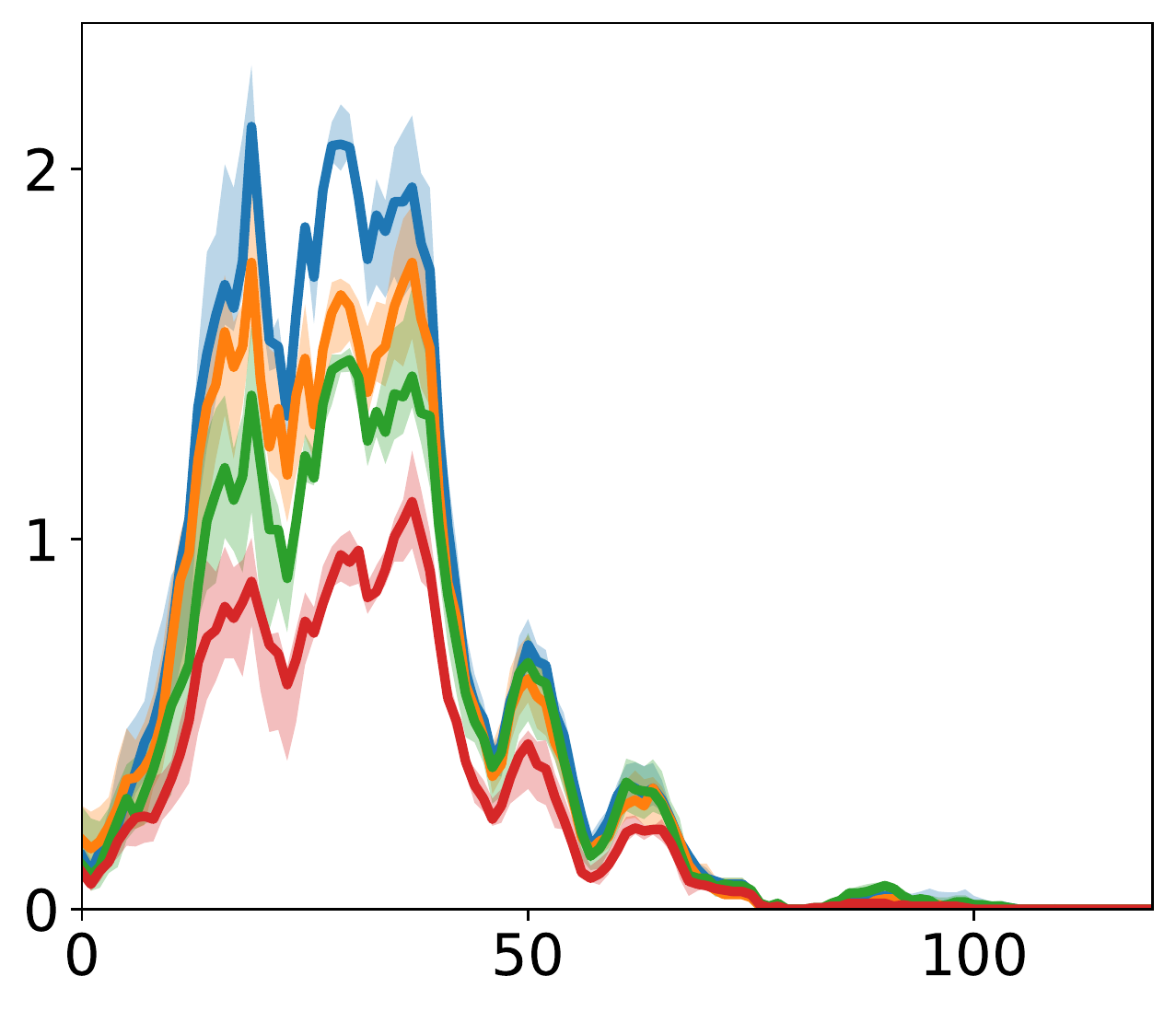}  &
\multirow{14}{*}{\includegraphics[height=2cm]{Fig/labels/growthrate.pdf}}&
\includegraphics[width=\fivesmallfig]{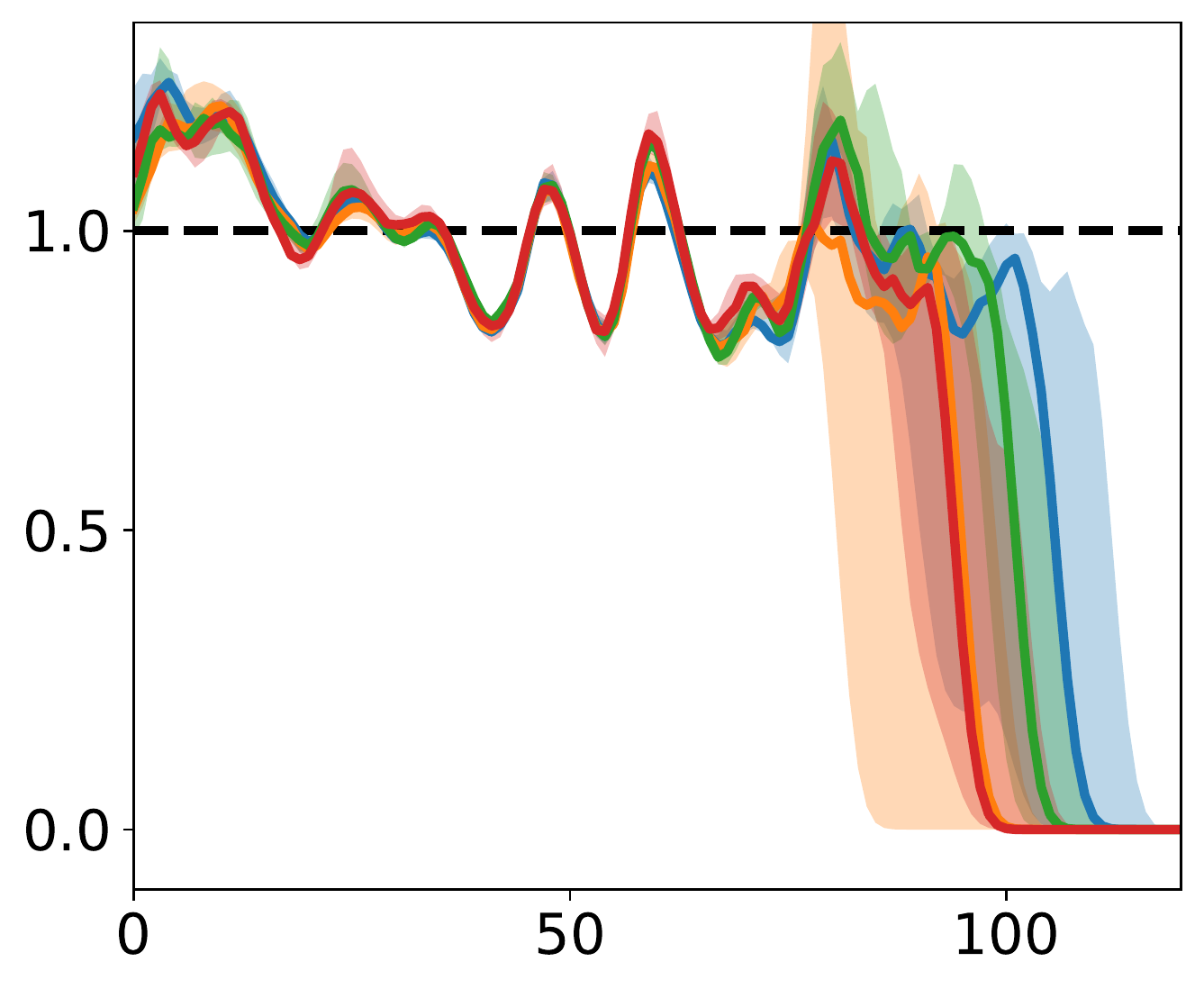} & &
\includegraphics[width=\fivesmallfig]{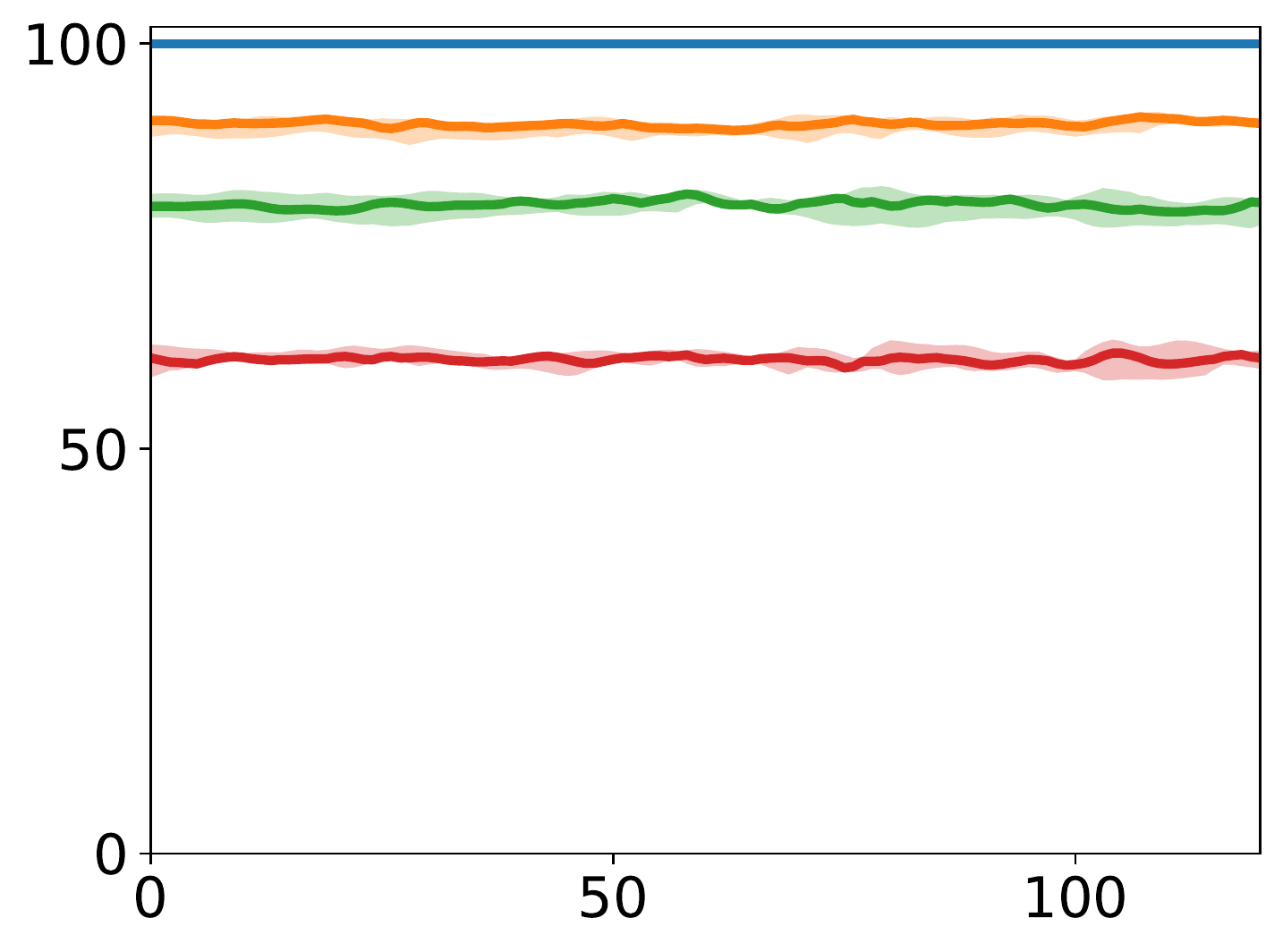} & 
\rotatebox{90}{NYC}
\\ [-0.25cm]
&
\includegraphics[width=\fivesmallfig]{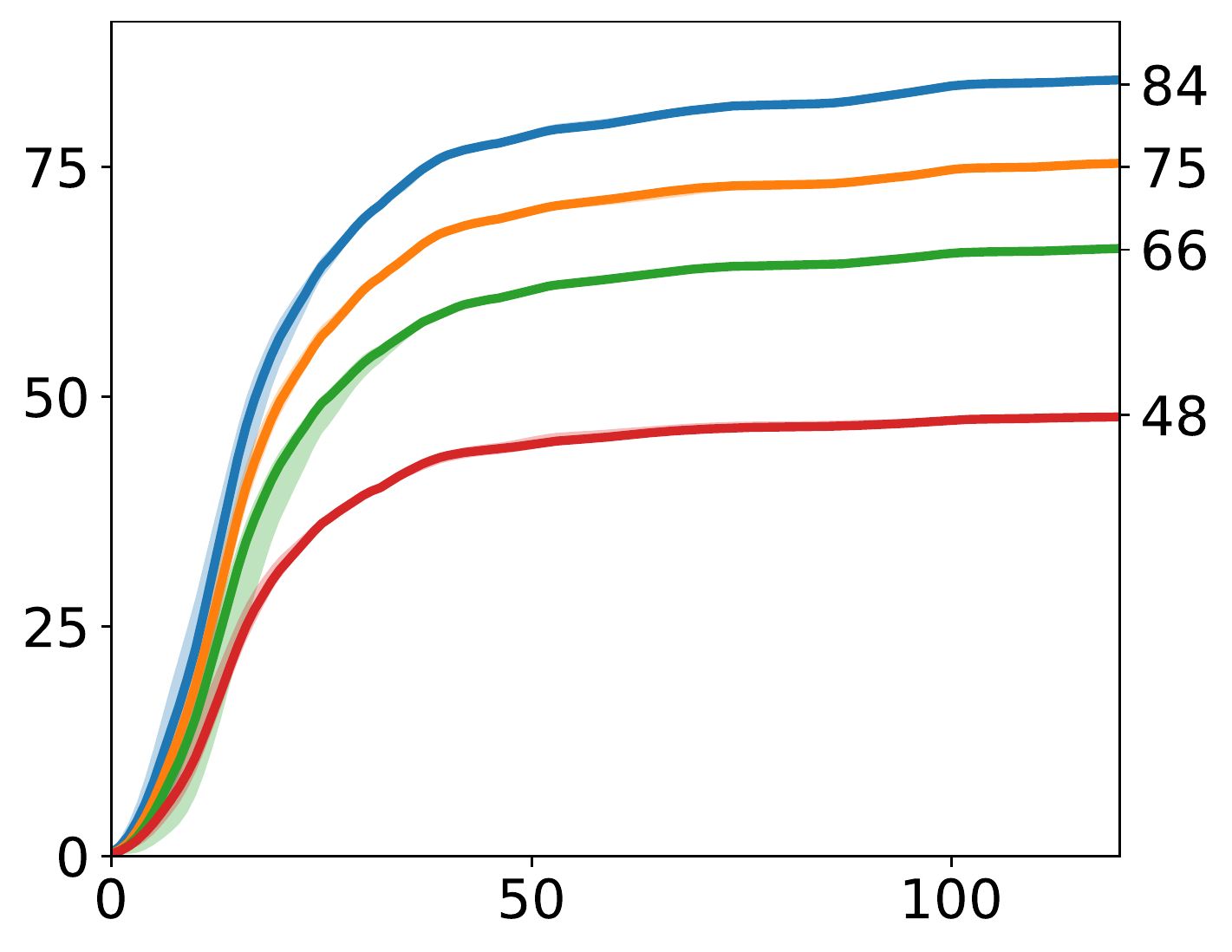} & &
\includegraphics[width=\fivesmallfig]{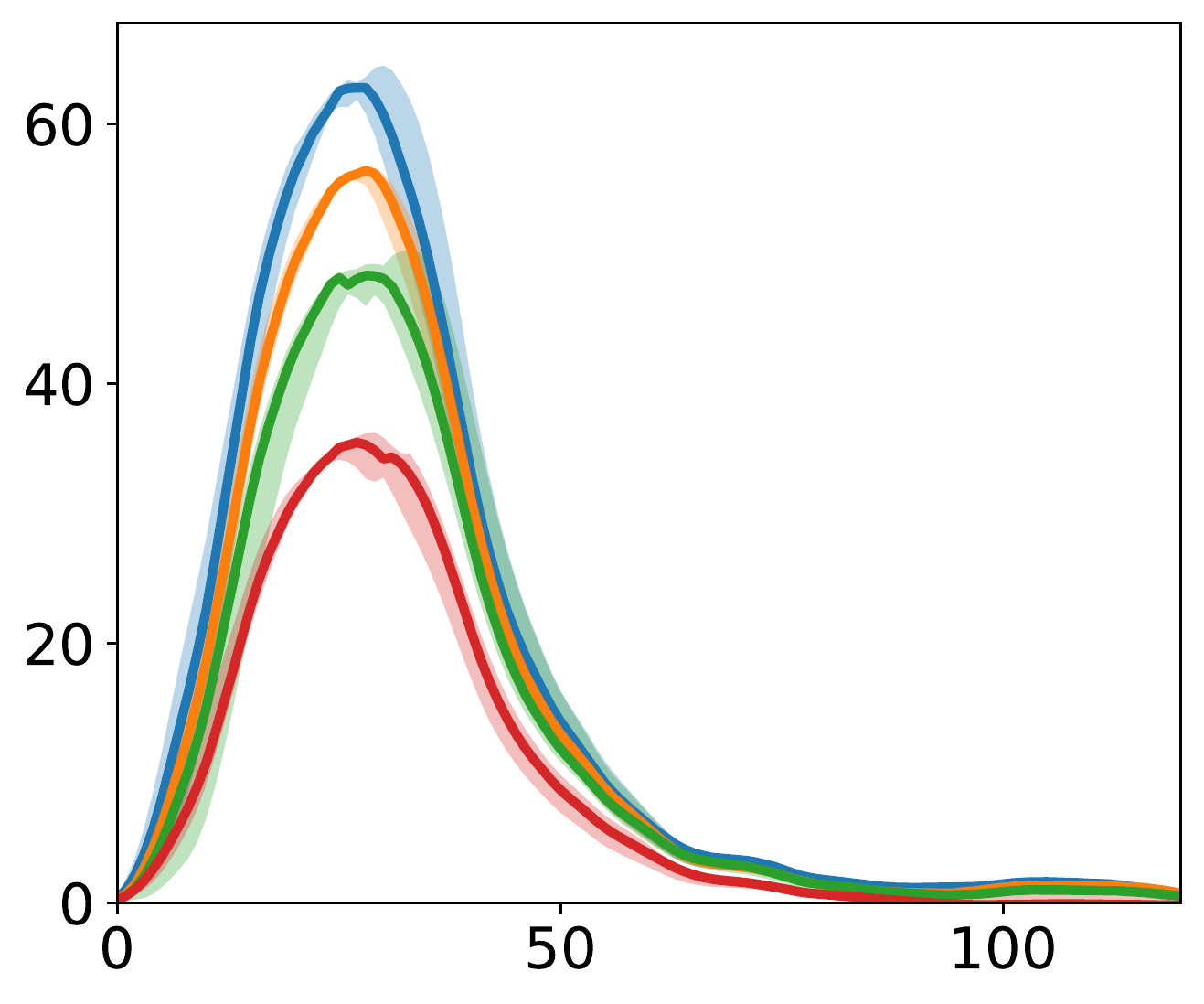}  & &
\includegraphics[width=\fivesmallfig]{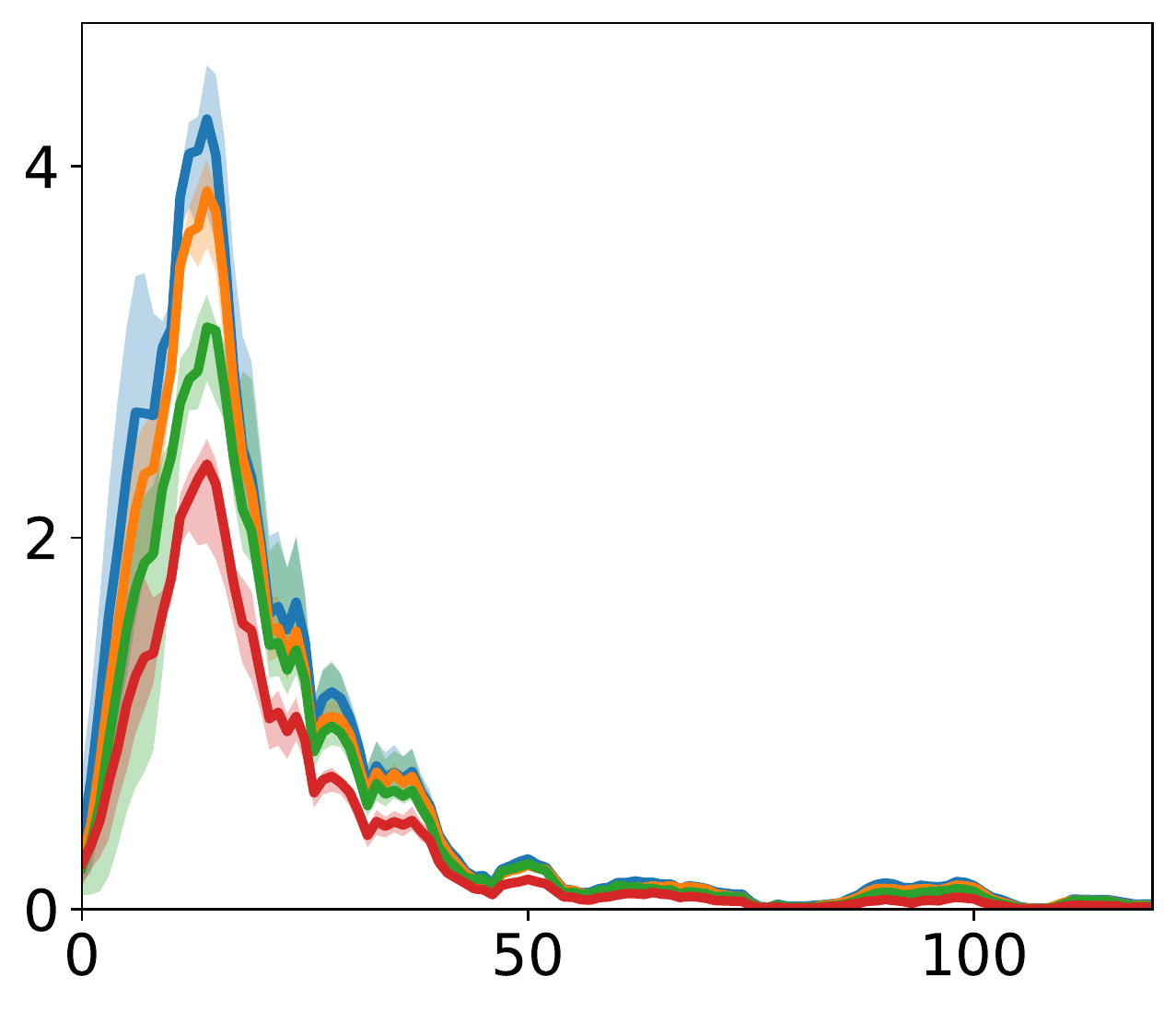}  & &
\includegraphics[width=\fivesmallfig]{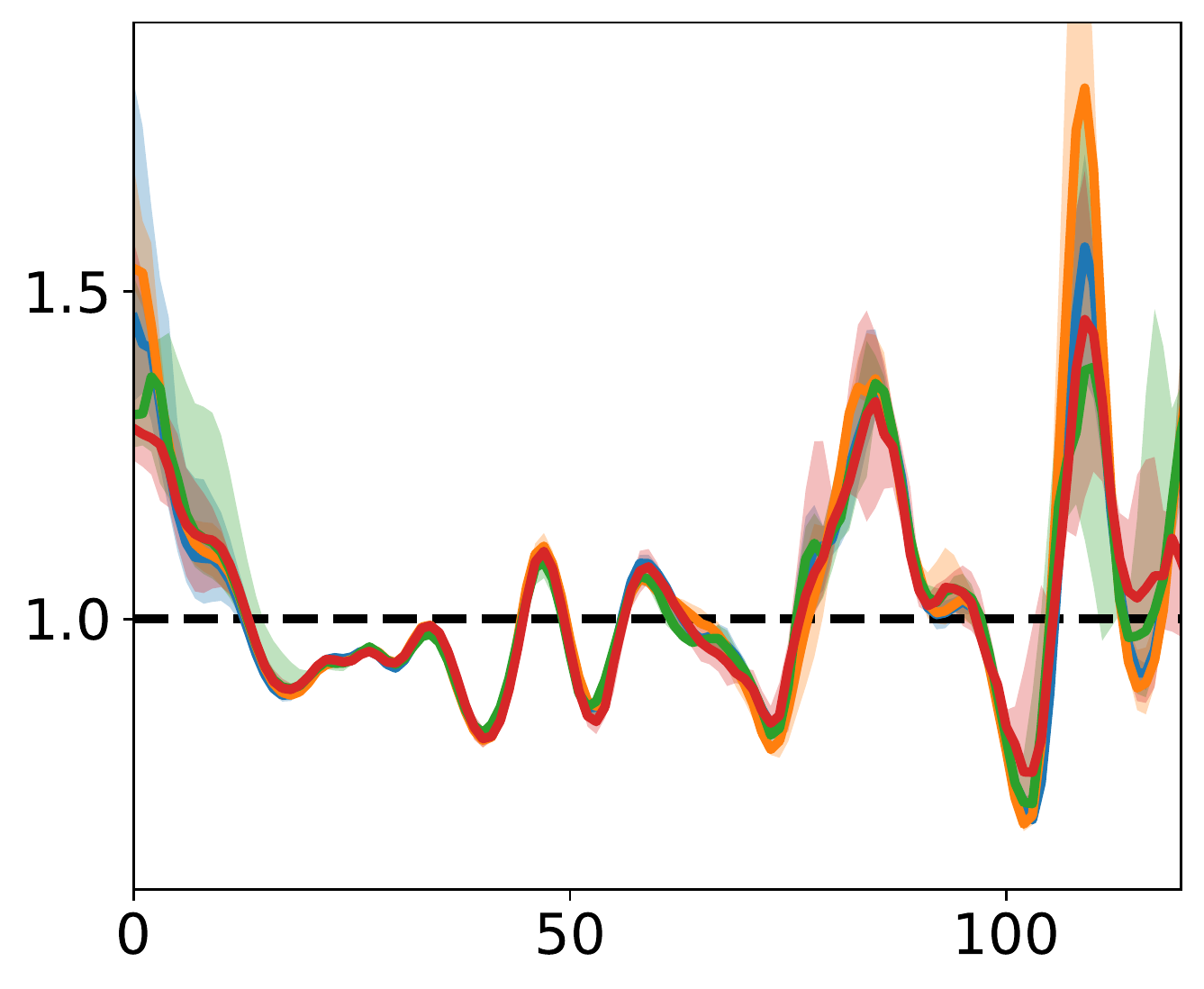} & &
\includegraphics[width=\fivesmallfig]{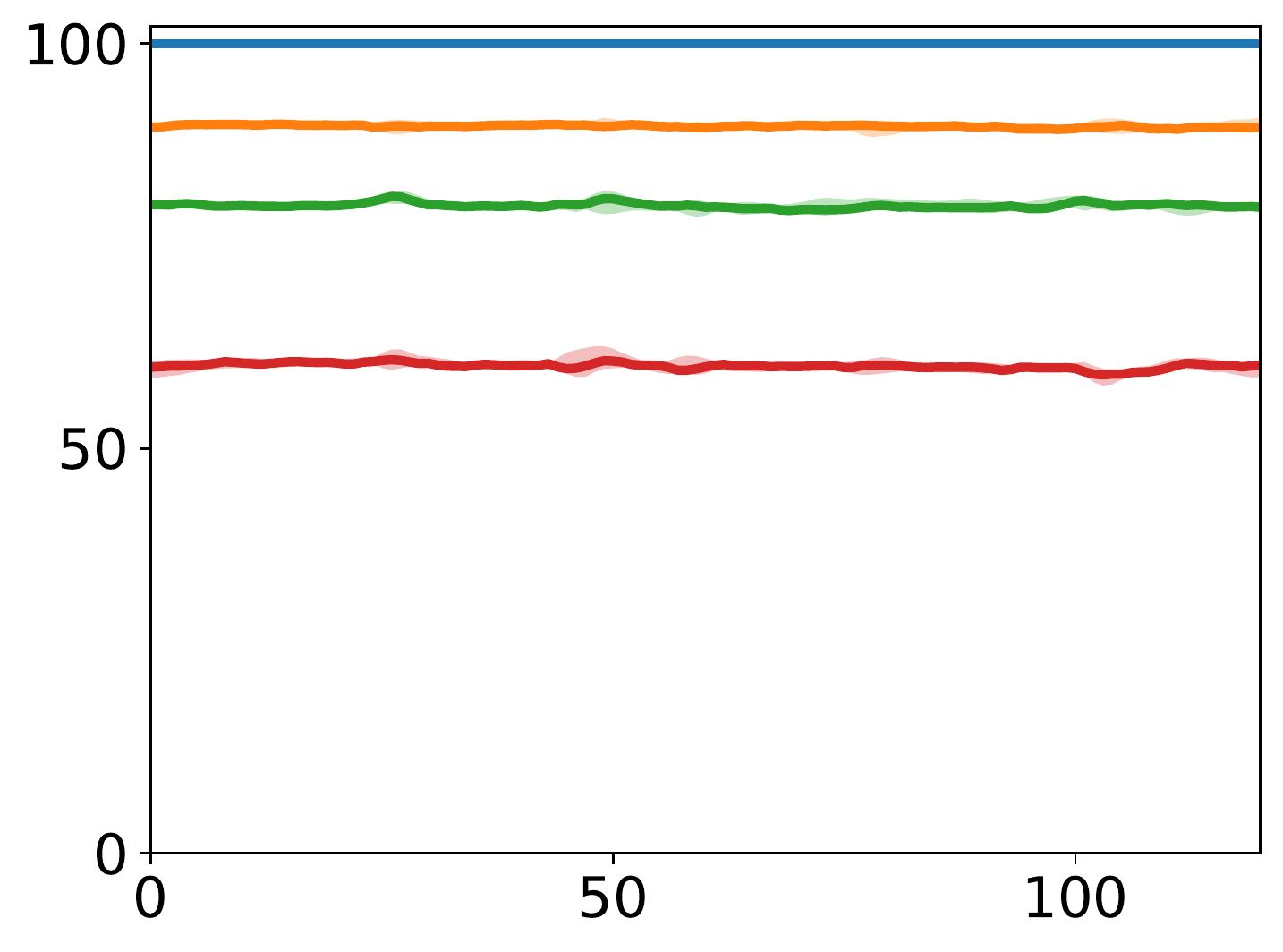} & 
\rotatebox{90}{Tokyo}
\\[-0.25cm]
&
\includegraphics[width=\fivesmallfig]{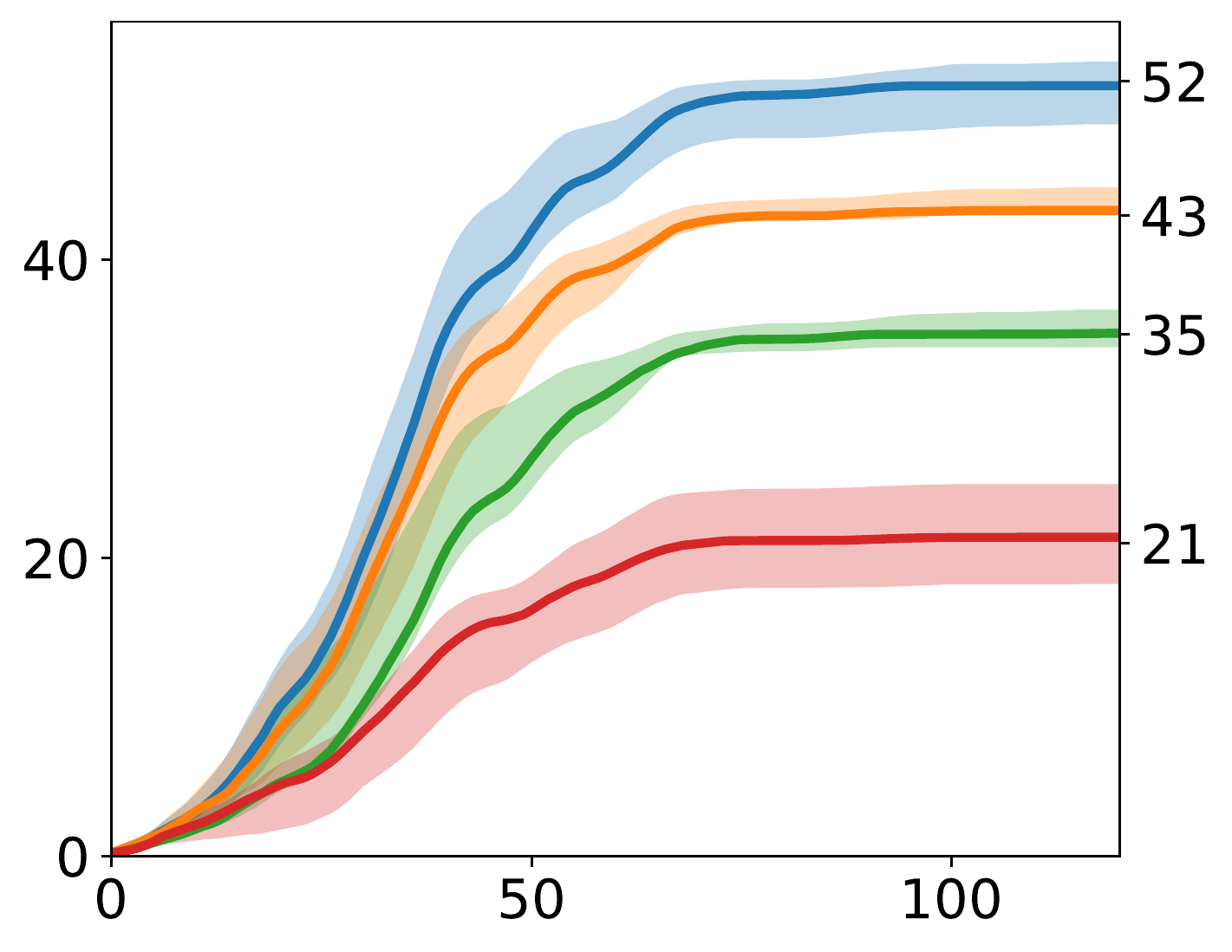} & &
\includegraphics[width=\fivesmallfig]{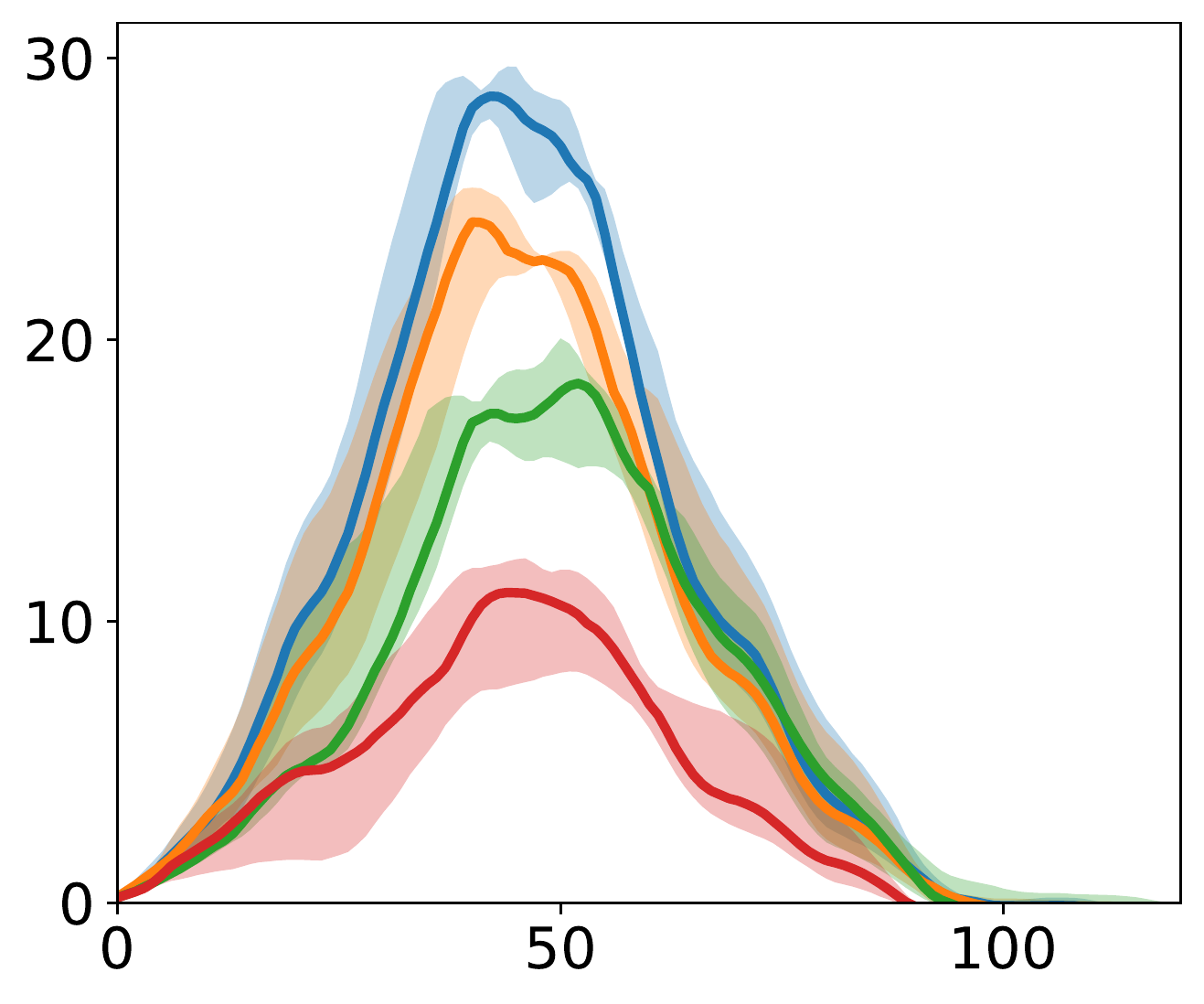}  & &
\includegraphics[width=\fivesmallfig]{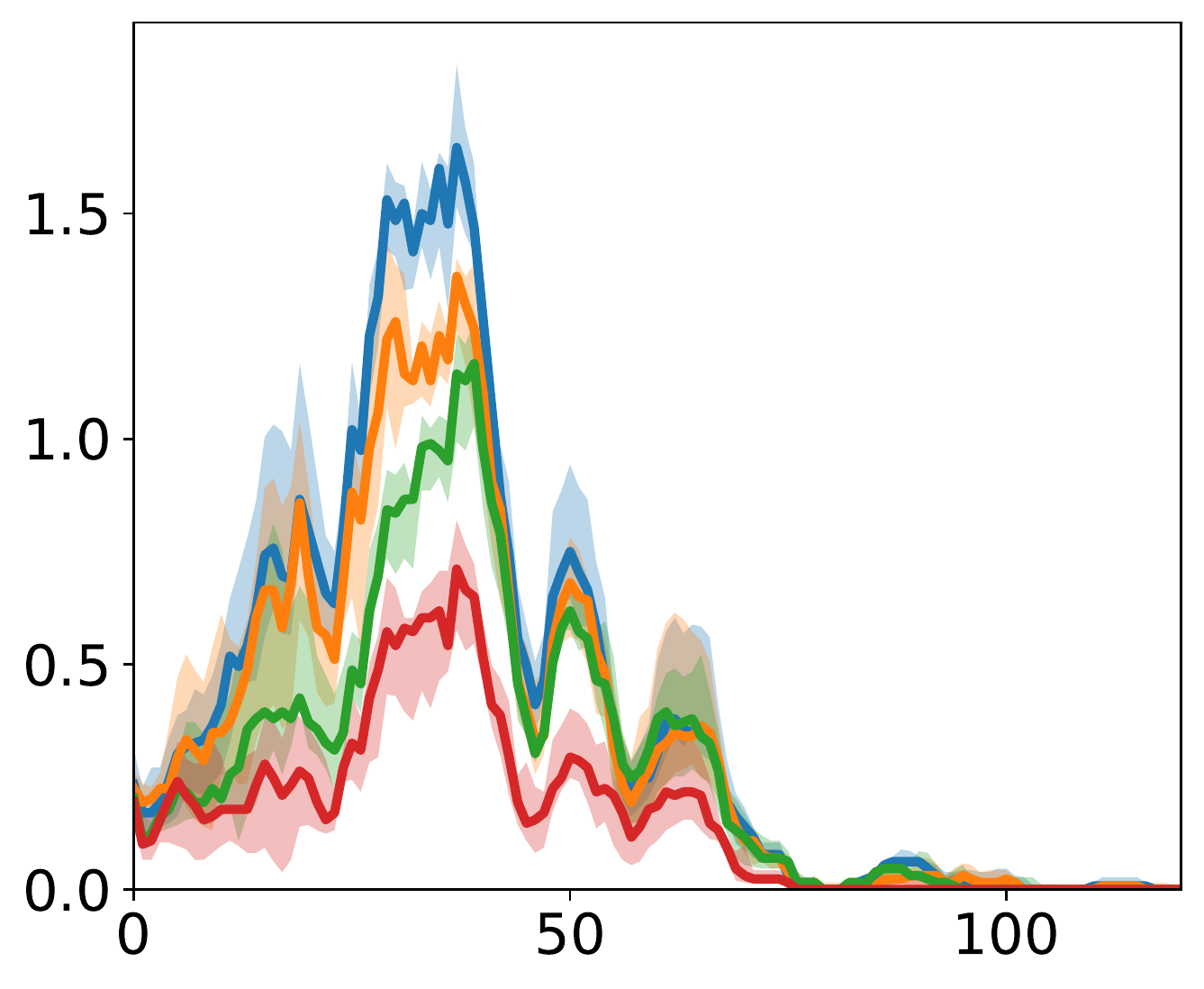}  & &
\includegraphics[width=\fivesmallfig]{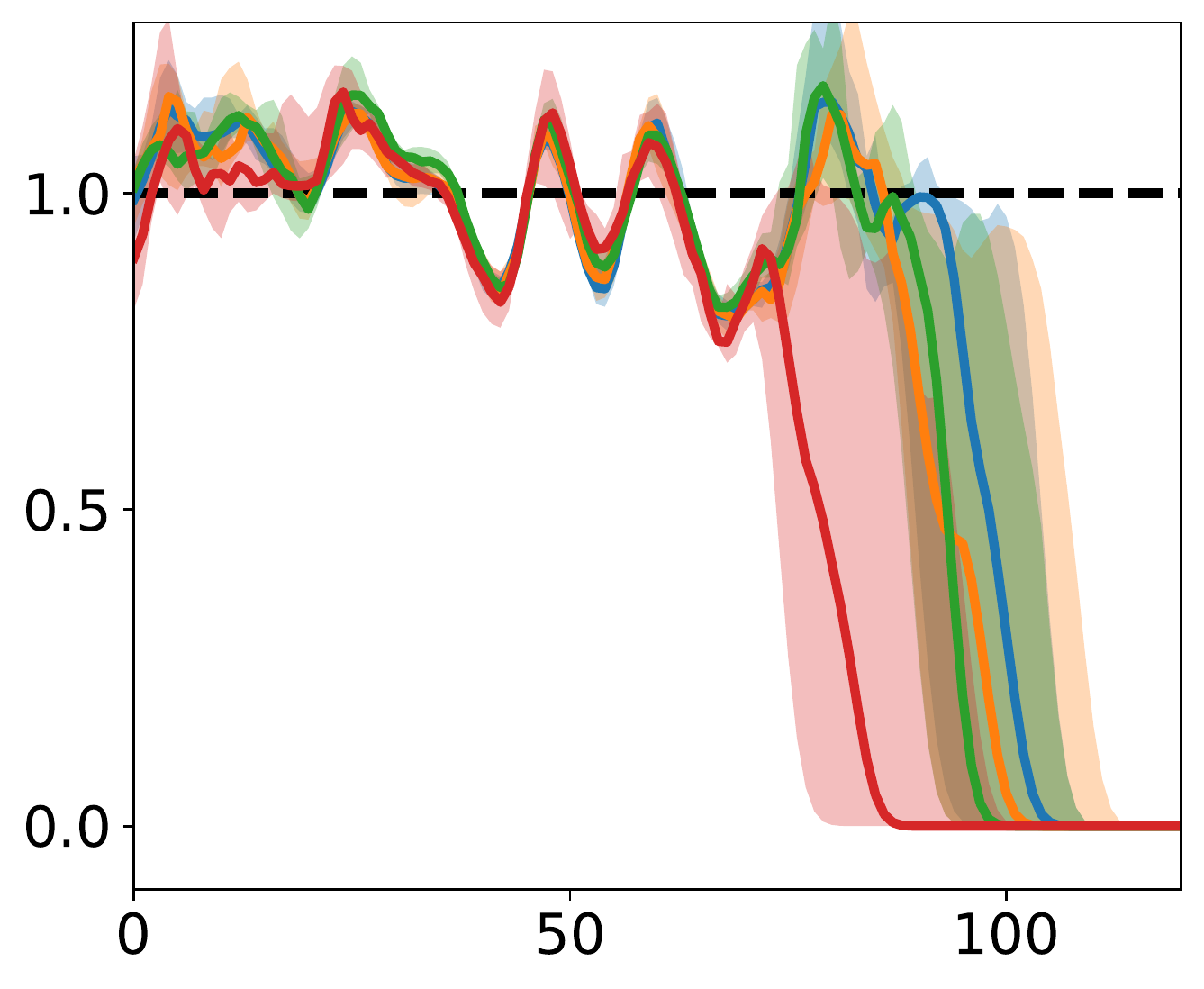}  & &
\includegraphics[width=\fivesmallfig]{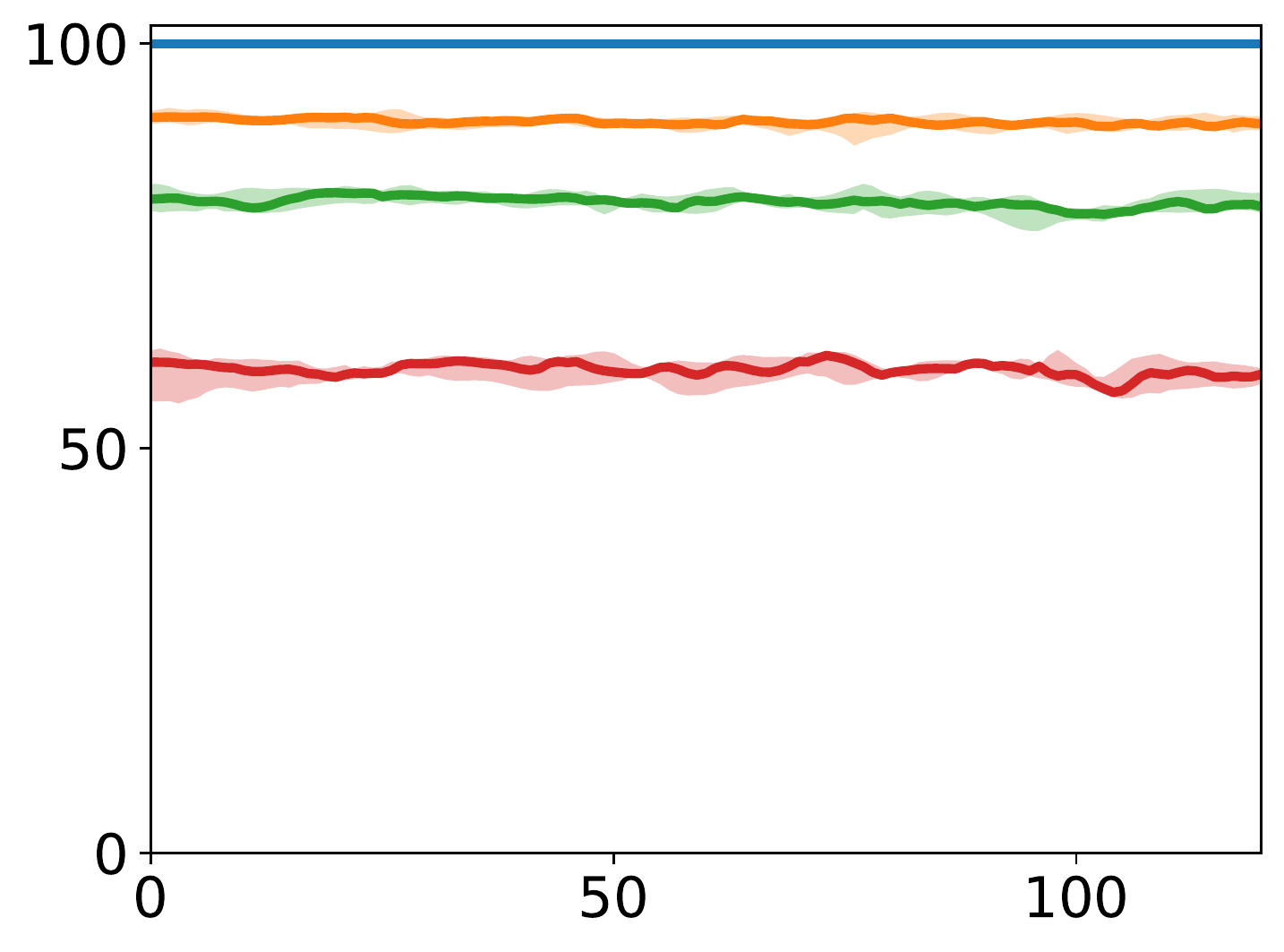}  &
\rotatebox{90}{Chicago}
\\ [-0.25cm]

&
\includegraphics[width=\fivesmallfig]{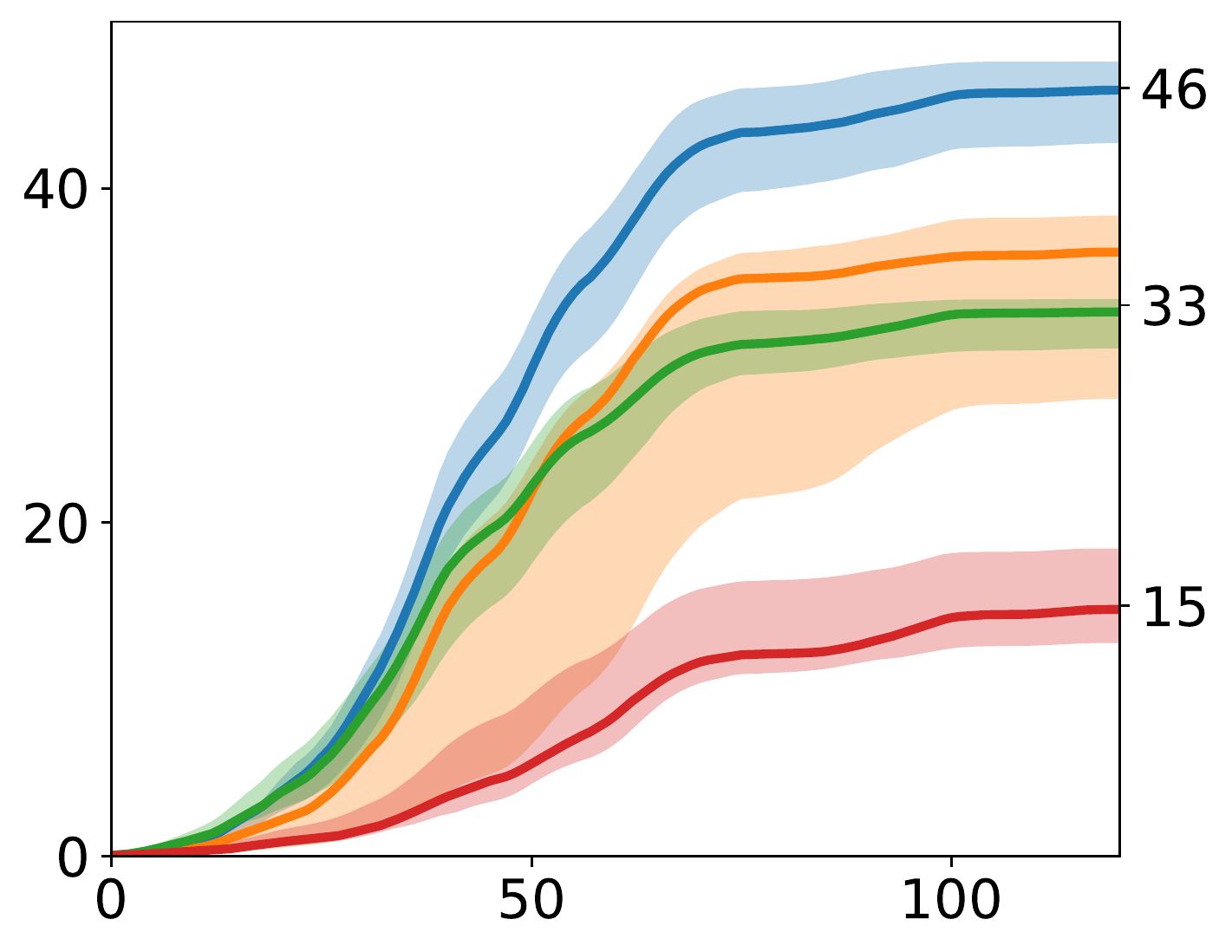} & &
\includegraphics[width=\fivesmallfig]{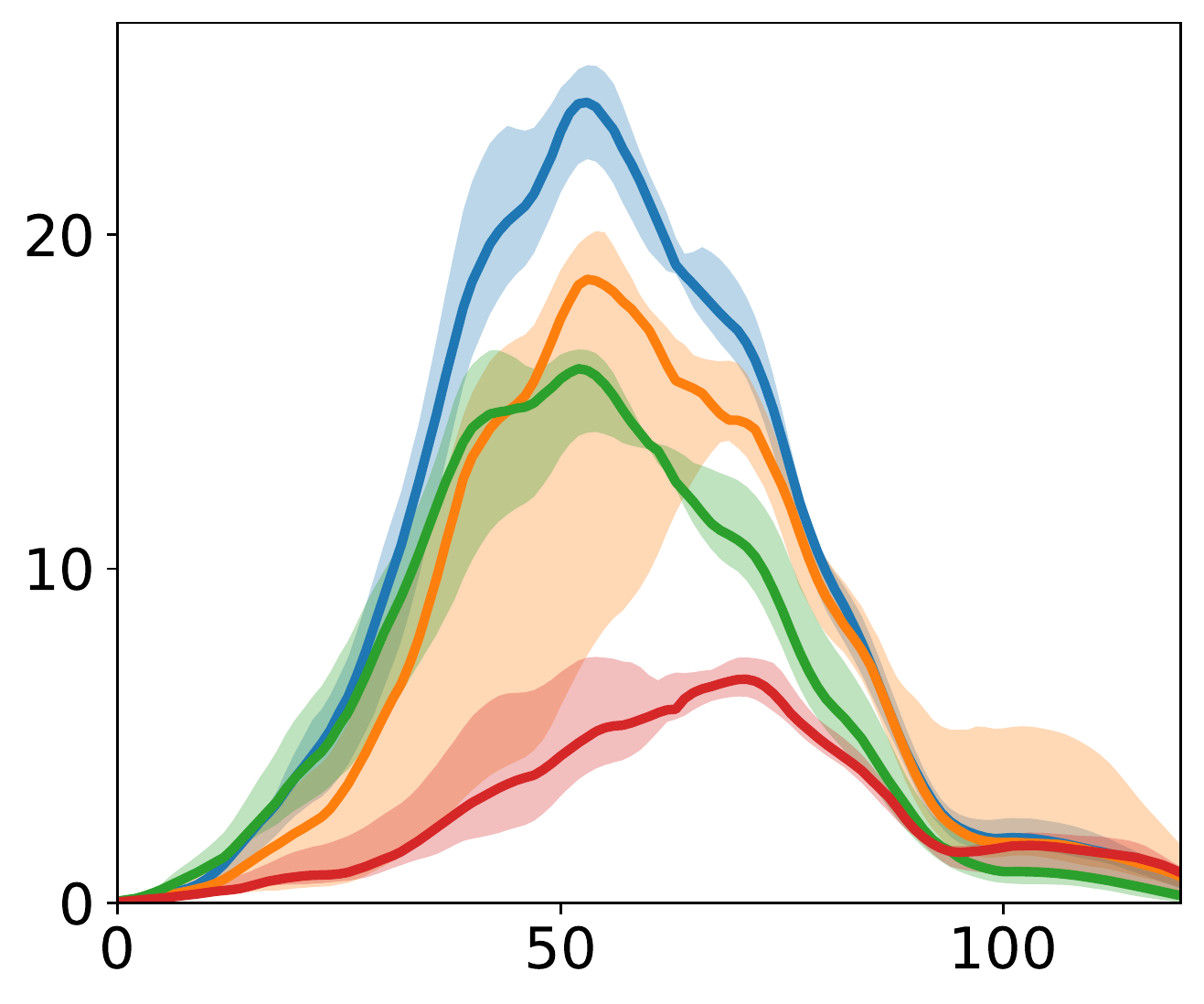} & &
\includegraphics[width=\fivesmallfig]{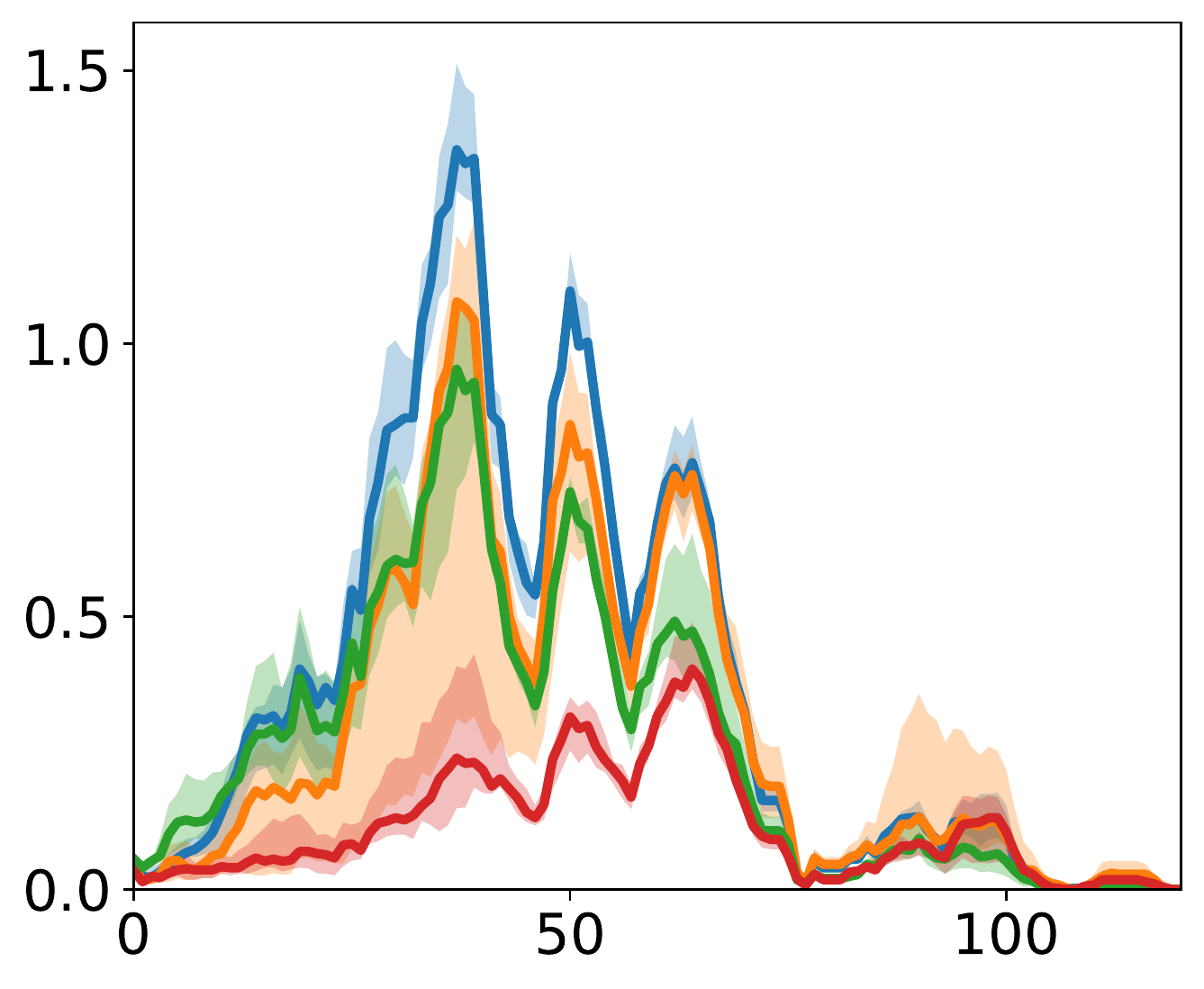}  & &
\includegraphics[width=\fivesmallfig]{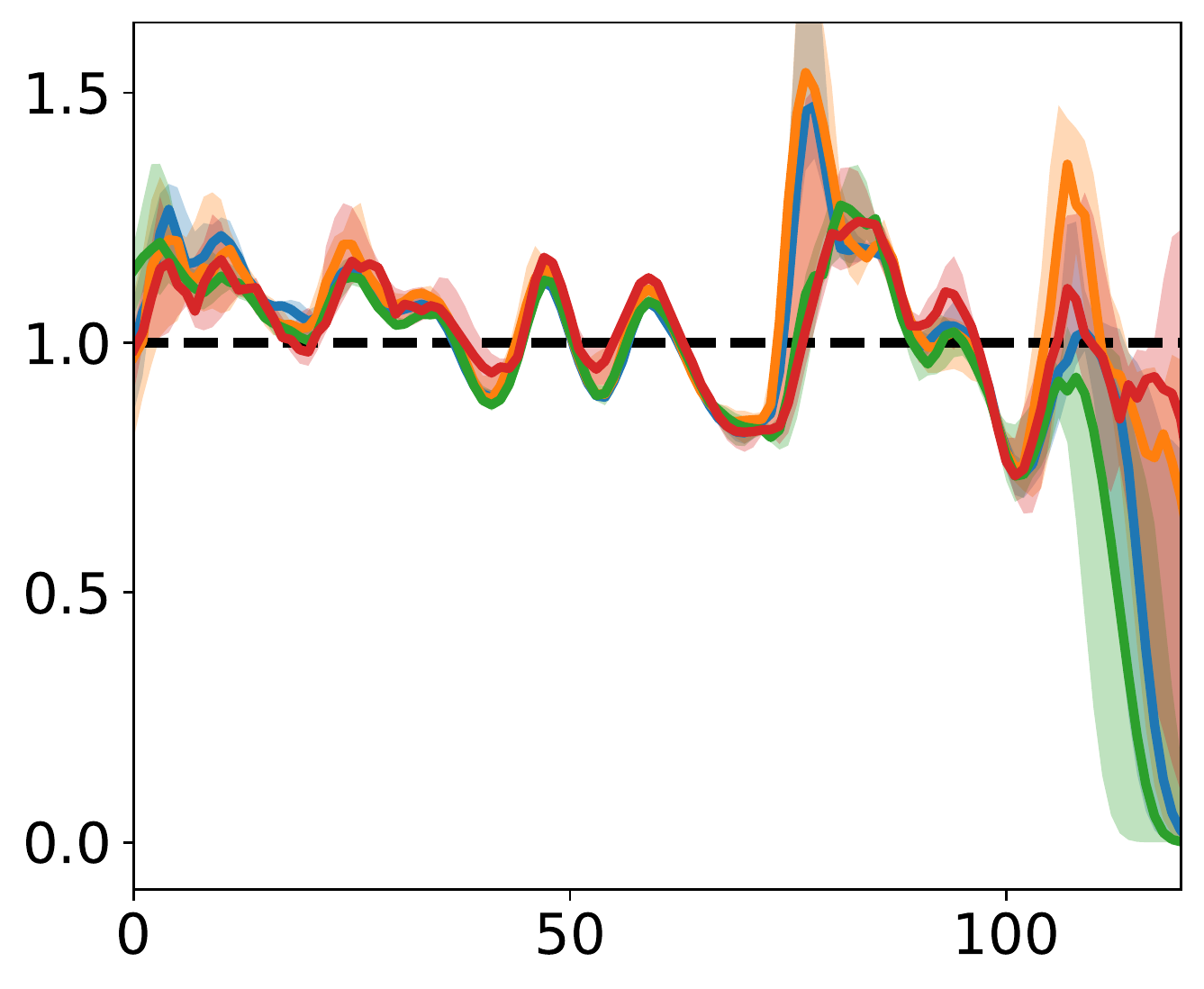} & &
\includegraphics[width=\fivesmallfig]{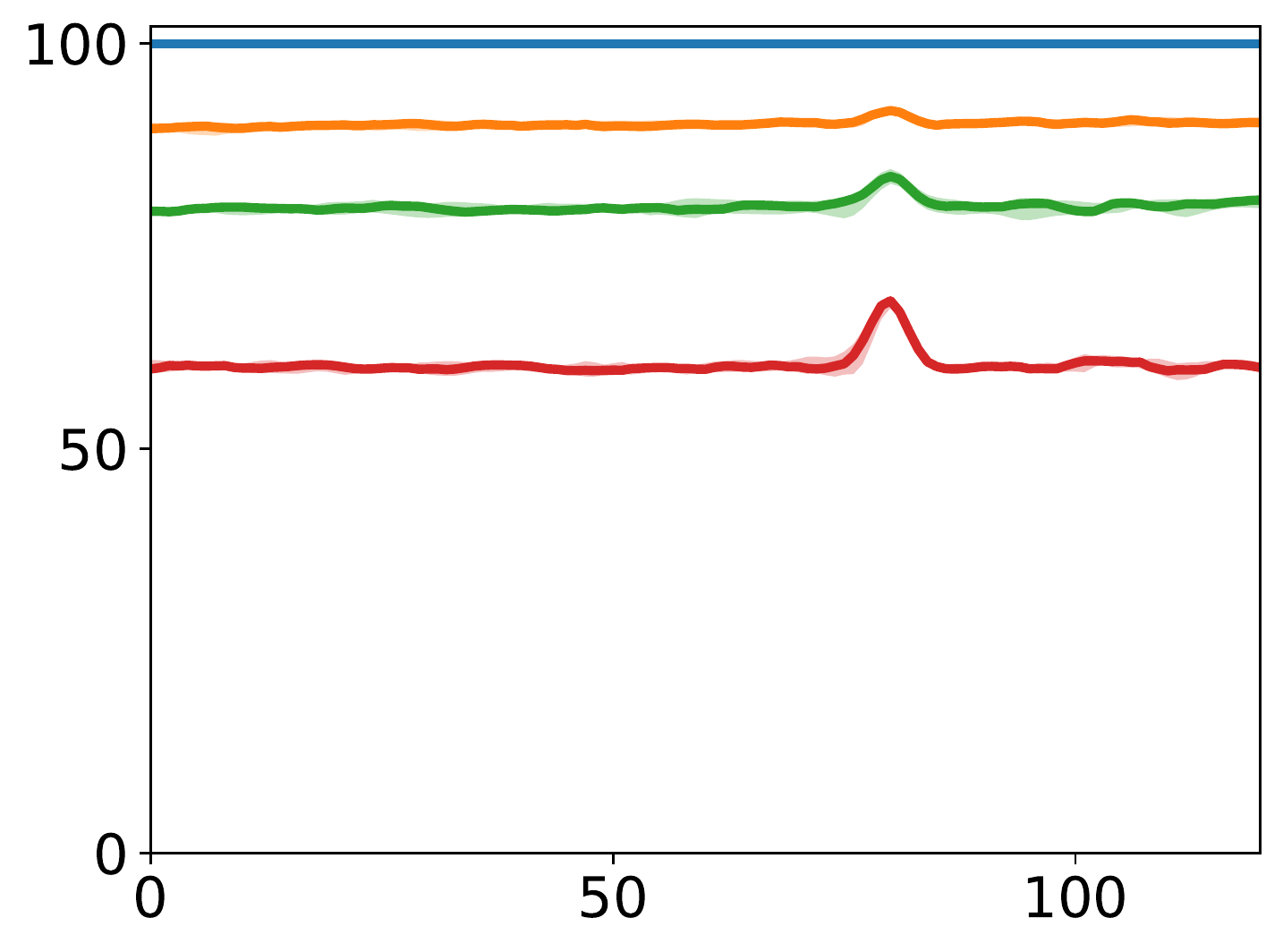} &
\rotatebox{90}{Los Angeles}
\\ [-0.25cm]

&
\includegraphics[width=\fivesmallfig]{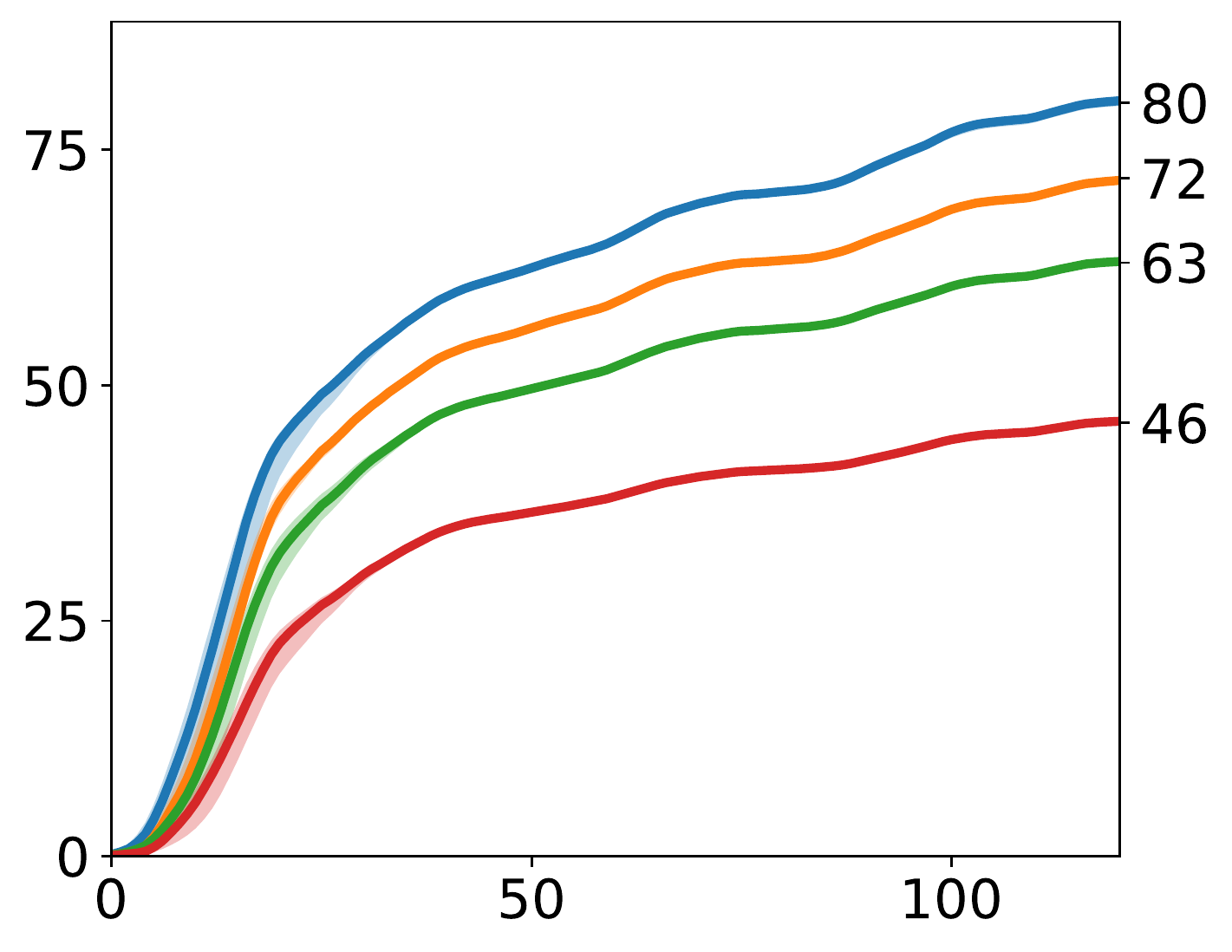} & &
\includegraphics[width=\fivesmallfig]{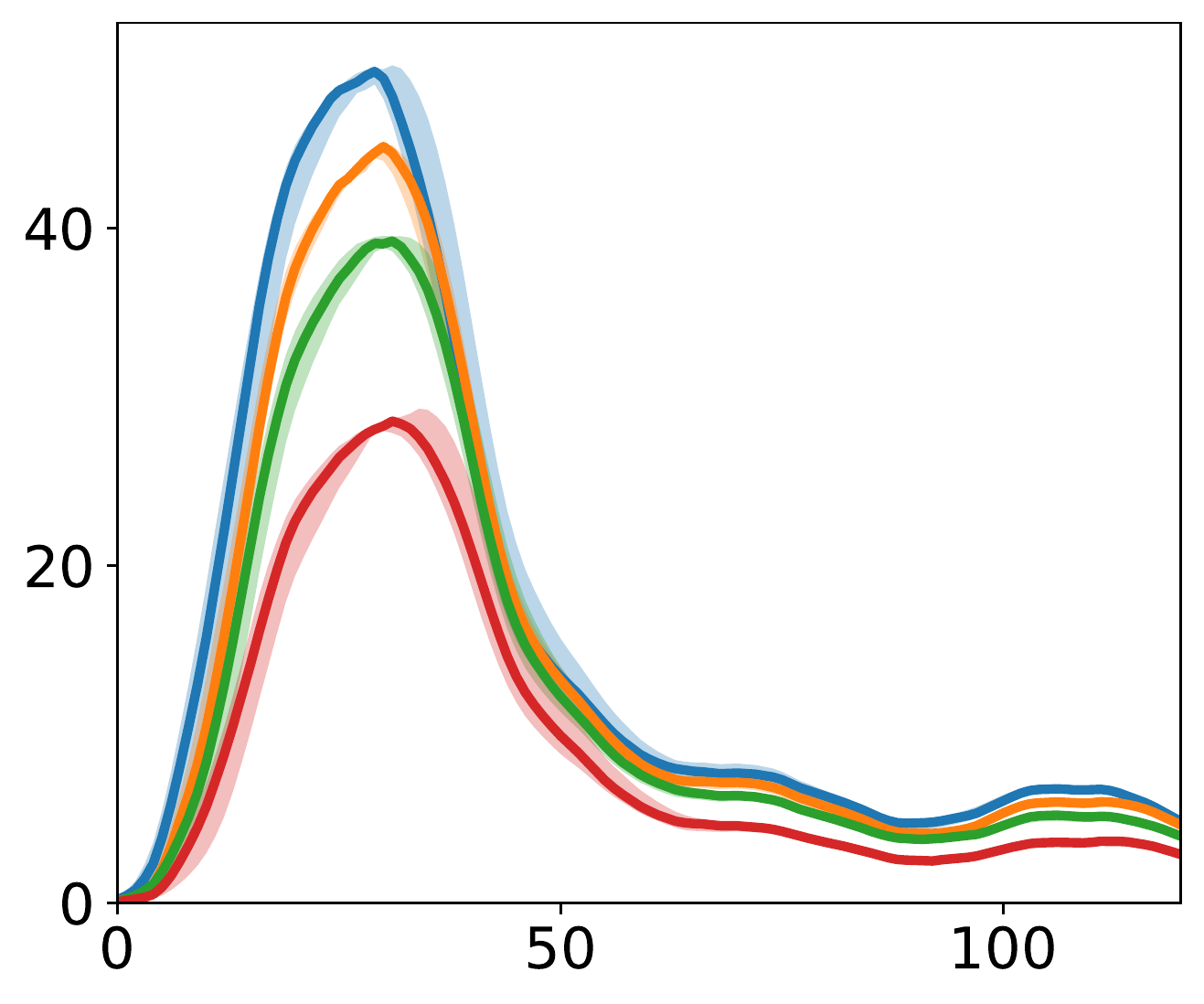}  &&
\includegraphics[width=\fivesmallfig]{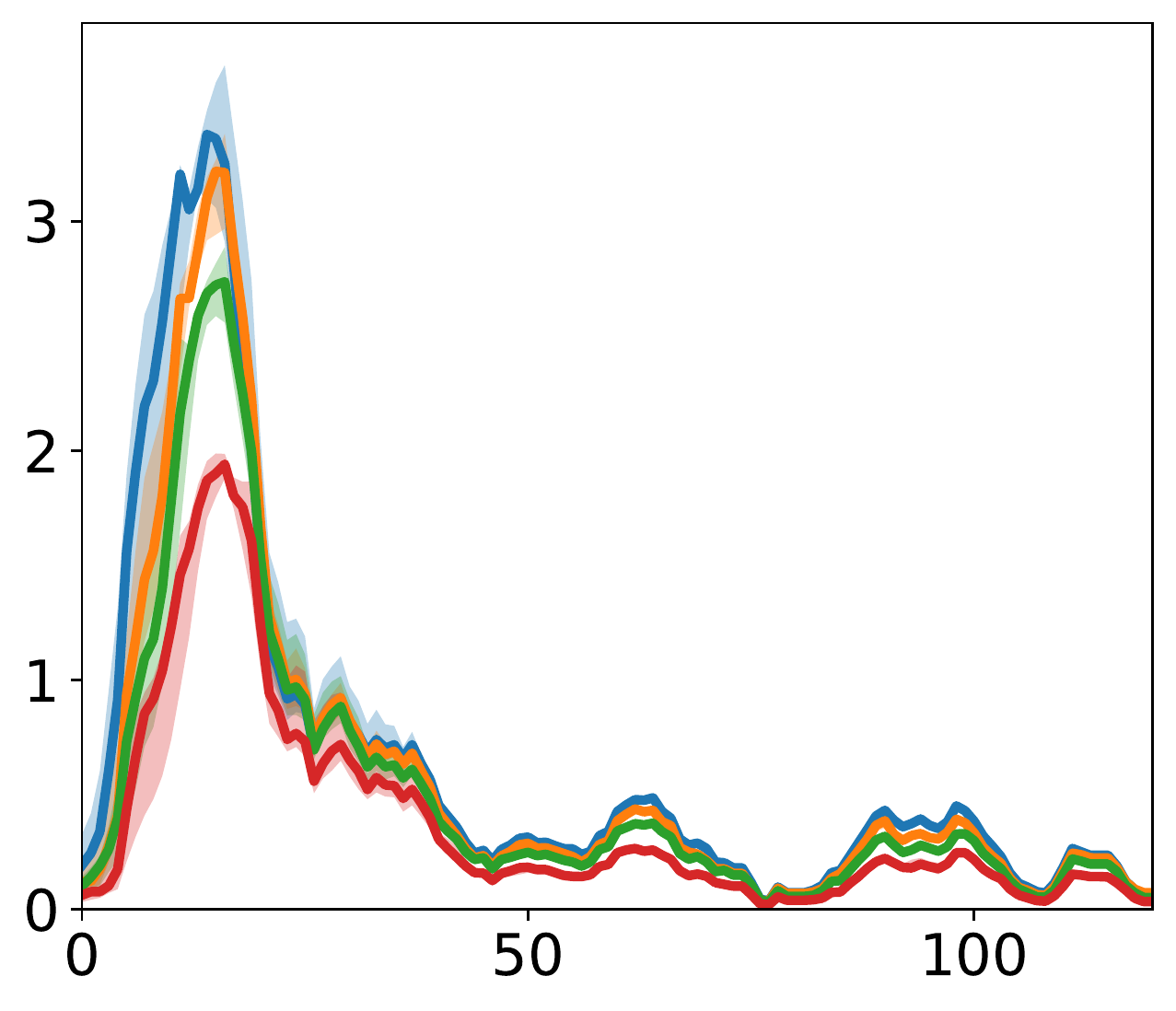}  &&
\includegraphics[width=\fivesmallfig]{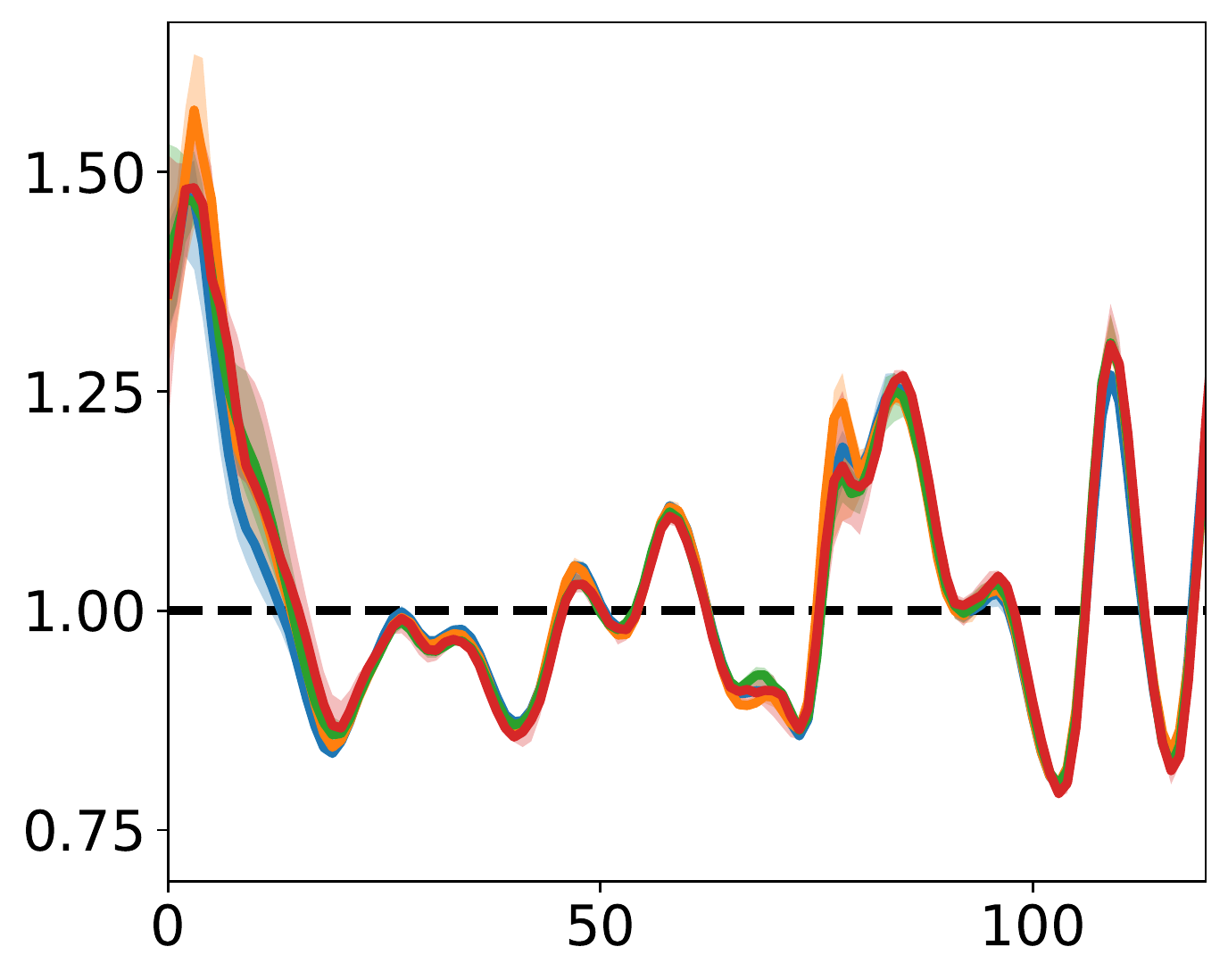} & &
\includegraphics[width=\fivesmallfig]{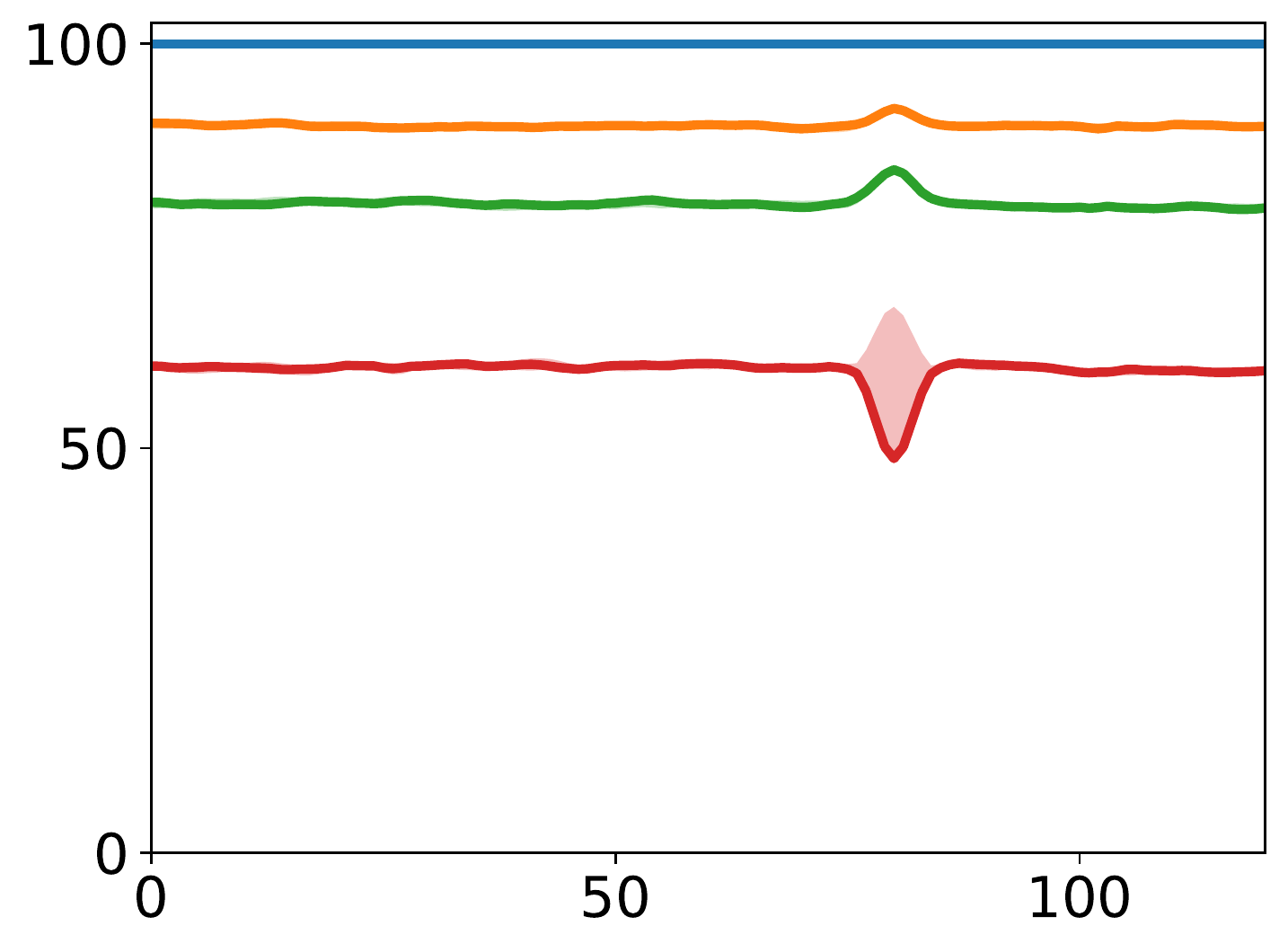} &
\rotatebox{90}{Istanbul}
\\ [-0.25cm]

&
\includegraphics[width=\fivesmallfig]{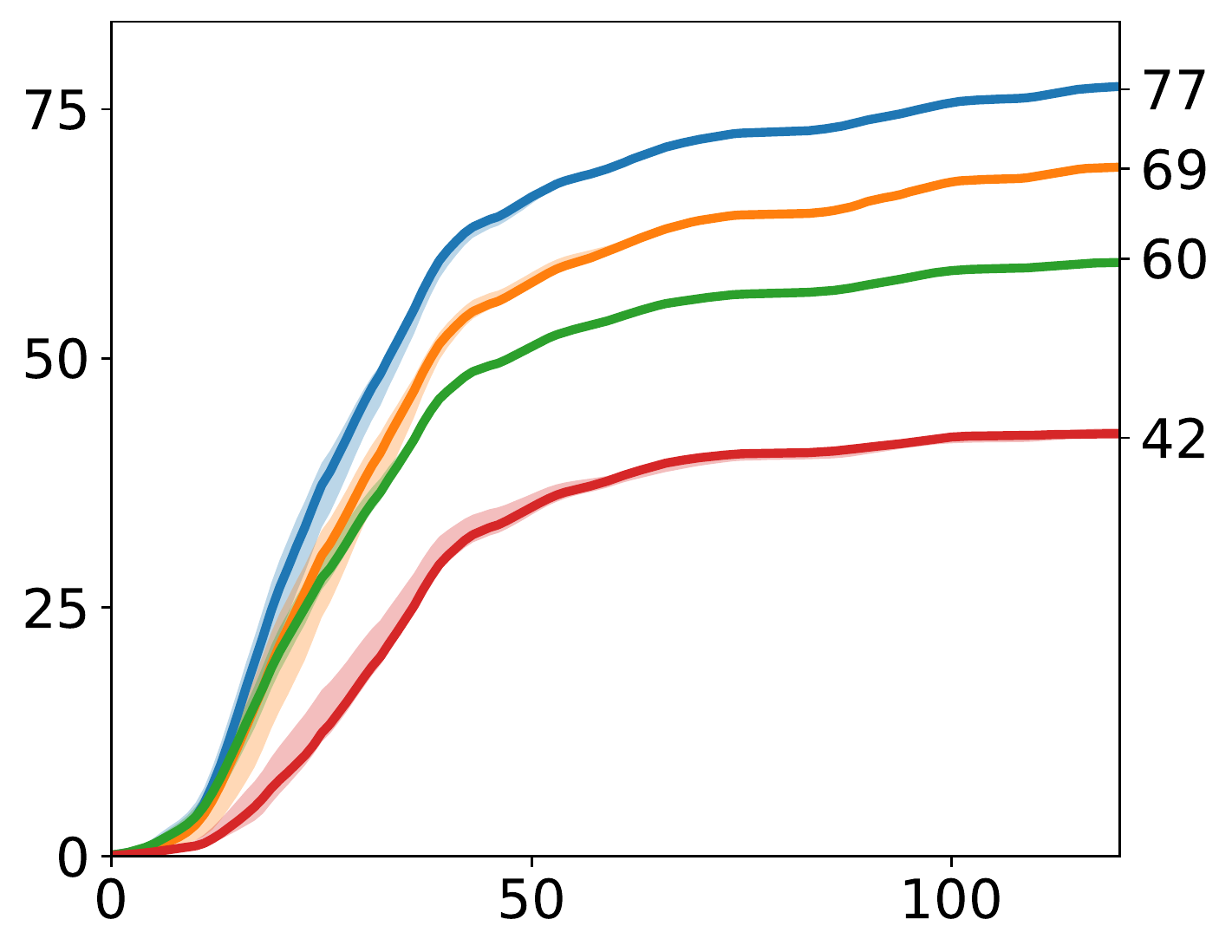} &&
\includegraphics[width=\fivesmallfig]{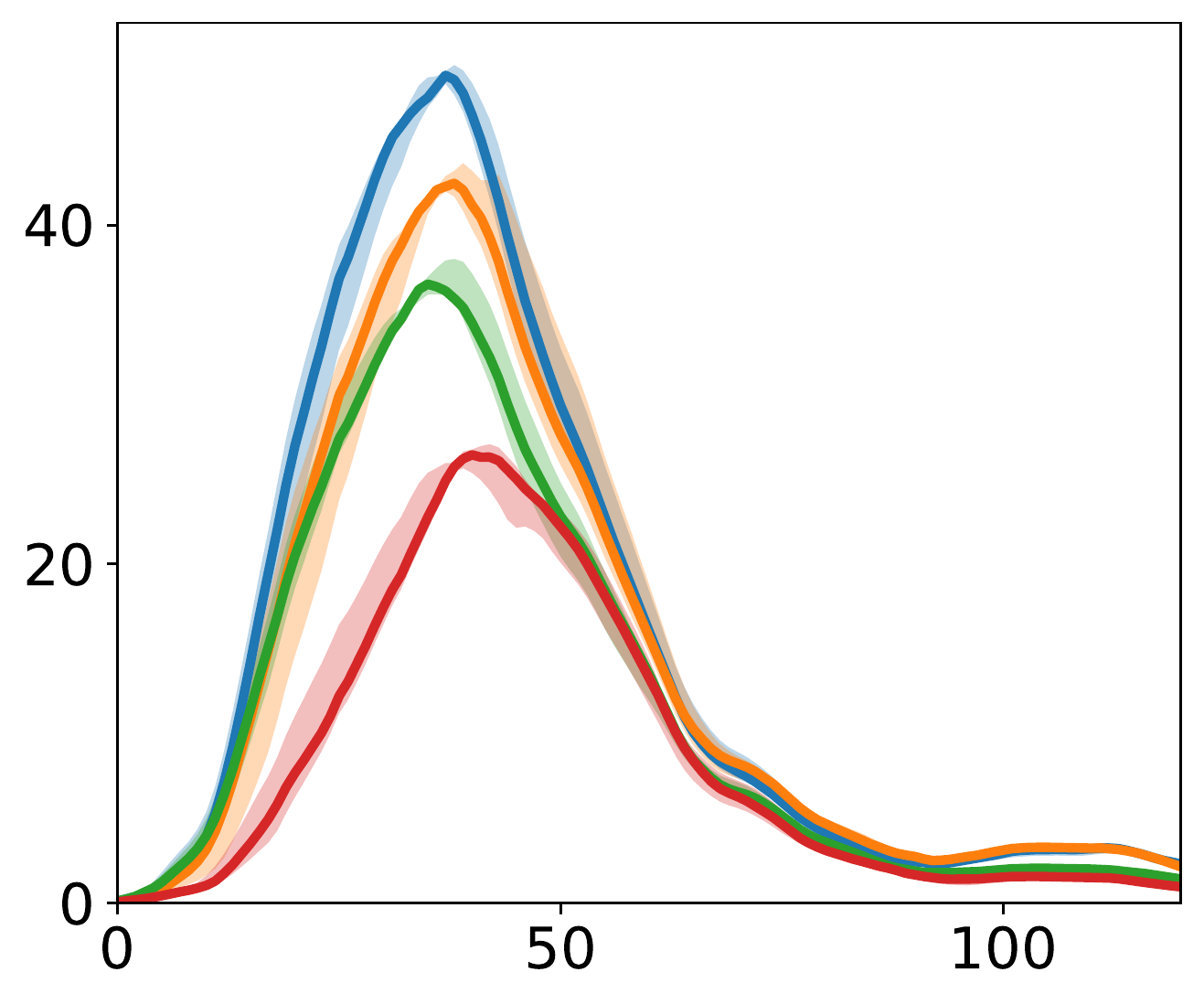}  &&
\includegraphics[width=\fivesmallfig]{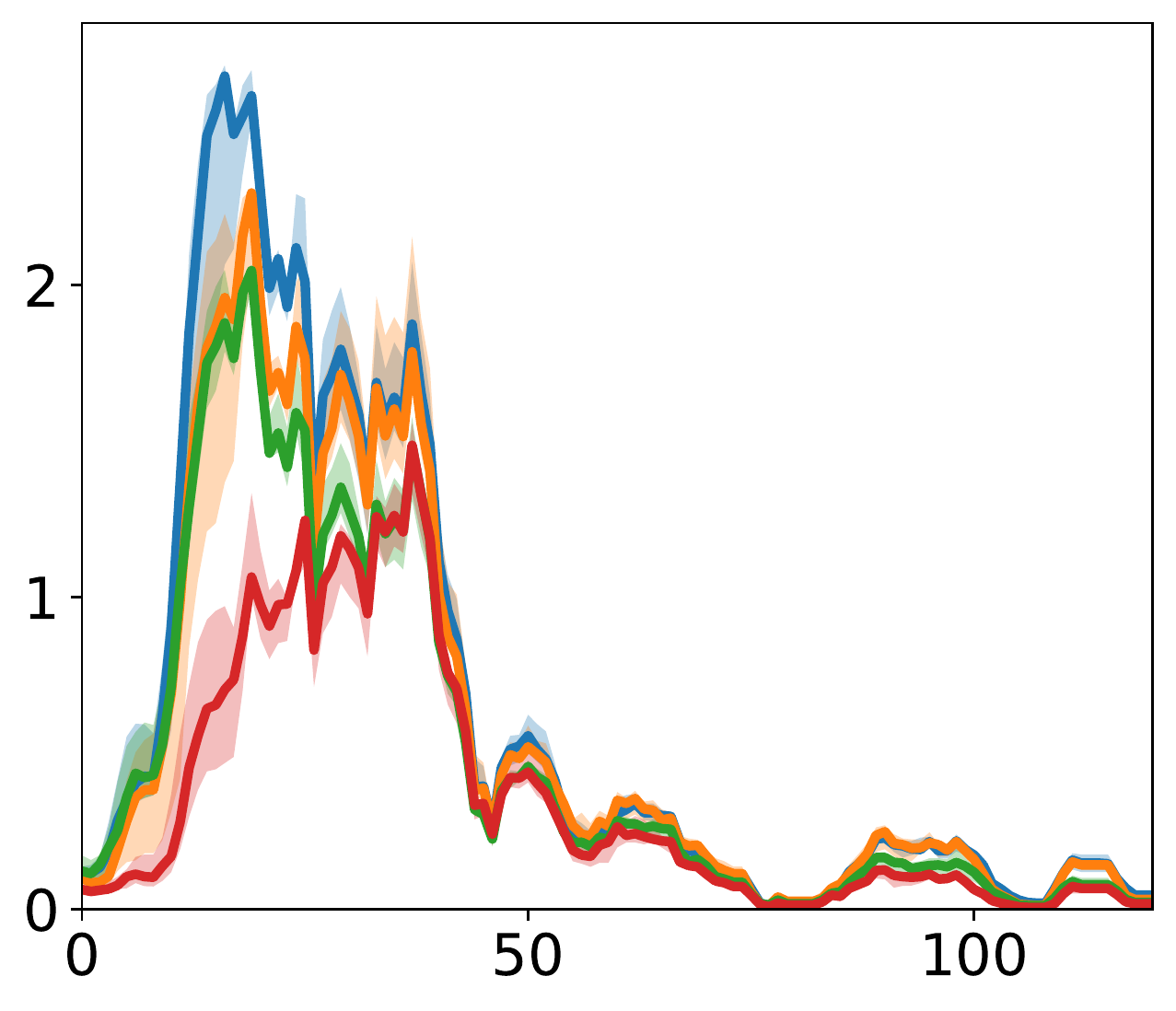}  &&
\includegraphics[width=\fivesmallfig]{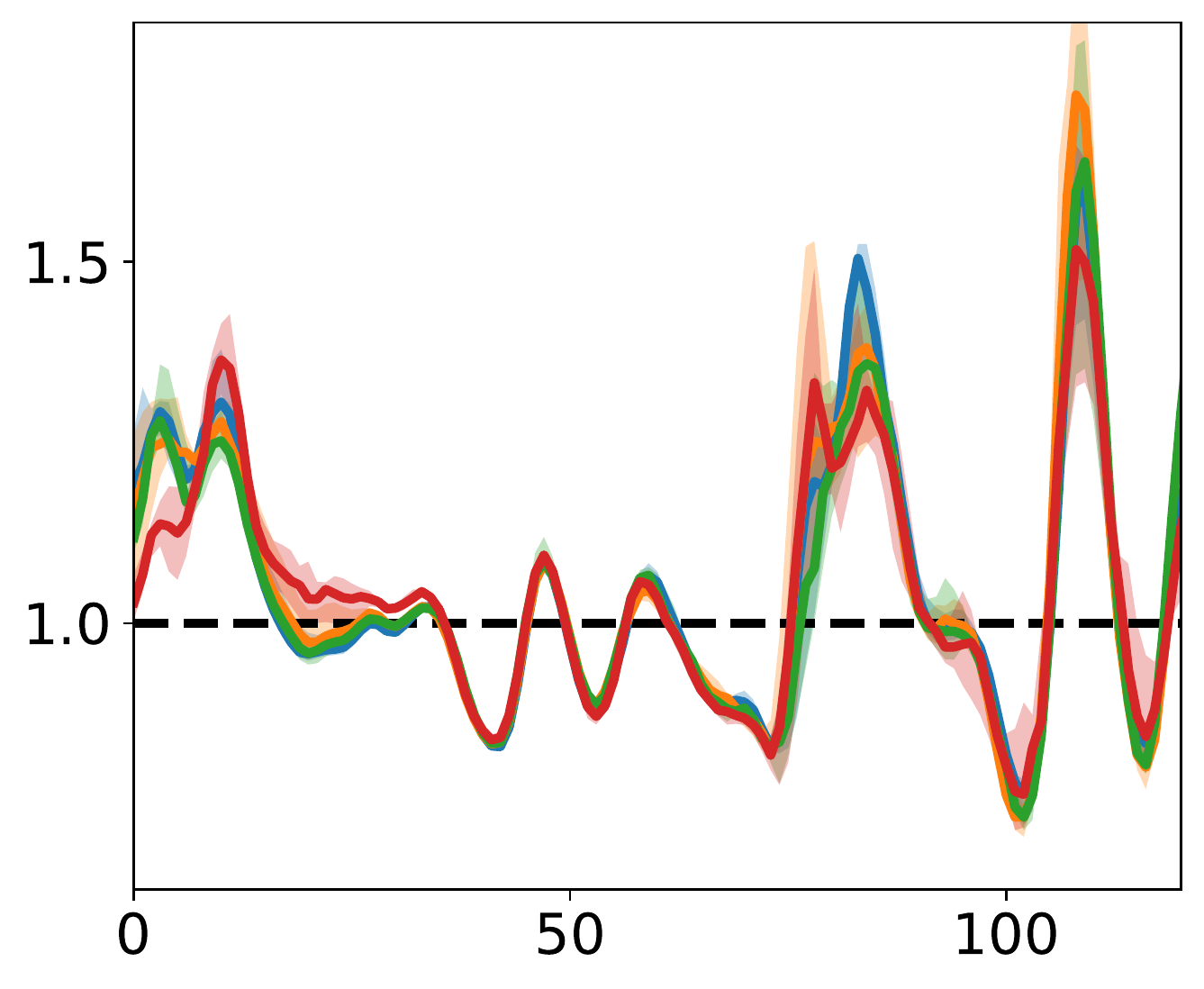} & &
\includegraphics[width=\fivesmallfig]{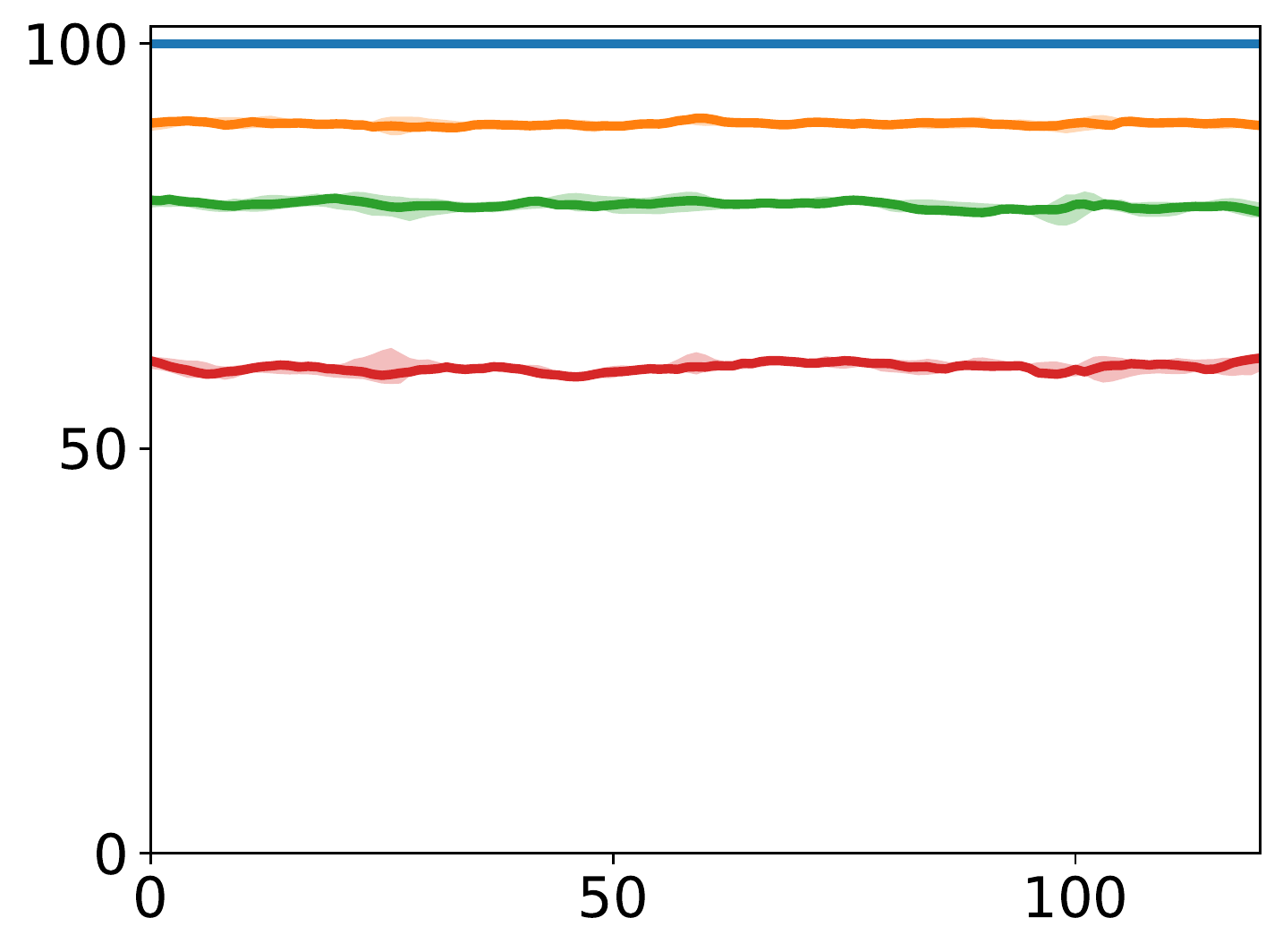} &
\rotatebox{90}{Jakarta}
\\ [-0.25cm]

&
\includegraphics[width=\fivesmallfig]{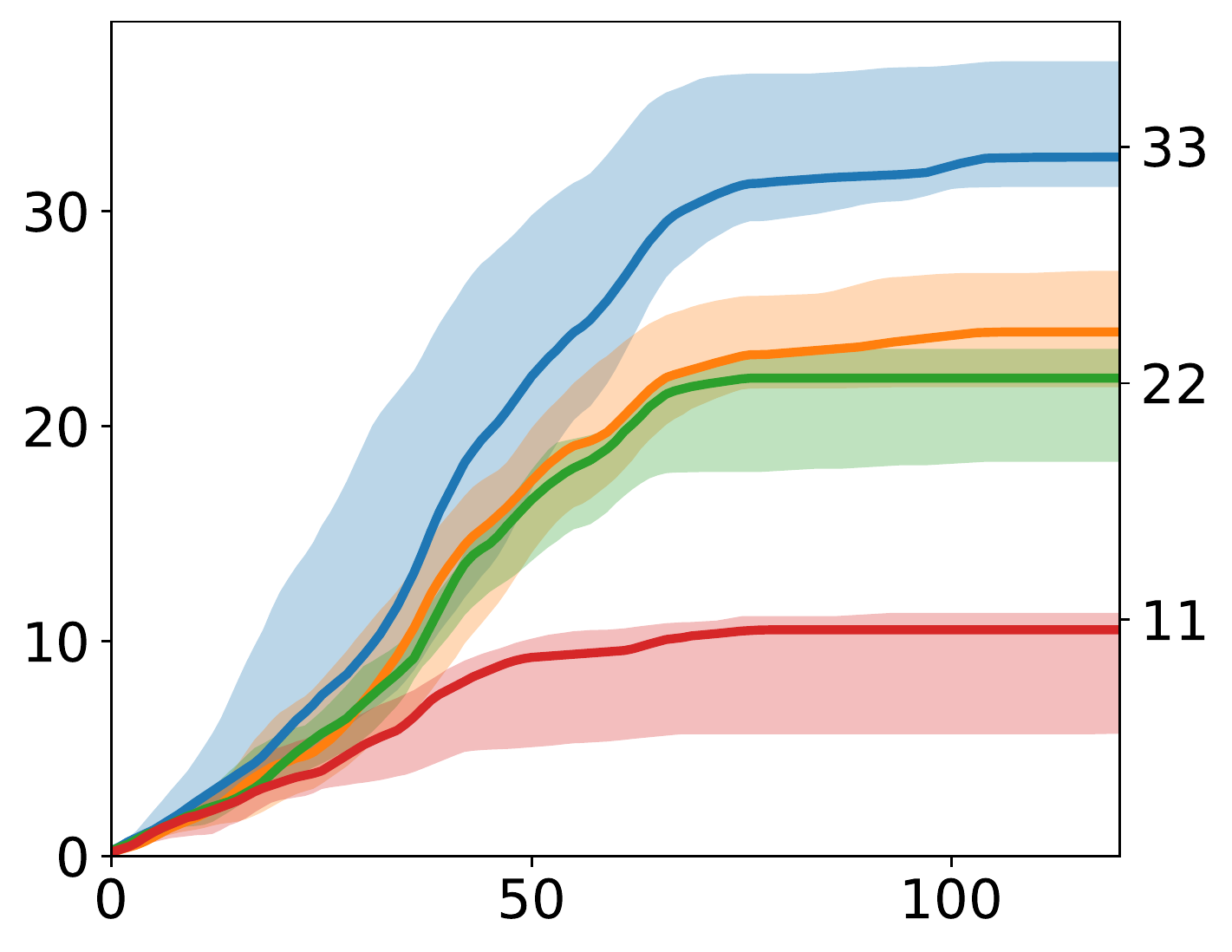} &&
\includegraphics[width=\fivesmallfig]{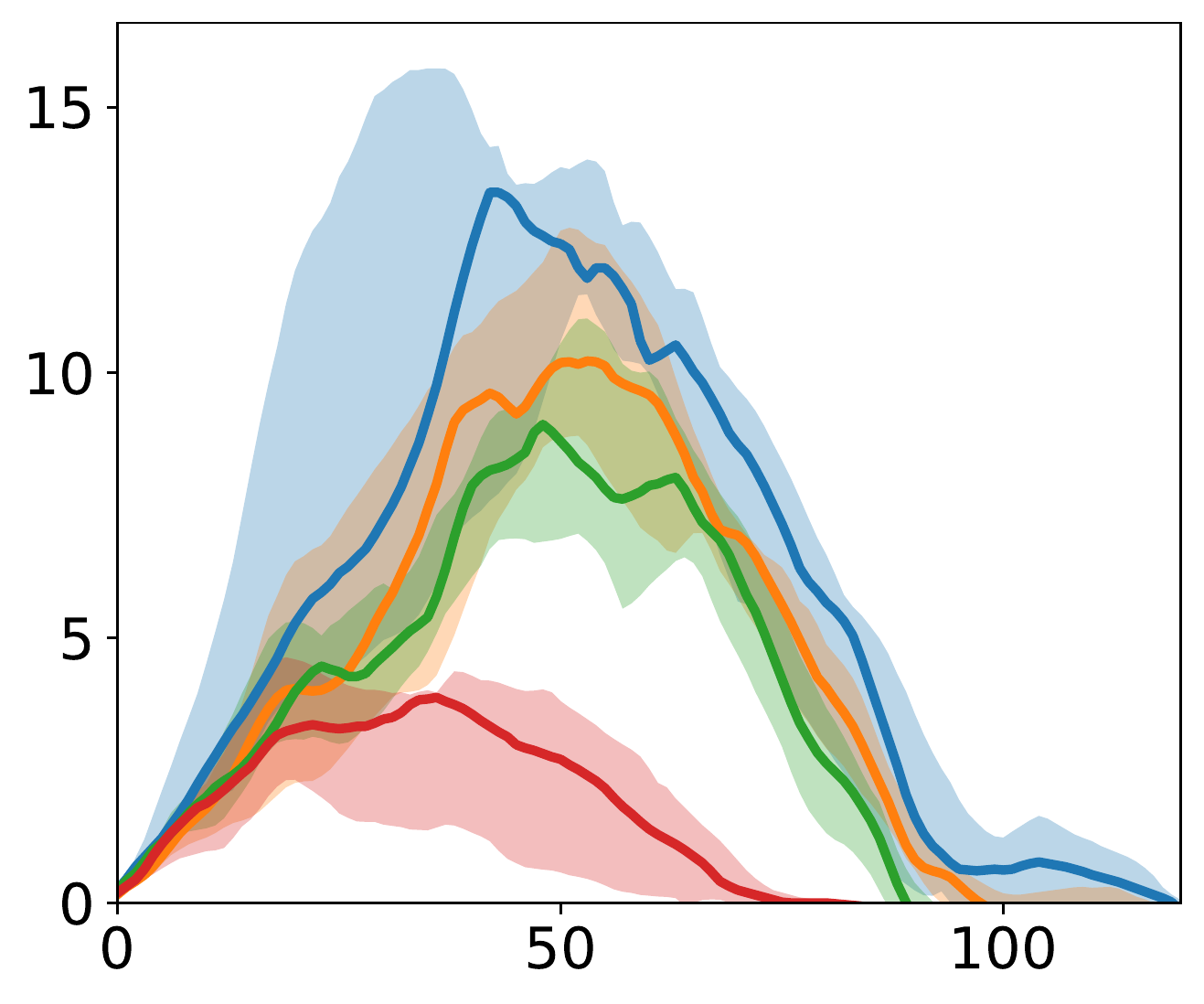}  &&
\includegraphics[width=\fivesmallfig]{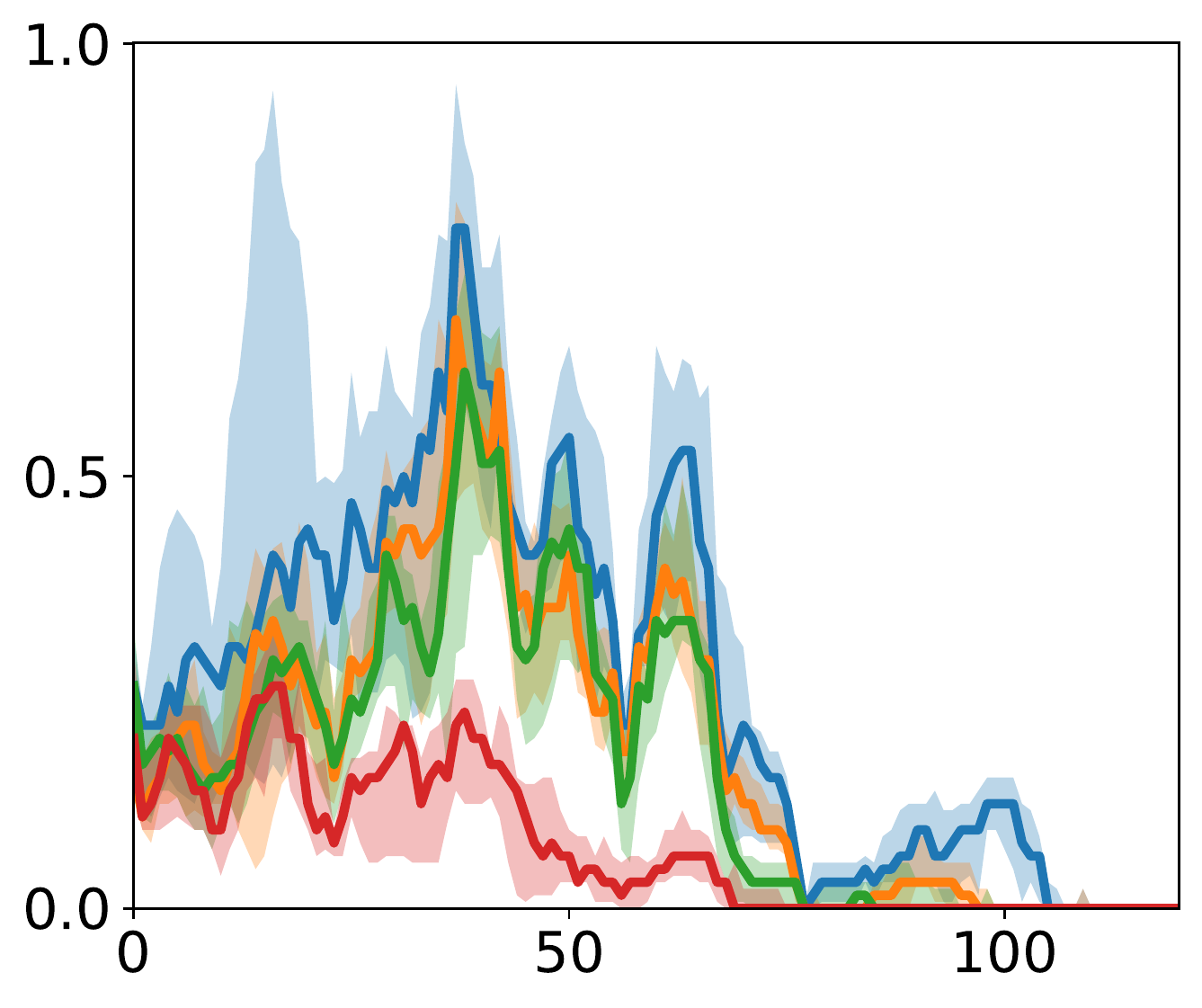}  &&
\includegraphics[width=\fivesmallfig]{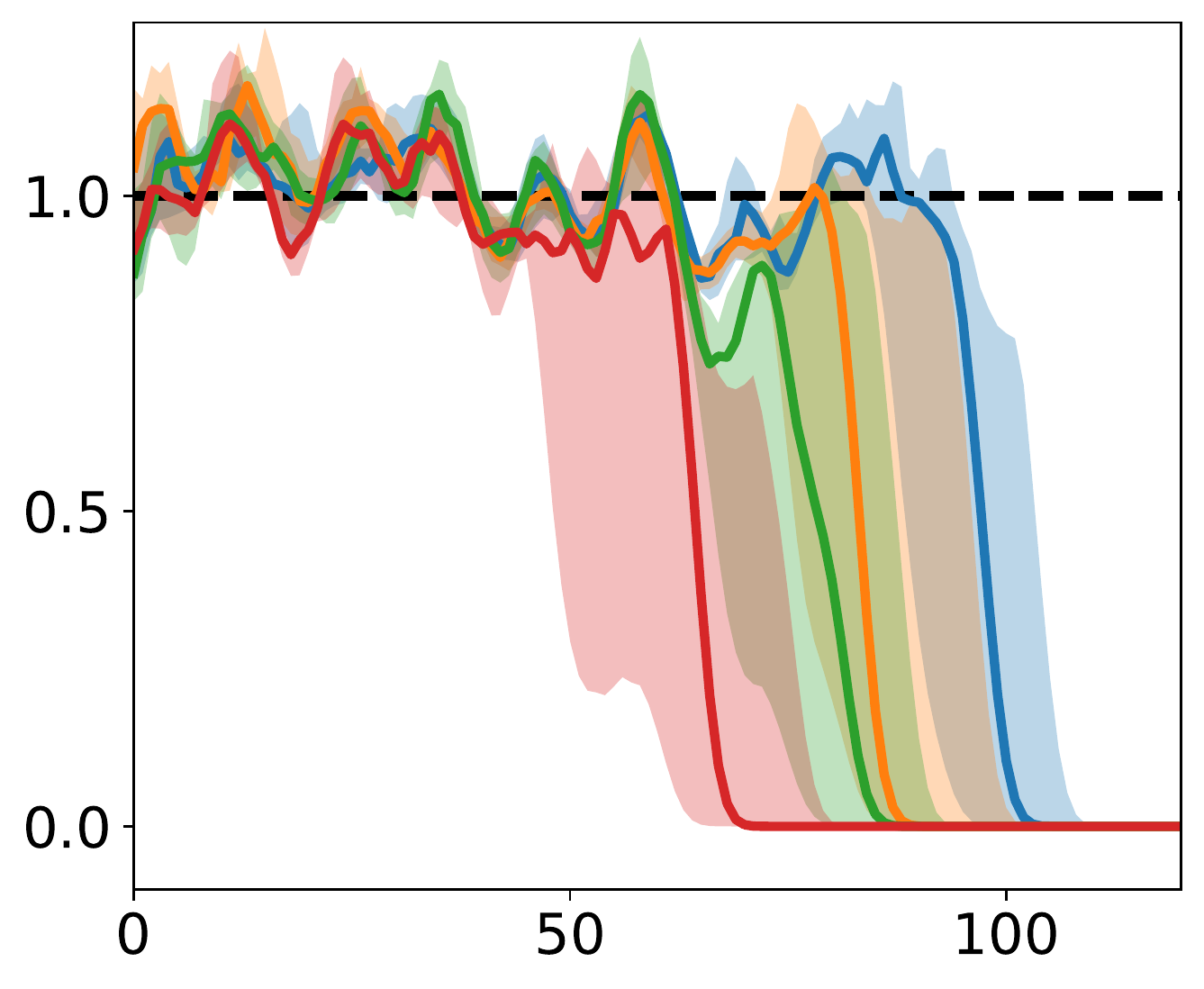} & &
\includegraphics[width=\fivesmallfig]{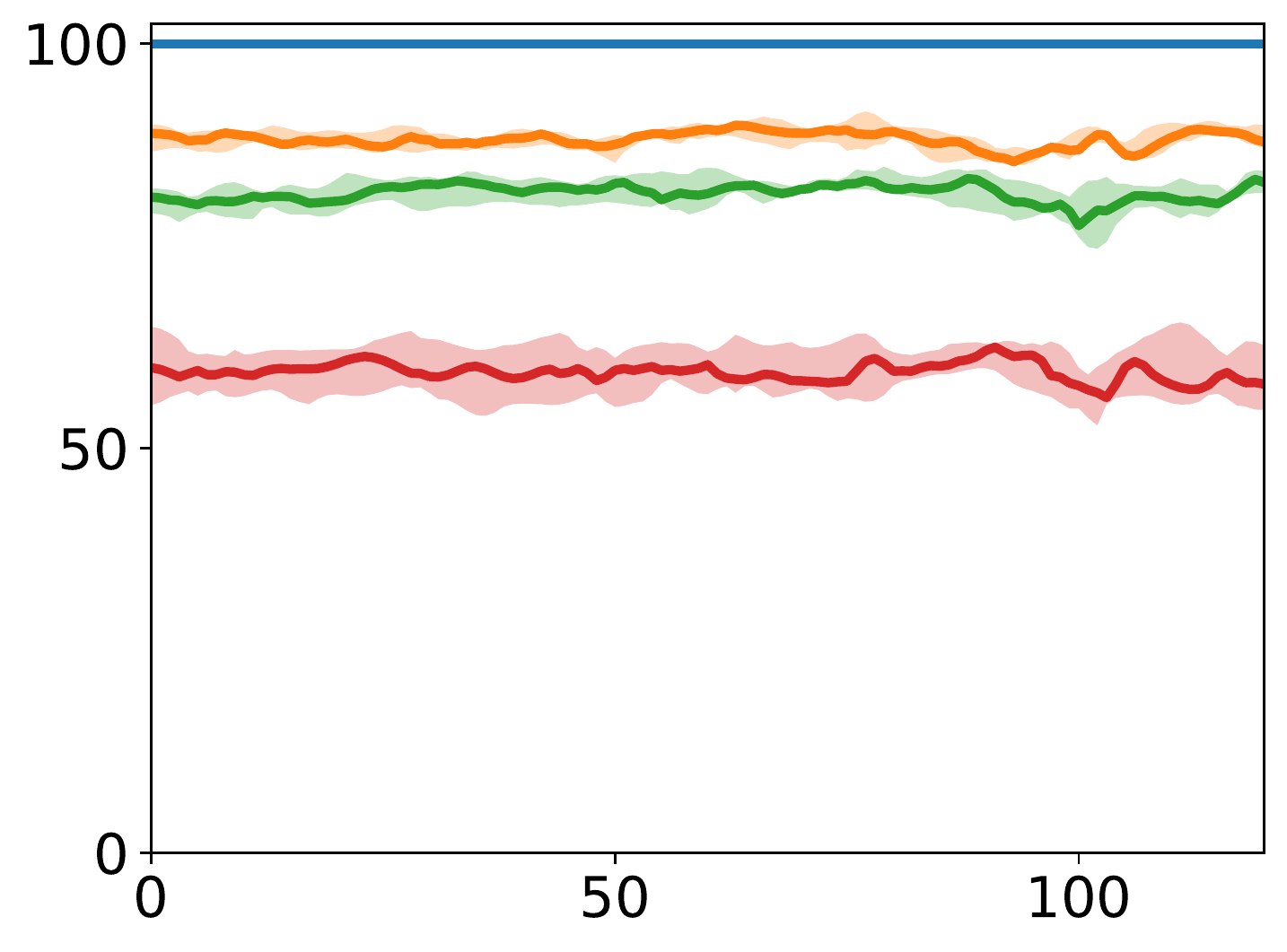} &
\rotatebox{90}{London}
\\ [-0.25cm]

&
\includegraphics[width=\fivesmallfig]{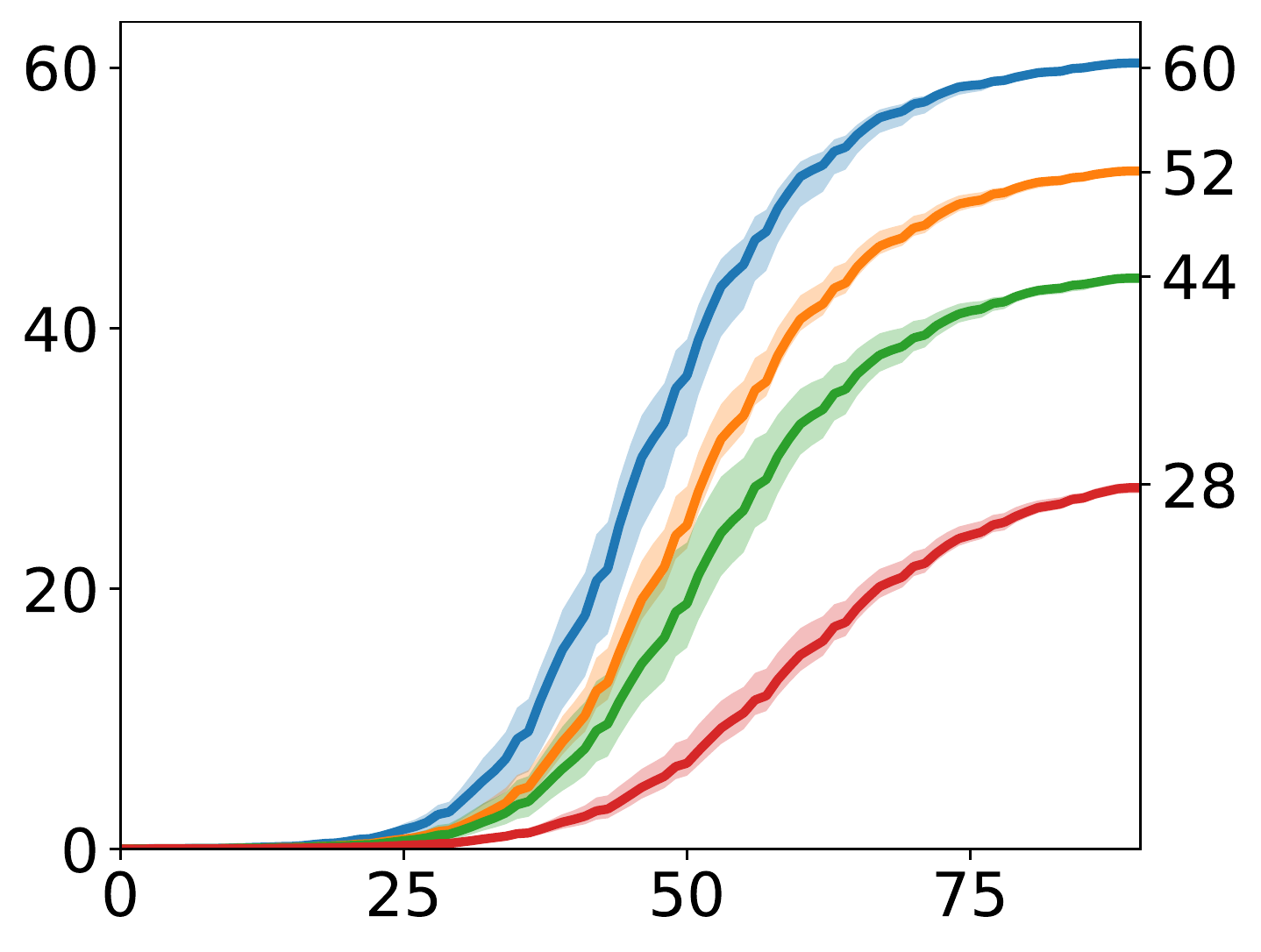} &&
\includegraphics[width=\fivesmallfig]{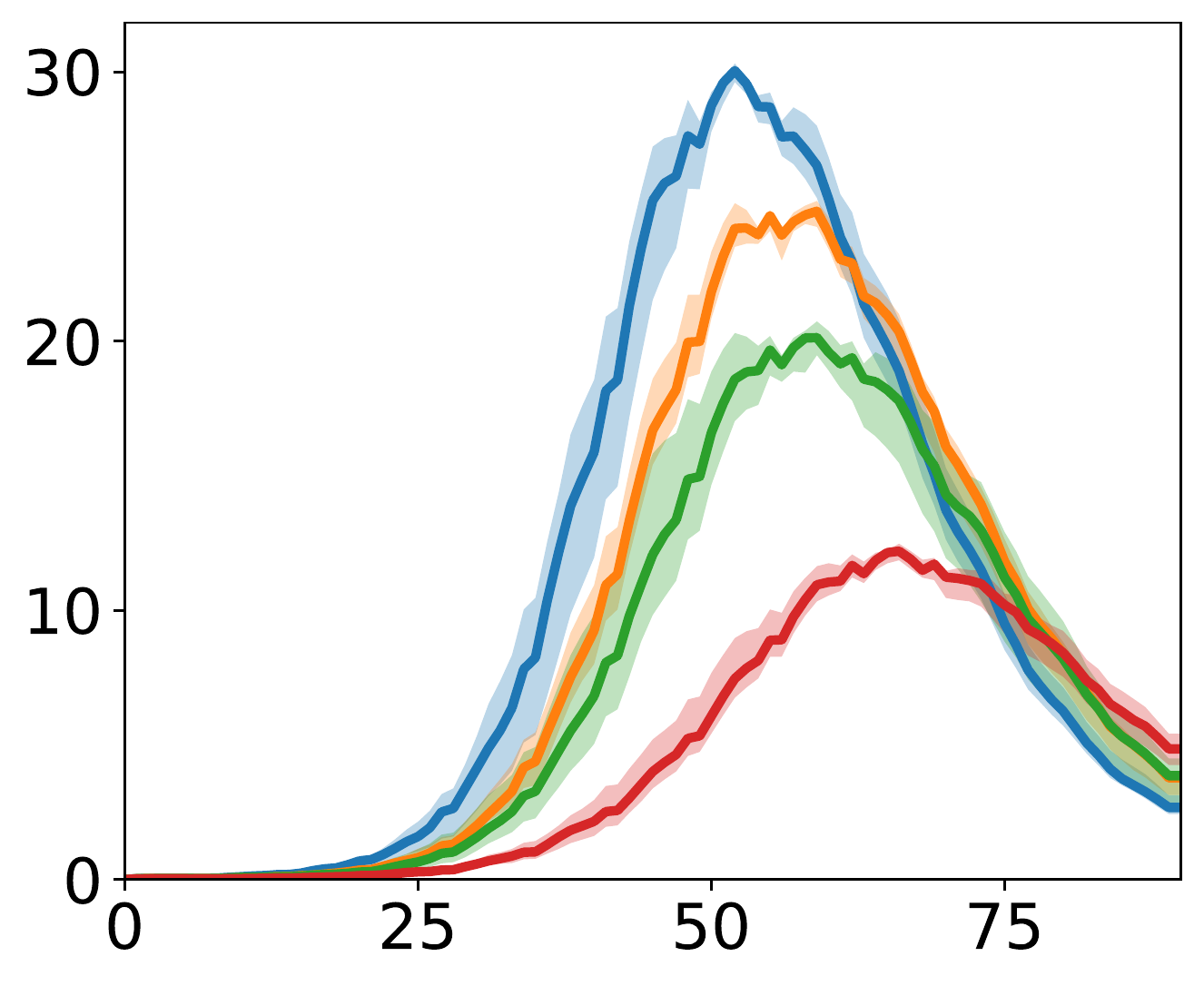}  &&
\includegraphics[width=\fivesmallfig]{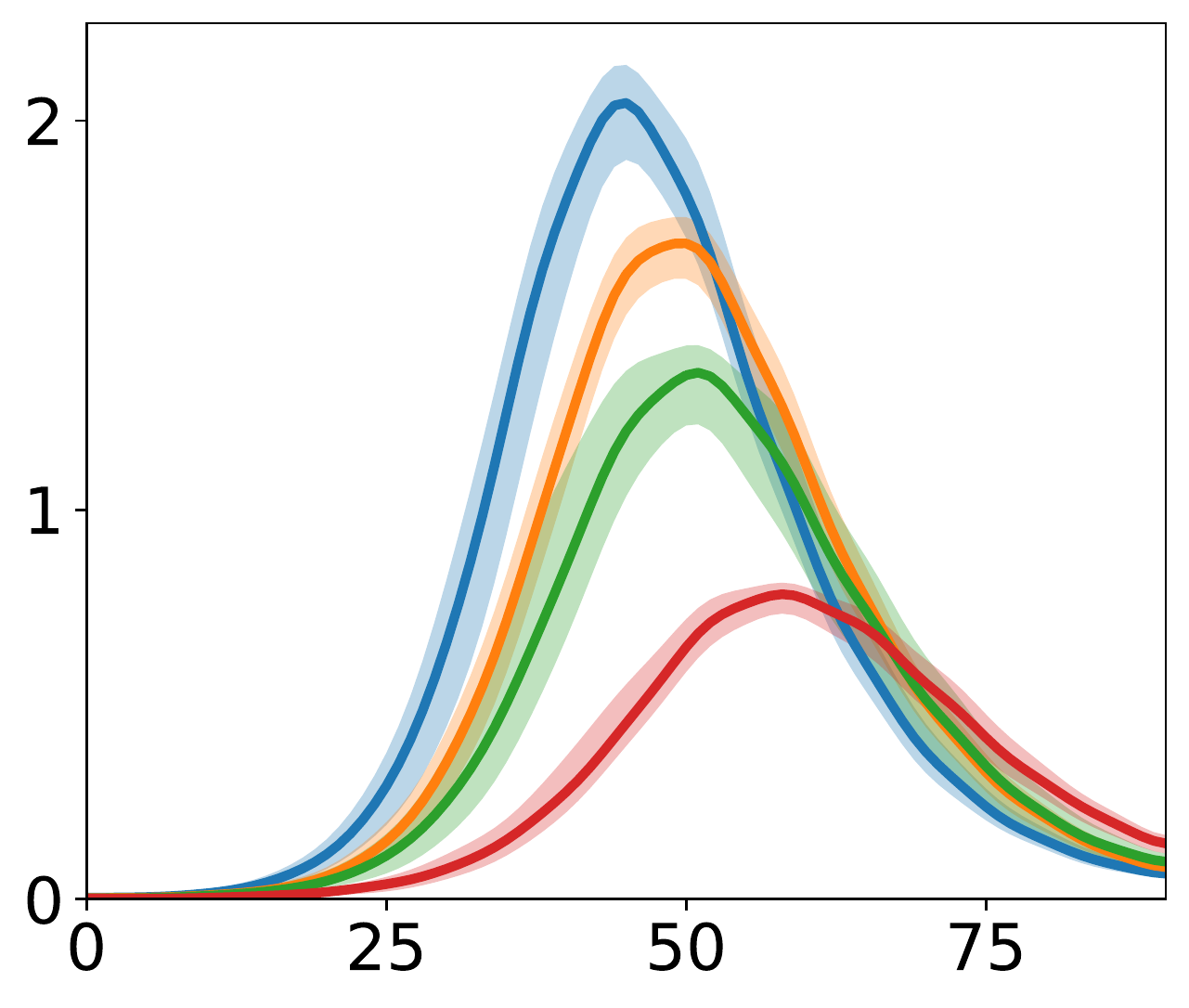}  &&
\includegraphics[width=\fivesmallfig]{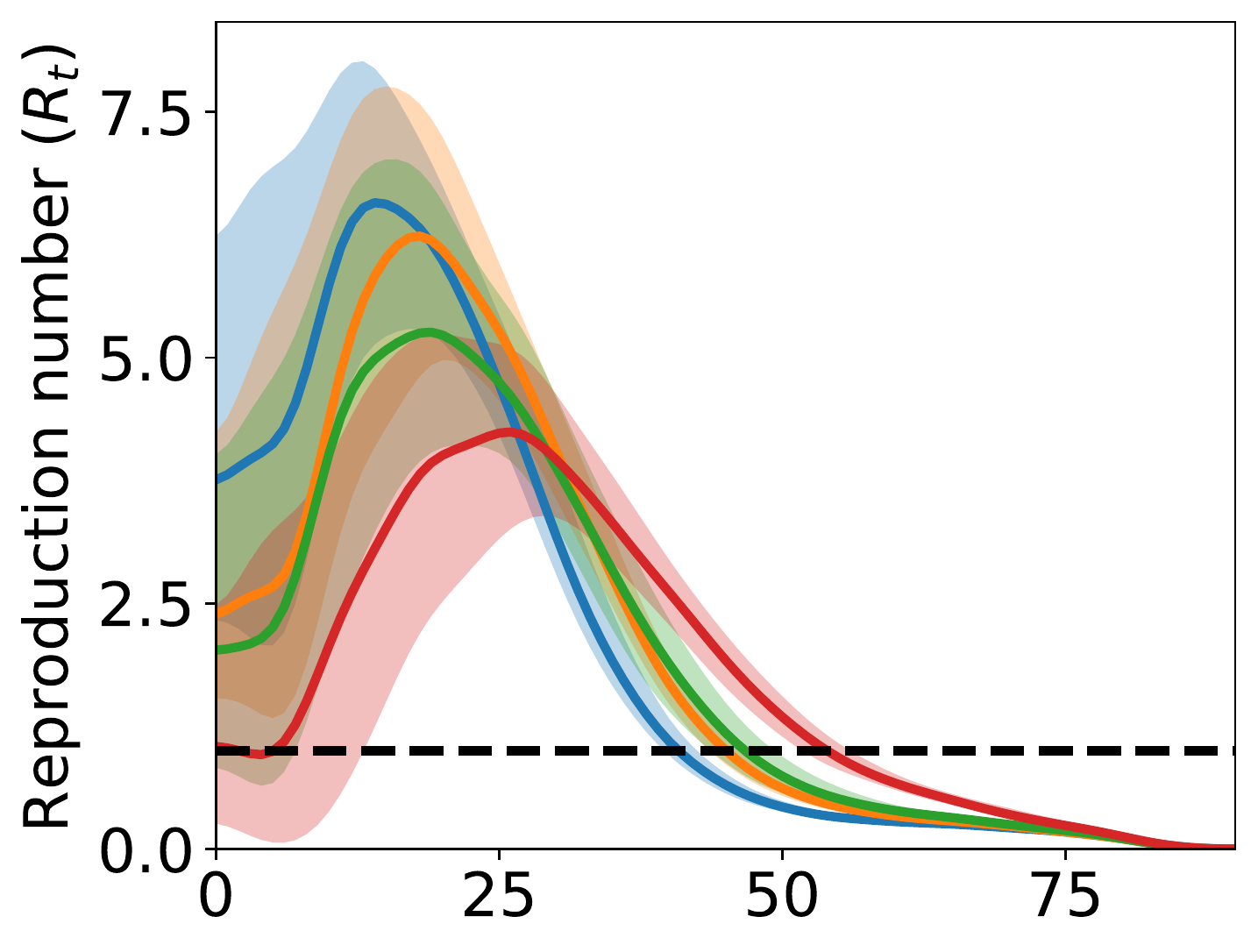} &&
\includegraphics[width=\fivesmallfig]{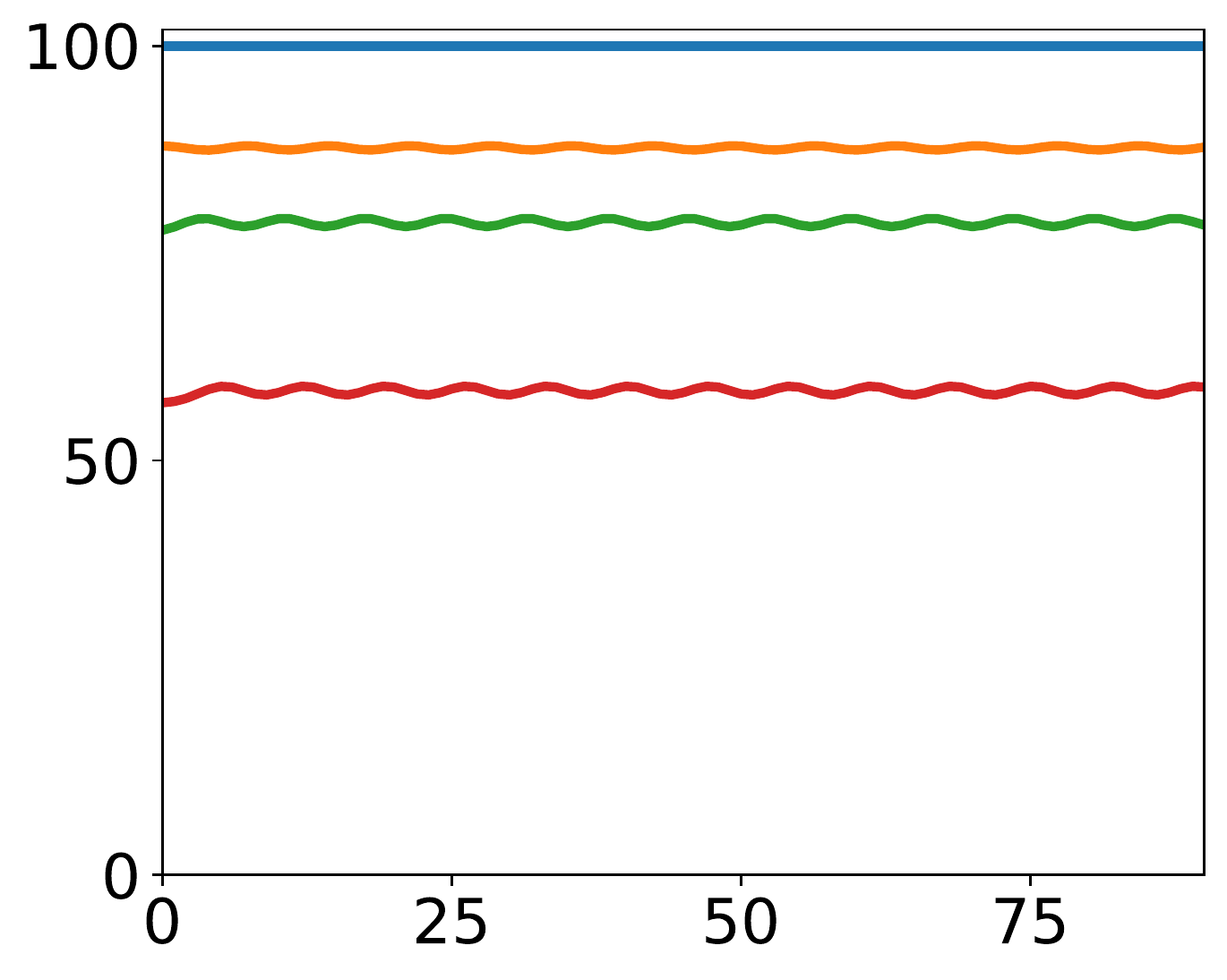} &
\rotatebox{90}{University}
\\ [-0.25cm]

&
\includegraphics[width=\fivesmallfig]{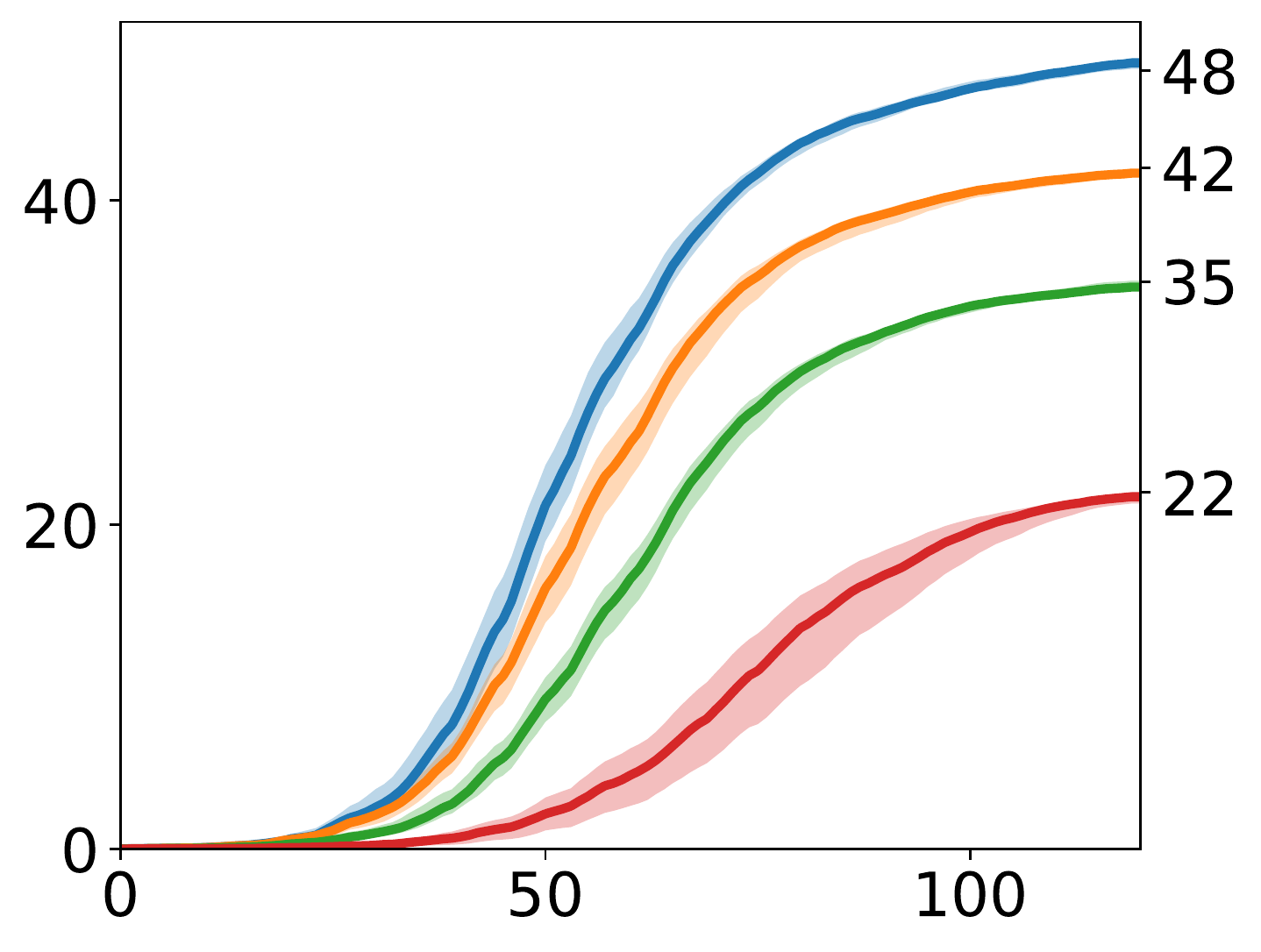} &&
\includegraphics[width=\fivesmallfig]{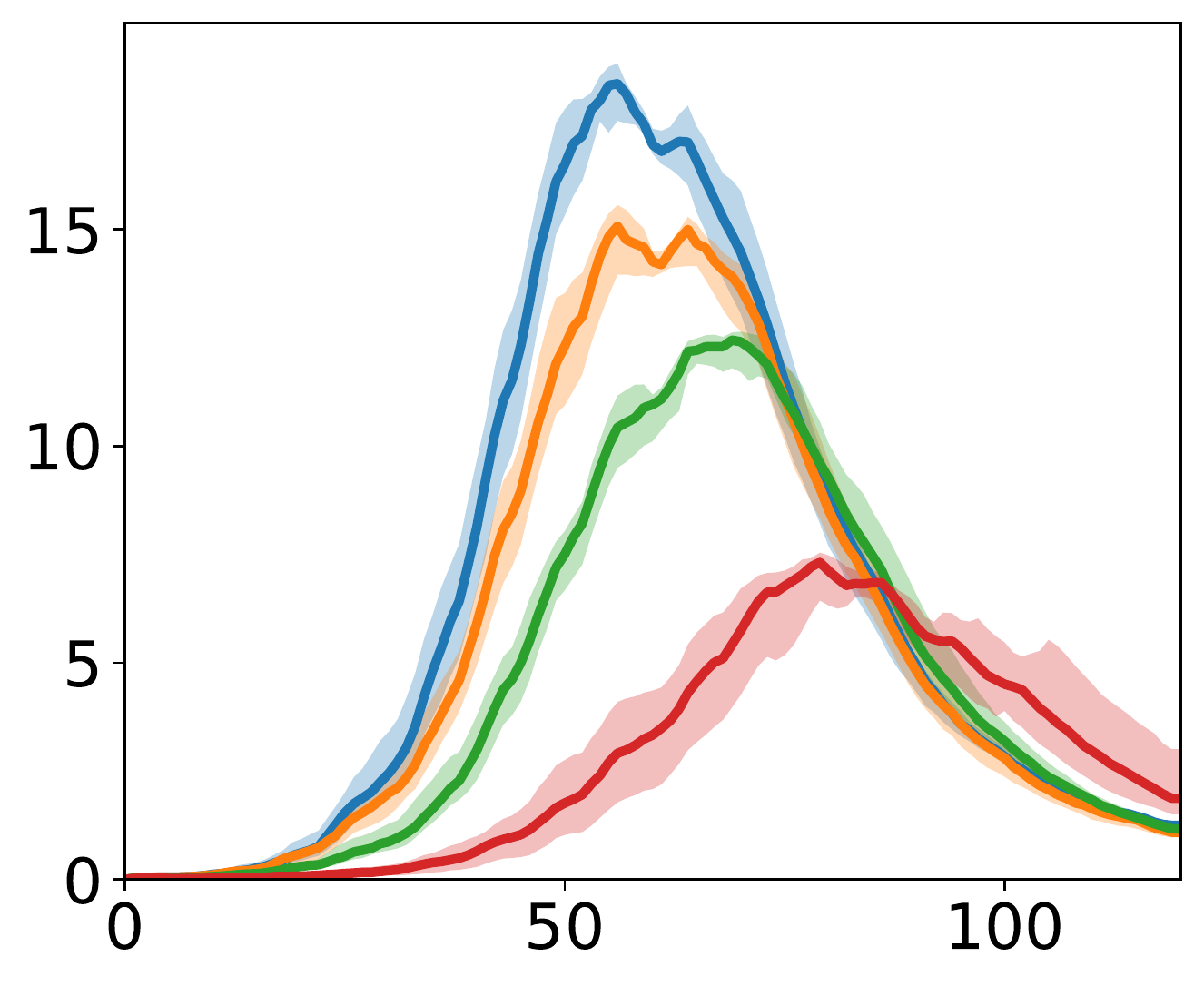}  &&
\includegraphics[width=\fivesmallfig]{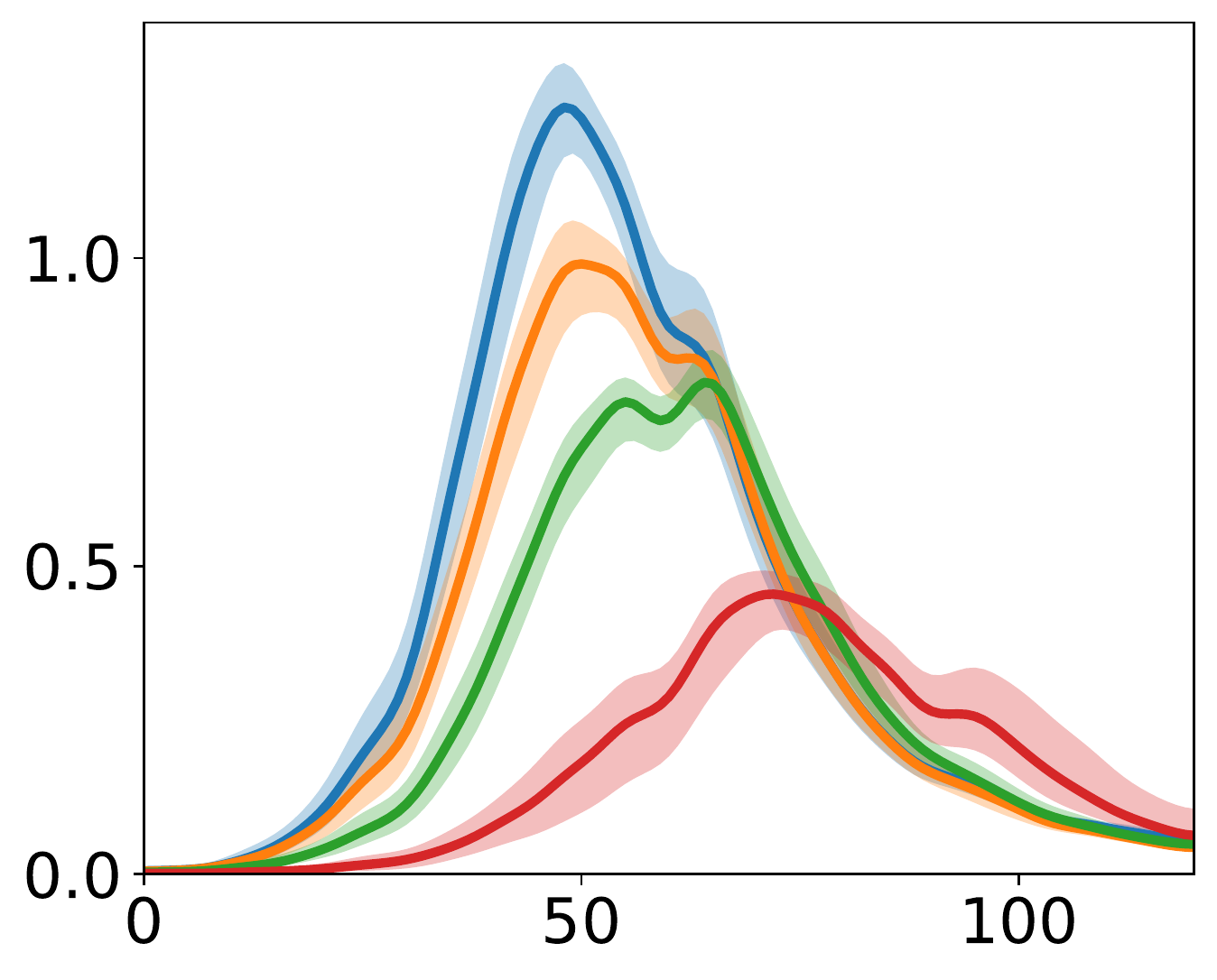}  &&
\includegraphics[width=\fivesmallfig]{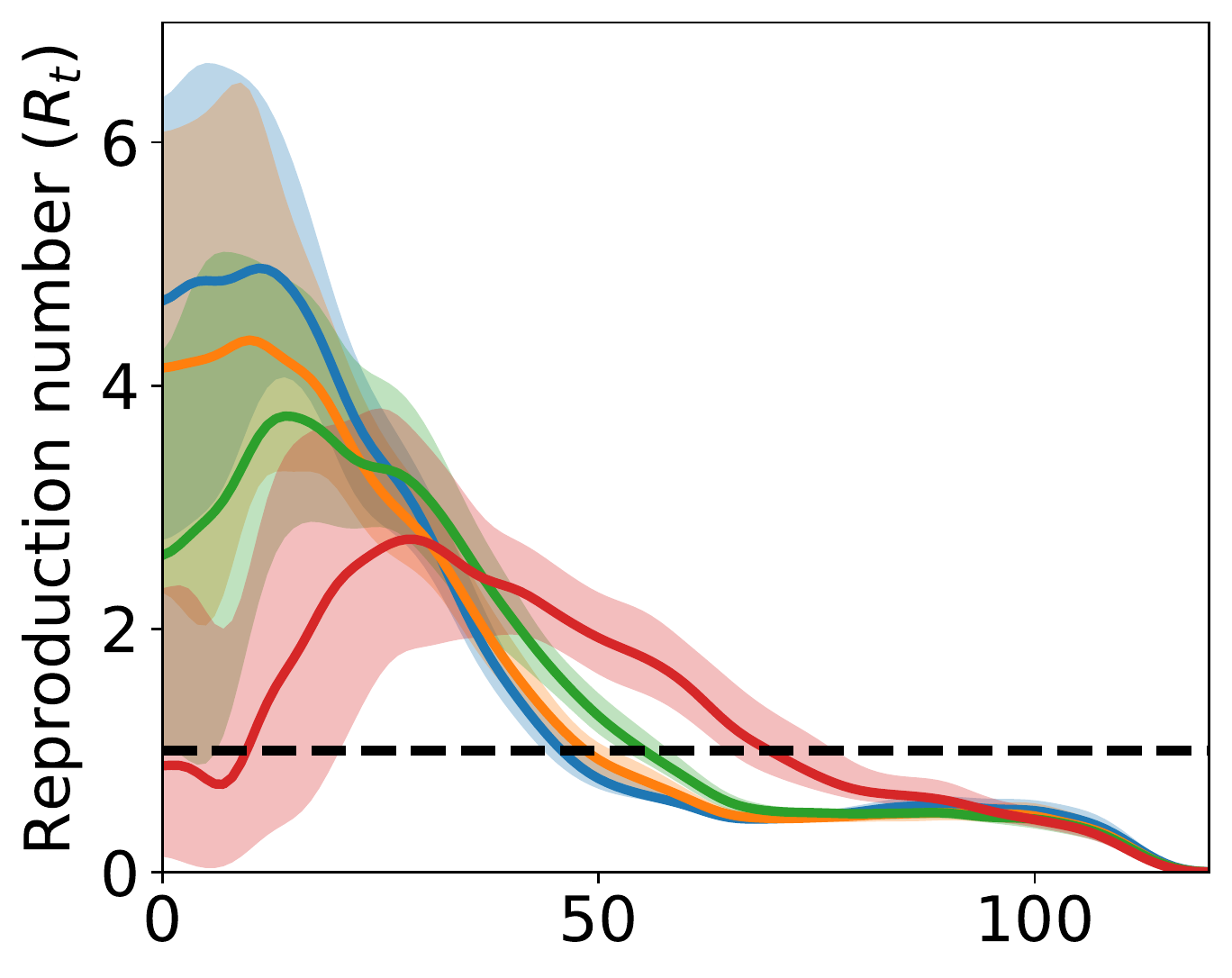} &&
\includegraphics[width=\fivesmallfig]{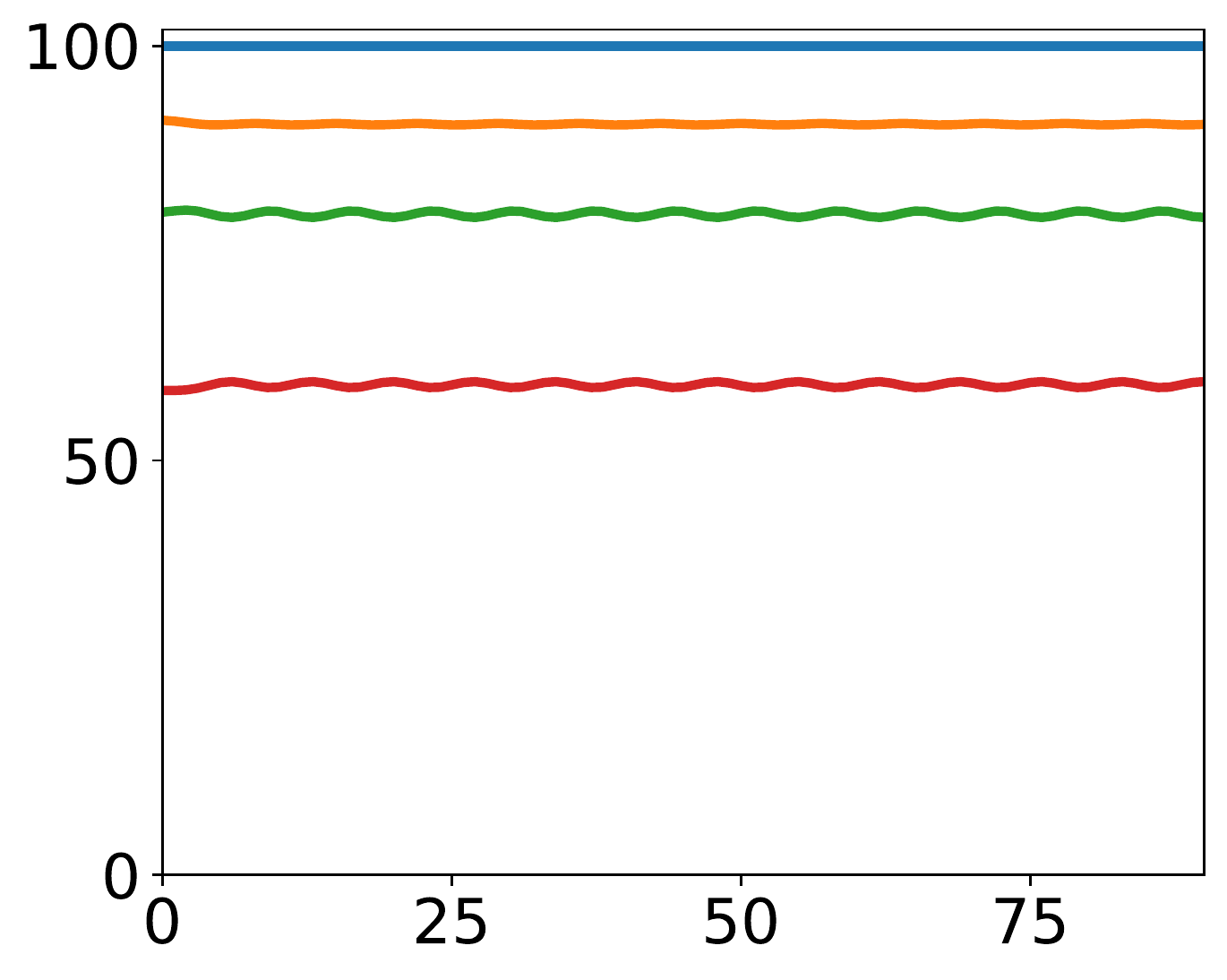} &
\rotatebox{90}{Bike}
\\ [-0.25cm]

\multicolumn{9}{c}{\includegraphics[width=0.15\textwidth]{Fig/labels/daysfromstart.pdf}}\\
\multicolumn{9}{c}{\includegraphics[width=0.6\textwidth]{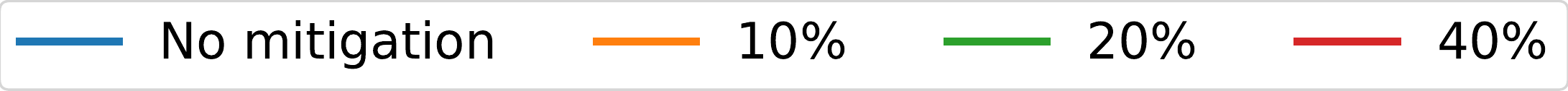}}\\

\end{tabular}
\caption{}
\label{figS:protect_random_people}
\end{figure}
\newpage
\noindent \textbf{Figure~\ref{figS:protect_random_people}:} Infection spreading with the intervention strategy of protecting random people. Compared with protecting most active people, this strategy needs to protect more people to achieve the similar health value. When $40\%$ people are protected, the total number of people infected can be reduced by about $20\%$, and the social value is reduced by $40\%$.

\begin{figure}[!hp]
    \centering
\begin{tabular}{m{0.3\textwidth}m{0.3\textwidth}m{0.3\textwidth}m{0.1cm}}

\includegraphics[width=0.3\textwidth]{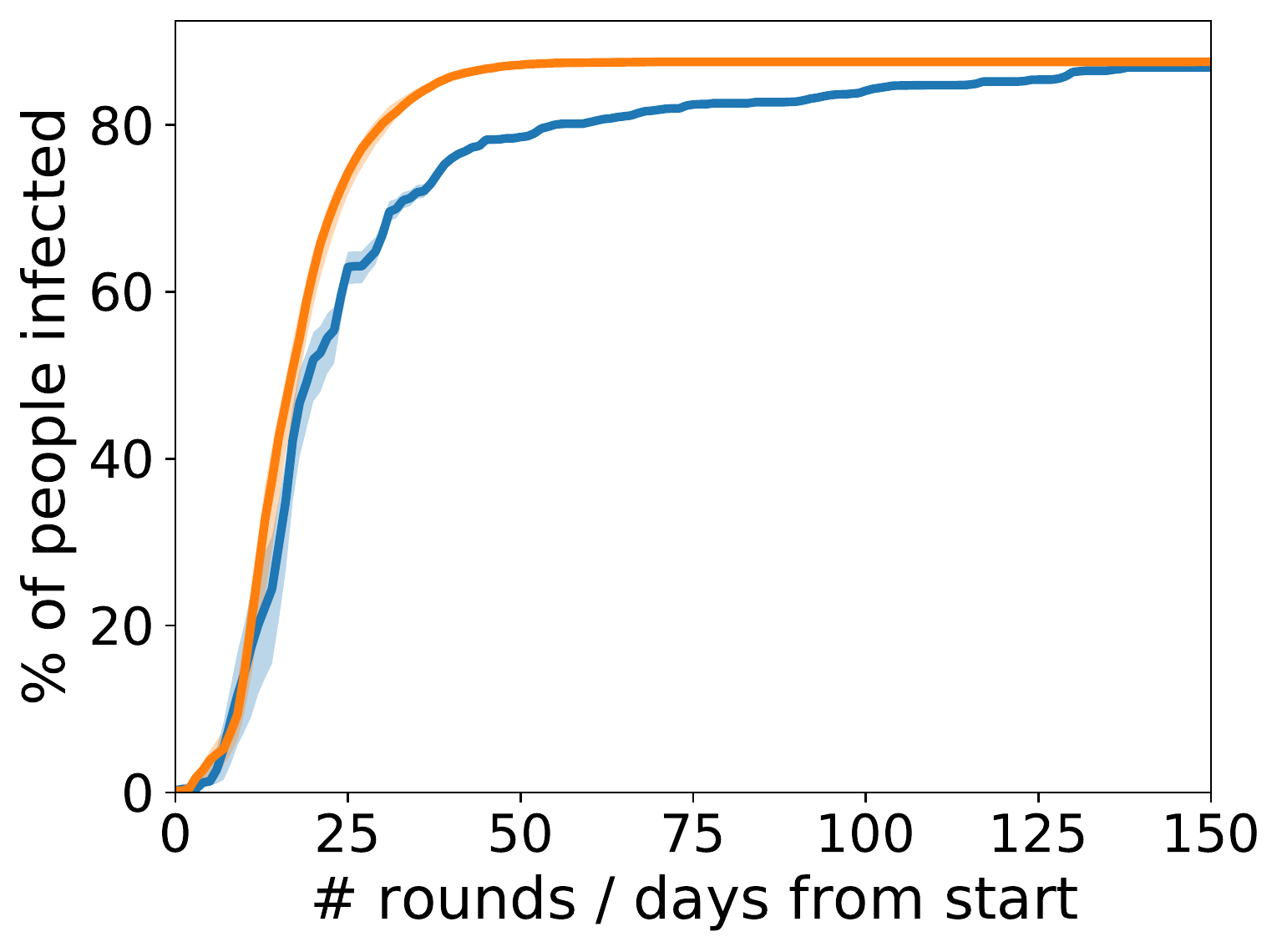} &
\includegraphics[width=0.3\textwidth]{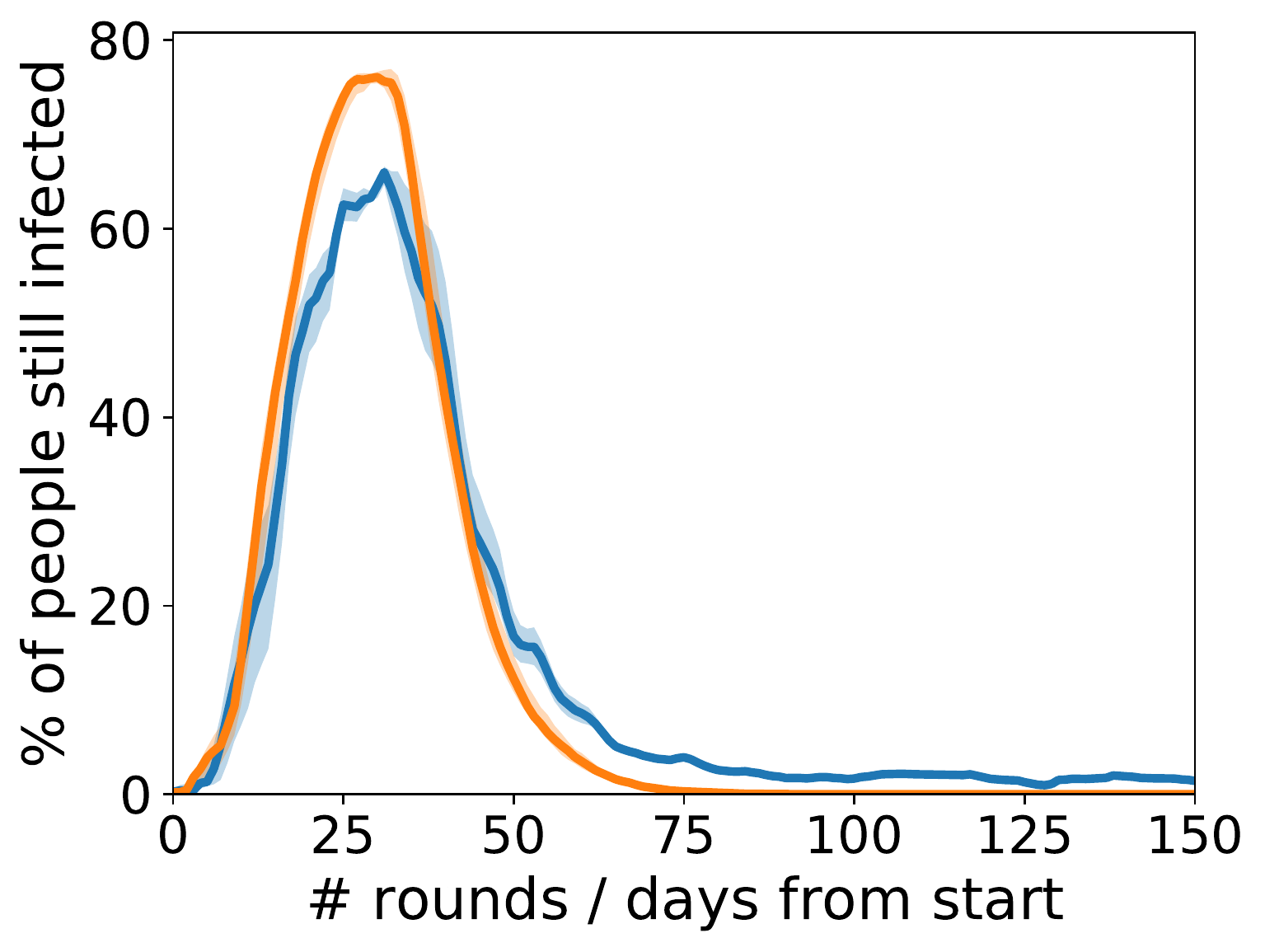}  &
\includegraphics[width=0.3\textwidth]{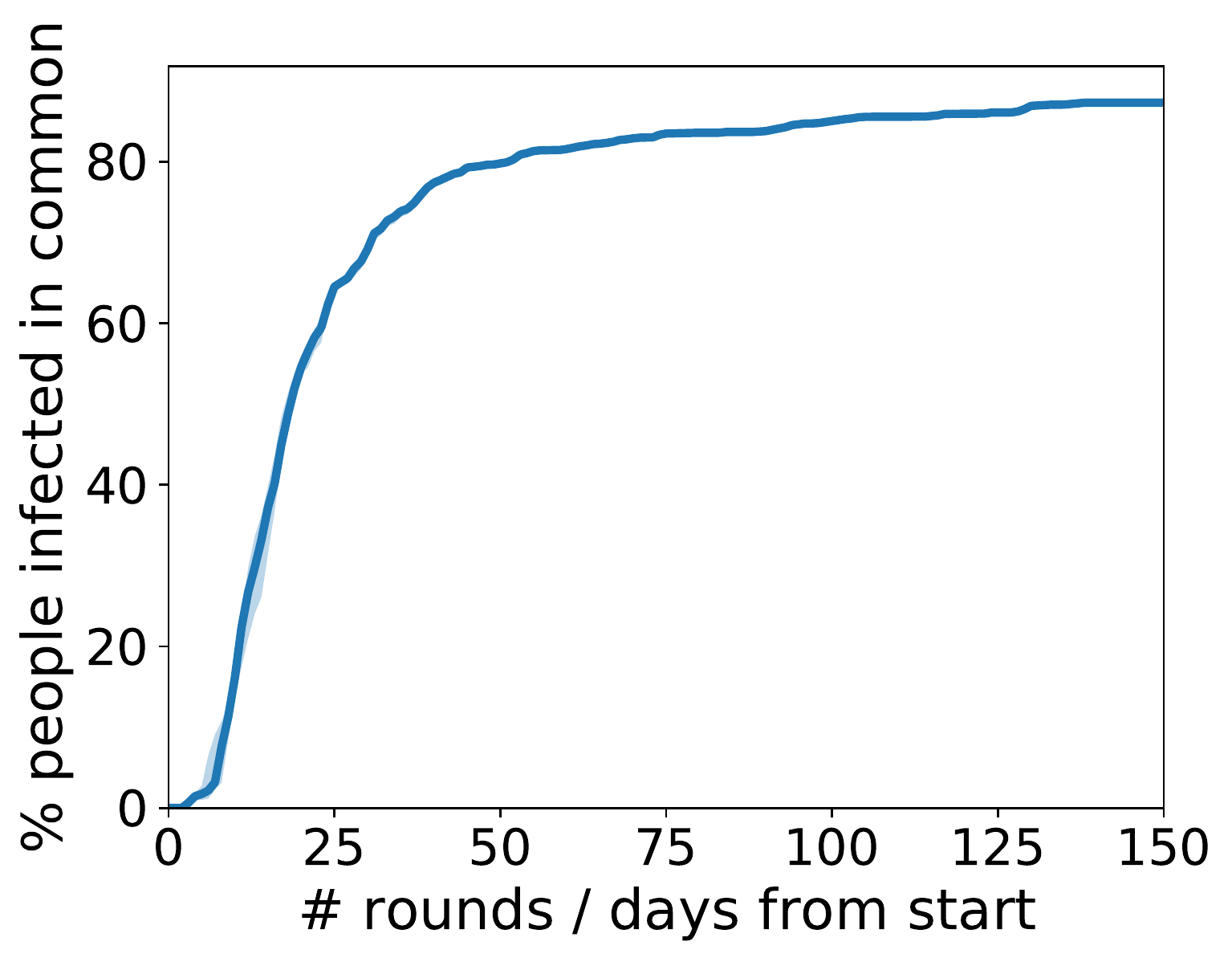}  &
\rotatebox{90}{Tokyo}
\\ [-0.25cm]

\includegraphics[width=0.3\textwidth]{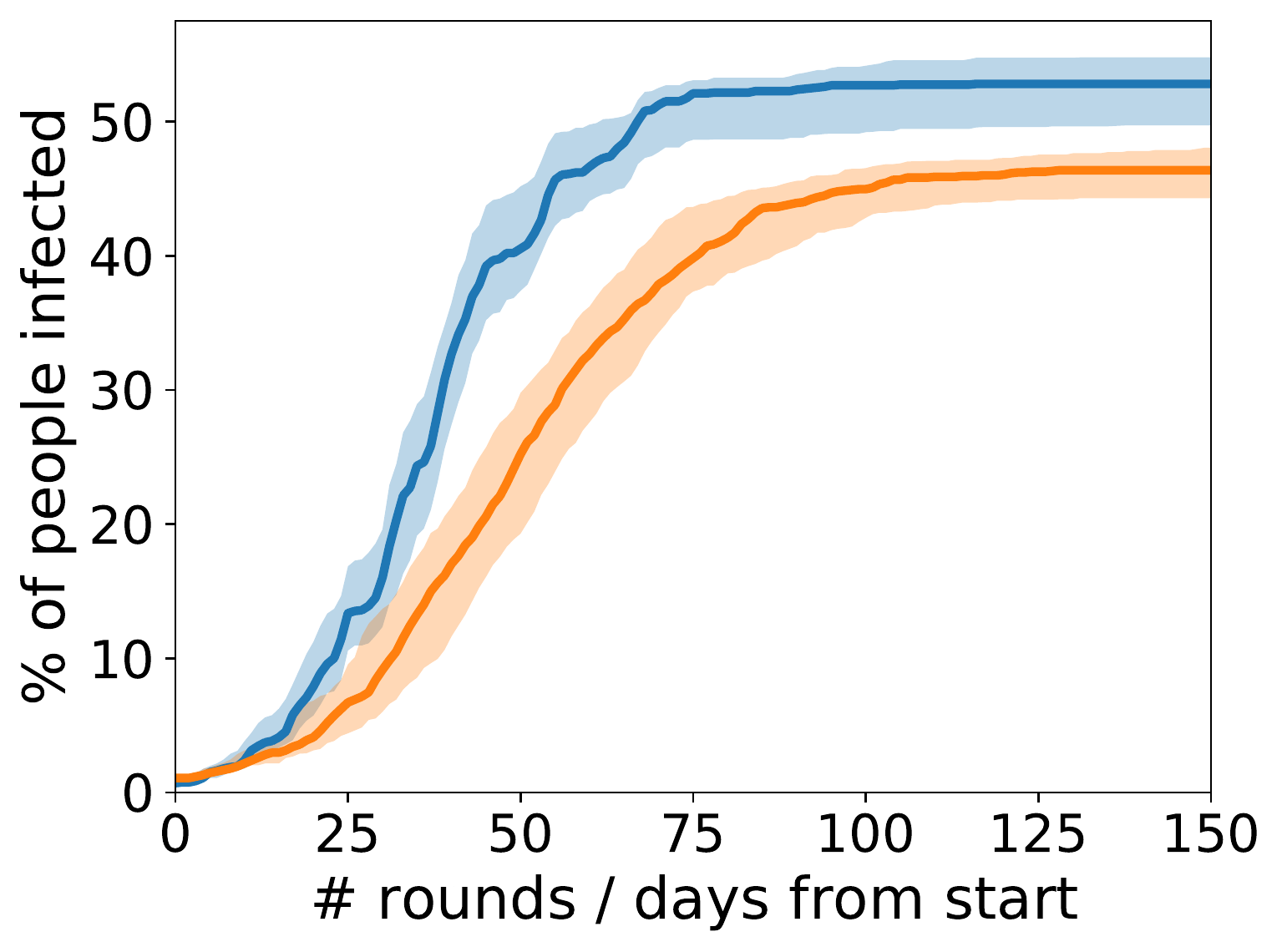} &
\includegraphics[width=0.3\textwidth]{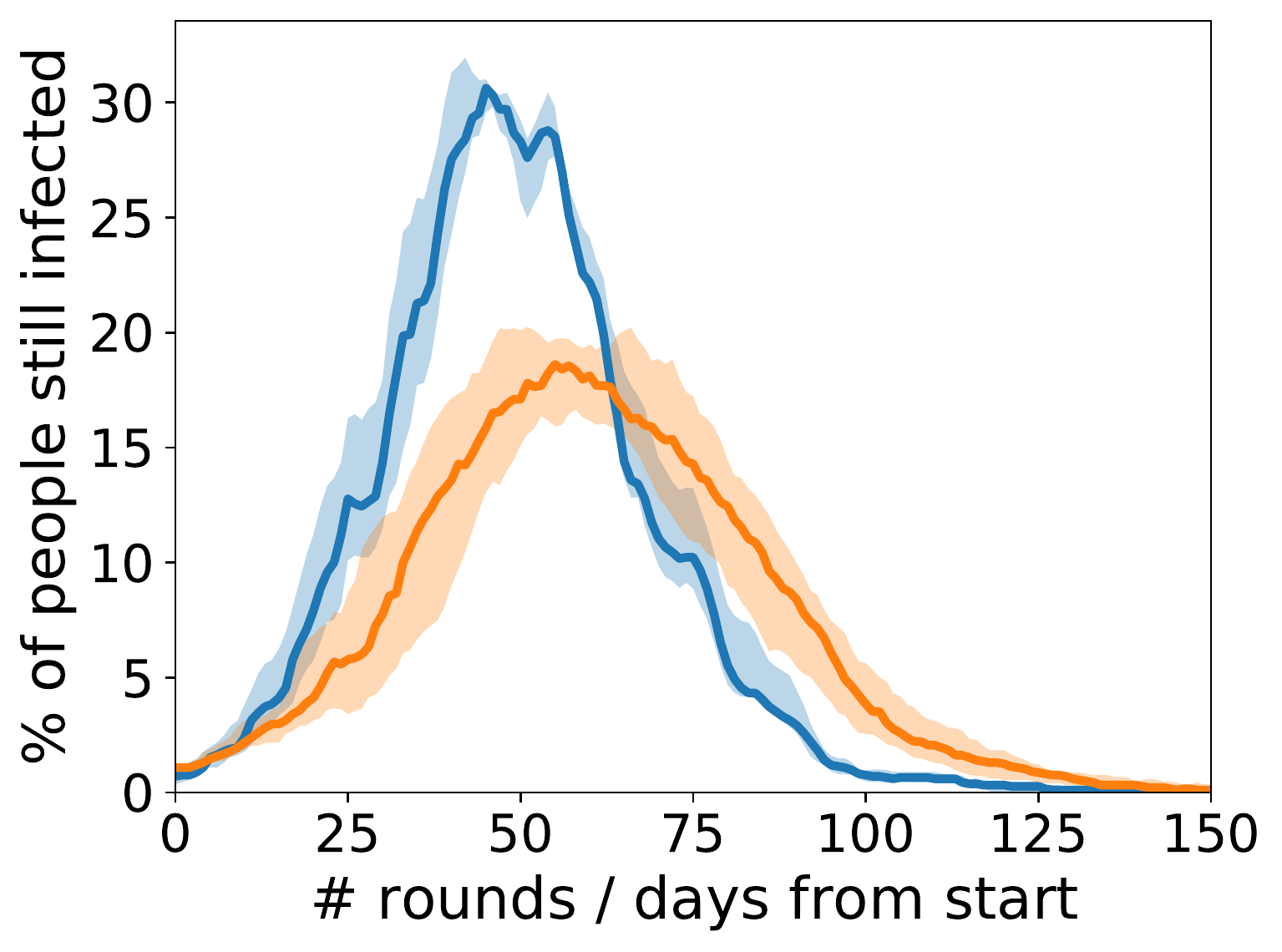}  &
\includegraphics[width=0.3\textwidth]{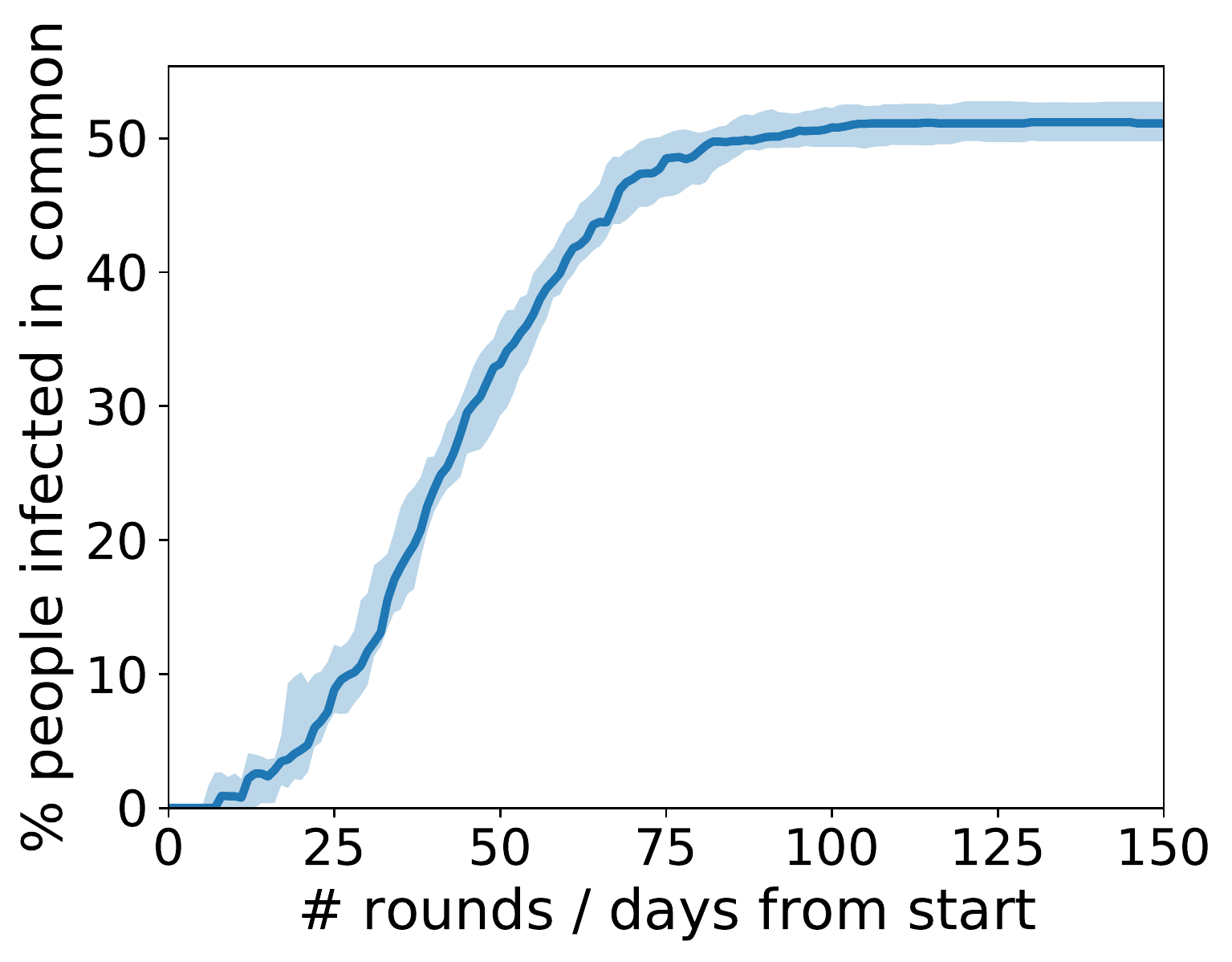}  &
\rotatebox{90}{Chicago}
\\ [-0.25cm]

\includegraphics[width=0.3\textwidth]{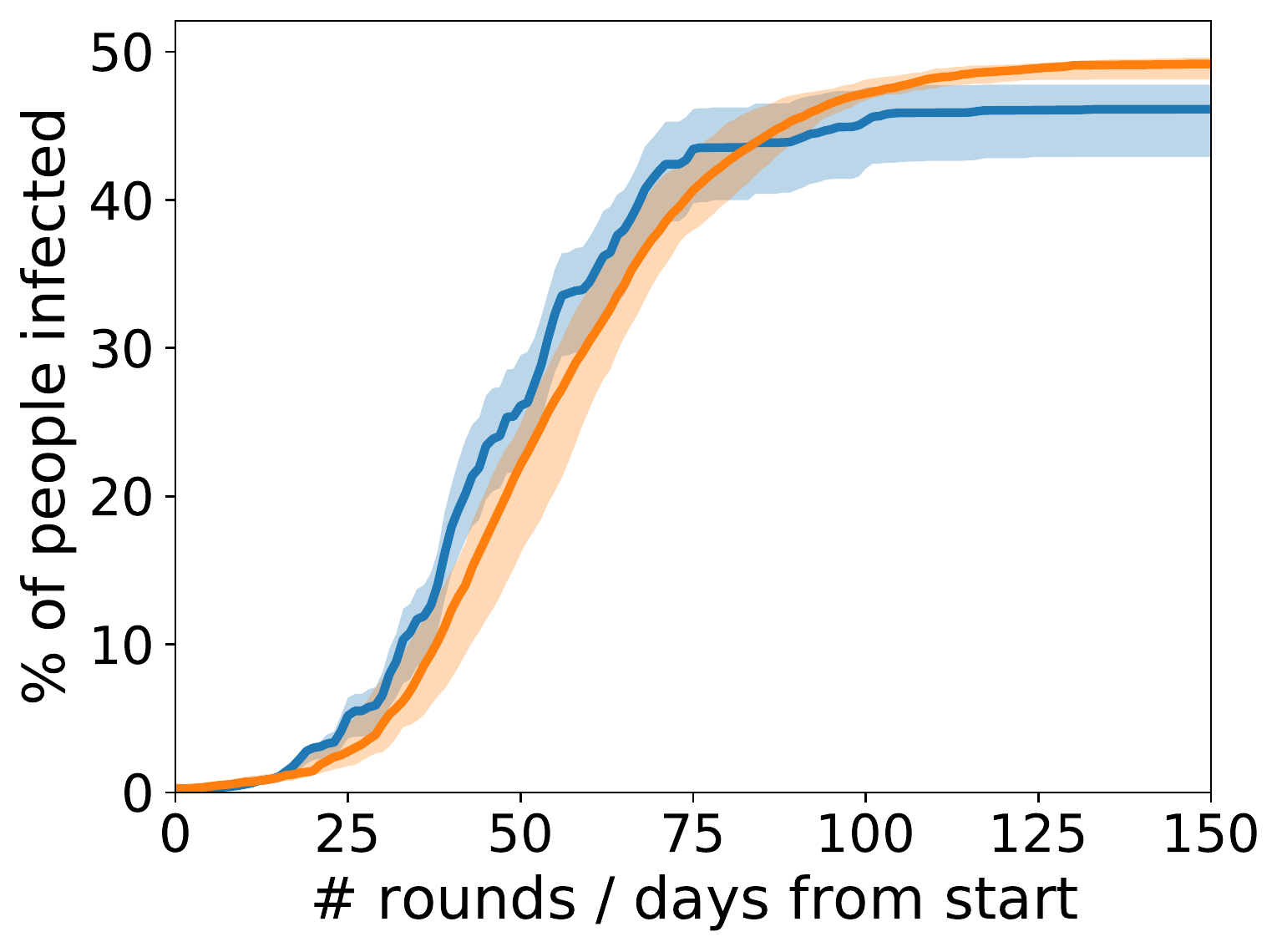} &
\includegraphics[width=0.3\textwidth]{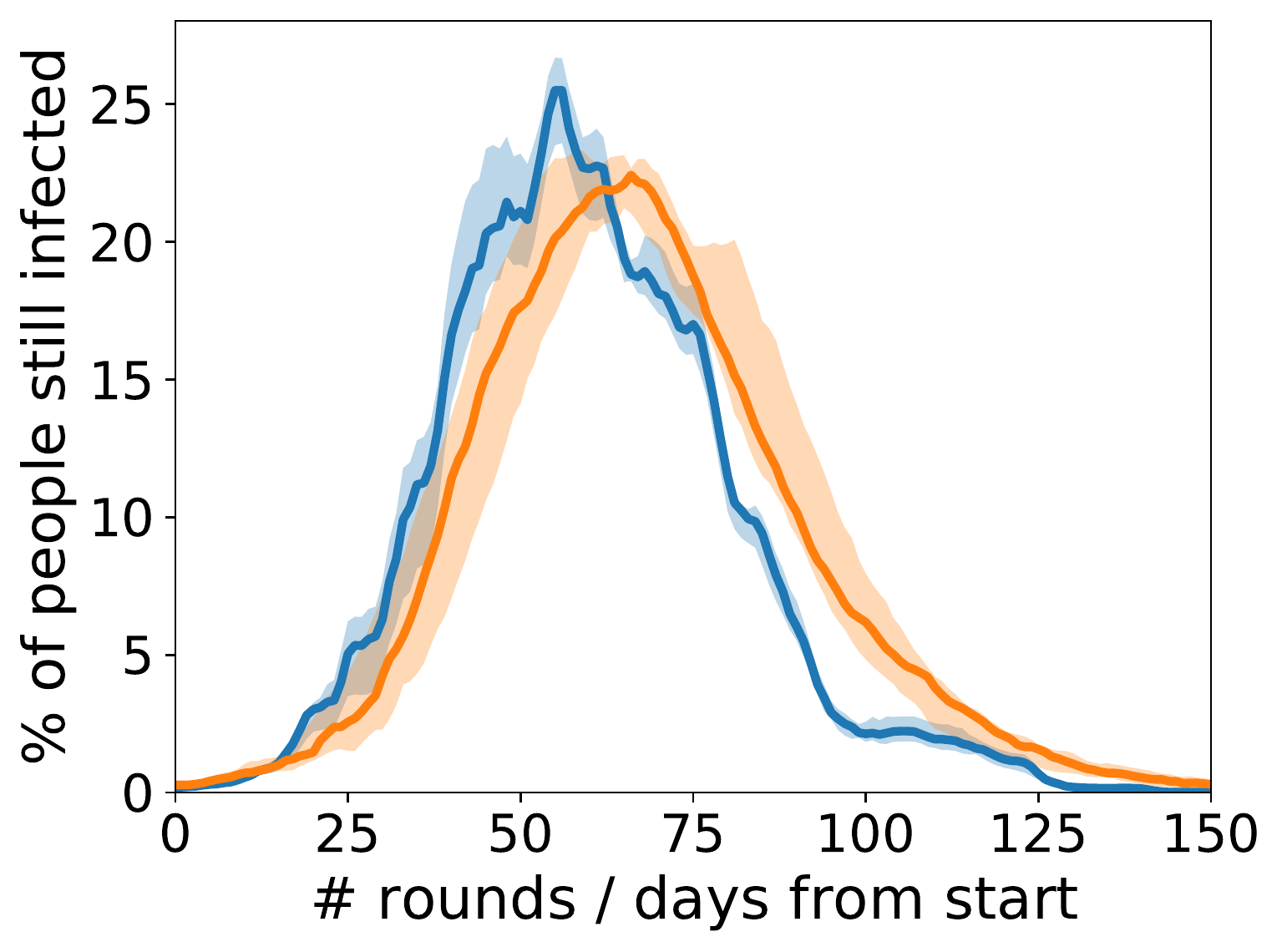}  &
\includegraphics[width=0.3\textwidth]{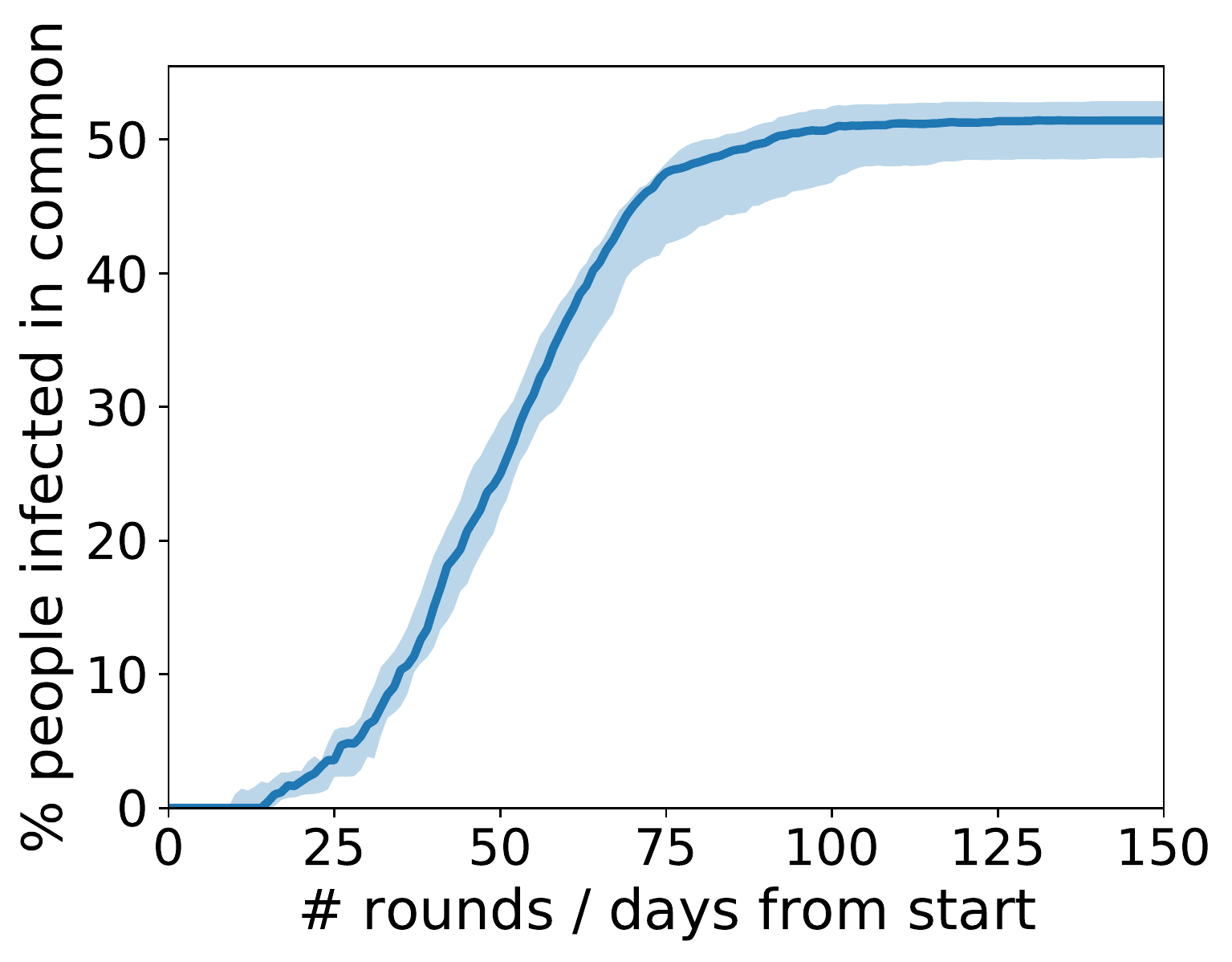}  &
\rotatebox{90}{Los Angeles}
\\ [-0.25cm]

\includegraphics[width=0.3\textwidth]{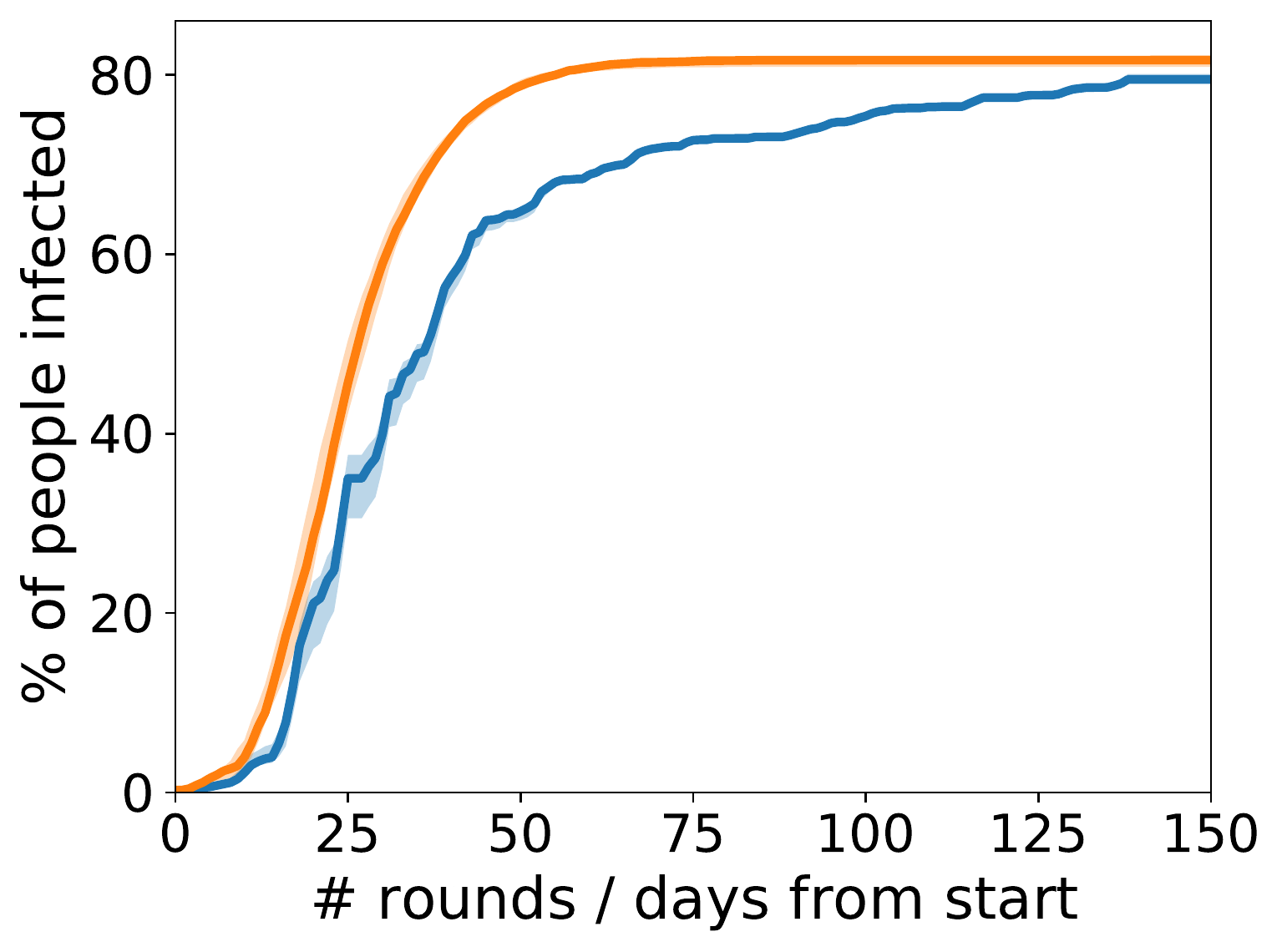} &
\includegraphics[width=0.3\textwidth]{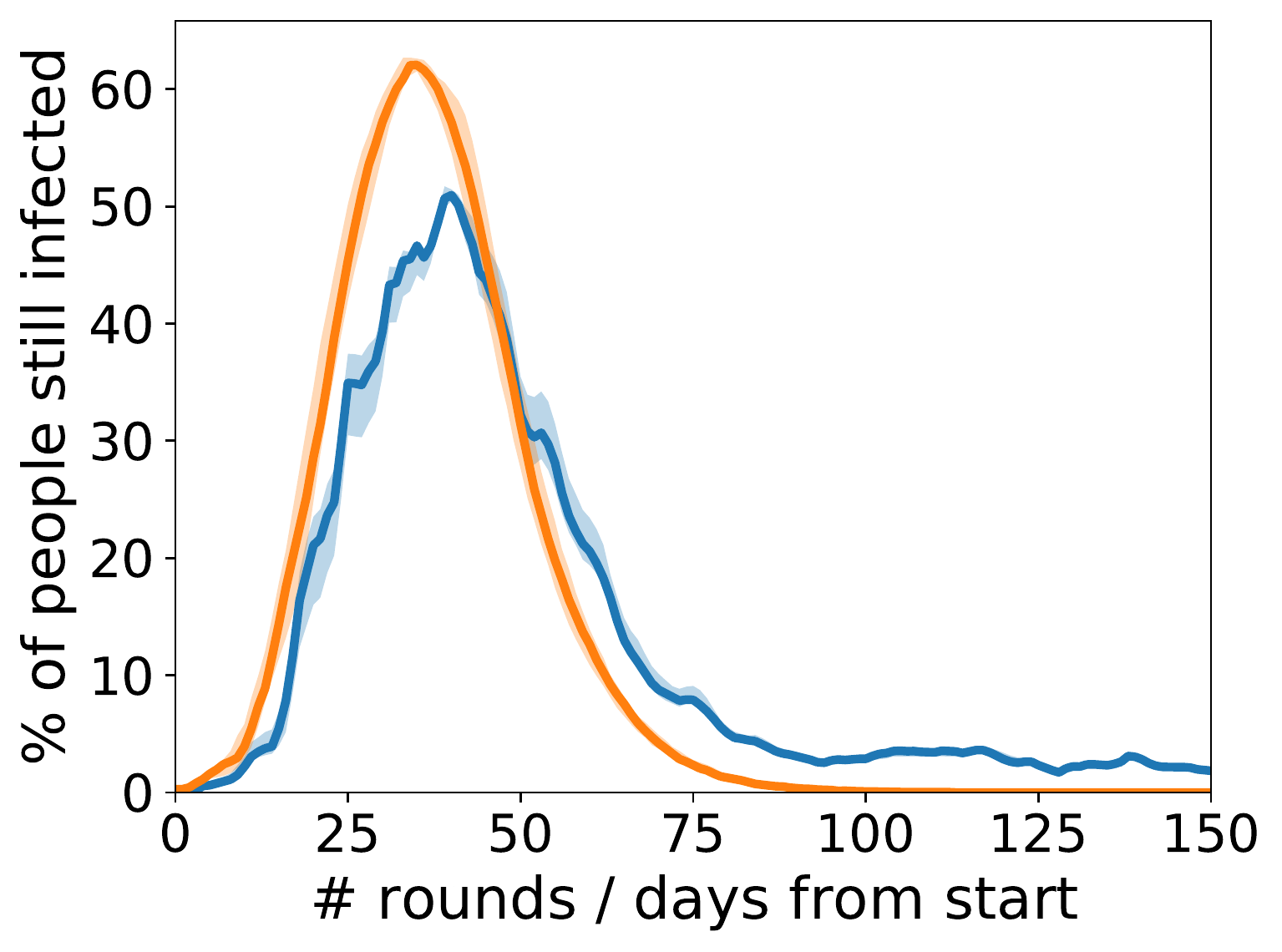}  &
\includegraphics[width=0.3\textwidth]{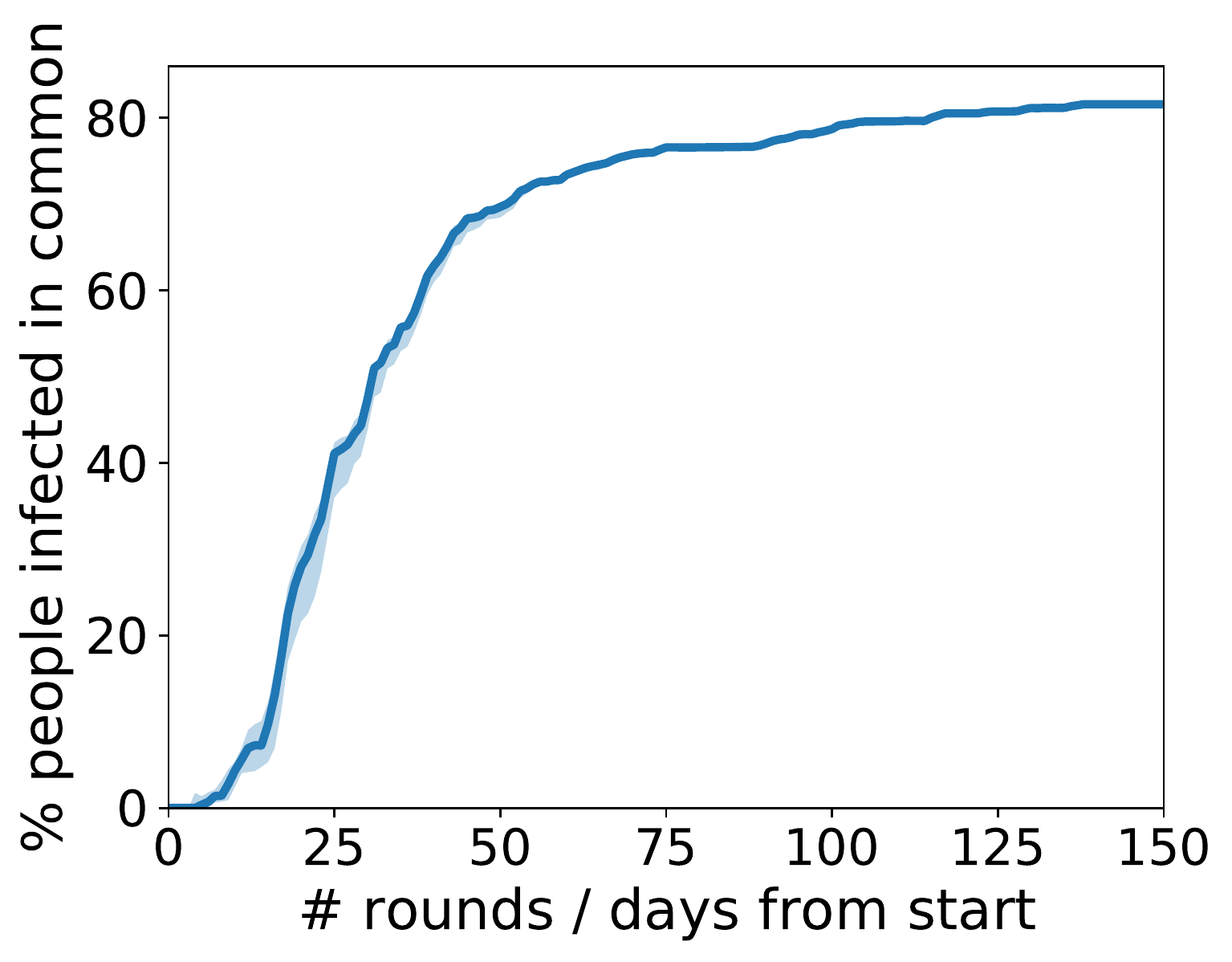}  &
\rotatebox{90}{Jakarta}
\\ [-0.25cm]

\includegraphics[width=0.3\textwidth]{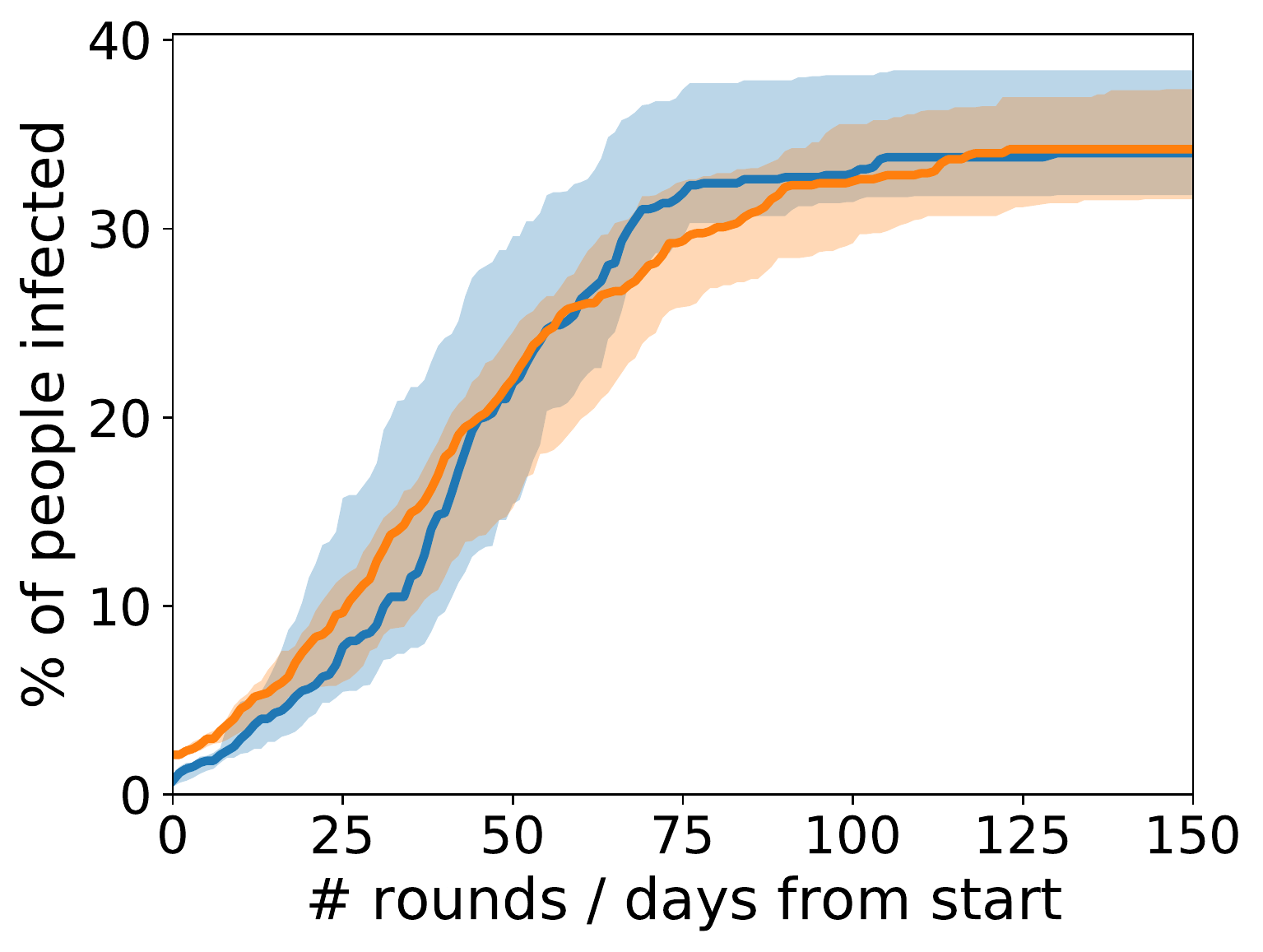} &
\includegraphics[width=0.3\textwidth]{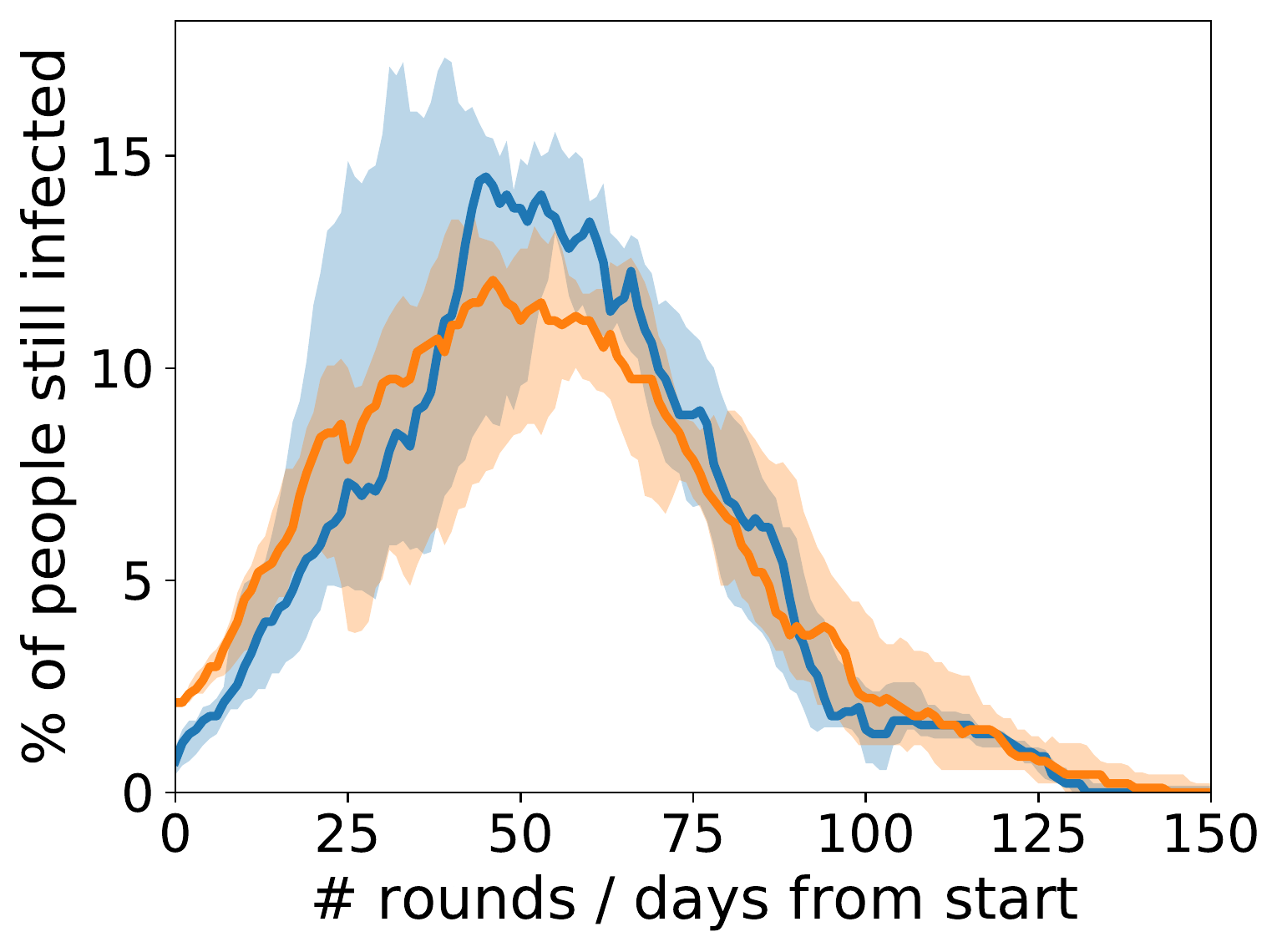}  &
\includegraphics[width=0.3\textwidth]{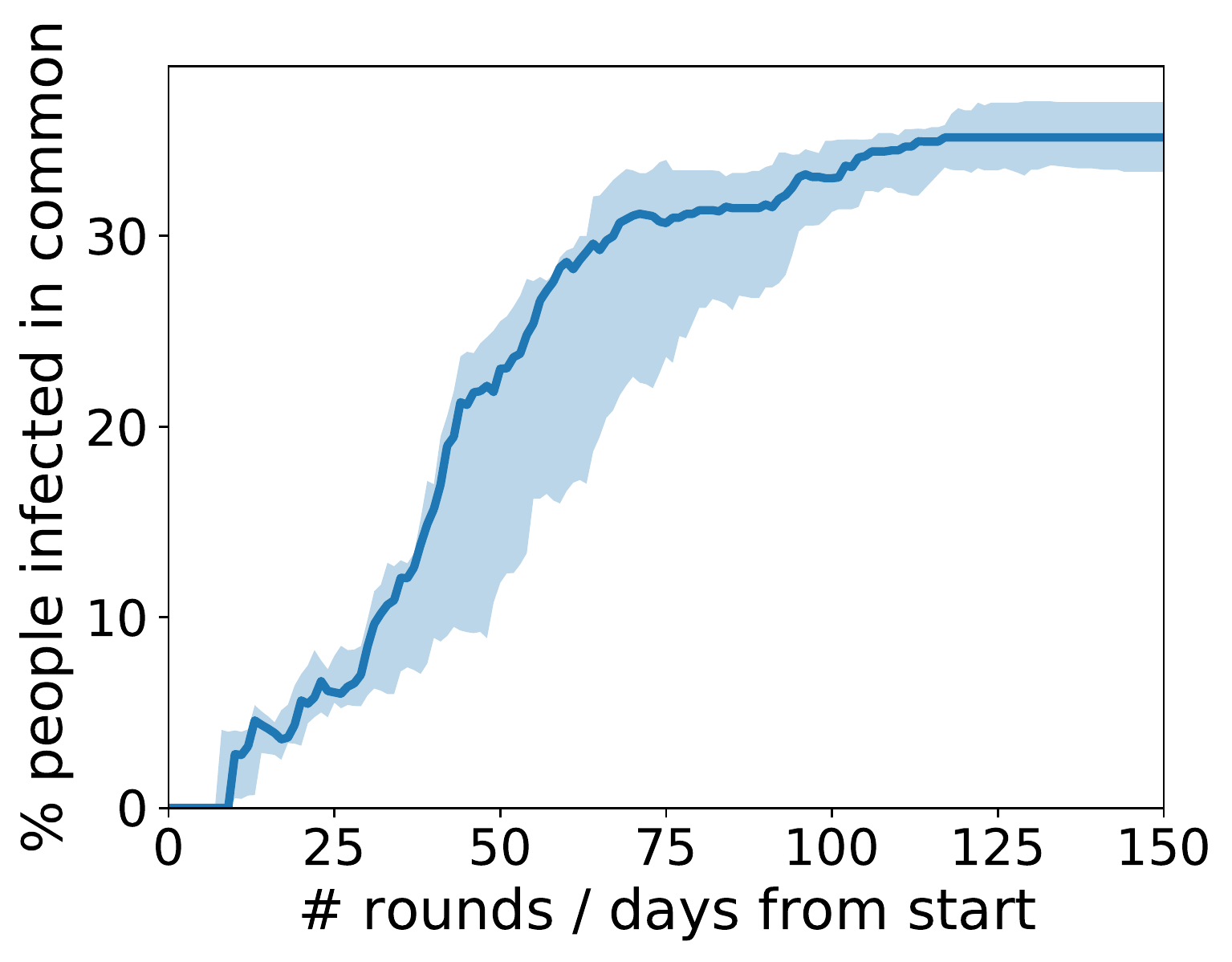}  &
\rotatebox{90}{London}
\\ [-0.25cm]

\end{tabular}
\caption{Infection spreading between mobility network and bipartite social networks. The infection curves between the two models are matched and the infection people in common also have a large percentage. }
\label{figS:comparison_social}
\end{figure}

\begin{figure}[!hp]
    \centering
    \includegraphics[width=0.83\textwidth]{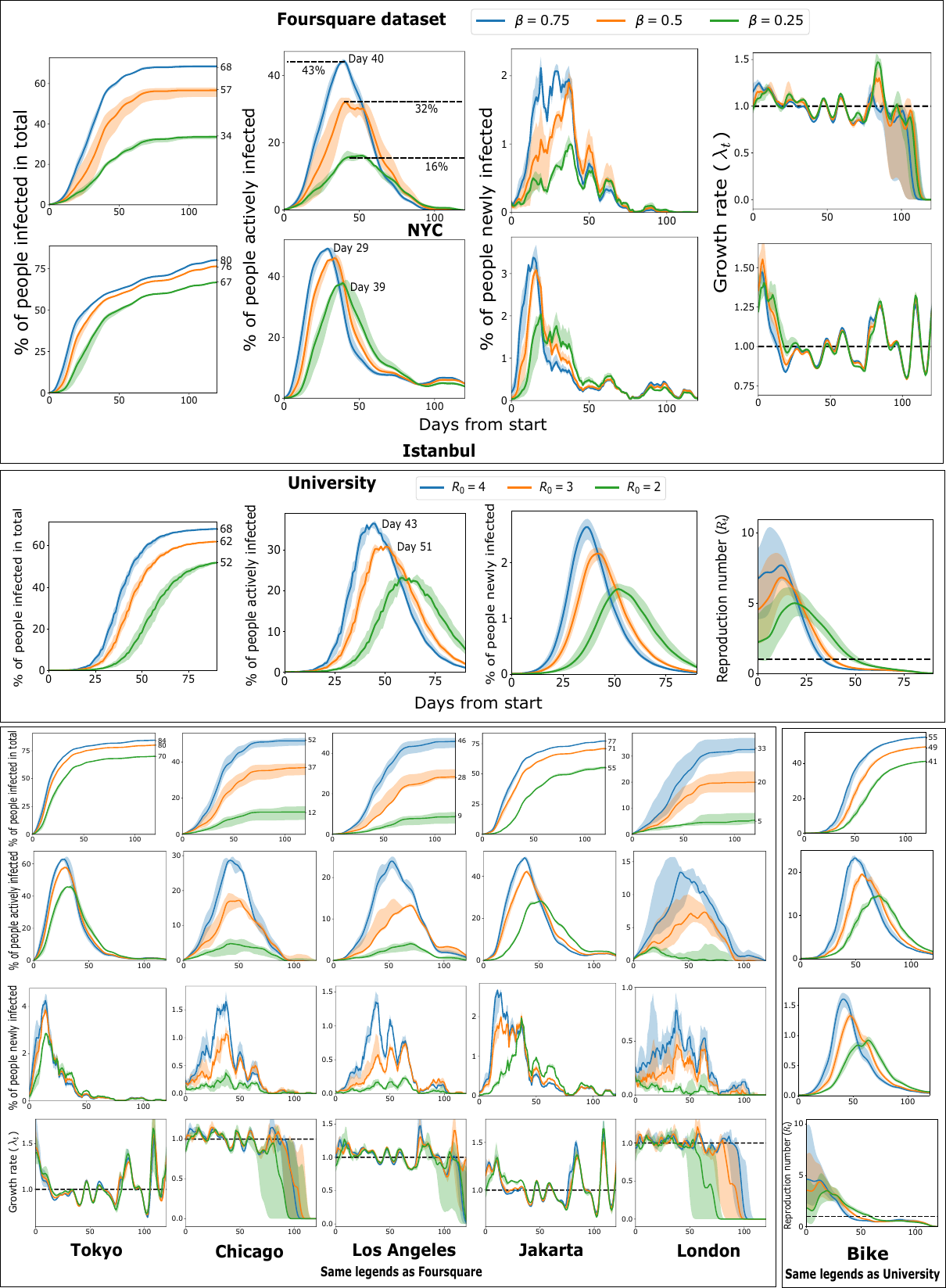}
    \caption{Infection spreading with varied transmission probabilities. In the Foursquare datasets, the transmission probabilities are provided directly. In the University and Bike datasets, the transmission probabilities are translated from $R_0$. With higher transmission probability values, the total number of people infected increase, and the peak of active cases get higher.}
    \label{figS:infection_probability}
\end{figure}
\newpage
\begin{figure}[H]
    \centering
    \includegraphics[width=0.85\textwidth]{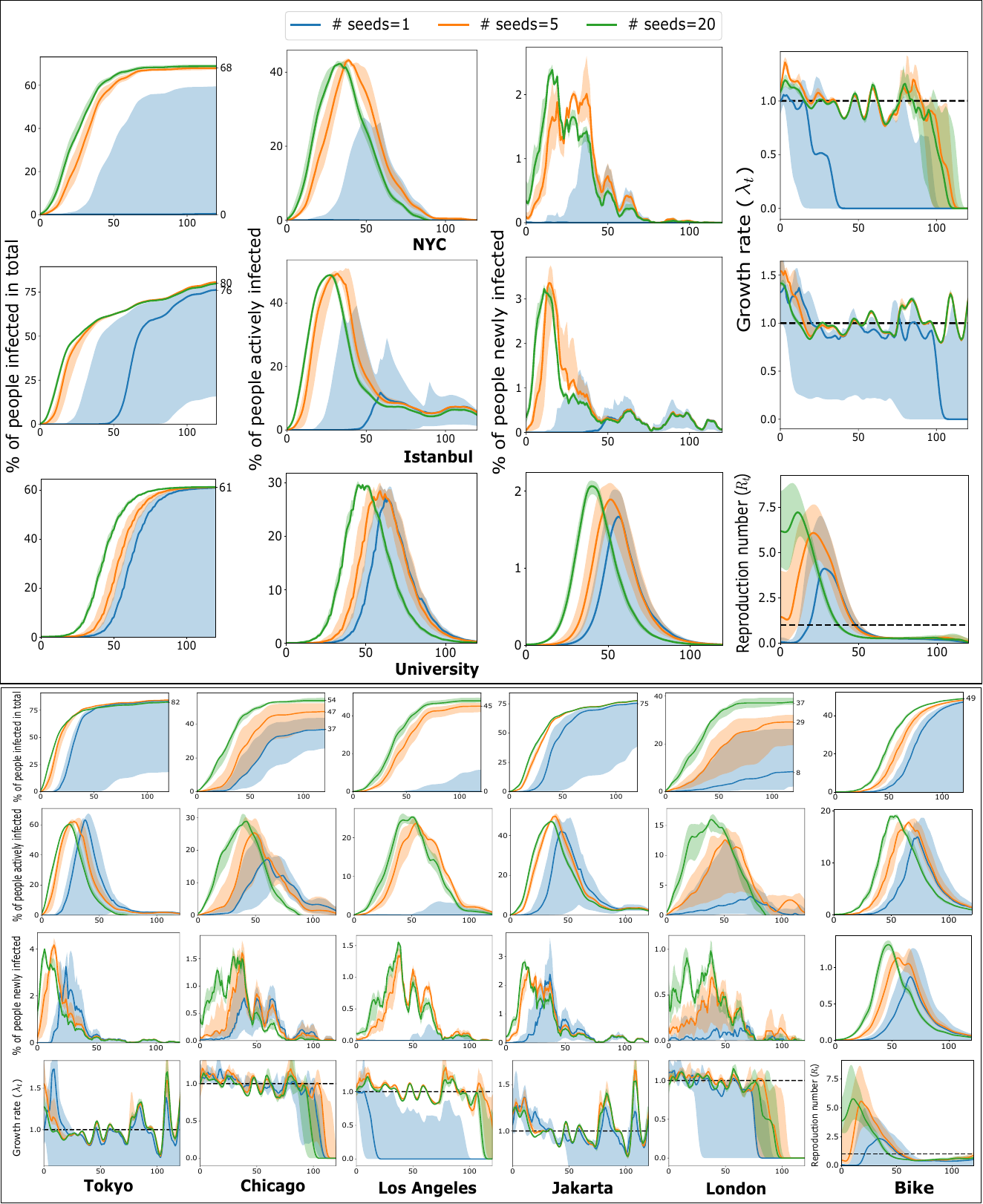}
    \caption{Infection spreading with a varied number of seeds. With a small number of initial seeds, the variance of infection is very large (shaded area). In most cases, when the number of seeds is enough, the total infection numbers and the peaks of people infected actively do not have a large difference.}
    \label{figS:number_seeds}
\end{figure}

\newpage
\begin{figure}[H]
    \centering
    \includegraphics[width=0.95\textwidth]{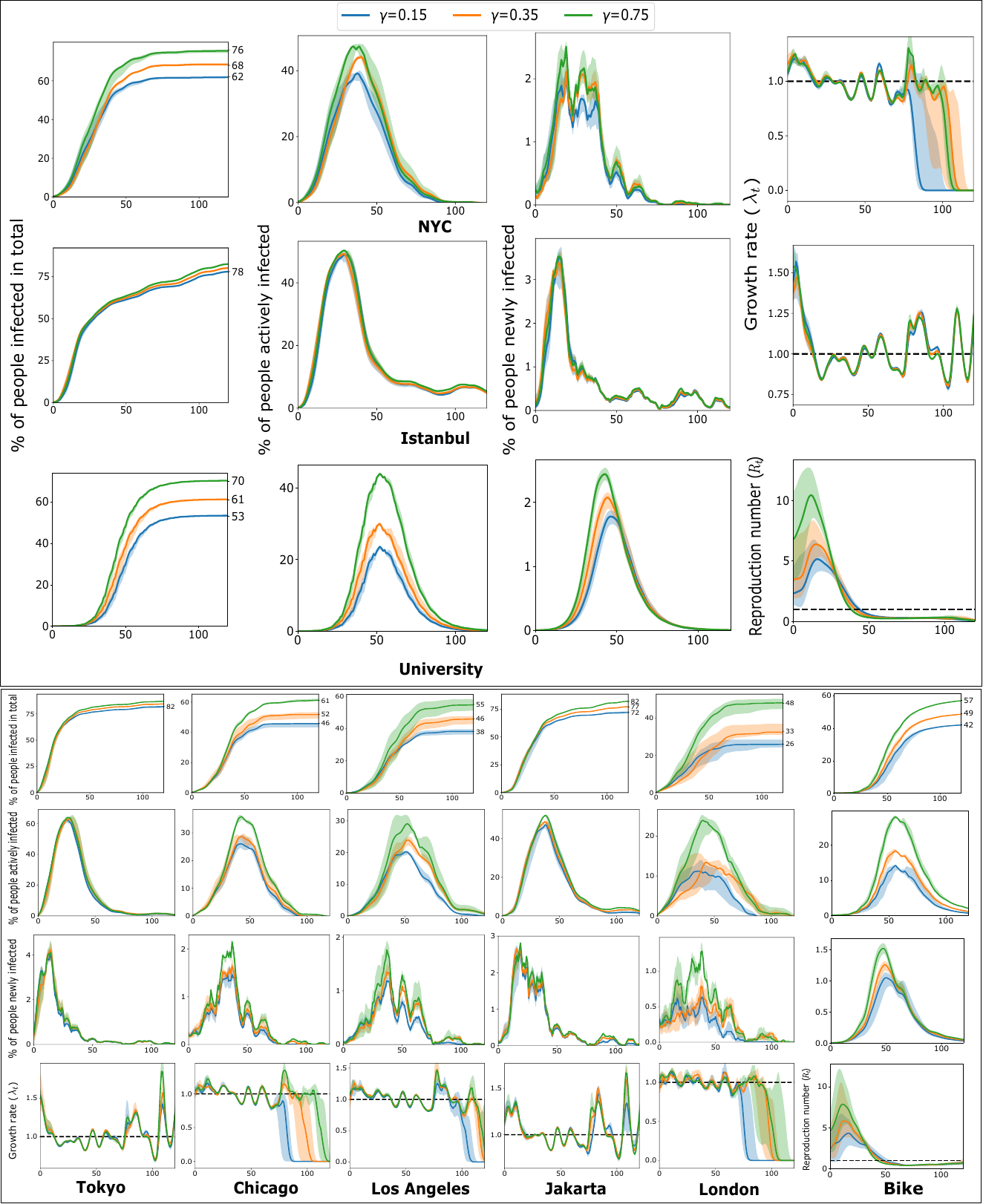}\\
    \caption{Infection spreading with the varying probabilities of being asymptomatic. With a high asymptomatic probability, there will be more infected people who have a longer duration to infect other susceptible people. It leads to a high percentage of people infected in total and a higher peak of people infected actively.}
    \label{figS:asymptotic_prob}
\end{figure}
\newpage

\begin{figure}[H]
    \centering
    \begin{tabular}{|m{0.1cm}>{\centering\arraybackslash}m{\threefig}|>{\centering\arraybackslash}m{\threefig}|}
    \hline
     & Protecting $x\%$ most active people & Closing $x\%$ most popular venues \\
     \hline
     &
     \includegraphics[width=\threefig]{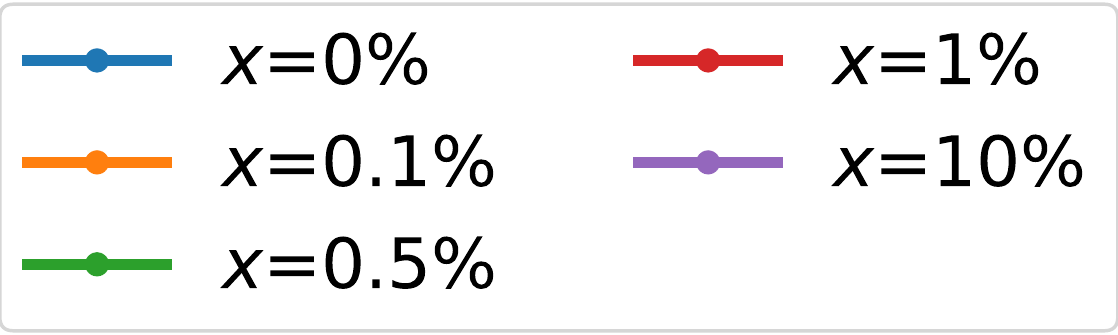} & 
     \includegraphics[width=\threefig]{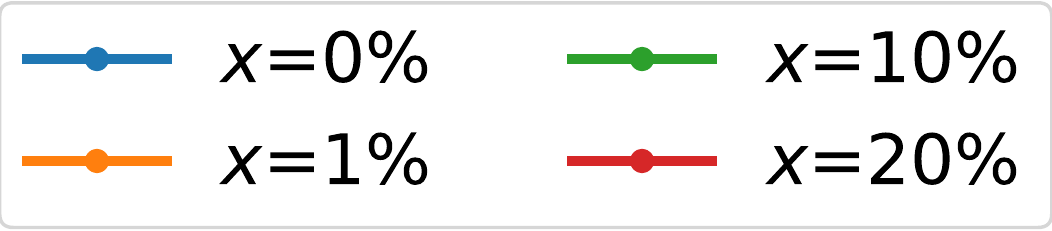}
     \\
     
    \rotatebox{90}{New York} &
    \includegraphics[width=\threefig]{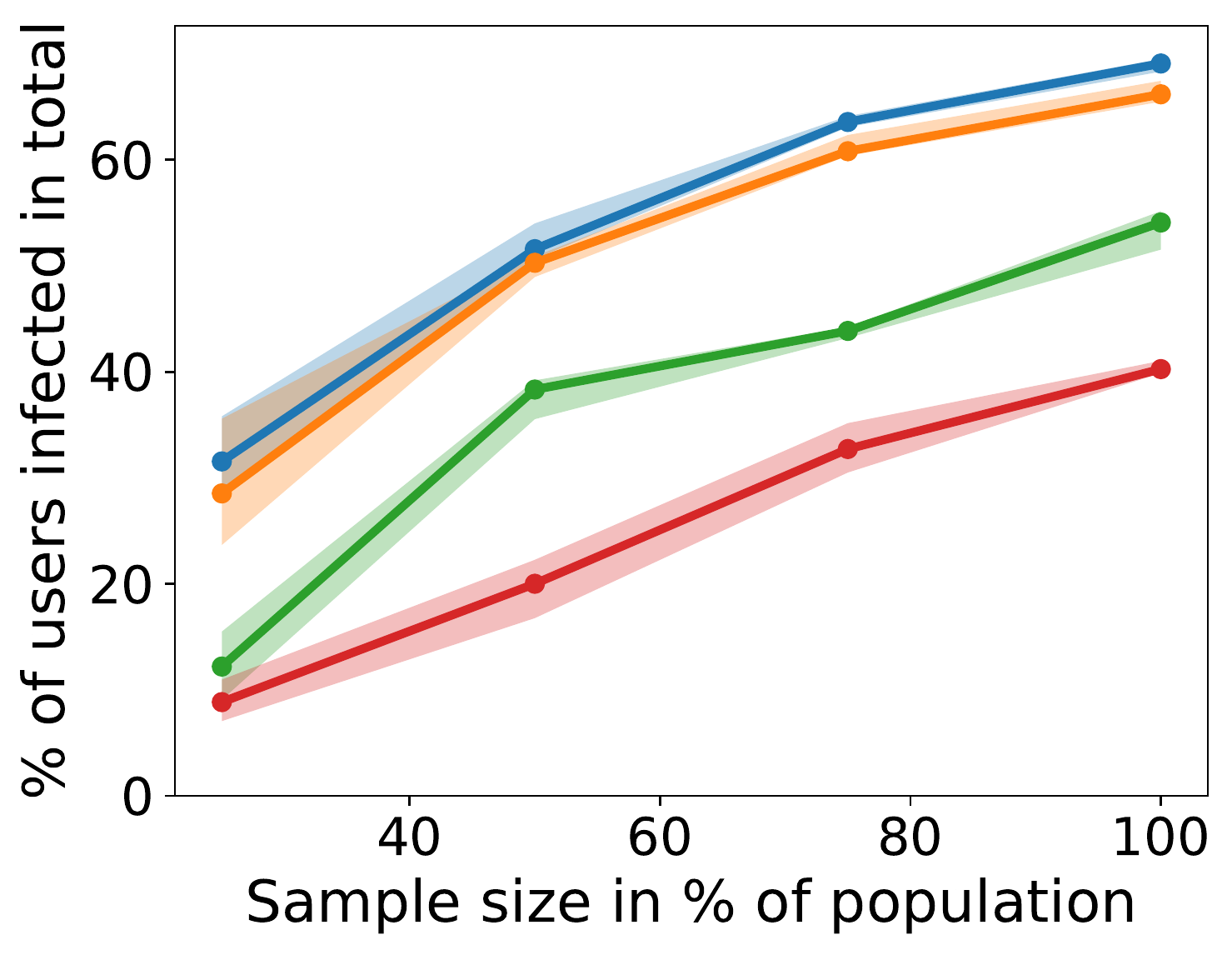} &
    \includegraphics[width=\threefig]{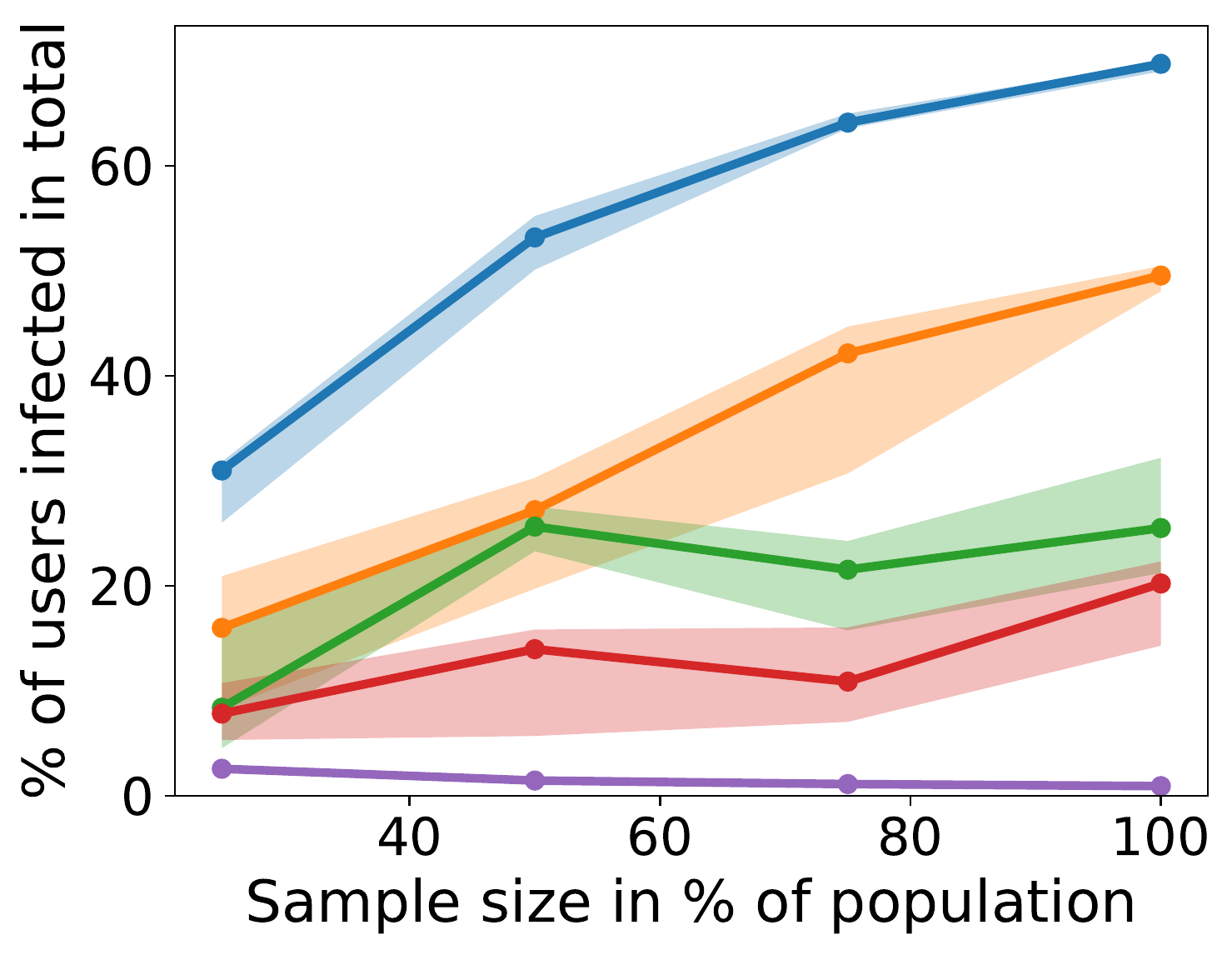} \\
    
    \rotatebox{90}{Istanbul} &
    \includegraphics[width=\threefig]{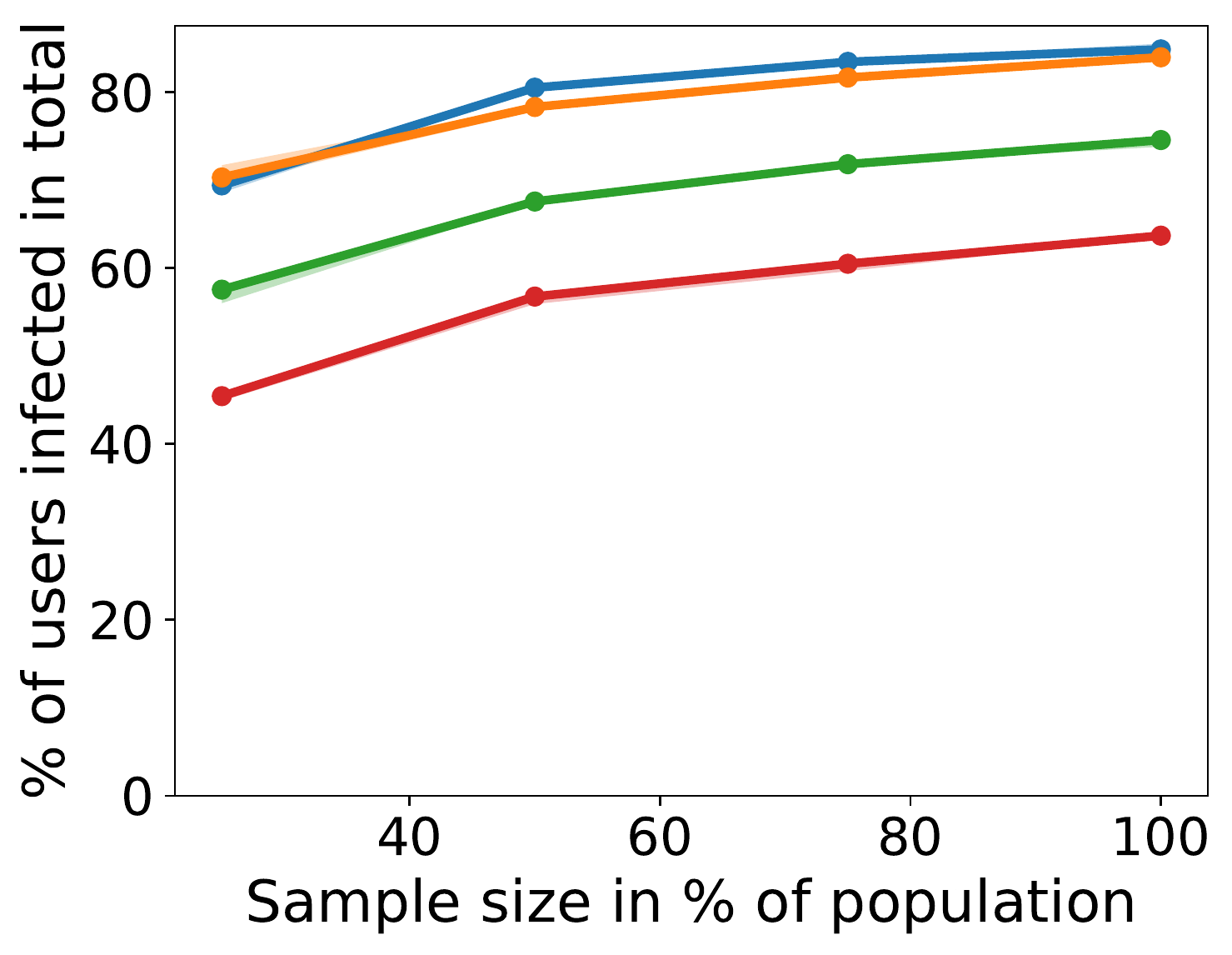} &
    \includegraphics[width=\threefig]{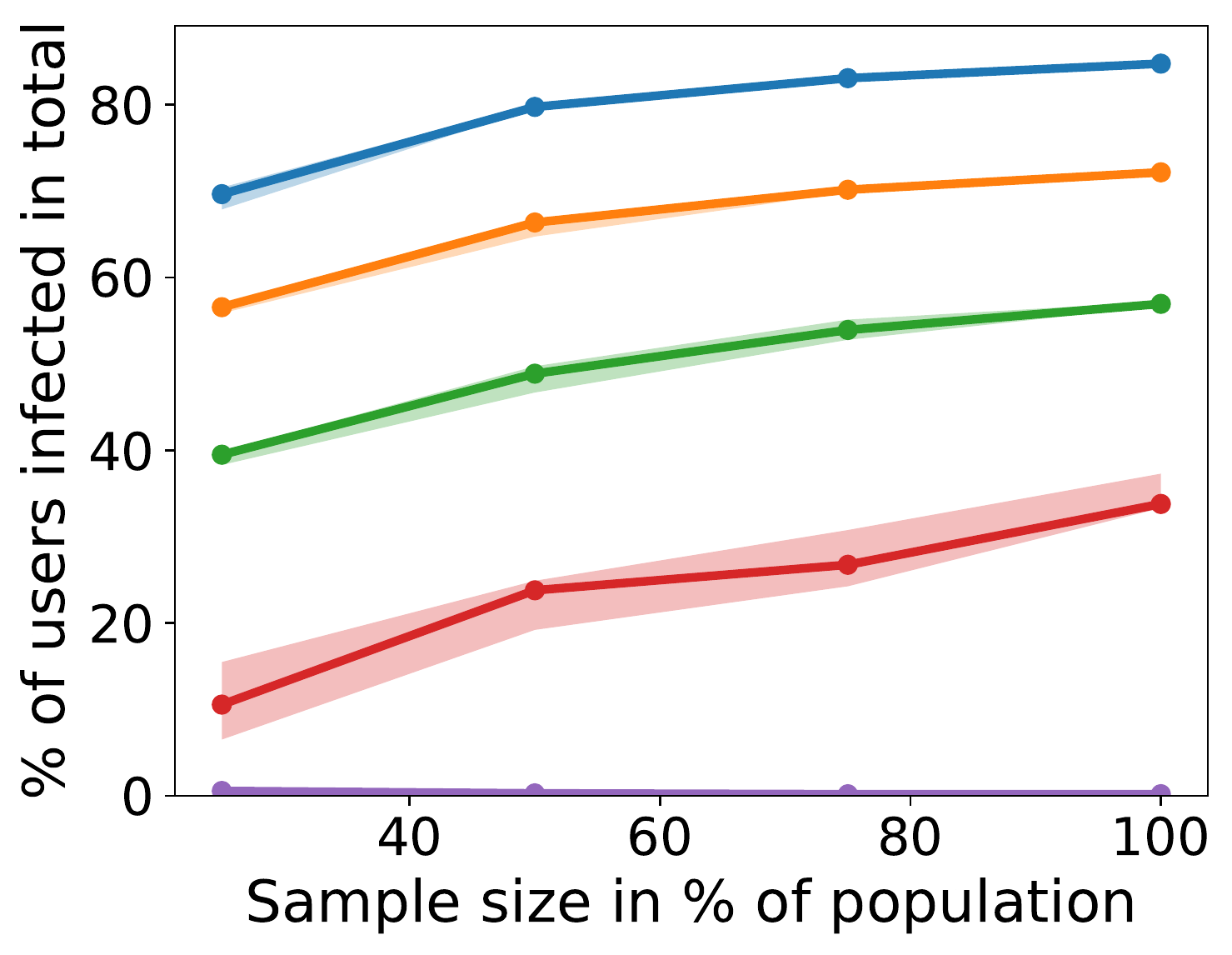} \\
    \hline
     & 
     \includegraphics[width=\threefig]{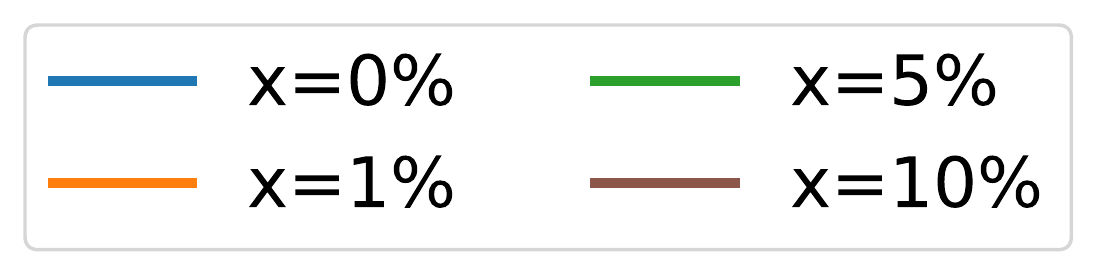} &
     \includegraphics[width=\threefig]{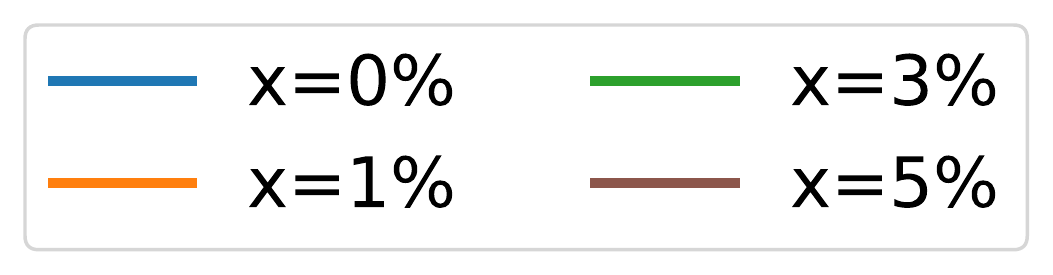} \\
    
    \rotatebox{90}{University} &
    \includegraphics[width=\threefig]{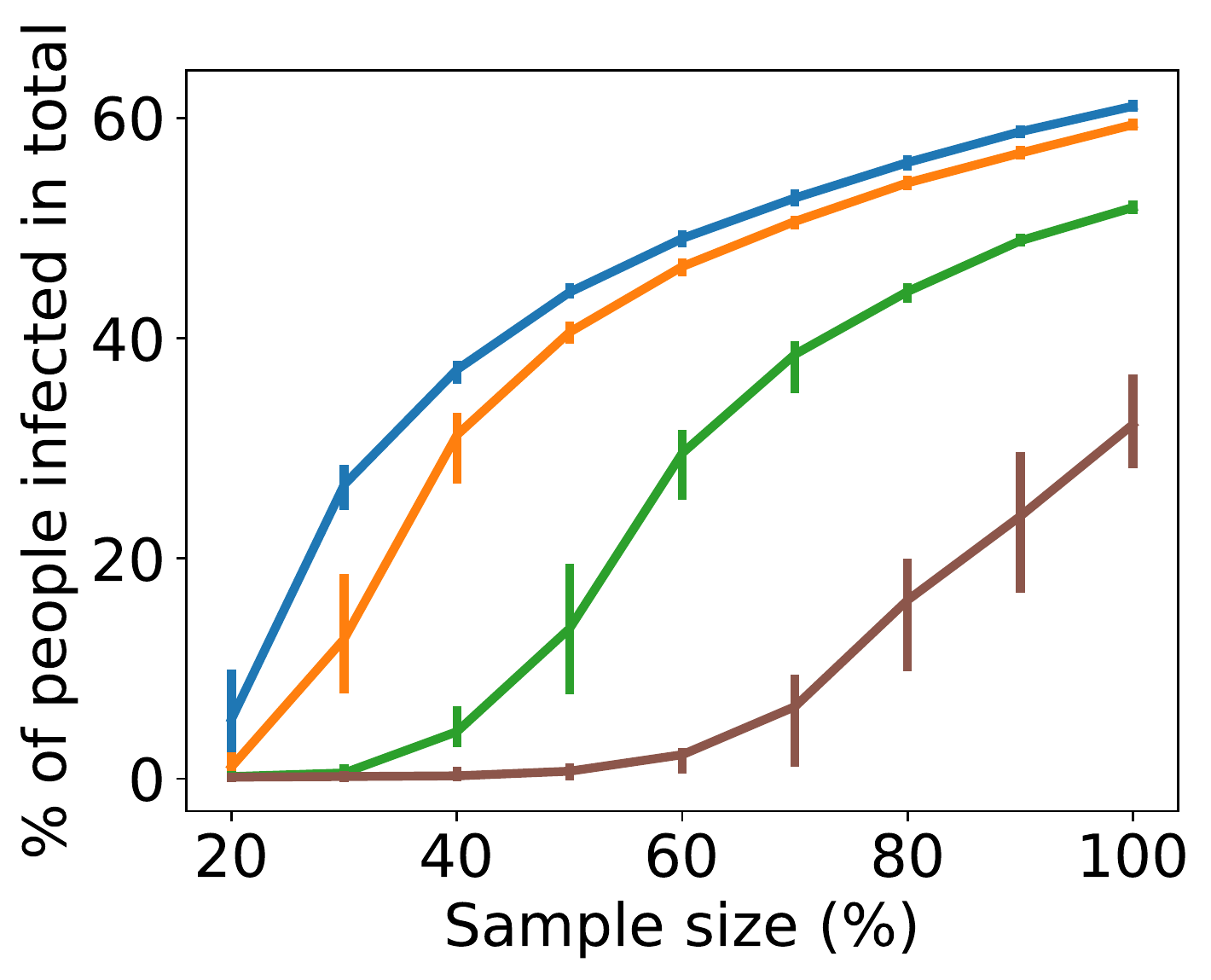}&
    \includegraphics[width=\threefig]{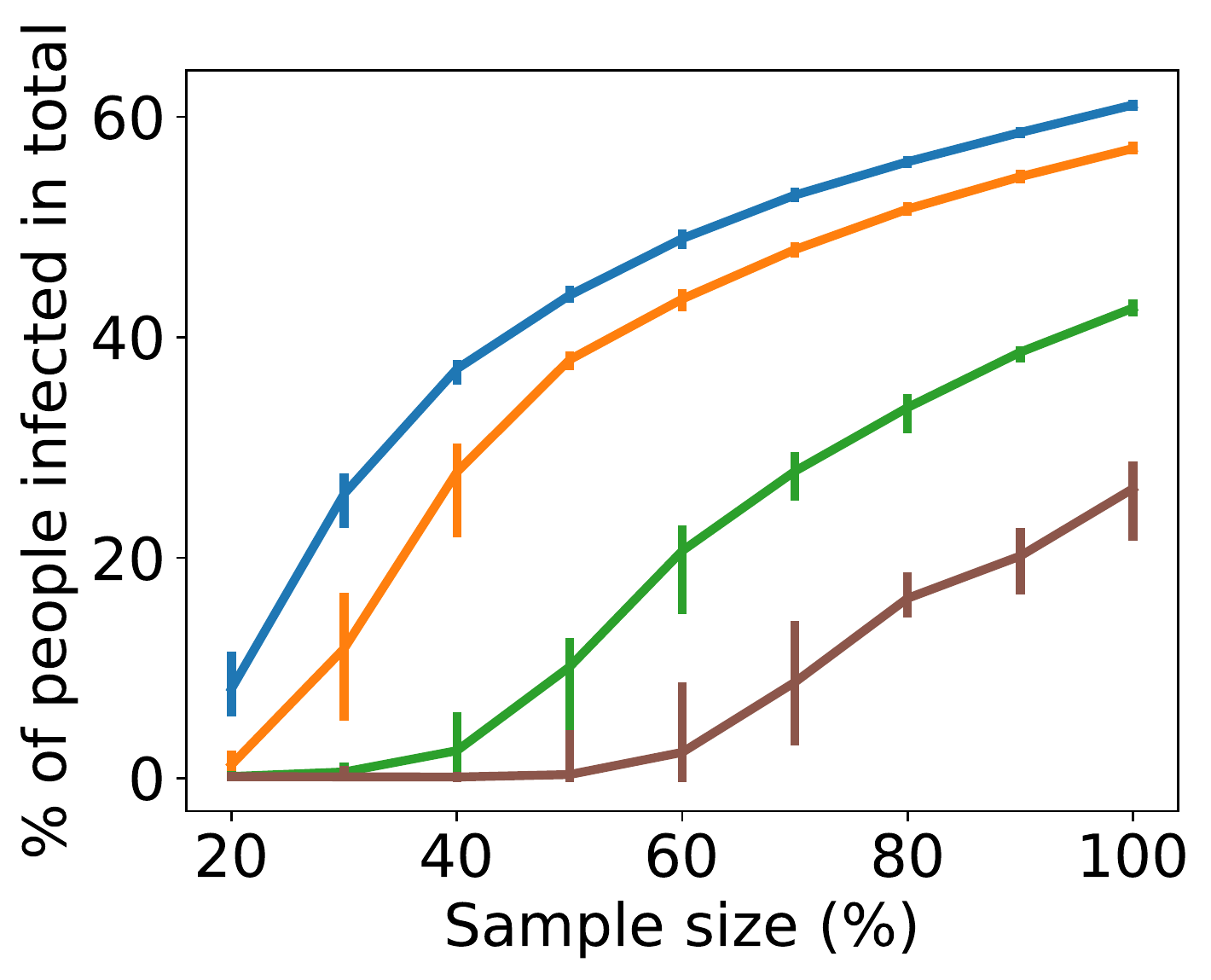} \\
    \hline
    \end{tabular}
    \caption{Infection spreading with the intervention strategy for different sizes of the dataset. The sample size is in the percentage of agents sampled. Using the intervention strategies to protect the most active people and close the most popular venues, the percent of infected agents decreases with sample size.}
    \label{figS:scale_size}
\end{figure}

\begin{figure}[H]
    \centering
    \begin{tabular}{cc}
      \includegraphics[width=0.45\textwidth]{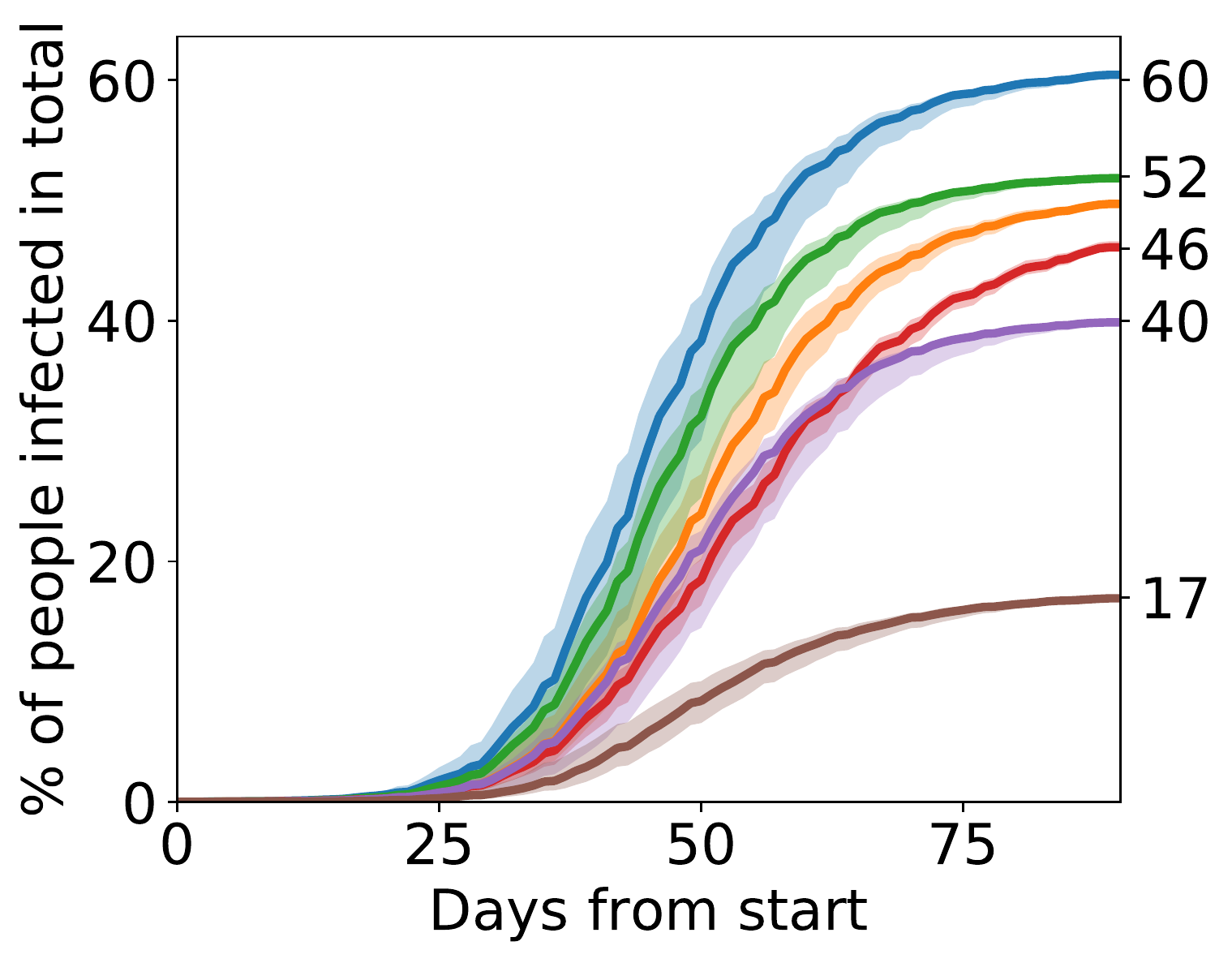}  & 
      \includegraphics[width=0.42\textwidth]{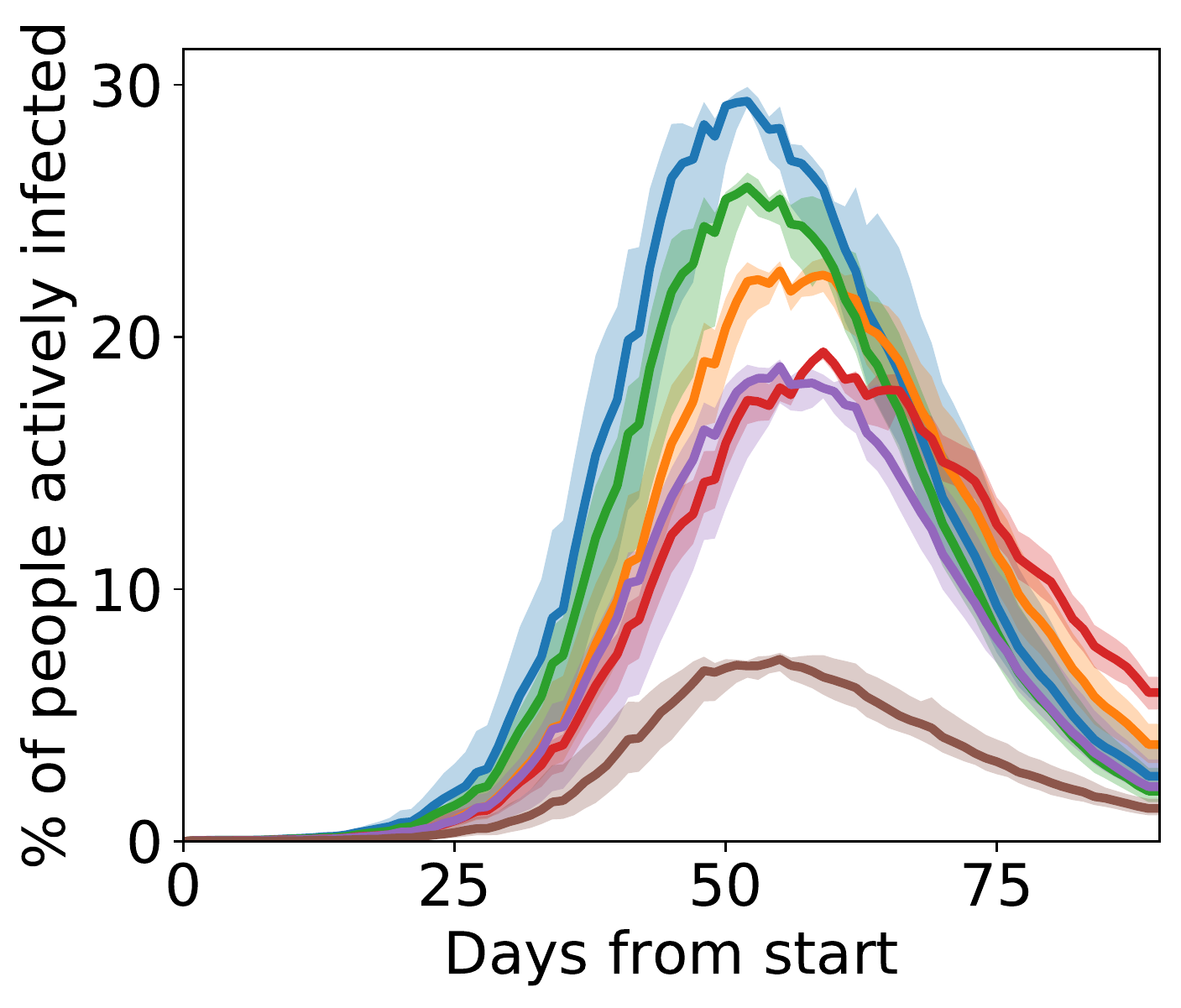}  \\
      \includegraphics[width=0.41\textwidth]{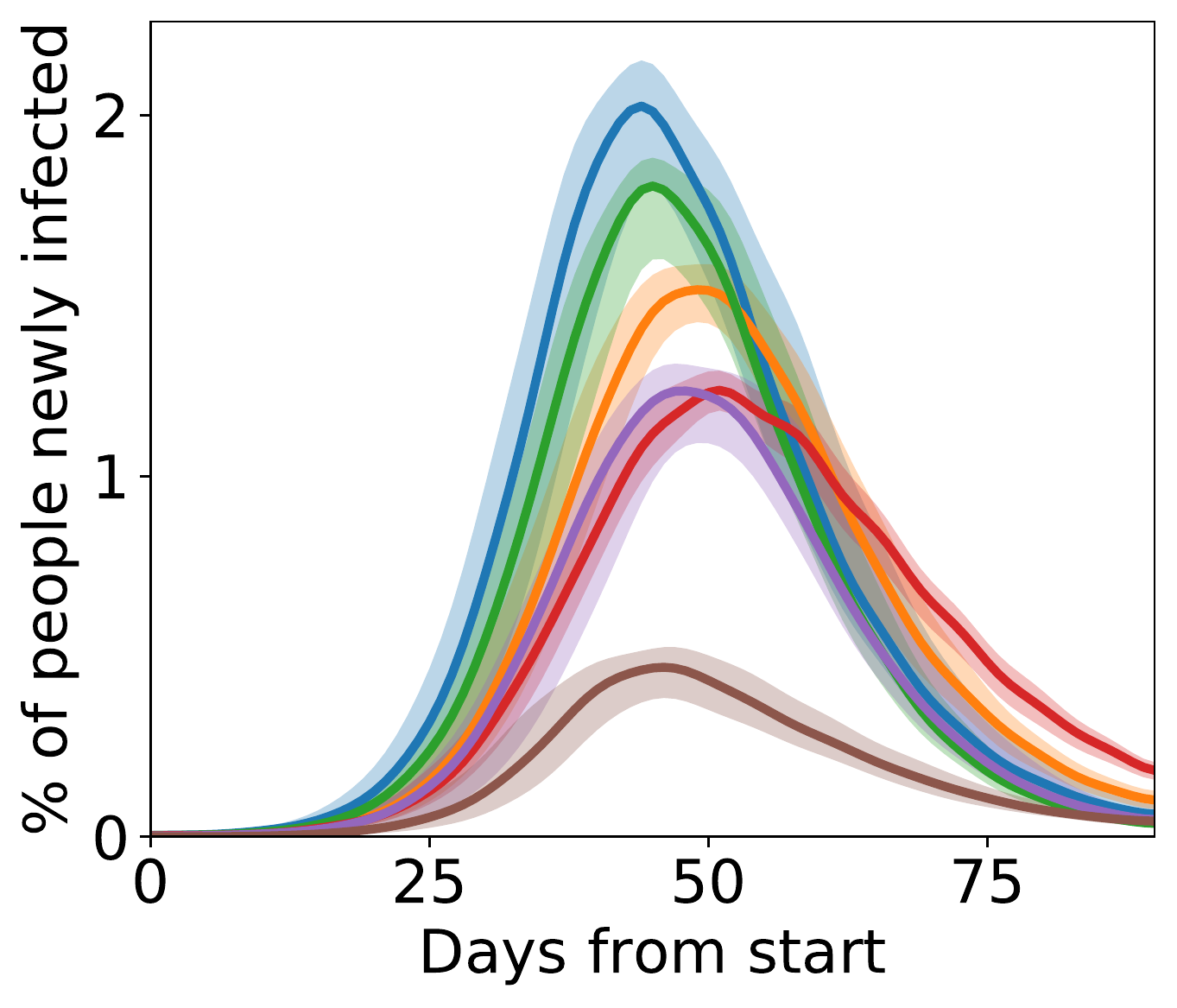}  & 
      \includegraphics[width=0.43\textwidth]{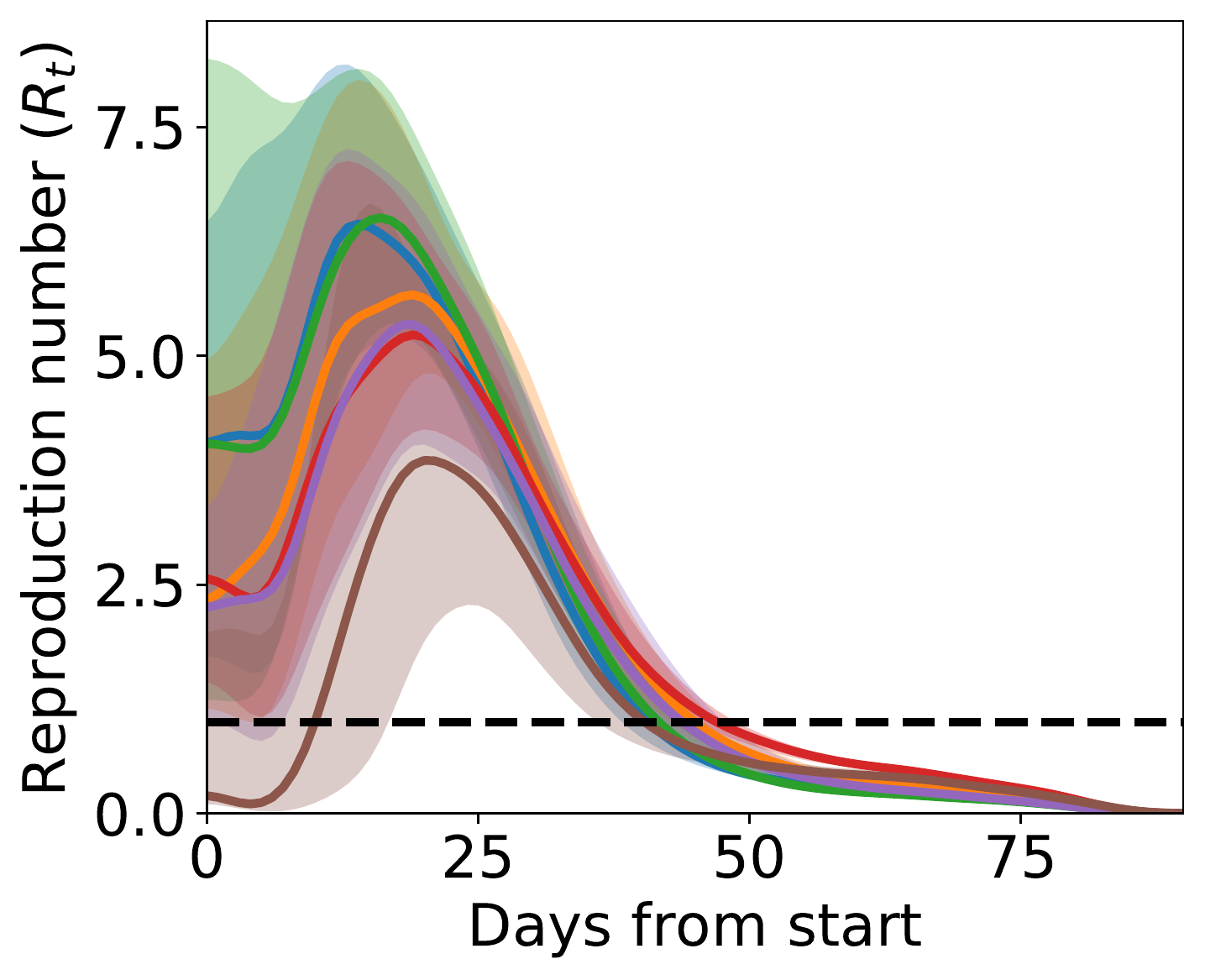}  \\
      \multicolumn{2}{c}{\includegraphics[width=0.9\textwidth]{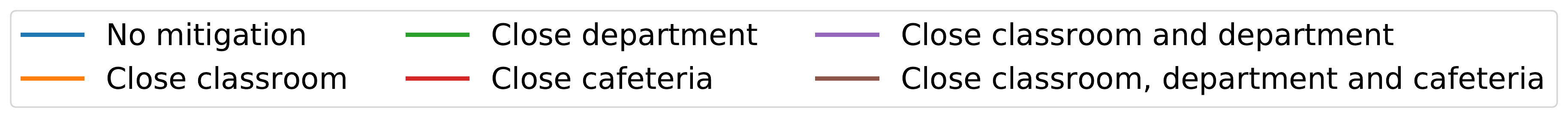}}
    \end{tabular}
    \caption{Closing some types of venues and constraining some activities are good intervention strategies in universities. In the university environment, there are many gathering events, especially taking courses and eating. In reality, many universities allow students to stay in the campus, but move the courses online and require students to take the meal back to dorms. So we test the spread in the university if some types of venues are closed. It shows that closing one type of venue is not enough to control the spreading. Only when we close all the classrooms and cafeteria, the total infection can be controlled under $20\%$, because students can also contact each other in other venues, due to the limited number of venues. 
    }
    \label{figS:close_type_venue}
\end{figure}


\begin{thebibliography}{10}

\bibitem{Flaxman2020-km}
S.~Flaxman, S.~Mishra, A.~Gandy, H.~J.~T.~Unwin, T.~A.~Mellan, H.~Coupland, C.~Whittaker, H.~Zhu, T.~Berah, J.~W.~Eaton, M.~Monod, Imperial College COVID-19 Response Team, A.~C.~Ghani, C.~A.~Donnelly, S.~Riley, M.~A.~C.~Vollmer, N.~M.~Ferguson, L.~C.~Okell, S.~Bhatt, Estimating the effects of non-pharmaceutical interventions on COVID-19 in Europe, {\it Nature\/}, 1-5 (2020). \href{https://doi.org/10.1038/s41586-020-2405-7}{doi:10.1038/s41586-020-2405-7}

\bibitem{Hsiang2020-ab}
S.~Hsiang, D.~Allen, S.~Annan-Phan, K.~Bell, I.~Bolliger, T.~Chong, H.~Druckenmiller L.~Y.~Huang A.~Hultgren, E.~Krasovich, P.~Lau, J.~Lee, E.~Rolf J.~Tseng, T.~Wu, Tiffany, The effect of large-scale anti-contagion policies on the {COVID-19} pandemic {\it Nature\/} (2020). \href{https://doi.org/10.1038/s41586-020-2404-8}{doi:10.1038/s41586-020-2404-8}

\bibitem{Chinazzi2020-pw}
M.~Chinazzi, J.~T.~Davis, M.~Ajelli, C.~Gioannini, M.~Litvinova, S.~Merler, A.~P.~Y.~Piontti, K.~Mu, L.~Rossi, K.~Sun, C.~Viboud, X.~Xiong, H.~Yu, M.~E.~Halloran, Jr I.~M.~Longini, A.~Vespignanio, The effect of travel restrictions on the spread of the 2019 novel coronavirus ({COVID-19}) outbreak, {\it Science\/} {\bf 368}, 395-400 (2020). \href{https://doi.org/10.1126/science.aba9757}{doi:10.1126/science.aba9757}

\bibitem{Kraemer2020-tq}
M.~U.~G.~Kraemer, C.~Yang, B.~Gutierrez, C.~Wu, B.~Klein, D.~M.~Pigott, Open COVID-19 Data Working Group†, L.~Plessis, N.~R. Faria, R.~Li, W.~P.~Hanage, J.~S.~Brownstein, M.~Layan, A.~Vespignani, H.~Tian, C.~Dye, O.~G.~Pybus, S.~V.~Scarpino, The effect of human mobility and control measures on the COVID-19 epidemic in China, {\it Science\/} {\bf 368}, 493-497 (2020). \href{https://doi.org/10.1126/science.abb4218}{doi:10.1126/science.abb4218}

\bibitem{Li2020-ej}
R.~Li, S.~Pei, B.~Chen, Y.~Song, T.~Zhang, W.~Yang, J.~Shaman, Substantial undocumented infection facilitates the rapid dissemination of novel coronavirus ({SARS-CoV-2}), {\it Science\/} {\bf 368}, 489-493 (2020). \href{https://doi.org/10.1126/science.abb3221}{doi:10.1126/science.abb3221}

\bibitem{badr2020association}
H.~S.~Badr, H.~Du, M.~Marshall, E.~Dong, M.~M.~Squire, L.~M.~Gardner, Association between mobility patterns and COVID-19 transmission in the USA: a mathematical modelling study, {\it The Lancet Infectious Diseases} (2020). \href{https://doi.org/10.1016/S1473-3099(20)30553-3}{doi:10.1016/S1473-3099(20)30553-3}

\bibitem{Metcalf368}
C.~J.~E.~Metcalf, D.~H.~Morris, S.~W.~Park, Mathematical models to guide pandemic response, {\it Science} {\bf 369}, 368--369 (2020). \href{https://science.sciencemag.org/content/369/6502/368.summary?casa_token=g0QrMYq4-0kAAAAA:2xKqEiAm3NSXBVOmOojZ6WuYUGTKGCthTbh9FrL2AedGyWmJS2qyGXr6aenNe0E2XBW-JPg4bfbO}{doi:10.1126/science.abd1668}

\bibitem{wang2020phase}
H.~Wang, Z.~Wang, Y.~Dong, R.~Chang, C.~Xu, X.~Yu, S.~Zhang, L.~Tsamlag, M.~Shang, J.~Huang, Y.~Wang, G.~Xu, T.~Shen, X.~Zhang, Y.~Cai, Phase-adjusted estimation of the number of coronavirus disease 2019 cases in Wuhan, China, {\it Cell discovery\/} {\bf 6}, 1-8 (2020). \href{https://doi.org/10.1038/s41421-020-0148-0}{doi:10.1038/s41421-020-0148-0}

\bibitem{prem2020effect}
K.~Prem, Y.~Liu, T.~W.~Russell, A.~J.~Kucharski, R.~M.~Eggo, N.~Davies, Centre for the Mathematical Modelling of Infectious Diseases COVID-19 Working Group, M.~Jit, P.~Klepac, The effect of control strategies to reduce social mixing on outcomes of the COVID-19 epidemic in Wuhan, China: a modelling study, {\it The Lancet Public Health\/}, (2020). \href{https://doi.org/10.1016/S2468-2667(20)30073-6}{doi:10.1016/S2468-2667(20)30073-6}

\bibitem{lai2020effect}
S.~Lai, N.~W.~Ruktanonchai, L.~Zhou, O.~Prosper, W.~Luo, J.~R.~Floyd, A.~Wesolowski, M.~Santillana, C.~Zhang, X.~Du, H.~Yu, A.~J.~Tatem, Effect of non-pharmaceutical interventions to contain COVID-19 in China, {\it Nature\/}, (2020). \href{https://doi.org/10.1038/s41586-020-2293-x}{doi:10.1038/s41586-020-2293-x}

\bibitem{covid19health}
I.~COVID, C.~J.~Murray, Forecasting COVID-19 impact on hospital bed-days, ICU-days, ventilator-days and deaths by US state in the next, medRxiv, March 20, 2020; \href{https://doi.org/10.1101/2020.03.27.20043752}{doi:10.1101/2020.03.27.20043752}.

\bibitem{Ferguson2020-kz}
N.~M.~Ferguson, D.~Laydon, G.~Nedjati-Gilani, N.~Imai, K.~Ainslie, M.~Baguelin, S.~Bhatia, A.~Boonyasiri, Z.~Cucunub{\'a}, G.~Cuomo-Dannenburg, others, Impact of non-pharmaceutical interventions ({NPIs}) to reduce {COVID-19} mortality and healthcare demand, {\it DOI} {\bf 10}, 77482 (2020). \href{https://doi.org/10.25561/77482}{doi:10.25561/77482}

\bibitem{yang2019revisiting} 
D.~Yang, B.~Qu, J.~Yang, P.~Cudre-Mauroux, Revisiting user mobility and social relationships in lbsns: a hypergraph embedding approach, {\it The World Wide Web Conference\/}, 4147--2157 (2019). \href{https://doi.org/10.1145/3308558.3313635}{doi:10.1145/3308558.3313635}

\bibitem{yang2020lbsn2vec++}
D.~Yang, B.~Qu, J.~Yang, P.~Cudre-Mauroux, LBSN2Vec++: Heterogeneous Hypergraph Embedding for Location-Based Social Networks, {\it IEEE Transactions on Knowledge and Data Engineering\/}, (2020). \href{https://doi.org/10.1109/TKDE.2020.2997869}{doi:10.1109/TKDE.2020.2997869}

\bibitem{sui2016characterizing}
K.~Sui, M.~Zhou, D.~Liu, M.~Ma, D.~Pei, Y.~Zhao, Z.~Li, T.~Moscibroda, Characterizing and improving wifi latency in large-scale operational networks, {\it Proceedings of the 14th Annual International Conference on Mobile Systems, Applications, and Services\/}, 347-360 (2016). \href{https://doi.org/10.1145/2906388.2906393}{doi:10.1145/2906388.2906393}

\bibitem{wang2020distributed}
H.~Wang, J.~Gao, Distributed Human Trajectory Sensing and Partial Similarity Queries, {\it 2020 19th ACM/IEEE International Conference on Information Processing in Sensor Networks (IPSN)\/}, 253--264 (2020). \href{https://doi.org/10.1109/IPSN48710.2020.00-43}{doi:10.1109/IPSN48710.2020.00-43}

\bibitem{wu2020characteristics}
Z.~Wu, J.~M. McGoogan, Characteristics of and important lessons from the coronavirus disease 2019 (COVID-19) outbreak in China: summary of a report of 72 314 cases from the Chinese Center for Disease Control and Prevention, {\it Jama\/} {\bf 323}, 1239--1242 (2020). \href{https://doi.org/10.1001/jama.2020.2648}{doi:10.1001/jama.2020.2648}

\bibitem{CDCparameter}
Centers for Disease Control and Prevention, ``COVID-19 Pandemic Planning Scenarios'', (2020); \href{https://www.cdc.gov/coronavirus/2019-ncov/hcp/planning-scenarios.html}{https://www.cdc.gov/coronavirus/2019-ncov/hcp/planning-scenarios.html}.


\bibitem{nishiura2020estimation}
H.~Nishiura, T.~Kobayashi, T.~Miyama, A.~Suzuki, S.~Jung, K.~Hayashi, R.~Kinoshita, Y.~Yang, B.~Yuan, A.~Akhmetzhanov, N.~M.~Linton, Estimation of the asymptomatic ratio of novel coronavirus infections (COVID-19), {\it International journal of infectious diseases\/} {\bf 94}, 154 (2020). \href{https://doi.org/10.1016/j.ijid.2020.03.020}{doi:10.1016/j.ijid.2020.03.020}

\bibitem{mizumoto2020estimating}
K.~Mizumoto, K.~Kagaya, A.~Zarebski, G.~Chowell, Estimating the asymptomatic proportion of coronavirus disease 2019 (COVID-19) cases on board the Diamond Princess cruise ship, Yokohama, Japan, 2020, {\it Eurosurveillance\/} {\bf 25}, 2000180 (2020). \href{https://doi.org/10.2807/1560-7917.ES.2020.25.10.2000180}{doi:10.2807/1560-7917.ES.2020.25.10.2000180}

\bibitem{gostic2020practical}
K.~M.~Gostic, L.~McGough, E.~Baskerville, S.~Abbott, K.~Joshi, C.~Tedijanto, R.~Kahn, R.~Niehus, J.~Hay, P.~de Salazar, J.~Hellewell, S.~Meakin, J.~Munday, N.~I.~Bosse, K.~Sherrat, R.~N.~Thompson, L.~F.~White, J.~S.~Huisman, J.~Scire, S.~Bonhoeffer, T.~Stadler, J.~Wallinga, S.~Funk, M.~Lipsitch, S.~Cobey, Practical considerations for measuring the effective reproductive number, Rt, medRxiv, June 23, 2020; \href{https://doi.org/10.1101/2020.06.18.20134858}{doi:10.1101/2020.06.18.20134858}


\bibitem{aktay2020google}
A.~Aktay, S.~Bavadekar, G.~Cossoul, J.~Davis, D.~Desfontaines, A.~Fabrikant, E.~Gabrilovich, K.~Gadepalli, B.~Gipson, M.~Guevara, C.~Kamath, M.~Kansal, A.~Lange, C.~Mandayam, A.~Oplinger, C.~Pluntke, T.~Roessler, A.~Schlosberg, T.~Shekel, S.~Vispute, M.~Vu, G.~Wellenius, B.~Williams, R.~J.~Wilson, Google COVID-19 Community Mobility Reports: Anonymization Process Description (version 1.0), arXiv:2004.04145, [Preprint]. April 9, 2020; \href{https://arxiv.org/abs/2004.04145v2}{arXiv:2004.04145}.



\bibitem{NJ2020Restart}
New Jersey Department of Education, "The Road Back: Restart and Recovery Plan for Education", (2020). \href{https://www.nj.gov/education/reopening}{https://www.nj.gov/education/reopening/}.

\bibitem{UNMC2020Covid}
J.~Lowe, A.~S.~Khan, J.~Herstein, J.~Lawler, B.~Grimm, B.~Rauner, D.~Brett-Major, M.~Schwedhelm, "COVID-19 Back to School PlayBook: Guiding Principles to Keep Students, Teachers, and Staff Safe in K-12 Schools", 2020; \href{https://www.unmc.edu/publichealth/news/UNMC_COPH_K-12_Playbookv1.pdf}{https://www.unmc.edu/publichealth/news/UNMC\_COPH\_K-12\_Playbookv1.pdf}.

\bibitem{Haushofer1063}
J.~Haushofer, C.~J.~E.~Metcalf, Which interventions work best in a pandemic?, {\it Science} {\bf 368}, 1063--1065 (2020). \href{https://doi.org/10.1126/science.abb6144}{doi:10.1126/science.abb6144}

\bibitem{Holtz202009522}
D.~Holtz, M.~Zhao, S.~G.~Benzell, C.~Cao, M.~A.~Rahimian, J.~Yang, J.~Allen, A.~Collis, A.~Moehring, T.~Sowrirajan, D.~Ghosh, Y.~Zhang, P.~Dhillon, C.~Nicolaides, D.~Eckles, S.~Aral, Interdependence and the cost of uncoordinated responses to COVID-19, OSF Preprints, May 27 (2020). \href{https://osf.io/b9psy/}{doi:10.31219/osf.io/b9psy}

\bibitem{abouk2020immediate}
R.~Abouk, B.~Heydari, The immediate effect of covid-19 policies on social distancing behavior in the united states, SSRN, April 15, 2020. \href{http://dx.doi.org/10.2139/ssrn.3571421}{doi:10.2139/ssrn.3571421}

\bibitem{gao2020mapping}
S.~Gao, J.~Rao, Y.~Kang, Y.~Liang, J.~Kruse, Mapping county-level mobility pattern changes in the United States in response to COVID-19, {\it SIGSPATIAL Special} {\bf 12}, 16--26 (2020). \href{https://arxiv.org/abs/2004.04544}{arXiv:2004.04544}

\bibitem{adams2020supporting}
J.~G.~Adams, R~M.~Walls, Supporting the Health Care Workforce During the COVID-19 Global Epidemic, {\it Jama} {\bf 323(15)}, 1439--1440 (2020). \href{https://doi.org/10.1001/jama.2020.3972}{doi:10.1001/jama.2020.3972}

\bibitem{chang2020protecting}
D.~Chang, H.~Xu, A.~Rebaza, L.~Sharma, C.~S~D.~Cruz, Protecting health-care workers from subclinical coronavirus infection, {\it The Lancet Respiratory Medicine} {\bf 8}, e13 (2020). \href{https://doi.org/10.1016/S2213-2600(20)30066-7}{doi:10.1016/S2213-2600(20)30066-7}

\bibitem{lancet2020covid}
Lancet, COVID-19: protecting health-care workers, {\it Lancet (London, England)} {\bf 395(10228)}, 922 (2020). \href{https://doi.org/10.1016/S0140-6736(20)30644-9}{doi:10.1016/S0140-6736(20)30644-9}

\bibitem{wang2020reasons}
J.~Wang, M.~Zhou, F.~Liu, Reasons for healthcare workers becoming infected with novel coronavirus disease 2019 (COVID-19) in China, {\it Journal of Hospital Infection}, (2020). \href{https://doi.org/10.1016/j.jhin.2020.03.002}{doi:10.1016/j.jhin.2020.03.002}

\bibitem{laborde2020covid}
D.~Laborde, W.~Martin, J.~Swinnen, R.~Vos, COVID-19 risks to global food security, {\it Science} {\bf 369}, 500--502 (2020). \href{https://doi.org/10.1126/science.abc4765}{doi:10.1126/science.abc4765}

\bibitem{stehle2011simulation}
J.~Stehlé, N.~Voirin, A.~Barrat, C.~Cattuto, V.~Colizza, L.~Isella, C.~Régis, J.~Pinton, N.~Khanafer, W.~V.~Broeck, P.~Vanhems, Simulation of an SEIR infectious disease model on the dynamic contact network of conference attendees, {\it BMC Medicine} {\bf 9}, 87 (2020). \href{https://doi.org/10.1186/1741-7015-9-87}{doi:10.1186/1741-7015-9-87}

\bibitem{githubcode}
Simulation Code; \href{https://github.com/SBUhaotian/Mobility_Contagion}{https://github.com/SBUhaotian/Mobility\_Contagion}

\end{thebibliography}
\end{document}